\documentclass[11pt]{cernrep}
\usepackage{psfig}
\usepackage{epsfig}
\usepackage{axodraw}
\usepackage{graphicx}
\usepackage{here}
\usepackage{a4}
\usepackage{graphics}
\usepackage{subfigure}
\usepackage{dcolumn}
\usepackage{amsmath}
\usepackage{psfrag}
\usepackage{cite}
\usepackage{rotating}
\usepackage{amssymb}
\usepackage{amsbsy}
\usepackage{latexsym,epsf}

\renewcommand{\textfraction}{-0.5}
\newcommand{\newc}{\newcommand}
\newc{\beq}{\begin{equation}}
\newc{\sgn}{\rm{sgn}\,}
\newc{\eeq}{\end{equation}}
\newc{\barr}{\begin{eqnarray}}
\newc{\earr}{\end{eqnarray}}
\newc{\ra}{\rightarrow}
\newc{\lam}{\lambda}
\newc{\eps}{\epsilon}
\newc{\half}{\frac{1}{2}}
\newc{\gev}{\mbox{~GeV}}
\newc{\etal}{{\it et al.}\ }
\newc{\nonum}{\nonumber}
\newc{\kap}{\kappa}
\newc{\rpv}{$/\!\!\!\!R_p $}
\newc{\nonr}{\nonumber}
\newc{\eq}[1]{(\ref{eq:#1})}
\newc{\eqs}[2]{(\ref{eq:#1},\ref{eq:#2})}
\newc{\lab}[1]{\label{eq:#1}}
\newc{\cw}{\cos\theta_w}
\newc{\ssw}{\sin^2\theta_w}
\newc{\upt}{\tilde{u}}
\newc{\elt}{\tilde{\ell}}
\newc{\mut}{\tilde{\mu}}
\newc{\nut}{\tilde{\nu}}
\newc{\dnt}{\tilde{d}}
\newc{\unt}{\tilde{u}}
\newc{\cht}{\tilde{\chi}}
\newc{\psb}{\bar{\psi}}
\newc{\ie}{{\it i.e.}} \newc{\eg}{{\it e.g.}}
\def\bit{\begin{itemize}}
\def\eit{\end{itemize}}
\def\etc{{\it etc.}}

\begin{document}

\setcounter{tocdepth}{0}

\bibliographystyle{revtex}
\pagestyle{myheadings}
\thispagestyle{empty}

\rightline{UCD-2002-05, SLAC-PUB-9183, UFIFT-HEP-02-11, BNL-HET-02/10}

\begin{center}
\vspace*{.5cm}

{\Large\sc \bf THE BEYOND THE STANDARD MODEL WORKING GROUP: }

\vspace*{0.3cm}

{\Large\sc \bf Summary Report}

\vspace*{.7cm}

Conveners: \\[0.2cm]
{\sc 
G.~Azuelos$^1$, 
J.~Gunion$^2$, 
J.~Hewett$^3$, 
G.~Landsberg$^4$,
K.~Matchev$^5$, 
F.~Paige$^6$, 
T.~Rizzo$^3$, 
L.~Rurua$^7$}
\vspace*{0.5cm}

Additional Contributors: \\[0.2cm]
{\sc 
S.~Abdullin$^8$, 
A.~Albert$^9$, 
B.~Allanach$^{10}$, 
T.~Blazek$^{11}$,
D.~Cavalli$^{12}$,
F.~Charles$^{9}$, 
K.~Cheung$^{13}$, 
A.~Dedes$^{14}$,
S.~Dimopoulos$^{15}$, 
H.~Dreiner$^{14}$, 
U.~Ellwanger$^{16}$,
D.S.~Gorbunov$^{17}$,
S.~Heinemeyer$^6$,
I.~Hinchliffe$^{18}$,
C.~Hugonie$^{19}$, 
S.~Moretti$^{10,19}$, 
G.~Polesello$^{20}$, 
H.~Przysiezniak$^{21}$,
P.~Richardson$^{22}$,
L.~Vacavant$^{18}$, 
G.~Weiglein$^{19}$ }
\vspace*{.5cm}

 Additional Working Group Members: \\[0.2cm]
{\sc 
S.~Asai$^{7}$, 
C.~Balazs$^{23}$, 
M.~Battaglia$^{7}$,
G.~Belanger$^{21}$, 
E.~Boos$^{24}$, 
F.~Boudjema$^{21}$, 
H.-C.~Cheng$^{25}$,
A.~Datta$^{26}$, 
A.~Djouadi$^{26}$, 
F.~Donato$^{21}$,
R.~Godbole$^{27}$, 
V.~Kabachenko$^{28}$, 
M.~Kazama$^{29}$, 
Y.~Mambrini$^{26}$,
A.~Miagkov$^{7}$, 
S.~Mrenna$^{30}$,
P.~Pandita$^{31}$, 
P.~Perrodo$^{21}$,
L.~Poggioli$^{21}$, 
C.~Quigg$^{30}$, 
M.~Spira$^{32}$, 
A.~Strumia$^{10}$,
D.~Tovey$^{33}$, 
B.~Webber$^{34}$ 
 }
\vspace*{.7cm}

Affiliations:
\vspace*{.5cm}

{\small
$^1$    Department of Physics, University of Montreal and TRIUMF, Canada. \\
$^2$    Department of Physics, University of California at Davis, Davis, CA, USA.\\
$^3$    Stanford Linear Accelerator Center, Stanford University, Stanford, CA, USA. \\
$^4$    Department of Physics, Brown University, Providence, RI, USA.\\
$^5$    Department of Physics, University of Florida, Gainesville, FL, USA.\\
$^6$    Brookhaven National Laboratory, Upton, NY, USA.\\
$^7$    EP Division, CERN, CH--1211 Geneva 23, Switzerland. \\
$^8$    I.T.E.P., Moscow, Russia. \\
$^9$    Groupe de Recherches en Physique des Hautes Energies, 
        Universit\'e de Haute Alsace, Mulhouse, France. \\
$^{10}$ TH Division, CERN, CH--1211 Geneva 23, Switzerland. \\
$^{11}$ Department of Physics and Astronomy, University of Southampton, Southampton, UK. \\ 
$^{12}$ INFN, Milano, Italy. \\
$^{13}$ National Center for Theoretical Science, National Tsing Hua University, Hsinchu, Taiwan. \\
$^{14}$ Physikalisches Institut der Universit\"at Bonn, Bonn, Germany. \\
$^{15}$ Physics Department, Stanford University, Stanford, CA, USA. \\
$^{16}$ Universit\'e de Paris XI, Orsay, Cedex, France. \\
$^{17}$ Institute for Nuclear Research of the Russian Academy of Sciences, Moscow, Russia. \\
$^{18}$ Lawrence Berkeley National Laboratory, Berkeley, CA, USA. \\
$^{19}$ Institute for Particle Physics Phenomenology, University of Durham, Durham, UK. \\
$^{20}$ INFN, Sezione di Pavia, Pavia, Italy. \\
$^{21}$ LAPP, Annecy, France. \\
$^{22}$ DAMTP, Centre for Mathematical Sciences and Cavendish Laboratory, Cambridge, UK. \\
$^{23}$ Department of Physics, University of Hawaii, Honolulu, HI, USA. \\
$^{24}$ INP, Moscow State University, Russia. \\
$^{25}$ Department of Physics, University of Chicago, Chicago, IL, USA. \\
$^{26}$ Lab de Physique Mathematique, Univ. de Montpellier II, Montpellier, Cedex, France. \\
$^{27}$ Center for Theoretical Studies, Indian Inst. of Science, Bangalore, Karnataka, India. \\
$^{28}$ IHEP, Moscow, Russia. \\
$^{29}$ Warsaw University, Warsaw, Poland. \\
$^{30}$ Fermilab, Batavia, IL, USA. \\
$^{31}$ Physics Dept., North-Eastern Hill Univ, HEHU Campus, Shillong, India. \\
$^{32}$ Paul Scherrer Institute, Villigen PSI, Switzerland. \\
$^{33}$ Dept. of Physics and Astronomy, Univ. of Sheffield, Sheffield, UK. \\
$^{34}$ Cavendish Laboratory, Cambridge, UK.
}
\vspace*{.5cm}

{\it Report of the ``Beyond the Standard Model'' working group for the Workshop \\[0.1cm]
``Physics at TeV Colliders", Les Houches, France, 21 May -- 1 June 2001.}
\end{center}


\tableofcontents
\newpage

\part{{\bf Preface}}
\label{preface}


\noindent
In this working group we have investigated 
a number of aspects of searches for new physics beyond the
Standard Model (SM) at the running or planned TeV-scale colliders.  
For the most part, we have considered hadron colliders, 
as they will define particle physics at the energy frontier 
for the next ten years at least. 
The variety of models for Beyond the Standard Model (BSM)
physics has grown immensely.  It is clear that only future
experiments can provide the needed direction to clarify
the correct theory.  Thus, our focus has been on exploring
the extent to which hadron colliders can discover
and study BSM physics in various models. We have placed
special emphasis on scenarios in which the new signal might
be difficult to find or of a very unexpected nature.
For example, in the context of supersymmetry (SUSY), we have considered:
\begin{itemize}
\item 
how to make fully precise predictions for the Higgs bosons as 
well as the superparticles of the Minimal Supersymmetric 
Standard Model (MSSM) (parts \ref{svensec} and \ref{allanachsec});
\item
MSSM scenarios in which most or all SUSY particles have
rather large masses (parts \ref{ian1sec} and \ref{ian2sec});
\item
the ability to sort out the many parameters of the MSSM
using a variety of signals and study channels (part \ref{arnauldsec});
\item
whether the no-lose theorem for MSSM Higgs discovery can
be extended to the next-to-minimal Supersymmetric Standard 
Model (NMSSM) in which an additional singlet superfield
is added to the minimal collection of superfields,
potentially providing a natural explanation of the 
electroweak value of the parameter $\mu$ (part \ref{nmssmhiggssec});
\item sorting out the effects of CP violation using Higgs plus squark
associate production (part \ref{dmsec});
\item
the impact of lepton flavor violation of various kinds (part \ref{blazeksec});
\item
experimental possibilities for the gravitino and its sgoldstino partner
(part \ref{gorbsec});
\item
what the implications for SUSY would be if the NuTeV signal
for di-muon events were interpreted as a sign of R-parity
violation (part \ref{ddrsec}).
\end{itemize}
Our other main focus was on the phenomenological implications of extra
dimensions.  There, we considered:
\begin{itemize}
\item constraints on Kaluza Klein (KK) excitations of the SM gauge
bosons from existing data (part \ref{landsberg2sec})
and the corresponding projected LHC reach (part \ref{dysec});
\item techniques for discovering and studying the radion field
which is generic in most extra-dimensional scenarios
(part \ref{azuelossec_new});
\item the impact of mixing between the radion and the Higgs sector,
a fully generic possibility in extra-dimensional models (part \ref{rizzo2sec});
\item
production rates and signatures of
universal extra dimensions at hadron colliders (part \ref{rizzo1sec});
\item
black hole production at hadron colliders, which would lead
to truly spectacular events (part \ref{landsberg1sec}).
\end{itemize}
The above contributions represent a tremendous amount of work
on the part of the individuals involved and represent the
state of the art for many of the currently most important phenomenological
research avenues. Of course, much more remains to be done. 
For example, one should continue to work
on assessing the extent to which the discovery reach
will be extended if one goes beyond the LHC to
the super-high-luminosity LHC (SLHC) or to a very large hadron
collider (VLHC) with $\sqrt s \sim 40$~TeV.  Overall, 
we believe our work shows 
that the LHC and future hadronic colliders will play
a pivotal role in the discovery and study of any kind
of new physics beyond the Standard Model. They provide
tremendous potential for incredibly exciting new discoveries. 

\noindent
{\bf Acknowledgments.} \\
We thank the organizers of this workshop for the friendly and
stimulating atmosphere during the meeting. We also thank our colleagues
of the QCD/SM and HIGGS working groups for the very constructive
interactions we had. We are grateful to the ``personnel'' of the Les
Houches school for providing an environment that enabled 
us to work intensively 
and especially for their warm hospitality during our stay.


\newpage

\def\tanb{\tan\beta}
\def\what{\widehat}
\def\vev#1{\langle #1 \rangle}
\def\lsim{\mathrel{\raise.3ex\hbox{$<$\kern-0.75em\lower1.1ex\hbox{$\sim$}}}}
\def\gsim{\mathrel{\raise.3ex\hbox{$>$\kern-0.75em\lower1.1ex\hbox{$\sim$}}}}

\part{{\bf  Theoretical Developments}\\[0.2cm]\hspace*{0.8cm}
{\it J.\,Gunion, J.\,Hewett, K.\,Matchev, T.\,Rizzo}}
\label{bsmtheorysec}


\begin{abstract}
Various theoretical aspects of physics beyond the 
Standard Model at hadron colliders are discussed.
Our focus will be on those issues that most immediately impact
the projects pursued as part of the BSM group at this meeting.
\end{abstract}

\section{Introduction}

The Standard Model (SM) has had a tremendous success describing
physical phenomena up to energies $\sim 100$ GeV.
Yet some of the deep questions of particle physics are still shrouded
in mystery - the origin of electroweak symmetry breaking 
(and the related hierarchy problem),
the physics of flavor and flavor mixing, $CP$-violation etc.
Any attempt to make further theoretical progress on any one of these 
issues necessarily requires new physics beyond the SM.

It is generally believed that the TeV scale will reveal 
at least some of this new physics. Throughout history,
we have never gone a whole order of magnitude up in energy without
seeing some new phenomenon. Further support is given by attempts to 
solve the gauge hierarchy problem. Either there is no Higgs
boson in the SM and then some new physics must appear around the 
TeV scale to unitarize $WW$ scattering, or the Higgs boson exists,
and one has to struggle to explain the fact that its mass is minute 
in (fundamental) Planck mass units. Very roughly, there are three
particularly compelling categories of new physics that are capable
of solving the hierarchy problem. 
\begin{itemize}
\item
{\bf Supersymmetry (SUSY):} 

Low energy supersymmetry eliminates the quadratic ultraviolet 
sensitivity of the Higgs boson mass, which arises through 
radiative corrections.
Supersymmetry guarantees that these contributions cancel between loops
with particles and those with their superpartners, making the weak scale
natural provided the superpartner masses are ${\cal O}(1\,{\rm TeV})$.

In its minimal version, a supersymmetrized standard model has only
one additional free parameter - the supersymmetric Higgs mass $\mu$. 
However, supersymmetry has to be broken, which leads to a proliferation
of the number of independent input parameters.
There are many different models on the market, 
differing only in the way SUSY breaking is communicated to
``our world''. Furthermore, one can go beyond the minimal
supersymmetric extension of the Standard Model (MSSM),
e.g. to the Next-to-Minimal Supersymmetric Standard Model (NMSSM)
where an extra singlet superfield is added to the MSSM
matter content. Then the so-called R-parity breaking models
introduce additional Yukawa-type couplings between the
SM fermions and their superpartners; there are 
models with multiple extra U(1) gauge groups, \etc\ 
(for a recent review, see \cite{Polonsky:2001pn}).
Garden varieties of all of these models have been extensively
studied. In this report, our focus will be on models which 
yield unusual signatures and/or make discovery/study of SUSY
more difficult.

\item
{\bf Technicolor (TC):}

Technicolor (for a recent review, see \cite{Hill:2002ap})
has made a resurgence through models where the heavy top quark
plays an essential role, such as the top-color assisted technicolor
model and models in which an extra heavy singlet quark 
joins with the top-quark to give rise to electroweak symmetry breaking
(EWSB).  Very little work was done on this class of models at this
workshop and so we will not discuss such models further.  It should,
however, be noted that in most of these models, an effective
low-energy Higgs sector emerges that typically is
equivalent to a  general two-Higgs-doublet model (2HDM). 
Light pseudo-Nambu-Goldstone bosons can also be present.

\item
{\bf Extra dimensions:}

Extra dimensions at or near the TeV$^{-1}$ scale may bring the 
relevant fundamental particle physics scale down to a TeV 
and thus eliminate the hierarchy 
problem~\cite{Arkani-Hamed:1998rs,Randall:1999ee}. If this scenario were
true, it would have a profound influence on all types of physics
at the LHC and other future colliders.  Extra dimensions 
impact the Higgs sector and can even give rise to EWSB.  
They can also lead to Kaluza Klein (KK) excitations of normal matter.
The production of small black holes at the LHC becomes a possibility.
Such black holes would promptly decay to multiple SM particles 
with a thermal distribution, giving striking signatures.
A number of the many possibilities and the related experimental
consequences were explored during this workshop and are reported here.

\end{itemize}
\section{SUSY and expectations for hadron colliders}
Even within the context of the minimal
supersymmetric model (MSSM) with R-parity conservation, 
there are 103 parameters beyond the usual Standard Model (SM) parameters.
Different theoretical ideas for soft-SUSY breaking can be used to motivate 
relations between these parameters, but as time progresses
more and more models are being proposed. In addition, one cannot
rule out the possibility that several sources of soft-SUSY breaking
are present simultaneously.

Typically, any theoretical model will provide predictions for 
the soft-SUSY breaking parameters at a high scale, such as the GUT scale. For example, in mSUGRA, the minimal supergravity model (sometimes also called the constrained MSSM -- cMSSM), 
the universal GUT-scale scalar mass $M_0$, 
the universal GUT-scale gaugino mass $M_{1/2}$, 
the universal trilinear term $A_0$,
the low-energy ratio $\tan\beta$ of Higgs
vacuum expectation values, and the sign of the $\mu$ parameter,
\begin{equation}
M_0, M_{1/2}, A_0, \tanb, {\rm sign}(\mu)
\label{msugrapardefs}
\end{equation}
fully specify all the soft-SUSY breaking parameters once
the renormalization group equations (RGE) 
are required to yield correct EWSB.  More generally,
the RGEs provide a link between the experimentally observed
parameters at the TeV scale and the fundamental physics 
at the high-energy scale. The amount of information
we can extract from experiment is therefore related to the precision
with which we can relate the values of the parameters at these two 
vastly different scales. Precise predictions require multi-loop
results for the RGE and the related threshold corrections, and a 
careful assessment of all systematic uncertainties. This is
the focus of a couple of the contributions to this report
(parts \ref{svensec} and \ref{allanachsec}). At the meeting,
there was also considerable discussion of the extent to which
a given set of low energy parameters could be ruled out or at least
discriminated against by virtue of constraints such as: requiring
that the LSP be the primary dark matter constitute; correct
$b\to s\gamma$; `correct' $g_\mu -2$; \etc\ 
Currently there are many programs available for evaluating the impact
of such constraints, and they tend to give diverse answers.
In some cases, numerically important effects have been left out,
\eg certain co-annihilation channels, higher-order terms in the 
RGE equations, and so forth. In the remaining cases,
the spread can be taken as an indication of the theoretical 
uncertainty involved in relating the TeV and unification scales.
While progress in this area has been made, as summarized in 
\cite{Allanach:2002nj}, no summary of the status was prepared for
this report.  However, one important conclusion from this effort is
clear.  There are regions of parameter space,
even for the conventional mSUGRA case of Eq.~(\ref{msugrapardefs}), 
for which very high sparticle masses could remain consistent
with all constraints. This observation led to renewed
focus on LHC sensitivity to SUSY models with very high mass scales
(parts \ref{ian1sec} and \ref{ian2sec}),
as possibly also preferred by coupling constant unification with
$\alpha_s(m_Z)<0.12$. For example, naturally heavy
squark masses are allowed in the focus point scenario 
\cite{Feng:1999mn} and would ameliorate any possible problems with
flavor-changing neutral currents (FCNC) related thereto \cite{Feng:2000bp}.

More generally, it would be unwise for the experimental
community to take too seriously
the predictions of any one theoretical model for soft-SUSY breaking.
It is important that convincing arguments be made that
TeV-scale SUSY (as needed to solve the hierarchy problem) 
can be discovered for all possible models.  Much work has been
done in recent years in this respect, and such efforts were
continued during the workshop and are reported on here.
In general, the conclusions are positive; TeV-scale SUSY discovery
at the LHC will be possible for a large class of models.  
Further, after the initial discovery, a multi-channel
approach, like the one presented in part \ref{arnauldsec}, 
can be used to determine the soft-SUSY-breaking parameters
with considerable precision.

An important aspect of verifying the nature of the SUSY model
will be a full delineation of its Higgs sector. 
In the MSSM, the Higgs sector is a strongly constrained 2HDM.
In particular, in the MSSM, there is a strong upper bound
on the mass of the lightest CP-even Higgs boson ($m_h<130\gev$)
and strong relations between its couplings and the CP-odd
Higgs mass parameter $m_A$.  As a result, there is a `no-lose'
theorem for MSSM Higgs discovery at the LHC (assuming that
Higgs decays to pairs of SUSY particles are not spread
out over too many distinct channels).  
However, if $m_A\gsim 300$~GeV and $\tanb$ has a moderate value
somewhat above 3, then existing analyses indicate that it
will be very hard to detect any Higgs boson other than the
light CP-even $h$ (which will be quite SM-like). 
The $H$, $A$ and $H^\pm$
(all of which will have similar mass) might well not be observable
at the LHC. Further 
work on extending the high-$\tanb$ $\tau$ signals for the $H,A,H^\pm$
to the lowest possible $\tanb$ values and on
finding new signals for them should be pursued.

However, an even bigger concern is the additional difficulties
associated with Higgs discovery if the MSSM 
is extended to include one or more additional singlet superfields
(leading to additional Higgs singlet scalar fields).
The motivation for such an extension is substantial.
First, such singlets are very typical of string models.
Second, it is
well-known that there is no convincing source for a weak-scale
value of the $\mu$
parameter of the MSSM.  The simplest and a very attractive model
for generating a weak-scale value for $\mu$ is the NMSSM
in which one singlet superfield is added to the MSSM.
The superpotential term $\lambda \what S\what H_d \what H_u$
(where $\what S$ is the singlet superfield and $\what H_d,\what H_u$
are the Higgs superfields whose neutral scalar component vevs give
rise to the down and up quark masses, respectively)
gives rise to a weak-scale value for $\mu$ provided $\lambda$
is in the perturbative domain and $\vev{S}={\cal O}\left(m_Z\right)$.
Both of these conditions can be naturally implemented in the NMSSM.
This simple and highly-motivated extension of the MSSM leads
to many new features for SUSY phenomenology at the LHC
and other future colliders.  However, its most dramatic impact
is the greatly increased difficulty of guaranteeing
the discovery of at least one of the NMSSM Higgs bosons 
(there now being 3 CP-even Higgs bosons, 2 CP-odd Higgs bosons
and a charged Higgs pair). Very substantial progress was made
as part of this workshop in filling previously identified
gaps in parameter space for which discovery could not be
guaranteed.  However, remaining additional dangerous parameter
regions, and the new relevant experimental discovery
channels, were identified. Substantial additional
effort on the part of the LHC community will be required in order
to demonstrate that Higgs discovery in these new channels will
always be possible.  Part \ref{nmssmhiggssec} of this report
discusses these issues in some depth.


In the simplest models of soft-SUSY-breaking, it is generally
assumed that the soft-SUSY-breaking parameters will not 
have phases (that cannot be removed by simple field redefinitions).
Even in the MSSM, the presence of  such phases would be
an essential complication for LHC SUSY phenomenology,
and most particularly for Higgs sector discovery and study.
In general, many things become more difficult.  An exception
would be if one can simultaneously produce a pair of squarks in association
with a Higgs boson. Such signals would allow a first determination of
the non-trivial phases of the theory, since the production of the CP-odd $A$
in association with two light top squarks, $A+\tilde t_1+\tilde t_1$,
is an unequivocal signal of non-trivial phases for the $\mu$ and
$A$ (soft tri-linear) parameters of the MSSM.  
Some aspects of this are explored in part \ref{dmsec}.
The experimental viability of such signals will
require further study.

In many SUSY models, lepton flavor violating (LFV) decays of various particles
can occur.   Lepton-flavor-violating interactions can easily arise
as a result of a difference between the flavor diagonalization in
the normal fermionic leptonic sector as compared to that in the slepton
sector.  Typically this is avoided by one of two assumptions:
a) a common leptonic flavor structure for the lepton and slepton sectors
(alignment) or b) flavor-blind mechanism of SUSY breaking, which yields
slepton mass matrices which are diagonal in flavor space.
No convincing GUT-scale motivation for either of these possibilities
has been expounded. In fact, many string models suggest quite the
contrary (see, \eg \cite{King:1998nv}).
Further, neutrino masses and mixing phenomenology could be
indicating the presence of lepton flavor violating interactions,
especially in the context of the see-saw mechanism.
In particular, as shown in part \ref{blazeksec} of
this report, expectations based on neutrino mixing
phenomenology lead to rates for
$\tau\to \mu \gamma$ decays at high $\tanb$
(which enhances these decays in the MSSM) that are very similar to existing
bounds on such decays, implying that they might be observed
in the next round of experiments.
If one wishes to suppress LFV decays in the most general case,
very large slepton masses would be required.
This would, of course, fit together with the large squark masses
needed for guaranteed suppression of FCNC decays.

One parameter that is not conventionally included in the 103 MSSM
SUSY parameters is the goldstino mass (which
determines the mass of the spin-3/2 gravitino). The gravitino mass
is related to the scale of SUSY breaking $F$ by
\begin{equation}
m_{3/2}=\sqrt{\frac{8\pi}{3}}\frac{F}{M_{Pl}}\,.
\end{equation}
Further, the interactions of the goldstino part of the gravitino (and 
of its spin zero sgoldstino partners)
are proportional to $1/F$.  (The masses of the goldstinos
$m_S,m_P$ are not determined.) 
In mSUGRA models and the like, $F$ is sufficiently large that
the goldstino and sgoldstino masses are so large, and their
interaction strengths so small, that they are not phenomenologically
relevant.  However, in some models of SUSY breaking $F$ 
is relatively small. A well-known example is
gauge-mediated SUSY breaking for which $F$ can be small enough
for the goldstino to be the
true LSP into which all more massive SUSY particles
ultimately decay.  In such a case, 
all of SUSY phenomenology changes dramatically.
The sgoldstinos might also be light,
with masses anywhere below $1$~TeV being reasonable.  
In this case, for $\sqrt F\lsim 1$~TeV,
they could yield some very significant experimental signals,
discussed in part \ref{gorbsec}.
For example, they might appear in rare decays of the $J/\psi$ and $\Upsilon$
or lead to FCNC interactions. For small enough $F$, direct
production of sgoldstinos becomes significant at the LHC for masses
up to about a TeV (in particular via a $gg\to S$ 
vertex of the form $\frac{m_{1/2}}{F}F_{\mu\nu}^a F^{\mu\nu\, a}S$)
and would yield some unique signatures.

The possibility of R-parity violation in SUSY models has been extensively
considered \cite{Allanach:1999bf}. There are three possible sets of
RPV couplings as specified in the superpotential:
\begin{equation}
\lam_{ijk} \what L_i\what L_j \what E_k+\lam'_{ijk}\what L_i Q_j D_k
+\lam^{\prime\prime}_{ijk} \what U_i \what D_j \what D_k\,,
\end{equation}  
where SU(2) and color-singlet structures are implied.
Here, $\lam_{ijk}$ ($\lam^{\prime\prime}_{ijk}$) 
must be antisymmetric under 
$i\leftrightarrow j$ ($j\leftrightarrow k$).
For proton stability, we require that either the $\lam^{\prime\prime}_{ijk}=0$
or that $\lam_{ijk}=\lam'_{ijk}=0$. One of the most under-explored
possibilities for the LHC is that one or more of the $\lam''$'s is non-zero.
This would imply that the neutralino ultimately decays to 3 jets
inside the detector.  There would be no missing energy.  If
the mass difference between the $\widetilde \chi_1^0$ 
and $\widetilde \chi_1^+$ is small (as possible, for example, 
for anomaly mediated SUSY-breaking and in some types of 
string-motivated boundary conditions) or if the leptonic branching
fractions of the charginos and heavier neutralinos are small,
then there might also be few hard leptons in the LHC events.  The main
SUSY signature would be extra events with large numbers of jets.
Whether or not such events can be reliably extracted from the large
QCD background, and especially the maximum SUSY particle mass
for which such extraction is possible, is a topic awaiting future
study.  The leptonic type of RPV would lead to very clear LHC
signals for SUSY, in which events would contain extra leptons as
well as some missing energy from the extra neutrinos that
would emerge from decays. For example, $\lam_{212}$
would lead to decays of the neutralino LSP 
such as $\widetilde \chi_1^0\to \mu\mu\nu$.

It is just possible that the NuTeV dilepton events~\cite{Formaggio:1999ne}
could be a first sign of R-parity violation.
The explanation proposed in part \ref{ddrsec} requires
$\lambda_{232}\neq 0$ (leading to the decays
$\widetilde \chi_1^0\to \mu^-_L\mu^+_R\nu_\tau$
and $\widetilde \chi_1^0\to \tau^-_L\mu^+_R\nu_\mu$, and conjugates
thereof). The explanation proposed for the Tevatron events,
in which the light neutralinos are produced in $B_d^0,B^+$ decays)
would also require the existence of  a mixed leptonic-hadronic
RPV coupling $\lambda_{113}^\prime$.  In general, the weakness of
the constraints on couplings involving the 3rd generation and 
the large size of the similar Yukawa couplings related to quark mass
generation both favor signals related to 3rd generation leptons
and quarks.


\def\nn{\noindent}

\def\Re{{\cal R \mskip-4mu \lower.1ex \hbox{\it e}\,}}
\def\Im{{\cal I \mskip-5mu \lower.1ex \hbox{\it m}\,}}
\def\ie{{\it i.e.}}
\def\eg{{\it e.g.}}
\def\etc{{\it etc}}
\def\etal{{\it et al.}}
\def\ibid{{\it ibid}.}
\def\sub#1{_{\lower.25ex\hbox{$\scriptstyle#1$}}}
\def\tev{\,{\ifmmode\mathrm {TeV}\else TeV\fi}}
\def\gev{\,{\ifmmode\mathrm {GeV}\else GeV\fi}}
\def\mev{\,{\ifmmode\mathrm {MeV}\else MeV\fi}} 
\def\MD{\ifmmode \overline M_{D}\else $\overline M_{D}$\fi}
\def\mpl{\ifmmode \overline M_{Pl}\else $\overline M_{Pl}$\fi}
\def\to{\rightarrow}
\def\slash{\not\!}
\def\subw{_{\rm w}}
\def\mh{\ifmmode m\sbl H \else $m\sbl H$\fi}
\def\mch{\ifmmode m_{H^\pm} \else $m_{H^\pm}$\fi}
\def\mt{\ifmmode m_t\else $m_t$\fi}
\def\mc{\ifmmode m_c\else $m_c$\fi}
\def\mz{\ifmmode M_Z\else $M_Z$\fi}
\def\mw{\ifmmode M_W\else $M_W$\fi}
\def\mws{\ifmmode M_W^2 \else $M_W^2$\fi}
\def\mhs{\ifmmode m_H^2 \else $m_H^2$\fi}   
\def\mzs{\ifmmode M_Z^2 \else $M_Z^2$\fi}
\def\mts{\ifmmode m_t^2 \else $m_t^2$\fi}
\def\mcs{\ifmmode m_c^2 \else $m_c^2$\fi}
\def\mchs{\ifmmode m_{H^\pm}^2 \else $m_{H^\pm}^2$\fi}
\def\ztwo{\ifmmode Z_2\else $Z_2$\fi}
\def\zone{\ifmmode Z_1\else $Z_1$\fi}
\def\mtwo{\ifmmode M_2\else $M_2$\fi}
\def\mone{\ifmmode M_1\else $M_1$\fi}
\def\tb{\ifmmode \tan\beta \else $\tan\beta$\fi}
\def\xw{\ifmmode x\subw\else $x\subw$\fi}
\def\ch{\ifmmode H^\pm \else $H^\pm$\fi}
\def\lum{\ifmmode {\cal L}\else ${\cal L}$\fi}
\def\inpb{\,{\ifmmode {\mathrm {pb}}^{-1}\else ${\mathrm {pb}}^{-1}$\fi}}
\def\infb{\,{\ifmmode {\mathrm {fb}}^{-1}\else ${\mathrm {fb}}^{-1}$\fi}}
\def\epem{\ifmmode e^+e^-\else $e^+e^-$\fi}
\def\ppb{\ifmmode \bar pp\else $\bar pp$\fi}
\def\bsg{\ifmmode B\to X_s\gamma\else $B\to X_s\gamma$\fi}
\def\bsll{\ifmmode B\to X_s\ell^+\ell^-\else $B\to X_s\ell^+\ell^-$\fi}
\def\bstt{\ifmmode B\to X_s\tau^+\tau^-\else $B\to X_s\tau^+\tau^-$\fi}
\def\lamt{\ifmmode \tilde\lambda\else $\tilde\lambda$\fi}
\def\shat{\ifmmode \hat s\else $\hat s$\fi}
\def\that{\ifmmode \hat t\else $\hat t$\fi}
\def\uhat{\ifmmode \hat u\else $\hat u$\fi}
\def\elli{\ell^{i}}
\def\ellj{\ell^{j}}
\def\ellk{\ell^{k}} 
\newskip\zatskip \zatskip=0pt plus0pt minus0pt
\def\matth{\mathsurround=0pt}
\def\lsim{\mathrel{\mathpalette\atversim<}}
\def\gsim{\mathrel{\mathpalette\atversim>}}
\def\atversim#1#2{\lower0.7ex\vbox{\baselineskip\zatskip\lineskip\zatskip
  \lineskiplimit 0pt\ialign{$\matth#1\hfil##\hfil$\crcr#2\crcr\sim\crcr}}}
\def\undertext#1{$\underline{\smash{\vphantom{y}\hbox{#1}}}$}

\section{Extra Dimensions}

An alternative to SUSY for explaining the hierarchy problem is that the
geometry of space-time is modified at scales much less than the Planck
scale, $M_{Pl}$. In such models, which
may still be regarded as rather speculative, but have attracted 
a lot of attention recently, the 3-spatial dimensions in which we 
live form a 3-dimensional `membrane', called `the wall', embedded 
in a much larger extra dimensional space, known as `the bulk', and that the 
hierarchy between the weak scale $\sim 10^3$ GeV and the 4-dimensional 
Planck scale $M_{Pl}\sim 10^{19}$ GeV
is generated by the geometry of the additional bulk dimensions.  
This is achievable either by compactifying all the extra dimensions on tori,
or by using strong curvature effects in the 
extra dimensions.  In the first case, Arkani-Hamed, Dimopoulos, and 
Dvali (ADD) \cite{Arkani-Hamed:1998rs,Antoniadis:1998ig,Arkani-Hamed:1998nn}
used this picture to generate the 
hierarchy by postulating a large volume for the extra dimensional space. 
In the latter case, the hierarchy can be established by a large curvature 
of the extra dimensions as demonstrated by Randall and Sundrum (RS) 
\cite{Randall:1999ee}.  It is the relation of these models to the hierarchy 
which yields testable predictions at the TeV scale. Such ideas have 
led to extra dimensional theories which have verifiable 
consequences at present and future colliders. 

There are three principal scenarios with predictions at the TeV scale, each of 
which has a distinct phenomenology. In theories with Large Extra Dimensions,  
proposed by ADD~\cite{Arkani-Hamed:1998rs,Antoniadis:1998ig,Arkani-Hamed:1998nn},
gravity alone propagates in the bulk where 
it is assumed to become strong near the weak scale.  Gauss' Law relates the 
(reduced) Planck scale \mpl\ of the effective 4d low-energy theory and the 
fundamental scale $\MD$, through the volume of the $\delta$ 
compactified dimensions, $V_\delta$, via 
$\mpl^2 = V_\delta \MD^{2+\delta}$. ~\mpl\ is thus no longer a fundamental 
scale as it is generated by the large volume of the higher dimensional space.
If it is  assumed that the extra dimensions are toroidal, then setting 
$\MD\sim$ TeV to eliminate the hierarchy between \mpl\ and the weak 
scale determines the compactification radius $R$ of the extra dimensions.
Under the further simplifying assumption that all radii are of equal size,
$V_\delta = (2\pi R)^\delta$, $R$ then ranges from a sub-millimeter to a few 
fermi for $\delta =2-6$. Note that the case of $\delta = 1$ is excluded as the 
corresponding dimension would directly alter Newton's law on solar-system 
scales. The bulk gravitons expand into a Kaluza-Klein (KK) tower of states, 
with the mass of each excitation state being given by $m_n^2=n^2/R^2$. 
With such large values of $R$ the KK mass spectrum appears almost 
continuous at collider energies. The ADD model has two important collider 
signatures: ($i$) the emission of real KK gravitons in a collision process 
leading to a final state with missing energy and ($ii$) the exchange of 
virtual KK graviton 
towers between SM fields which leads to effective dim-8 contact interactions. 
Except for the issue of Black Hole (BH) production to be discussed 
below, we will say no more about the ADD scenario as work was not performed 
on this model at this workshop. 

A second possibility is that of Warped Extra Dimensions; in the simplest form 
of this scenario \cite{Randall:1999ee} gravity propagates in a 5d bulk of finite extent 
between two $(3+1)$-dimensional branes which have opposite tensions.  
The Standard Model fields are assumed 
to be constrained to one of these branes which is called the TeV brane.  
Gravity is localized on the 
opposite brane which is referred to as the Planck brane. This configuration 
arises from the metric $ds^2 =e^{-2ky}\eta_{\mu\nu}dx^\mu dx^\nu - dy^2$ 
where the exponential function, or warp factor, multiplying the usual 4d 
Minkowski term produces a non-factorizable geometry, and $y \in [0,\pi R]$ is 
the coordinate of the extra dimension. The Planck (TeV) brane is placed at 
$y=0 (\pi R)$. The space between the two branes is 
thus a slice of $AdS_5$: 5d anti-deSitter space. The original extra dimension 
is compactified on a circle $S^1$ so that the wave functions in the extra 
dimension are periodic and then orbifolded by a single discrete 
symmetry $Z_2$ forcing the KK graviton 
states to be even or odd under $y \to -y$. Here, 
the parameter $k$ describes the curvature scale, which together with 
$\MD$ ($D=5$) is assumed \cite{Randall:1999ee} to be of order \mpl, with the relation 
$\mpl^2=\MD^3/k$ following from the integration over the 5d action. 
Note that that there are no hierarchies amongst these mass parameters. 
Consistency of the low-energy description requires that the 5d curvature, 
$R_5=-20k^2$, be small in magnitude in comparison to $\MD$, which implies 
$k/\mpl \lsim 0.1$. We note that mass scales which are 
naturally of order $\mpl$ on the $y=0$ brane will appear to be of order the TeV 
scale on the $y=\pi R$ brane due to the exponential warping provided that 
$\pi R \simeq 11-12$. This leads to a solution of the hierarchy problem. 

The 4d phenomenology of the RS model is governed by two parameters, 
$\Lambda_\pi=\mpl e^{-kR\pi}$, which is of order a TeV, and $k/\mpl$. 
The masses of the bulk graviton KK 
tower states are $m_n=x_nke^{-kR\pi} = x_n\Lambda_\pi k/\mpl$ with the $x_n$ 
being the roots of the first-order Bessel function $J_1$. The KK states are 
thus not evenly spaced. For typical values of the parameters, the mass of the 
first graviton KK excitation is of order a TeV. The interactions of the bulk 
graviton KK tower with the SM fields are \cite{Davoudiasl:1999jd}
\begin{equation} 
\Delta{\cal L}=-{\frac{1}{M_{\rm Pl}}}T^{\mu\nu}(x)h^{(0)}_{\mu\nu}(x)
-\frac{1}{\Lambda_\pi}T^{\mu\nu}(x)\sum_{n=1}^\infty h^{(n)}_{\mu
\nu}(x)\,,
\end{equation} 
where $T^{\mu\nu}$ is the stress-energy tensor of the SM fields, 
$h^{(0)}_{\mu\nu}$ is the 
ordinary graviton and $h^{(n)}_{\mu\nu}$ are the KK graviton tower fields. 
Experiment can determine or constrain the masses $m_n$ and the coupling 
$\Lambda_\pi$. 
In this model KK graviton resonances with spin-2 can be produced in a number 
of different reactions at colliders. 
Extensions of this basic model allow for the SM fields to propagate
in the bulk \cite{Davoudiasl:1999tf,Pomarol:1999ad,Davoudiasl:2000wi,%
Gherghetta:2000qt,Hewett:2002fe}. In this case, 
the masses of the bulk fermion, gauge, and graviton KK states are 
related.  A third parameter, associated with the fermion bulk mass, 
is introduced and governs the 4d phenomenology. In this case, KK excitations 
of the SM fields may also be produced at colliders. 
  
One important aspect of the RS model is the need to stabilize the separation 
of the two branes with $kR\simeq 11-12$ in order to solve the hierarchy 
problem. This 
can be done in a natural manner {\cite{Goldberger:1999uk}} but leads to the existence of a 
new, relatively light scalar field with a mass significantly less than 
$\Lambda_\pi$ called the radion. This is most likely the lightest new state in 
the RS scenario. The radion has a flat wavefunction in the bulk and is a 
remnant of orbifolding and of the graviton KK decomposition. This field 
couples to the trace of the stress-energy tensor, 
$\sim T^\mu_\mu/\Lambda_\pi$,  and is thus Higgs-like in its 
interactions with SM fields. In addition, it may mix with the SM Higgs 
altering the couplings of both fields. Searches for the radion and its 
influence on the SM Higgs couplings will be discussed below.

The possibility of TeV$^{-1}$-sized extra dimensions arises in braneworld 
models \cite{Antoniadis:1990ew}. By themselves, they 
do not allow for a reformulation of the 
hierarchy problem but they may be incorporated into a larger structure in 
which this problem is solved.  In these scenarios, the Standard Model fields 
may propagate in the bulk. This allows for a wide number of model building 
choices: 
\begin{itemize}
\item all, or only some, of the SM gauge fields are present in the bulk; 
\item the Higgs field(s) may be in the bulk or on the brane; 
\item the SM fermions may be confined to the brane or
to specific locales in the extra dimensions. 
\end{itemize}
If the Higgs field(s) propagate 
in the bulk, the vacuum expectation value (vev) of the Higgs zero-mode, the 
lowest lying KK state, generates spontaneous symmetry breaking.  In this case, 
the gauge boson KK mass matrix is diagonal with the excitation masses given by
$[M_0^2+\vec n\cdot\vec n /R^{2}]^{1/2}$, where $M_0$ is the vev-induced mass 
of the gauge zero-mode and $\vec n$ labels the KK excitations in $\delta$ 
extra dimensions. However, if the Higgs is confined to the brane, its 
vev induces off-diagonal elements in the mass matrix generating mixing amongst 
the gauge KK states of order $(M_0 R)^2$.  For the case of 1 extra dimension,
the coupling strength of the bulk KK gauge states to the SM fermions on the 
brane is $\sqrt 2 g$, where $g$ is the corresponding SM coupling. The fermion 
fields may  (a) be constrained to the $(3+1)$-brane, in which case they are 
not directly affected by the extra dimensions; (b) be localized at
specific points in the TeV$^{-1}$ dimension, but not on a rigid brane.
Here the zero and excited mode KK fermions obtain narrow Gaussian-like wave 
functions in the extra dimensions with a width much smaller than $R^{-1}$. 
This possibility may suppress the rates for a number of dangerous processes 
such as proton decay \cite{Arkani-Hamed:1999dc}. (c) The SM fields may also 
propagate in the bulk.  This scenario is known as universal extra dimensions 
\cite{Appelquist:2000nn}. $(4+\delta)$-dimensional momentum is then 
conserved at tree-level, and KK parity, $(-1)^n$, is conserved to all orders.   
TeV extra dimensions lead to an array of collider signatures some of which 
will be discussed in detail below. 

Theories with extra dimensions and a low effective Planck scale ($\MD$) offer 
the exciting possibility that black holes (BH) somewhat 
more massive than $\MD$ can be produced with large rates at future colliders. 
Cross sections of order 100 pb at the LHC have been advertised 
in the analyses presented by Giddings and Thomas~\cite{Giddings:2001bu} 
and by Dimopoulos and Landsberg~\cite{Dimopoulos:2001hw}. 
These early analyses and discussions of the production of BH at colliders 
have been elaborated upon by several groups of authors
{\cite {Giddings:2001ih,Cheung:2001ue,Landsberg:2001sj,Hofmann:2001pz,%
Eardley:2002re,Hossenfelder:2001dn,Ahn:2002mj}} and the 
production of BH by cosmic rays has also been considered 
{\cite {Kowalski:2002gb,Anchordoqui:2001cg,Uehara:2001yk,Emparan:2001kf,%
Alfaro:2001gk,Kazanas:2001ep,Anchordoqui:2001ei,Feng:2001ib}}. A  
most important question to address is whether or not the BH cross sections are 
actually this large or, at the very least, large enough to lead to visible 
rates at future colliders. 

The basic idea behind the original collider BH papers is as follows: 
consider the collision of two high energy SM partons which are 
confined to a 3-brane, as they are in both the ADD and RS models. 
In addition, gravity is free to propagate in $\delta$ extra 
dimensions with the $4+\delta$ dimensional Planck scale assumed to be 
$\MD \sim 1$ TeV. The curvature of the space is assumed to be small compared 
to the energy scales involved in the collision process so that quantum 
gravity effects can be neglected. When these partons have a center of 
mass energy in excess of $\sim \MD$ and the impact parameter of the 
collision is less than the Schwarzschild radius, $R_S$, associated with this 
center of mass energy, a 
$4+\delta$-dimensional BH is formed with reasonably high efficiency. It is 
expected that a very large fraction of the collision energy 
goes into the BH formation process so that $M_{BH}\simeq \sqrt s$. 
The subprocess cross section for the production of a non-spinning BH 
is thus essentially geometric for {\it each} pair of initial partons: 
$\hat \sigma \simeq \epsilon \pi R_S^2$, where $\epsilon$ is a factor that 
accounts for finite impact parameter and angular momentum corrections and 
is expected to be $\lsim 1$. Note that the $4+\delta$-dimensional 
Schwarzschild radius scales as 
$R_S \sim \Big[\frac{M_{BH}}{\MD^{~2+\delta}}\Big]^\frac{1}{1+\delta}$, 
apart from an overall $\delta$- and {\it convention-dependent} numerical 
prefactor. 
This approximate geometric subprocess cross section expression is claimed to 
hold when the ratio $M_{BH}/\MD$ is ``large", \ie, 
when the system can be treated semi-classically and quantum gravitational 
effects are small.

Voloshin {\cite {Voloshin:2001vs,Voloshin:2001fe}} 
has provided several arguments which suggest that 
an additional exponential suppression factor must be included which 
presumably damps the pure geometric cross section for this process even in 
the semi-classical case. This issue remains somewhat controversial. 
Fortunately it has been shown {\cite {Rizzo:2002kb}} 
that the numerical influence of this suppression, if present,  
is not so great as to preclude BH production at significant 
rates at the LHC. These objects will decay promptly and yield  
spectacular signatures. A discussion of BH production at future 
colliders is presented in one of the contributions. 

\section{Acknowledgments}
JFG is supported in part by the U.S. Department
of Energy contract No. DE-FG03-91ER40674 and by the Davis
Institute for High Energy Physics.
The work of JH and TR is supported by the Department of Energy, 
Contract DE-AC03-76SF00515.

\setcounter{figure}{0}
\setcounter{table}{0}
\setcounter{section}{0}
\setcounter{equation}{0}
\clearpage

\def\msbar{$\overline{\rm{MS}}$}
\def\drbar{$\overline{\rm{DR}}$}
\def\tev{\,\, \mathrm{TeV}}
\def\gev{\,\, \mathrm{GeV}}
\def\mev{\,\, \mathrm{MeV}}
\def\onel{one-loop}
\def\twol{two-loop}
\def\drbarm{\overline{\rm{DR}}}
\def\als{\alpha_s}
\def\KL{\left(}
\def\KR{\right)}
\def\KKL{\left[}
\def\KKR{\right]}
\def\KKKL{\left\{}
\def\KKKR{\right\}}
\def\BL{\lbrack}
\def\BR{\rbrack}
\def\BC{begin{center}}
\def\EC{\end{center}}
\def\BEA{\protect\begin{eqnarray}}
\def\BEAnn{\protect\begin{eqnarray*}}
\def\EEA{\protect\end{eqnarray}}
\def\EEAnn{\protect\end{eqnarray*}}
\def\non{\nonumber}
\def\oaas{{\cal O}(\alpha\alpha_s)}

\part{{\bf  {FeynSSG v.1.0}: Numerical Calculation 
of the mSUGRA and Higgs spectrum} \\[0.2cm] \hspace*{0.8cm}
{\it A.~Dedes, S.~Heinemeyer, G.~Weiglein}}
\label{svensec}

\begin{abstract}
{\tt FeynSSG v.1.0} is a program for the numerical evaluation of the
Supersymmetric (SUSY) particle spectrum and Higgs boson masses in the
Minimal Supergravity (mSUGRA) scenario. 
We briefly present the physics behind the program and as an 
example we calculate the SUSY and Higgs spectrum for a set of sample
points. 
\end{abstract}


In the Minimal Supersymmetric Standard Model (MSSM) no specific
assumptions are made about the underlying 
SUSY-breaking mechanism, and a parameterization of all 
possible soft SUSY-breaking terms is used. This gives rise to the huge
number of more than 100 new parameters in addition to the SM, 
which in principle can be chosen
independently of each other. A phenomenological analysis of this model
in full generality would clearly be very involved, and one usually
restricts to certain benchmark scenarios, see Ref.~\cite{Allanach:2002nj} for a
detailed discussion.
On the other hand, models in which all the low-energy parameters are
determined in terms of a few parameters at the Grand Unification (GUT)
scale (or another high-energy scale), 
employing a specific soft SUSY-breaking scenario, are
much more predictive. The most prominent scenario at present
is the minimal Supergravity (mSUGRA) 
scenario~\cite{Nilles:1982ik,Nilles:1983xx,Chamseddine:1982jx,%
Barbieri:1982eh,Nilles:1983dy,Cremmer:1983vy,Ferrara:1983qs,%
Hall:1983iz,Soni:1983rm,Nath:1983aw}.

\smallskip
In this note we present the Fortran code {\tt FeynSSG} for the
evaluation of the low-energy mSUGRA spectrum, including a precise
evaluation for the MSSM Higgs sector. The high-energy input parameters
(see below) are related to the low-energy SUSY parameters via 
renormalization group (RG) running (taken from the program 
{\tt SUITY}~\cite{Dedes:1996sb,Dedes:1998wn}), 
taking into account contributions up
to two-loop order. The low-energy parameters are then used as input
for the program {\tt FeynHiggs}~\cite{Heinemeyer:1998yj} for the evaluation of
the MSSM Higgs sector.

\bigskip
The simplest possible choice for an underlying theory is to take at
the GUT scale all scalar particle masses equal to a common mass parameter
$M_0$, all gaugino masses are chosen to be equal to the parameter
$M_{1/2}$ and all trilinear couplings flavor blind and equal to
$A_0$. This situation 
can be arranged in Gravity Mediating SUSY breaking Models by imposing 
an appropriate symmetry in the K\"ahler 
potential~\cite{Nilles:1982ik,Nilles:1983xx,Chamseddine:1982jx,%
Barbieri:1982eh,Nilles:1983dy,Cremmer:1983vy,Ferrara:1983qs,%
Hall:1983iz,Soni:1983rm,Nath:1983aw}, called
the minimal Supergravity (mSUGRA) scenario. 
In order to solve the minimization conditions of the Higgs potential,
i.e.\ in order to impose the constraint of REWSB, one needs as input
$\tan\beta(M_Z)$ and ${\rm sign}(\mu)$. 
The running soft SUSY-breaking parameters in the Higgs potential, 
$m_{H_1}$ and $m_{H_2}$, are defined at the EW scale after their
evolution from the GUT scale where we assume that they also have the common
value $M_0$. Thus, apart from
the SM parameters (determined by experiment)
4~parameters and a sign are required to define the mSUGRA scenario:
%
\begin{eqnarray}    
\{\; M_0\;,\;M_{1/2}\;,\;A_0\;,\;\tan\beta \;,\; {\rm sign}(\mu)\; \} \;.
\label{mSUGRAparams}
\end{eqnarray}
%
In the numerical procedure we employ a two-loop 
renormalization group running for all parameters involved, i.e.\ all
couplings, dimensionful parameters and VEV's. We start with the \msbar\
values for the gauge couplings at the scale $M_Z$, where for the strong
coupling constant $\alpha_s$ a trial input value in the vicinity 
of 0.120 is used. The \msbar\ values are converted into the corresponding
\drbar\ ones~\cite{Martin:1993yx}. 
The \msbar\ running $b$ and $\tau$ masses are run down to 
$m_b = 4.9 \gev$, $m_\tau = 1.777 \gev$ 
with the ${\rm SU(3)}_C \times {\rm U(1)}_{em}$
RGE's~\cite{Arason:1992ic} to derive
the running bottom and tau masses (extracted from their pole masses).
This procedure includes all SUSY corrections at the \onel\ level and
all QCD corrections at the \twol\ level as given in \cite{Pierce:1997zz}.
Afterwards by making use of the \twol\ RGE's
for the running masses $\overline{m}_b$, $\overline{m}_\tau$,
we run upwards to derive their \msbar\ values at $M_Z$, which
are subsequently converted to the corresponding \drbar\ values. 
This procedure provides the bottom and tau Yukawa couplings at the
scale $M_Z$. The top Yukawa coupling is derived from the 
top-quark pole mass, $m_t = 175 \gev$, which is
subsequently converted to the \drbar\ value, $\overline{m_t}(m_t)$,
where the top Yukawa coupling is defined.
The evolution of all couplings from $M_Z$ running upwards to high
energies now determines the unification scale
$M_{\rm GUT}$ and the value of the unification coupling $\alpha_{\rm GUT}$ by
%
\begin{eqnarray}
\alpha_1(M_{\rm GUT})|_{\drbarm}=\alpha_2(M_{\rm GUT})|_{\drbarm}
=\alpha_{\rm GUT} \;.
\label{unification}
\end{eqnarray}
%
At the GUT scale we set the boundary conditions for the 
soft SUSY breaking parameters, i.e.\ the values for $M_0$, $M_{1/2}$
and $A_0$ are chosen, and also $\alpha_3(M_{\rm GUT})$ is set equal to
$\alpha_{\rm GUT}$.
All parameters are run down again from $M_{\rm GUT}$ to $M_Z$.
For the calculation of the soft SUSY-breaking
masses at the EW scale we use the ``step function
approximation''~\cite{Dedes:1996sb,Dedes:1998wn}. Thus, if
the equation employed is the RGE for a particular running
mass $m(Q)$, then $Q_0$ is the corresponding physical mass 
determined by the condition $m(Q_0) =Q_0$. 
After running down to $M_Z$, the trial input value for $\als$ has
changed. 
At this point the value for $\tan\beta$ is chosen and fixed.
The parameters $|\mu|$ and $B$
are calculated from the minimization conditions 
\begin{eqnarray}
\label{min1} 
\mu^2(Q) \ &=& \ \frac{\bar{m}_{H_1}(Q)^2-\bar{m}_{H_2}(Q)^2 \tan^2\beta(Q)}
                      {\tan^2\beta -1}
                -\frac{1}{2} M_Z^2(Q) \;, \\
B(Q) \ &=& \ -\frac{(\bar{m}_{1}(Q)^2+\bar{m}_{2}(Q)^2)\sin2\beta(Q)}
                   {2\mu(Q)} \;.
\label{min2}
\end{eqnarray}
Only the sign of the $\mu$-parameter is
not automatically fixed and thus chosen now.
This procedure is iterated several times until convergence is reached.

In (\ref{min1}),(\ref{min2})
$Q$ is the renormalization scale. It is chosen
such that radiative corrections to the effective potential are rather
small compared to other scales.
In (\ref{min1}),(\ref{min2}) $\tan\beta \equiv v_2/v_1$ 
is the ratio of the two vacuum 
expectation values of the Higgs fields  $H_2$ and $H_1$ responsible 
for giving masses to the up-type and down-type quarks, respectively. In 
(\ref{min1}),(\ref{min2}), $\tan\beta$ is evaluated at the scale $Q$,
from the scale $M_Z$, where it is considered as an input parameter%
\footnote{
See
for example the discussion in the Appendix of \cite{Drees:1992mx}. 
}%
. 
By $\bar{m}_{H_i}^2 = m_{H_i}^2+\Sigma_{v_i}$ in (\ref{min1}),(\ref{min2}) 
we denote the radiatively corrected ``running '' Higgs soft-SUSY
breaking masses and  
%
\begin{eqnarray}
\bar{m}_{i}^2 = {m}_{H_i}^2 + \mu^2 + \Sigma_{v_i} 
\equiv \bar{m}_{H_i}^2 + \mu^2 \;\; (i=1,2) \;,
\label{mbar}
\end{eqnarray}
%
where $\Sigma_{v_i}$ are the one-loop corrections
based on  the 1-loop Coleman-Weinberg effective potential $\Delta V$, 
$\Sigma_{v_i}=\frac{1}{2v_i}\frac{\partial \Delta V}{\partial v_i}$,
%
\begin{eqnarray}
\Sigma_{v_i} = \frac{1}{64\pi^2} \sum_a (-)^{2J_a} (2J_a+1) 
               C_a \Omega_a
               \frac{M_a^2}{v_i} 
               \frac{\partial M_a^2}{\partial v_i} 
               \KKL \ln \frac{M_a^2}{Q^2} - 1 \KKR \;.
\label{effpot}
\end{eqnarray}
%
Here $J_a$ is the spin of the particle $a$, $C_a$ are the color degrees of
freedom, and $\Omega_a=1(2)$ for real scalar (complex scalar), 
$\Omega_a=1(2)$ for Majorana (Dirac) fermions. $Q$ is the energy scale 
and the $M_a$ are the field dependent mass matrices. Explicit formulas
of the $\Sigma_{v_i}$ are given in the Appendices of \cite{Pierce:1997zz,Barger:1994gh}.
In our analyses contributions from all SUSY particles at the \onel\ level
are incorporated%
\footnote{
The corresponding \twol\ corrections are now available for a general
renormalizable softly broken SUSY theory~\cite{Martin:2001vx}. Assuming
the size of these higher-order corrections to be of the same
size as for the Higgs-boson mass matrix,
the resulting values of $\mu$ and $B$ could change by
$\sim 5-10\%$.  The possible changes would hardly
affect the results in the 
Higgs-boson sector but could affect to some extent the analysis of SUSY
particle spectra, especially when $M_0$ and $M_{1/2}$ are lying in
different mass regions.
}%
.
With $M_Z^2$ here we denote
the tree level ``running'' $Z$~boson mass, 
$M_Z^2(Q)=\frac{1}{2}(g_1^2+g_2^2)v^2$ ($v^2 \equiv v_1^2 + v_2^2$),
extracted at the scale $Q$ from
its physical pole mass $M_Z = 91.187 \gev$. The REWSB
is fulfilled,
if and only if there is 
a solution to the conditions (\ref{min1}),(\ref{min2})%
\footnote{
Sometimes in the
literature, the requirement of the REWSB is
described by the inequality $m_1^2(Q)m_2^2(Q)-|\mu(Q) B(Q)|^2 <0$. This
relation is automatically satisfied here from (\ref{min1}),(\ref{min2})
and from the fact that the physical squared Higgs masses must
be positive.
}%
.

\bigskip
For the predictions in the MSSM Higgs sector we use the code 
{\tt FeynHiggs}~\cite{Heinemeyer:1998yj}, which is implemented as a subroutine
into {\tt FeynSSG}. The code is based on the evaluation of the
low-energy Higgs sector parameters 
in the Feynman-diagrammatic (FD) 
approach~\cite{Heinemeyer:1998jw,Heinemeyer:1998kz,Heinemeyer:1998np} within the
on-shell renormalization scheme. Details about the conversion of the
low-energy results from the RG running, obtained in the
$\overline{\rm{DR}}$ scheme, to the on-shell scheme can be found in
Ref.~\cite{Ambrosanio:2001xb}.  
In the FD approach the masses of the two CP-even
Higgs bosons, $m_h$ and $m_H$, are derived beyond tree level 
by determining the poles of the $h-H$-propagator
matrix, which is equivalent to solving the equation
\begin{equation}
\left[q^2 - m_{h,{\rm tree}}^2 + \hat\Sigma_{h}(q^2) \right]
\left[q^2 - m_{H,{\rm tree}}^2 + \hat\Sigma_{H}(q^2) \right] - 
\left[\hat\Sigma_{hH}(q^2)\right]^2 = 0 ,
\end{equation}
where $\hat\Sigma_s, s = h, H, hH$ denotes the renormalized Higgs
boson self-energies. Their evaluation consists of 
the complete \onel\ result combined with the dominant \twol\ contributions of
$\oaas$~\cite{Heinemeyer:1998jw,Heinemeyer:1998kz,Heinemeyer:1998np} and further subdominant
corrections~\cite{Brignole:2001jy,Espinosa:2000df}, see Refs.~\cite{Heinemeyer:1998jw,%
Heinemeyer:1998kz,Heinemeyer:1998np,Frank:2002qf}
for details.


\bigskip
An analysis employing {\tt FeynSSG} for the constraints on the mSUGRA
scenario from the Higgs boson search at LEP2 and the corresponding
implications for SUSY searches at future colliders has been presented
in Refs.~\cite{Ambrosanio:2001xb,Dedes:2001ew}. As another example we present here the
results of the low-energy SUSY spectrum for some sample
points~\cite{Battaglia:2001jn}. (Some of these sample points are now included
in the ``SPS'' (Snowmass
Points and Slopes)~\cite{Allanach:2002nj} that have recently been proposed as new
benchmark scenarios for SUSY searches at current and future colliders.)

The sample points are presented in Table~1.
For these results we have set the 1-loop corrections 
$\Sigma_{v_i}$ equal to zero and all the thresholds are switched on.
Thus for the points considered here a one loop improved tree level
analysis is done.
If we switch on the full 1-loop corrections $\Sigma_{v_i}$, then the 
points  E,F,H,J,K, and M, fail to satisfy electroweak symmetry breaking,
$\mu^2$ from (\ref{min2}) is negative. In addition, the weak mixing angle, 
$\sin^2\theta_W(M_Z)$, has been set to $0.2315$. An updated version
which employs 
the effective weak mixing angle as a boundary condition at the
electroweak scale is under way (in fact such an analysis had been done 
in the past using the program {\tt SUITY}, 
see \cite{Dedes:1996sb,Dedes:1998wn,Dedes:1998kq}). 
It is intended to regularly update {\tt FeynSSG} with the upcoming
new versions of the {\tt SUITY} and {\tt FeynHiggs} programs.

%
\begin{table}[htb!]
\small
\centering
\begin{tabular}{|c||r|r|r|r|r|r|r|r|r|r|r|r|r|}
\hline
Model          & A   &  B  &  C  &  D  &  E  &  F  &  G  &  H  &  I  &  J  &  K  &  L  &  M   \\ 
\hline
$m_{1/2}$      & 624 & 258 & 415 & 549 &  315& 1090& 390 & 1585.5& 364 & 785 & 1006& 471 & 1600 \\
$m_0$          & 137 & 100 &  90 & 120 & 1500& 2970& 123 & 459 & 188 & 320 & 1000& 330 & 1500 \\
$\tan{\beta}$  & 5   & 10  & 10  & 10  & 10  & 10  & 20  & 20  & 35  & 35  &40.3 & 45  & 48   \\
sign($\mu$)    & $+$ & $+$ & $+$ & $-$ & $+$ & $+$ & $+$ & $+$ & $+$ & $+$ & $-$ & $+$ & $+$  \\ 
$A_0$          & 0   & 0   &  0  &  0  &  0  &  0  &  0  &  0  & 0   & 0   & 0   &  0  &  0   \\
$m_t$          & 175 & 175 & 175 & 175 & 175 & 175 & 175 & 175 & 175 & 175 & 175 & 175 & 175  \\ \hline
Masses         &     &     &     &     &     &     &     &     &     &     &     &     &      \\
 \hline
$|\mu |$       & 811 & 362 & 551 & 705 & --- & 941 & 515 & 1719 & 480 & 936 & --- & 595 & 1660  \\
 \hline
h$^0$          & 114 & 113 & 116 & 116 & --- & 118 & 117 & 121 & 117 & 121 & --- & 119 & 123  \\
H$^0$          & 947 & 414 & 629 & 769 & --- & 3171 & 580 & 2065 & 502 & 1003 & --- & 578 & 1709  \\
A$^0$          & 947 & 414 & 629 & 769 & --- & 3171 & 580 & 2065 & 502 & 1003 & --- & 578 & 1709  \\
H$^{\pm}$      & 939 & 420 & 625 & 789 & --- & 3151 & 569 & 1920 & 472 & 867 & --- & 461 & 818  \\ 
\hline
$\chi^0_1$     & 260 & 101 & 169 & 229 & --- & 475 & 158 & 693 & 148 & 332 & ---  & 196 & 705  \\
$\chi^0_2$     & 484 & 185 & 314 & 429 & --- & 853 & 295 & 1273 & 274 & 618 & --- & 363 & 1293  \\
$\chi^0_3$     & 813 & 368 & 555 & 707 & --- & 942 & 520 & 1720 & 485 & 938 & --- & 599 & 1661  \\
$\chi^0_4$     & 827 & 387 & 570 & 713 & --- & 985 & 534 & 1728 & 499 & 948 & --- & 611 & 1670  \\
$\chi^{\pm}_1$ & 483 & 185 & 314 & 429 & --- & 852 & 295 & 1273 & 274 & 618 & --- & 362 & 1293  \\
$\chi^{\pm}_2$ & 826 & 387 & 570 & 715 & --- & 985 & 534 & 1728 & 500 & 948 & --- & 612 & 1670  \\ 
\hline
$\tilde{g}$    & 1382 & 619 & 953 & 1228 & --- & 2371 & 901 & 3266 & 847 & 1713 & --- & 1074 & 3301  \\ 
\hline
$e_L$, $\mu_L$ & 437 & 206 & 295 & 386 & --- & 3038 & 292 & 1127 & 311 & 610 & --- & 456 & 1818  \\
$e_R$, $\mu_R$ & 273 & 146 & 184 & 241 & --- & 2991 & 195 & 744 & 236 & 435 & --- & 376 & 1609  \\
$\nu_e$, $\nu_{\mu}$
               & 431 & 190 & 284 & 378 & --- & 3037 & 281 & 1125 & 300 & 605 & --- & 449 & 1816  \\
$\tau_1$       & 271 & 137 & 176 & 234 & --- & 2966 & 168 & 702 & 165 & 351 & --- & 261 & 1228  \\
$\tau_2$       & 438 & 209 & 297 & 387 & --- & 3026 & 299 & 1118 & 322 & 602 & --- & 449 & 1673  \\
$\nu_{\tau}$   & 430 & 189 & 283 & 377 & --- & 3025 & 277 & 1112 & 289 & 584 & --- & 419 & 1666  \\ 
\hline
$u_L$, $c_L$   & 1261 & 575 & 874 & 1122 & --- & 3546 & 831 & 2958 & 794 & 1581 & --- & 1028 & 3293  \\
$u_R$, $c_R$   & 1216 & 559 & 845 & 1082 & --- & 3507 & 805 & 2835 & 770 & 1524 & --- & 997 & 3183  \\
$d_L$, $s_L$   & 1264 & 581 & 877 & 1125 & --- & 3547 & 835 & 2959 & 798 & 1583 & --- & 1031 & 3294  \\
$d_R$, $s_R$   & 1211 & 559 & 843 & 1078 & --- & 3503 & 803 & 2820 & 768 & 1517 & --- & 994 & 3169  \\
$t_1$          & 971 & 419 & 663 & 874 & --- & 2465 & 630 & 2340 & 596 & 1237 & --- & 779 & 2534  \\
$t_2$          & 1211 & 604 & 864 & 1076 & --- & 3077 & 820 & 2735 & 772 & 1457 & --- & 953 & 2826  \\ 
$b_1$          & 1167 & 531 & 807 & 1037 & --- & 3071 & 754 & 2711 & 686 & 1393 & --- & 859 & 2739  \\
$b_2$          & 1211 & 560 & 842 & 1075 & --- & 3481 & 799 & 2772 & 752 & 1460 & --- & 941 & 2833  \\
\hline
\end{tabular}
\caption{
Mass spectra in GeV for mSUGRA points
calculated with program {\tt FeynSSG v1.0} (see text for details).
Points (E) and (K) fail to pass the Radiative Electroweak Breaking
requirement, i.e., $\mu^2 < 0$. Points (F) and (M) exhibit
instability, i.e., the program reaches a poor convergence.
The charged Higgs Boson ($H^\pm$) mass is given at tree level.} 
\end{table}

\setcounter{figure}{0}
\setcounter{table}{0}
\setcounter{section}{0}
\setcounter{equation}{0}
\clearpage

\part{\noindent
{\bf  Theoretical Uncertainties in Sparticle Mass Predictions and SOFTSUSY
} \\[0.5cm]\hspace*{0.8cm}
{\it B.C. Allanach}}
\label{allanachsec}


\begin{abstract}
We briefly introduce the SOFTSUSY calculation of sparticle masses and
mixings and illustrate the output with post-LEP benchmarks.
We contrast the sparticle spectra obtained from 
ISASUGRA7.58, SUSPECT2.004 with those obtained from SOFTSUSY1.3
along SNOWMASS model lines in minimal supersymmetric standard model (MSSM)
parameter space. 
From this we gain an idea of the uncertainties involved with sparticle spectra
calculations.
\end{abstract}

Supersymmetric phenomenology is notoriously complicated. 
Even if one assumes the particle spectrum of the minimal supersymmetric
standard model (MSSM),
fundamental patterns of supersymmetry (SUSY) breaking are numerous.
It seems that there is currently nothing to strongly favor one particular scenario
above all others.
In ref.~\cite{Allanach:2001qe}, it was shown that measuring two ratios of sparticle masses to
$1\%$ could be enough to discriminate different SUSY breaking scenarios 
(in that case, mirage, grand-unified or intermediate scale type I
string-inspired unification). Thus,
in order to discriminate high energy models of supersymmetry
breaking, it will be necessary to have better than 1$\%$ accuracy in
both the experimental {\em and} theoretical determination of some superparticle
masses. 
An alternative bottom-up approach~\cite{Blair:2000gy} is to evolve soft
supersymmetry breaking
parameters from the weak scale to a high scale once they are `measured'. 
The parameters of the high-scale theory are then inferred, and theoretical
errors involved in the calculation will need to be minimized.

We now briefly introduce {\tt SOFTSUSY1.3}~\cite{Allanach:2001kg}, a tool to
calculate the masses and
mixings of MSSM sparticles. 
It can be downloaded from the URL
\begin{quote}
{\tt http://allanach.home.cern.ch/allanach/softsusy.html}
\end{quote}
It is valid for the R-parity conserving MSSM with
real couplings and includes full 3-family particle or sparticle mixing.
The manual~\cite{Allanach:2001kg} can be consulted for a more complete description of approximations
and the algorithm used. Low energy data (together with $\tan \beta(M_Z)$) set
the Standard Model gauge couplings and Yukawa couplings: $G_F$, $\alpha$,
$\alpha_S(M_Z)$ and the fermion masses and CKM matrix elements. The user
provides a high-energy unification scale and supersymmetry breaking boundary
conditions at that scale. The program derives the MSSM spectrum consistent with
both of these constraints and radiative electroweak symmetry
breaking at a scale $M_{SUSY}=\sqrt{m_{{\tilde t}_1} m_{{\tilde t}_2}}$.
Below $M_Z$, three-loop QCD$\otimes$one-loop QED is used to evaluate the
$\overline{MS}$ Yukawa couplings and gauge couplings at $M_Z$. These are then
converted into the $\overline{DR}$ scheme, including finite and logarithmic 
corrections coming from sparticle loops. All one-loop corrections are added to
the top mass and gauge couplings, while the other Standard Model couplings
receive approximations to the full one-loop result. The radiative electroweak
symmetry breaking constraint incorporates full one-loop tadpole corrections.
The gluino, stop and sbottom masses receive full one-loop (logarithmic and
finite) corrections, with approximations being employed in the one-loop
corrections to the other sparticles. In the CP-even Higgs sector, the
calculation is {\tt FEYNHIGGSFAST}-like~\cite{Heinemeyer:1999be,Haber:1997fp},
with additional two-loop top/stop corrections. The other Higgs' receive full
one-loop radiative corrections,
except for the charged Higgs, which is missing a self-energy correction.
Currently, the MSSM renormalization group equations (used above $M_Z$) 
are two-loop order except for the scalar masses and scalar trilinear couplings,
which are all one-loop order equations.

\small\begin{table}[p] 
\centering
\begin{tabular}{|c||r|r|r|r|r|r|r|r|r|r|r|r|r|}
\hline
Model          & A   &  B  &  C  &  D  &  E  &  F  &  G  &  H  &  I  &  J  &  K  &  L  &  M   \\ 
\hline
$m_{1/2}$      & 624 & 258 & 415 & 549 &  315& 1090& 390 & 1585.5& 364 & 785 & 1006& 471 & 1600 \\
$m_0$          & 137 & 100 &  90 & 120 & 1500& 2970& 123 & 459 & 188 & 320 & 1000& 330 & 1500 \\
$\tan{\beta}$  & 5   & 10  & 10  & 10  & 10  & 10  & 20  & 20  & 35  & 35  &40.3 & 45  & 48   \\
sign($\mu$)    & $+$ & $+$ & $+$ & $-$ & $+$ & $+$ & $+$ & $+$ & $+$ & $+$ & $-$ & $+$ & $+$  \\ 
$m_t$          & 175 & 175 & 175 & 175 & 175 & 175 & 175 & 175 & 175 & 175 & 175 & 175 & 175  \\ \hline
Masses         &     &     &     &     &     &     &     &     &     &     &     &     &      \\
 \hline
$|\mu(M_Z)|$ & 738 & 322 & 494 & 632 & - & - & 461 & 1579 & 429 & 847 & - & 531 & -\\
$h^0$ & 118 & 114 & 119 & 119 & - & - & 119 & 126 & 118 & 123 & - & 119 & -\\
$H^0$ & 877 & 379 & 575 & 708 & - & - & 528 & 1884 & 452 & 905 & - & 440 & -\\
$A^0$ & 863 & 365 & 558 & 721 & - & - & 495 & 1779 & 392 & 792 & - & 289 & -\\
$H^\pm$ & 869 & 376 & 566 & 727 & - & - & 506 & 1791 & 410 & 813 & - & 331 & -\\ \hline 
$\chi^0_1$ & 252 & 99 & 165 & 221 & - & - & 154 & 654 & 144 & 319 & - & 187 & -\\
$\chi^0_2$ & 465 & 176 & 301 & 411 & - & - & 282 & 1211 & 262 & 593 & - & 347 & -\\
$\chi^0_3$ & 740 & 328 & 498 & 636 & - & - & 465 & 1582 & 433 & 847 & - & 530 & -\\
$\chi^0_4$ & 756 & 351 & 516 & 644 & - & - & 482 & 1591 & 450 & 859 & - & 546 & -\\
$\chi^\pm_1$ & 465 & 175 & 300 & 411 & - & - & 282 & 1211 & 262 & 593 & - & 347 & -\\
$\chi^\pm_2$ & 755 & 351 & 515 & 646 & - & - & 483 & 1590 & 450 & 859 & - & 546 & -\\ \hline 
$\tilde{g}$ & 1372 & 617 & 945 & 1216 & - & - & 894 & 3194 & 840 & 1684 & - & 1063 & -\\ \hline 
$e_L$, $\mu_L$ & 427 & 202 & 287 & 376 & - & - & 283 & 1072 & 300 & 584 & - & 464 & -\\
$e_R$, $\mu_R$ & 269 & 144 & 181 & 238 & - & - & 190 & 703 & 227 & 414 & - & 391 & -\\
$\nu_e$, $\nu_{\mu}$ & 420 & 186 & 277 & 368 & - & - & 272 & 1069 & 290 & 579 & - & 458 & -\\
$\tau_1$ & 427 & 205 & 289 & 376 & - & - & 289 & 1063 & 310 & 576 & - & 444 & -\\
$\tau_2$ & 267 & 137 & 174 & 232 & - & - & 166 & 665 & 161 & 335 & - & 240 & -\\
$\nu_{\tau}$ & 420 & 186 & 277 & 368 & - & - & 272 & 1069 & 290 & 579 & - & 458 & -\\ \hline 
$u_L$, $c_L$ & 1252 & 570 & 864 & 1111 & - & - & 822 & 2904 & 784 & 1553 & - & 1021 & -\\
$u_R$, $c_R$ & 1200 & 551 & 830 & 1066 & - & - & 791 & 2767 & 756 & 1487 & - & 985 & -\\
$d_L$, $d_L$ & 1254 & 576 & 867 & 1114 & - & - & 825 & 2905 & 788 & 1555 & - & 1024 & -\\
$d_R$, $d_R$ & 1193 & 550 & 827 & 1060 & - & - & 787 & 2748 & 753 & 1479 & - & 981 & -\\
$t_1$ & 1174 & 583 & 834 & 1044 & - & - & 791 & 2632 & 742 & 1397 & - & 903 & -\\
$t_2$ & 949 & 415 & 649 & 856 & - & - & 617 & 2252 & 583 & 1192 & - & 755 & -\\
$b_1$ & 1146 & 523 & 790 & 1018 & - & - & 740 & 2632 & 672 & 1353 & - & 884 & -\\
$b_2$ & 1190 & 548 & 822 & 1053 & - & - & 776 & 2692 & 722 & 1400 & - & 811 &
-\\\hline 
\end{tabular}
\caption{Post-LEP Benchmark points.
Mass spectra in GeV for minimal SUGRA models 
calculated with program {\tt SOFTSUSY1.3} and unification scale $M_X=1.9 \times
10^{16}$ GeV, $A_0=0$.
Columns with dashes for spectra indicate points which did not
break electroweak symmetry correctly. All massive parameters are quoted in
units of GeV.
}\label{postlep}
\end{table}\normalsize
A series of points in MSSM universal supersymmetry breaking parameter space
were identified~\cite{Battaglia:2001zp} as being relevant for study, taking the
results of the LEP2 collider searches (and dark matter considerations) into
account. For this workshop, the parameters of each 
benchmark were changed until the output of {\tt ISASUGRA7.51} matched that of 
ref.~\cite{Battaglia:2001zp}. 
The standard of these parameters is used to compare the output of
several codes in these proceedings.
We illustrate the {\tt SOFTSUSY1.3} calculation
by presenting its output of these modified ``post-LEP benchmark points'' in
table~\ref{postlep}. We use
$\alpha_s(M_Z)^{\overline MS}=0.119$, $m_t=175$ GeV, $m_b(m_b)^{\overline
MS}=4.2$ GeV.
We note that four of these points do not break the
electroweak symmetry consistently.
However, many of the points were picked specifically in order to be close to 
the electroweak symmetry-breaking boundary and so this feature is perhaps not
so surprising.

Studies of the ability of future colliders to search for and measure
supersymmetric 
parameters have often focused on isolated `bench-mark' model
points~\cite{Allanach:2000kt,Battaglia:2001zp,atlastdr3} such as the post-LEP
benchmarks. This approach, while  
being a start, is not ideal because one is not sure how many of the
features used in the analyses will apply to other points of parameter space.
Collider signatures typically rely upon identifying decay products
of produced sparticles through cascade decay chains.
The resulting signatures of different scenarios of SUSY breaking are not only
highly dependent upon the scenario that is assumed, but also upon any model
parameters~\cite{atlastdr3}. 
As a supersymmetry breaking parameter is changed, the ordering of
sparticle masses can change, switching various sparticle decay branches on
and off. 
In an attempt to cover more of the available parameter space, the 
{\em Direct Investigations of SUSY Subgroup} of {\em SNOWMASS 2001} has
proposed eight bench-mark model {\em lines} for study~\cite{Allanach:2002nj}.

The lines were defined to have the spectrum output from the {\tt
ISASUGRA}~program (part of the {\tt ISAJET7.51} package~\cite{Baer:1999sp})
for $m_t=175$ GeV. 
Knowledge of the uncertainties in this calculation will be important when data
is confronted with theory, i.e. when information upon a high-energy SUSY
breaking sector is sought from low-energy data.
Here, we intend to investigate the theoretical uncertainties in sparticle
mass determination. To this end, we contrast the sparticle masses predicted by
three modern up-to-date publicly available and supported codes: {\tt
ISASUGRA7.58*}, {\tt SOFTSUSY1.3}~\cite{Allanach:2001kg} and {\tt
SUSPECT2.004}~\cite{Djouadi:1998di}. The asterisk indicates a changed version
of {\tt ISASUGRA7.58}, as detailed below.

Each of the three packages calculates sparticle masses in a similar way, but
with different approximations~\cite{stefano}. In certain model line
scenarios, we calculate the fractional difference for some sparticle $s$
\begin{equation}
f_s^{\mbox{\tiny CODE}} = \frac{m_s^{\mbox{\tiny SOFTSUSY1.3}} - m_s^{\mbox{\tiny
CODE}}}{m_s^{\mbox{\tiny SOFTSUSY1.3}}}, \label{fractions}
\end{equation}
where {\small CODE}~refers to {\tt ISASUGRA7.58*}, or {\tt
SUSPECT2.004}. $f_s^{\mbox{\tiny CODE}}$  
then gives the 
fractional difference of the mass of sparticle $s$ between the predictions of
{\small CODE} and {\tt SOFTSUSY1.3}. A positive value of $f_s^{\mbox{\tiny CODE}}$ then
implies that $s$ is heavier in {\tt SOFTSUSY1.3}~ than in {\small CODE}. 

We focus upon model lines in scenarios which are currently supported by all
three packages, i.e. supergravity mediated supersymmetry breaking (mSUGRA).
At a high unification scale $M_{GUT}\equiv 1.9 \times 10^{16}$, 
the soft-breaking scalar masses are set to be all equal to $m_0$, the
universal scalar trilinear coupling to $A_0$ and each gaugino mass $M_{1,2,3}$ 
is set. $\tan \beta$ is set at $M_Z$. 
The three choices of model lines are displayed in Table~\ref{:tabmodels}.
 \begin{table}
\begin{center}
 \caption{Model lines in mSUGRA investigated here. $m_t=175$ GeV, $M_{GUT}=1.9 \times 10^{16}$ GeV and
$\alpha_s(M_Z)^{\overline{MS}} = 0.119$ are used.}
 \label{:tabmodels}
 \begin{tabular}{|c|ccccccc|} \hline
Model line & $\tan \beta$ & $A_0$ & $M_1$ & $M_2$ & $M_3$ & $m_0$ & sgn$\mu$
\\ \hline
A          & 10 & -0.4$M_{1/2}$ & $M_{1/2}$ & $M_{1/2}$ & $M_{1/2}$ &
0.4$M_{1/2}$ & + \\
B & 10 & 0 & 1.6$M_2$ & $M_2$ & $M_2$ & $M_2/2$ & + \\
F & 10 & 0 &  $M_{1/2}$ & $M_{1/2}$ & $M_{1/2}$ & $2M_{1/2}+800$ GeV & + \\ \hline
 \end{tabular}
\end{center}
 \end{table}
Model line A displays gaugino mass dominance, ameliorating the SUSY flavor
problem. Model line B has non-universal gaugino masses and 
model line F corresponds to focus-point supersymmetry~\cite{Feng:1999mn},
close to the electroweak symmetry breaking boundary.

The differences in the output between three earlier versions of the codes has
already been discussed~\cite{Allanach:2001hm}. 
Ref.~\cite{Allanach:2001hm}
showed significant order 1$\%$ 
numerical round-off error in the gluino and
squark masses. Even worse, along model line F there were 10$\%$,
3$\%$ numerical round-off errors in the lightest neutralino and chargino
masses respectively. These numerical round-off errors were due to the {\tt
ISASUGRA} calculation, but this was not obvious because {\tt ISASUGRA} was
used for the normalization in the equivalent of eq.~\ref{fractions}.
Stop masses were not examined.
The lightest stop mass could be very important for SUSY searches, for example
at the Tevatron collider.
We now perform the comparison again, with the following differences: 
the output of {\tt SOFTSUSY} is used for the normalization,
up-to-date and bug-fixed versions of each code are used, 
we include the lightest stop mass in the comparison and
the {\tt ISASUGRA7.58*} package is hacked to provide better accuracy
in the renormalization 
group evolution\footnote{We re-set two parameters in subroutine {\tt SUGRA} to 
{\tt DELLIM=2.0e-3} and {\tt NSTEP=2000}.}

\begin{figure}
\unitlength=1in
\begin{picture}(6,2.5)
\put(0,0){\includegraphics[scale=0.75]{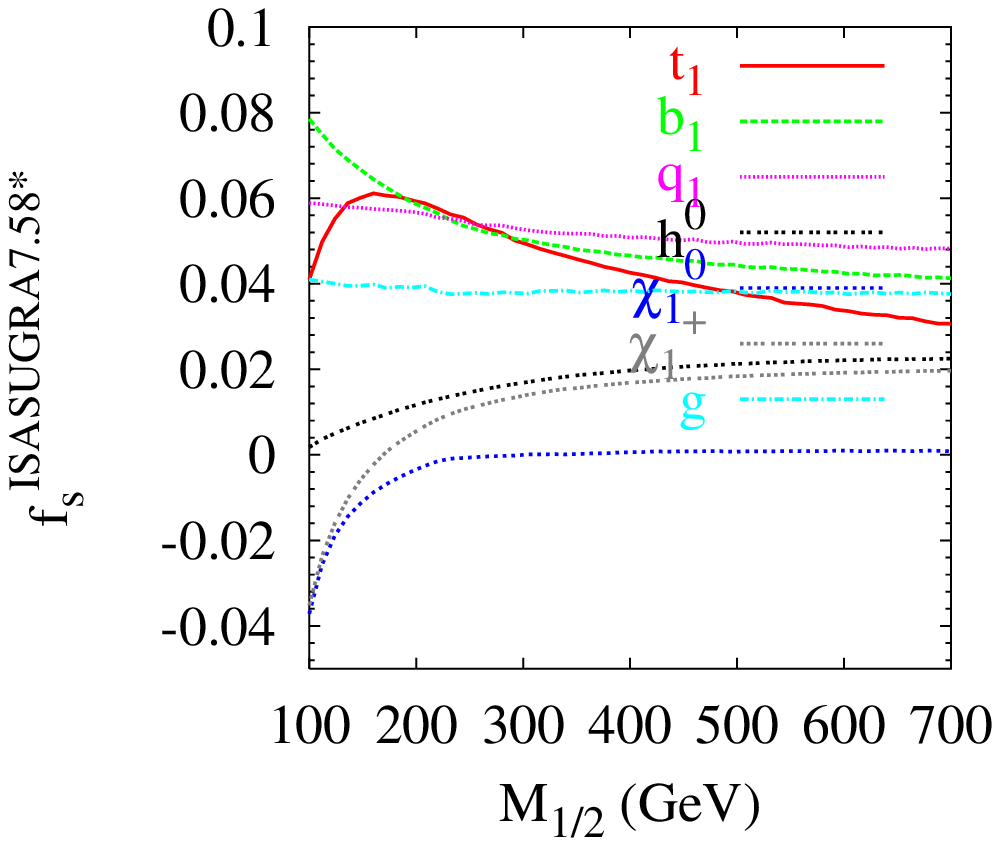}}
\put(3,0){\includegraphics[scale=0.75]{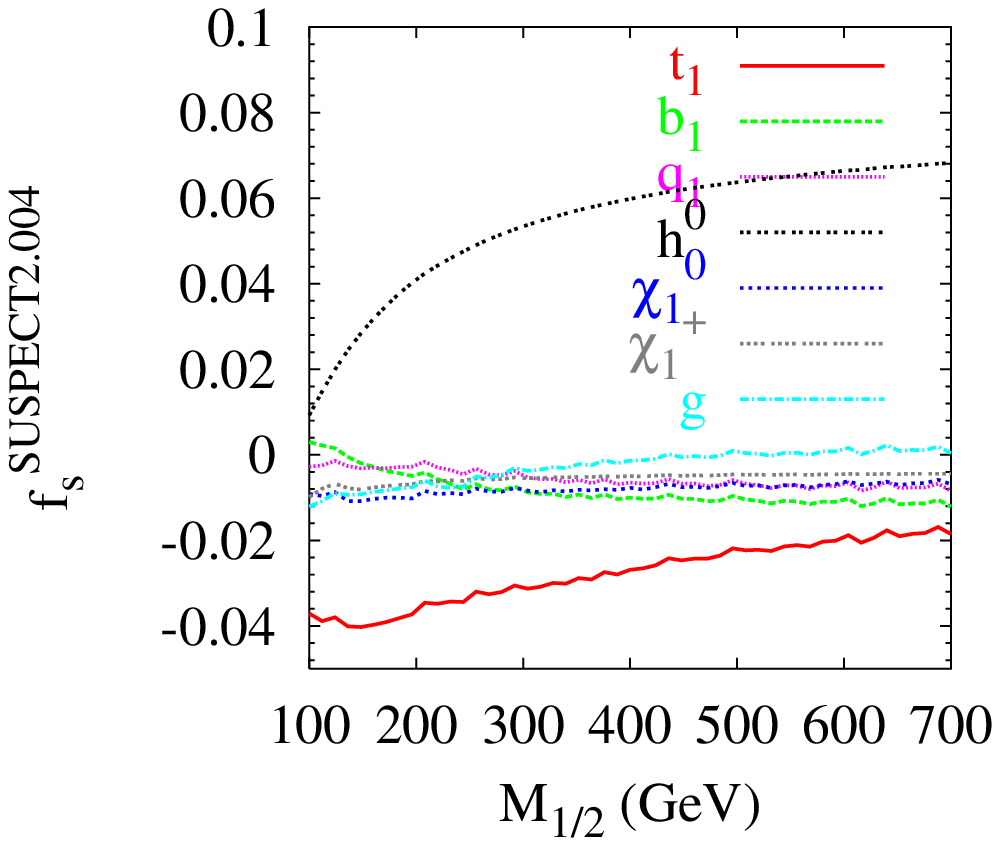}}
\put(0.,2.4){(a)}
\put(3.0,2.4){(b)}
\end{picture}
 \caption{Fractional differences between the spectra predicted for mSUGRA
model line A}
 \label{:fig:modelA}
 \end{figure}
\begin{figure}
\unitlength=1in
\begin{picture}(6,2.5)
\put(0,0){\includegraphics[scale=0.75]{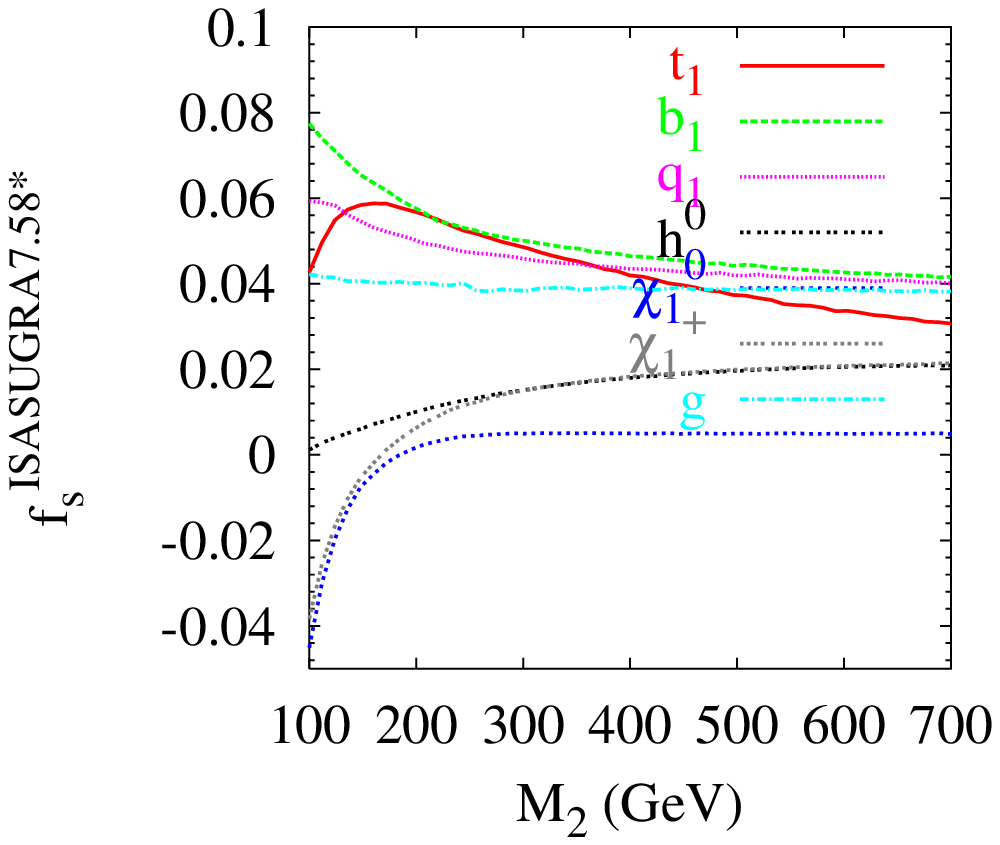}}
\put(3,0){\includegraphics[scale=0.75]{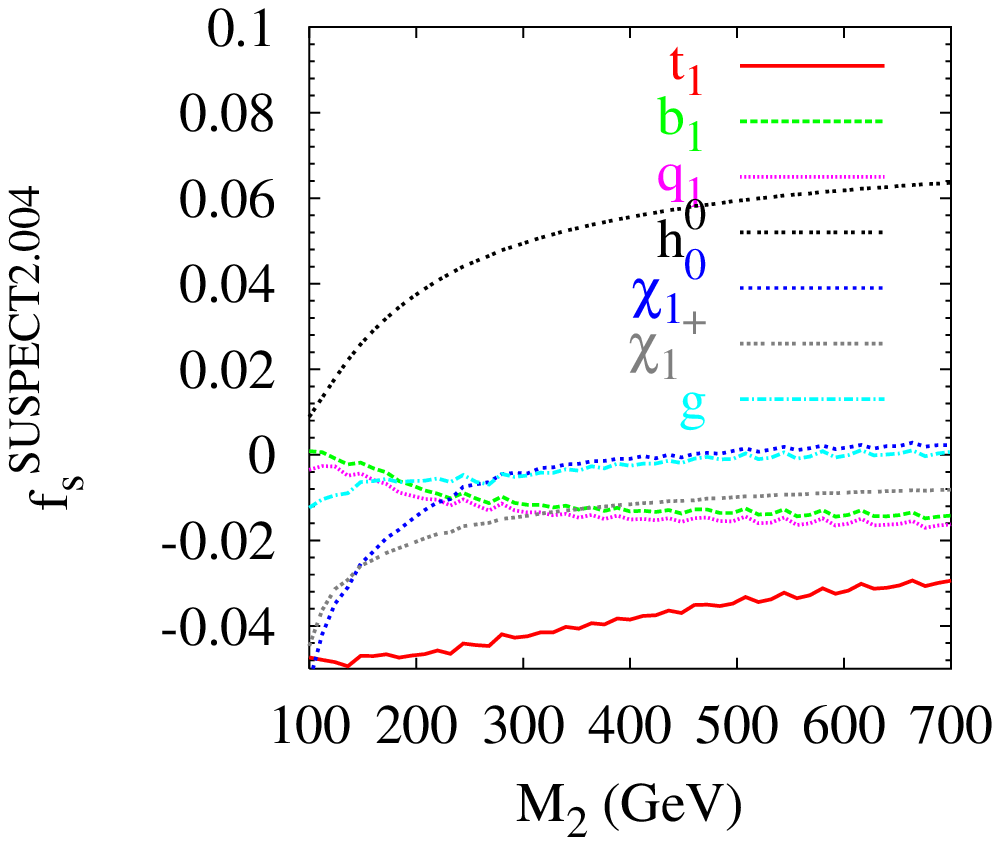}}
\put(0.,2.4){(a)}
\put(3.0,2.4){(b)}
\end{picture}
 \caption{Fractional differences between the spectra predicted for
model line B}
 \label{:fig:modelB}
 \end{figure}

We pick various sparticle masses that show a large difference in their
prediction between the
three calculations. For model line A,
Fig.~\ref{:fig:modelA}a shows
$f_{t_1,b_1,q_1,h^0,\chi_1^0,\chi_1^+,g}^{\mbox{\tiny ISASUGRA7.58*}}$
(the lightest stop, sbottom, squark, neutral Higgs, neutralino, chargino and
gluino mass difference fractions respectively). Fig.~\ref{:fig:modelA}b
shows 
the equivalent results for the output of {\tt SUSPECT}. 
Model line B differences are shown in Fig.~\ref{:fig:modelB}.
Jagged curves in the figures are a result of numerical error in the {\tt
SUSPECT} calculation, and are at an acceptable per-mille level
level for squarks, gluinos and the lightest neutralino. 
The 
lightest Higgs and lightest chargino do not display any appreciable numerical
error. 

Figs.~\ref{:fig:modelA},\ref{:fig:modelB} share some common
features. 
The largest discrepancies occur mostly for low $M_{1/2}$, where the
super-particle spectrum is lightest.
The gluino and squark masses are consistently around 5$\%$ lower in
{\tt ISASUGRA}~ than the other two codes, which agree with each other to
better than $1\%$ with the exception of the lightest stop, which {\tt SUSPECT}
finds to be less than 4$\%$ heavier than {\tt SOFTSUSY}. 
We note here that this uncertainty is not small,
a $3\%$ error on the lightest stop mass at $M_{1/2}=700$ GeV in model line A
corresponds to an error of 35 GeV, for example.
The lightest CP-even Higgs is predicted to be heaviest in {\tt SOFTSUSY},
{\tt SUSPECT} gives a value up to 6$\%$ lighter for large $M_{1/2}$,
whereas {\tt ISASUGRA} gives a value up to 2$\%$ lighter (again for large
$M_{1/2}$).
This could be to some degree due to the fact that {\tt SOFTSUSY} uses a
{\tt FEYNHIGGSFAST} calculation of the neutral Higgs
masses with important two-loop effects added~\cite{Heinemeyer:1999be}, which
predicts masses that tend to be higher 
than the one-loop calculation (as used in {\tt ISASUGRA} or {\tt SUSPECT}).
The gaugino masses display differences between the output of
each of these two codes and {\tt SOFTSUSY}, up to 4$\%$ at the lighter end of
the model lines.

\begin{figure}
\unitlength=1in
\begin{picture}(6,2.5)
\put(0,0){\includegraphics[scale=0.75]{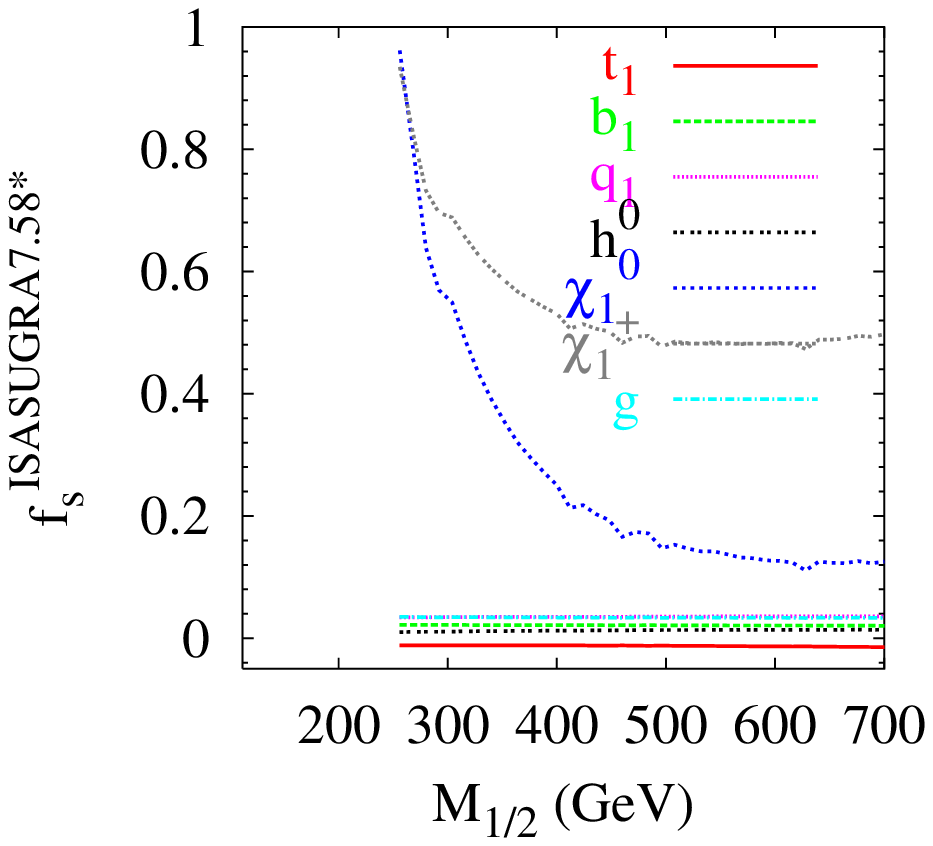}}
\put(3,0){\includegraphics[scale=0.75]{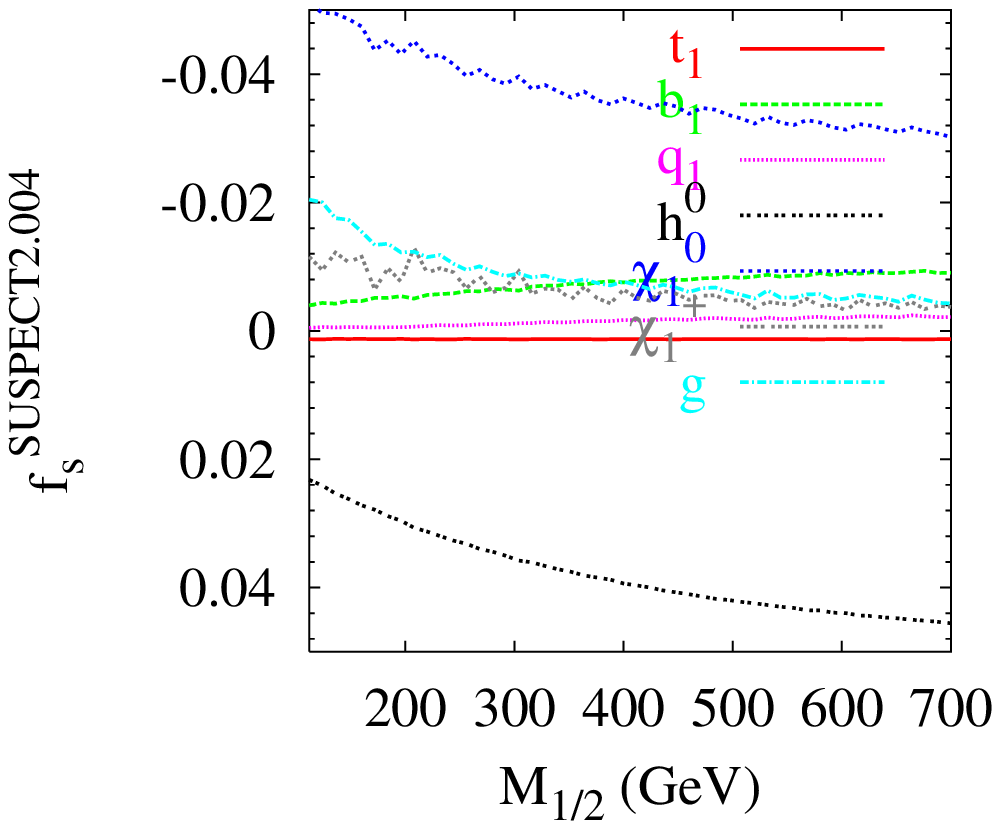}}
\put(0.,2.4){(a)}
\put(3.0,2.4){(b)}
\end{picture}
 \caption{Fractional differences between the spectra predicted for 
model line F}
 \label{:fig:modelF}
 \end{figure}

The focus-point scenario (model line F) is displayed in
Fig.~\ref{:fig:modelF}.
Fig.~\ref{:fig:modelF}a is cut off for low $M_{1/2}$ because {\tt ISASUGRA}
does not find a consistent solution that breaks electroweak symmetry there,
contrary to the other two codes.
The overall view of spectral differences is similar to that in model lines A
and B except for the masses of the lightest chargino and neutralino.
They display large 10-100$\%$ differences in Fig.~\ref{:fig:modelF}. 
In focus point supersymmetry, the bilinear Higgs mass parameter $\mu$ is close
to zero and is very sensitive to threshold corrections to
$m_t$~\cite{Allanach:2000ii}. For small $\mu < M_Z$, the lightest chargino and
neutralino masses become sensitive to its value. The predicted
value of $\mu(M_Z)$ differs by 10$\%$-100$\%$ between {\tt ISASUGRA} and the
other two codes' output. {\tt SUSPECT} and {\tt SOFTSUSY} have closer
agreement, the largest differences being that the chargino is predicted to be 
4$\%$ lighter at low $M_{1/2}$ and the lightest CP even Higgs to be 4$\%$
heavier in {\tt SOFTSUSY}.
Only a few of the threshold corrections to $m_t$ are included 
in the {\tt ISASUGRA}~ calculation, whereas {\tt SOFTSUSY}, for example,
includes all one-loop corrections with sparticles in the loop. {\tt SUSPECT}
also adds many of the sparticle loop corrections to $m_t$. Because model line
F has heavy scalars, another
possibility for the large discrepancy with {\tt ISASUGRA} could potentially be
that {\tt ISASUGRA} employs two-loop renormalization group equations for 
scalar masses, whereas the other two codes use one-loop order for them.
This explanation seems unlikely because of the relative agreement observed in
the scalar masses, which ought to be more sensitive to this effect.

To summarize, with the current technology, we do not yet have the desired accuracy
for 
discrimination of supersymmetry breaking models or
measurement of their parameters from the sparticle spectrum.
We note that possible future linear colliders could determine some
sparticle masses at the per-mille level~\cite{Aguilar-Saavedra:2001rg}.
An increase in accuracy of the theoretical predictions of sparticle masses by
about a factor 10 will be necessary. 


\setcounter{figure}{0}
\setcounter{table}{0}
\setcounter{section}{0}
\setcounter{equation}{0}
\clearpage

\def\etal{{\it et al.}}
\def\abs#1{\left| #1\right|}
\def\sgn{\mathop{\rm sgn}}
\def\gtap{\raisebox{-.4ex}{\rlap{$\sim$}} \raisebox{.4ex}{$>$}}  
\def\etmiss{\slashchar{E}_T}
\def\fb{{\rm fb}}
\def\ltap{\raisebox{-.4ex}{\rlap{$\sim$}} \raisebox{.4ex}{$<$}}
\def\tG{{\tilde G}}
\def\ns{{\rm ns}}
\def\tell{{\tilde\ell}}
\def\ttau{{\tilde\tau}}
\def\fbi{{\rm fb}^{-1}}
\def\Meff{M_{\rm eff}}
\def\Msusy{M_{\rm SUSY}}
\def\lsp{{\tilde\chi_1^0}}
\def\ra{\rightarrow}
\def\GeV{{\rm GeV}}
\def\TeV{{\rm TeV}}
\def\mhalf{m_{1/2}}
\def\tchi{\tilde\chi}
\def\tg{\tilde g}
\def\tq{\tilde q}
\def\Cgrav{C_{\rm grav}}
\let\badcite=\cite
\def\cite{~\badcite}
\def\thefootnote{\fnsymbol{footnote}}
\def\Frac#1#2{{\displaystyle#1\over\displaystyle#2}}
\def\cmsec{{\rm cm}^{-2}{\rm s}^{-1}}
\def\hc{{\rm h.c.}}
\def\jet{{\rm jet}}
\def\jets{{\rm jets}}

\def\slashchar#1{\setbox0=\hbox{$#1$}           
   \dimen0=\wd0                                 
   \setbox1=\hbox{/} \dimen1=\wd1               
   \ifdim\dimen0>\dimen1                        
      \rlap{\hbox to \dimen0{\hfil/\hfil}}      
      #1                                        
   \else                                        
      \rlap{\hbox to \dimen1{\hfil$#1$\hfil}}   
      /                                         
   \fi}                                         %


\catcode`@=11
\newdimen\vbigd@men                             

\def\vbigl{\mathopen\vbig}
\def\vbigm{\mathrel\vbig}
\def\vbigr{\mathclose\vbig}

\def\vbig#1#2{{\vbigd@men=#2\divide\vbigd@men by 2%
   \hbox{$\left#1\vbox to \vbigd@men{}\right.\n@space$}}}
\catcode`@=12

\def\simge{
    \mathrel{\rlap{\raise 0.511ex
        \hbox{$>$}}{\lower 0.511ex \hbox{$\sim$}}}}
\def\simle{
    \mathrel{\rlap{\raise 0.511ex 
        \hbox{$<$}}{\lower 0.511ex \hbox{$\sim$}}}}

\catcode`@=11
\def\citenum#1{\csname b@#1\endcsname}
\catcode`@=12

\part{{\bf  High-Mass Supersymmetry with High Energy Hadron Colliders
} \\[0.5cm]\hspace*{0.8cm}
{\it I. Hinchliffe and F.E. Paige}}
\label{ian1sec}


\begin{abstract}
While it is natural for supersymmetric particles to be well within the
 mass range of the large hadron collider, it is possible that the
 sparticle masses could be very heavy.
Signatures are examined at a very high energy hadron collider and a
very high luminosity option for the Large Hadron Collider 
 in such scenarios.
\end{abstract}


\section{Introduction}

If supersymmetry is connected to the hierarchy problem, it is
expected \cite{Anderson:1995dz,Barbieri:1988fn} 
that sparticles will be sufficiently light that at least
some of them will be observable at the Large Hadron Collider (LHC) or
even at the Tevatron. 
However it is not possible to set a rigorous bound on the sparticle
masses.  As the
sparticle masses rise, the fine tuning problem of the standard model
reappears, but the sparticle masses become large enough so that they are
difficult to observe at LHC.

\begin{table}[htb]
\caption{Benchmark SUGRA points and masses from Ref.~\cite{Battaglia:2001zp}
\label{points}}
\medskip
\renewcommand{\arraystretch}{0.95}
\begin{centering}
\small
\begin{tabular}{|c||r|r|r|r|r|r|r|r|r|r|r|r|r|}
\hline
Model          & A   &  B  &  C  &  D  &  E  &  F  &  G  &  H  &  I  &  J  &  K  &  L  &  M   \\ 
\hline
$m_{1/2}$      & 600 & 250 & 400 & 525 &  300& 1000& 375 & 1500& 350 & 750 & 1150& 450 & 1900 \\
$m_0$          & 140 & 100 &  90 & 125 & 1500& 3450& 120 & 419 & 180 & 300 & 1000& 350 & 1500 \\
$\tan{\beta}$  & 5   & 10  & 10  & 10  & 10  & 10  & 20  & 20  & 35  & 35  & 35  & 50  & 50   \\
sign($\mu$)    & $+$ & $+$ & $+$ & $-$ & $+$ & $+$ & $+$ & $+$ & $+$ & $+$ & $-$ & $+$ & $+$  \\ 
$\alpha_s(m_Z)$& 120 & 123 & 121 & 121 & 123 & 120 & 122 & 117 & 122 & 119 & 117 & 121 & 116 \\
$m_t$          & 175 & 175 & 175 & 175 & 171 & 171 & 175 & 175 & 175 & 175 & 175 & 175 & 175  \\ \hline
Masses         &     &     &     &     &     &     &     &     &     &     &     &     &      \\ \hline
h$^0$          & 114 & 112 & 115 & 115 & 112 & 115 & 116 & 121 & 116 & 120 & 118 & 118 &  123 \\
H$^0$          & 884 & 382 & 577 & 737 &1509 &3495 & 520 &1794 & 449 & 876 &1071 & 491 & 1732 \\
A$^0$          & 883 & 381 & 576 & 736 &1509 &3495 & 520 &1794 & 449 & 876 &1071 & 491 & 1732 \\
H$^{\pm}$      & 887 & 389 & 582 & 741 &1511 &3496 & 526 &1796 & 457 & 880 &1075 & 499 & 1734 \\ \hline
$\chi^0_1$     & 252 &  98 & 164 & 221 & 119 & 434 & 153 & 664 & 143 & 321 & 506 & 188 &  855 \\
$\chi^0_2$     & 482 & 182 & 310 & 425 & 199 & 546 & 291 &1274 & 271 & 617 & 976 & 360 & 1648 \\
$\chi^0_3$     & 759 & 345 & 517 & 654 & 255 & 548 & 486 &1585 & 462 & 890 &1270 & 585 & 2032 \\
$\chi^0_4$     & 774 & 364 & 533 & 661 & 318 & 887 & 501 &1595 & 476 & 900 &1278 & 597 & 2036 \\
$\chi^{\pm}_1$ & 482 & 181 & 310 & 425 & 194 & 537 & 291 &1274 & 271 & 617 & 976 & 360 & 1648 \\
$\chi^{\pm}_2$ & 774 & 365 & 533 & 663 & 318 & 888 & 502 &1596 & 478 & 901 &1279 & 598 & 2036 \\ \hline
$\tilde{g}$    &1299 & 582 & 893 &1148 & 697 &2108 & 843 &3026 & 792 &1593 &2363 & 994 & 3768 \\ \hline
$e_L$, $\mu_L$ & 431 & 204 & 290 & 379 &1514 &3512 & 286 &1077 & 302 & 587 &1257 & 466 & 1949 \\
$e_R$, $\mu_R$ & 271 & 145 & 182 & 239 &1505 &3471 & 192 & 705 & 228 & 415 &1091 & 392 & 1661 \\
$\nu_e$, $\nu_{\mu}$
               & 424 & 188 & 279 & 371 &1512 &3511 & 275 &1074 & 292 & 582 &1255 & 459 & 1947 \\
$\tau_1$       & 269 & 137 & 175 & 233 &1492 &3443 & 166 & 664 & 159 & 334 & 951 & 242 & 1198 \\
$\tau_2$       & 431 & 208 & 292 & 380 &1508 &3498 & 292 &1067 & 313 & 579 &1206 & 447 & 1778 \\
$\nu_{\tau}$   & 424 & 187 & 279 & 370 &1506 &3497 & 271 &1062 & 280 & 561 &1199 & 417 & 1772 \\ \hline
$u_L$, $c_L$   &1199 & 547 & 828 &1061 &1615 &3906 & 787 &2771 & 752 &1486 &2360 & 978 & 3703 \\
$u_R$, $c_R$   &1148 & 528 & 797 &1019 &1606 &3864 & 757 &2637 & 724 &1422 &2267 & 943 & 3544 \\
$d_L$, $s_L$   &1202 & 553 & 832 &1064 &1617 &3906 & 791 &2772 & 756 &1488 &2361 & 981 & 3704 \\
$d_R$, $s_R$   &1141 & 527 & 793 &1014 &1606 &3858 & 754 &2617 & 721 &1413 &2254 & 939 & 3521 \\
$t_1$          & 893 & 392 & 612 & 804 &1029 &2574 & 582 &2117 & 550 &1122 &1739 & 714 & 2742 \\
$t_2$          &1141 & 571 & 813 &1010 &1363 &3326 & 771 &2545 & 728 &1363 &2017 & 894 & 3196 \\ 
$b_1$          &1098 & 501 & 759 & 973 &1354 &3319 & 711 &2522 & 656 &1316 &1960 & 821 & 3156 \\
$b_2$          &1141 & 528 & 792 &1009 &1594 &3832 & 750 &2580 & 708 &1368 &2026 & 887 & 3216 \\ 
\hline
\end{tabular}
\end{centering}
\end{table}

It is also possible that SUSY is the solution to the dark matter
problem \cite{Ellis:1984ew,Gabutti:1996qd,Baer:1996nc},  the stable,
lightest supersymmetric particle (LSP) being the particle that
pervades the universe. This constraint can be applied to the minimal
SUGRA \cite{Alvarez-Gaume:1983gj,Ibanez:1982ee,Ellis:1983wr,Inoue:1982pi,Chamseddine:1982jx} 
model and used to constrain the masses of the other sparticles.
Recently sets of parameters in  the minimal SUGRA 
model have been proposed~\cite{Battaglia:2001zp}
that satisfy existing constraints, including the dark matter
constraint and the one from the precise measurement of the
anomalous magnetic
moment of the muon~\cite{Brown:2001mg},  but do not impose any fine tuning
requirements. This set of points is not a random sampling of the  available
parameter space but is rather intended to illustrate the possible
experimental consequences. These points and their mass spectra are shown in
Table~\ref{points}. Most of the allowed parameter space corresponds to
cases for which the sparticles have masses less than 1 TeV or so and
is accessible to LHC. Indeed 
some of these points are quite similar to ones studied in earlier LHC
simulations \cite{Abdullin:1998pm,AtlasPhysTDR}. 
Points A, B, C, D, E, G, J and L fall into this category.
As the masses of the sparticles are increased, the LSP contribution to
dark matter rises and typically violates the experimental
constraints. However there are certain regions of parameter space
where the annihilation rates for the LSP can be increased and the
relic density of LSP's lowered sufficiently. In these
narrow regions, the sparticle masses can be much larger. Points F, K, H
and M illustrate these regions. This paper considers Point K, H and M 
at the LHC with a luminosity upgrade to $1000\,\fbi$ per year (SLHC)
and at a possible higher energy hadron collider (VLHC). 
We assume an energy of $40\,\TeV$ for the VLHC
and use the identical analysis for both machines. 
Point F has  similar phenomenology to Point K except that the
squark and slepton masses are much larger and consequently more
difficult to observe.
For the purposes of this simulation, the detector
performance at $10^{35}\,\cmsec$ and at the VLHC 
is assumed to be the same as that of ATLAS for the LHC design 
luminosity. In particular, the additional pileup present at higher
luminosity is taken into account only by raising some of the
cuts.  Isajet~7.54\cite{ISAJET,Baer:1999sp}  is used for the event
generation. Backgrounds from $t\overline{t}$, gauge boson pairs, large
$p_T$ gauge boson production and QCD jets are included.

\section{Point K}

\begin{figure}[tb]
\includegraphics[width=3.0in]{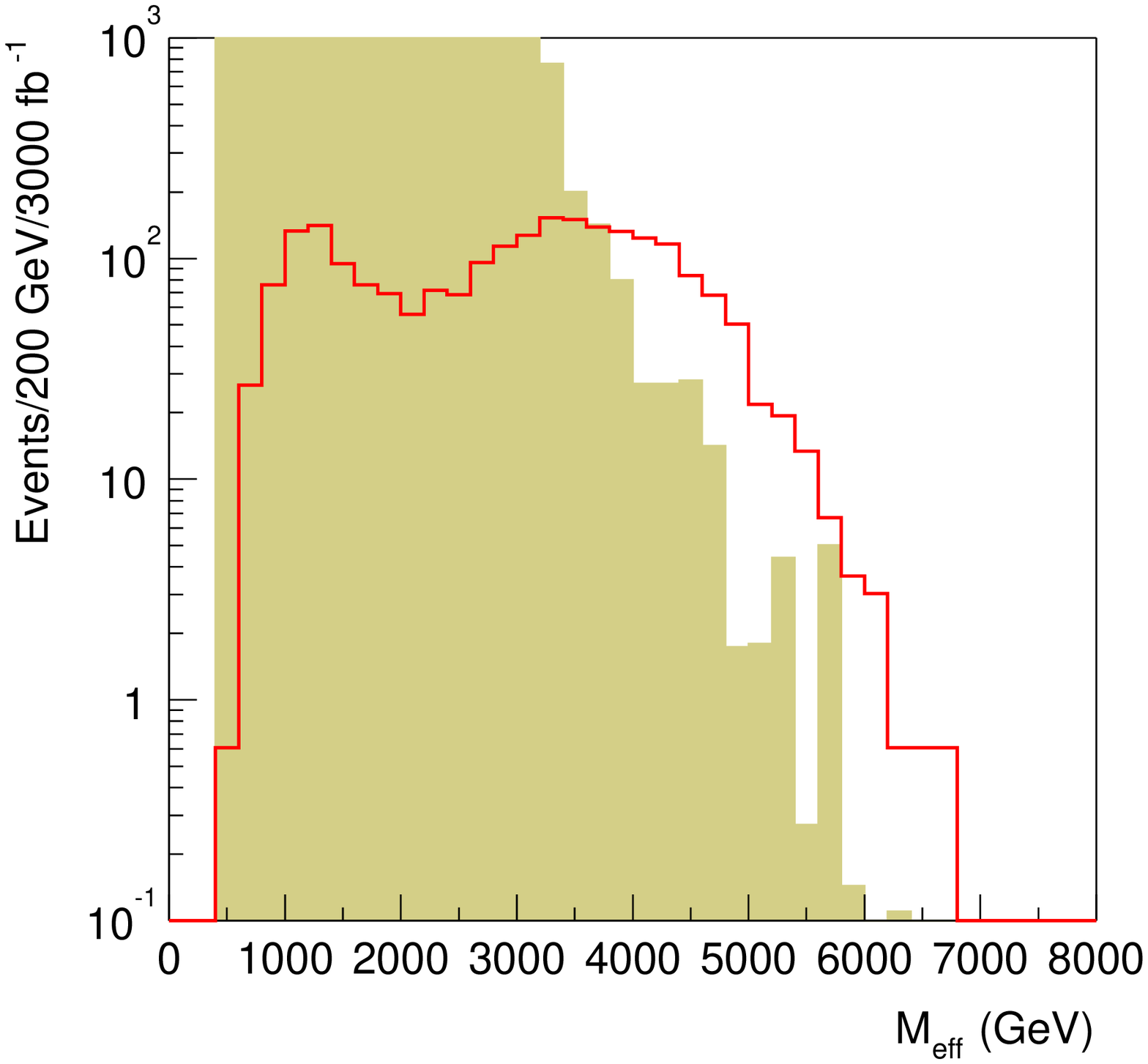}
\includegraphics[width=3.0in]{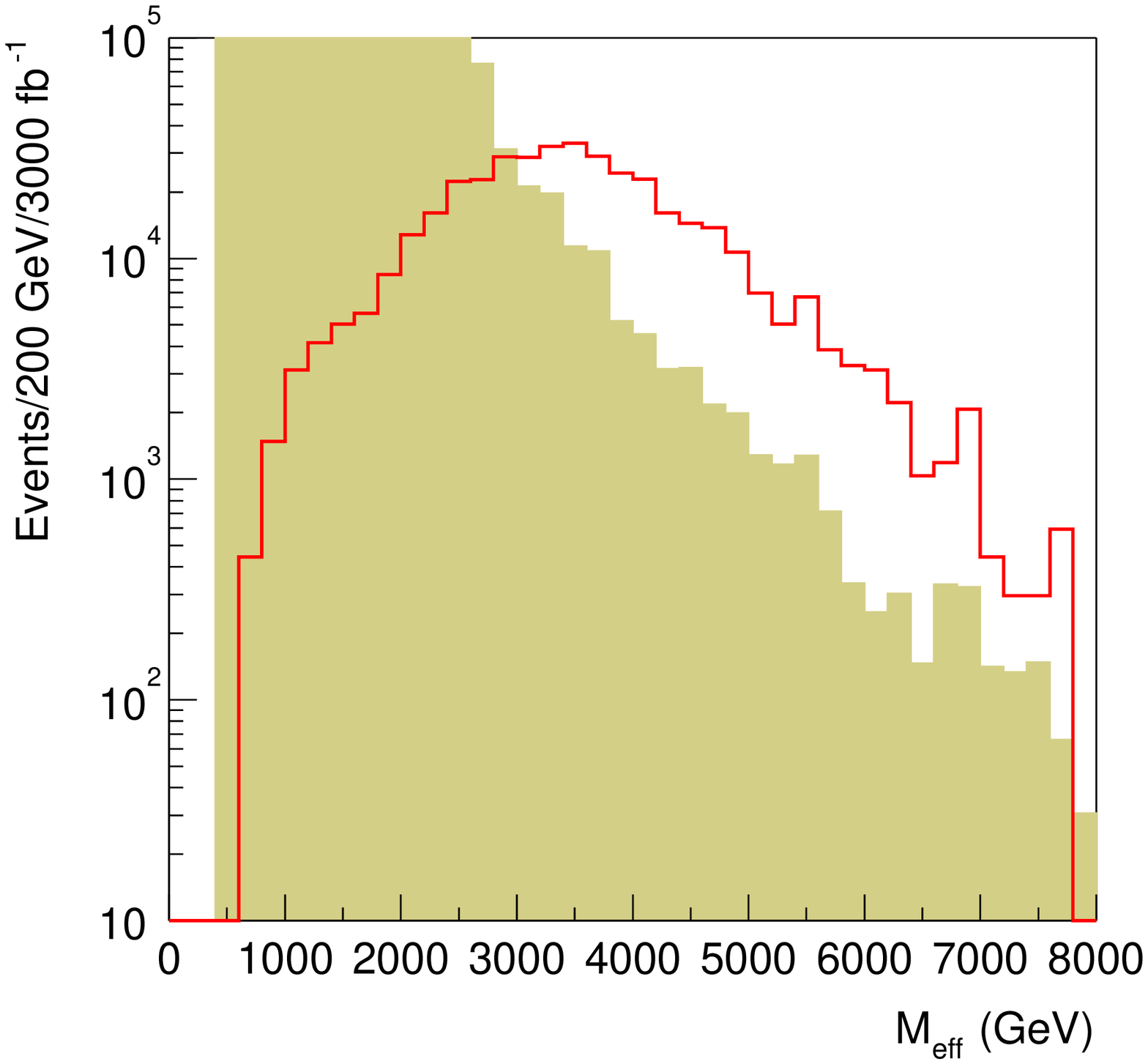}
\caption{$\Meff$ distribution for SLHC (left) and VLHC (right) for Point
K. Solid: signal. Shaded: SM background. \label{pointkmeff}}
\end{figure}

Point K has $M_A \approx 2M_\lsp$ and gluino and squark masses above
$2\,\TeV$. The strong production is dominated by valance squarks,
which have the characteristic decays
$\tq_L \to \tchi_1^\pm q, \tchi_2^0 q$ and $\tilde q_R \to \lsp q$.
The signal can be observed in the inclusive effective mass
distribution. 
Events are selected with hadronic jets and missing $E_T$, and the
following scalar quantity is formed:
$$ 
M_{eff} = \etmiss + \sum_{jets}E_{T,jet} +\sum_{leptons}E_{T,lepton}
$$
where the sum runs over all jets with $E_T> 50 $ GeV and
$\abs{\eta}<5.0$ and isolated leptons with  $E_T> 15 $ GeV and$\abs{\eta}<2.5$ . 
The following further selection was then made: events were selected 
with at least two jets
with $p_T > 0.1\Meff$, $\etmiss > 0.3\Meff$, $\Delta\phi(j_0,\etmiss) <
\pi-0.2$, and $\Delta\phi(j_0,j_1)<2\pi/3$. These cuts help to
optimize the signal to background ratio.
The distributions in  $ M_{eff}$ for signal and
background are shown in Figure~\ref{pointkmeff}. It can be seen that the
signal emerges from the background at large values of $M_{eff}$.
 The LHC  with 3000 $\fbi$ of integrated luminosity has a
signal of 510 events on a background of 108 for
$\Meff>4000\,\GeV$. These rates are sufficiently large so that a
discovery could be made with the standard integrated luminosity of 300
$\fbi$. However the limited data samples available will restrict
detailed studies.
\begin{figure}[h!t]
\includegraphics[width=3.0in]{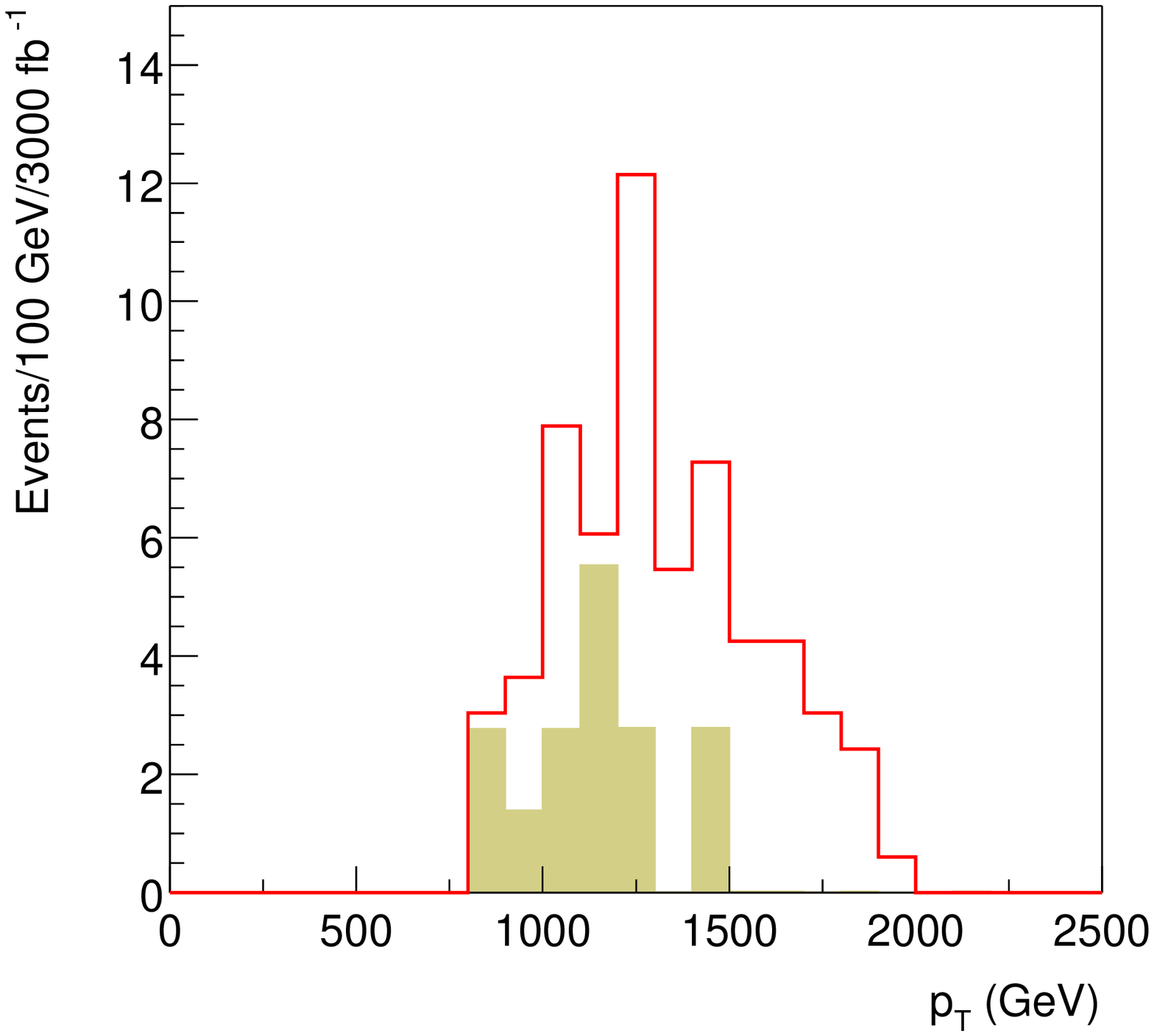}
\includegraphics[width=3.0in]{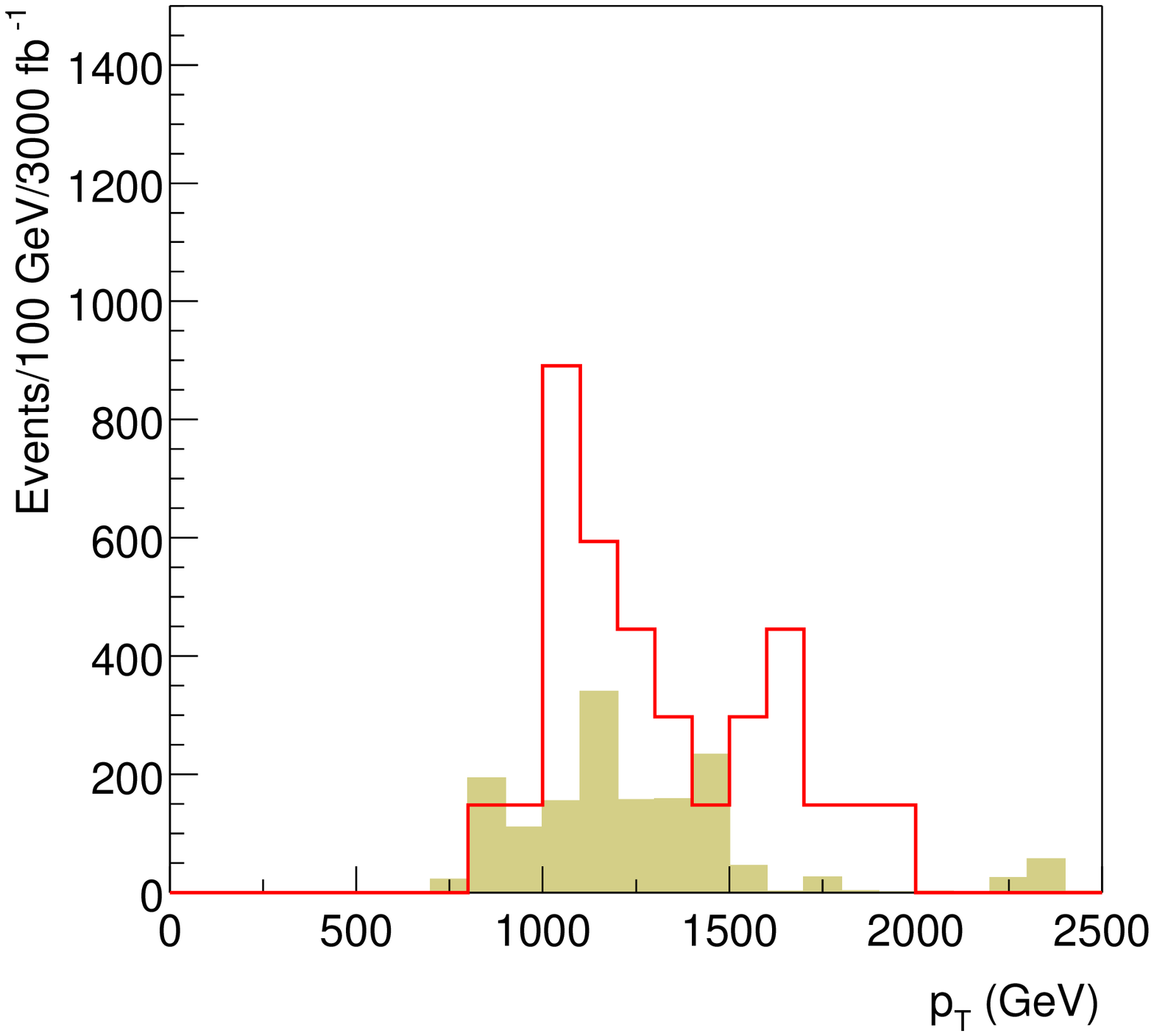}
\caption{$p_T$ distribution of hardest jet in $2\jet+\etmiss$ events for
SLHC (left) and VLHC (right) for Point K. \label{pointkptfroid}}
\end{figure}
\begin{figure}[h!t]
\includegraphics[width=3.0in]{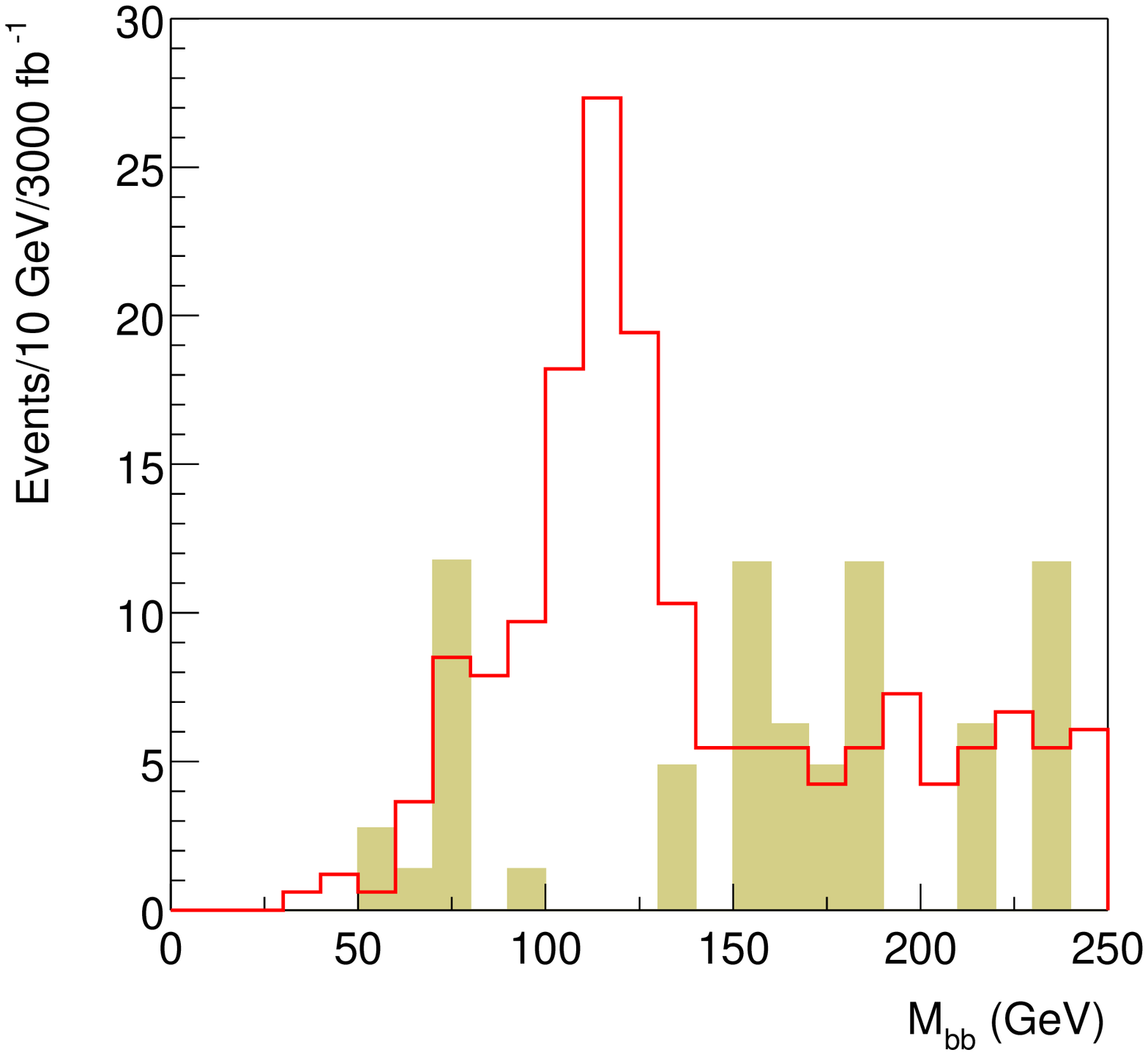}
\includegraphics[width=3.0in]{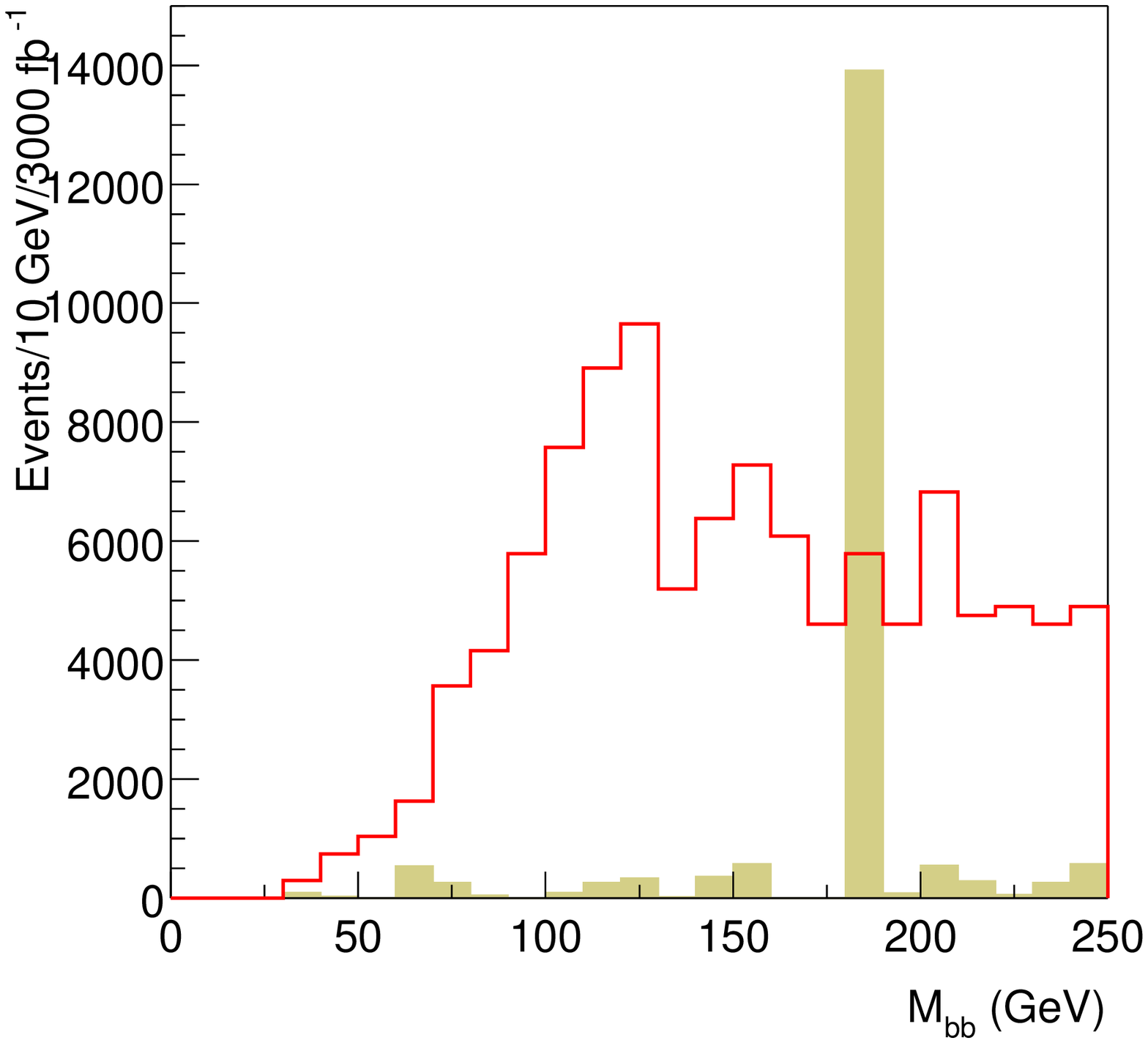}
\caption{$M_{bb}$ distribution for SLHC (left) and VLHC (right) for Point
K. \label{pointkmbb}}
\end{figure}

Production of $\tq_R\tq_R$ followed by the decay of each squark to
$q\lsp$ gives a dijet signal accompanied by missing $E_T$. 
In order to  extract this from the standard model  background, 
hard cuts on the jets and $\etmiss$ are needed.
Events were required
to have two jets with $p_T>700\,\GeV$, $\etmiss>600\,\GeV$, and
$\Delta\phi(j_1,j_2)<0.8\pi$. The resulting distributions are shown in
Figure~\ref{pointkptfroid}. Only a few events survive at the LHC with
3000 $\fbi$. The transverse momentum of the hardest jet is 
 sensitive to the
$\tq_R$ mass\cite{AtlasPhysTDR}. The mass determination will be
limited by the available statistics.

The decay $\tchi_2^0 \to \lsp h$ is dominant so we should expect to
see Higgs particles in the decay of $\tq_L$ ($\tq_L\to \tchi_2^0 q \to
\lsp h q $). The Higgs signal can  be observed as a peak in the
$b\overline{b}$ mass distributions. In order to do this, it is
essential that $b-$jets can be tagged with good efficiency and
excellent 
rejection against light quark jets. 
There is a large
background from $t \bar t$ that must be overcome using topological cuts.
 Events were selected to have at least three
jets with $p_T > 600,300,100\,\GeV$, $\etmiss>400\,\GeV$,
$\Meff>2500\,\GeV$, $\Delta\phi(j_1,\etmiss)<0.9\pi$, and
$\Delta\phi(j_1,j_2)<0.6\pi$. The distributions are shown in
Figure~\ref{pointkmbb} assuming the same $b$-tagging performance as for
standard luminosity, {\it i.e.,}  that shown in  Figure~9-31 of
Ref.~\cite{AtlasPhysTDR} which corresponds to an efficiency of 60\%
and a rejection factor against light quark jets of $\sim 100$.
This $b-$tagging performance may be optimistic in the very high
luminosity environment. However our event 
selection is only $\sim 10\%$ efficient at SLHC and might be improved.
There is much less standard model background at VLHC. However,  there is
significant SUSY background from  $\tg \to \tilde b_i \bar b, \tilde
t_1 \bar t$ which becomes more important at the higher energy.  
At the VLHC and possibly a the SLHC, it should be
possible to extract information on the mass of $\tq_L$ by combining
the Higgs with a jet and probing the decay chain $\tq_L \to  \tchi_2^0
q \to q h \lsp $ (see {e.g.} \cite{Hinchliffe:1997iu}).

\section{Point M}

Point M has squark and gluino masses around 3.5 TeV and is beyond the
reach of the standard LHC. Only 375 SUSY events of
all types are produced for $1000\,\fbi$ at LHC, mainly valence squarks
($\tilde u_L, \tilde d_L, \tilde u_R, \tilde d_R$) and gauginos
($\tchi_1^\pm,\tchi_2^0$). The VLHC cross section is a factor of 200
larger. About half of the SLHC SUSY events are from electro weak 
gaugino pair production mostly $\tchi_2^0$ and $\tchi_1^\pm$ . The
dominant decays of these are  $\tchi_2^0 \to \lsp h$ and $\tchi_1^\pm \to \lsp
W^\pm$. Rates are so small that no signal
 close to the Standard Model  backgrounds could be  found for the
SLHC.
\begin{figure}[tb]
\includegraphics[width=3.0in]{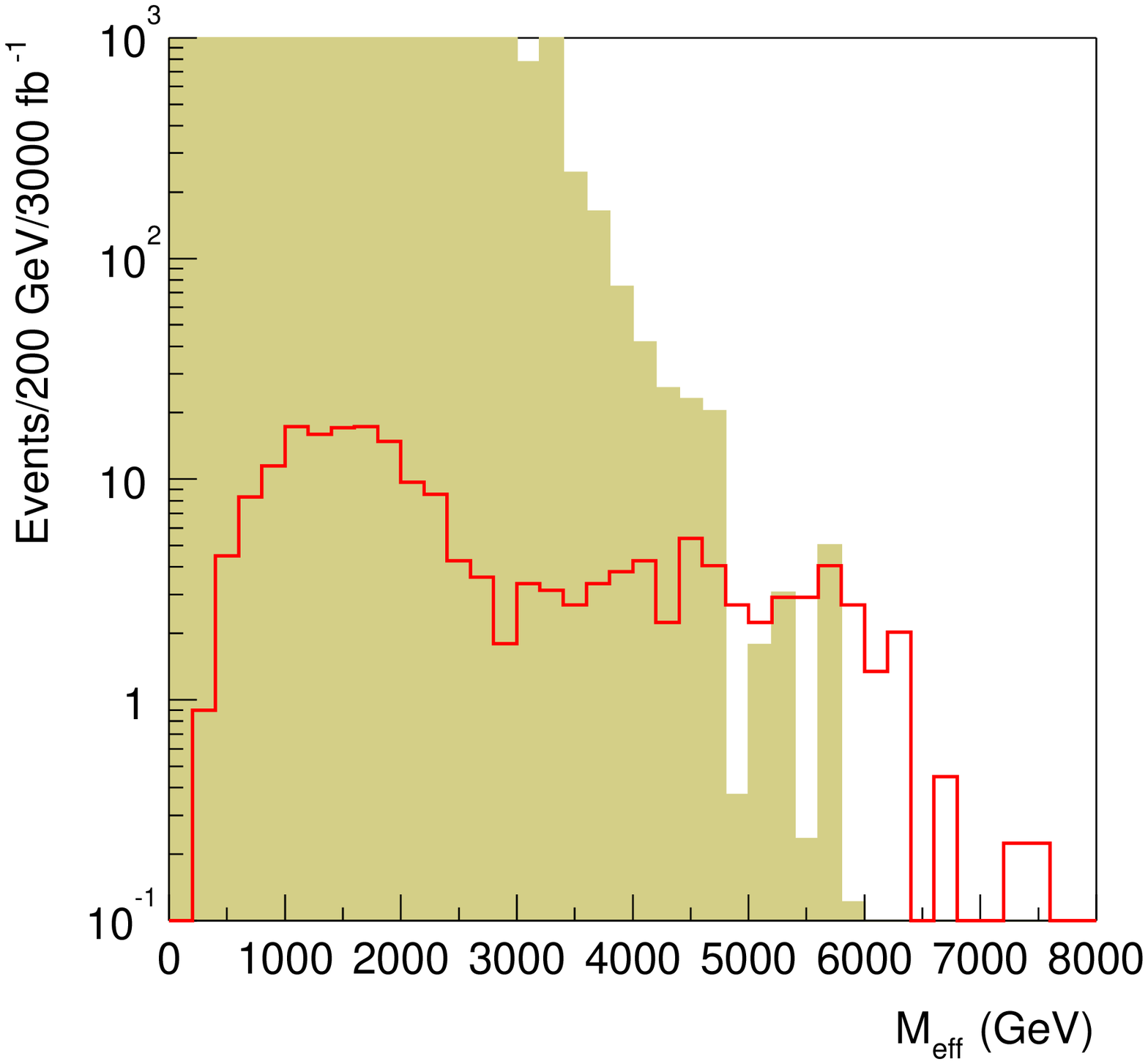}
\includegraphics[width=3.0in]{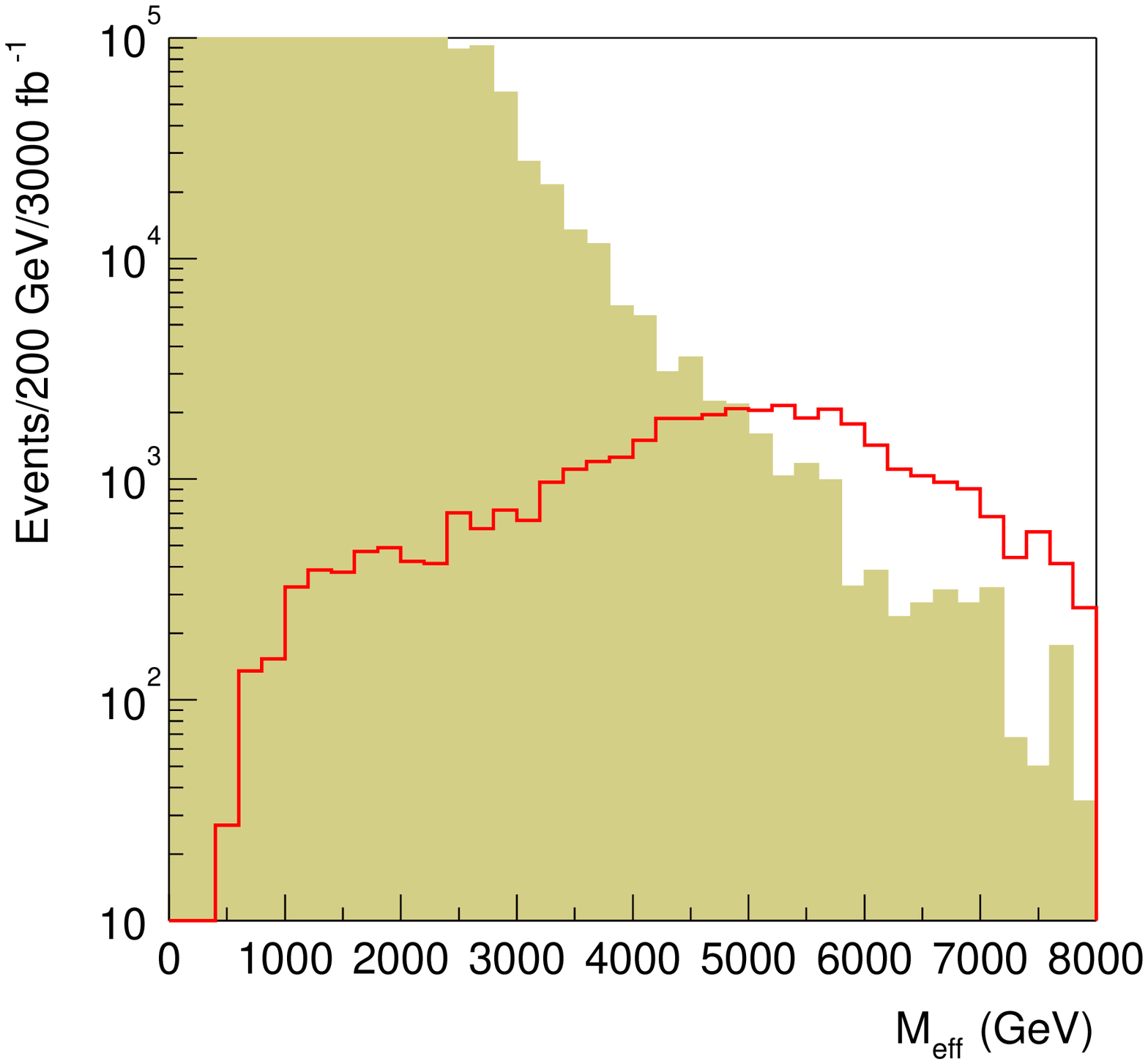}
\caption{$\Meff$ distribution for SLHC (left) and VLHC (right) for
Point M. \label{pointmmeff}}
\end{figure}
\begin{figure}[h!t]
\includegraphics[width=3.0in]{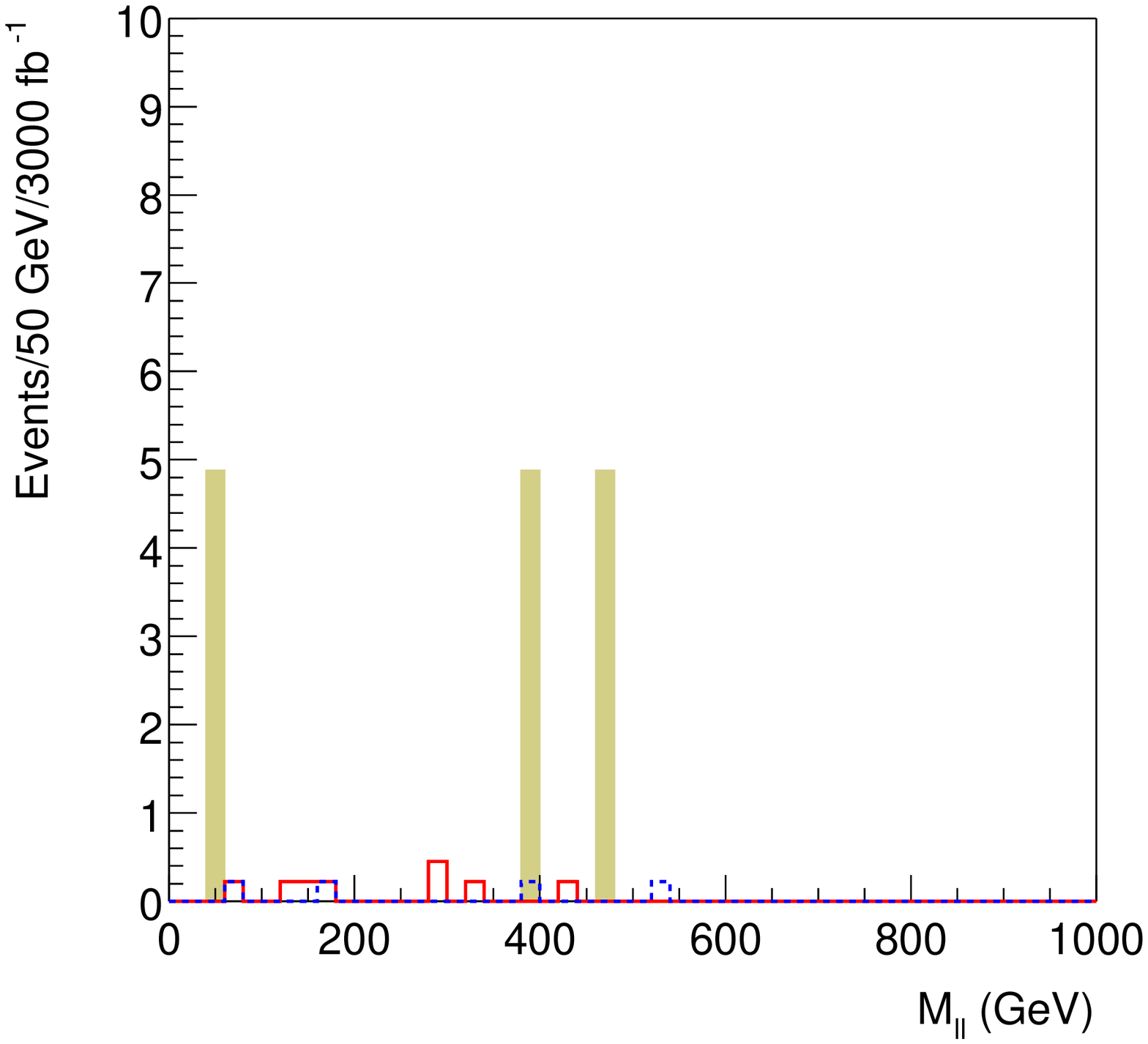}
\includegraphics[width=3.0in]{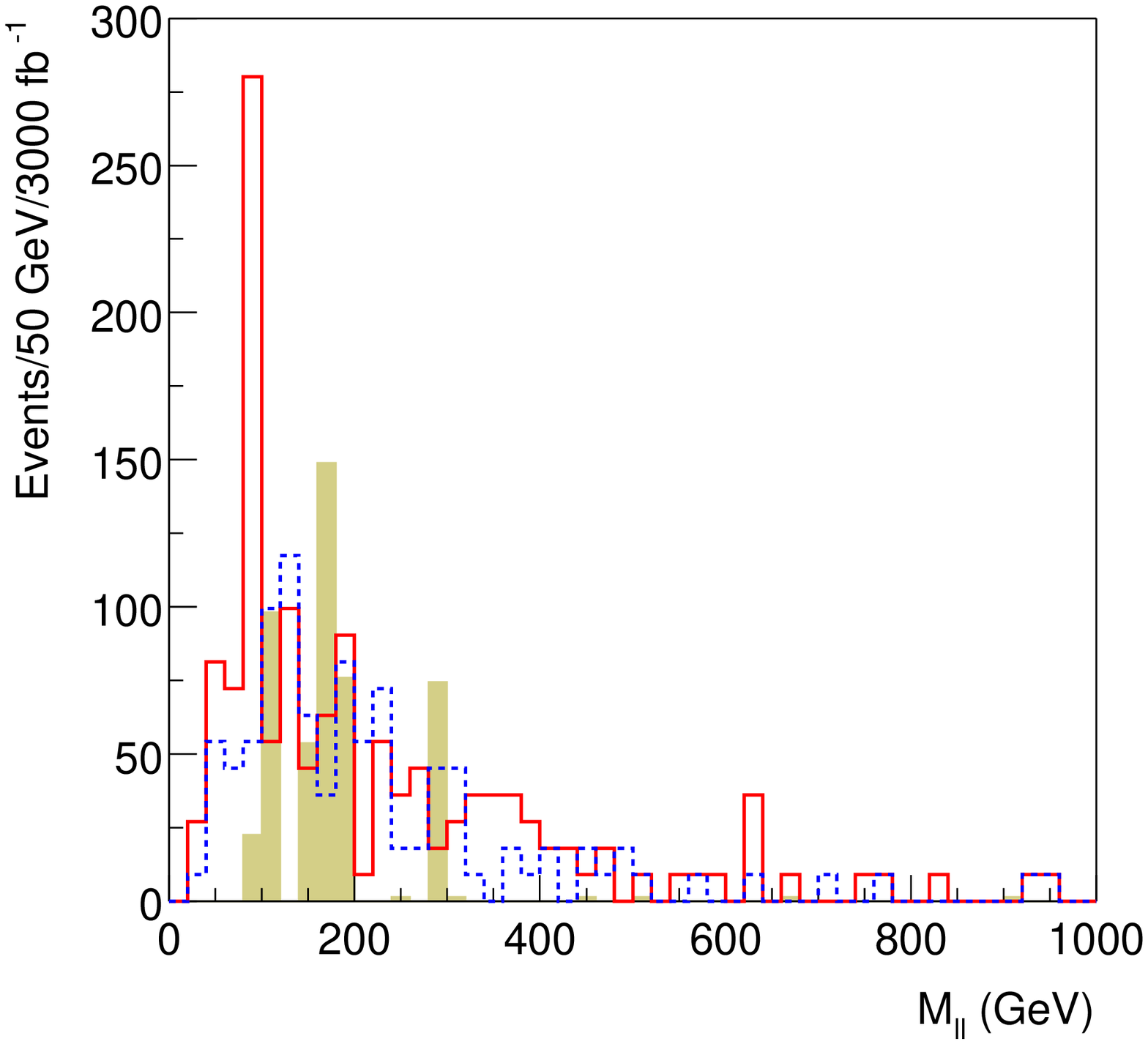}
\caption{Dilepton mass distribution for SLHC (left) and VLHC (right) for
Point M. Solid: $\ell^+\ell^-$. Dashed: $e^\pm\mu^\mp$.
\label{pointmmll}}
\end{figure}

The effective mass distributions for Point M at SLHC and VLHC are shown
in Figure~\ref{pointmmeff} using the same cuts as for Point K. As
expected, the SLHC
signal is very marginal: there are only 20 signal events with 10
background events for $\Meff > 5000\,\GeV$ and $3000 \fbi$. 
Several attempts to optimize
the cuts did not give any improvement. Requiring a lepton, a hadronic
$\tau$, or a tagged $b$ jet  did not help.  We are forced to conclude
that it is unlikely that a signal of any type could be observed.
The VLHC signal is
clearly visible and could be further optimized. 

The dilepton rates are shown in Fig~\ref{pointmmll}. Events are
selected that have $M_{eff}>3000 \GeV$ $\etmiss>0.2M_{eff}$ and two
isolated leptons with $p_T>15 \GeV$ and the mass distribution of the
dilepton pair is shown.  As expected, nothing is visible
at SLHC. The distribution at VLHC is dominated by two independent
decays ({\it e.g.} $\tchi_1^\pm \tchi_1^\mp \to \lsp W^\pm \lsp
W^\mp$),  so that $e^+e^-+\mu^+\mu^-$ and $e^\pm\mu^\mp$ rates are almost
identical except for the $Z$ peak in the former which arises mainly from $\tq
\to q \tchi_2^\pm \to q \tchi_1^\pm Z$.

On the basis of this preliminary, 
 study it seems unlikely that Point M can
be detected at $14\,\TeV$ even with $3000\,\fbi$. Higher energy
would be required.

\section{Point H}

Point H is
able to accommodate very heavy  sparticles  without
overclosing the universe as the destruction rate for the $\lsp$
 is enhanced  by coannihilation  with a stau. This implies a very
small splitting between the $\ttau_1$ and the $\lsp$. In this
particular case,  $\ttau_1
\not\to \lsp \tau$, so it must decay by second order weak processes,
$\ttau_1 \to \lsp e\bar\nu_e\nu_\tau$, giving it a long lifetime. The
dominant SUSY rates arise from the  strong production of  valance squarks, with
$\tq_L \to \tchi_1^\pm q, \tchi_2^0 q$ and $\tilde q_R \to \lsp q$. The
staus which are produced from cascade decays of these squarks, then
exit the detector with a signal similar to a ``heavy muon''.

\begin{figure}[tb]
\includegraphics[width=3.0in]{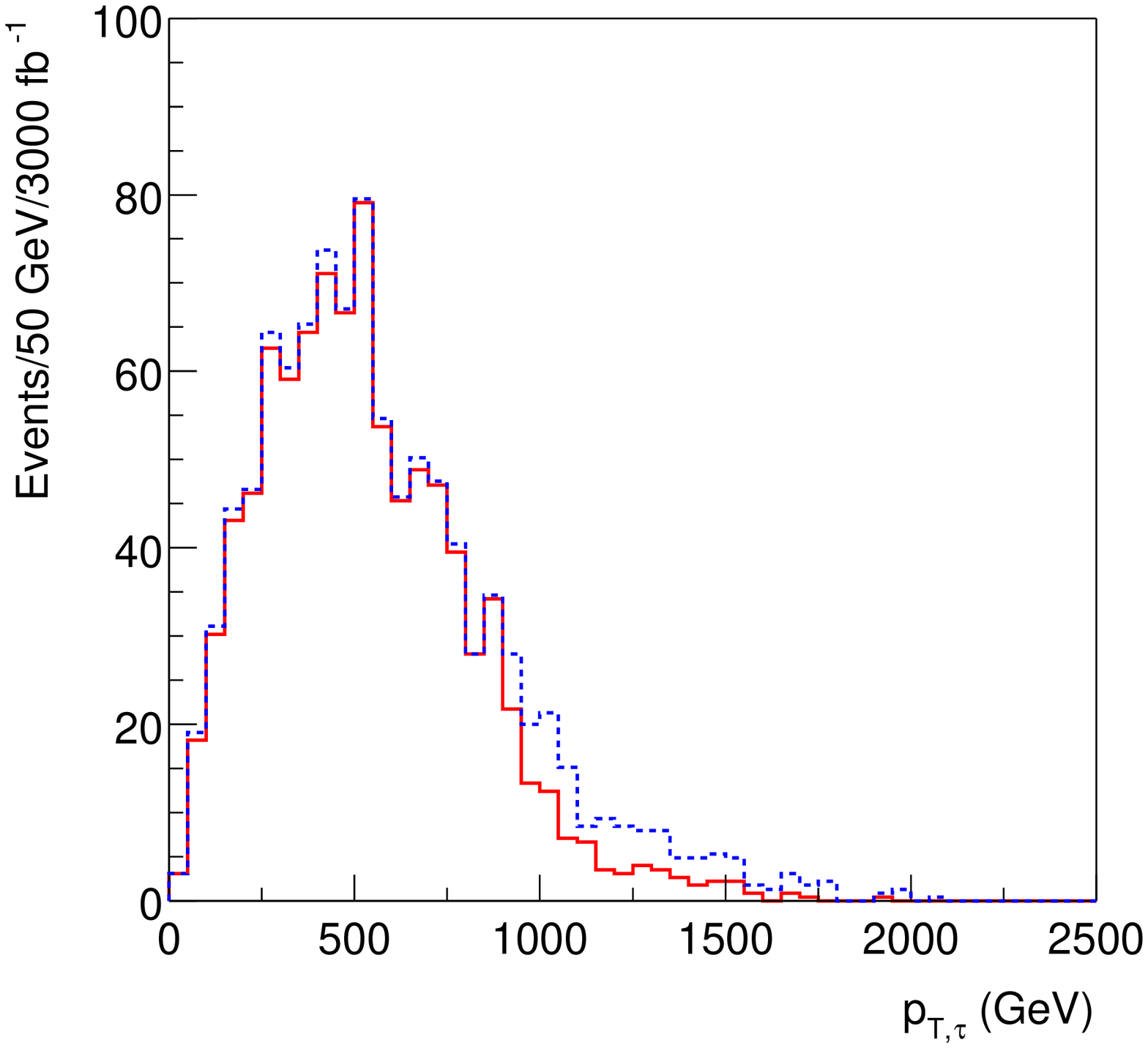}
\includegraphics[width=3.0in]{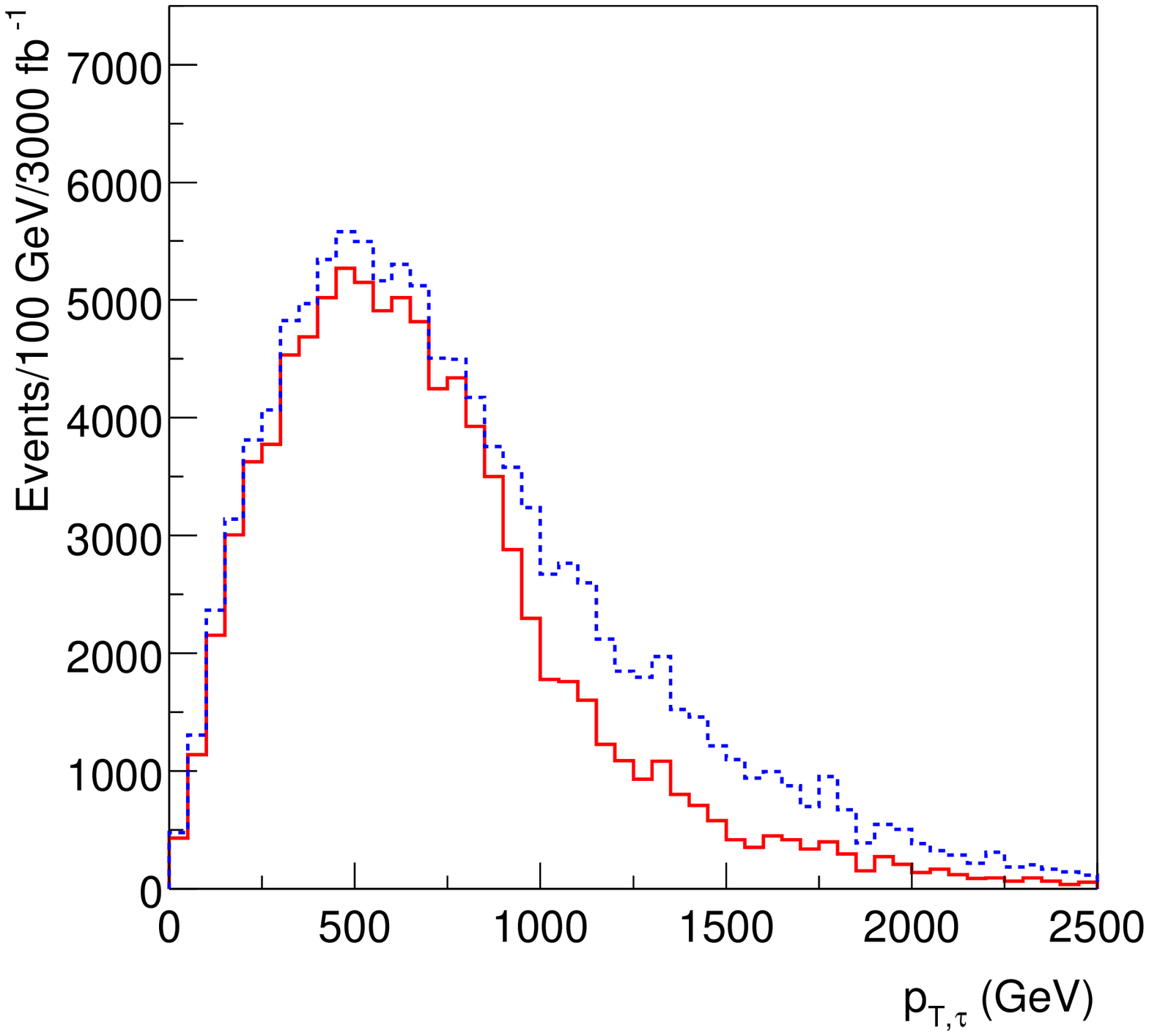}
\caption{$p_T$ distribution of $\ttau_1$ at SLHC (left) and VLHC (right)
for Point H. Dashed: all $\ttau_1$. Solid: $\ttau_1$ with $\Delta t >
7{\,\rm ns}$ \label{pointhptstau}}
\end{figure}
\begin{figure}[h!t]
\includegraphics[width=3.0in]{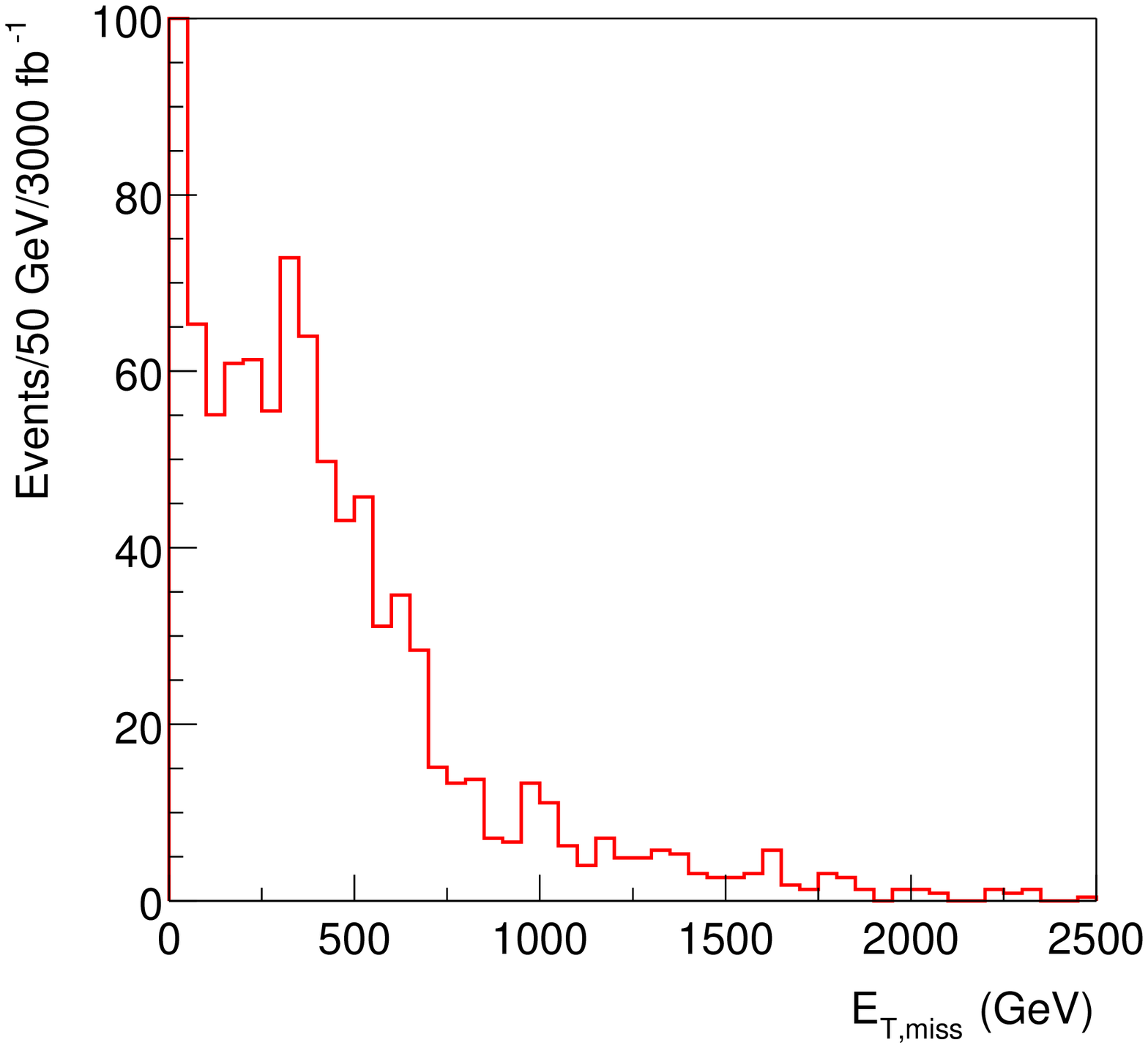}
\includegraphics[width=3.0in]{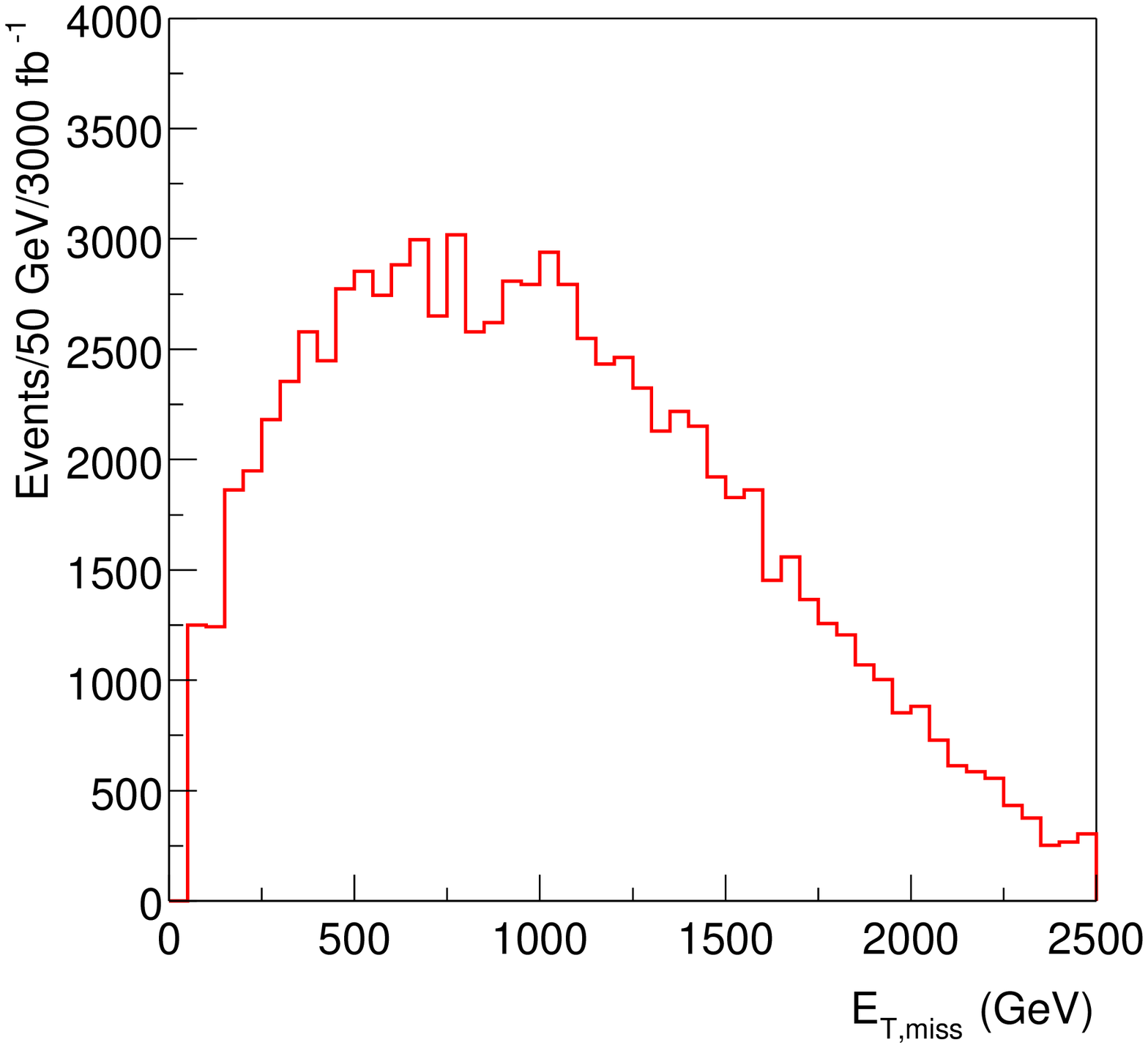}
\caption{Calorimetric $\etmiss$ distributions in $\ttau_1$ events for
SLHC (left) and VLHC (right) for Point H. \label{pointhetxstau}}
\end{figure}
The $p_T$ spectrum of these quasi-stable $\ttau_1$ for $1000\,\fbi$ is
shown in Figure~\ref{pointhptstau}. The ATLAS muon system
\cite{AtlasPhysTDR} has a time
resolution of about 0.7~ns for time of flight over a cylinder of radius
10~m and half-length 20~m. The spectrum with a time delay $\Delta t >
10\sigma(7\,{\rm ns})$ is also shown. Notice that this signal could be
observed at
the LHC with $\sim 300\,\fbi$.
Triggering on a slow $\ttau_1$ may be a problem since the time-window  for the
trigger chambers is limited. 
However, the $\etmiss$ in SUSY events as measured by the calorimeter
is quite large as shown in Figure~\ref{pointhetxstau}.
 It probably is possible to trigger just on
jets plus $\etmiss$, the distribution for which is shown in
Figure~\ref{pointhetxstau}. The mass of the stable stau can be measured
by exploiting the time of flight measurements in the muon measurement
system.
Studies of such quasi stable particles at somewhat smaller masses
carried out at the ATLAS detector showed a mass resolution of
approximately 3\% given sufficient statistics (see Section 20.3.4.2 of
Ref~\cite{AtlasPhysTDR}).
A precision of this order should be achievable with 3000 $\fbi$ at
either the LHC or VLHC. One can then build on the stable stau to
reconstruct the decay chain using techniques similar to those used for
the GMSB studies \cite{AtlasPhysTDR,Hinchliffe:1998ys}. This is
not pursued here.

\begin{figure}[tb]
\includegraphics[width=3.0in]{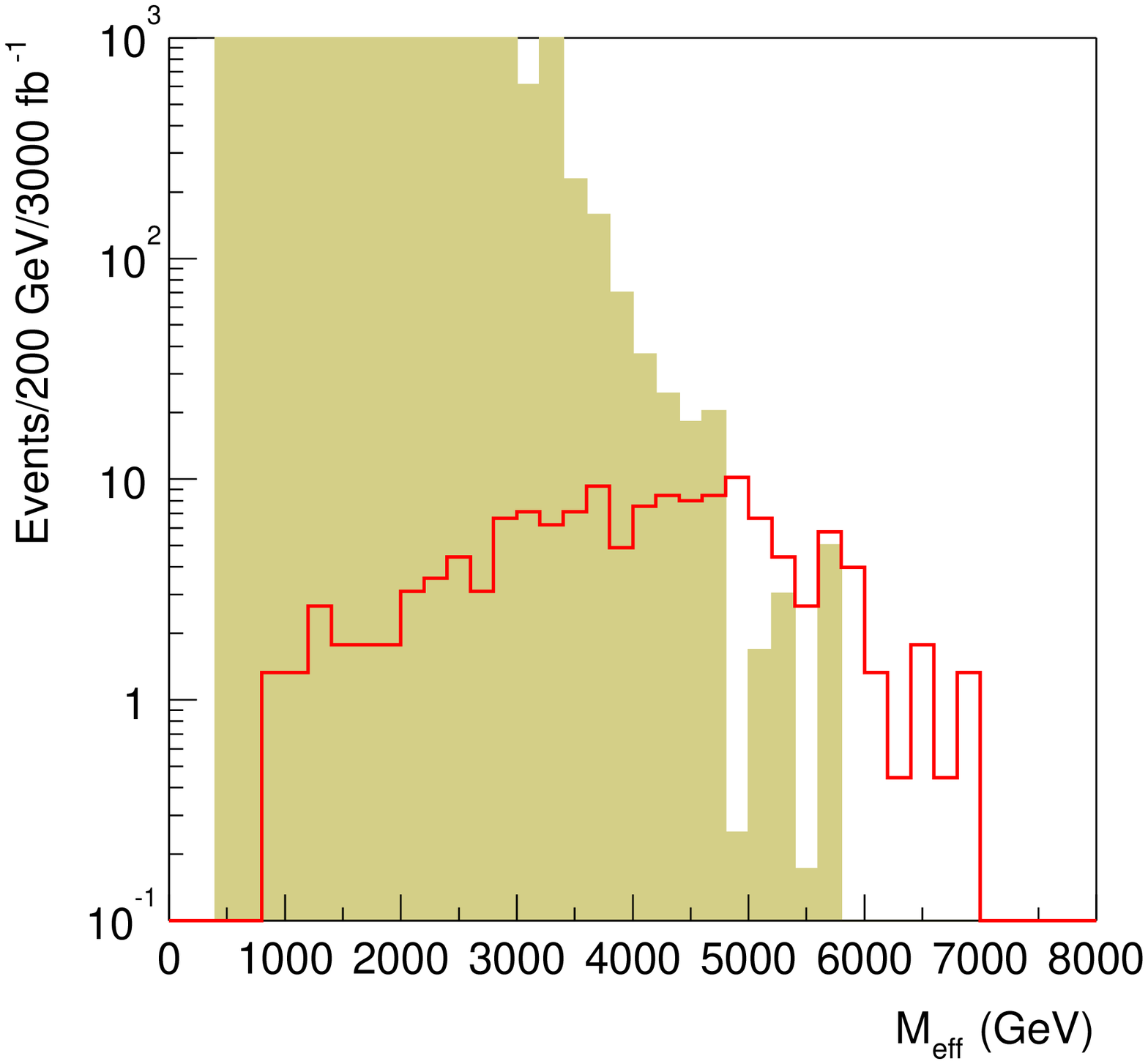}
\includegraphics[width=3.0in]{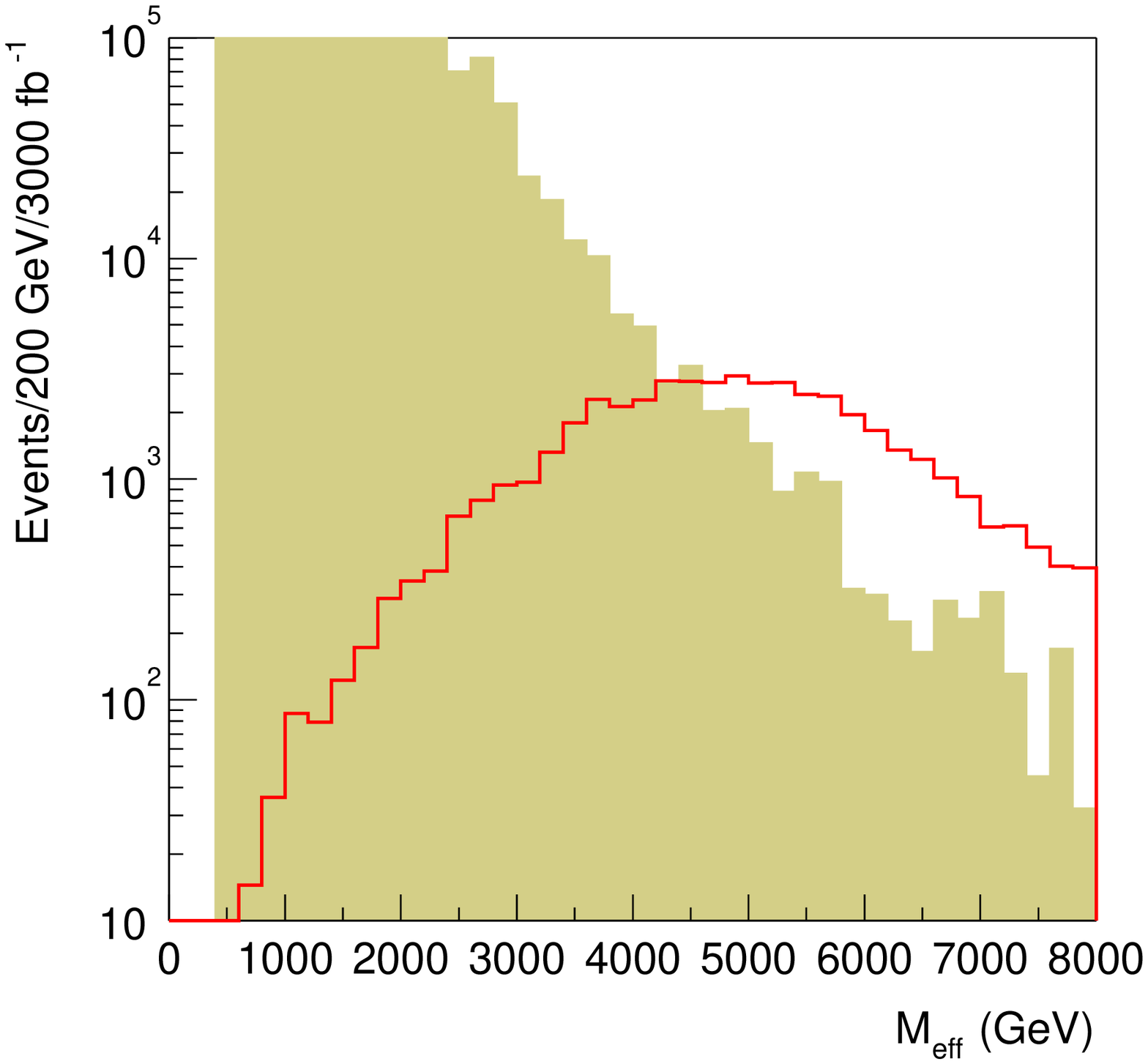}
\caption{$\Meff$ distribution for SLHC (left) and VLHC (right) for Point
H. Solid: signal. Shaded: SM background. \label{pointhmeff}}
\end{figure}

The stable $\tau_1$ signature is somewhat exceptional so we explore
other signatures that do not require it and would be present if the
stau decayed inside the detector. For such high
masses the strong production is mainly of $\tilde u$ and $\tilde d$.
Events are selected with hadronic jets and missing $E_T$ and the
effective mass formed as in the case of Point K.
 To optimize this signature, events were further 
selected with at least two jets
with $p_T > 0.1\Meff$, $\etmiss > 0.3\Meff$, $\Delta\phi(j_0,\etmiss) <
\pi-0.2$, and $\Delta\phi(j_0,j_1)<2\pi/3$. The $\Meff$ distributions
after these cuts for the SLHC and the VLHC are shown in
Fig~\ref{pointhmeff}. 
Note that at the SLHC the number of events in
the region where $S/B>1$ is very small. 
Given the uncertainties in the modeling of
the standard model backgrounds via the shower Monte Carlo, 
it is not possible to claim that the
SLHC could see a signal using this global variable.
 The VLHC should have no difficulty as there are several thousand
 events for $M_{eff}>5$ TeV.

\begin{figure}[tb]
\includegraphics[width=3.0in]{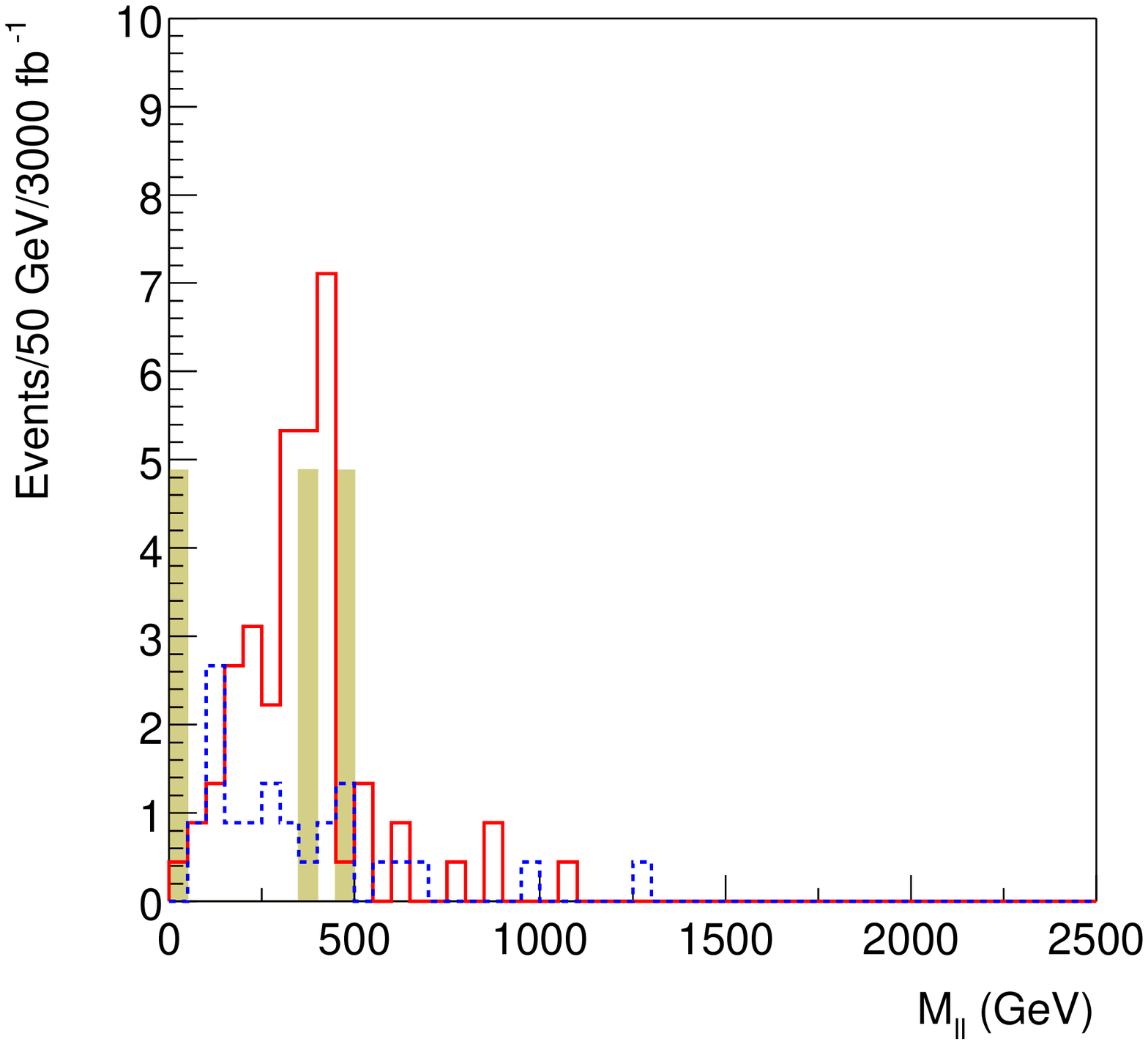}
\includegraphics[width=3.0in]{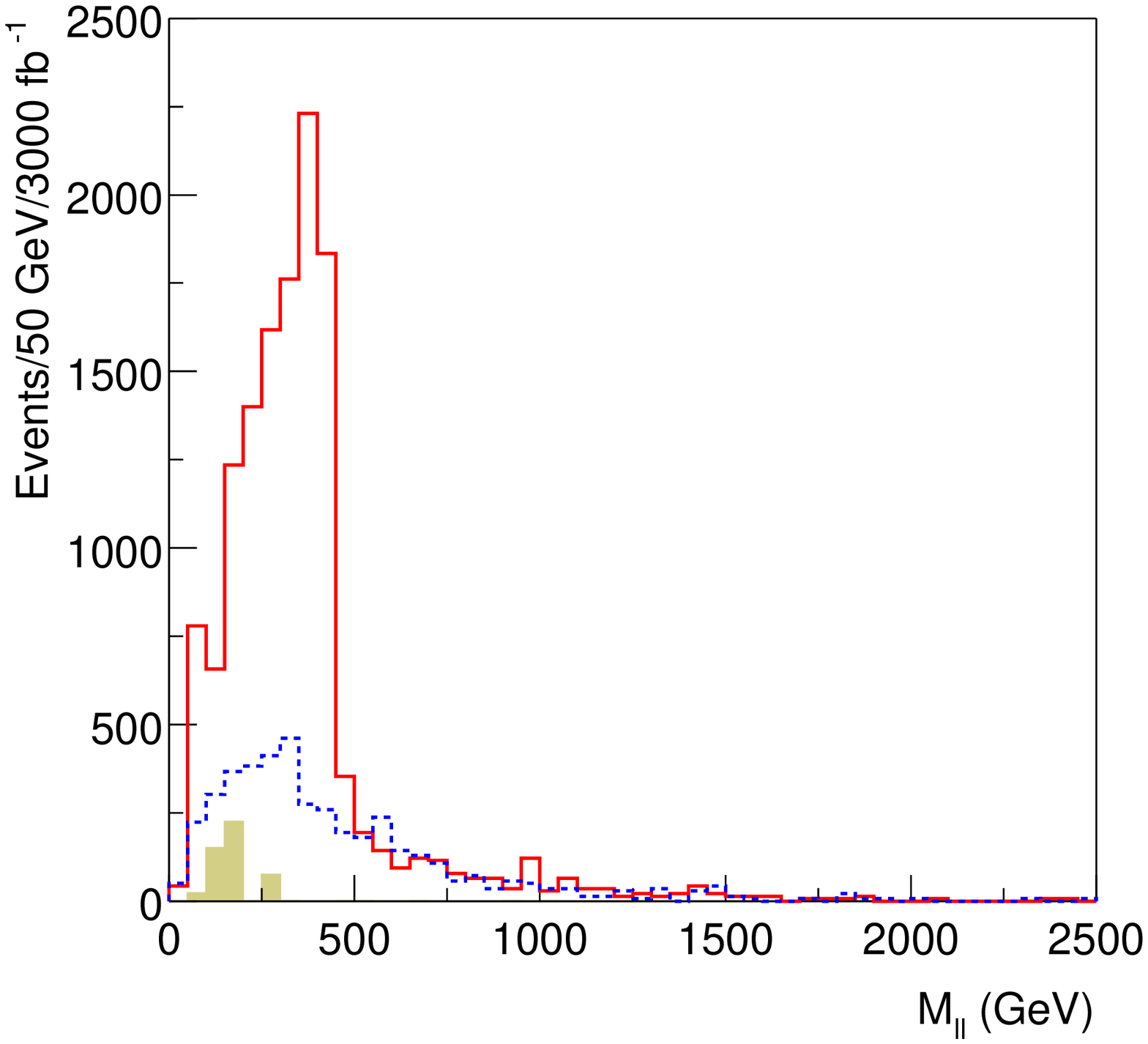}
\caption{$M_{\ell\ell}$ distribution for SLHC (left) and VLHC (right)
for Point H.  Solid: $\ell^+\ell^-$. Dashed: $e^\pm\mu^\mp$.
\label{pointhmll}}
\end{figure}
\begin{figure}[h!t]
\includegraphics[width=3.0in]{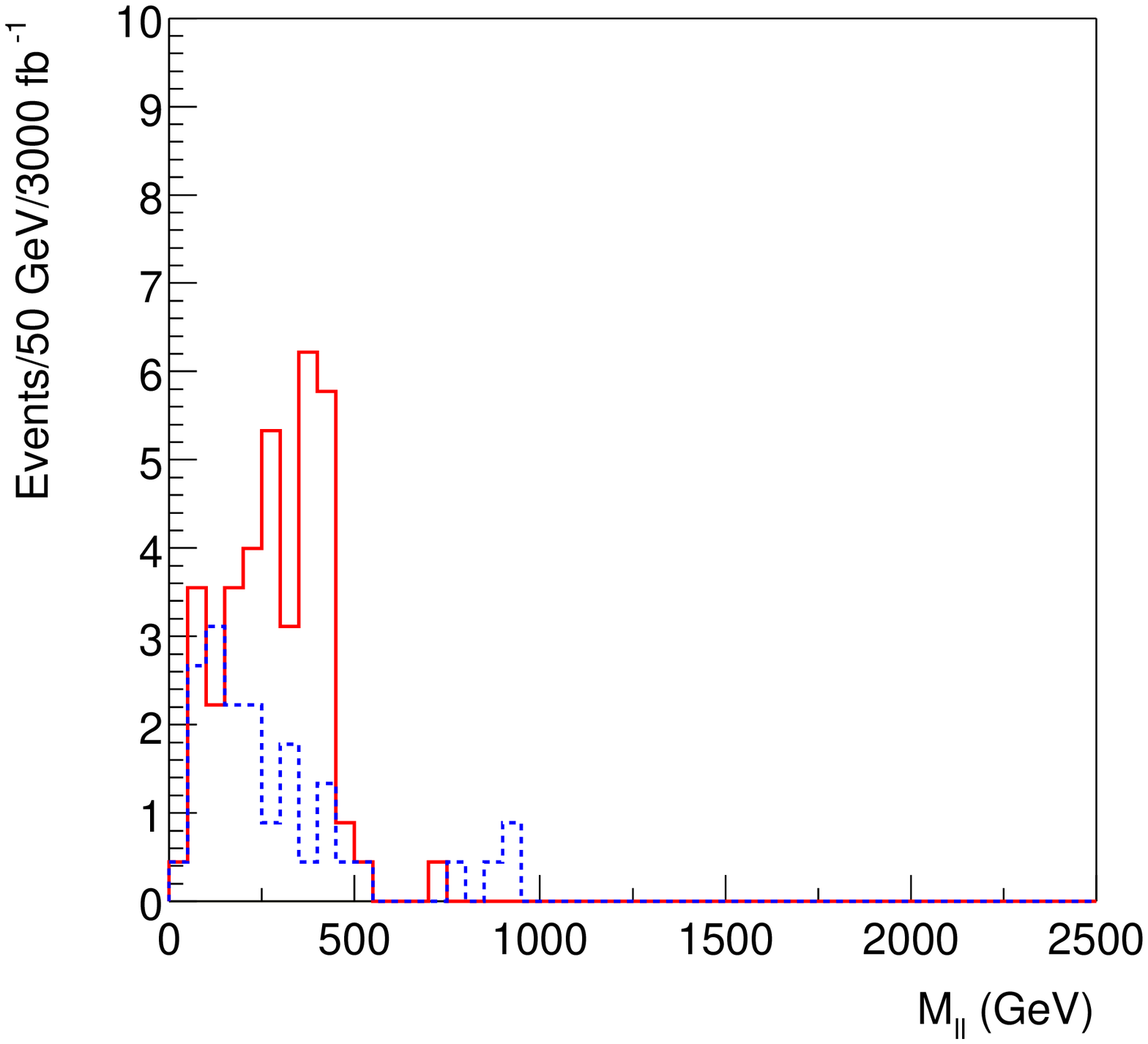}
\includegraphics[width=3.0in]{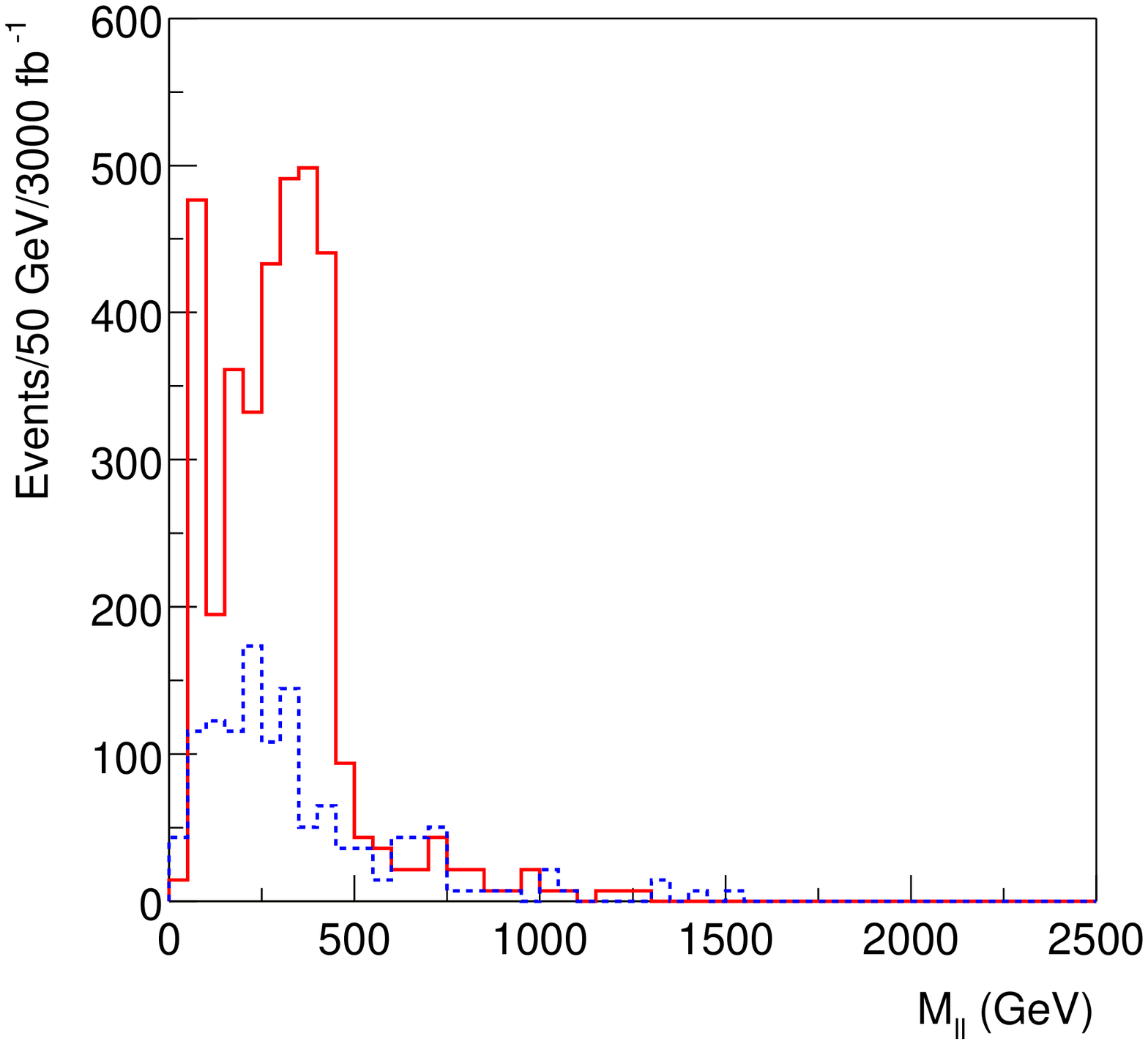}
\caption{$M_{\ell\ell}$ distribution for SLHC (left) and VLHC (right)
for events containing a $\ttau_1$ for Point H.  Solid: $\ell^+\ell^-$.
Dashed:  $e^\pm\mu^\mp$. \label{pointhmllstau}}
\end{figure}

Dileptons arise from the cascade $\tq_L\to q \tchi_2^0\to
q\ell^+\ell^- \lsp$, 
The dilepton mass distributions should have a kinematic endpoint
corresponding to this decay. Figure~\ref{pointhmll} shows the
distribution for same flavor and different flavor lepton pairs.
Events were required to have $\Meff>3000\,\GeV$
and $\etmiss>0.2\Meff$ and to have two isolated opposite sign leptons
with $E_T>15$ GeV and $\abs{\eta}<2.5$.
 The structure at the VLHC is clear;
the edge comes mainly from $\tchi_2^0 \to
\tell_L^\pm \ell^\mp$, which has  a branching ratio of 15\% per flavor. This
gives an endpoint at
$$
\sqrt{\frac{(M_{\tchi_2^0}^2-M_{\tell_L}^2)(M_{\tell_L}^2-M_{\lsp}^2)}
               {M_{\tell_L}^2}} = 447.3\,\GeV
$$
consistent with the observed endpoint in Figure~\ref{pointhmll}. Of course
this plot does not distinguish $\tell_L$ and $\tell_R$. In the case of
the upgraded LHC, the signal may be observable, but it should be noted
that the background is uncertain as only three generated events passed
the cuts.

If the stable stau is used then the situation improves considerably.
The dilepton mass for events containing a $\ttau_1$ with a time delay $7
< \Delta t < 21.5\,{\rm ns}$ is shown in Figure~\ref{pointhmllstau}.
Since $\Delta t>10\sigma$, the standard model  background is expected  to be
negligible. The SLHC signal is  improved and a measurement should   be possible.
The acceptance for VLHC is somewhat worse than the inclusive sample, but
having the correlation of the dileptons with the $\ttau_1$ should be
useful.

The VLHC gives a gain of $\sim 100$ in statistics over the LHC for the
same luminosity at this point, which is at the limit of observability at
the LHC. If the VLHC luminosity were substantially lower, the
improvement provided by it would be rather marginal. The cross section
increases by another factor of $\sim100$ at $200\,\TeV$.

\section{Conclusions}

We have surveyed the signals at hadron colliders for the SUGRA models
proposed by \cite{Battaglia:2001zp} concentrating on the cases where
the sparticle masses are very large.
While the masses of the sparticles at Point K are such that SUSY would
be discovered at the baseline LHC, the event rates are small and
detailed SUSY studies will not be possible.
The reach of the LHC would be improved by
higher luminosity where the extraction of specific final states will
become possible.  The cross section at a 
$40\,\TeV$ VLHC is approximately $100$ times larger than that at LHC.
 This leads to a substantial gain, but it is important to
emphasize that this gain 
requires  luminosity at least as large as that ultimately reached by
the LHC
and detectors capable of exploiting it. Point H has a special feature
in that the stau is quasi-stable. This feature would enable a signal
to be extracted at SLHC. If the tau mass were raised slightly so that
its lifetime were short, then only the VLHC could observe it. The
masses in the case of Point M are so large that the VLHC would be
required for discovery. Point F has a gluino mass of order 2 TeV and
should be observable at the LHC exploiting the production of gluinos
followed by the decays to $\tchi_i$ and hence to leptons.

The Points A, B, C, D,  G, I, and L  which are much less fine tuned
have  similar phenomenology to  the ``Point 5'' or
``Point 6'' 
analysis of \cite{AtlasPhysTDR} in that lepton structure from the
decay $\tchi_2^0\to \tell_R \ell \to \ell^+\ell^- \lsp$ and/or
$\tchi_2^0\to \tilde{\tau}\tau \to \tau^+\tau^- \lsp$ is present.
In most cases
decay
$\tchi_2^0\to \tell_L\ell$ is also allowed, so that a more complicated
dilepton mass spectrum is observable. This should enable the
extraction of $m_{\ell_L}$ in addition
(for an example see Fig 20-53 of \cite{AtlasPhysTDR}). Points A, D
and L have higher squark/gluino masses and will require more
integrated luminosity. Nevertheless one can have confidence that the
baseline LHC will make many measurements in all of these cases.

\section{Acknowledgments}

This work was supported in part by the Director, Office of Energy
Research, Office of High Energy and Nuclear Physics of 
the U.S. Department of Energy under Contracts
DE--AC03--76SF00098 and DE-AC02-98CH10886.  Accordingly, the U.S.
Government retains a nonexclusive, royalty-free license to publish or
reproduce the published form of this contribution, or allow others to
do so, for U.S. Government purposes.

\setcounter{figure}{0}
\setcounter{table}{0}
\setcounter{section}{0}
\setcounter{equation}{0}
\clearpage

\overfullrule=20pt

\def\abs#1{\left| #1\right|}
\def\topfraction{1.0}
\def\bottomfraction{1.0}
\def\textfraction{0.0}
\def\etal{{\it et al.}}
\def\sgn{\mathop{\rm sgn}}
\def\gtap{\raisebox{-.4ex}{\rlap{$\sim$}} \raisebox{.4ex}{$>$}}  
\def\etmiss{\slashchar{E}_T}
\def\fb{{\rm fb}}
\def\ltap{\raisebox{-.4ex}{\rlap{$\sim$}} \raisebox{.4ex}{$<$}}
\def\tG{{\tilde G}}
\def\ns{{\rm ns}}
\def\tell{{\tilde\ell}}
\def\ttau{{\tilde\tau}}
\def\fbi{{\rm fb}^{-1}}
\def\Meff{M_{\rm eff}}
\def\Msusy{M_{\rm SUSY}}
\def\lsp{{\tilde\chi_1^0}}
\def\ra{\rightarrow}
\def\GeV{{\rm GeV}}
\def\mhalf{m_{1/2}}
\def\tchi{\tilde\chi}
\def\tg{\tilde g}
\def\tq{\tilde q}
\def\Cgrav{C_{\rm grav}}
\def\thefootnote{\fnsymbol{footnote}}
\def\Frac#1#2{{\displaystyle#1\over\displaystyle#2}}
\def\cmsec{{\rm cm}^{-2}{\rm s}^{-1}}
\def\hc{{\rm h.c.}}
\def\jet{{\rm jet}}
\def\jets{{\rm jets}}

\def\slashchar#1{\setbox0=\hbox{$#1$}           
   \dimen0=\wd0                                 
   \setbox1=\hbox{/} \dimen1=\wd1               
   \ifdim\dimen0>\dimen1                        
      \rlap{\hbox to \dimen0{\hfil/\hfil}}      
      #1                                        
   \else                                        
      \rlap{\hbox to \dimen1{\hfil$#1$\hfil}}   
      /                                         
   \fi}                                         %


\catcode`@=11
\newdimen\vbigd@men                             

\def\simge{
    \mathrel{\rlap{\raise 0.511ex
        \hbox{$>$}}{\lower 0.511ex \hbox{$\sim$}}}}
\def\simle{
    \mathrel{\rlap{\raise 0.511ex 
        \hbox{$<$}}{\lower 0.511ex \hbox{$\sim$}}}}

\def\dofig#1#2{\centerline{\epsfxsize=#1\epsfbox{#2}}}
\def\dofigs#1#2#3{\centerline{\epsfxsize=#1\epsfbox{#2}%
  \epsfxsize=#1\epsfbox{#3}}}

\part{{\bf  SUSY with Heavy Scalars at LHC
} \\[0.5cm]\hspace*{0.8cm}
{\it I. Hinchliffe and F.E. Paige
}}
\label{ian2sec}


\begin{abstract}
Signatures at the LHC are examined for a SUSY model in which all the
squarks and sleptons are heavy.
\end{abstract}

\section{Introduction\label{sec:intro}}

SUSY models may give new contributions to flavor changing neutral
currents, $CP$ violation, etc., through loops involving squarks and
sleptons. These effects are reduced if the scalars are heavy. The
``inverted hierarchy''\cite{Bagger:1999ty} and ``focus
point''\cite{Feng:2000zg} scenarios provide examples of ways in which
heavy scalars could be accommodated naturally.

This note examines the LHC signatures for a minimal SUGRA model with
$$
m_0=1500\,\GeV,\ \mhalf=300\,\GeV,\ A_0=0,\ \tan\beta=10,\ \sgn\mu=+.
$$
The gaugino masses are similar to those considered previously, e.g.,
$$
M(\lsp)\approx109\,\GeV,\ M(\tchi_1^+)\approx161\,\GeV,\ M(\tchi_2^+)
\approx289\,\GeV,\ M(\tg)\approx782\,\GeV.
$$
Most of the scalars have masses around $1500\,\GeV$; the light Higgs
mass is $116\,\GeV$.

The SUSY production cross section at the LHC is dominated by gluino
pairs. The two largest branching ratios are
$$
B(\tg \to \tchi_1^- t \bar b + \hc) \approx
B(\tg \to \tchi_2^- t \bar b + \hc) \approx 23\%.
$$
However, decays into both charginos and all four neutralinos with all
allowed combinations of quarks are significant. This leads to many
complex signatures.

ISAJET~7.51 was used to generate events for the signal and for all the
Standard Model (SM) backgrounds. The detector response to these events
was simulated using  a parameterized simulation with
parameters appropriate to the ATLAS detector. Jets were found using
a simple cone algorithm with $R=0.4$. Lepton
identification efficiency and $b$ and $\tau$ jet tagging and
misidentification were included with parameterized efficiencies and
backgrounds based on full simulation of ATLAS. A micro-DST was saved and
subsequently analyzed using Root as a framework. The statistics for the
signal correspond to approximately $100\,\fbi$. The statistics for the
largest SM background samples correspond to a much smaller luminosity
but are sufficient to show that the Standard Model backgrounds are
small after cuts.

\section{Effective Mass Distribution\label{sec:meff}}

An inclusive signature based on multiple jets plus missing energy
$\etmiss$ was useful at many of the SUSY points considered previously
and remains so here. Since the jet multiplicity is higher here, the
effective mass was defined to include all jets and leptons, not just the
four hardest jets:
$$
\Meff = \etmiss + \sum_{i=1}^{N_{\rm jet}} p_{T,i}^{\rm jet}
+ \sum_{i=1}^{N_{\rm lep}} p_{T,i}^{\rm lep}
$$

\begin{figure}[tb]
\dofigs{3in}{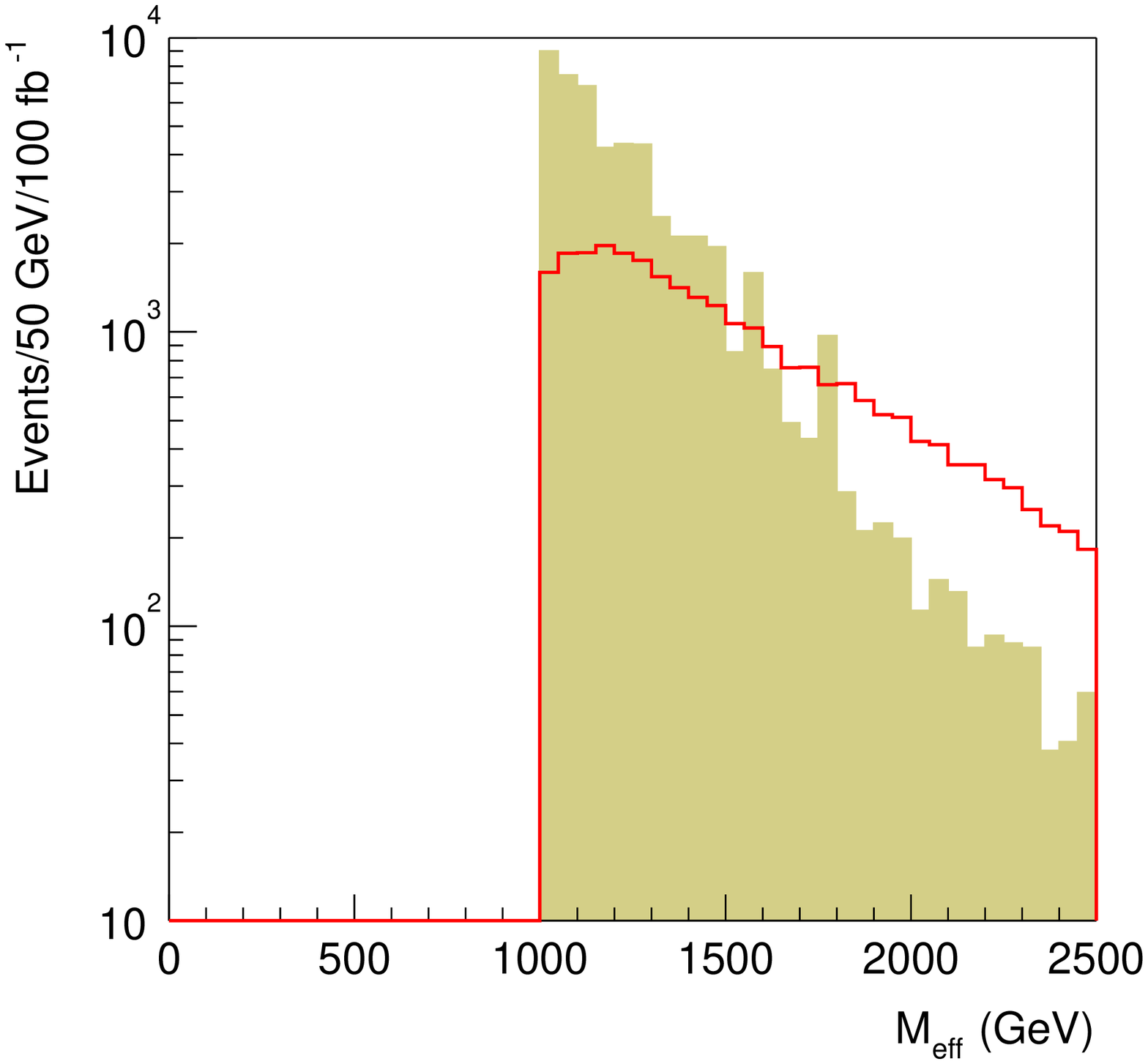}{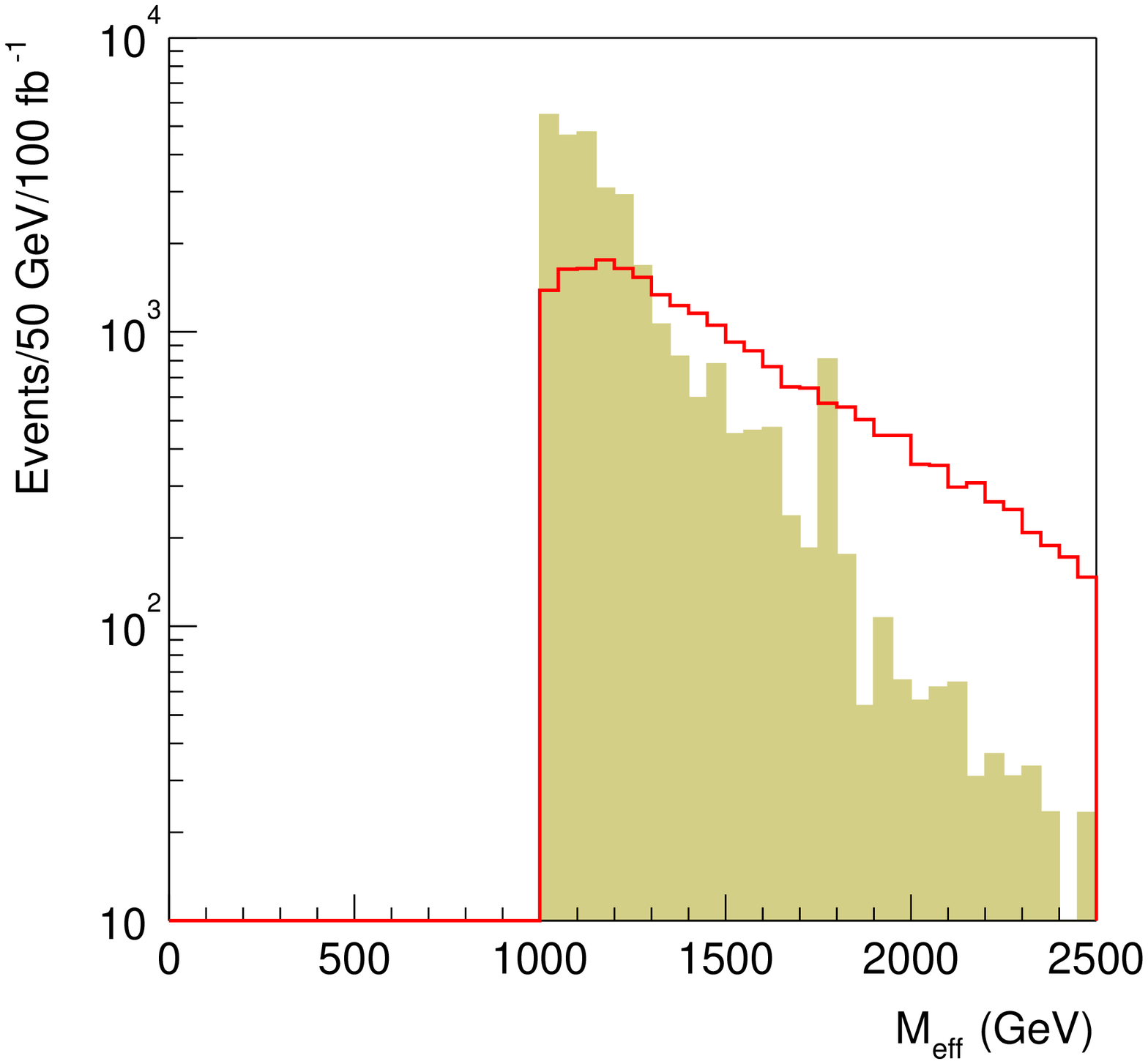}
\caption{Left: $\Meff$ distribution for signal (curve) and SM background
(shaded). Right: Same with $\ge1$ $b$ tags. \label{c1500meff}}
\end{figure}

Events were selected to have
\begin{itemize}
\item   At least six jets with $p_T>100,50,30,30,30,30\,\GeV$;
\item   $\etmiss > \max(100\,\GeV,0.2\Meff)$;
\item   Transverse sphericity $S_T>0.2$;
\item   $\Meff>1000\,\GeV$.
\end{itemize}
The resulting $\Meff$ distribution for signal and background is shown in
Figure~\ref{c1500meff}. In contrast to many previous cases, the signal
emerges from the SM background well past its peak, but nevertheless the
$S/B$ ratio is large for large enough $\Meff$. Thus, discovery of a
deviation from the SM is easy, although not quite so easy as in earlier
cases. 

This signal can be improved by requiring at least one $b$ jet. A $b$
tagging probability of 70\% was chosen, and the corresponding light jet
rejection was taken from full simulation results for ATLAS. This
distribution is also shown in Figure~\ref{c1500meff}. As expected, the
$S/B$ ratio is improved with only a small loss of signal.

\section{Top Reconstruction\label{sec:top}}

Given the large branching ratios for $\tg \to \tchi_i^- t \bar b + \hc$,
it is natural to try to reconstruct hadronic top decays. If everything
decays hadronically, the jet multiplicity from each gluino is 6 for
$\tchi_1^\pm$ and 8 for $\tchi_2^\pm$, giving a total of 12 to 16 jets
without any gluon radiation. This produces a severe combinatorial
background; lepton-based signatures are considerably easier. Work on top
reconstruction is continuing. A more sophisticated jet algorithm might
work better for these complex events.

\section{\boldmath $\ell^+\ell^-$ Signature\label{sec:ellell}}

\begin{figure}[tb]
\dofigs{3in}{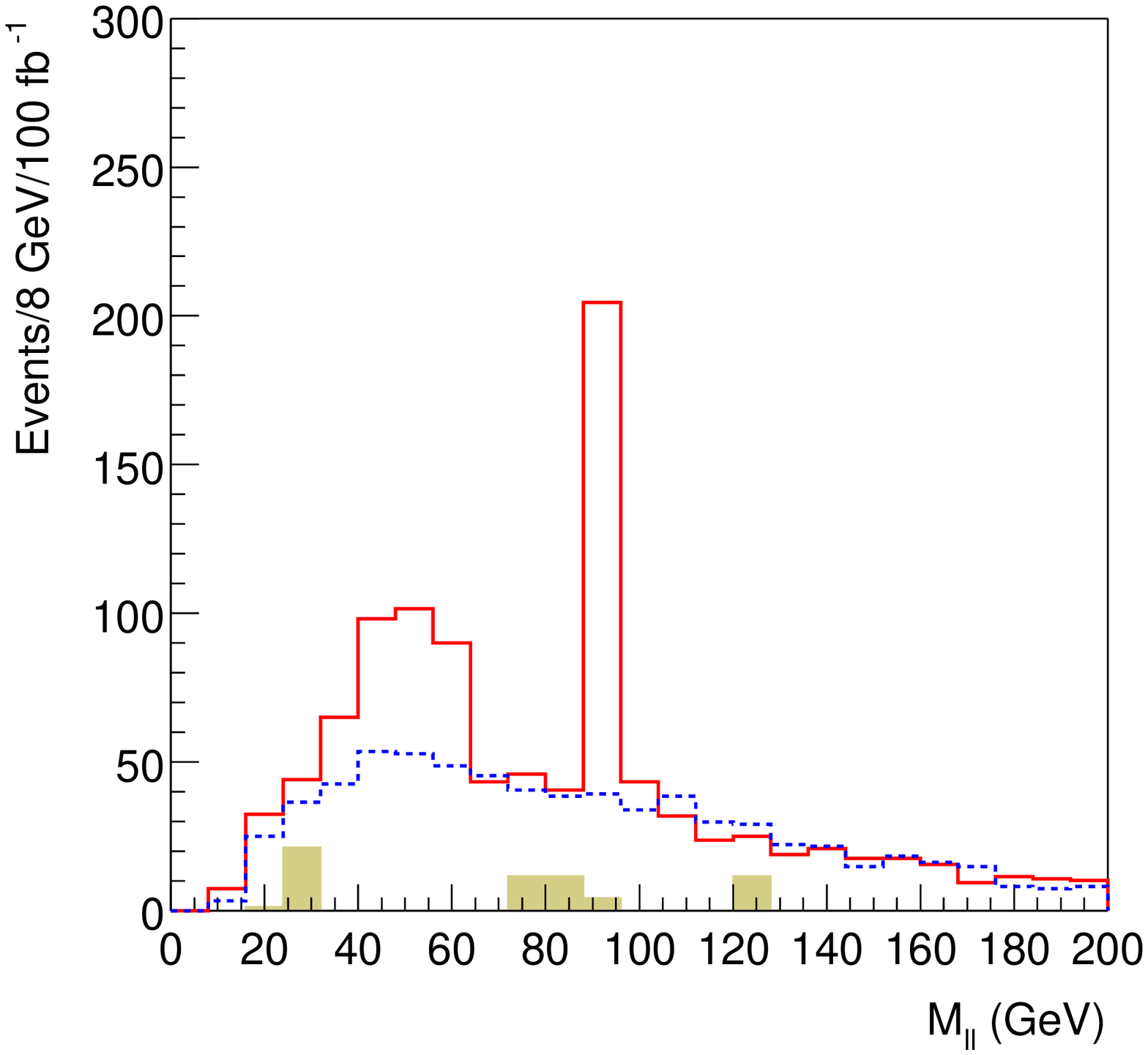}{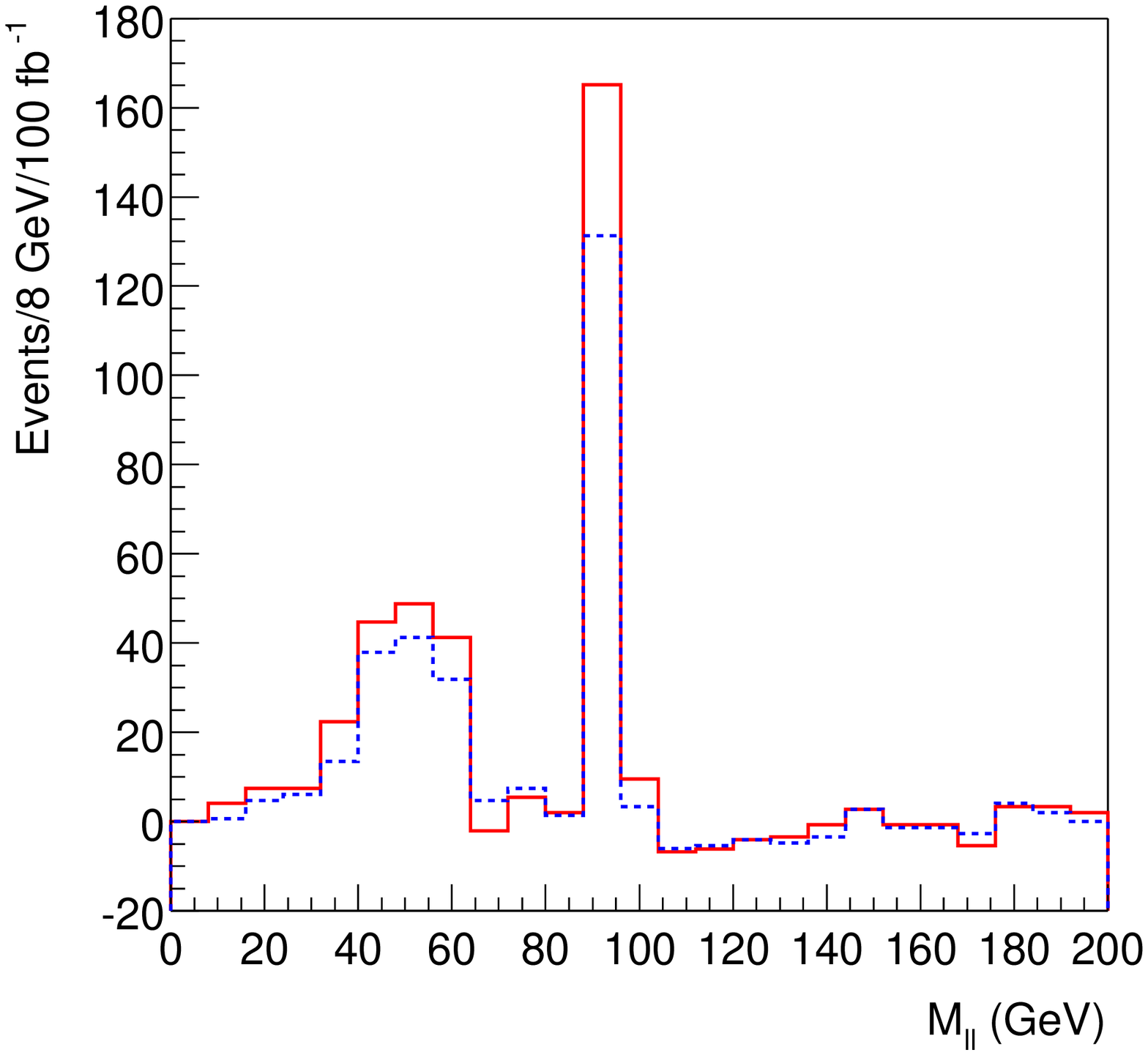}
\caption{Left: Dilepton mass distributions for $\ell^+\ell^-$ from
signal (solid), $e^\pm\mu^\pm$ from signal (dashed), and $\ell^+\ell^-$
from SM background (shaded) after cuts. Right: $\ell^+\ell^- -
\ell^\pm\ell^\pm$ distribution (solid) and with $\ge1$ $b$ tag (dashed).
\label{c1500mll}}
\end{figure}

The three-body decay $\tchi_2^0 \to \lsp \ell^+\ell^-$ has a kinematic
endpoint at $M(\tchi_2^0)-M(\lsp)=61.5\,\GeV$. Events satisfying the
cuts given in Section~\ref{sec:meff} were required to have two OS,SF
leptons with $p_T>15\,\GeV$ and $|\eta|<2.5$. The reconstruction
efficiency was assumed to be 90\% for both $e$ and $\mu$.
Figure~\ref{c1500mll} shows the resulting $\ell^+\ell^-$ and
$e^\pm\mu^\mp$ mass distributions. Any contribution from two independent
decays should cancel in the difference of these. This difference shows
both a continuum with an endpoint at the expected place and a $Z$ peak
coming from decays of heavy gauginos.

The largest single source of $\tchi_2^0$ is $\tg \to \tchi_2^0 t \bar
t$; while the largest sources of heavy gauginos are $\tg \to \tchi_2^\pm
t \bar b$ and $\tg \to \tchi_3^0 t \bar t$. Thus one expects a large
fraction of dileptons to be accompanied by a $b$. Figure~\ref{c1500mll}
also shows the subtracted distribution without and with at least one $b$
tag.

\section{\boldmath $e^\pm\mu^\mp - e^\pm\mu^\pm$ Signature}

\begin{figure}[tb]
\dofig{3in}{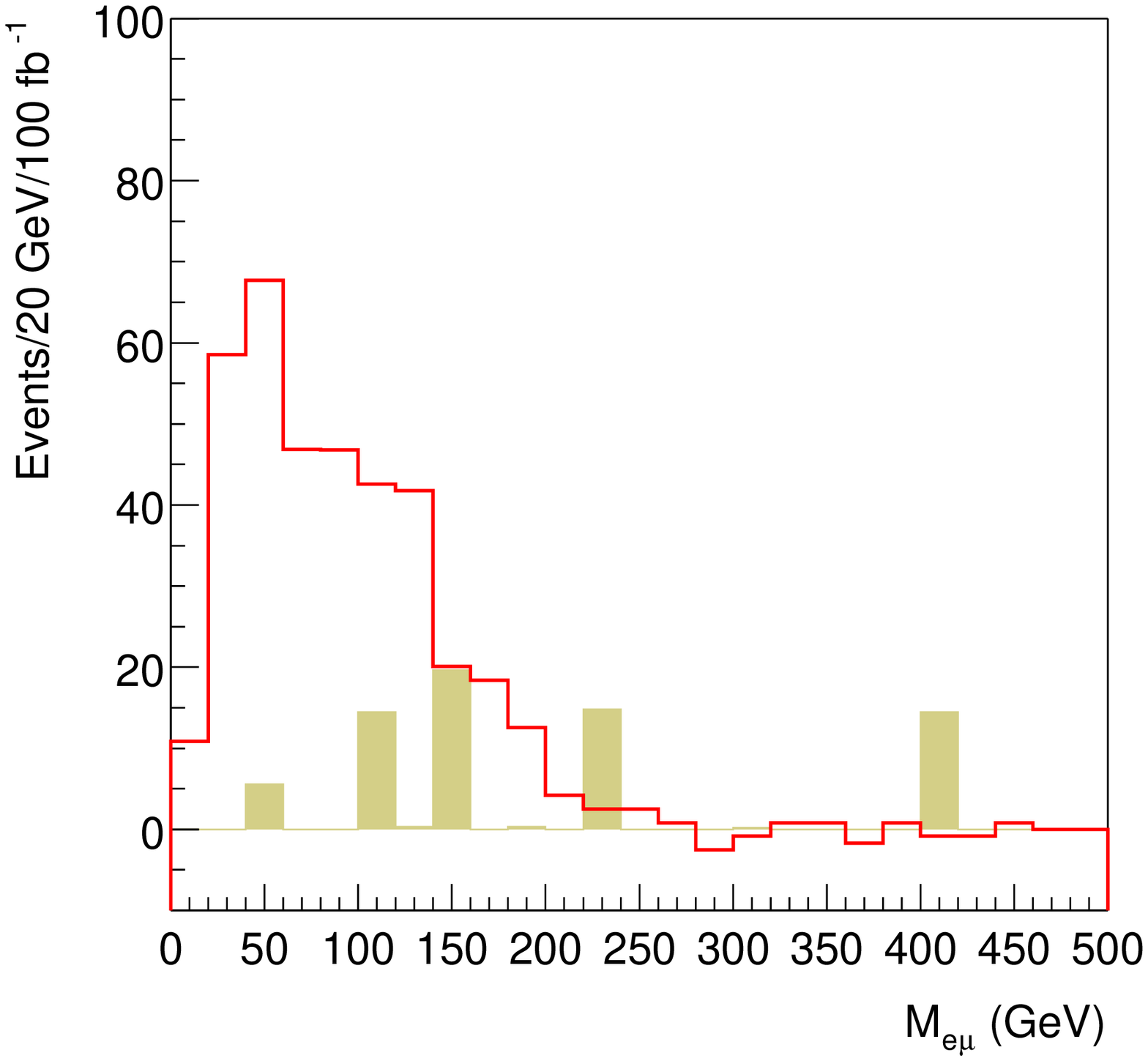}
\caption{$e^\pm\mu^\mp - e^\pm\mu^\pm$ distribution for signal (solid)
and SM background (shaded). \label{c1500mgemu}}
\end{figure}

Two independent leptonic decays of the same gluino, e.g., $\tg \to
\tchi_i^- t \bar b$ with $\tchi_i^- \to e^- X$ and $t \to \mu^+ X$,
gives an OS dilepton signature. Since the $\tg$ is a Majorana fermion,
any contribution from leptonic decays involving both gluinos will cancel
in the combination $e^\pm\mu^\mp - e^\pm\mu^\pm$. (Equal acceptance for
$e$ and $\mu$ is assumed here. In reality one would have to correct for
acceptance; this correction can be checked using $Z \to
e^+e^-,\mu^+\mu^-$ data.) The resulting distribution using the cuts
described above is shown in Figure~\ref{c1500mgemu}.

\begin{figure}[htb]
\dofigs{3in}{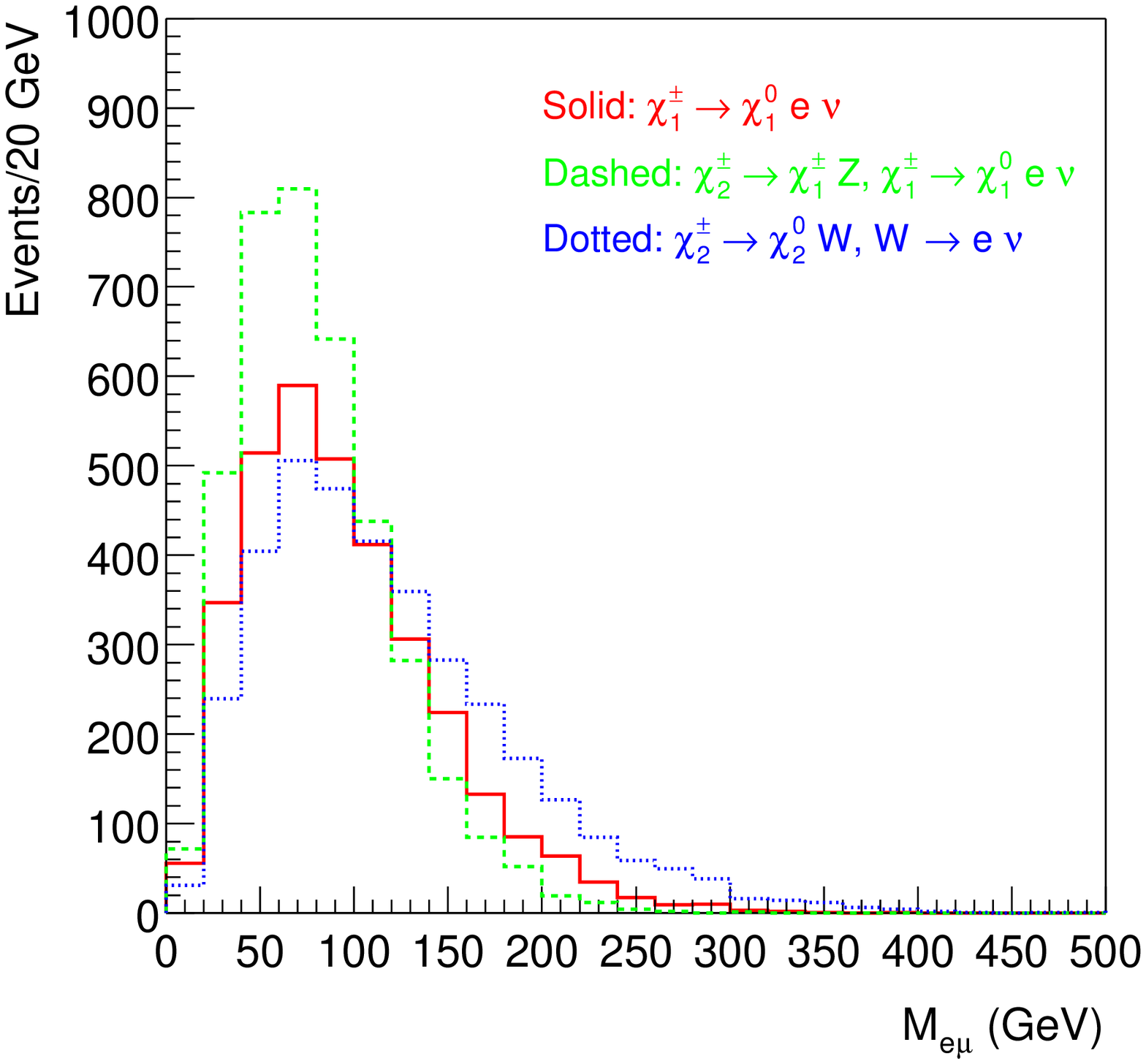}{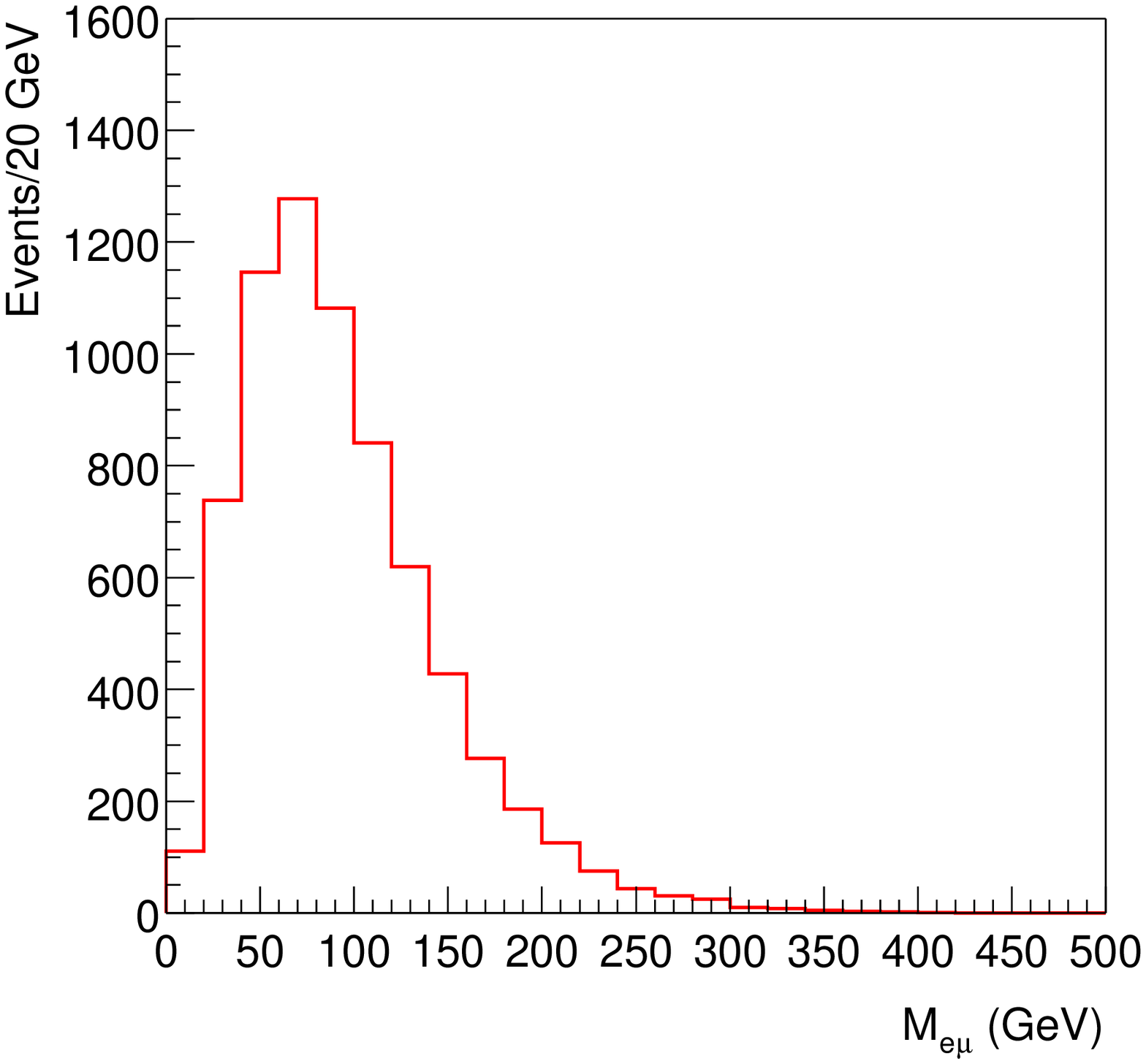}
\caption{Left: $e^\pm\mu^\mp - e^\pm\mu^\pm$ mass distribution for three
possible signal contributions. Right: Sum weighted by branching ratios.
\label{c1500mue}}
\end{figure}

\begin{figure}[htb]
\dofigs{3in}{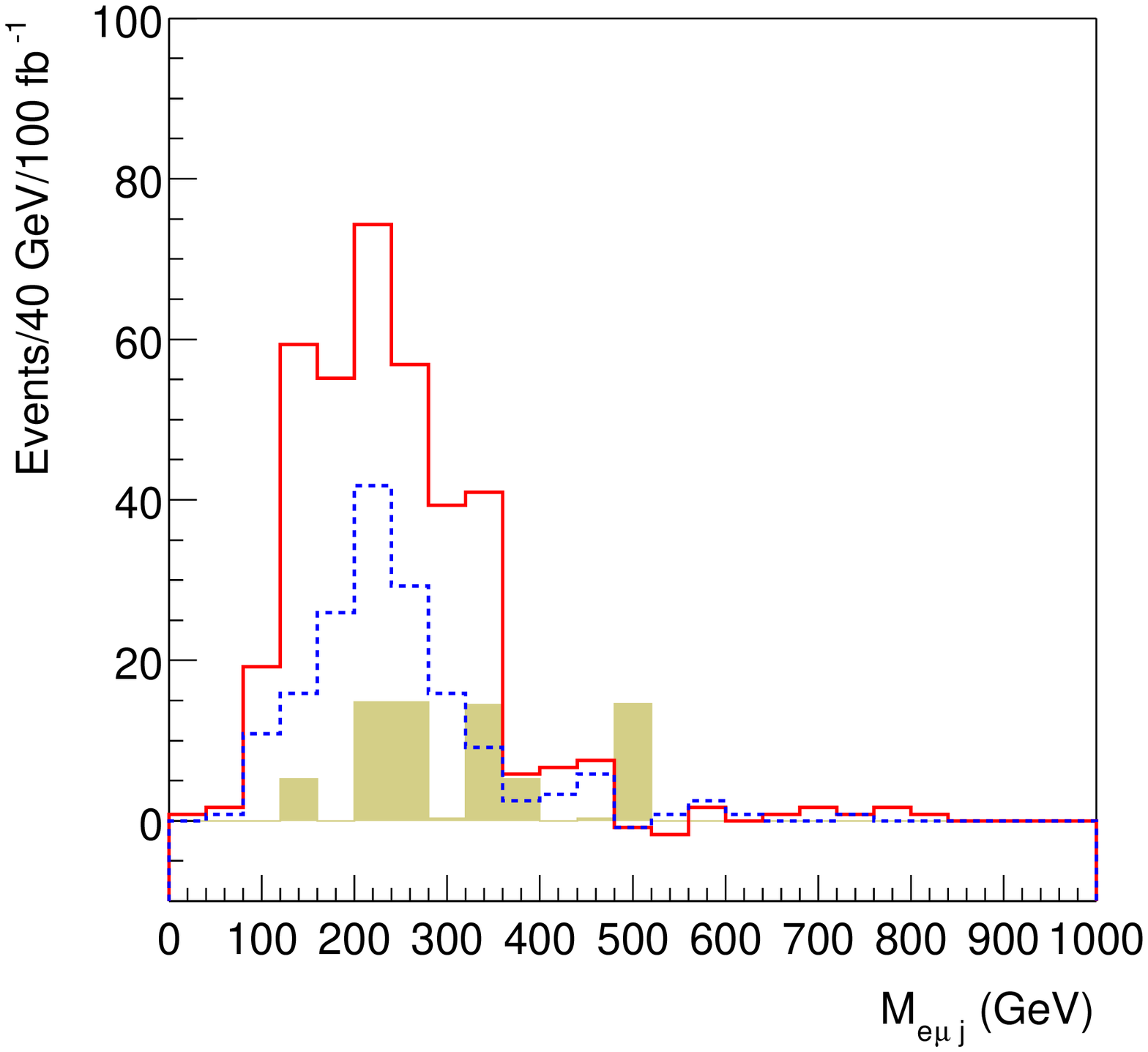}{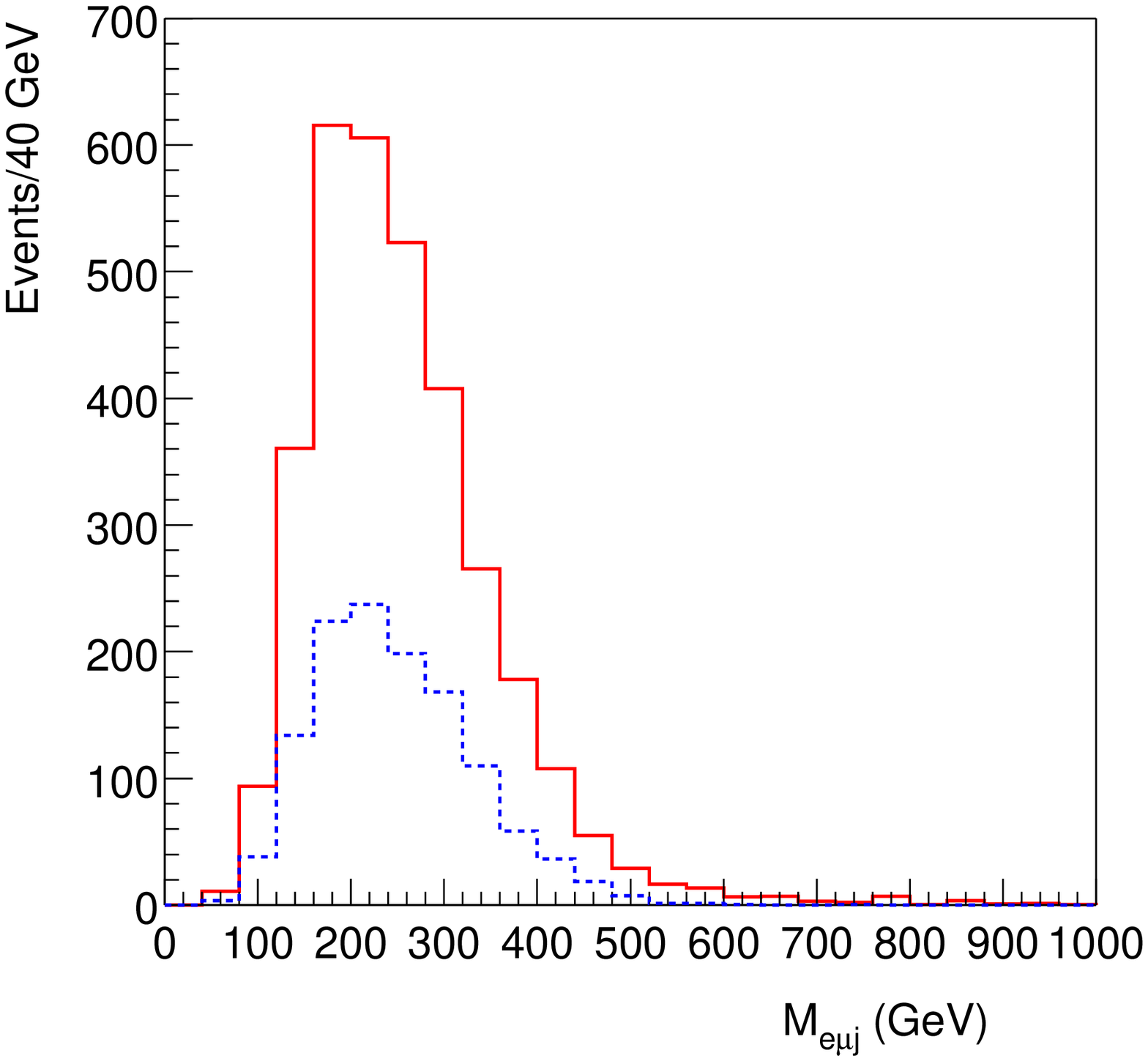}
\caption{Left: Minimum $e\mu j$ mass formed from the sign-subtracted $e\mu$
pair, Figure~\protect\ref{c1500mue} and one of the three hardest jets
with $p_T>100\,\GeV$. Solid: all jets. Dashed: with $b$ tag. Shaded: SM
background. Right: Same for $\tg+\lsp$ events
with forced decays. \label{c1500mgemuj}}
\end{figure}

\begin{figure}[htb]
\dofig{3in}{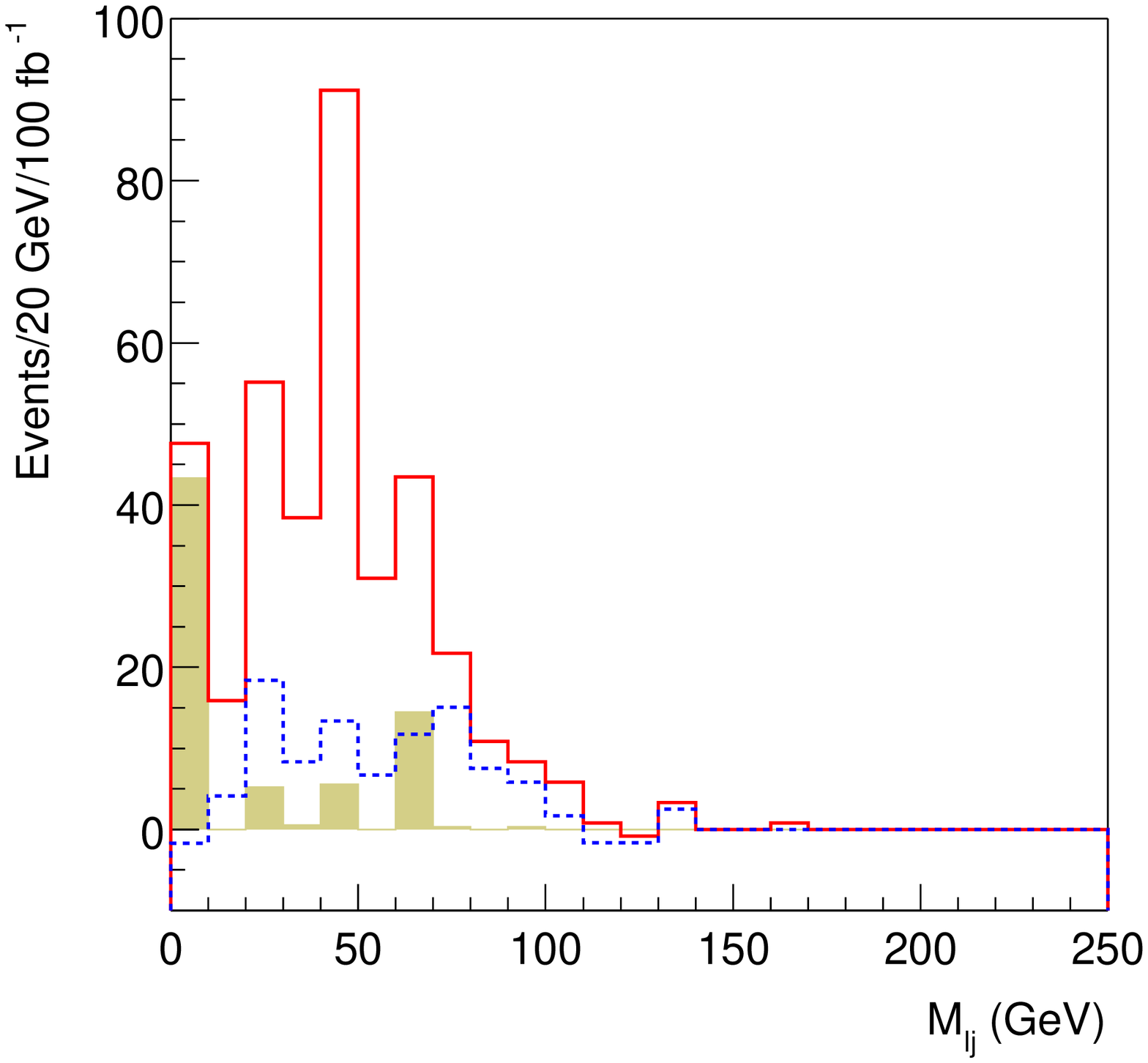}
\caption{Minimum $\ell j$ mass from sample in
Figure~\protect\ref{c1500mgemu}. Solid: all jets. Dashed: with $b$ tag.
\label{c1500mglj}}
\end{figure}

While the $e^\pm\mu^\mp - e^\pm\mu^\pm$ dilepton distribution should
have a true kinematic endpoint corresponding to the maximum possible
mass from gluino decay, this is not useful because many particles are
unobservable ($\nu$, $\lsp$) or not included (jets). The largest
contributions to this channel should come 
from $\tg \to \tchi_{1,2}^\pm t b$ 
with $\tchi_{1,2}^\pm \to \lsp X$ and $t \to \ell\nu b$. Three
samples of $\tg\lsp$ events with $200<p_T<600\,\GeV$ (the typical range
for the gluino $p_T$) and with one of the three decay chains forced
were generated. The
same analysis was applied except that the required number of jets was
reduced from 6 to 3. The mass distributions for the three possibilities
are shown in Figure~\ref{c1500mue}. All three are qualitatively similar
to that shown in Figure~\ref{c1500mgemu}. The shapes are somewhat
different and presumably could be distinguished with sufficient
statistics after a detailed analysis. This has not yet been attempted.

The sign-subtracted $e\mu$ pair was next combined with each of the three
hardest jets (with $p_T>100\,\GeV$) in the event. The distribution of
the minimum of the three masses is shown in Figure~\ref{c1500mgemuj}.
The distribution in the case that the jet giving the minimum is tagged
as a $b$ is also shown in the Figure. If one of the jets is from the
same gluino as the dilepton pair, then this distribution should have a
kinematic endpoint related to the gluino mass. The choice of three jets
is a compromise between including the right jet and including too many.
The expected shape from a single gluino was again determined using the
$\tg+\lsp$ sample; this is also shown in Figure~\ref{c1500mgemuj}. A similar
analysis combining the $e\mu$ with two jets found too much combinatorial
background.

If one of the leptons is from $t \to W b$, then the smallest $\ell j$
mass should be less than the kinematic limit for this decay, 
$\sqrt{(m_t^2-m_W^2)/2}=110\,\GeV$. This minimum mass is plotted in
Figure~\ref{c1500mglj} and has the expected shape.
However, a rather small fraction of the jets so selected are tagged as
$b$'s, while the $b$ tagging efficiency is about 60\%. 

\section{\boldmath $\ell^\pm\ell^\pm + \jets$ Signature}

If both gluinos decay via $\tg \to \tchi_1^\pm q \bar q$ with
$\tchi_1^\pm \to \lsp\ell^\pm\nu$, the signature is four hard jets plus
two leptons. Requiring the leptons to be the same sign causes the loss
of half of the signal but greatly reduces the SM background. Events were
selected to have
\begin{itemize}
\item   Four jets with $p_T>40\,\GeV$, the first with $p_T>100\,\GeV$;
\item   $\Meff>500\,\GeV$;
\item   $\etmiss>\max(100\,\GeV,0.1\Meff)$;
\item   $S_T>0.2$;
\item   At least 2 leptons;
\item   Less than 2 tagged $b$ jets.
\end{itemize}

\begin{figure}[htb]
\dofigs{3in}{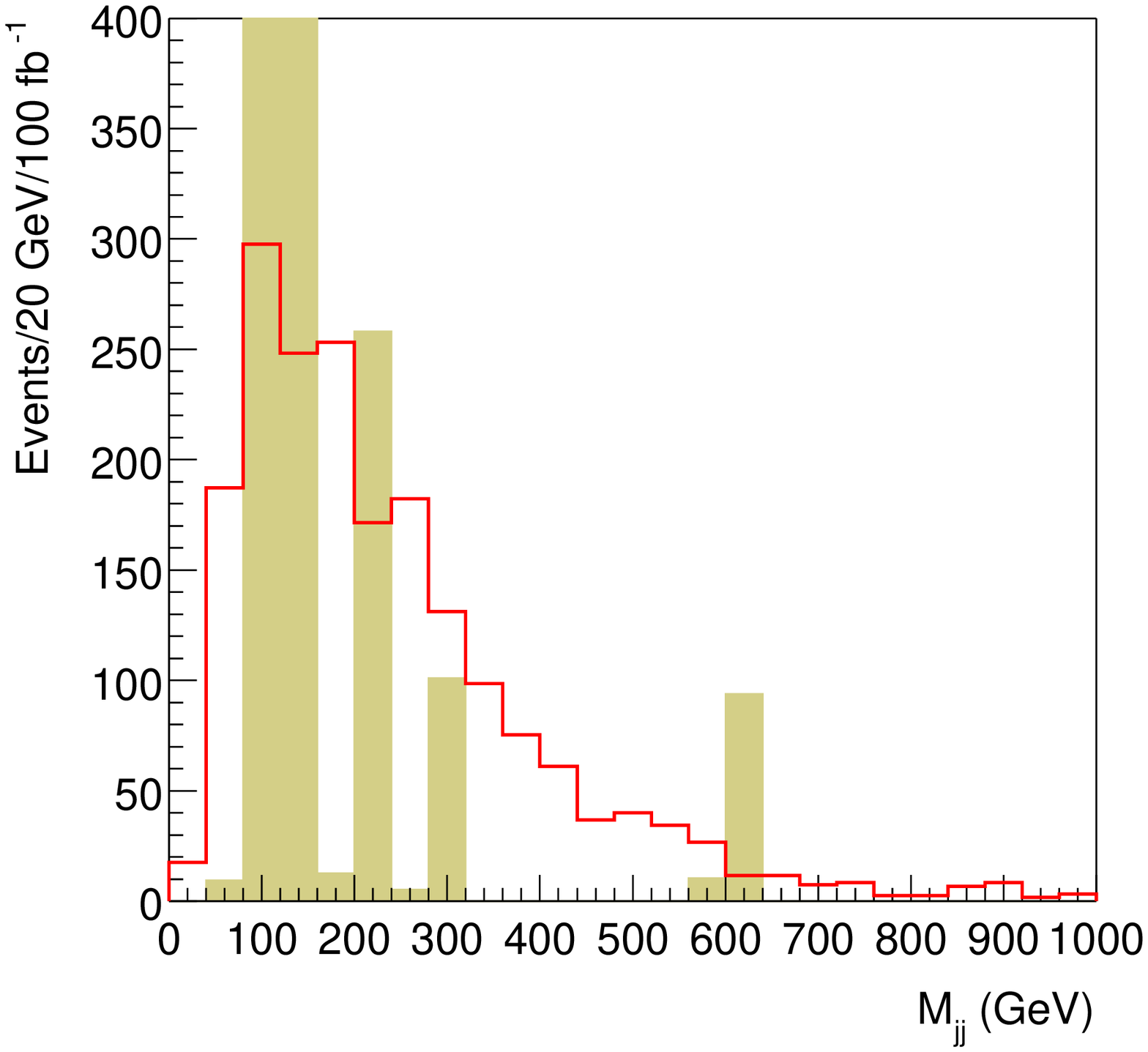}{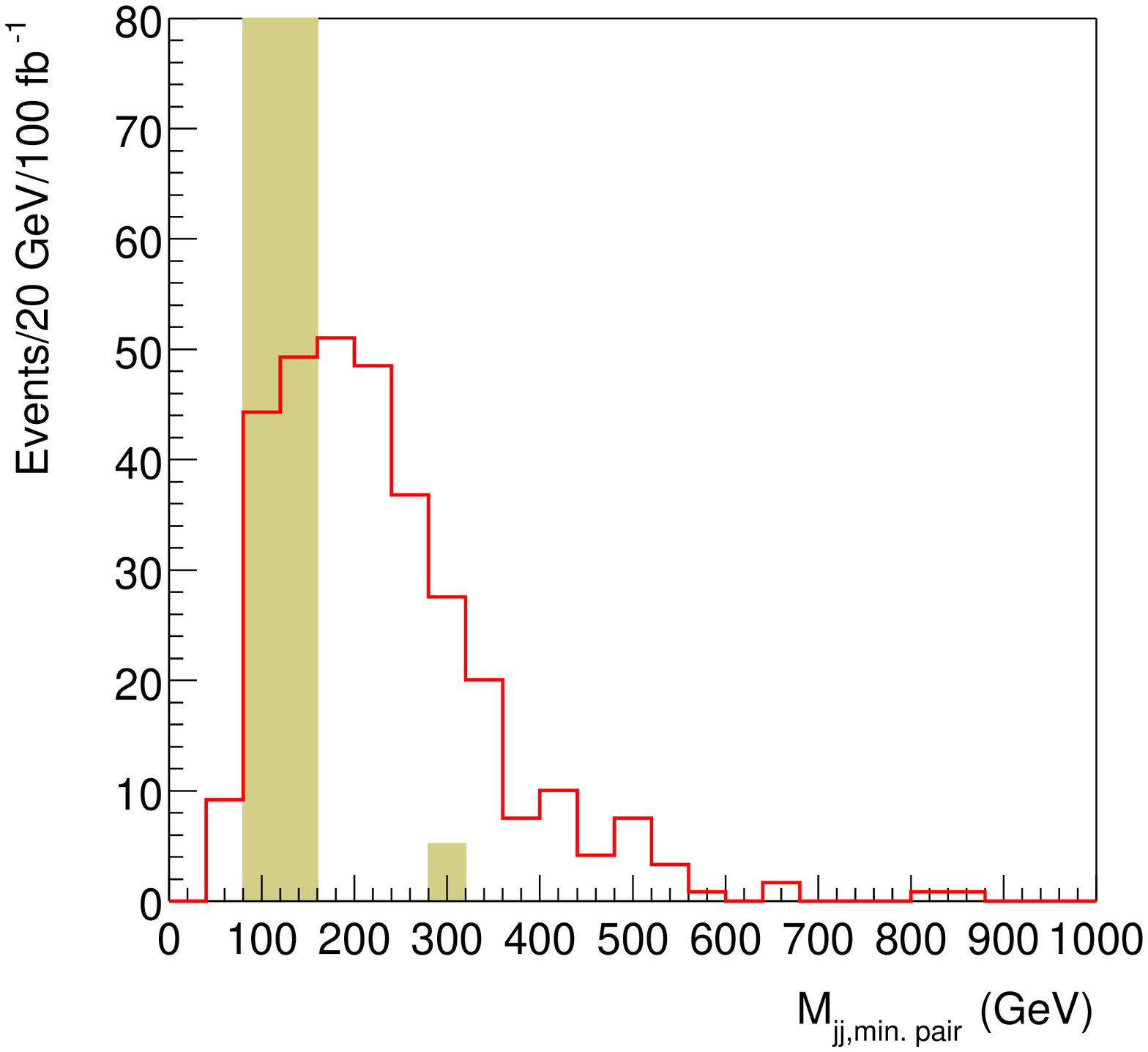}
\caption{Dijet mass distributions for like-sign dilepton events for
signal (solid) and SM background (shaded). Left:  all three
combinations.  Right: minimum combination. \label{c1500mjj2l}}
\end{figure}

\begin{figure}[htb]
\dofig{3in}{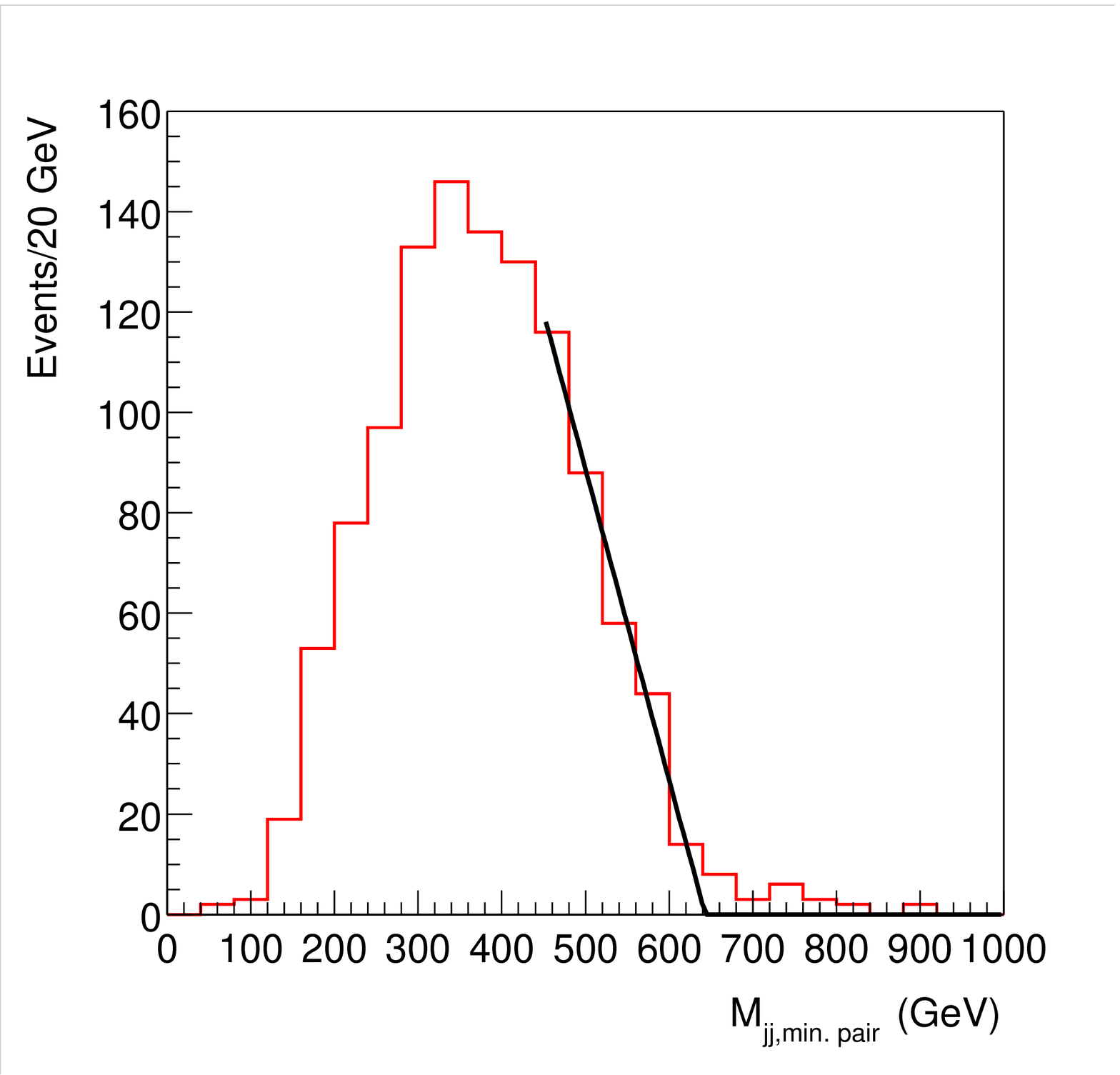}
\caption{Same as Figure~\protect\ref{c1500mjj2l} for forced decays.
\label{c1500mjj2lforce}}
\end{figure}

\begin{figure}[htb]
\dofig{3in}{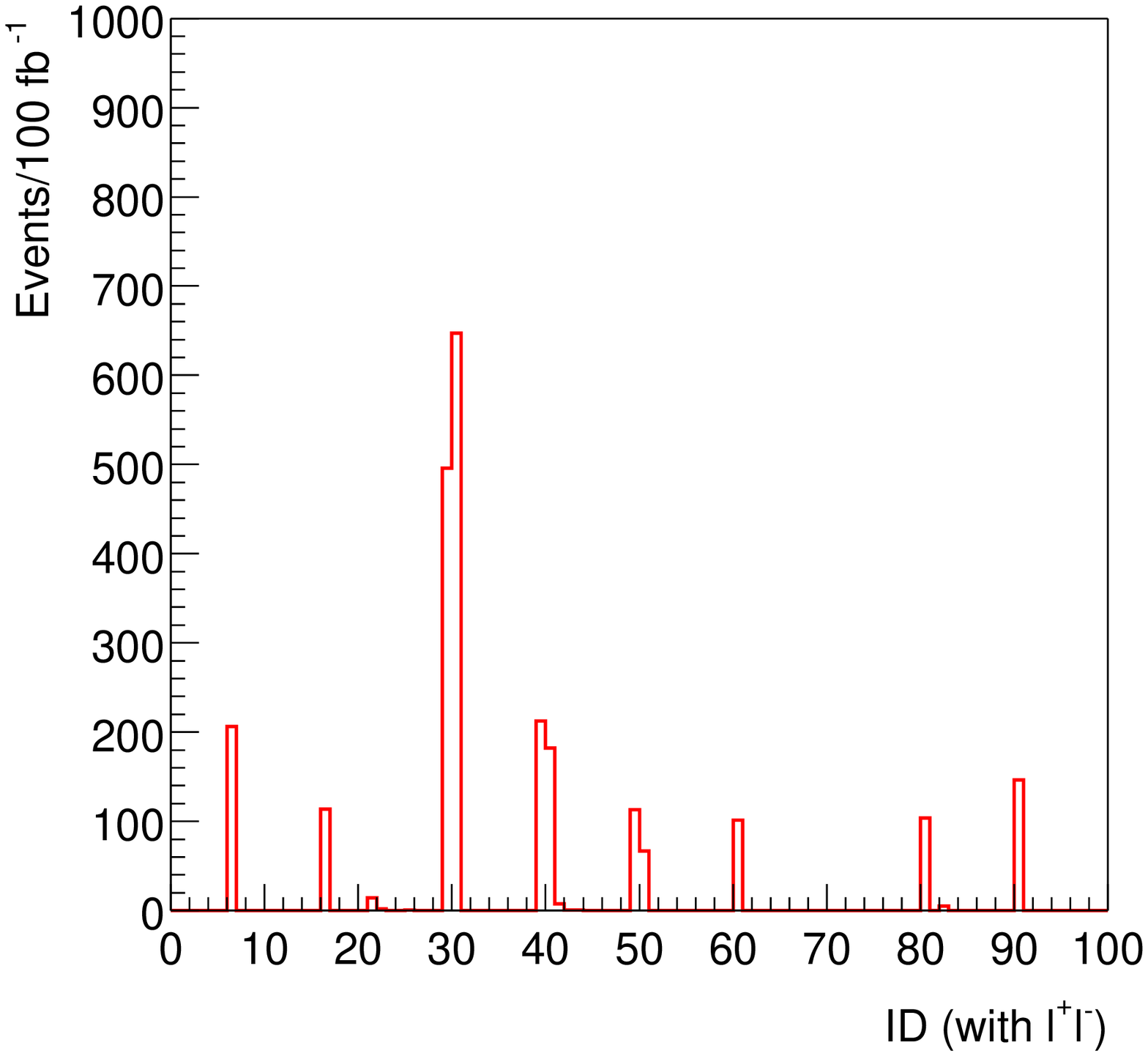}
\caption{SUSY particle content for events selected in
Figure~\protect\ref{c1500mjj2l} using the ISAJET particle numbering
scheme. \label{c1500id2l}}
\end{figure}

The two hardest leptons were required to be the same sign. For each of
the three possible ways of pairing the jets, the larger of the dijet
masses was taken, so the minimum of the
three masses should be less than the dijet endpoint for gluino decay,
$M(\tg)-M(\tchi_1^\pm)=620\,\GeV$. The distributions for all three and
for the minimum are shown in Figure~\ref{c1500mjj2l}. The expected
distribution obtained by forcing the decays $\tg \to \tchi_1^+ \bar u d$
and $\tchi_1^+ \to \lsp e^+\nu$ is shown in
Figure~\ref{c1500mjj2lforce}. The endpoint has the expected value. 
However, the sample is not very pure:  Figure~\ref{c1500id2l}
shows that there are other contributions.
Harder cuts on extra jets did not help significantly to improve the purity.

\begin{figure}[htb]
\dofigs{3in}{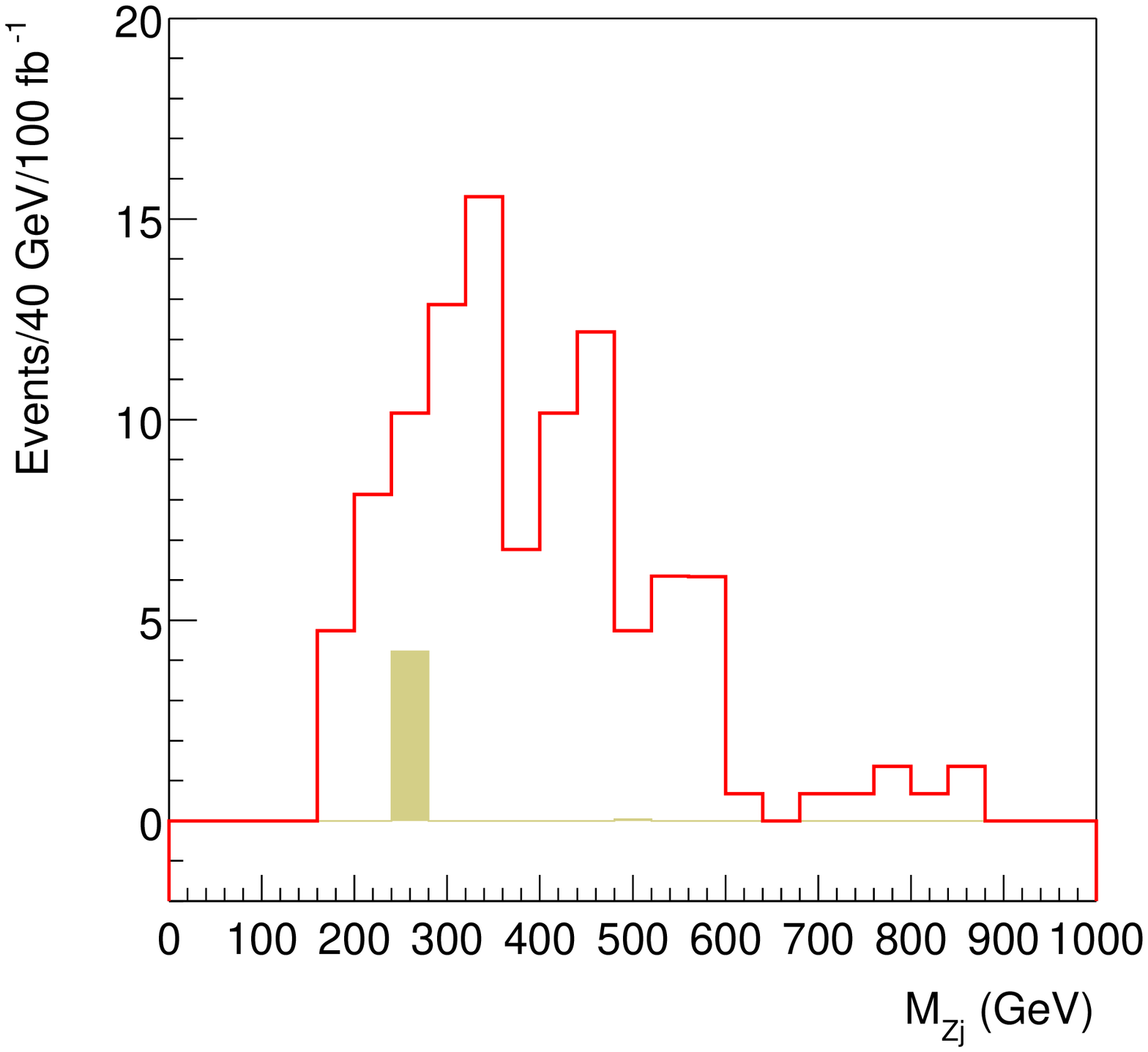}{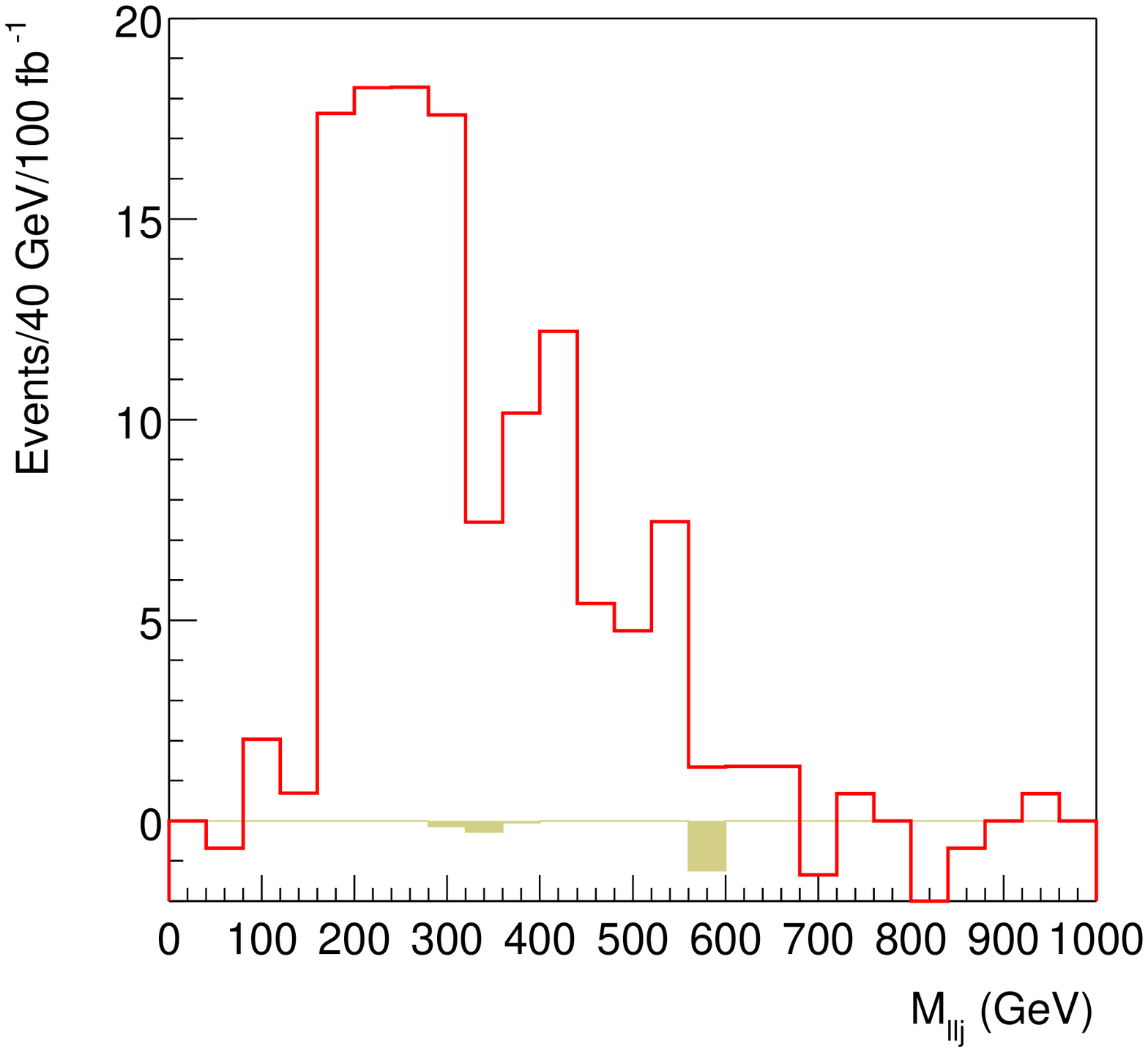}
\caption{Minimum mass of $e^+e^- + \mu^+\mu^- - e^\pm\mu^\mp$ combined
with jet with $p_T>200\,\GeV$. Left: $Z$ peak. Right: Below $\tchi_2^0
\to \lsp \ell^+\ell^-$ endpoint. \label{c1500mzj}}
\end{figure}

\section{\boldmath $\ell^+\ell^-j$ Signature from $\tg \to \tchi_i^0 g$}

The decays $\tg \to \tchi_2^0 g, \tchi_3^0 g$ have branching ratios of
$\sim1\%$ and $\sim2\%$ respectively at this point; $\tchi_3^0 \to \lsp
Z$ has a branching ratio of $\sim100\%$. The $\ell^+\ell^-$ pair was
combined with any jet with $p_T>200\,\GeV$ not tagged as a $b$.
Figure~\ref{c1500mzj} shows the resulting $(e^+e^- + \mu^+\mu^- -
e^\pm\mu^\mp)+j$ mass distributions for the $Z$ peak and for the
$\tchi_2^0 \to \lsp \ell^+\ell^-$ continuum. The $Z$ distribution should
have an endpoint at $602.3\,\GeV$ that can be calculated in terms of the
masses involved. The continuum is more complicated since the dilepton
mass also has a distribution. 

\begin{figure}[ht]
\dofigs{3in}{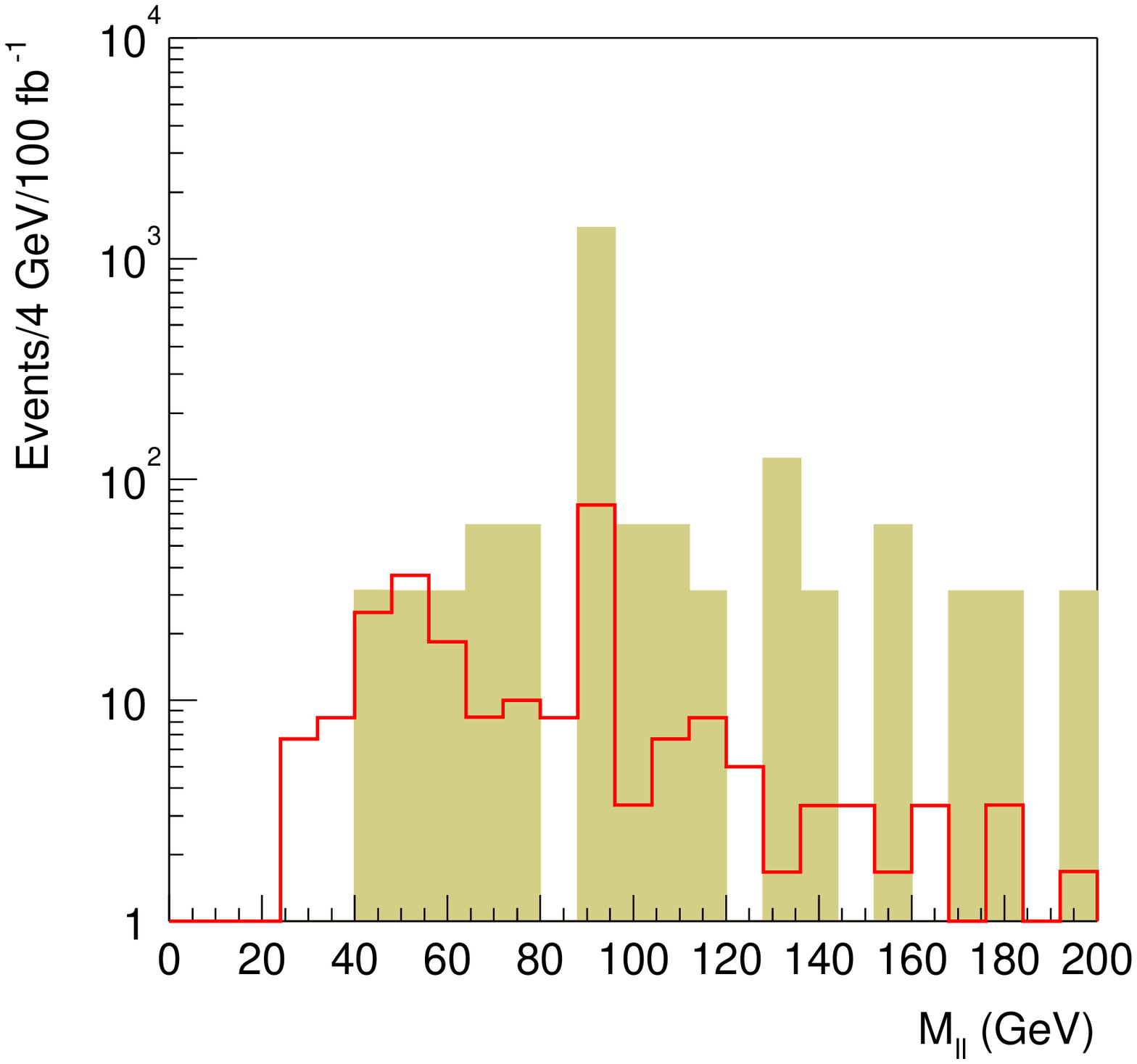}{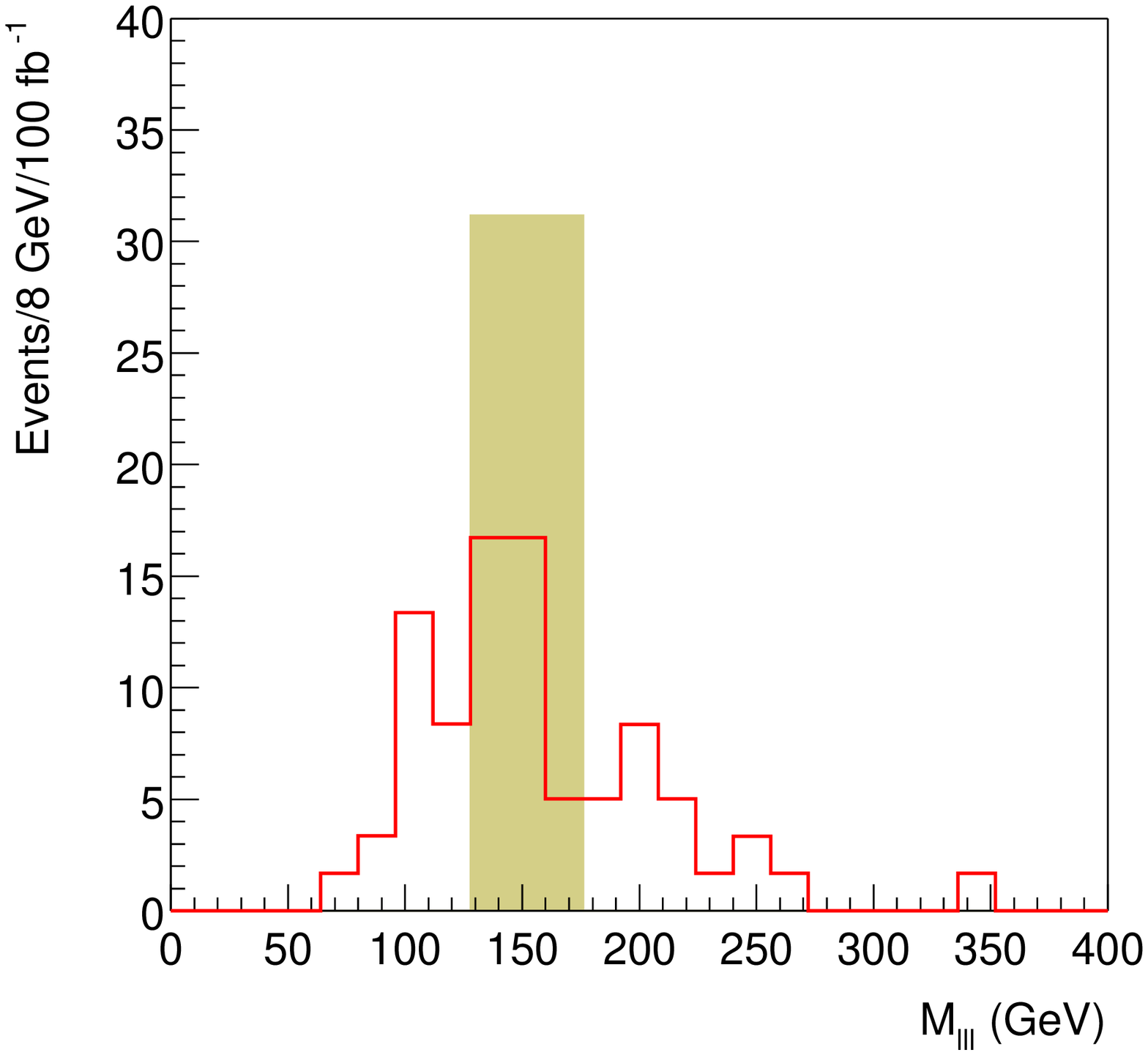}
\caption{Mass distributions for $\ell^+\ell^-$ (left) and
$\ell^+\ell^-\ell^\pm$ (right) with jet veto for signal (curves) and SM
background (shaded). \label{c1500mwzll}}
\end{figure}

\section{\boldmath $\tchi_1^\pm + \tchi_2^0$ Signature}

Direct production of gauginos is only a factor of $\sim10$ smaller than
$\tg\tg$ production at this point. Events were selected to have
three leptons with $p_T>50,20,20\,\GeV$. A jet veto of $30\,\GeV$ was
imposed. The OS,SF dilepton mass distribution is shown in
Figure~\ref{c1500mwzll}. The dilepton endpoint is known from
Figure~\ref{c1500mll}. Requiring an OS,SF pair below this endpoint gives
the $\ell^+\ell^-\ell^\pm$ distribution in the same figure. Clearly the
SM background is still comparable to the signal; it would be worse if
the $Z$ width in SM $WZ$ events were properly taken into account. Also,
the effect of pileup on the jet veto has not been included. Thus this
channel does not seem very promising.

\section{Acknowledgments}

This work was supported in part by the Director, Office of Energy
Research, Office of High Energy and Nuclear Physics of 
the U.S. Department of Energy under Contracts
DE--AC03--76SF00098 and DE-AC02-98CH10886.  Accordingly, the U.S.
Government retains a nonexclusive, royalty-free license to publish or
reproduce the published form of this contribution, or allow others to
do so, for U.S. Government purposes.

\setcounter{figure}{0}
\setcounter{table}{0}
\setcounter{section}{0}
\setcounter{equation}{0}
\clearpage



\part{{\bf Inclusive study of MSSM in CMS
} \\[0.5cm]\hspace*{0.8cm}
{\it 
S.~Abdullin,
A.~Albert,
F.~Charles
}}
\label{arnauldsec}


  

  \begin{abstract}
   The Minimal Supersymmetric Standard Model is an extension of the Standard Model,
the most economical one in terms of new particles and new couplings. 
Many studies have been performed on the
observation of supersymmetry, but mostly limited to the mSUGRA model. 
Here we consider the possibility of a broader test of SUSY, 
using a less constrained model than mSUGRA, the pMSSM 
(phenomenological MSSM). This study is made in an inclusive way in the 
framework of the CMS experiment. We first show the ability of 
CMS to discover SUSY in a large domain of pMSSM parameter values. 
We then attempt to estimate the uncertainties in the determination of MSSM 
parameter values using essentially kinematical measurements.
  \end{abstract}

\section{Aim}
The MSSM is a good candidate for the new Physics
expected at the TeV scale. Experiments at both LEP
\cite{lep_susy} and Tevatron \cite{teva_susy} 
have been looking for evidence of SUSY, but for the
moment no signal has been observed. The results of the
searches give limits on sparticle masses. Some indirect
measurements like the $ b \rightarrow s \gamma$  
branching ratio, the anomalous muon magnetic moment g-2
\cite{Battaglia:2001zp,Ellis:2001yu}, dark matter searches \cite{Ellis:2001zk} 
or the Z width also provide constraints on SUSY
parameter values. But all these results still leave  
a large MSSM parameter space unexplored.

The goal of this study is to evaluate the ability of the 
CMS detector \cite{cmstdr} to observe signals of
supersymmetry in a large domain of MSSM parameter values. 
The mSUGRA \cite{Gunion:1998dh,Djouadi:1998di} 
(minimal SUper GRAvity) model, with its only 
five free parameters ($m_{0}$, $m_{1/2}$, $A_{0}$, 
$\tan \beta$ and ${\rm sign}(\mu)$) is very popular 
and has been the subject of many studies up to now. 
A study of mSUGRA, performed in a similar way \cite{Abdullin:1998nv}
to the work presented here, concludes that for both 
low and high values of $ \tan \beta$, and for both 
positive and negative $ \mu$, the mass reach
for gluinos and squarks is up to $ \approx 2.5 $ to $ 2.7\, $ TeV 
for $100 {\rm fb}^{-1}$.

Two reasons motivate us to extend the mSUGRA study to a less 
constrained model, the ``phenomenological" MSSM
(pMSSM). On the one hand, mSUGRA is a rather constrained model, 
very specific and not illustrative of the variety
of all possible supersymmetric models. On the other hand, 
contrary to mSUGRA, the pMSSM has no fixed hierarchy
of masses. In this case, some extreme mass hierarchies could 
show a significantly different kinematical behavior 
than in the case of mSUGRA, which could prevent 
the discovery of supersymmetry even for relatively low
values of the sparticle masses. Moreover, in the case of 
pMSSM we have various types of cascades which produce 
many types of final states, with similar signatures, 
but not the same types of particles. For example, are 
we going to be able to observe supersymmetry with a final state  
containing multiple jets; taus instead of muons and electrons;
and jets produced by c quarks instead of b quarks? The
type of particles produced is really important in order 
to discover supersymmetry in CMS and to identify the SUSY
scenario at work.

In this pMSSM framework, we are going to show that supersymmetry could 
be discovered over a large scale of masses
in the $m_{\tilde{q}}$ versus $m_{\tilde{g}}$ plane. Next we show that 
there are ways to estimate the values of the
MSSM parameters using kinematical quantities measured by the CMS 
detector and event rates. An advantage of this approach
is its model independence -- the only dependence 
comes from the hierarchy of masses \cite{Bityukov:2000zg}. 
Finally, we estimate the statistical uncertainties
due to this method of extraction of the MSSM parameter values.

\section{Theoretical framework}

\subsection{MSSM}

Supersymmetry (SUSY) is a symmetry between fermions and bosons. 
Some of the motivation for SUSY has been reviewed in the Introduction.
The MSSM \cite{Gunion:1998dh,Djouadi:1998di}
is the Supersymmetric extension of the 
Standard Model which introduces the minimal number 
of new particles (only one per SM particle and 4 additional 
Higgs bosons) and no new couplings. The MSSM contains 124 
independent parameters, including the 19 ones of the Standard Model.

\subsubsection{pMSSM (phenomenological MSSM)}

Some phenomenological constraints allow to reduce the number of MSSM free parameters:

\begin{itemize}
\item no new sources of CP violation,
\item no Flavor Changing Neutral Current effects, 
\item universality of the first two generations. 
\end{itemize}
These three constraints leave only 19 free parameters : 

\begin{itemize}
\item $ \tan \beta  $ : ratio of the vacuum expectation values of the two Higgs
doublets fields,
\item $ M_{A} $ : mass of the pseudoscalar Higgs boson,
\item $ \mu  $ : SUSY preserving Higgs mass parameter,
\item $ M_{1} $, $ M_{2} $, $ M_{3} $ : bino, wino and gluino mass parameters,
\item $ M_{\tilde{q}} $ $ M_{\tilde{u_{R}}} $,$ M_{\tilde{d_{R}}} $,
$ M_{\tilde{l}} $, $ M_{\tilde{e_{R}}} $ : unified first and second generation
sfermion masses,
\item $ M_{\tilde{t_{R}}} $, $ M_{\tilde{b_{R}}} $, $ M_{\tilde{Q}} $,
$ M_{\tilde{L}} $, $ M_{\tilde{\tau _{R}}} $ : third generation
sfermion masses,
\item $ A_{t} $, $ A_{b} $, $ A_{\tau } $: third generation trilinear couplings.
\end{itemize}

\subsubsection{A restricted pMSSM }

The model used in our study is a pMSSM, but with a further reduction 
in the number of free parameters. It is an
intermediate model between mSUGRA and the pMSSM, 
relaxing the constraints of mSUGRA but still more constrained than the
pMSSM. Reference \cite{Djouadi:1998di} gives some examples of such models, 
taking into account more constraints than
pMSSM. We take into account, respectively, the mass unification 
of squarks and sleptons (universality of the
three generations of sparticles) assuming that the mixing is 
not too large for the third generation. We also
consider the unification of trilinear coupling $ A_{t} = A_{b} = A_{\tau }$.  
This leads to 9 free parameters:
$ \tan \beta  $, $ M_{A} $, $ \mu  $, $ M_{1} $, $ M_{2} $,
$ M_{\tilde{g}} $, $ M_{\tilde{q}} $, $M_{\tilde{l}} $, 
$ A_{3} $. This constrained model allows us to perform simpler simulation, 
while keeping the diversity of signatures of MSSM events.

\subsection{Examples of signal events}

\subsubsection{An example of MSSM cascade}

\begin{figure}[tb]
\begin{center}
\resizebox*{8cm}{6cm}{\includegraphics{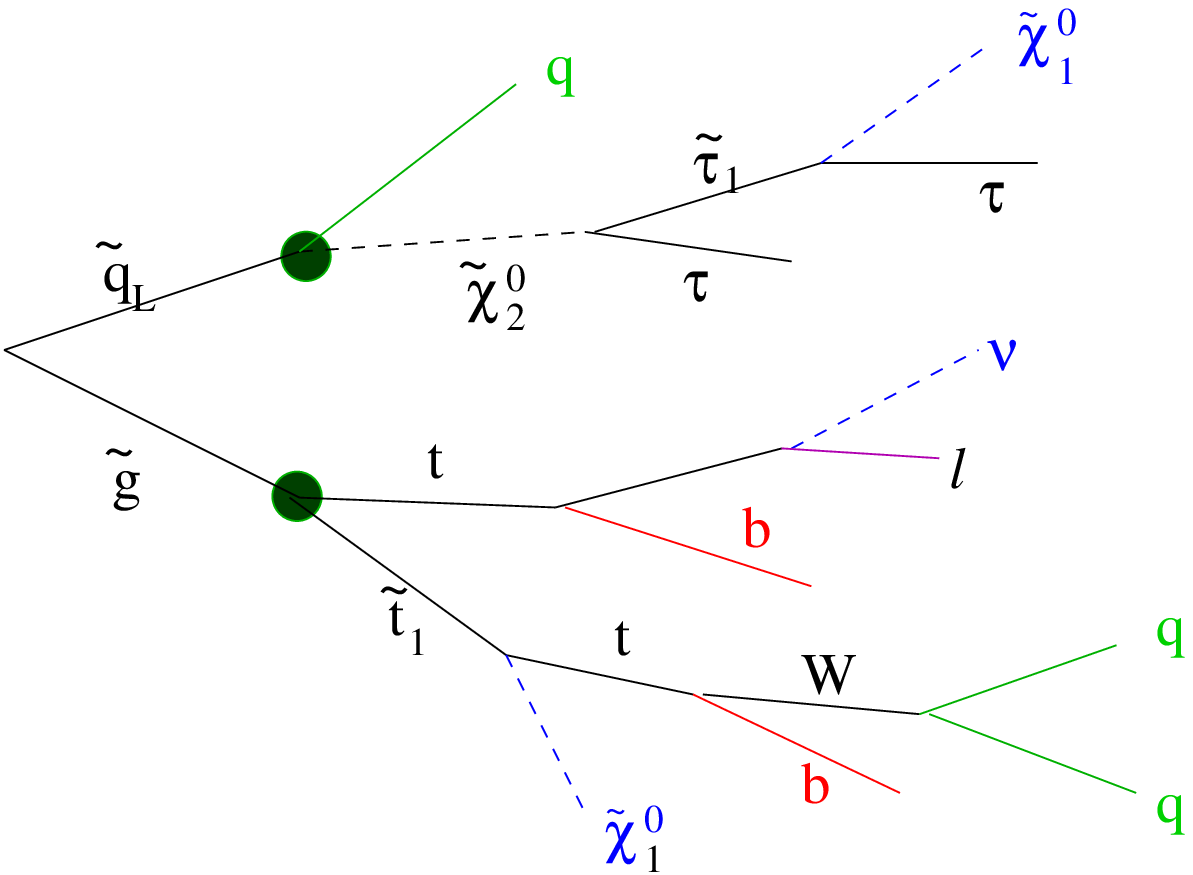}}
\caption{\label{fig: cas1} An example of an MSSM cascade
resulting from $\tilde g \tilde q_L$ production.}
\end{center}
\end{figure}

Figure \ref{fig: cas1} shows an MSSM event of the type 
$ gq\rightarrow \tilde{q_{L}}\tilde{g} $,
with 5 jets including 2 b quark jets and 3 leptons 
including 2 $\tau$'s in the final state. The 2
neutralinos $\chi_1^0$ produce missing transverse 
energy $E_t^{miss}$.

\subsubsection{A more detailed mSUGRA example}

\begin{figure}[tb]
\subfigure[\label{fig: cas2}  Another example of an mSUGRA cascade.]
{\includegraphics*[width=8cm,height=6cm]{pic_hbb.ps}}
\subfigure[\label{fig: geant} GEANT output for the mSUGRA event
shown in Fig.~\ref{fig: cas2}.]
{\includegraphics*[width=8cm,height=8cm]{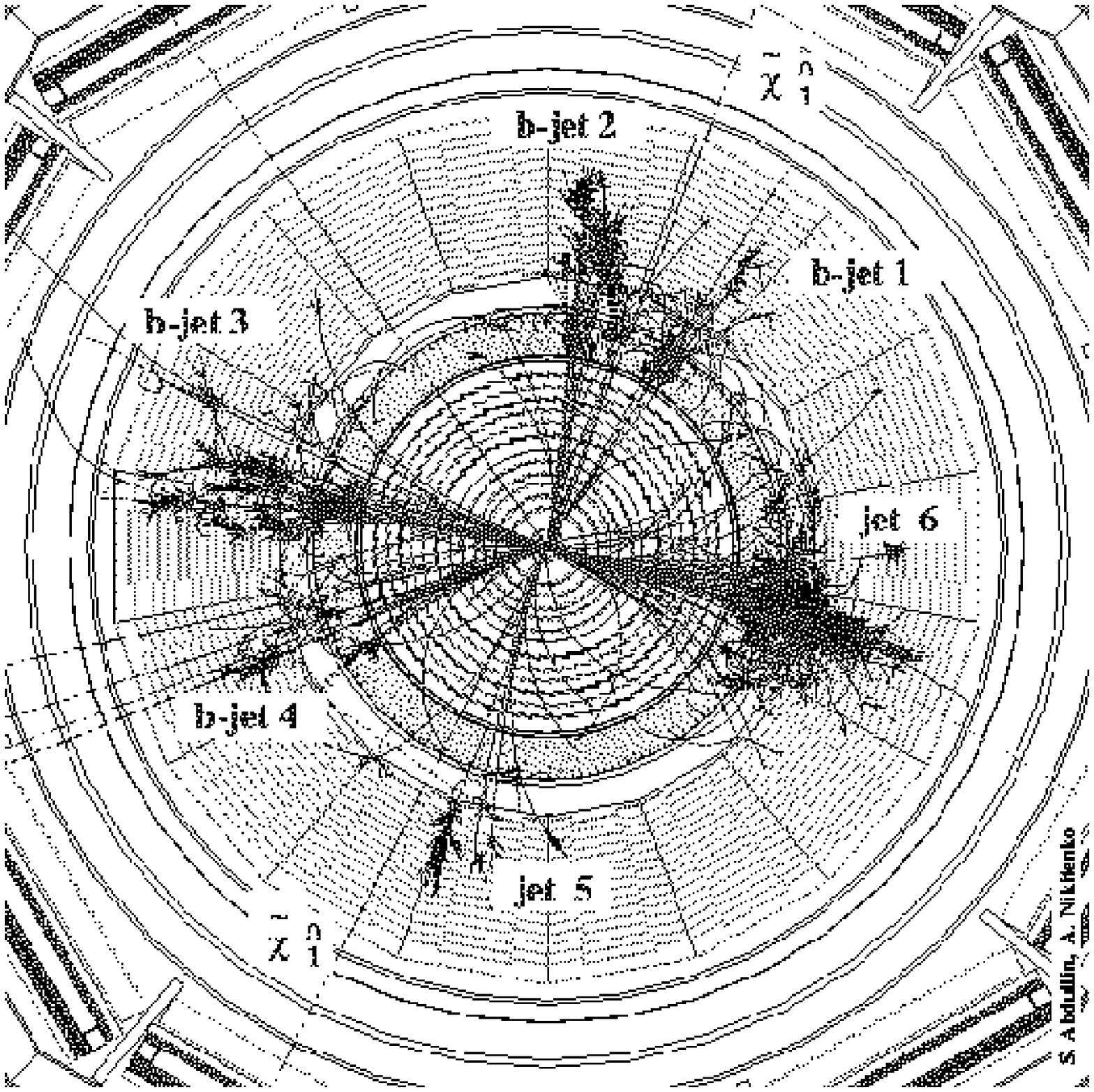}} 
\caption{Another detailed example of an mSUGRA event.}
\end{figure}

An mSUGRA event of the type $ gq \rightarrow \tilde{g}\tilde{q} $
is shown in figure \ref{fig: cas2}, while figure \ref{fig: geant} shows
the corresponding event display in CMS obtained from GEANT\cite{geant}
for this event. We used the following parameter values:

$ m_{0}=1000 $ GeV, $ m_{1/2}=500$ GeV, $ A_{0}=0$, $\tan\beta =35$, $\mu >0 $.

The final state is made of 6 jets including 4 b-quark
jets, and 2 neutralinos which produce $E_t^{miss}$.

\section{Simulation procedure}

\subsection{Signal production}

We use a model with 9 parameters, which make up a hyperspace in 9 dimensions.
To simplify the analysis we use a discretization of the parameter values. The
choice of the number of values for each parameter depends on the parameter sensitivity.
We used a grid for squark and gluino masses with 9 values evenly spaced between
600 and 3000 GeV, because the event characteristics at LHC depend primarily
on these two masses. On the other hand, many observables are not very 
strongly dependent on the parameter $\tan\beta$. We thus use only two values,
to distinguish the behavior at large and small values of this parameter. 
The values selected for each parameter in this analysis are the following: 

\begin{itemize}
\item $ M_{\tilde{q}} $ : $ 600,\, 900,\, 1200,\, 1500,\, 1800,\, 2100,\, 2400,\, 2700,\, 3000 $
 GeV 
\item $ M_{\tilde{g}} $ : $ 600,\, 900,\, 1200,\, 1500,\, 1800,\, 2100,\, 2400,\, 2700,\, 3000 $
 GeV 
\item $ M_{\tilde{l}} $ : $ 200,\, 1000,\, 3000 $  GeV 
\item $ M_{1} $ : $ 100,\, 500,\, 1000,\, 2000 $ GeV 
\item $ M_{2} $ : $ 100,\, 500,\, 1000,\, 2000 $ GeV 
\item $ M_{A} $ : $ 200,\, 1000,\, 3000 $ GeV 
\item $ A_{3} $ : $ 0,\, 2000 $  GeV 
\item $ \mu  $ : $ 200,\, 500,\, 2000 $ GeV 
\item $ \tan \beta  $ : $ 2,\, 50 $
\end{itemize}

We end up with a total of 140000 different sets of parameter values. For each
set, we generate 1000 events, a compromise between the limits imposed by the
handling of the data flow and sufficiently small statistical errors. The theoretical
and experimental constraints make it possible to reduce the number of combinations
to a total of $ 17.10^{3} $. 

These imposed constraints are the following:

\begin{itemize}
\item constraint on the Higgs mass  $ > \, 100\,$ GeV, 
\item lightest chargino mass   $ > \,100\,$ GeV, 
\item lightest neutralino mass $ > \, 50\,$ GeV,
\item the lightest neutralino is the LSP.
\end{itemize}

\begin{figure}[tb]
\begin{center}
\resizebox*{10cm}{12cm}{\includegraphics{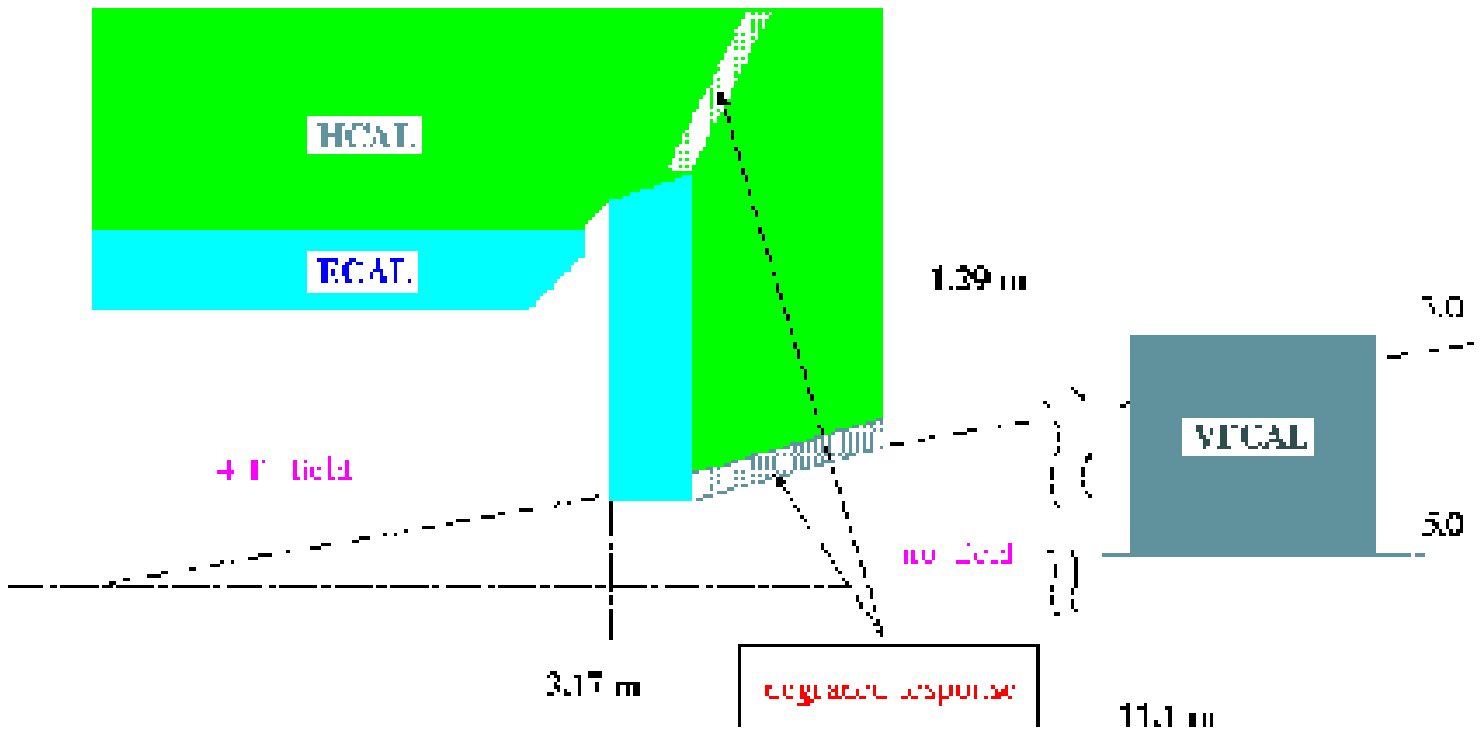}}
\vspace{-2cm}
{
\unitlength=1.5 pt
\SetScale{1.5}
\SetWidth{0.7}      
\normalsize    
{} \qquad\allowbreak
\begin{picture}(200,70)(0,30)
\CBoxc(25,50)(30,20){Black}{ProcessBlue}
\CBoxc(75,60)(25,15){Black}{ProcessBlue}
\CBoxc(75,40)(25,15){Black}{ProcessBlue}
\ArrowLine(40,50)(62.5,60)
\ArrowLine(40,50)(62.5,40)
\Text(25,50)[]{ECAL}
\Text(75,60)[]{barrel}
\Text(75,40)[]{endcap}
\Text(100,60)[l]{$\sigma/E = 5.7\%/\sqrt{E}+0.55\%+0.915/E$} 
\Text(130,50)[l]{\footnotesize no projective crack} 
\Text(100,40)[l]{$\sigma/E = 2.7\%/\sqrt{E}+0.55\%+0.210/E$} 
\end{picture}
{} \allowbreak
\vspace{-1cm}
\begin{picture}(200,70)(0,30)
\CBoxc(25,50)(30,20){Black}{Green}
\CBoxc(75,60)(25,15){Black}{Green}
\CBoxc(75,40)(25,15){Black}{Green}
\ArrowLine(40,50)(62.5,60)
\ArrowLine(40,50)(62.5,40)
\Text(25,50)[]{HCAL}
\Text(75,60)[]{hadr.}
\Text(75,40)[]{crack}
\Text(100,50)[l]{$\sigma/E = f(\eta)\quad$ HCAL TDR, Fig.~1.43, p.60 } 
\end{picture}
{} \allowbreak
\vspace{-1cm}
\begin{picture}(200,70)(0,30)
\CBoxc(25,50)(30,20){Black}{Gray}
\CBoxc(75,60)(25,15){Black}{Gray}
\CBoxc(75,40)(25,15){Black}{Gray}
\ArrowLine(40,50)(62.5,60)
\ArrowLine(40,50)(62.5,40)
\Text(25,50)[]{VFCAL}
\Text(75,60)[]{hadr.}
\Text(75,40)[]{$e,\gamma$}
\Text(100,60)[l]{$\sigma/E = 182\%/\sqrt{E}+9\%$} 
\Text(110,50)[l]{\footnotesize D.~Litvintsev, CMS Internal Note/1996-03} 
\Text(100,40)[l]{$\sigma/E = 138\%/\sqrt{E}+5\%$} 
\end{picture}
}
\caption{Details of CMSJET characteristics.}
\label{fig:cmsjet}
\end{center}
\end{figure}

Signal events were generated using the ISAJET 
program \cite{ISAJET}.  The CMS detector
response was obtained from the fast Monte-Carlo 
code (non-Geant) CMSJET 4.51 \cite{cmsjet}.
Characteristics of CMSJET software are given in figure \ref{fig:cmsjet}. 

\subsection{Background production}
The background production to this $\tilde{g}$, $\tilde{q}$ 
SUSY search was estimated using Standard Model events
leading to similar signatures as the MSSM events. 
The background was produced using  PYTHIA \cite{PYTHIA}. We
consider here SUSY signals with the following event characteristics: 

\begin{itemize} 
\item production from 0 to n isolated leptons (electrons or muons),  
\item a large value for the average missing transverse energy ($>200$ GeV),  
\item more than 2 jets with large transverse energy ($> 40 $ GeV).  
\end{itemize}

\begin{figure}[tb]
\begin{center} 
\resizebox*{17cm}{6cm}{\includegraphics{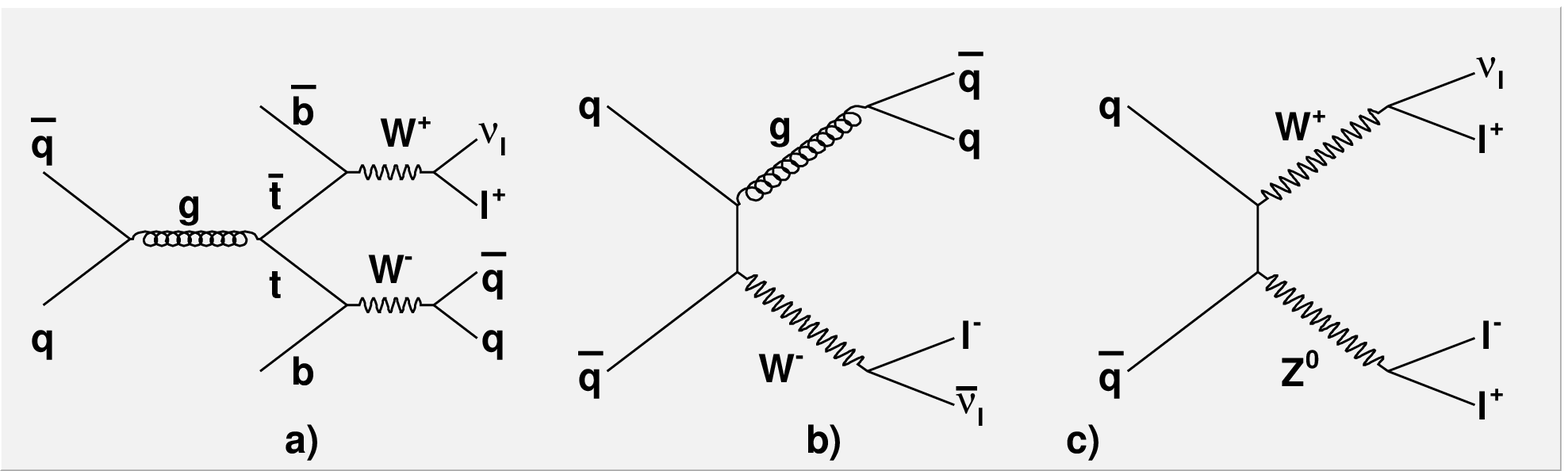}} 
\caption{Feynman diagrams of a few background examples. \label{fig:fe}}
\end{center}
\end{figure}

Therefore we must consider as potential Standard Model backgrounds 
all processes with large rest masses which yield
large transverse missing energy, producing energetic jets, 
and possible isolated leptons. Thus the backgrounds
we consider are the following: 
$ pp\, \rightarrow \, t\overline{t},W+jet,Z+jet,WW,ZZ,ZW $. 
We also consider $
QCD $ events with several high energy jets, including  
heavy flavors (b and c). The missing transverse energy in
this type of event originates either from semi-leptonic 
b,c decay or from imperfections and fluctuations in the
response of the detector which may fake missing transverse 
energy.   Some of the background mechanisms are shown
in figure \ref{fig:fe}. The primary contributions to the background
after cuts is at large values of the transverse momentum 
$\hat p_T$ of the produced particles in the 2-body final state.
To obtain sufficient statistics, we generate events independently 
for several intervals of $\hat p_T $
(see table \ref{fig: table1}). A total of two hundred million 
events have been generated. 

\begin{table}[tb]
\begin{center}
\begin{tabular}{|c|c|c|c|c|}
\hline 
processes&  $\hat p_T$ interval (GeV)& $ \sigma  $ (pb)& N$ _{ev} $ generated&
\% of needed for $100\ {\rm fb}^{-1} $\\
\hline 
\hline 
$ t\overline{t} $ 
& $ \begin{array}{c}
0-100\\
100-200\\
200-400\\
400-800\\
>800
\end{array} $&
$ \begin{array}{c}
267\\
240\\
80.7\\
6.3\\
0.163
\end{array} $&
$ \begin{array}{c}
1.461\cdot 10^{7}\\
6.638\cdot 10^{6}\\
6.864\cdot 10^{6}\\
6.484\cdot 10^{5}\\
1.630\cdot 10^{4}
\end{array} $&
$ \begin{array}{c}
54.7\\
27.7\\
85.1\\
102.9\\
100.0
\end{array} $\\
\hline 
\hline 
$ Wj $ &
$ \begin{array}{c}
50-100\\
100-200\\
200-400\\
400-800\\
>800
\end{array} $&
$ \begin{array}{c}
7140\\
1470\\
155\\
9.5\\
0.33
\end{array} $&
$ \begin{array}{c}
2.753\cdot 10^{7}\\
8.618\cdot 10^{6}\\
6.424\cdot 10^{6}\\
9.909\cdot 10^{5}\\
3.300\cdot 10^{4}
\end{array} $&
$ \begin{array}{c}
3.9\\
5.9\\
41.4\\
104.3\\
100.0
\end{array} $\\
\hline 
\hline 
$ Zj $&
$ \begin{array}{c}
50-100\\
100-200\\
200-400\\
400-800\\
>800
\end{array} $&
$ \begin{array}{c}
2670\\
580\\
64.0\\
4.0\\
0.137
\end{array} $&
$ \begin{array}{c}
1.554\cdot 10^{7}\\
9.998\cdot 10^{6}\\
4.455\cdot 10^{6}\\
4.927\cdot 10^{5}\\
1.370\cdot 10^{4}
\end{array} $&
$ \begin{array}{c}
5.8\\
17.2\\
71.2\\
123.2\\
100.0
\end{array} $\\
\hline 
\hline 
$ QCD $&
$ \begin{array}{c}
100-200\\
200-400\\
400-800\\
800-1200\\
>1200
\end{array} $&
$ \begin{array}{c}
1.37\cdot 10^{6}\\
7.15\cdot 10^{4}\\
2740\\
60.0\\
4.8
\end{array} $&
$ \begin{array}{c}
6.000\cdot 10^{7}\\
3.229\cdot 10^{7}\\
3.259\cdot 10^{7}\\
6.033\cdot 10^{6}\\
4.947\cdot 10^{5}
\end{array} $&
$ \begin{array}{c}
0.04\\
0.45\\
11.9\\
100.0\\
103.1
\end{array} $\\
\hline 
\hline 
total&
&
&
$ 2.342\cdot 10^{8} $&
\multicolumn{1}{|c|}{}\\
\hline 
\end{tabular}
\caption{Background repartition by $\hat{p}_{T}$ 
interval. For each $\hat{p}_{T}$
interval we give the process cross-section in pb 
and the number of generated events. 
The last column shows the percentage of events we have
generated compared to the expected number of events 
for $ 100\ {\rm fb}^{-1} $ of integrated luminosity.
\label{fig: table1}}

\end{center}
\end{table} 

The QCD event sample generated for low $\hat{p}_T$ is tiny compared 
to the required one but fortunately, there is correlation between 
$\hat{p}_T$ and the maximal produced $E_T^{miss}$ value, 
so one does not expect high values of $E_T^{miss}$ for low $\hat{p}_T$.
To be confident in the simulation, we apply some 
preliminary cuts during the generation:

\begin{itemize}
\item $E_T^{miss} > 200 $ GeV limit, below 
which the QCD jet background become dominant;
\item at least 2 jets with $E_T^{jet} > 40 $ GeV in $|\eta^{jet}| < 3$.
\end{itemize} 

The isolation of the leptons  is given by the following requirement
\begin{itemize}
\item  muon with $p_T^{\mu} > 10 $ GeV within the 
muon acceptance or electron with $p_T^e > 20 $ GeV within $|\eta^e| <
2.4 $;
\item no charged particles with $p_T > 2 $ GeV in a 
cone of $R = 0.3$ around the direction of the lepton;
\item $\sum E_T^{cell}$ in a cone ring $0.05 < R < 0.3$ 
around the lepton impact point has to be less than  10\% of the
lepton transverse energy.
\end{itemize}

\subsubsection{Pile-up}

\begin{figure}[tb]
\begin{center}
\resizebox*{12cm}{12cm}{\includegraphics{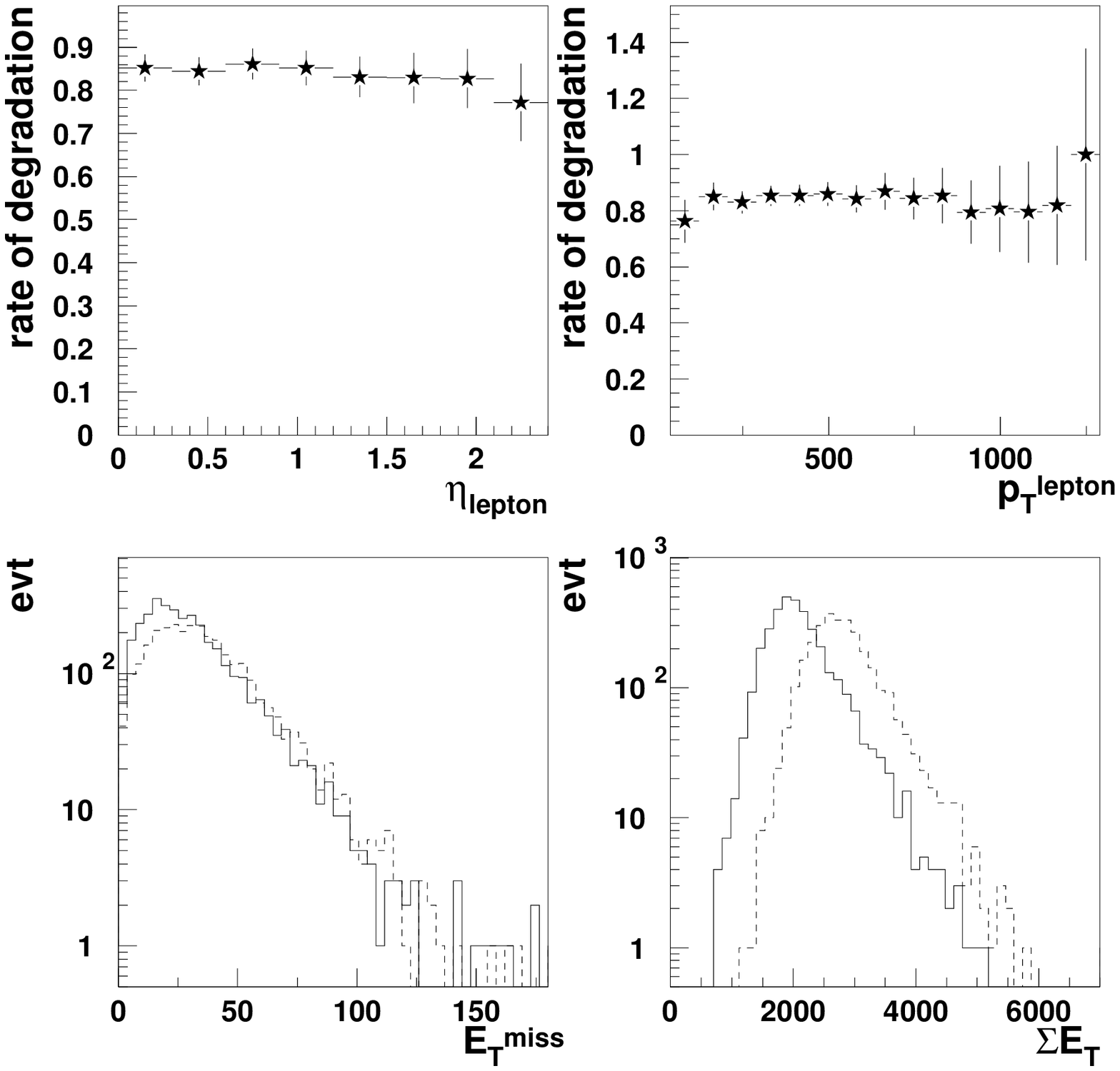}} 
\caption{Importance of pile-up for lepton isolation 
and energy measurements. \label{fig: pile up}}
\end{center}
\end{figure}

We also take into account event pile-up, i.e. 25 inelastic 
$pp$ events on average per bunch crossing with a
Poisson distribution. The two upper graphs in figure \ref{fig: pile up} 
illustrate the ratio between  the
lepton isolation efficiency with and without pileup, 
as a function of pseudorapidity, and transverse
momentum (the definition of lepton isolation is given 
in \cite{cmsjet}). The efficiency is reduced  to 85\% due
to the multiplicity of particles produced in each bunch crossing. 
In the two lower plots giving the event
missing transverse energy, and the scalar sum of the event 
transverse energy, the  solid curves
are without pile-up and the dashed ones with pile-up. 
Pile-up does not make a very significant difference for total 
missing transverse energy, but increases the total transverse energy.

\subsection{Different selection criteria}

To distinguish signal from background, we are led to apply kinematical cuts
on the observables we extract from the CMS detector.

Table~\ref{tab:cut} gives all the different values we use for the selection.
This yields approximately 10000 combinations of cuts to optimize 
signal to background ratio.

\subsection{Signal significance estimator}

We make a systematic search for all sets of parameters and thus define limits
of discovery through calculations of the significance of the signals. The definition
of the significance we use is the
following one:
\begin{equation}
\mbox{significance }=\frac{S}{\sqrt{B}},
\end{equation}
where $ S $ is the number of signal events and $ B $ the number of background
events. A significance exceeding 5 indicates that the corresponding set of MSSM
parameter values is experimentally accessible.  
In order to optimize the significance, we used some cuts which are listed in the Table
\ref{tab:cut}.

\begin{table}[!htb]
\centering
\small
\begin{tabular}{|c|c|}
\hline
parameter & different cut values \\
\hline
Number of jets &  2, 3, 4, 5, 6, 7, 8, 9, 10 jets \\
(with a minimal transverse energy of 40 GeV) & \\ 
\hline
Transverse momentum of the highest energy jet &  40, 150, 300, 400, 500,
600, 700, 800, 900, 1000 GeV  \\
\hline
Transverse momentum of the second highest energy jet &  40, 80, 200,
300, 400, 500 GeV  \\
\hline
Missing transverse momentum & 200, 300, 400, 500, 600, 700, 800, \\
& 900, 1000, 1100, 1200, 1300, 1400 GeV  \\
\hline
Total transverse momentum & 700, 900, 1100, 1300, 1500, 1700, 1900,
2100, \\
&2300, 2500, 2700, 2900, 3100, 3300, 3500 GeV  \\
\hline
Angle $ \Phi  $ between the missing transverse & 0, 20 degrees \\
momentum and the momentum of the isolated lepton & \\ 
\hline
\end{tabular}
\caption{All sets of cut values for significance optimization.}
\label{tab:cut}
\end{table}

\begin{table}[!htb]
\begin{center}
\begin{tabular}{|c|c|c|}
\hline
& before cuts & after cuts \\
\hline
number of signal events & 6152 & 431 \\
\hline
number of background events & 240 $10^6$ & 124 \\
\hline
significance & 0.397 & 38.614 \\
\hline
\end{tabular}
\caption{An example of the effect of cuts on the number of signal and background
events. We give also the significance obtained. The MSSM parameter values are
the following:
$M_{\tilde{l}} = 1000$ GeV, $M_1 = 500$ GeV, $M_2= 1000$ GeV, $M_{\tilde{g}}
= 2700$ GeV, $M_{\tilde{q}} = 900$ GeV, $M_{A} = 200$ GeV, $\tan \beta =
50$, $\mu = 500$ GeV, $A_3 = 0$ GeV.}
\label{tab:compsig1}
\end{center}
\end{table}

\begin{table}[!htb]
\begin{center}
\begin{tabular}{|c|c|c|}
\hline
& before cuts & after cuts \\
\hline
number of signal events & 6121 & 355 \\
\hline
number of background events & 240 $10^6$ & 112 \\
\hline
significance & 0.395 & 33,5 \\
\hline
\end{tabular}
\caption{An example of the effect of cuts on the number of signal and background
events. We give also the significance obtained. The MSSM parameter values are
the following:
$M_{\tilde{l}} = 1000$ GeV, $M_1 = 100$ GeV, $M_2= 500$ GeV, $M_{\tilde{g}}
= 2100$ GeV, $M_{\tilde{q}} = 2100$ GeV, $M_{A} = 1000$ GeV, $\tan \beta = 50$, 
$\mu = 2000$ GeV, $A_3 = 2000$ GeV, which correspond to Fig.~\ref{fig: S/N 2}.}
\label{tab:compsig2}
\end{center}
\end{table}

To show the importance of cuts to separate the signal from the background, tables
\ref{tab:compsig1} and \ref{tab:compsig2} give examples of  the number of events for
signal and background, before and after cuts. We notice the very important
effect of the cuts, the number of background events decreasing from $2.4\,10^6$
to $\approx 100$, i.e. by a factor roughly equal to 
$2\times10^6$, and the significance increasing by a factor $\approx$ 100.

\section{Analysis}
\subsection{Calculation of significance}
\subsubsection{Illustration of the analysis for some specific parameter values}

\begin{figure}[tbp]
\vskip -1.8cm
\subfigure[Sample without specific hierarchy. $ M_{\tilde l}=1000 $ GeV,  
$ M_{1}=100$ GeV, $ M_{2}=500$ GeV, $ M_{\tilde{q}}=2100$ GeV, 
$ M_{\tilde{g}}=2100$ GeV, $ M_{A}=200$ GeV, $ \tan \beta =50,$  
$\mu =2000$ GeV, $ A_{3}=2000$ GeV. \label{fig: S/N 2}]
{\includegraphics*[width=8.0cm,height=10cm]{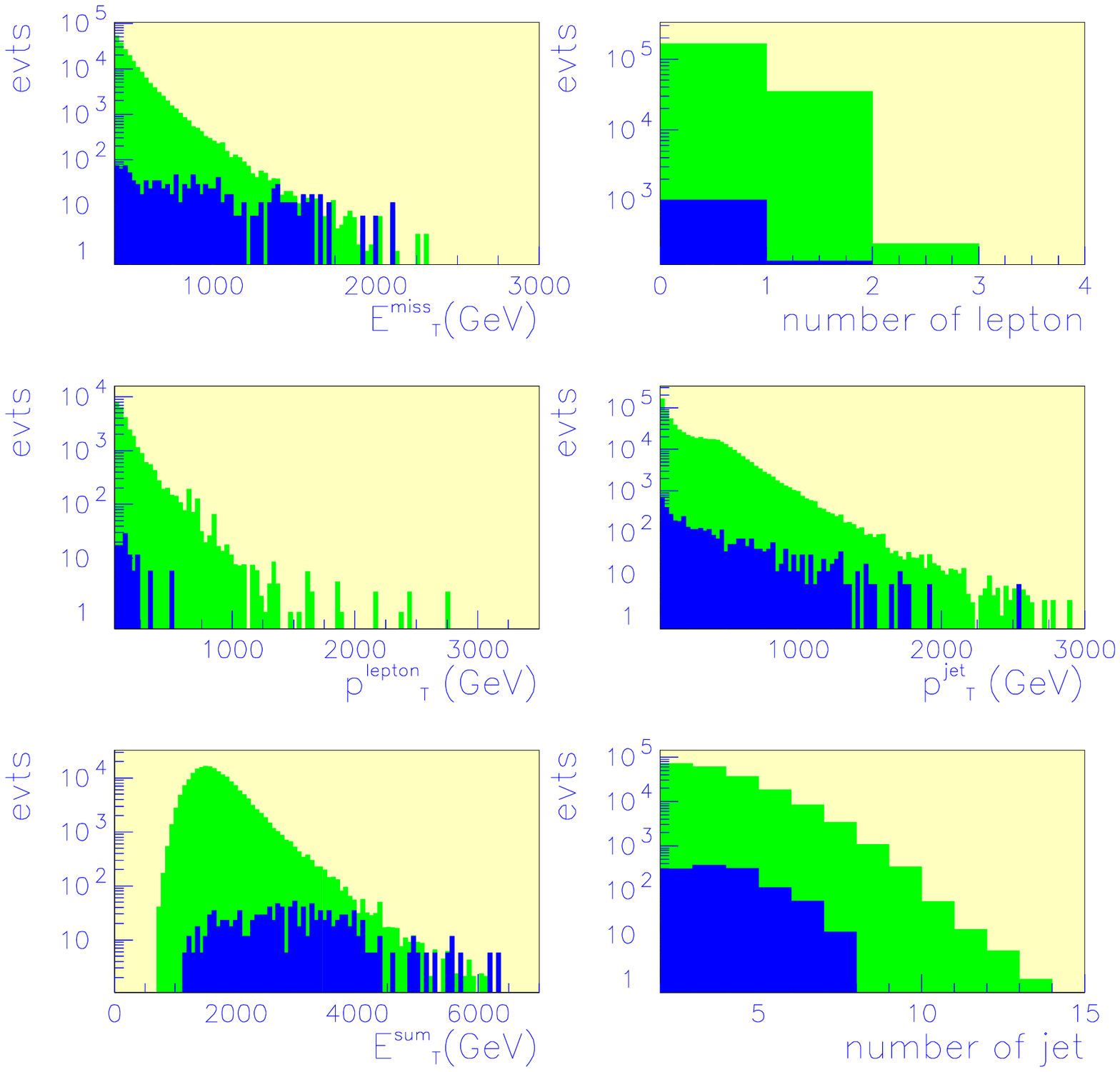}} 
\subfigure[Sample with broad hierarchy of masses. 
$ M_{\tilde{l}}=2000$ GeV, $ M_{1}=500 $ GeV $ M_{2}=500 $ GeV, 
$ M_{\tilde{g}}=2000 $ GeV, $ M_{\tilde{q}}=2000$ GeV 
$ M_{A}=1000$ GeV,  $ \tan \beta =50, $ $ \mu =200$ GeV, 
$ A_{3}=0$ GeV. \label{fig: masse 1}]
{\includegraphics[width=8.0cm,height=9.5cm]{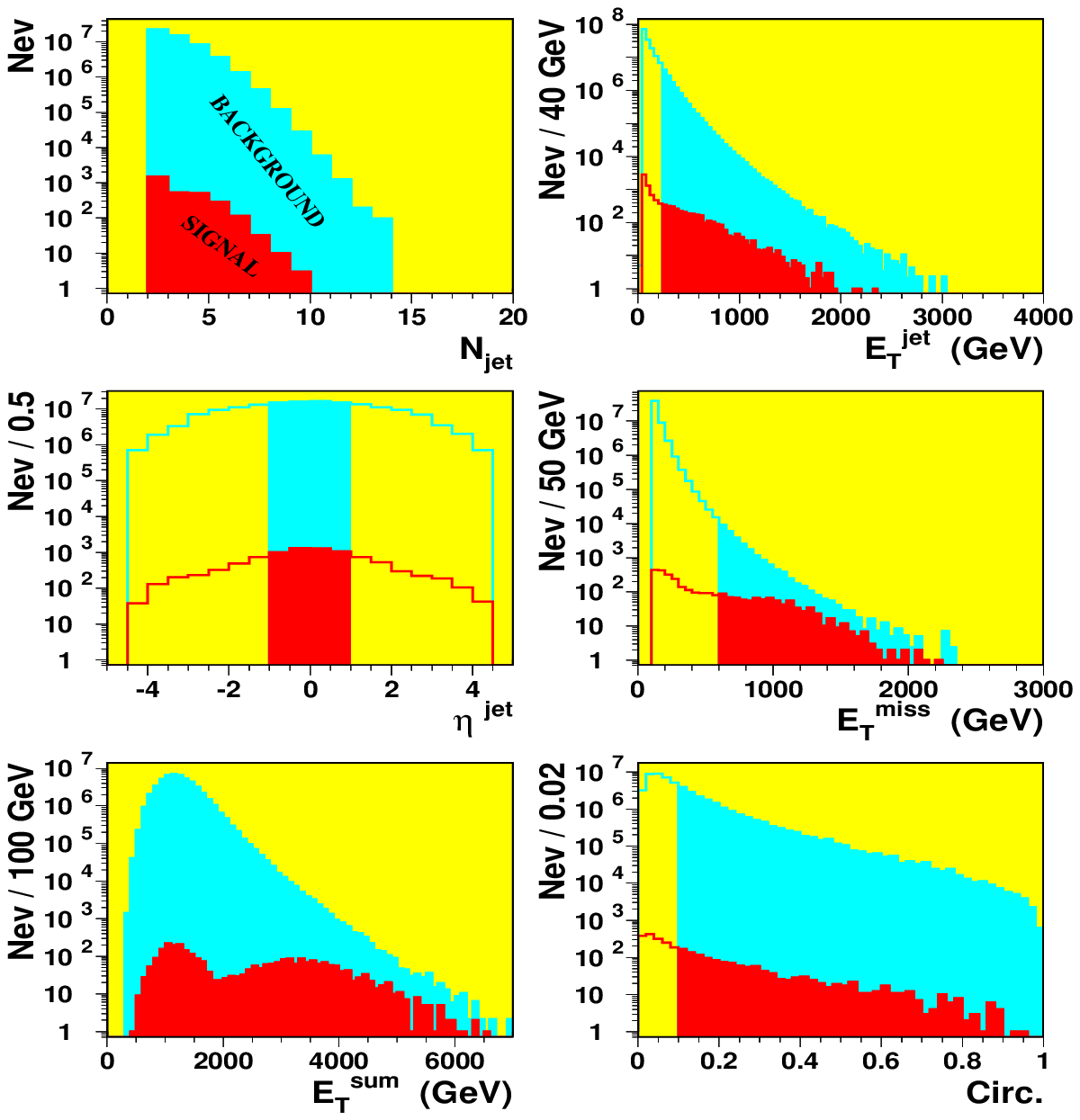}} 
\vskip -1.0cm
\subfigure[Sample with narrow hierarchy of masses, 
low value. $ M_{\tilde{l}}=1500$ GeV,  $ M_{1}=940$ GeV, 
$ M_{2}=2000$ GeV, $ M_{\tilde{g}}=1000$ GeV, 
$ M_{\tilde{q}}=1020$ GeV, $ M_{A}=1000$ GeV, 
$ \tan \beta =50,$ $ \mu =1050$ GeV, $ A_{3}=0$ GeV. 
\label{fig: masse 2}]{\includegraphics[width=8.0cm,height=6.5cm]{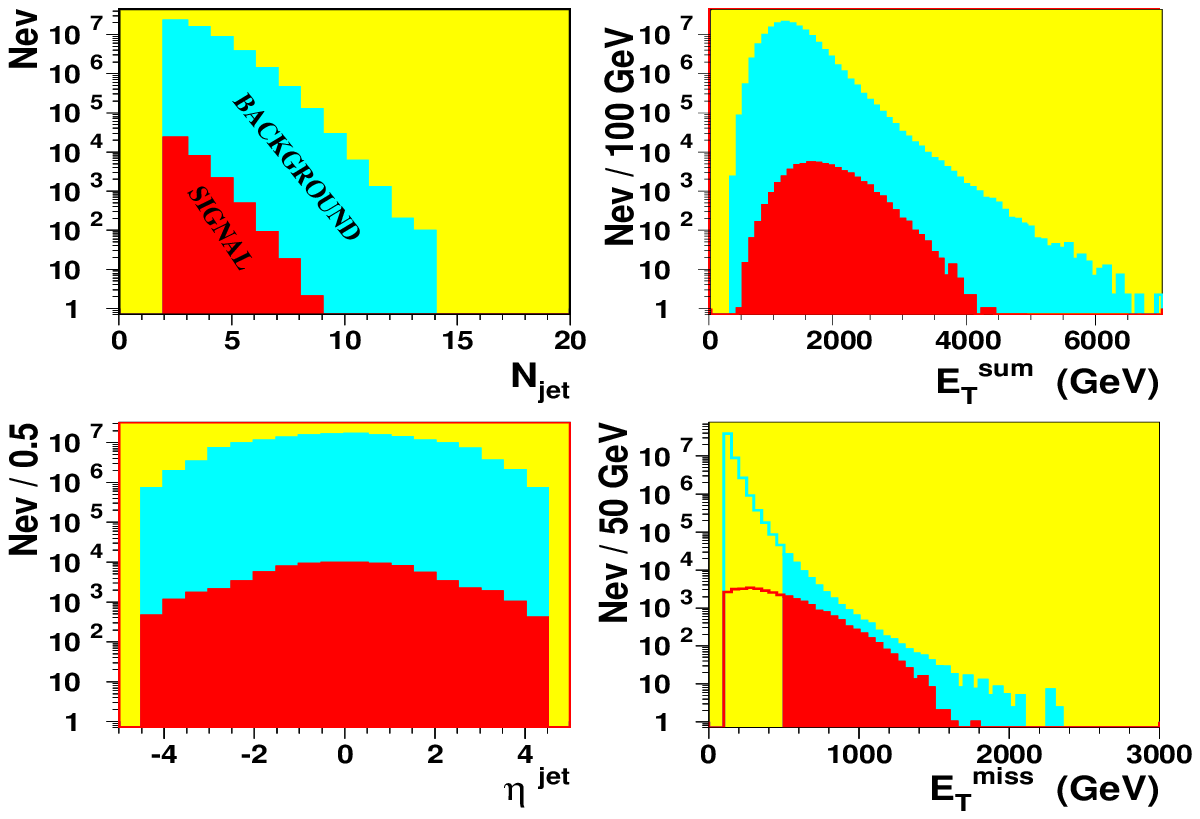}}
\subfigure[Sample with narrow hierarchy of masses, 
medium value. $ M_{\tilde{l}}=1520$ GeV, 
$ M_{1}=1450$ GeV, $ M_{2}=2000$ GeV, 
$ M_{\tilde{g}}=1500$ GeV, 
$ M_{\tilde{q}}=1520$ GeV, $ M_{A}=1000$ GeV, 
$\tan \beta =50,$ $ \mu =1500$ GeV, $ A_{3}=0$ GeV. 
\label{fig:masse3}]{\includegraphics[width=8.2cm,height=9.5cm]{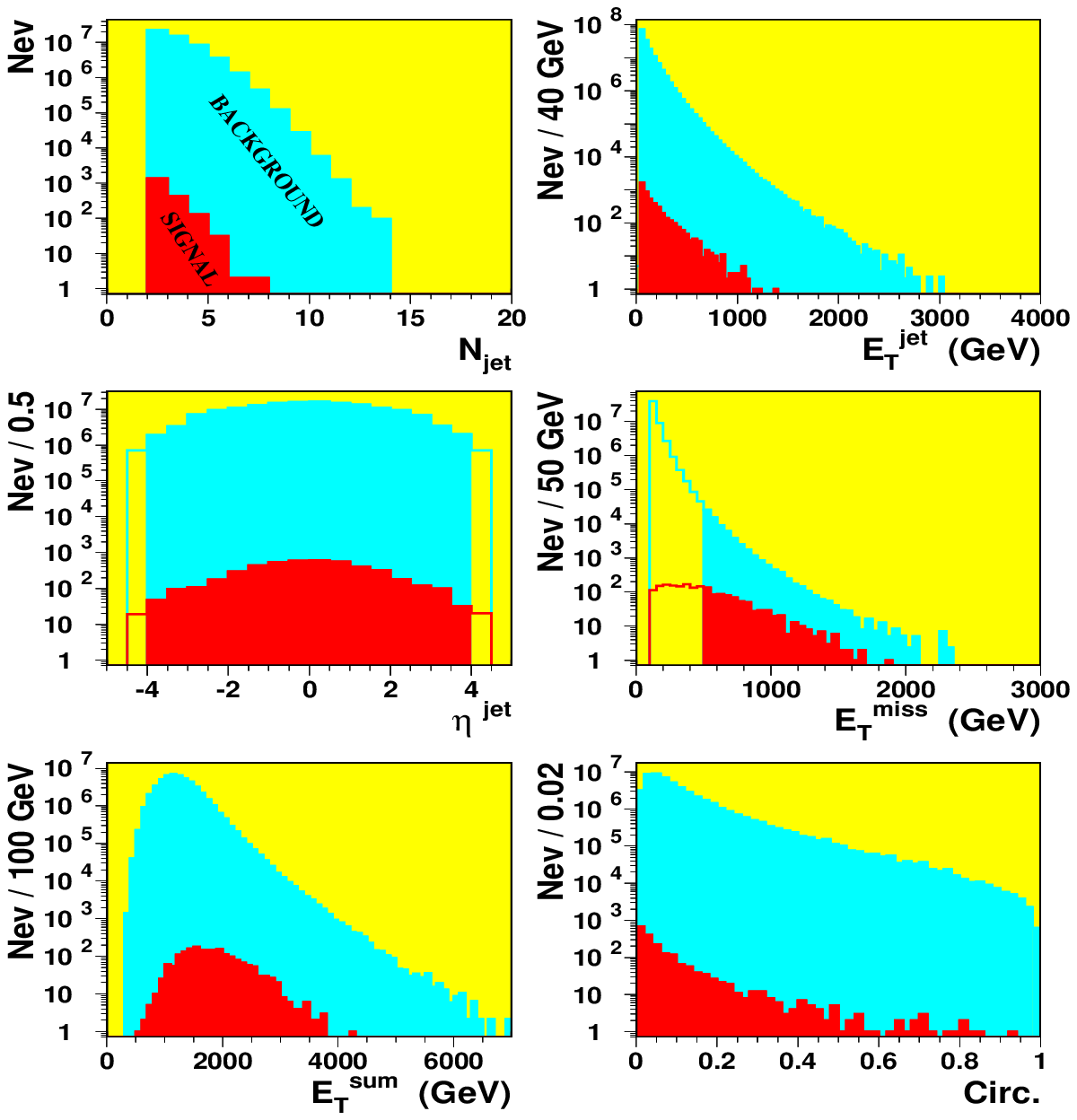}} 
\caption{Distribution of signal and background 
for different observables before cuts.} 
\end{figure}

Figure \ref{fig: S/N 2}  gives distributions of signal
(in black) and backgrounds (in gray) for some kinematical 
quantities before any cuts are applied; the 
signal is not easily distinguishable from the background 
at this stage, as the cross section is too much
smaller. But quantities such as $E_t^{miss}$ and $E_t^{sum}$
have a very different shape, thus cutting on these
variables would greatly enhance the signal to background ratio. 

The specific example shown in figure \ref{fig: S/N 2}
corresponds to the following
values of MSSM parameters:
\begin{eqnarray}
&&M_{\tilde l}=1000\ \gev, M_{1}=100\ \gev, 
M_{2}= 500\ \gev, M_{\tilde{q}}=2100\ \gev,
M_{\tilde{g}}=2100\ \gev, \nonumber \\
&&M_{A}=200\ \gev,  \mu =2000\ \gev, 
A_{3}=2000\ \gev, \tan \beta =50.
\end{eqnarray}
The cross section for this set of parameters is $ \sigma =58\ {\rm fb} $, and the
significance after applying all cuts is equal to 33.

\subsubsection{Example with either very broad or very narrow hierarchy of masses}

We have investigated in some detail one of the major 
points of difference  between mSUGRA and pMSSM, namely
the non fixed hierarchy of masses in case of pMSSM. 
In the first example (figure \ref{fig: masse 1}), with a very broad mass
spectrum, the masses of neutralinos are chosen to be 
much lower than the masses of squarks, gluinos
and sleptons. Sparticles production is therefore 
dominated by neutralinos and charginos.
The specific parameter values are the following:
\begin{eqnarray}
&&M_{\tilde{l}}=2000\ \gev, 
M_{1}=500\ \gev,
M_{2}=500\ \gev, 
M_{\tilde{g}}=2000\ \gev,
M_{\tilde{q}}=2000\ \gev, \nonumber \\
&& M_{A}=1000\ \gev,  \mu =200\ \gev, A_{3}=0\ \gev, \tan \beta =50. 
\end{eqnarray}
The cross section for this set of parameters is $ \sigma =1.22$ pb  and
despite the abundance of neutralinos and the low production rate of gluinos and
squarks, one is still able to obtain after appropriate cuts 
(discussed in the following section) a significance equal to $ 10.2 $. 


As a second example (figure \ref{fig: masse 2}) we chose 
a case where the masses of the neutralinos, 
gluinos, squarks and sleptons are comparable. 
The specific parameter values are the following: 
\begin{eqnarray}
&&M_{\tilde{l}}=1500\ \gev,  
M_{1}=940\ \gev, 
M_{2}=2000\ \gev, 
M_{\tilde{g}}=1000\ \gev, 
M_{\tilde{q}}=1020\ \gev, \nonumber \\
&&M_{A}=1000\ \gev, 
\mu =1050\ \gev, A_{3}=0\ \gev, \tan \beta =50.
\end{eqnarray}
SUSY production now mostly proceeds via gluinos 
and squarks with a cross section $ \sigma =2.014$ pb 
and a significance equal to $ 36.3 $ after selection cuts.


For the third example (figure \ref{fig:masse3}) 
the parameter values are the following: 
\begin{eqnarray}
&&M_{\tilde{l}}=1520\ \gev, 
M_{1}=1450\ \gev, 
M_{2}=2000\ \gev, 
M_{\tilde{g}}=1500\ \gev, 
M_{\tilde{q}}=1520\ \gev, \nonumber \\
&&M_{A}=1000\ \gev, 
\mu =1500\ \gev, A_{3}=0\ \gev,
\tan \beta = 50.
\end{eqnarray}
The masses of the neutralinos, gluinos, squarks 
and sleptons are similar again but heavier
than in the previous example. The main sparticle 
production proceeds still via gluinos and
squarks, with, in this case, a cross section $ \sigma =0.126$ pb and a
significance equal to $ 3.2 $ after cuts are applied.


Even for the sets of parameter values 
which would seem difficult (either very
similar masses or on the contrary very broad span of masses), 
applying cuts allows 
to obtain good results. However, in the case of a small 
spread of sparticle masses, the discovery limit is
about 1.5 TeV instead of 2.5 TeV as obtained with the usual 
mSUGRA-type mass hierarchies.

\subsubsection{Discovery limits}

We now generalize this study to determine a limit of discovery for
the MSSM. For each MSSM point, we are looking for the set of cuts
which gives the maximum value for the significance.
With these collected values of the maximal significance, 
we can draw the discovery limits (estimated as the isocurve 
of a significance equal to 5).
The significances are calculated for an integrated luminosity 
$ \int Ldt=100\ {\rm fb}^{-1}$ corresponding to one year of LHC 
at high luminosity. Our first result  
is the isocurve of significance equal to 5, given in the plane
$(M_{\tilde{q}},M_{\tilde{g}})$,
which are the two most important parameters. The 7 other
parameters have the same fixed value for all the 
MSSM points of the plane. 
\begin{figure}[tb]
\begin{center}
\resizebox*{10cm}{10cm}{\includegraphics{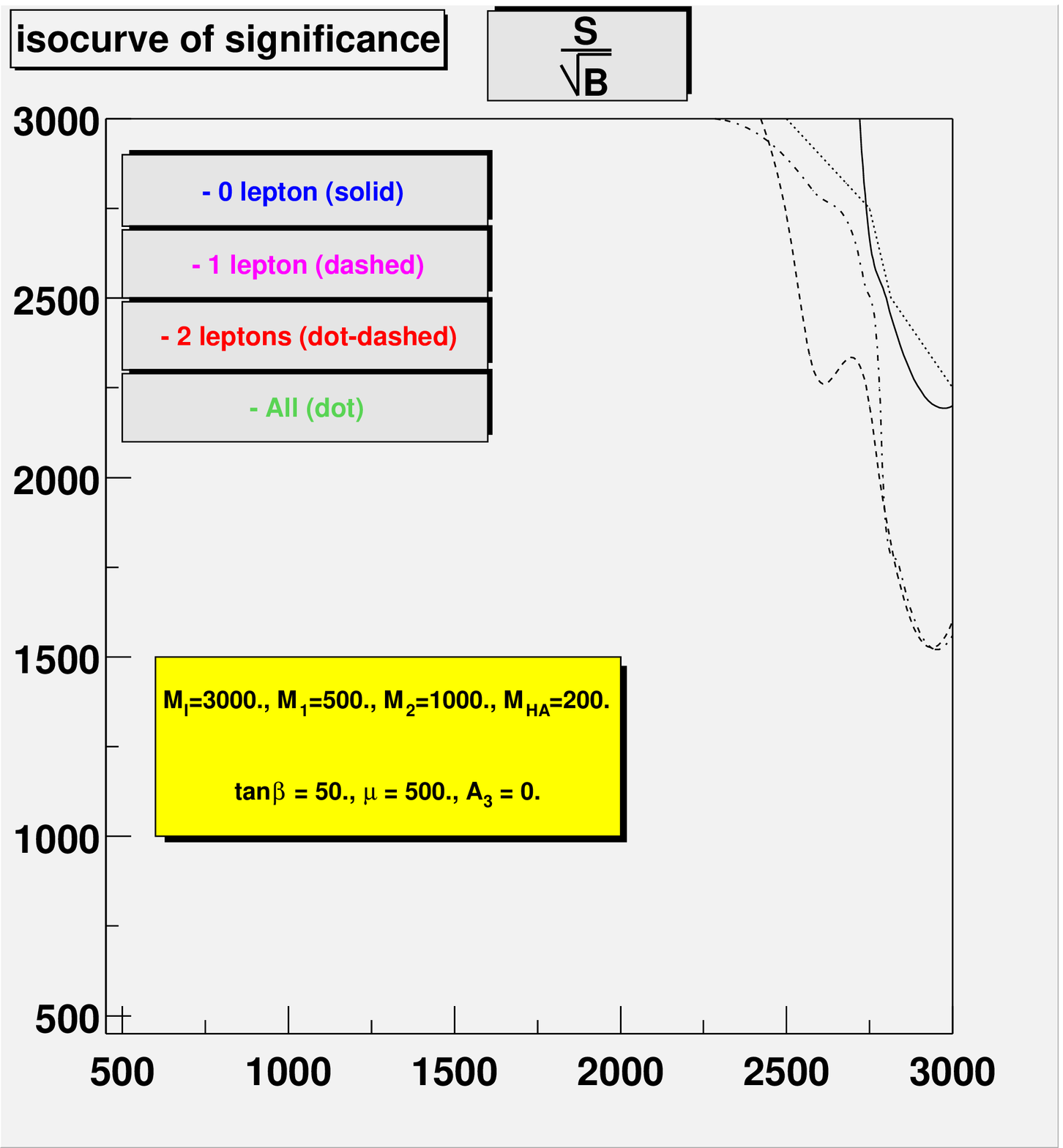}}
\caption{Graph of discovery limits in the 
$ M_{\tilde{q}}$ versus 
$ M_{\tilde{g}} $ plane.}
\label{fig: exclu1}
\end{center}
\end{figure}
One example of such a result 
is given by Figure \ref{fig: exclu1}
with the following values of the other parameters: 
\begin{eqnarray}
&&M_{\tilde{l}}=3000\ \gev, 
M_{1}=500\ \gev, 
M_{2}=1000\ \gev, 
M_{A}=200\ \gev, \nonumber\\
&& 
\mu =500\ \gev, A_{3}=0\ \gev, \tan \beta =50.
\end{eqnarray}

Four isocurves are given in this figure, corresponding to a 
specific event topology selection according to the
number of isolated  leptons produced. The curves labelled 
``0 lepton",  ``1 lepton", ``2 leptons" correspond to a calculation of
significance using only events with respectively  
$ 0 $, $ 1 $, $ 2 $ lepton(s). The curves labelled ``all" 
use all events to
calculate the significance. For each point and each type 
of lepton selection, we manage to find the set of cuts which gives the 
largest significance.

\begin{figure}[tb]
\begin{center}
\resizebox*{10cm}{10cm}{\includegraphics{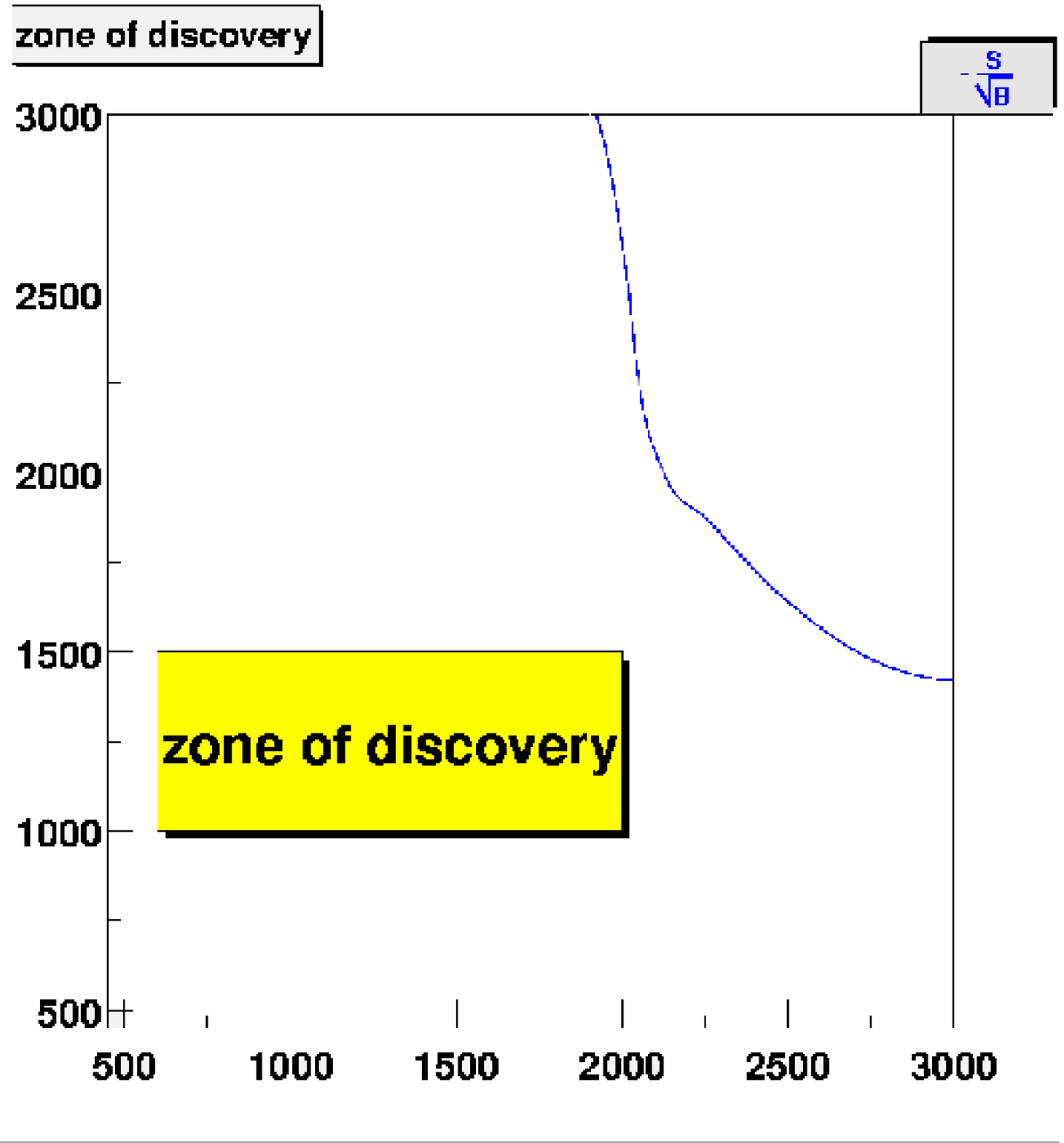}}
\caption{Discovery limits of the MSSM in the $(M_{\tilde{q}},M_{\tilde{g}})$ plane.}
\label{fig: exclu3}
\end{center}
\end{figure}

We can now try to compile all these results in one characteristic limit.
We could first combine our results in a conservative fashion,
establishing the region in the $(M_{\tilde{q}},M_{\tilde{g}})$ plane
in which any set of the orthogonal parameters will be accessible
by the CMS detector. In other words, for each point in
the $(M_{\tilde{q}},M_{\tilde{g}})$ plane above the reach curve
there exist at least one set of the 7 other parameter which has
a significance under 5.  Figure~\ref{fig: exclu3} 
shows this pessimistic mass reach in
the $(M_{\tilde{q}},M_{\tilde{g}})$ plane.
In terms of statistics, there are a total of 2,962 combinations 
(out of the original 35,000 combinations of parameters) which
don't pass the cut on significance ($>5$).  
We have also found a single point under the curve which
does not pass the significance cut:
\begin{eqnarray}
&&M_{\tilde{l}}=3000\ \gev,
M_{1}=2000\ \gev, 
M_{2}=2000\ \gev, 
M_{\tilde{g}}=1500\ \gev, 
M_{\tilde{q}}=1800\ \gev,\nonumber \\
&&M_{A}=3000\ \gev, \mu =200\ \gev, A_{3}=2000\ \gev, \tan \beta =50.
\end{eqnarray}
This particular point exhibits a narrow hierarchy of masses just 
like in the previous example, and provides a lower limit of discovery.
In a typical configuration we have a discovery limit of about 2.5 TeV. 
This exact reach in parameter space depends on the  magnitude of the
background cross section within the kinematical cuts. Here, we assume 
that the PYTHIA cross section are correct. This is clearly
invalid, as higher order QCD corrections to 
$t\, \bar t$, W+jets, Z+jets are not incorporated. 
This is an aspect of systematic uncertainties to be addressed in a later study.

\subsection{From kinematical observables to MSSM parameters}

\subsubsection{Choice of observables}

In a second exercise, we reverse the problem and try to see whether 
on the basis of event kinematical variables and event rates 
it would be possible to determine the MSSM parameter values. 
A total of 11 observables are used to separate the different sets
of MSSM parameter values. 

\begin{itemize}
\item average number of leptons per event $ \left\langle N_{l}\right\rangle$,
\item average number of jets per event $ \left\langle N_{j}\right\rangle$,
\item mean value of jet momenta  $ \left\langle P_{tj}\right\rangle$,
\item mean value of lepton momenta $ \left\langle P_{tl}\right\rangle$,
\item mean value of missing transverse energy $ \left\langle E^{miss}_{t}\right\rangle$,
\item number of events $ N_{tot}$,
\item number of events with 0,1,2 or 3 leptons $ N_{0} $, $ N_{1} $, $ N_{2}$,
$ N_{3}$,
\item mean value of total transverse energy $ \left\langle E^{sum}_{t}\right\rangle$.
\end{itemize}

These observables are characteristic of the measurement to be done with the
CMS detector, and correspond to signatures of MSSM events. In particular, there
are correlations between these observables and the MSSM parameters.

\begin{figure}[tb]
\begin{center}
\resizebox*{13cm}{13cm}{\includegraphics{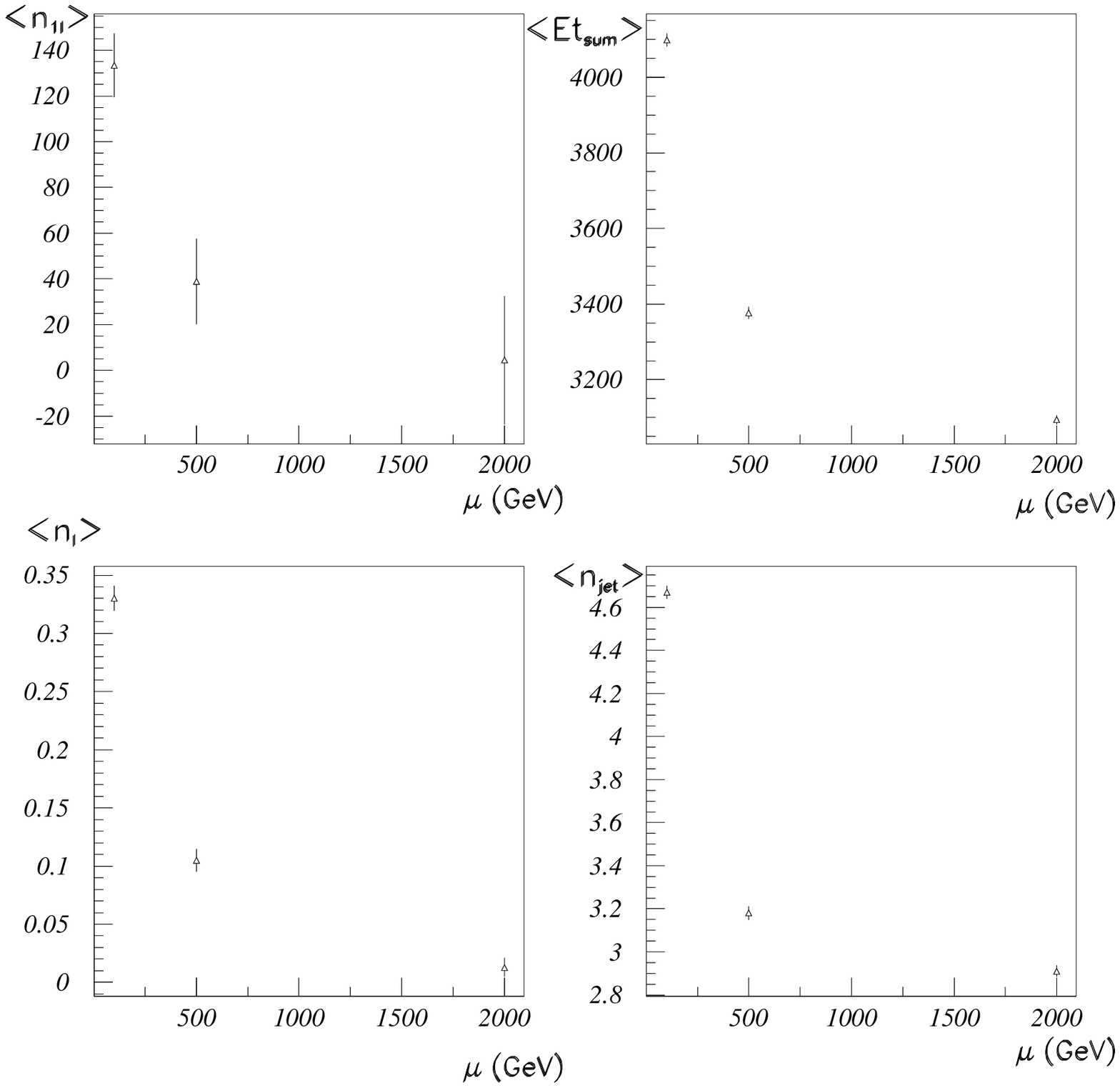}} 
\caption{Relationship between some observables and the $ \mu  $ parameter  for 
$M_{\tilde{l}}=3000$ GeV,  $ M_{1}=500$ GeV, $ M_{2}=500$ GeV, $ M_{\tilde{q}}=2100$ GeV, 
$M_{\tilde{g}}=2400$ GeV, $ M_{A}=200$ GeV, $ \tan \beta =50, $ $ A_{3}=2000$ GeV.}
\label{fig: par mu}
\end{center}
\end{figure}

Figure \ref{fig: par mu} illustrates some of these correlations between 
observables and pMSSM parameters,
more specifically, between $ \left\langle E^{sum}_{t}\right\rangle  $, $ N_{1} $, 
$ \left\langle N_{l}\right\rangle  $, $ \left\langle N_{j}\right\rangle  $
and the parameter $ \mu  $. 
The values of the other parameters are: 
\begin{eqnarray}
&&M_{\tilde{l}}=3000\ \gev, 
M_{1}=500\ \gev, 
M_{2}=500\ \gev, 
M_{\tilde{q}}=2100\ \gev,
M_{\tilde{g}}=2400\ \gev, \nonumber\\ 
&& M_{A}=200\ \gev,
A_{3}=2000\ \gev, 
\tan \beta =50.
\end{eqnarray}
It can be seen that the value of all these kinematical quantities decreases with increasing $ \mu  $.
This shows that these observables are sensitive to the value of $ \mu  $. We have
obtained the same behavior with the other pMSSM parameters, which is an argument to use pMSSM 
instead of mSUGRA. 

\subsubsection{Separation of parameters }

After optimizing the cuts to achieve maximal significance, we extract
the values of all observables for signal and background, and we calculate the
statistical uncertainties $ \sigma _{e} $ associated with each observable;
the uncertainties $ \sigma _{e} $ are equal to the ratio of the standard
deviation of the distribution and the number of events, thus :

\begin{itemize}
\item $ \sigma _{e}\propto \sqrt{N} $ for $ N $  number of events
(for example the number of events with 0 leptons);
\item $ \sigma _{e}\propto \frac{\sigma _{i}}{\sqrt{N}} $ in general and for mean value
( like $ \left\langle E^{miss}_{t}\right\rangle  $) with $ \sigma _{i} $ the root mean square. 
\end{itemize}
In this way we take into account the uncertainties on the
averages derived from the small number of 1000 events generated for each MSSM point. Can
these values of the observables be linked to the values of the pMSSM parameters? In
other words, are we able to distinguish two different sets of MSSM parameter values using only
the values of the observables for each set? If this discrimination is possible, we associate each
set of parameter values with the corresponding values of kinematical quantities, and also we
use interpolation when we turn to the continuous case. Discrimination is carried out in the
following way: one MSSM point is considered distinguishable from another one when the difference
between values of at least one of the observables for the two points is greater than 5 standard
deviations of the considered observable $ \sigma _{e} $ , calculated for the point we take
as reference, \textit{i.e.} if, for example, $N_{jet}^{ref} - N_{jet}^{i} > 5
\sigma_{e}^{ref}$ where $N_{jet}^{ref}$ is the number of jets for the reference point,
$N_{jet}^{i}$ is the number of jets for another points and $\sigma_{e}^{ref}$ is the
statistical uncertainty on the average number of jets of the reference point. By calculating, for
each MSSM point, the difference between the MSSM reference point and any other point for each
observable and by expressing these variations in terms of the value of respective
uncertainties calculated for the reference point, we can discriminate between them.
\begin{figure}[tb]
\begin{center}
\resizebox*{13cm}{13cm}{\includegraphics{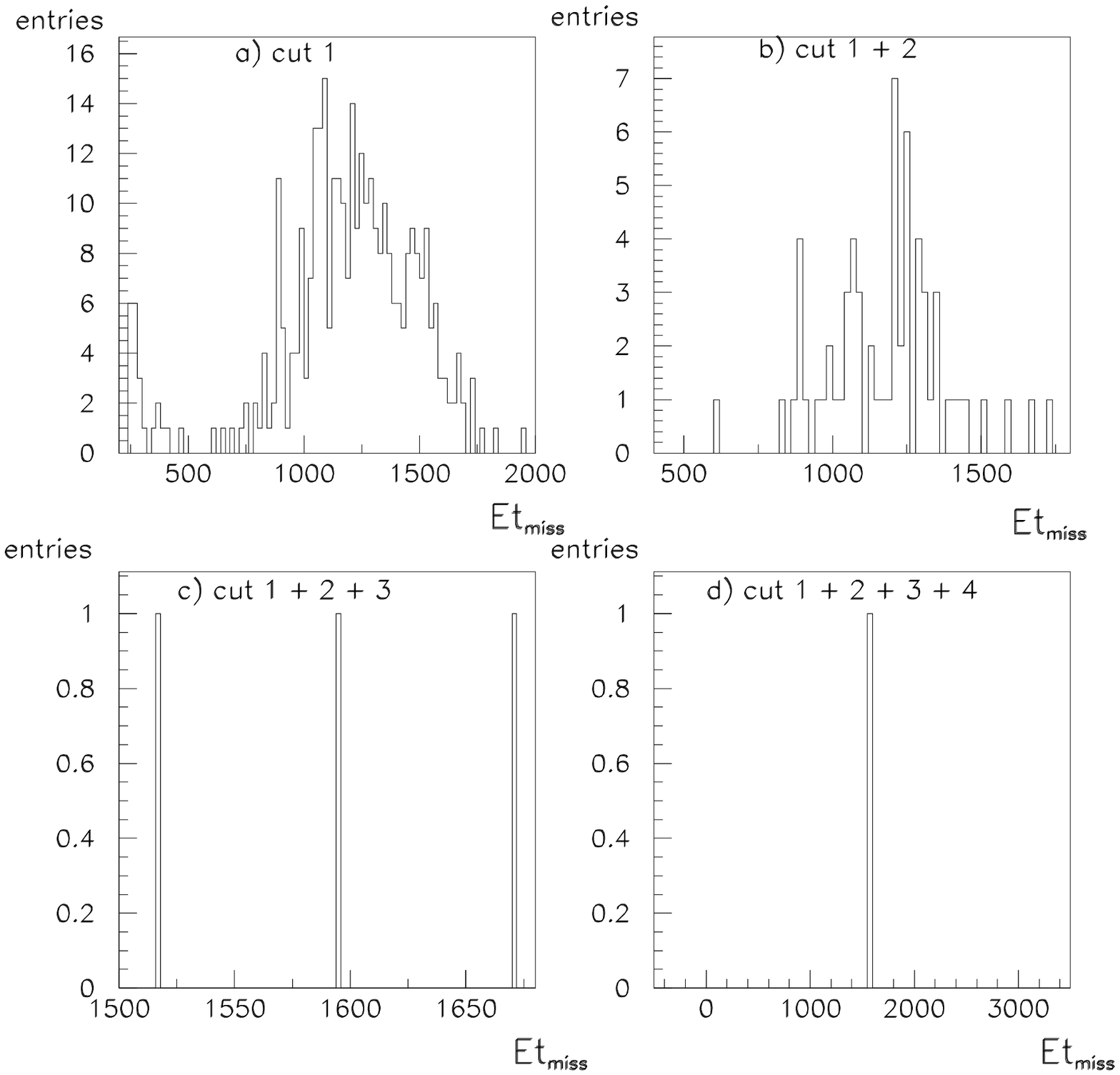}} 
\caption{ Separation of MSSM points \label{fig: mssm 13} (each entry corresponding
to a set of MSSM parameter values):
a) sets remaining after one cut ( $ \left\langle n_{lep}\right\rangle  $
variation $ <5\sigma _{e} $);
b) sets remaining after two cuts ( $ \left\langle n_{jet}\right\rangle  $
variation $ <5\sigma _{e} $);
c) sets remaining after three cuts ( $ E^{miss}_{t} $ variation
$ <5\sigma _{e} $ ) and
d) sets remaining after four cuts ( $ E^{sum}_{t} $ variation
$ <5\sigma _{e} $ ). }
\end{center}
\end{figure}
Figure
\ref{fig: mssm 13} shows an example of discrimination for the following values of parameters: 
\begin{eqnarray}
&&M_{\tilde{l}}=3000\ \gev, 
M_{1}=100\ \gev,
M_{2}=1000\ \gev, 
M_{\tilde{q}}=2700\ \gev, 
M_{\tilde{g}}=2400\ \gev,
\nonumber \\
&&M_{A}=200\ \gev,
\mu =500\ \gev, 
A_{3}=0\ \gev, 
\tan \beta =50.
\end{eqnarray}
By applying successively a cut at 5 standard deviations on the following observables: 
average number of leptons, of jets, missing transverse energy and total transverse
energy, we manage to separate the reference point from all the other ones.
This method works for all MSSM points ( the number of cuts needed for the
separation varies from point to point).

\subsubsection{Evaluation of the statistical uncertainties}

We showed in the previous section that there is a possibility to distinguish 
one set of parameter values from
the others. We are now going to estimate the statistical uncertainties for each
MSSM parameter value. We estimate these uncertainties using the uncertainties
calculated for each observable. For that, we consider one set of MSSM parameter values
defined as the reference point. Then for any chosen parameter, we measure the number of
standard deviations for each observable between the reference point
and the MSSM point having the same parameter values except for the one
in question. The value of this parameter has to be different by one unit 
on the grid of values. For example, if we want to obtain the statistical 
uncertainties on the $\mu$ parameter
for fixed values of the other parameters, we consider the point which has 
the same values for the other parameters and the next higher value 
for $\mu$ in the grid of parameters and then we calculate
\begin{equation}
N = \frac{|E_{T1}^{Miss} - E_{T2}^{Miss}|}{\sigma_{E_{T1}^{Miss}}}
\end{equation}
where $E_{T1}^{Miss}$ is the missing transverse energy at the point 
we want to calculate the resolution, $E_{T2}^{Miss}$ 
the missing transverse energy at the other point and $\sigma_{E_{T1}^{Miss}}$ the statistical
uncertainty estimate for $E_{T1}^{Miss}$. We assumed that
\begin{equation}
N = \frac{|\mu_1 - \mu_2|}{\sigma_{\mu_1}}
\end{equation}
where $\sigma_{\mu_1}$ is the statistical uncertainty we want to 
estimate. We could also take another observable
or a linear combination of observables to calculate the uncertainties.

In the following two tables, we have used $ E_{T}^{miss} $
as the observable for the calculation of the resolution. 
\begin{table}[tb]
{\centering \begin{tabular}{|c|c|c|c|}
\hline 
parameter &
value&
$ \sigma _{+} $&
$ \sigma _{-} $\\
\hline 
\hline 
M($ \tilde{L} $) (GeV)&
$ 3000. $&
$ 9.899 $&
$ 9.899 $\\
\hline 
M1 (GeV)&
$ 500. $&
$ 3.035 $&
$ 4.259 $\\
\hline 
M2 (GeV)&
$ 500. $&
$ 5.451 $&
$ 5.451 $\\
\hline 
M($ \tilde{Q} $) (GeV)&
$ 1800. $&
$ 12.384 $&
$ 21.092 $\\
\hline 
M($ \tilde{G} $) (GeV)&
$ 1800. $&
$ 2.481 $&
$ 1.901 $\\
\hline 
A (GeV)&
$ 200. $&
$ 4.749 $&
$ 4.749 $\\
\hline 
$ \tan \beta  $&
$ 50. $&
$ 0.575 $&
$ 0.575 $\\
\hline 
$ \mu  $ (GeV)&
$ 2000. $&
$ 17.331 $&
$ 17.331 $\\
\hline 
A$ _{3} $ &
$ 2000. $&
$ 8.534 $&
$ 8.534 $\\
\hline 
\end{tabular}\par}
\caption{Resolutions of the MSSM parameter values for intermediate masses  
M($ \tilde{q} $)
and M($ \tilde{g} $) (large statistics) \label{tab: resolMSSM1}
and significance = 30.}
\end{table} 

\begin{table}[!ht]
{\centering \begin{tabular}{|c|c|c|c|}
\hline 
parameter &
value&
$ \sigma _{+} $&
$ \sigma _{-} $\\
\hline 
\hline 
M($ \tilde{L} $) (GeV)&
$ 1000. $&
$ 146.491 $&
$ 146.491 $\\
\hline 
M1 (GeV)&
$ 2000. $&
$ 55.387 $&
$ 55.387 $\\
\hline 
M2 (GeV)&
$ 500. $&
$ 23.740 $&
$ 23.740 $\\
\hline 
M($ \tilde{Q} $) (GeV)&
$ 2700. $&
$ 18.868 $&
$ 16.289 $\\
\hline 
M($ \tilde{G} $) (GeV)&
$ 2700. $&
$ 42.142 $&
$ 40.814 $\\
\hline 
A (GeV)&
$ 3000. $&
$ 118.390 $&
$ 118.390 $\\
\hline 
$ \tan \beta  $&
$ 50. $&
$ 4.411 $&
$ 4.411 $\\
\hline 
$ \mu  $ (GeV)&
$ 500. $&
$ 136.504 $&
$ 136.504 $\\
\hline 
A$ _{3} $ &
$ 2000. $&
$ 95.730 $&
$ 95.730 $\\
\hline 
\end{tabular}\par}
\caption{Resolution of the MSSM parameter values for high masses 
M($ \tilde{q} $) and M($ \tilde{g} $) (low statistics) 
\label{tab: resolMSSM2} and significance = 6.}
\label{resolMSSM2}
\end{table} 

\begin{figure}[!htb]
{\centering \resizebox*{10cm}{10cm}{\includegraphics{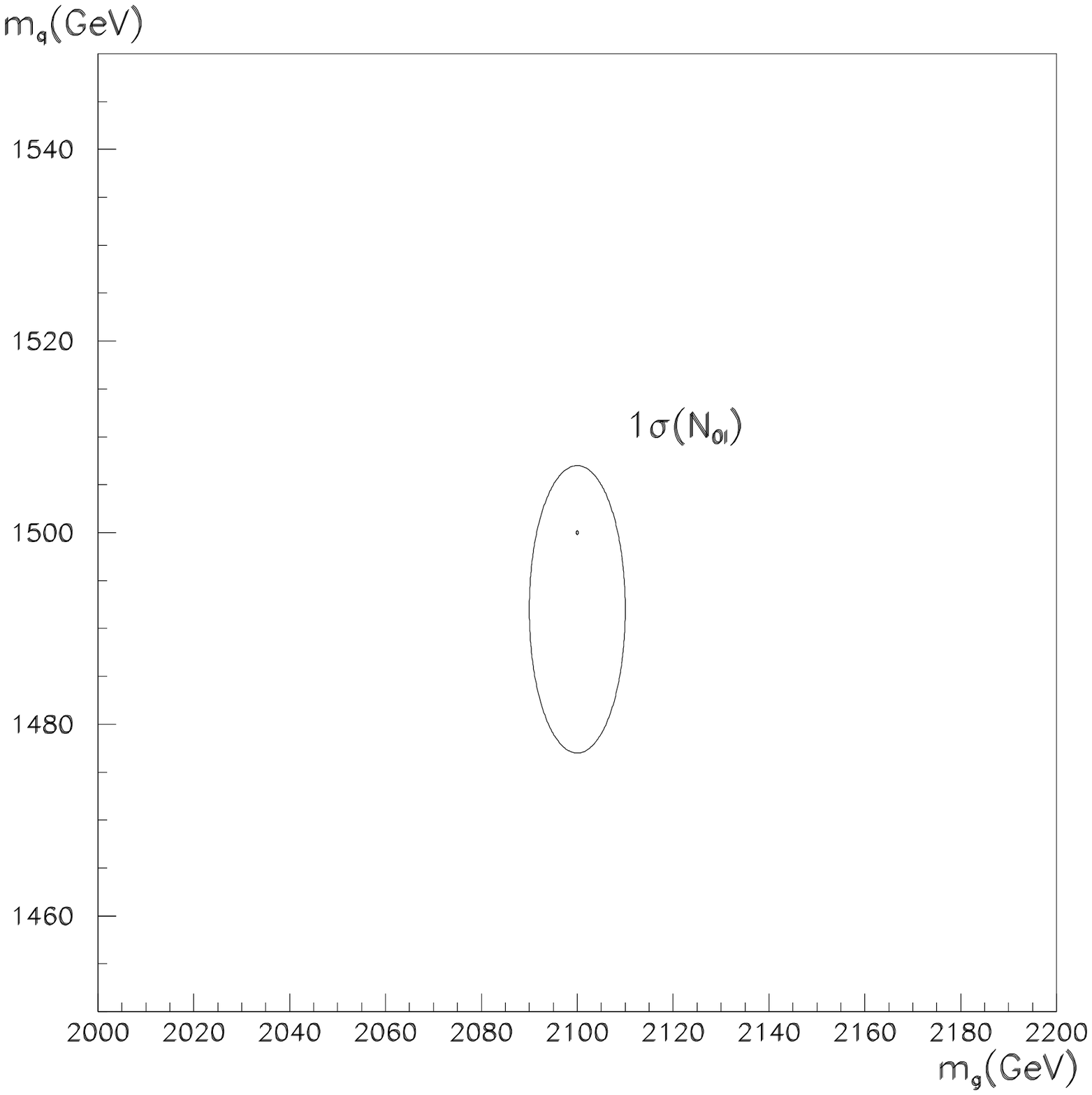}} \par}
\caption{An example of a resolution in the plane $ M_{\tilde{q}} $
versus $ M_{\tilde{g}}$ (the curve represents 1 $\sigma$).
\label{fig: reso5}}
\end{figure} 

Tables \ref{tab: resolMSSM1} and \ref{tab: resolMSSM2} show some 
examples of resolution obtained by this
method. These values are small since this is only the statistical error. 
The resolution should degrade after including systematic errors. Table 
\ref{tab: resolMSSM1} shows the calculation of
statistical resolution for medium squark and gluino masses, 
where statistics are large. Table \ref{tab: resolMSSM2}
shows a similar calculation for large masses for squarks and gluinos, 
where statistics are low and the uncertainties are thus much
more important. 

Figure \ref{fig: reso5}
shows an example of resolution in the plane $ m_{\tilde{q}} $ versus $ m_{\tilde{g}} $
for the following reference point: 
\begin{eqnarray}
&&
M_{\tilde{l}}=200\ \gev, 
M_{1}=100\ \gev, 
M_{2}=500\ \gev, 
M_{\tilde{q}}=1500\ \gev,
M_{\tilde{g}}=2100\ \gev, \nonumber \\
&&
M_{A}=200\ \gev, 
\mu =200\ \gev, 
A_{3}=2000\ \gev,
\tan \beta =50.
\end{eqnarray}
We have, for this point, a resolution of about 5 to 10 GeV 
(using the number of events with 0 leptons, $ N_{0l} $, as the discriminator).

\section{Conclusion}

We demonstrated the possibility to discover a phenomenological MSSM using an
inclusive study in the MSSM parameters space. Once we have discovered SUSY, 
using the kinematical observables for parameter determination proved to be an efficient
method. Statistical uncertainties obtained are relatively small ($< 50 $ GeV for
squark and gluino masses). We could note, at the end of our study, that there
was little difference between mSUGRA and the pMSSM. The discovery limit
is to a large extent determined by the total cross section 
(and is around 2.7 TeV at CMS). The only difference appears
for some points having a specific mass hierarchy. As an example, in the case of
a compact hierarchy of masses, the limit we expect is about 1.5 TeV.



\setcounter{figure}{0}
\setcounter{table}{0}
\setcounter{section}{0}
\setcounter{equation}{0}
\clearpage

\part{{\bf Establishing a No-Lose Theorem for NMSSM 
Higgs Boson Discovery at the LHC}
 \\[0.5cm]\hspace*{0.8cm}
{\it U. Ellwanger, J.F. Gunion, C. Hugonie
}}
\label{nmssmhiggssec}


\def\sq{\hbox {\rlap{$\sqcap$}$\sqcup$}}
\def\lsim{\raise0.3ex\hbox{$<$\kern-0.75em\raise-1.1ex\hbox{$\sim$}}}
\def\gsim{\raise0.3ex\hbox{$>$\kern-0.75em\raise-1.1ex\hbox{$\sim$}}}
\def\noi{\noindent}
\def\to{\rightarrow}
\def\mw{m_W}
\def\ie{{\it i.e.}}
\def\nn{\nonumber}
\def\noi{\noindent}
\def\beeq{\begin{eqnarray}} \def\eeeq{\end{eqnarray}}
\def\what{\widehat}
\def\vev#1{\langle #1 \rangle}
\def\tev{~{\rm TeV}}
\def\gev{~{\rm GeV}}
\def\fbi{~{\rm fb}^{-1}}
\def\abi{~{\rm ab}^{-1}}
\def\anti{\overline}
\def\tanb{\tan\beta} 
\def\br{BR}
\def\gam{\gamma}
\def\nsd#1{$N_{SD}(#1)$}


\begin{abstract}
  We scan the parameter space of the NMSSM for the observability of at
  least one Higgs boson at the LHC with $300~\mbox{fb$^{-1}$}$ integrated
  luminosity, taking the present LEP2 constraints into account. We
  restrict the scan to those regions of parameter space for which
  Higgs boson decays to other Higgs bosons and/or supersymmetric
  particles are kinematically forbidden.  We find that if $WW$-fusion
  detection modes for a light Higgs boson are not taken into account,
  then there are still significant regions in the scanned portion of
  the NMSSM parameter space where no Higgs boson can be observed at
  the $5\sigma$ level, despite the recent improvements in ATLAS and
  CMS procedures and techniques and even if we combine all non-fusion
  discovery channels. However, if the $WW$-fusion detection modes are
  included using the current theoretical study estimates, then we find
  that for all scanned points at least one of the NMSSM Higgs bosons
  will be detected. If the estimated $300~\mbox{fb$^{-1}$}$ significances for ATLAS
  and CMS are combined, one can also achieve $5\sigma$ signals after
  combining just the non-$WW$-fusion channels signals. We present the
  parameters of several particularly difficult points, and discuss the
  complementary roles played by different modes. We conclude that the
  LHC will discover at least one NMSSM Higgs boson unless there are
  large branching ratios for decays to SUSY particles and/or to other
  Higgs bosons.
\end{abstract}

\section{Introduction}

Supersymmetric extensions of the standard model generally predict relatively
light Higgs bosons. One of the most important tasks of the LHC is the search
for Higgs bosons \cite{atlastdr3,newCMSTDR,Abdullin:1998pm}.
An important milestone in understanding the
potential of the LHC was the demonstration that at least one Higgs boson of
the minimal supersymmetric standard model (MSSM) would be detectable at the
$\geq 5\sigma$ level throughout all of the MSSM parameter space so long as top
squark masses do not exceed 1.5 to 2 TeV and so long as large branching
fractions to decay channels containing supersymmetric particles are not
substantial.

In \cite{Ellwanger:2001iw}, we studied, subject to these same and a few other
simplifying restrictions, the detectability of Higgs bosons in the
next-to-minimal supersymmetric standard model (NMSSM). 
This short note presents the most relevant procedures and conclusions
of \cite{Ellwanger:2001iw}. In the NMSSM, one Higgs
singlet superfield, $\what S$, is added to the MSSM in order to render
unnecessary the bilinear superpotential term $\mu \what H_1 \what H_2$ by
replacing it with $\lambda \what S\what H_1 \what H_2 $, where the vacuum
expectation value of the scalar component of $\what S$, $\vev{S}$, results in
an effective bilinear Higgs mixing with $\mu=\lambda\vev{S}$. The detectability
of the NMSSM Higgs bosons was first considered in a contribution to Snowmass 96
\cite{Gunion:1996fb}. The result, using the experimentally established modes and
sensitivities available at the time, was that substantial regions in the
parameter space of the NMSSM were found where none of the Higgs bosons would
have been observable either at LEP2 or at the LHC even with an integrated
luminosity of $600~\mbox{fb$^{-1}$}$ (two detectors with $L=300~\mbox{fb$^{-1}$}$ each). 

Since then, progress has been made both on the theoretical and the experimental
sides. On the theoretical side, the dominant two-loop corrections to the
effective potential of the model have been computed \cite{Yeghian:1999kr,Ellwanger:1999ji}.
These lead to
a modest decrease in the mass of the lightest Higgs scalar, holding fixed the
stop sector parameters. Inclusion of the two-loop corrections thus increases
somewhat the part of the NMSSM parameter space excluded by LEP2 (and accessible
at the Tevatron) \cite{Ellwanger:1999ji}, but is of less relevance for the LHC. On the
experimental side the expected statistical significances have been improved
since 1996 \cite{atlastdr3,newCMSTDR,Abdullin:1998pm}.
Most notably, associated $t {\anti t} h$ production
with $h \to b{\anti b}$ (originally discussed in \cite{Dai:1993gm}), which in the SM
context is particularly sensitive to $m_h\ \lsim\ 120$ GeV, has been added by
ATLAS and CMS to the list of Higgs boson detection modes
\cite{atlastdr3,newCMSTDR,Abdullin:1998pm,Richter-Was:1999sa,Sapinski:2001tn,%
Drollinger:2001xm,CMSnote2001-39}. 
Analysis of this mode was recently extended
\cite{thankstosapinski} to $m_h=140\gev$, which, though not relevant in the SM case due to
the decline in the $b\anti b$ branching ratio as the $WW^*$ mode increases, is
highly relevant for points in our searches for which the $WW^*$ mode is
suppressed in comparison to the SM prediction. In addition, techniques have
been proposed \cite{Rainwater:1998kj,Plehn:1999xi,Rainwater:1999sd,%
Kauer:2000hi,Zeppenfeld:2000td,zepp} 
for isolating signals for $WW$ fusion to a light
Higgs boson which decays to $\tau\anti\tau$ or $WW^{(*)}$.

It turns out that adding in just the $t\anti t h$ process renders the
no-Higgs-discovery parameter choices described and plotted in \cite{Gunion:1996fb},
including the ``black point'' described in detail there, visible \cite{Gunion:2001pg}.
In~\cite{Ellwanger:2001iw}, we searched 
for any remaining parameter choices for which no
Higgs boson would produce a $\geq 5\sigma$ signal. 
In this search, we performed a
scan over nearly all of the parameter space of the model, the only parameter
choices not included being those for which there is sensitivity to the highly
model-dependent decays of Higgs bosons to other Higgs bosons and/or
superparticles. The outcome is that, for an integrated luminosity of $300~\mbox{fb$^{-1}$}$
at the LHC, there are still regions in the parameter space with $<5 \sigma$
expected statistical significance (computed as $N_{SD}=S/\sqrt B$ for a given
mode) for all Higgs detection modes so far studied in detail by ATLAS and CMS,
\ie\ including the $t\anti t h\to t\anti t b\anti b$ mode but not the
$WW$-fusion modes. On the other hand, the expected statistical significance for
at least one of these detection modes is always above $3.6 \sigma$ at
$300~\mbox{fb$^{-1}$}$, and the statistical significance obtained by combining (using the
naive Gaussian procedure) all the non-$WW$-fusion modes is at least $4.8
\sigma$. However, we find that all such cases are quite observable (at $\geq
10.1 \sigma$) in one of the $WW$-fusion modes (using theoretically estimated
statistical significances for these modes). For all points in the scan of
parameter space, statistical significances obtained by combining all modes,
including $WW$-fusion modes, are always $\gsim 10.7 \sigma$. Thus, NMSSM Higgs
discovery by just one detector with $L=300~\mbox{fb$^{-1}$}$ is essentially guaranteed for
those portions of parameter space for which Higgs decays to other Higgs bosons
or supersymmetric particles are kinematically forbidden. This represents
substantial progress towards guaranteeing LHC discovery of at least one of the
NMSSM Higgs bosons.

In order to clarify the nature of the most difficult points in those portions
of parameter space considered, we present, in sect. 4, examples of particularly
difficult bench mark points for the Higgs sector of the NMSSM. Apart from the
``bare'' parameters of the model, we give the masses and couplings of all Higgs
scalars, their production rates and branching ratios to various channels
(relative to the SM Higgs) and details of the statistical significances
predicted for each Higgs boson in each channel. The latter will allow an
assessment of exactly what level of improvement in statistical significance
will be required in the various different detection modes in order to render
marginal modes visible. Of course, our estimates of the expected statistical
significances are often somewhat crude (e.g. their dependence on the
accumulated integrated luminosity). We believe that our procedures always err
in the conservative direction, leading to statistical significances that might
be a bit small. Thus, the LHC procedures for isolating Higgs boson signals
could provide even more robust signals for NMSSM Higgs boson detection than we
estimate here.

The detection modes, which serve for the searches for standard model or MSSM
Higgs bosons, include (using the notation $h$, $a$ for CP-even, CP-odd Higgs
bosons, respectively):\par
\noi 1) $gg \to h \to \gamma \gamma$;\par
\noi 2) associated $W h$ or $t \bar{t}h$ production with $\gamma \gamma
\ell^{\pm}$ in the final state;\par
\noi 3) associated $t {\anti t}h$ production with $h \to b{\anti b}$;\par
\noi 4) $gg \to h/a$ or associated $b {\anti b}h/a$ production with $h/a \to
\tau {\anti \tau}$;\par
\noi 5) $gg \to h \to ZZ^{(*)} \to$ 4 leptons;\par
\noi 6) $gg \to h \to WW^{(*)} \to l^+ l^- \nu {\anti \nu}$;\par
\noi 7) LEP2 $e^+e^- \to Zh$ and $e^+e^- \to ha$;\par
\noi 8) $WW\to h \to \tau\anti\tau$;\par
\noi 9) $WW\to h\to WW^{(*)}$,\par
\noi where 8) and 9) are those analyzed at the 
theoretical level in \cite{Rainwater:1998kj,Plehn:1999xi,Rainwater:1999sd,%
Kauer:2000hi,Zeppenfeld:2000td,zepp}
and included in the NMSSM analysis for the first time in this paper. The above
detection modes do not employ the possibly important decay channels i)~$h \to
hh$, ii)~$h \to aa$, iii)~$h \to h^+h^-$, iv)~$h \to aZ$, v)~$a \to ha$, vi)~$a
\to hZ$, vii)~$h,a \to h^\pm W^\mp$, viii)~$h,a \to t\anti t$ and ix)~$t \to
h^+b$. The decay modes i)-vii) give high multiplicity final states and
deserve a dedicated study \cite{wip}, while the existing analyses of the
$t\anti t$ final state signatures are not very detailed. Further, when
kinematically allowed, the $t\to h^+b$ signal would be easily observed
according to existing analyzes. Thus, in this paper we restrict our scan over
NMSSM parameter space to those parameters for which none of these decays are
present. In addition, we take the constraints of LEP2 [via the mode 7)] into
account, and only accept points for which $5\sigma$ discovery at LEP2 would not
have been possible \cite{LEP0103,LEP0104}.

The Higgs sector of the NMSSM consists of 3 scalars, denoted $h_1, h_2, h_3$
with $m_{h_1} < m_{h_2} < m_{h_3}$, 2 pseudo-scalars, denoted $a_1, a_2$ with
$m_{a_1} < m_{a_2}$, and a charged Higgs pair, denoted $h^{\pm}$. Mixing of the
neutral doublet fields with the gauge singlet fields in the scalar and in the
pseudo-scalar sector can be strong. The scalar mixing can lead to a
simultaneous suppression of the couplings of all the $h_i$ to gauge bosons, and
hence to a suppression of many of the detection modes above. (Of course, the
$a_i$ have no tree-level couplings to gauge boson pairs and the one-loop
couplings are too small to yield useful event rates.) The couplings of the
Higgs bosons to t- or b-quarks can be amplified, reduced or even change sign
with respect to the standard model couplings. Hence negative interferences can
occur among the (loop-) diagrams contributing to $gg \to h_i$ and $h_i \to
\gamma \gamma$, leading again to suppressions of the above detection modes. A
complete simultaneous annulation of all detection modes is not possible, but
simultaneous reduction of all detection modes is possible and it is for such
parameter choices that NMSSM Higgs boson discovery is most difficult.

In the next section, we define the class of models we are going to consider,
and the way we perform the scan over the corresponding parameter space. In
section 3 we describe our computations of the expected statistical
significances of the detection modes 1) -- 9) above. In section 4, we present 
six particularly difficult bench mark points (in table \ref{table_1}) and details
regarding their statistical significances in channels 1)-9) in table \ref{table_2}, with a
summary of overall statistical significances in table \ref{table_3}. Using these tables, we
give a discussion of the properties of these points.

\begin{table}[p]
\begin{center}
\small
\begin{tabular} {|l|l|l|l|l|l|l|} 
\hline
Point Number & 1 & 2 & 3 & 4 & 5 & 6  \\
\hline \hline
Bare Parameters &\multicolumn{6}{c|}{} \\
\hline
$\lambda$            & 0.0340 & 0.0450 & 0.0230 & 0.0230 & 0.1330 & 0.0230 \\
\hline
$\kappa$             & 0.0198 & 0.0248 & 0.0129 & 0.0069 & 0.1459 & 0.0114 \\
\hline
$\tan\beta$          &   6.00 &   5.25 &   -5.5 &   5.75 &     -8 &     -6 \\
\hline
$\mu_{\rm eff} (GeV)$&    140 &   -110 &    115 &   -235 &    100 &    150 \\
\hline
$A_{\lambda} (GeV)$  &    -35 &     25 &    -95 &     40 &   -135 &   -100 \\
\hline
$A_{\kappa} (GeV)$   &   -150 &     70 &    -90 &     80 &    -75 &   -110 \\
\hline \hline
Scalar Masses and Couplings &\multicolumn{6}{c|}{} \\
\hline \hline
$m_{h_1}$ (GeV)      &    115 &    100 &    103 &    113 &    114 &    112 \\
\hline
$c_V $               &  -0.66 &   0.32 &  -0.34 &   0.67 &  -0.87 &  -0.71 \\
\hline
$c_t $               &  -0.65 &   0.30 &  -0.31 &   0.65 &  -0.81 &  -0.66 \\
\hline
$c_b $               &  -1.07 &   0.66 &  -1.27 &   1.16 &  -4.50 &  -2.40 \\
\hline
gg Production Rate   &   0.39 &   0.08 &   0.08 &   0.39 &   0.56 &   0.36 \\
\hline
$\br\gamma \gamma$   &   0.43 &   0.26 &   0.09 &   0.38 &   0.05 &   0.11 \\
\hline
$\br b\anti 
b=\br\tau\anti\tau$  &   1.12 &   1.08 &   1.10 &   1.12 &   1.18 &   1.15 \\
\hline
$\br WW^{(*)}$       &   0.42 &   0.25 &   0.08 &   0.37 &   0.04 &   0.10 \\
\hline \hline

$m_{h_2}$ (GeV)      &    125 &    114 &    114 &    126 &    144 &    122 \\
\hline
$c_V $               &  -0.74 &  -0.83 &   0.79 &  -0.73 &   0.46 &   0.59 \\
\hline
$c_t $               &  -0.72 &  -0.74 &   0.70 &  -0.71 &   0.57 &   0.54 \\
\hline
$c_b $               &  -1.49 &  -3.28 &   3.46 &  -1.47 &  -6.66 &   2.24 \\
\hline
gg Production Rate   &   0.46 &   0.44 &   0.40 &   0.45 &   1.18 &   0.23 \\
\hline
$\br\gamma \gamma$   &   0.33 &   0.08 &   0.07 &   0.34 &   0.01 &   0.10 \\
\hline
$\br b\anti 
b=\br\tau\anti\tau$  &   1.30 &   1.18 &   1.18 &   1.32 &   3.06 &   1.31 \\
\hline
$\br WW^{(*)}$       &   0.32 &   0.08 &   0.06 &   0.33 &   0.01 &   0.09 \\
\hline \hline

$m_{h_3}$ (GeV)      &    205 &    153 &    148 &   201 &     202 &    155 \\
\hline
$c_V $               &  -0.14 &  -0.46 &  -0.51 &  -0.15 &   0.18 &  -0.39 \\
\hline 
$c_t $               &  -0.30 &  -0.63 &  -0.67 &  -0.32 &   0.17 &  -0.55 \\
\hline
$c_b $               &   5.80 &   4.17 &   4.20 &   5.53 &   0.68 &   5.12 \\
\hline
gg Production Rate   &   0.31 &   0.84 &   0.95 &   0.33 &   0.02 &   0.80 \\
\hline
$\br\gamma \gamma$   &   0.13 &   0.05 &   0.05 &   0.15 &   0.98 &   0.03 \\
\hline
$\br b\anti 
b=\br\tau\anti\tau$  & 308.66 &   5.83 &   3.92 & 274.41 &  13.97 &   8.12 \\
\hline
$\br WW^{(*)}$       &   0.18 &   0.07 &   0.06 &   0.21 &   0.96 &   0.05 \\
\hline \hline

Pseudo-Scalar Masses and Couplings &\multicolumn{6}{c|}{} \\
\hline \hline
$m_{a_1}$ (GeV)      &    191 &    112 &    130 &    130 &    113 &    145 \\
\hline
$c_t $               &   0.03 &  -0.03 &  -0.10 &  -0.01 &  -0.10 &  -0.16 \\
\hline
$c_b $               &   1.16 &  -0.83 &  -2.95 &  -0.19 &  -6.55 &  -5.77 \\
\hline
gg Production Rate   &   0.00 &   0.00 &   0.03 &   0.00 &   0.31 &   0.08 \\
\hline \hline

$m_{a_2} $(GeV)      &    206 &    141 &    137 &    198 &    174 &    158 \\
\hline
$c_t $               &   0.16 &   0.19 &  -0.15 &   0.17 &  -0.07 &  -0.05 \\
\hline
$c_b $               &   5.89 &   5.18 &  -4.64 &   5.75 &  -4.59 &  -1.65 \\
\hline
gg Production Rate   &   0.02 &   0.07 &   0.06 &   0.02 &   0.03 &   0.00 \\
\hline \hline

Charged Higgs Mass   &\multicolumn{6}{c|}{} \\
\hline
$m_{c} $(GeV)        &    221 &    162 &    157 &    213 &    157 &    167 \\
\hline
\end{tabular}
\caption{We tabulate the input bare model parameters, the
corresponding Higgs masses, and the corresponding Higgs couplings, relative to
SM Higgs boson coupling strength, for 6 bench mark points. Also given for the
CP-even $h_i$ are ratios of the $gg$ production rate and various branching
fractions relative to the values found for a SM Higgs of the same mass. For
the CP-odd $a_i$, ``gg Production Rate'' refers to the value relative to what
would be found if both the $b\anti b$ and the $t\anti t$ $\gamma_5$ couplings
had SM-like strength.
\label{table_1}}
\end{center}
\end{table}

\begin{table}[p]
\begin{center}
\small
\begin{tabular} {|l|l|l|l|l|l|l|} 
\hline
Point & 1 & 2 & 3 & 4 & 5 & 6   \\
\hline \hline
Channel & \multicolumn{6}{c|}{$h_1$ Higgs boson} \\
\hline
\nsd1  &  3.74 &  0.35 &  0.13 &  3.18 &  0.62 &  0.83 \\
\nsd2  &  4.37 &  0.59 &  0.22 &  3.92 &  0.85 &  1.22 \\
\nsd3  &  2.79 &  0.85 &  0.85 &  3.03 &  4.83 &  3.30 \\
\nsd4  &  0.08 &  0.07 &  0.76 &  0.09 &  4.52 &  0.40 \\
\nsd5  &  0.83 &  0.00 &  0.00 &  0.64 &  0.12 &  0.16 \\
\nsd6  &  1.10 &  0.09 &  0.03 &  0.90 &  0.16 &  0.22 \\
\nsd7  &  0.00 &  3.37 &  3.40 &  3.29 &  0.00 &  4.79 \\
\nsd8  &  9.29 &  1.22 &  1.59 &  8.93 & 16.78 & 10.08 \\
\nsd9  &  2.39 &  0.00 &  0.00 &  1.74 &  0.41 &  0.49 \\
$\sqrt{\sum_{i=1}^6 [N_{SD}(i)]^2}$ 
       &  6.54 &  1.09 &  1.17 &  5.99 &  6.69 &  3.65 \\
$\sqrt{\sum_{i=1}^7 [N_{SD}(i)]^2}$ 
       &  6.54 &  3.55 &  3.59 &  6.84 &  6.69 &  6.02 \\
$\sqrt{\sum_{i=1-6,8,9} [N_{SD}(i)]^2}$ 
       & 11.61 &  1.64 &  1.97 & 10.89 & 18.07 & 10.73 \\
$\sqrt{\sum_{i=1}^9 [N_{SD}(i)]^2}$ 
       & 11.61 &  3.75 &  3.93 & 11.38 & 18.07 & 11.75 \\
\hline \hline
Channel & \multicolumn{6}{c|}{$h_2$ Higgs boson} \\
\hline
\nsd1  &  3.69 &  0.83 &  0.61 &  3.62 &  0.22 &  0.55 \\
\nsd2  &  4.01 &  1.25 &  0.92 &  3.93 &  0.05 &  0.74 \\
\nsd3  &  2.49 &  3.95 &  3.58 &  2.30 &  0.99 &  1.77 \\
\nsd4  &  0.16 &  2.76 &  2.93 &  0.16 &  3.62 &  2.99 \\
\nsd5  &  1.84 &  0.16 &  0.11 &  1.94 &  0.56 &  0.20 \\
\nsd6  &  1.44 &  0.22 &  0.16 &  1.46 &  0.38 &  0.18 \\
\nsd7  &  0.00 &  0.00 &  3.31 &  0.00 &  0.00 &  0.00 \\
\nsd8  & 15.39 & 15.17 & 13.46 & 15.05 &  7.41 &  9.89 \\
\nsd9  &  5.79 &  0.63 &  0.44 &  6.05 &  0.19 &  0.82 \\
$\sqrt{\sum_{i=1}^6 [N_{SD}(i)]^2}$ 
       &  6.44 &  5.05 &  4.76 &  6.31 &  3.82 &  3.61 \\
$\sqrt{\sum_{i=1}^7 [N_{SD}(i)]^2}$ 
       &  6.44 &  5.05 &  5.80 &  6.31 &  3.82 &  3.61 \\
$\sqrt{\sum_{i=1-6,8,9} [N_{SD}(i)]^2}$ 
       & 17.65 & 16.00 & 14.28 & 17.40 &  8.34 & 10.56 \\
$\sqrt{\sum_{i=1}^9 [N_{SD}(i)]^2}$ 
       & 17.65 & 16.00 & 14.66 & 17.40 &  8.34 & 10.56 \\
\hline \hline
Channel  & \multicolumn{6}{c|}{$h_3$ Higgs boson} \\
\hline
\nsd1  &  0.00 &  0.59 &  0.66 &  0.01 & 0.00 &  0.32 \\
\nsd2  &  0.00 &  0.21 &  0.25 &  0.00 & 0.00 &  0.08 \\
\nsd3  &  0.00 &  0.00 &  1.13 &  0.00 & 0.00 &  0.00 \\
\nsd4  &  3.79 &  3.43 &  3.62 &  3.56 & 1.55 &  4.86 \\
\nsd5  &  3.65 &  2.51 &  2.07 &  4.46 & 1.54 &  1.66 \\
\nsd6  &  0.80 &  2.13 &  1.52 &  1.17 & 0.38 &  1.55 \\
\nsd7  &  0.00 &  0.00 &  0.00 &  0.00 & 0.00 &  0.00 \\
\nsd8  &  0.00 &  0.00 &  9.06 &  0.00 & 0.00 &  0.00 \\
\nsd9  &  0.00 &  0.77 &  0.79 &  0.00 & 0.00 &  0.43 \\
$\sqrt{\sum_{i=1}^6 [N_{SD}(i)]^2}$ 
       &  5.32 &  4.80 &  4.64 &  5.83 & 4.76 &  5.37 \\
$\sqrt{\sum_{i=1}^7 [N_{SD}(i)]^2}$ 
       &  5.32 &  4.80 &  4.64 &  5.83 & 4.76 &  5.37 \\
$\sqrt{\sum_{i=1-6,8,9} [N_{SD}(i)]^2}$ 
       &  5.32 &  4.86 & 10.21 &  5.83 & 4.76 &  5.39 \\
$\sqrt{\sum_{i=1}^9 [N_{SD}(i)]^2}$ 
       &  5.32 &  4.86 & 10.21 &  5.83 & 4.76 &  5.39 \\
\hline
\end{tabular} 
\caption[]{Scalar Higgs statistical significances, $N_{SD}=S/\sqrt B$,
in various channels for the 6 bench mark points. For each individual Higgs, we
give (in order): $N_{SD}$ for the channels 1) -- 9) described in the text;
Gaussian combined $N_{SD}$ for non-$WW$-fusion LHC channels; combined $N_{SD}$
for non-$WW$-fusion LHC channels plus LEP2; combined $N_{SD}$ for all LHC
channels, including the fusion channels $WW\to h\to \tau\anti\tau$ and $WW\to
h\to WW^{(*)}$ channels; and combined $N_{SD}$ for all LHC channels plus LEP2.
\label{table_2}}
\end{center}
\end{table}

\begin{table}
\begin{center}
\small
\hspace*{-.5cm}
\begin{tabular} {|l|l|l|l|l|l|l|}
\hline
Point Number & 1 & 2 & 3 & 4 & 5 & 6   \\
\hline
Best non-$WW$ fusion $N_{SD}$ 
&  4.37 ($h_1$) &  3.95 ($h_2$) &  3.62 ($h_3$) &  4.46 ($h_3$) &  4.83 ($h_1$)  &  4.86 ($h_3$) \\
\hline Best $WW$ fusion $N_{SD}$ 
& 15.39 ($h_2$) & 15.17 ($h_2$) & 13.46 ($h_2$) & 15.05 ($h_2$) & 16.78 ($h_1$)  & 10.08 ($h_1$) \\
\hline \begin{minipage}{3.8cm}{\baselineskip=0pt Best combined $N_{SD}$
w.o.\\ $WW$-fusion modes\vspace*{.15cm}}\end{minipage} 
&  6.54 ($h_1$) &  5.05 ($h_2$) &  4.76 ($h_2$) &  6.31 ($h_2$) &  6.69 ($h_1$)  &  5.37 ($h_3$) \\
\hline \begin{minipage}{3.8cm}{\baselineskip=0pt Best combined $N_{SD}$
with\\ $WW$-fusion modes\vspace*{.15cm}}\end{minipage}
& 17.65 ($h_2$) & 16.00 ($h_2$) & 14.28 ($h_2$) & 17.40 ($h_2$) & 18.07 ($h_1$)  & 10.73 ($h_1$) \\
\hline
\end{tabular}
\caption[]{Summary for all Higgs bosons. The entries are: maximum
non-$WW$ fusion LHC $N_{SD}$; maximum LHC $WW$ fusion $N_{SD}$; best combined
$N_{SD}$ after summing over all non-$WW$-fusion LHC channels; and best
combined $N_{SD}$ after summing over all LHC channels. The Higgs boson for
which these best values are achieved is indicated in the parenthesis. One
should refer to the preceding table in order to find which channel(s) give the
best values.
\label{table_3}}
\end{center}
\end{table}

\section{NMSSM Parameters and Scanning Procedure}

In this paper, we consider the simplest version of the NMSSM
\cite{Ellis:1989er,Derendinger:1984bz,Drees:1989fc,Espinosa:1992gr,%
Binetruy:1992mk,Ellwanger:1993xa,Ellwanger:1995ru,Ellwanger:1997gw,%
Kamoshita:1994iv,Franke:1995xn,King:1995vk,King:1996ys,Ham:1997sf,Krasnikov:1998nh}, 
where the term $\mu \what H_1 \what H_2$ in the
superpotential of the MSSM is replaced by (we use the notation $\what A$ for
the superfield and $A$ for its scalar component field)
\begin{equation}
\label{2.1r}
\lambda \what H_1 \what H_2 \what S\ + \ \frac{\kappa}{3} \what S^3 \ \ ,
\end{equation}
so that the superpotential is scale invariant. We make no assumption on
``universal'' soft terms. Hence, the five soft supersymmetry breaking terms
\begin{equation}
\label{2.2r}
m_{H_1}^2 H_1^2\ +\ m_{H_2}^2 H_2^2\ +\ m_S^2 S^2\ +\ \lambda
A_{\lambda}H_1 H_2 S\ +\ \frac{\kappa}{3} A_{\kappa}S^3
\end{equation}
are considered as independent. The masses and/or couplings of sparticles are
assumed to be such that their contributions to the loop diagrams inducing Higgs
production by gluon fusion and Higgs decay into $\gamma \gamma$ are negligible.
In the stop sector, which appears in the radiative corrections to the Higgs
potential, we chose the soft masses $m_Q = m_T \equiv M_{susy}= 1$ TeV, and
varied the stop mixing parameter 
\begin{equation}
\label{2.4r}
X_t \equiv 2 \ \frac{A_t^2}{M_{susy}^2+m_t^2} \left ( 1 -
\frac{A_t^2}{12(M_{susy}^2+m_t^2)} \right ) \ .
\end{equation}
As in the MSSM, the value $X_t = \sqrt{6}$ -- so called maximal mixing --
maximizes the radiative corrections to the Higgs masses, and we found that it
leads to the most challenging points in the parameter space of the NMSSM. 

Assuming that the Higgs sector is CP conserving, the independent parameters of 
the model are thus: $\lambda, \kappa, m_{H_1}^2, m_{H_2}^2, m_S^2, A_{\lambda}$
and $A_{\kappa}$. For purposes of scanning and analysis, it is more convenient
to eliminate $m_{H_1}^2$, $m_{H_2}^2$ and $m_S^2$ in favor of $M_Z$,
$\tan\beta$ and $\mu_{\rm eff} = \lambda \langle S \rangle$ through the three
minimization equations of the Higgs potential (including the dominant 1- and
2-loop corrections \cite{Ellwanger:1999ji}) and to scan over the six independent parameters 
\begin{equation}
\label{2.5r}
\lambda, \kappa, \tan\beta, \mu_{\rm eff}, A_{\lambda}, A_{\kappa}\ .
\end{equation}
We adopt the convention $\lambda,\kappa>0$, in which $\tanb$ can have either
sign. The absence of Landau singularities for $\lambda$ and $\kappa$ below the
GUT scale ($\sim 2\times10^{16}$ GeV) imposes upper bounds on these couplings
at the weak scale, which depend on the value of $h_t$ and hence of $\tanb$
\cite{Ellis:1989er,Derendinger:1984bz,Drees:1989fc,Espinosa:1992gr,Binetruy:1992mk}. 
Using $m_{top}^{pole} = 175$ GeV, one finds $\lambda_{\rm
max}\sim 0.69$ and $\kappa_{\rm max}\sim 0.62$ for intermediate values of
$\tanb$.

For each point in the parameter space, we diagonalize the scalar and
pseudo-scalar mass matrices and compute the scalar, pseudo-scalar and charged
Higgs masses and couplings taking into account the dominant 1- and 2-loop
radiative corrections \cite{Ellwanger:1999ji}. We then demand that the Higgs scalars satisfy
the LEP2 constraints on the $e^+e^- \to Zh_i$ production mode (taken from
\cite{LEP0103}, fig. 10), which gives a lower bound on $m_{h_i}$ as a function of
the $ZZh_i$ reduced coupling. We also impose LEP2 constraints on $e^+e^- \to
h_ia_j$ associated production (from \cite{LEP0104}, fig. 6), yielding a lower
bound on $m_{h_i} + m_{a_j}$ as a function of the $Zh_ia_j$ reduced coupling.

In order to render the above-mentioned processes i) -- ix) kinematically
impossible, we require the following inequalities among the masses:
\begin{eqnarray}
& m_{h_3} < 2m_{h_1}, \ 2m_{a_1}, \ 2m_{h^\pm}, \ m_{a_1}+M_Z, \ m_{h^\pm}+M_W;
\nn \\
& m_{a_2} < m_{h_1}+m_{a_1}, \ m_{h_1}+M_Z, \ m_{h^\pm}+M_W; \quad m_{h^\pm} >
155 \mbox{GeV}. \nn
\end{eqnarray}
\noi In addition we require $|\mu_{\rm eff}|\ >\ 100$ GeV; otherwise a light
chargino would have been detected at LEP2. (The precise lower bound on
$|\mu_{\rm eff}|$ depends somewhat on $\tan\beta$ and the precise experimental
lower bound on the chargino mass; however, our subsequent results do not depend
on the precise choice of the lower bound on $|\mu_{\rm eff}|$.) We further note
that for the most challenging parameter space points that we shall shortly
discuss, $|\mu_{\rm eff}|\ >\ 100\gev$ is already sufficient to guarantee that
the NMSSM Higgs bosons cannot decay to chargino pairs so long as the SU(2)
soft-SUSY-breaking parameter $M_2$ is also large. In fact, in order to avoid
significant corrections to $\gam\gam h_i$ and $\gam\gam a_i$ couplings coming
from chargino loops it is easiest to take $M_2\gg \mu_{\rm eff}$ (or vice
versa). This is because the $h_i \widetilde \chi^+_i \widetilde \chi^-_i$
coupling is suppressed if the $\widetilde \chi^+_i$ is either pure higgsino or
pure gaugino. Since the parts of parameter space that are challenging with
regard to Higgs detection typically have $|\mu|\sim 100 - 200\gev$, the
validity of our assumptions requires that $M_2$ be large and that the chargino
be essentially pure higgsino.

Using a very rough sampling, we determined, as expected from previous work
\cite{Gunion:1996fb}, that it is only for moderate values of $\tanb$ that $<5\sigma$
signals might possibly occur. From this sampling, we determined the most
difficult parameter space regions and further refined our scan to the
following:
\begin{itemize}
\itemsep=0in
\item
$4.5<|\tanb|<8$ (both signs) in steps of 0.25;
\item
$0.001<\lambda<{\rm min}[0.21,\lambda_{\rm max}]$, using 20 points;
\item
$0.001<\kappa<{\rm min}[0.24,\kappa_{\rm max}]$, using 20 points; 
\item
$100\gev<|\mu_{\rm eff}|<300\gev$ (both signs), in steps of 5 GeV;
\item
$0<|A_\lambda|<160\gev$, with $A_\lambda$ opposite in sign to $\mu_{\rm eff}$,
using steps of 5 GeV;
\item
$25\gev<|A_\kappa|<170\gev$, with $A_\kappa$ opposite in sign to $\mu_{\rm
eff}$, using steps of 5 GeV.
\end{itemize}
For those points sampled in this final scan which satisfy all the constraints
detailed earlier, we compute the expected statistical significances for the
processes 1) to 9) listed in section 1, as described in the next section. As a
rough guide, from the $\sim 10^9$ points detailed in the above list, we find
about $250,000$ that pass all constraints and have $N_{SD}<5$ (for $L=300~\mbox{fb$^{-1}$}$)
in each of the individual discovery modes 1) -- 7). We shall tabulate a number
of representative points taken from this final set in section 4.

\section{Expected Statistical Significances}

>From the known couplings of the NMSSM Higgs scalars to gauge bosons and
fermions it is straightforward to compute their production rates in gluon-gluon
fusion and various associated production processes, as well as their partial
widths into $\gamma \gamma$, gauge bosons and fermions, either relative to a
standard model Higgs scalar or relative to the MSSM $H$ and/or $A$. This allows
us to apply ``NMSSM corrections'' to the processes 1) -- 9) above.

These NMSSM corrections are computed in terms of the following ratios. For the
scalar Higgs bosons, $c_V$ is the ratio of the coupling of the $h_i$ to vector
bosons as compared to that of a SM Higgs boson (the coupling ratios for $h_iZZ$
and $h_iWW$ are the same), and $c_t$, $c_b$ are the corresponding ratios of
the couplings to top and bottom quarks (one has $c_\tau=c_b$). Note that we
always have $|c_V| < 1$, but $c_t$ and $c_b$ can be larger, smaller or even
differ in sign with respect to the standard model. For the CP-odd Higgs bosons,
$c_V$ is not relevant since there is no tree-level coupling of the $a_i$ to the
$VV$ states; $c_t$ and $c_b$ are defined as the ratio of the $i\gamma_5$
couplings for $t\anti t$ and $b\anti b$, respectively, relative to SM-like
strength.

We emphasize that our procedure implicitly includes QCD
corrections to the Higgs production processes at precisely the same
level as the experimental collaborations.
First, the ATLAS and CMS collaborations employed Monte Carlo programs
such as ISAJET \cite{ISAJET} and PYTHIA \cite{PYTHIA} in obtaining
results for the (MS)SM. These programs include many QCD corrections to
Higgs production in a leading-log sense. This is the best that can
currently be done to implement QCD corrections in the context of
experimental cuts and neural-net analyses. Clearly the more exact NNLO
results for many of the relevant processes will slowly be implemented
in the Monte Carlo programs and increased precision for Higgs discovery
expectations will result. Since our goal is to obtain NMSSM results
that are completely analogous to the currently available (MS)SM
results, we have proceeded by simply rescaling the available (MS)SM
experimental analyses. 
In doing the rescaling of the Higgs branching ratios we have included
all relevant higher-order QCD corrections \cite{Djouadi:1996gt} using an
adapted version of the FORTRAN code HDECAY \cite{Djouadi:1998yw}. 
Details regarding our rescaling procedures can
be found in \cite{Ellwanger:2001iw}.  
Using the rescaling  procedures, 
for each point in the parameter space of the NMSSM
we obtain the statistical significances predicted for an integrated luminosity
of $100~\mbox{fb$^{-1}$}$ for each of the detection modes 1) -- 9). In order to obtain the
statistical significances for the various detection modes at $300~\mbox{fb$^{-1}$}$, we
multiply the $100~\mbox{fb$^{-1}$}$ statistical significances by $\sqrt{3}$ in the cases
1), 2), 3), 5) and 6), but only by a factor of $1.3$ in the cases 4), 8) and
9). That such a factor is appropriate for mode 4), see, for example, fig. 19-62
in \cite{atlastdr3}. Use of this same factor for modes 8) and 9) is simply a
conservative guess.

\section{Difficult Points}

As stated in the introduction we still find ``black spots'' in the parameter
space of the NMSSM, where the expected statistical significances for all Higgs
detection modes 1) -- 7) are below $5 \sigma$ at $300~\mbox{fb$^{-1}$}$. The reasons for
this phenomenon have been described above; see also the corresponding
discussion in \cite{Gunion:1996fb}. However, after including the modes 8) and 9), the
points that provide the worst 1) -- 6) statistical significances typically
yield robust signals in one or the other of the $WW$-fusion modes 8) and 9).

In order to render the corresponding suppression mechanisms of the detection
modes reproducible, we present the detailed properties of several difficult
points in the parameter space in table \ref{table_1}. The notation is as follows: The bare
parameters are as in eq. (2.5), with $m_{H_1}^2$, $m_{H_2}^2$ and $m_S^2$ fixed
implicitly by the minimization conditions. (As noted earlier, with the
convention $\lambda$, $\kappa > 0$ in the NMSSM, the sign of $\tan\beta$ can no
longer be defined to be positive.) For the reasons discussed below eq.
(\ref{2.4r}) we chose in the stop sector $m_Q = m_T \equiv M_{susy}= 1$ TeV and
$X_t=\sqrt{6}$ for all of the points (1 -- 6). We have also fixed
$m_{top}^{pole}=175$ GeV. For both scalar and pseudoscalar Higgs bosons, ``gg
Production Rate'' denotes the ratio of the gluon-gluon production rate with
respect to that obtained if $c_t=c_b=1$, keeping the Higgs mass fixed. For
scalar $h_i$, this is the same as the ratio of the $gg$ production rate
relative to that predicted for a SM Higgs boson of the same mass. For the
scalar $h_i$, $\br\gamma \gamma$ denotes the ratio of the $\gamma \gamma$
branching ratio with respect to that of a SM Higgs boson with the same mass. (A
verification of the reduced gluon-gluon production rates or $\gamma \gamma$
branching ratios would sometimes require the knowledge of the couplings to
higher precision than given, for convenience, in table \ref{table_1}.) Also given for the
scalar $h_i$ are the ratios $\br b\anti b$ and $\br WW^*$ of the $b\anti b$ and
$WW^*$ branching ratios relative to the SM prediction (as noted above, one has
$\br\tau\anti\tau=\br b\anti b$). 

In table \ref{table_2}, we tabulate the statistical significances for the $h_i$ in all the
channels 1) -- 9); production of the CP-odd $a_i$ turns out to be relevant only
when they add to the $h_i$ signals in process 4). Also note that, all these
problematical points are such that $m_{h_1}+m_{a_1}>206\gev$, so that
$e^+e^-\to h_1+a_1$ followed by $h_1,a_1 \to b\anti b$ would have been
kinematically forbidden at the highest LEP2 energy. Hence, for LEP2 mode 7) we
only give the statistical significance for $e^+e^- \to Z h_i$. Also tabulated
in table \ref{table_2} are four statistical significances obtained by combining various
channels. This combination is done in the Gaussian approximation:
\begin{equation}
N_{SD}^{\rm combined}= \left[\sum_{i} \left(N_{SD}^i\right)^2\right]^{1/2} \, ,
\nn
\end{equation}
where $\sum_i$ runs over the channels $i$ being combined. We give results for
the following combinations:
\begin{description}
\item{a)} $N_{SD}$ obtained by combining LHC channels 1) -- 6); 
\item{b)} $N_{SD}$ obtained by combining LHC channels 1) -- 6) and LEP2; 
\item{c)} $N_{SD}$ obtained by combining LHC channels 1) -- 6) with the
$WW$-fusion channels 8) and 9);
\item{d)} $N_{SD}$ obtained by combining all LHC channels and LEP2, \ie\ by
combining all channels 1) -- 9).
\end{description}
In those cases where there is no LEP2 signal, a)=b) and c)=d). In addition, in
our point selection we have required a mass difference of at least 10 GeV
between scalar Higgses, so that they yield well separated signals and no
statistical significance combination of two different scalar Higgses is needed.
All parameter choices for which Higgs boson masses differ by less than
10 GeV yield stronger signals than the cases retained. 
(The increased net signal
strength of overlapping Higgs signals in those
channels with limited mass resolution arises as a result
of $N_{SD}^{\rm eff}(1+2)\sim (S_1+S_2)/\sqrt B>\sqrt{S_1^2+S_2^2}/\sqrt B$.) 

As summarized in table \ref{table_3}, all of the tabulated ``bench mark points'' have
statistical significances below $5 \sigma$ for all of the detection modes 1) --
6) at $300~\mbox{fb$^{-1}$}$ and 7) at LEP2. In more detail, as tabulated in table \ref{table_2} and
summarized in table \ref{table_3}, the best signals in the modes 1) -- 6) for the points
\#1 -- \#6 at the LHC are:
\begin{itemize}
\item
point \#1, $N_{SD}=$4.37 for mode 2) and $h_1$;
\item
point \#2, $N_{SD}=$3.95 for mode 3) and $h_2$;
\item
point \#3, $N_{SD}=$3.62 for mode 4) and $h_3$; 
\item
point \#4, $N_{SD}=$4.46 for mode 5) and $h_3$;
\item 
point \#5, $N_{SD}=$4.83 for mode 3) and $h_1$; 
\item
point \#6, $N_{SD}=$4.86 for mode 4) and $h_3$;
\end{itemize}
Further, for point \#3, the combined statistical significance of modes 1) -- 6)
(also tabulated in table \ref{table_3}) would still be below $5$ for any one $h_i$, 
although $\sqrt 2 N_{SD}^{1-6}>5$ (as is likely to be relevant by combining
ATLAS and CMS data once each detector has accumulated $L=300~\mbox{fb$^{-1}$}$) for at least
one of the $h_i$. However, for all these ``difficult'' points the $WW$-fusion
modes 8) and/or 9) provide (according to theoretical estimates) a decent
(sometimes very strong) signal.

The points \#1 -- \#4 differ as to which of the modes 1) -- 6) and which $h_i$
yields the largest statistical significance should the $WW$-fusion mode 8) not
provide as strong a signal as suggested by the theoretical estimates. To render
these points observable without the $WW$-fusion mode 8) would require
improvements of all detection modes 2) -- 5). 

As in \cite{Gunion:1996fb}, we find that difficult points in the parameter space generally
have $|\tan\beta| \sim 5$. This is the region of $\tan\beta$ for which the
$b\anti b h,b\anti b a$ signals are still not very much enhanced but yet the 
$gg\to h,a$ and $t\anti t h,t\anti t a$ signals have been suppressed somewhat.
In a few cases, however, difficulties also arise for $|\tan\beta|$ as large as
8, as shown in the case of point \#5. Also as in \cite{Gunion:1996fb}, the most difficult
points are those in which the masses of the $h_i$ and $a_i$ are relatively
close in magnitude, typically clustered in a $\sim 60\gev$ interval above $\sim
105\gev$. Such clustering maximizes the mixing among the different Higgs bosons
and thereby minimizes the significance of the discovery channels for any one
Higgs boson. In particular, it is for strong mixing among the $h_i$ that the
statistical significance for discovery modes based on a large $VV$ coupling for
any one $h_i$ are most easily suppressed.

Finally, for point \#6, we have minimized the statistical significances for the
$WW$-fusion modes over the parameter space, while keeping the statistical
significances of modes 1) -- 6) below 5. One can see that it still gives a
strong $10.1 \sigma$ signal in mode 8). 
[Smaller $N_{SD}$ for mode 8) would have been possible if we had allowed 
stronger signals in modes 1) -- 6), in particular had we allowed smaller
mass separation, $<10\gev$, between the two lightest Higgs bosons.]
In addition, for point \#6 $m_{h_1} =
112$ GeV and the $ZZ$ coupling of $h_1$ is sufficiently large that it would
have yielded a $4.8 \sigma$ signal at LEP2. Had we taken a top quark mass
slightly larger, $m_{top}^{pole}=178$ GeV, we would have found a very similar
point with a $h_1$ mass of $\sim 115$ GeV, which could have been responsible
for the excess observed at LEP2 \cite{Barate:2000ts,Abreu:2000fw,%
Acciarri:2000hv,Abbiendi:2000ac}.

\section{Discussion and Conclusions}

In this paper, we have addressed the question of whether or not it would be
possible to fail to discover any of the Higgs bosons of the NMSSM using
combined LEP2 and LHC data, possibly resulting in the erroneous conclusion that
Higgs bosons with masses below 200 GeV have been excluded. We have demonstrated
that, assuming that the decay channels i) -- ix) are either kinematically
disallowed or render a Higgs boson observable, this is unlikely (at the
$>5\sigma$ level) to happen. Certainly, there are points in NMSSM parameter
space for which the statistical significances for the individual detection
modes 1) -- 6) (\ie\ those analyzed in detail by ATLAS and CMS) are all well
below $5 \sigma$ for integrated luminosity of $300~\mbox{fb$^{-1}$}$. However, by combining
several of the modes 1) -- 6) and $300~\mbox{fb$^{-1}$}$ data from both ATLAS and CMS, a
$>5\sigma$ signal can be achieved based just on modes 1) -- 6). Further, we
have found that throughout all of the NMSSM parameter space (scanned subject to
the earlier listed restrictions) for which such weak signals in modes 1) -- 6)
are predicted, the theoretical estimates for the $WW$-fusion modes indicate
that an easily detected $WW\to h\to \tau\anti \tau$ signal should be present.
Thus, our conclusion is that for all of the parameter space of the NMSSM
compatible with reasonable boundary conditions for the parameters at the GUT
scale (with, of course, non-universal soft terms in general) and such that
Higgs pair and SUSY pair decays of the Higgs bosons are kinematically
forbidden, at least one of the NMSSM Higgs bosons will be detected at the LHC.
This is a big improvement over the results from the earlier Snowmass 1996 study
which was somewhat negative without the inclusion of the $t\anti t h\to t\anti
t b\anti b$ mode 3), and the $WW$-fusion modes 8) and 9).

It is amusing to note that all of our bench mark points for which Higgs
discovery is most difficult at the LHC include a Higgs scalar with mass close
to 115 GeV (with, however, reduced couplings to the $Z$ boson), which could be
responsible for the excess observed at LEP2 \cite{Barate:2000ts,Abreu:2000fw,%
Acciarri:2000hv,Abbiendi:2000ac}.

Another important point that appears from our analysis is the fact that the
full $L=300~\mbox{fb$^{-1}$}$ of integrated luminosity (per detector) is needed in order to
have robust NMSSM Higgs discovery in the portion of parameter space considered
here. Of course, as in the MSSM, it is very possible that only one of the
CP-even NMSSM Higgs bosons might be detected at the LHC but that, as studied by
Kamoshita et al. in \cite{Ellwanger:1993xa,Ellwanger:1995ru,Ellwanger:1997gw,%
Kamoshita:1994iv,Franke:1995xn,King:1995vk,King:1996ys,Ham:1997sf,Krasnikov:1998nh}, 
the observation of all the CP-even Higgs bosons
of the NMSSM would be possible at the LC by virtue of all having some
non-negligible level of $ZZ$ coupling and not having very high masses. Even at
the LC, the CP-odd Higgs bosons might escape discovery, although this would not
be the case for the parameter choices that we have found which make LHC
discovery of even one NMSSM Higgs bosons most challenging. This is because, for
such parameters, the $a_i$ are relatively light and could be readily seen at
the LC in the processes $e^+e^-\to h_i a_j$, $e^+e^-\to \nu\anti \nu a_ia_i$
and $e^+e^-\to Z^*\to Z a_ia_i$, assuming an integrated LC luminosity of
$1000~\mbox{fb$^{-1}$}$ and energy $\sqrt s\geq 500\gev$ \cite{Farris:2002ny}.

This study makes clear the importance of continuing to expand the sensitivity
of existing modes and continuing to develop new modes for Higgs detection at
the LHC in order not to have to wait for construction of a linear $e^+e^-$
collider for detection of at least one of the SUSY Higgs bosons. In particular,
study of modes i) -- ix) and SUSY pair channels should all be pushed. The
problematical points that we have emphasized here are unlikely to be
substantially influenced by $t\anti t $ or SUSY decays since all the Higgs
masses are below $\sim 200\gev$ so that $t\anti t$ decays will be kinematically
highly suppressed (one of the top quarks would have to be virtual) and SUSY
pair decays are quite unlikely to be significant given LEP2 limits on the
masses of SUSY particles. However, by allowing Higgs (in particular,
pseudoscalar) masses such that one or more of the channels i)-vii) are
kinematically allowed we have found points for which discovery in modes 1)-9)
will not be possible \cite{wip}. Thus, a full ``no-lose'' theorem for NMSSM
Higgs boson discovery at the LHC will require exploring additional discovery
modes sensitive to those portions of parameter space for which Higgs decays to
other Higgs bosons are important, and might necessitate combining results from
both the ATLAS and CMS detectors and/or accumulating more integrated
luminosity.


\section{Acknowledgments}
We wish to acknowledge helpful discussions with M. Sapinski and D. Zeppenfeld.
C.H. would like to thank the late theory group of the Rutherford Appleton
Laboratory and the Theoretical Physics Department of Oxford, where part of this
work was achieved, for their kind hospitality. This work was supported in part
by the U.S. Department of Energy.

\setcounter{figure}{0}
\setcounter{table}{0}
\setcounter{section}{0}
\setcounter{equation}{0}
\clearpage

\def\squaresm{\lower0.085ex\hbox{$\square$}}
\def\be{\begin{equation}}
\def\ee{\end{equation}}
\def\br{\begin{eqnarray}}
\def\er{\end{eqnarray}}
\def\ba{\begin{array}}
\def\ea{\end{array}}
\def\bi{\begin{itemize}}
\def\ei{\end{itemize}}
\def\bn{\begin{enumerate}}
\def\en{\end{enumerate}}
\def\bc{\begin{center}}
\def\ec{\end{center}}
\def\ul{\underline}
\def\ol{\overline}
\def\eps{\epsilon}

\def\Ord{\buildrel{\scriptscriptstyle <}\over{\scriptscriptstyle\sim}}
\def\OOrd{\buildrel{\scriptscriptstyle >}\over{\scriptscriptstyle\sim}}

\def\gsim{\lower.7ex\hbox{$\;\stackrel{\textstyle>}{\sim}\;$}}
\def\lsim{\lower.7ex\hbox{$\;\stackrel{\textstyle<}{\sim}\;$}}

\def\zpc#1 #2 #3{Z.\ Phys.\ {\bf C#1} (19#2) #3}
\def\rmp#1 #2 #3{Rev.\ Mod.\ Phys.\ {\bf C#1} (19#2) #3}
\def\plb#1 #2 #3{Phys.\ Lett.\ {\bf B#1} (19#2) #3}
\def\plold#1 #2 #3{Phys.\ Lett.\ {\bf #1B} (19#2) #3}
\def\npb#1 #2 #3{Nucl.\ Phys.\ {\bf B#1} (19#2) #3}
\def\prd#1 #2 #3{Phys.\ Rev.\ {\bf D#1} (19#2) #3}
\def\prdr#1 #2 #3{Phys.\ Rev.\ {\bf D#1} (19#2) R#3}
\def\prl#1 #2 #3{Phys.\ Rev.\ Lett.\ {\bf #1} (19#2) #3}
\def\prep#1 #2 #3{Phys.\ Rep.\ {\bf C#1} (19#2) #3}
\def\niam#1 #2 #3{Nucl.\ Instr.\ and Meth.\ {\bf #1} (19#2) #3}
\def\mpl#1 #2 #3{Mod.\ Phys.\ Lett.\ {\bf A#1} (19#2) #3}
\def\cpc#1 #2 #3{Comp.\ Phys.\ Commun.\ {\bf #1} (19#2) #3}

\def\preprint{{\it preprint}}

\def\sign{\mathrm{sign}}
\def\ppb{p\bar{p}}
\def\as{\alpha_s}
\def\gl{\tilde{g}}
\def\sq{\tilde{q}}
\def\sqb{{\tilde{q}}^*}
\def\qb{\bar{q}}
\def\sqL{\tilde{q}_{_L}}
\def\sqR{\tilde{q}_{_R}}
\def\ms{m_{\tilde q}}
\def\mg{m_{\tilde g}}
\def\Gs{\Gamma_{\tilde q}}
\def\Gg{\Gamma_{\tilde g}}
\def\md{m_{-}}
\def\eps{\varepsilon}
\def\Ce{C_\eps}
\def\dnq{\frac{d^nq}{(2\pi)^n}}
\def\DR{$\overline{DR}$\,\,}
\def\MS{$\overline{MS}$\,\,}
\def\DRm{\overline{DR}}
\def\MSm{\overline{MS}}
\def\ghat{\hat{g}_s}
\def\shat{\hat{s}}
\def\sihat{\hat{\sigma}}
\def\Li{\text{Li}_2}
\def\bs{\beta_{\sq}}
\def\xs{x_{\sq}}
\def\xsa{x_{1\sq}}
\def\xsb{x_{2\sq}}
\def\bg{\beta_{\gl}}
\def\xg{x_{\gl}}
\def\xga{x_{1\gl}}
\def\xgb{x_{2\gl}}
\def\lsp{\tilde{\chi}_1^0}
\def\as{\alpha_{\mbox{\tiny S}}}

\def\gluino{\mathaccent"7E g}
\def\mgluino{m_{\gluino}}
\def\squark{\mathaccent"7E q}
\def\M{ \overline{|\mathcal{M}|^2} }
\def\utm{ut-M_a^2M_b^2}
\def\MiLR{M_{i_{L,R}}}
\def\MiRL{M_{i_{R,L}}}
\def\MjLR{M_{j_{L,R}}}
\def\MjRL{M_{j_{R,L}}}
\def\tiLR{t_{i_{L,R}}}
\def\tiRL{t_{i_{R,L}}}
\def\tjLR{t_{j_{L,R}}}
\def\tjRL{t_{j_{R,L}}}
\def\tg{t_{\gluino}}
\def\uiLR{u_{i_{L,R}}}
\def\uiRL{u_{i_{R,L}}}
\def\ujLR{u_{j_{L,R}}}
\def\ujRL{u_{j_{R,L}}}
\def\ug{u_{\gluino}}
\def\utot{u \leftrightarrow t}
\def\ar{\to}

\def\cpmtwo{\mbox{$ {\chi}^{\pm}_{2}                    $}}
\def\cpmone{\mbox{$ {\chi}^{\pm}_{1}                    $}}

\part{{\bf  Effects of Supersymmetric Phases on Higgs Production  
in Association with Squark Pairs 
in the Minimal Supersymmetric Standard Model}\\[0.5cm] \hspace*{0.8cm}
{\it A. Dedes, S. Moretti}}
\label{dmsec}


\begin{abstract}{
We show how  the Supersymmetric (SUSY) CP-violating phases 
can induce new final states in associated production of Higgs bosons with
squark pairs of identical flavor (for the $A^0$) as well as modify 
substantially those
already present when the soft SUSY parameters are real (in the case
of $H^0$ and $h^0$). Hence these processes, particularly for light
stop squarks, $\tilde t_1$, are good candidates 
for phenomenological investigation, in order to confirm or disprove 
the existence of complex soft SUSY
parameters. We illustrate this in the context of a general
Minimal Supersymmetric Standard Model (MSSM), 
for a choice of SUSY parameters  accessible 
at the Large Hadron Collider (LHC).
}\end{abstract}

\noindent
It has recently been shown \cite{Dedes:1999sj,Dedes:1999zh,Choi:1999aj,%
Asakawa:2000es,Akeroyd:2001kt,Choi:2001iu,Arhrib:2001pg} that, if 
CP-violating effects are manifestly inserted into the MSSM Lagrangian, by 
allowing the Higgsino mass term, $\mu$, and the trilinear couplings,
$A$\footnote{For simplicity, we assume $A\equiv A_u = A_d$ at the electroweak (EW)
scale, i.e., ${\cal O}(M_Z)$,
where $u$ and $d$ refer to all flavors of 
up- and down-type (s)quarks.}, to be 
complex, thereby introducing  two independent CP-violating 
phases \cite{Dugan:1985qf,Dimopoulos:1996kn}, $\phi_\mu$ and $\phi_A$, such that
$e^{i\phi_\mu} \ = {\mu}/{|\mu|}$ and 
$e^{i\phi_A  } \ =   {A}/{|A|}$, then the strength of the
Higgs couplings to (s)particles can drastically be modified, inducing
sizable effects, e.g., in the dominant  production mode
of neutral Higgs bosons at the LHC, i.e.,
$gg\to \Phi^0$ (where $\Phi^0=h^0,H^0$ and $A^0$), through the
squark-squark-Higgs vertices involving stops and sbottoms. 
These effects are a consequence
of large values attained by $\phi_\mu$ and/or $\phi_A$
consistent with cancellations taking place in the SUSY contributions to
the Electric Dipole Moments (EDMs) of neutron and
electron~\cite{Ibrahim:1998gj,Falk:1998pu,Ibrahim:1998je,Brhlik:1998zn}.
 
\noindent
These same interactions also affect the associated production of Higgs bosons
with third generation squark pairs.  Since this process is expected
to be accessible at the LHC, see 
Refs.~\cite{Djouadi:1998xx,Djouadi:1999dg,Dedes:1998yt,%
Dedes:1999ku,Belanger:1999pv,Dedes:1999yr,Djouadi:2000gu},
we investigate here some aspects of its phenomenology in the
presence of complex parameters in the MSSM Lagrangian.
Schematically, the production mechanism is the following:
\begin{equation}
g + g~,~q + \bar q ~\longrightarrow~
{\tilde{q}}_{\chi} + {\tilde{q}}^{*}_{\chi'} + \Phi^0,
\label{proc}
\end{equation}
where $q=t,b$, $\chi^{(')}=1,2$ and $\Phi^0=h^0,H^0,A^0$, in all
possible combinations, as appropriate in the MSSM. 
Notice that in such processes the existence of CP-violating effects in the
SUSY Lagrangian would
immediately be manifest from the detection of three particle
final states involving a pseudoscalar Higgs boson and two identical 
squarks. In fact, if $\phi_\mu=\phi_A=0$, even in presence of
mixing between the third-generation squarks, the 
${\tilde{q}}_{\chi}{\tilde{q}}_{\chi} A^0$ couplings, 
with $q=t,b$ and $\chi=1,2$,  are identically zero \cite{Dedes:1998yt}.
Depending on the relative value of the final state masses in (\ref{proc}), 
$m_{{\tilde{q}}_{\chi}}$, $m_{{\tilde{q}}_{\chi'}}$ and
$M_{\Phi^0}$, the production of Higgs particles can be regarded
as taking place either via a (anti)squark decay 
or via a Higgs-strahlung. 

\noindent
We work in the theoretical framework provided by the `complex' MSSM, the
latter including explicitly
the two CP-violating phases, $\phi_\mu$ and $\phi_A$,
and assuming universality of the soft gaugino
masses only at the Grand Unification (GUT) scale.
 We define its fundamental parameters
without making any assumptions about the structure of the
SUSY breaking dynamics
at the Planck scale, whether driven by Supergravity (SUGRA),  
gauge/anomaly mediated (GMSB/AMSB) 
or proceeding via other mechanisms as we
treat the MSSM as a low-energy effective theory.  
Among the possible setups of the MSSM parameter space compatible
at one-loop with the EDM data, we choose here the one
presented in Table~\ref{tab:setup}, in terms
of $\mu$, $A$,  $M_{A^0}$ (the mass of the pseudoscalar Higgs
state), $\tan\beta$ (the  ratio of the vacuum expectation values of the two
Higgs doublet fields), 
$M_{\tilde{q}_{1,2,3}}$ (the soft squark masses of the 
three squark generations) and $M_{\tilde{g}}$ (the soft gluino mass).
For this choice of MSSM inputs, the derived masses of the $h^0$
scalar are barely consistent with the latest bounds on $M_{h^0}$ from LEP2,
of about 91 GeV \cite{limitH} ($m_{{\tilde t}_1}$ is instead set
above the Tevatron limit from Run 1, around 140 GeV \cite{Tevatron} for
our $\tan\beta$). This means that our specific parameter point will readily
be probed at the LHC, as in correspondence of the inputs given in Table~\ref{tab:setup}
and by varying both $\phi_\mu$ and $\phi_A$ between 0 and $\pi$, one has
that the lightest Higgs mass spans from 91 to 100 GeV and the 
lightest stop one is between 140 and 240 GeV or so. Notice that 
the MSSM setup given in Table~\ref{tab:setup} should be regarded 
as one possible example of the rich phenomenology that can be
induced by the CP-violating phases in the MSSM with a low
$\tan\beta$ and squark/Higgs masses small enough to be 
produced at detectable rates via process (\ref{proc}). 
In fact, we have found many others but refrained from showing them here
for reasons of space.

\begin{figure}[!t]
\begin{center}
\epsfig{file=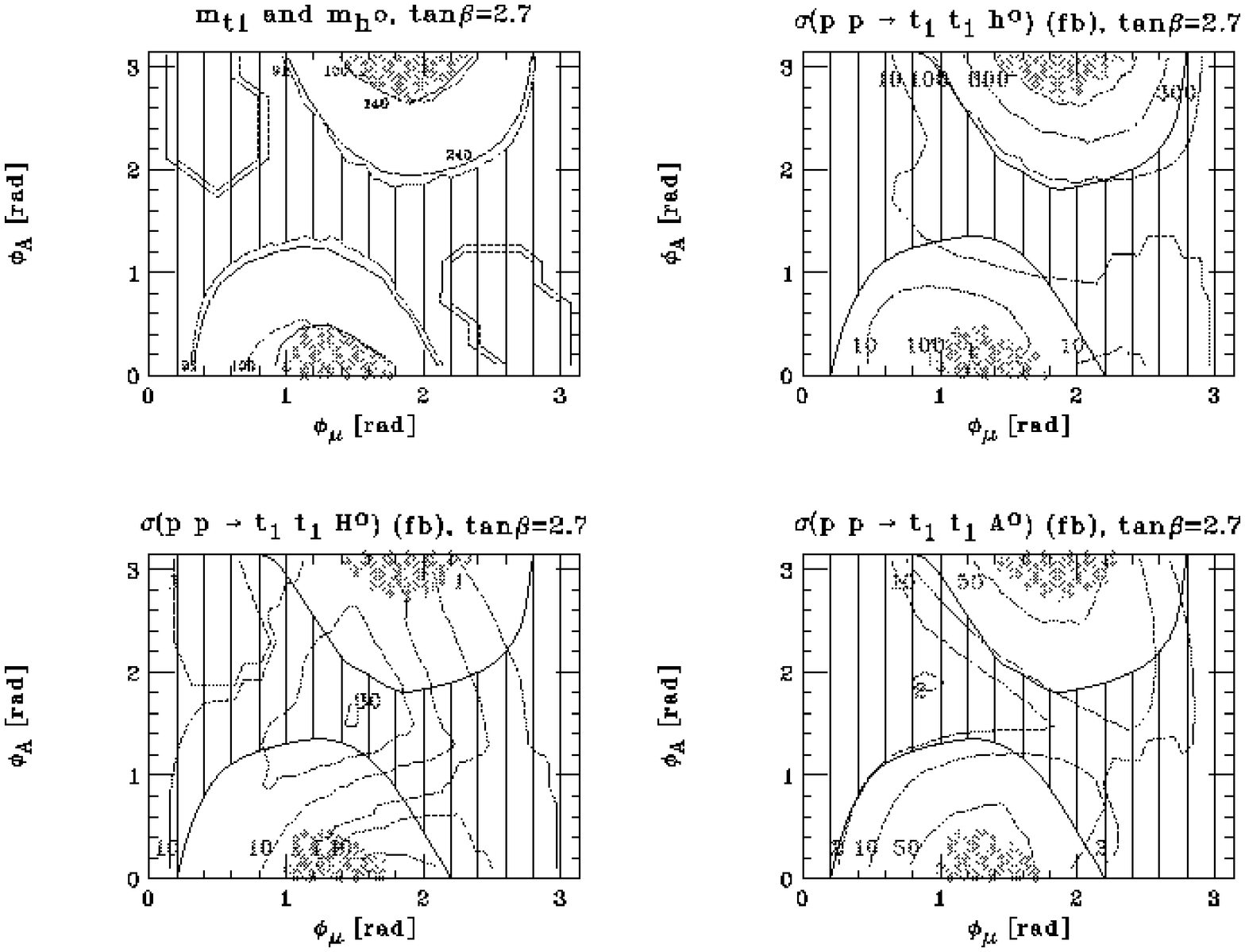,angle=90,height=12.cm}
\end{center}
\caption{\small Contour plots illustrating the $\phi_\mu$ and $\phi_A$
dependence of: 
the lightest Higgs and stop masses (top-left), in red and blue, respectively;
the cross sections for
$gg,q\bar q\to {\tilde{t}}_{1} {\tilde{t}}^{*}_{1}  (h^0)[H^0]\{A^0\}$
(top-right)[bottom-left]\{bottom-right\}, in red.
The meaning of the black shaded area and of
the colored green/magenta symbols is given in the text.
For the cross sections, we 
have used the CTEQ(4L) \cite{Lai:1997mg} Parton Distribution Functions
with $Q=\sqrt{\hat{s}}$ as the factorisation scale 
(also in  $\alpha_s$), the latter evolved 
at two loops with all relevant 
(s)particle thresholds onset within the MSSM
(as described in \cite{Dedes:1996sb,Dedes:1998wn}).}

\label{fig}
\end{figure}

In the remainder, we will denote the regions of the ($\phi_\mu,\phi_A$) plane 
excluded from Higgs and squark direct searches by a shaded area.
In addition, 
the inputs in Table~\ref{tab:setup} comply with the constraints
deduced from the two-loop Barr-Zee type contributions to the
fermionic EDMs \cite{Chang:1998uc,Pilaftsis:1999td} 
(green ``$\diamond\hskip-0.160cm{\cdot}$'' symbols in the following) for most 
choices
of $\phi_\mu$ and $\phi_A$.
Finally, some ($\phi_\mu,\phi_A$) points will further be neglected following
 the requirement of positive definiteness of the squared squark 
masses: see eqs. (5)--(6) of Ref.~\cite{Choi:1999aj,%
Asakawa:2000es,Akeroyd:2001kt,Choi:2001iu,Arhrib:2001pg}
(magenta ``$\times$''      symbols in the forthcoming plots).
The top-left corner of Fig.~\ref{fig}
shows the phase dependence of $M_{h^0}$ and $m_{{\tilde t}_1}$, outside
such experimentally excluded areas.

In the course of our discussion, we shall make only one simplification, 
which will not alter the conclusions of our work. That is, we will neglect
one-loop mixing effects among the three neutral Higgs bosons 
of the MSSM \cite{Demir:1999hj,Demir:1999zb,Pilaftsis:1998dd,Pilaftsis:1998pe,%
Pilaftsis:1999qt,Carena:2000yi}, on the ground that for our
choice of parameters they turn out to be of the order of a few
percent at most (see discussion in Refs.~\cite{Dedes:1999sj,%
Dedes:1999zh,Choi:1999aj,%
Asakawa:2000es,Akeroyd:2001kt,Choi:2001iu,Arhrib:2001pg}). Indeed, much
larger effects will remain unaccounted for, such as the higher-order QCD
corrections to the production process (\ref{proc}), which are likely
to induce $K$-factors of the order 1.5--2 and whose calculation
is presently not available.

\begin{table}[htb]
\begin{center}
\begin{tabular}{|c|c|c|c|c|c|}
\hline
$|\mu|$ & $M_{\tilde{q}_{1,2}}$ & $M_{\tilde{q}_3}$
& $M_{\tilde{g}}$ & $M_{A^0}$ & $\tan\beta$   \\[2mm] \hline
600 & 2500 & 300 & 1000 & 200 & 2.7 \\ \hline
\end{tabular}

\vspace*{2mm}

\caption{\small One possible parameter setup of the MSSM
satisfying the one-loop EDM constraints
(all quantities in GeV, apart from the dimensionless $\tan\beta$)
and yielding cross sections for process (\ref{proc})
 manifestly dependent on the CP-violating phases
$\phi_\mu$ and $\phi_A$.}
\label{tab:setup} 
\end{center}
\end{table}

We start our numerical analysis by referring to Fig.~1 of 
Ref.~\cite{Dedes:1999sj,Dedes:1999zh}, where one can find
the contour plots for the minimum values of
the modulus of the common trilinear coupling, $|A|$, 
above which the mentioned 
cancellations work at a level which is compatible with the
experimental accuracy achieved in both the neutron and electron
EDM measurements, i.e.,
$|d_n| \le 6.3 \times 10^{-26} \; e \, {\rm cm}$ \cite{Harris:1999jx} and 
$|d_e| \le 4.3 \times 10^{-27} \; e \, {\rm cm}$ \cite{Commins:1994gv},
for the choice of parameters in Table~I and as a function of $\phi_\mu$
and $\phi_A$. 
These $A$ values are those entered in the production vertices
of the processes we considered, alongside the two discrete quantities
$|\mu|$ and $\tan\beta$. 
As for the Higgs masses, we keep $M_{A^0}$
fixed at 200 GeV and derive the values of $M_{h^0}$ and $M_{H^0}$ at 
two-loop level (see the discussion in Refs.~\cite{Dedes:1999sj,%
Dedes:1999zh,Choi:1999aj,%
Asakawa:2000es,Akeroyd:2001kt,Choi:2001iu,Arhrib:2001pg} concerning
the residual theoretical error on the latter). 

Among the processes of the type (\ref{proc}), much emphasis has
been put on the case in which both the squark and Higgs scalar
states are the lightest, that is, on the mechanism
$gg,q\bar q\to {\tilde{t}}_{1}  {\tilde{t}}^{*}_{1} h^0$ 
\cite{Djouadi:1998xx,Djouadi:1999dg}. In fact, if the typical $A$ scale is in the TeV
regime, then two concurrent effects take place, that render
light Higgs production in association with the lightest scalar top quarks a
more favorable Higgs discovery channel than the corresponding SM-like one,
$gg,q\bar q\to t\bar t h^0$ \cite{Kunszt:1984ri,Marciano:1991qq}.
On the one hand, since the mixing angle $\theta_{\tilde t}$
is proportional to $m_{t}A$, the ${\tilde t}_1$ squark
becomes much lighter than the $t$-quark and all 
other squarks. On the other hand, by looking at the
expression of the ${\tilde t}_1{\tilde t}_1h^0$ vertex 
(e.g., see eq.~(3) of Ref.~\cite{Djouadi:1998xx}), for large 
$A$ values, it is clear that
its strength can overcome that of the top-Higgs Yukawa
coupling entering the SM-like reaction. 

We thus continue our investigation of squark-squark-Higgs production
by considering this particular final state. Our choice of $\tan\beta$
in Table~I reflects the remark made in Refs.~\cite{Djouadi:1998xx,Djouadi:1999dg}
that the ${\tilde t}_1{\tilde t}_1h^0$ production rates are larger
at smaller $\tan\beta$ (see, e.g., Fig.~3 of Ref~\cite{Djouadi:1998xx}), 
this being the consequence of the increase of $M_{h^0}$ for large values of 
the latter \cite{Choi:1999aj,Asakawa:2000es,Akeroyd:2001kt,%
Choi:2001iu,Arhrib:2001pg}. (As representative of the low $\tan\beta$ regime,
the two values $\tan\beta=2,3$ were used in Refs.~\cite{Djouadi:1998xx,Djouadi:1999dg}.) 
As for the $A$ spectrum, we are bound
to those values guaranteeing the EDM cancellations 
(again, see Fig.~1 in \cite{Dedes:1999sj,Dedes:1999zh}): 
that is, between 100 and 700 GeV,
depending on $\phi_\mu$ and $\phi_A$.  The top-right corner of Fig.~\ref{fig}
presents the cross sections for $gg,q\bar q\to {\tilde{t}}_{1}  
{\tilde{t}}^{*}_{1} h^0$ as a contour plot over the ($\phi_\mu,\phi_A$) plane.
We see a strong dependence on the SUSY phases, as
the production rates vary over several orders of magnitude.
 The maximum of the cross section occurs at large
$\phi_A$ values, when $\phi_\mu$ is slightly above $\pi/2$.
This can be understood in the following terms. Here, it is where the
lightest stop mass reaches its allowed minimum.
However, the fact that $\sigma({\tilde t}_1{\tilde t}_1^*h^0)$ does not grow
similarly at small $\phi_A$ values,  when $\phi_\mu$ is slightly below 
$\pi/2$ -- the other region of the ($\phi_\mu,\phi_A$) plane
where $m_{{\tilde t}_1}$ is minimal -- implies that the
${\tilde t}_1{\tilde t}_1h^0$ vertex too plays an important r\^ole in
determining the actual size of the cross section. 
The fact that this coupling is maximal where the stop mass is minimal
can easily be understood by noticing the $\cos\phi_\mu$
dependence  of eqs.~(5) and (A.53) of Ref.~\cite{Choi:1999aj,%
Asakawa:2000es,Akeroyd:2001kt,Choi:2001iu,Arhrib:2001pg} when 
$\phi_A\to\pi$. Besides, a large value of the ${\tilde t}_1{\tilde t}_1h^0$
coupling combined with a small value of $m_{{\tilde t}_1}$ implies
that the two-loop contributions of the Barr-Zee type graphs
to the EDMs can be sizable, so that
it is not surprising to see that 
larger rates for $gg,q\bar q\to {\tilde{t}}_{1}  {\tilde{t}}^{*}_{1} h^0$ 
accumulate towards the correspondingly excluded area. 
However, in areas not yet removed through the EDM measurements,
 one can find production rates as large as 800 fb. 
Finally, notice that --
for the choice of parameters in Table~I and 100 GeV $\lsim A\lsim$
700 GeV as in Fig.~1 of \cite{Dedes:1999sj,Dedes:1999zh} -- in the `phaseless' limit,
i.e., the standard MSSM case, $\phi_\mu,\phi_A\to 0$, the yield is
several orders of magnitude smaller, indeed well below detection level.

A strong hierarchy exists,
$M_{h^0}\ll M_{H^0}\approx M_{A^0}$, among the neutral Higgs masses,
for our setup of the MSSM. This should naturally allow one
to disentangle in the experimental samples 
${\tilde{t}}_{1}  {\tilde{t}}^{*}_{1} h^0$ from
${\tilde{t}}_{1}  {\tilde{t}}^{*}_{1} H^0$ $+$
${\tilde{t}}_{1}  {\tilde{t}}^{*}_{1} A^0$ events.
Furthermore, under the assumption that the lightest
scalar quark will promptly be discovered at the LHC
from some other source of SUSY events than those
in (\ref{proc}), and its mass measured,
we believe that final states with two heavy objects of
{\sl identical} mass $m_{{\tilde t}_1}$
recoiling against a rather central one
with mass $M_{h^0}$ (that, again, we assume to be known
and reconstructed through $b\bar b$ and/or $\gamma\gamma$ decays),
should be distinguishable from other
${{\tilde q}_\chi}{{\tilde q}_{\chi'}}^*\Phi^0$ channels ($q=t,b$),
in which $\chi$ or $\chi'\ne1$, e.g., in the transverse mass distributions
of the visible ${\tilde{t}}_{1}$ and ${\tilde{t}}^{*}_{1}$ decay products.

Even in such circumstances though, ${\tilde{t}}_{1}  {\tilde{t}}^{*}_{1} H^0$
and ${\tilde{t}}_{1}  {\tilde{t}}^{*}_{1} A^0$ events would still
obey  a sort of degeneracy in their appearance  
(especially after accounting for large
 detector resolutions in reconstructing masses),
that could render non-trivial the operation of
separating data containing one type
from those induced by the other. (Our
 choice of MSSM parameters, producing $M_{H^0}\approx M_{A^0}$, ought to be 
representative also of such extreme experimental conditions.)
Nonetheless, we see that the maxima and minima
of $\sigma({\tilde{t}}_{1}  {\tilde{t}}^{*}_{1} H^0)$ and 
   $\sigma({\tilde{t}}_{1}  {\tilde{t}}^{*}_{1} A^0)$ 
occur in very different regions of the ($\phi_\mu,\phi_A$) plane. Moreover,
in the allowed areas, the difference between the production
rates of the two processes can even be a factor of 10 or more.
In other
terms, although it could always be possible to attempt the above `separation'
on the basis of the different decay patterns of the two Higgs bosons
(and/or their topology), this might not be needed after all.
In fact, as long as $\mu$, $A$ and $\tan\beta$ have been constrained
to some extent through some other measurements, then
for some specific values of $m_{{\tilde t}_1}$, a clear excess (i.e., well
above the size of the uncertainties induced by unknown higher order QCD 
effects) of 
${\tilde{t}}_{1}  {\tilde{t}}^{*}_{1} \Phi^0$ events, with 200 GeV
$\lsim M_{\Phi^0}\lsim$ 210 GeV, above the  $\Phi^0=H^0$ rates,
could only be explained if 
${\tilde{t}}_{1}  {\tilde{t}}^{*}_{1} A^0$ events have indeed been 
produced.

By comparing the bottom-left to the bottom-right plots in 
 Fig.~\ref{fig}, one realises that this separation
can happen over a large portion of the ($\phi_\mu,\phi_A$) plane.
One should also note the very different shapes of the 
two contours, such that 
${\tilde{t}}_{1}  {\tilde{t}}^{*}_{1} A^0$ rates are largest
where the 
${\tilde{t}}_{1}  {\tilde{t}}^{*}_{1} H^0$ ones are smallest. 
For example, 
just outside the areas excluded by the EDMs (when $\phi_A\to0$ or $\pi$
and $\phi_\mu\approx\pi/2$), the pseudoscalar
Higgs channel can reach the 100 fb level, whereas the scalar
Higgs rates are always around 10 fb. For 
 $\phi_A\to0$ and $\phi_\mu\approx\pi/6$ (or, quite symmetrically,
for $\phi_A\to\pi$ and $\phi_\mu\approx\pi-\pi/6$), the two process
rates are of the same order, about 10 fb.
Finally, in the  limit $\phi_\mu,\phi_A\to0$, 
$\sigma({\tilde{t}}_{1}  {\tilde{t}}^{*}_{1} H^0)$ is about 10 fb
and $\sigma({\tilde{t}}_{1}  {\tilde{t}}^{*}_{1} A^0)$ is, of course,
zero.

In our numerical simulations, we also have considered the large
$\tan\beta$ scenario. However, in this case, once the EDM constraints
were taken into account, we have found that both the theoretical plausibility 
of the MSSM and the phenomenological impact
of the CP-violating phases were much reduced.
On the one hand, in order to obtain the mentioned cancellations also for 
$\tan\beta\gsim10$, one would need to have the soft squark masses 
$M_{{\tilde q}_{1,2}}$ as large as 6 TeV or more and the
gluino one $M_{\tilde g}\gsim3$ TeV, that is, a quite `unnatural'
hierarchy in the soft SUSY breaking sector, if one aims to maintain 
$M_{{\tilde q}_{3}}$ around 300
GeV (so that ${{\tilde t}_1}{{\tilde t}_1^*}\Phi^0$ 
final states remain within the reach of the LHC energy).  
On the other hand, for large $\tan\beta$ values, the not yet excluded
(by direct searches, two-loop effects in the EDMs and positive
definiteness of the squark masses) 
area of the $(\phi_\mu,\phi_A)$ plane is much smaller 
(see Fig.~3 of Ref.~\cite{Choi:1999aj,%
Asakawa:2000es,Akeroyd:2001kt,Choi:2001iu,Arhrib:2001pg}). 
(Besides, also mixing effects
among neutral Higgs states start becoming relevant for $\tan\beta\gsim10$).
Thus, although some sporadic points over the allowed ($\phi_\mu,\phi_A$)
regions can still be found, these yielding cross sections significantly
different from those obtained in the phaseless case, 
we would conclude that only the $2\lsim\tan\beta\lsim10$ region 
is relevant in the experimental analysis of squark-squark-Higgs production.

To summarise, we have shown that the LHC production rates of the lightest
Higgs boson of the MSSM in association with a pair of lightest
stop scalars are strongly affected by the presence of complex parameters
in the soft sector of the SUSY Lagrangian, even when
EDM constraints are taken into account. As a matter
of fact, the $gg,q\bar q\to {{\tilde t}_1}{{\tilde t}_1^*} h^0$
mechanism has recently been advocated as a new possible discovery 
mode of the $h^0$ boson, at least for certain combinations of the MSSM 
parameters, that we have emulated here to some extent. Thus,
our results in this case have a twofold meaning.
On the one hand, they emphasise that more inputs than those pertaining
to a phaseless MSSM could be needed to describe the phenomenology
of ${{\tilde t}_1}{{\tilde t}_1^*} h^0$ events
(further recall that we have limited ourselves to the case of only
two independent phases, $\phi_\mu$ and $\phi_A$,
those associated to the Higgsino mass and the 
universal trilinear couplings, respectively). On
 the other hand, they make the point that such a mechanism can be
useful in assessing whether or not soft CP-violating phases are present.

In this last respect, however, it would be even more intriguing to detect
final states involving the pseudoscalar Higgs boson in place of the
lightest scalar one at the LHC. In fact, no matter the actual setup of the MSSM
parameters, the detection of a ${{\tilde t}_1}{{\tilde t}_1^*} A^0$ state 
would unequivocally mean that $\phi_\mu$ and/or $\phi_A$ are finite.
In fact, the corresponding interaction is prohibited at tree-level in a MSSM
with real masses and couplings in the soft sector and EW
effects at the one-loop level are unlikely to yield~
${{\tilde t}_1}{{\tilde t}_1^*} A^0$
production rates as large as those shown here: up to 10000 events
per year for some phase combinations at high collider luminosity. 
Further notice that, in 
our analysis, we deliberately have chosen a small value of
$\tan\beta$, so that mixing effects among the three neutral Higgs states
are very small even when the values of $M_{H^0}$ and $M_{A^0}$ are rather 
close. 
Finally, despite this mass degeneracy between the heavy scalar
and pseudoscalar Higgs bosons, the relative production rates
of ${{\tilde t}_1}{{\tilde t}_1^*} H^0$ and
${{\tilde t}_1}{{\tilde t}_1^*} A^0$ events are very different, both
in size and in shape,  over most of the ($\phi_\mu,\phi_A$) plane,
so that the two samples could even be separated experimentally.

\setcounter{figure}{0}
\setcounter{table}{0}
\setcounter{section}{0}
\setcounter{equation}{0}
\clearpage

%
%
%
%

\part{{\bf  Study of the Lepton Flavor Violating Decays of Charged Fermions 
 in SUSY GUTs
} \\[0.5cm]\hspace*{0.8cm}
{\it T. Bla\v{z}ek 
}}
\label{blazeksec}


Neutrino oscillations clearly show that individual lepton flavor 
is violated in nature. Here we present results of a study of lepton
flavor violating decays $\ell^\pm\to \ell^{\prime \:\pm}\,\gamma$
 in a class of SUSY GUT models assuming that the third right-handed 
neutrino couples
equally to the second and third lepton doublets with a large coupling. 
This corresponds to the neutrino Yukawa matrix (in the left-right
basis) of the form 
\begin{equation}
   Y_{\nu} \sim 
               \left( \begin{array}{ccc}
                       0 & 0 & 0    \\
                       0 & 0 & 1    \\
                       0 & 0 & 1   
                      \end{array}
               \right).
\label{eq:Yuknu}
\end{equation}
The large 23 entry, responsible for the large atmospheric neutrino
mixing angle, induces potentially large lepton flavor violating 
effects in the low-energy effective theory.
The see-saw mechanism yields a physical neutrino with a mass
about $5\times 10^{-2}$ eV consistent with the SuperKamiokande 
observation providing the third right-handed neutrino mass is
$M_{R3} \approx 3\times 10^{14}$ GeV, much greater than the
masses of the other two right-handed neutrinos. 

We focus on the large 
tan$\beta\approx 50$ regime of these models with 
$(Y_e)_{33}=(Y_\nu)_{33}$. The rate of a decay 
$\ell^\pm\to \ell^{\prime \:\pm}\,\gamma$
is enhanced by $(\tan\,\beta)^2$ and can be, approximately, rescaled
by this factor for lower values of tan$\beta$. Thus our study provides
for the upper estimate of the lepton flavor violating decay rates
of charged fermions.
In Figure 1a-b we present the results for the branching ratio 
$\tau\to\mu\,\gamma$ in a typical model of this class. Besides 
(\ref{eq:Yuknu}) all other Yukawa matrices are hierarchical with
small off-diagonal entries (more details can be found in \cite{Blazek:2001zm}). 
The model assumes the Pati-Salam $SU(4)\times SU(2)_L \times SU(2)_R$ 
symmetry at the GUT scale $M_G\sim 3\times 10^{16}\,$GeV 
but the results in the leptonic sector would be quite similar
for models with different unifying gauge group leading to Eq.~(\ref{eq:Yuknu}).
The two plots were obtained for two different values of the 
$\mu$ parameter fixed to $120\,$GeV in plot (a) and $300\,$GeV in plot (b).
$m_F$ is the soft universal scalar mass and $M_{1/2}$ is its gaugino analogue,
both at $M_G$. The $\tau\to\mu\,\gamma$ contour lines in Figure 1a-b
should be compared to the experimental upper bound  
$BR(\tau\to\mu + \gamma) < 1.1\cdot 10^{-6}$.
The allowed region overlaps with the region preferred by the data on
the muon $g-2$, as shown in plots 1c-d. 
Clearly, this decay rate should be very close to the present limits
and thus presents an exciting opportunity to observe
lepton flavor violation or to constrain substantially this class of models.

The decays $\mu\to e\,\gamma$ and $\tau\to e\,\gamma$ are much more
model dependent since small Yukawa entries are necessarily involved.
In the model studied in \cite{Blazek:2001zm} they are found to be
well below the experimental
limit. Thus our main result is the correlation between the
maximal atmospheric mixing angle and large $\tau\to\mu\,\gamma$ 
branching ratio related through the large off-diagonal entry in the
Yukawa matrix in (\ref{eq:Yuknu}).

\begin{figure}[ht]
\centering
\epsfysize=6.5truein
\epsffile{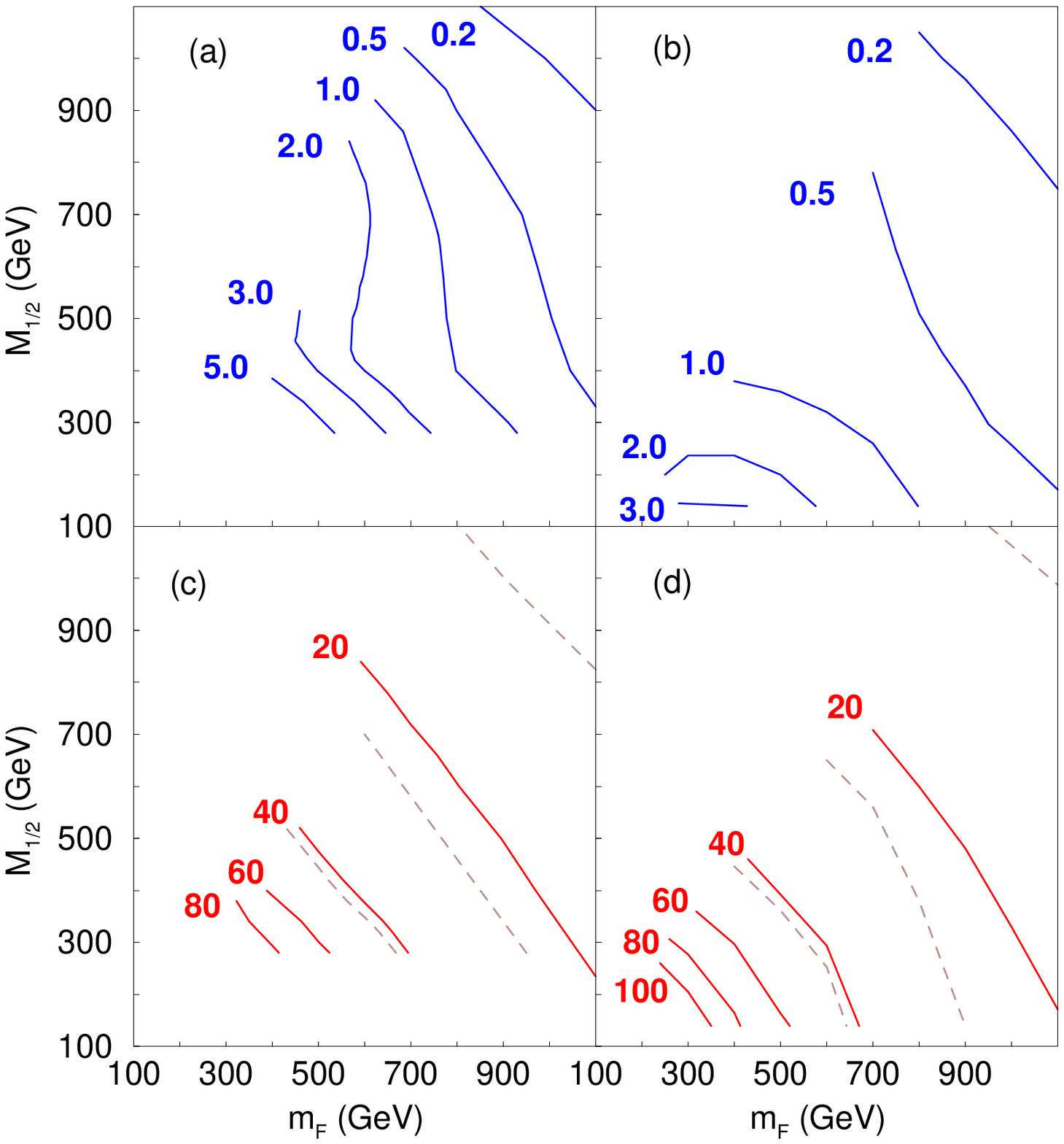}
\caption{
Contour lines of $\:BR(\tau\to\mu\gamma)\times 10^{6}$
and $\:\delta a_\mu({SUSY})\times 10^{10}$
for two different values of $\mu$.
(a) $\:BR(\tau\to\mu\gamma)\times 10^{6}$,      for $\mu=120$GeV. $\;$
(b) $\:BR(\tau\to\mu\gamma)\times 10^{6}$,      for $\mu=300$GeV. $\;$
(c) $\delta a_\mu({SUSY})\times 10^{10}$, for $\mu=120$GeV. $\;$
(d) $\delta a_\mu({SUSY})\times 10^{10}$, for $\mu=300$GeV. $\;$
In (c) and (d) the dashed curves mark the central value for
$a_\mu$ not accounted for by the Standard Model
and the borderlines of 1-sigma region for this quantity.
The experimental upper limit on $\:BR(\tau\to\mu\gamma)$ 
is $1.1\times 10^{6}$.
In all plots $\tan\beta=50$ and soft trilinear parameter $A=0$.
}
\label{f:tmg_amu}
\end{figure}


\setcounter{figure}{0}
\setcounter{table}{0}
\setcounter{section}{0}
\setcounter{equation}{0}
\clearpage

\def\const{\mbox{const}}
\def\e{{\rm e}}
\def\al{\alpha}
\def\eps{\epsilon}
\def\d{\partial}
\def\l{\left(}
\def\r{\right)}
\def\la{\langle}
\def\ra{\rangle}
\def\S{{\cal S}}
\def\tg{\mathop{\rm tg}\nolimits}
\renewcommand{\ln}{\mathop{\rm ln}\nolimits}
\def\sm#1{{\scriptscriptstyle \rm #1}}
\def\Tr{{\rm Tr}}
\def\eps{\epsilon}
\def\half{\frac{1}{2}}
\def\eq#1{(\ref{#1})}
\def\goe{\gtrsim}
\def\loe{\lesssim}

\part{{\bf  Interactions of the Goldstino Supermultiplet with Standard Model
Fields
} \\[0.5cm]\hspace*{0.8cm}
{\it D.S. Gorbunov
}}
\label{gorbsec}


\begin{abstract}
In a set of supersymmetric extensions of the Standard Model the masses of
the sgoldstinos are of the order of the electroweak scale. Thus 
sgoldstinos are
expected to be produced at future colliders. The sgoldstino interactions
with the fermions and gauge bosons of the Standard Model are determined by 
the MSSM soft mass
terms and the scale of supersymmetry breaking. These interactions have
been included into the CompHEP package. On the other hand, the
sgoldstino couplings to Higgs bosons depend on the parameters of 
the hidden
sector responsible for mediation of supersymmetry breaking. The
measurement of these coupling constants would offer
a unique probe of the hidden sector. 
\end{abstract}

\section{Goldstino supermultiplet}
In any supersymmetric extension of the Standard Model (SM) of particle
physics spontaneous supersymmetry breaking occurs due to a
non-zero vacuum expectation value of an auxiliary component 
of some chiral or vector superfield. 
As a simple case, let us consider a model where 
\begin{equation}
\S=s+\sqrt{2}\theta\psi+\theta^2F_s
\label{spurion}
\end{equation}
is the only chiral superfield which obtains a non-zero vacuum 
expectation value $F$ for its auxiliary component, 
\begin{equation}
\la F_s\ra\equiv F\;.
\label{spurion-1}
\end{equation}
Then $\psi$ is a two-component Goldstone fermion, {\it goldstino},  
and its superpartners 
\begin{equation}
S\equiv\frac{1}{\sqrt{2}}(s+s^*)\;,~~~~
P\equiv\frac{1}{i\sqrt{2}}(s-s^*)\;,
\label{definition}
\end{equation}
are respectively a scalar and a pseudoscalar {\it sgoldstino}. 

In the framework of supergravity $\partial_\mu\psi$ becomes the longitudinal
component of the gravitino, due to the super-Higgs effect. 
As a result, the gravitino acquires a mass $m_{3/2}$ which 
in realistic models with a vanishing cosmological constant
is completely determined by the supersymmetry breaking parameter $F$: 
\begin{equation}
m_{3/2}=\frac{\sqrt{8\pi}}{\sqrt{3}}\frac{F}{M_{Pl}}\;.
\label{gravitino-mass}
\end{equation} 
The sgoldstinos remain massless at tree level and become massive due to
corrections from higher order terms in the K\"ahler potential. If
these terms are sufficiently suppressed, the sgoldstinos are light and may
appear in particle collisions at high energy colliders. Such pattern
emerges in a number of non-minimal supergravity
models~\cite{Ellis:1984kd,Ellis:1985xe}
as well as in gauge mediation models if supersymmetry is broken via
a non-trivial superpotential (see, e.g.~\cite{Giudice:1998bp,Dubovsky:1999xc} 
and references therein). Here we shall consider the sgoldstino masses
$m_S$ and $m_P$ as free parameters.

The gravitinos and sgoldstinos interact with the 
MSSM fields and the corresponding
coupling constants are inversely proportional to the
supersymmetry breaking parameter $F$. Eq.~(\ref{gravitino-mass}) 
then implies that the gravitino has to be very
light, otherwise the gravitino as well as
the sgoldstinos are effectively decoupled from the MSSM fields at the
energy scale of the colliders of the near-term future.

\section{Effective Lagrangian}

The effective Lagrangian for the gravitino $\tilde{G}_\mu$ is obtained
from N=1 supergravity~\cite{Cremmer:1983en}, and may be
used to calculate scattering processes involving any of the helicity
components of the massive gravitino. Meanwhile the energy scale
attainable at the present and the nearest future generation of
accelerators favors the longitudinal component of gravitino as the
most promising to be studied in collision experiments. In
this case the longitudinal component of the gravitino,
$\tilde{G}_\mu\sim i\d_\mu\psi/m_{3/2}$, 
effectively behaves as a
two-component fermion $\psi$, the {\it goldstino}, 
and the interaction between the goldstino and the other fields is
plainly given by the Goldberger--Treiman relation
\[
{\cal L}_{GT}=\frac{1}{F}J^{\mu}_{ SUSY}\partial_\mu\psi\;,
\] 
with $J^{\mu}_{SUSY}$ being a supercurrent. 

In order to obtain the low-energy effective Lagrangian for sgoldstinos
one can use the spurion method~\cite{Brignole:1997fn}. It 
exploits the fact that, by definition, sgoldstinos are scalar
components of the very supermultiplet $\S$ whose auxiliary component
acquires a non-zero vacuum expectation value due to supersymmetry
breaking. Then one can consider a simple supersymmetric model with
non-renormalizable interactions between $\S$ and the MSSM superfields,
which yield the MSSM soft terms when a non-zero $\la F_s\ra$
is generated. Consequently, the corresponding coupling 
constants are fixed by the ratios of the soft terms and $F$.

In general, the supersymmetry breaking part of the Lagrangian has the form
\begin{eqnarray}
-{\cal L}_{breaking}=\sum_km_k^2|\tilde{\phi}_k|^2+\l\half
\sum_\alpha M_\alpha~\Tr\lambda^\alpha\lambda^\alpha+h.c.\r\nonumber
\\\label{soft-terms}
-\epsilon_{ij}\l 
Bh^i_Dh^j_U+
A^L_{ab}\tilde{l}_a^j\tilde{e}_b^ch_D^i+
A_{ab}^D\tilde{q}_a^j\tilde{d}_b^ch_D^i+
A_{ab}^U\tilde{q}_a^i\tilde{u}_b^ch_U^j+h.c.\r\;,
\end{eqnarray} 
where $k$ ($\alpha$) runs over all scalar $\tilde{\phi}_k$ (gaugino
$\lambda_\alpha$) fields. Since supersymmetry is broken {\it spontaneously},
Eq.~(\ref{soft-terms}) implies the following effective interaction
between the MSSM superfields and the goldstino supermultiplet $\S$:
\[
{\cal L}_{\S-MSSM}={\cal L}_{\S-K\ddot{a}hler}+{\cal L}_{\S-gauge}+{\cal L}_{\S-superpotential}\;,
\]
where 
\begin{eqnarray*}
{\cal L}_{\S-K\ddot{a}hler}&=&-\int\! d^2\theta~ d^2\bar{\theta}~\S^\dag\S\cdot\!\!\!\!\!\!\!\!\sum_{all~matter\atop and~Higgs~fields}{m_k^2\over F^2}~\!\Phi_k^\dag 
~\!e^{g_1V_1+g_2V_2+g_3V_3}~\!\Phi_k\;,
\\
{\cal L}_{\S-gauge}&=&{1\over 2}\int\!\!d^2\theta~
\S\cdot \!\!\!\!\!\!\sum_{all~gauge\atop fields}{M_\alpha\over F}~\!Tr W^\alpha W^\alpha+h.c.\;,
\\
{\cal L}_{\S-superpotential}&=&\int\!\!d^2\theta~\S\cdot\epsilon_{ij}\!
\l{B\over F}~\!H_D^iH_U^j+{A^L_{ab}\over F}~\! L_a^jE_b^cH_D^i+
{A_{ab}^D\over F}~\!Q_a^jD_b^cH_D^i+{A_{ab}^U\over F}~\!Q_a^iU_b^cH_U^j\r+h.c.
\end{eqnarray*}
These terms emerge if the fields from the hidden sector, where
supersymmetry breaking occurs, are integrated out. The only remnant of
the hidden sector is the goldstino supermultiplet, which may remain light.

Integrating over $\theta$, $\bar{\theta}$ and taking into account
Eqs.~(\ref{spurion}), (\ref{spurion-1}) we obtain the soft supersymmetry
breaking terms~(\ref{soft-terms}) as well as the interactions between the
components of the goldstino supermultiplet and the components of the MSSM
superfields:
\begin{eqnarray}
{\cal L}_{\S-K\ddot{a}hler}&=&-\!\!\!\!\!\!\sum_{all~matter\atop and~Higgs~fields}\l{m_k^2\over F}s\cdot\tilde{\phi}_k^\dag F_k+h.c.\r\;,
\label{S-kahler}
\\
{\cal L}_{\S-gauge}&=&\!\!\!\!\sum_{all~gauge\atop fields}\biggl(
-i{M_\alpha\over F}s\cdot\lambda^\alpha_a\sigma^\mu D_\mu\bar{\lambda}^\alpha_a+{M_\alpha\over2F}s\cdot D^\alpha_a D^\alpha_a
\nonumber
\\&-&
{M_\alpha\over4F}s\cdot F_{a~\mu\nu}^\alpha F_a^{\alpha~\mu\nu}-i{M_\alpha\over 8F}s\cdot F_{a~\mu\nu}^\alpha \epsilon^{\mu\nu\lambda\rho}F_{a~\lambda\rho}^\alpha+h.c.\biggr)\;,
\label{S-gauge}
\\
{\cal L}_{\S-superpotential}&=&-\epsilon_{ij}\biggl( 
{B\over F}s\cdot\chi_D^i\chi_U^j
-{B\over F}s\cdot\l h_D^iF_{H_U}^j+F_{H_D}^ih_U^j\r
\nonumber
\\&+&
{A^L_{ab}\over F}
\l
l_a^je_b^c\cdot sh_D^i
+l_a^j\chi_D^i\cdot s\tilde{e}_b^c
+s\tilde{l}_a^j\cdot e_b^c\chi_D^i
-sF_{L_a}^j\tilde{e}_b^ch_D^i
-s\tilde{l}_a^jF_{E_b^c}h_D^i
-s\tilde{l}_a^j\tilde{e}_b^cF_{H_D}^i
\r
\nonumber
\\&+&
{A_{ab}^D\over F}
\l
q_a^jd_b^c\cdot sh_D^i
+s\tilde{q}_a^j\cdot d_b^c\chi_D^i
+s\tilde{d}_b^c\cdot q_a^j\chi_D^i
-sF_{Q_a}^j\tilde{d}_b^ch_D^i
-s\tilde{q}_a^jF_{D_b^c}h_D^i
-s\tilde{q}_a^j\tilde{d}_b^cF_{H_D}^i
\r
\nonumber
\\&+&
{A_{ab}^U\over F}
\l
q_a^iu_b^c\cdot sh_U^j
+s\tilde{q}_a^i\cdot u_b^c\chi_U^j
+s\tilde{u}_b^c\cdot q_a^i\chi_U^j
-sF_{Q_a}^i\tilde{u}_b^ch_U^j
-s\tilde{q}_a^iF_{U_b^c}h_U^j
-s\tilde{q}_a^i\tilde{u}_b^cF_{H_U}^j
\r
\nonumber
\\
&+&h.c.\biggr)\;.
\label{S-superpotential}   
\end{eqnarray}
Here we presented only the leading order in $1\over F$ terms; the
convention for the Levi--Civita tensor is $\epsilon^{0123}=-1$.

Eliminating the auxiliary fields, one obtains the low-energy effective
Lagrangian for the interactions between the components of the goldstino
supermultiplets and the MSSM fields.  

\section{Phenomenology of the model}

Until now, there is no experimental evidence for a gravitino or sgoldstinos. 
The study of their phenomenology places bounds on their coupling constants. 
Note that all
sgoldstino coupling constants introduced in the previous sections are
completely determined by the MSSM soft terms and the
supersymmetry breaking parameter $F$, while the sgoldstino masses 
$m_S$ and $m_P$ remain arbitrary. 
Depending on the values of $m_S$ and $m_P$, the sgoldstinos
may show up in different experiments. The phenomenologically interesting
models can be separated into four classes:
\begin{itemize}
\item The sgoldstino masses are of order the electroweak scale, while
$\sqrt{F}\sim1$~TeV --- sgoldstinos may then be produced at high-energy 
colliders~\cite{Perazzi:2000id,Perazzi:2000ty} (see section 3 of
Ref.~\cite{Gorbunov:2001pd} for a sketch of the sgoldstino
collider phenomenology).
\item The sgoldstino masses $m_S,m_P\sim1$~MeV$\div1$~GeV, while
$\sqrt{F}\sim1$~TeV --- sgoldstinos may then emerge as products of rare
meson decays~\cite{Dicus:1990su,Gorbunov:2000th}, such as
$\Upsilon\to S(P)\gamma$, $J/\psi\to S(P)\gamma$.
\item Models with flavor violation in the soft
trilinear couplings, $A_{ab}\neq A\delta_{ab}$ --- sgoldstino
interactions then lead to flavor violating processes. In particular, 
sgoldstinos may contribute to FCNC (mass differences 
and/or CP-violation in the neutral meson 
systems)~\cite{Brignole:2000wd,Gorbunov:2000cz}, 
and, if kinematically allowed, sgoldstinos appear in rare
decays such as $t\to cS(P)$~\cite{Gorbunov:2000ht}, $\mu\to eS(P)$,
$K\to\pi S~$\cite{Gorbunov:2000th}, etc.
\item The sgoldstinos are lighter than 1~MeV --- these models may be tested
in low energy experiments~\cite{Gorbunov:2000th}, such as reactor
experiments, conversion in a magnetic field, etc. Sgoldstinos may
also play a very important role in astrophysics and
cosmology~\cite{Nowakowski:1995ag,Grifols:1997hi,Gherghetta:1998zq,Gorbunov:2000th}:
they may change the predictions of Big Bang Nucleosynthesis, distort
the CMB spectrum, affect SN explosions and the cooling rate of stars, etc.
\end{itemize}

\section{Incorporation into CompHEP}

To leading order in $1/F$ and to zero order in the
MSSM gauge and Yukawa coupling constants, the interactions between the
components of the goldstino supermultiplet and the MSSM fields are
derived in Ref.~\cite{Gorbunov:2001pd}. 
They correspond to the processes most attractive for collider studies --
where only one of these {\it new} particles appears in a
final state. In this case the light gravitino behaves exactly as
a goldstino. For sgoldstinos, as they are R-even, 
only the sgoldstino-goldstino and sgoldstino-SM fields couplings
have been included 
as the most interesting phenomenologically. All
new coupling constants between the components of the goldstino
supermultiplet and the MSSM fields are completely determined by the ratios
of the soft supersymmetry breaking parameters and $F$. This Lagrangian 
has been incorporated (see Ref.~\cite{Gorbunov:2001pd} for details) 
into the CompHEP software 
package~\footnote{The CompHEP package~\cite{Pukhov:1999gg} 
automatically calculates tree-level particle decay rates and
cross sections and is aimed to improve the accuracy, to cut
down the efforts and to shorten the time usually required 
for studying high-energy collision processes.}. This package 
may be used in the calculation of 
any tree-level process with one on-shell gravitino or sgoldstino. 
The universality of the Lagrangian makes it possible to apply the package in
studying the phenomenology of any supersymmetric extension of the Standard
Model. Being included into CompHEP, this model should be regarded as an
additional option allowed within the framework of the CompHEP/SUSY
model. Currently accessible versions of this package operate only with
real parameters and coupling constants. Likewise the trilinear soft
couplings are assumed to be proportional to the corresponding Yukawa
couplings, $A_{ab}={\cal A}_{ab}y_{ab}$. The sgoldstino Lagrangian
transformed in accordance with these rules reads
\begin{eqnarray} 
{\cal L}_{S}&=&-\sum_{all~gauge\atop fields}{M_\alpha\over2\sqrt{2}F}S\cdot F_{a~\mu\nu}^\alpha F_a^{\alpha~\mu\nu}
-{{\cal A}^L_{ab}\over\sqrt{2} F}y^L_{ab}\cdot S
\bigl(\epsilon_{ij} l_a^je_b^c h_D^i +h.c.\bigr)
\nonumber
\\&-&
{{\cal A}_{ab}^D\over\sqrt{2} F}y_{ab}^D\cdot S
\bigl( \epsilon_{ij} q_a^jd_b^c h_D^i+h.c.\bigr)
-{{\cal A}_{ab}^U\over\sqrt{2} F}y_{ab}^U\cdot S
\bigl( \epsilon_{ij} q_a^iu_b^ch_U^j+h.c.\bigr)\;,
\label{scalar}
\\
{\cal L}_{P}&=&\sum_{all~gauge\atop fields}{M_\alpha\over 4\sqrt{2}F}P\cdot F_{a~\mu\nu}^\alpha \epsilon^{\mu\nu\lambda\rho}F_{a~\lambda\rho}^\alpha
-i{{\cal A}^L_{ab}\over\sqrt{2} F}y^L_{ab}\cdot P
\bigl(\epsilon_{ij} l_a^je_b^c h_D^i -h.c.\bigr)
\nonumber
\\&-&
i{{\cal A}_{ab}^D\over\sqrt{2} F}y_{ab}^D\cdot P
\bigl( \epsilon_{ij} q_a^jd_b^c h_D^i-h.c.\bigr)
-i{{\cal A}_{ab}^U\over\sqrt{2} F}y_{ab}^U\cdot P
\bigl( \epsilon_{ij} q_a^iu_b^ch_U^j-h.c.\bigr)\;.
\label{pseudoscalar}
\end{eqnarray}

\section{Remarks}

It is worth noting that the independent direct measurement of the
MSSM soft supersymmetry
breaking terms and the gravitino or/and sgoldstino couplings offers a
unique possibility to estimate the supersymmetry breaking scale
$\sqrt{F}$. 

The sgoldstino couplings to
superpartners become relevant for models with heavy sgoldstinos,
where the sgoldstinos also decay to SM superpartners. In the spurion
approach the corresponding coupling constants are not 
completely determined by the MSSM
soft breaking terms, but may depend on new parameters originating from
the hidden sector. A similar situation happens with the sgoldstino
interaction terms proportional to the MSSM gauge or Yukawa couplings. For
instance, accounting for the $F$ and $D$ auxiliary fields (see
Eqs.~(\ref{S-kahler}), (\ref{S-gauge}), (\ref{S-superpotential}))
may cause sgoldstino-Higgs mixing. The corresponding
coefficients also depend on new parameters from the hidden sector. Indeed,
the additional interaction between the sgoldstino and the Higgs bosons arises 
from the effective superpotential 
\[
{\cal L}=\tilde{\mu}\int\!\!d^2\theta\S^2\cdot\epsilon_{ij}H^i_DH^j_U+h.c.
\]
with $\tilde{\mu}$ being a new dimensionful constant. 
Thus the sgoldstino couplings to Higgs bosons allow us to probe the hidden sector.
The sgoldstino decay mode into the lightest MSSM
Higgs boson becomes important in models with fairly light sgoldstinos
($m_{S(P)}\gtrsim250$~GeV) and the measurement of this partial width 
constrains some combination of hidden sector parameters,
if the supersymmetry breaking scale is known (i.e., from the sgoldstino
partial width into two photons and a measurement of the gaugino masses).

\setcounter{figure}{0}
\setcounter{table}{0}
\setcounter{section}{0}
\setcounter{equation}{0}
\clearpage


\def\npb#1 #2 #3 #4 {Nucl.~Phys. B {\bf #1}, #2 (#3)#4 }
\def\plb#1 #2 #3 #4 {Phys.~Lett. B {\bf #1}, #2 (#3)#4 }
\def\prd#1 #2 #3 #4 {Phys.~Rev.  D {\bf #1}, #2 (#3)#4 }
\def\prl#1 #2 #3 #4 {Phys.~Rev.~Lett. {\bf #1}, #2 (#3)#4 }

\part{{\bf  Attempts at Explaining the NuTeV Observation of Di-Muon Events
} \\[0.5cm]\hspace*{0.8cm}
{\it A. Dedes, H. Dreiner, and P. Richardson}
}
\label{ddrsec}


\begin{abstract}
\noindent
The NuTeV Collaboration has observed an excess in their di-muon channel, 
possibly corresponding to a long-lived neutral particle with only weak 
interactions and which decays to muon pairs. We show that this can {\it not} be 
explained by pair production of neutralinos as suggested in the literature.
In the parameter region allowed by LEP
the event rate is far too small. We propose instead a new neutralino production 
method via $B$-mesons, which can fully explain the observation.
\end{abstract}

\section{Introduction}

The NuTeV Collaboration has searched for long-lived neutral particles
($N^0$) with mass $M_{N^0}\geq2.2\gev$ and small interaction rates
with ordinary matter
\cite{Formaggio:1999ne,Adams:2000rd,Adams:2001sk}.  They look for the
decay of the neutral particles in a detector which is $1.4\,\rm{km}$
downstream from the production point and observe 3 $\mu\mu$ events
where they only expect to see a background of $0.069\pm0.010$ events.
The probability that this is a fluctuation of this specific channel is
 about $4.6\,\sigma$. 
The simple supersymmetric scenarios discussed previously can
not lead to an excess at NuTeV, since the decisive supersymmetric
parameter range has been excluded by LEP. We propose instead the production of
light neutralinos via $B$-mesons which could give a measurable excess.
We briefly present the two possible models and then discuss them
quantitatively.
We have also considered the production rate for neutral
heavy leptons but this is too low and does not lead to a viable
explanation \cite{Dedes:2001zi}.

\section{The $R_p$ Violating Model}
\label{sec:model}
Only couplings, $\lam_{212}$ and $\lam_{232}$ give a di-muon
signature.
For $\lam_{212}$ the neutralino will decay with equal probability to
$e\mu\nu$ and $\mu\mu\nu$. No $e\mu$-events are observed, we therefore
propose one dominant R-parity violating coupling $\lambda_{232}$.
This coupling corresponds 
to the two neutralino decay modes $\chi^0_1\rightarrow\mu^-_L\mu^+_R\nu_\tau$ and 
$\chi^0_1\rightarrow\tau^-_L\mu^+_R\nu_\mu$,
as well as their complex conjugates. For a light neutralino 
the $\tau\mu$ decays
are sufficiently phase space suppressed to give an expectation below
one event. For the light neutralino production we shall only consider
single neutralino production in the decay of bottom hadrons. The
bottom hadrons are formed following the production of a $b\bar{b}$
pair. These hadrons can then decay via the R-parity couplings
$\lambda'_{i13}$, ($i=1,2,3$). We will only consider the decays of the
$B^0_{d}$ and $B^+$ via R-parity violation \ie\ 
 $B^0_d\longrightarrow \bar{\nu}_i \cht^0_1$ and 
$B^+ \longrightarrow \ell^+_i \cht^0_1$.
This mechanism allows one to produce light neutralinos via a strong
interaction process.

\begin{figure}[t]
\begin{center}
\includegraphics[width=0.5\textwidth,angle=90]{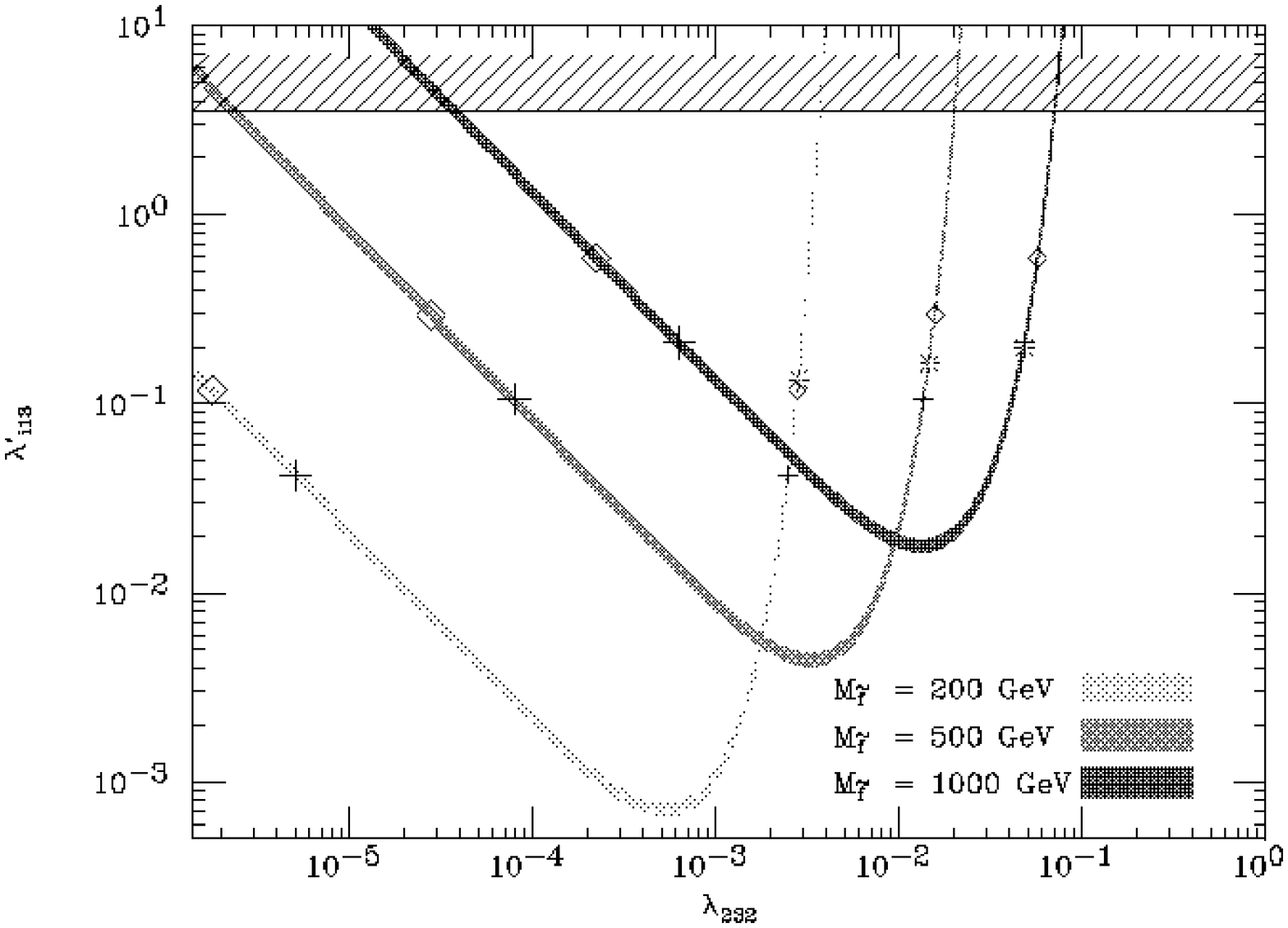}
\caption{Regions in $\lambda_{232},\lambda'_{i13}$ parameter space in which
  we would expect $3\pm1$ events to be observed in the NuTeV detector.
  The limits \cite{Allanach:1999ic} on the
couplings $\lam'_{113}$ (crosses) and $\lam'_{213}$ (diamonds) allow
solutions between the two points.  The
region above the stars is ruled out for the coupling $\lam'_{213}$ by
the limit on the product of the couplings $\lam_{232}\lam'_{213}$.
The hatched region shows the bound on the coupling
$\lam'_{i13}$ from perturbativity.}
\label{fig:parameter}
\end{center}
\end{figure}

Using results for the RPV branching ratios of the B mesons and the
neutralino lifetime we can find regions in $(\lam_{232},\lam'_{113})$
parameter space, for a given sfermion mass, in which there are $3\pm1$
events inside the NuTeV detector, this is shown in
Fig.\,\ref{fig:parameter}. 

This model can be tested at other experiments.
At the NOMAD experiment for the same $B^0$-meson
branching ratio we obtain about an order of
magnitude more events than at NuTeV. Thus our model can be {\it
  completely} tested by the NOMAD data! 
For neutralino production we are relying on a rare
B-meson decay
which can possibly be observed at a present or future B-factory
 although this may be difficult as the leptons produced will be very soft.

\section{Conclusions}
We have reconsidered the NuTeV di-muon observation in the light of
supersymmetry with broken R-parity and neutral heavy leptons. We have
shown that it is not possible to obtain the observed event rate with
pair production of light neutralinos or via the production of neutral
heavy leptons. However, we have introduced a new production method of
neutralinos via $B$-mesons. Due to the copious production of
$B$-mesons in the fixed target collisions the observed di-muon event
rate can be easily obtained for allowed values of the R-parity
violating couplings.

The model we have proposed can be completely tested using current
NOMAD data.
If the NOMAD search is negative our model is ruled out and the NuTeV
observation is most likely not due to physics beyond the Standard
Model.

\section{Acknowledgments}
A.D. would like to acknowledge financial support from the Network
RTN European Program HPRN-CT-2000-00148 ``Physics Across the Present
Energy Frontier: Probing the Origin of Mass''.

\setcounter{figure}{0}
\setcounter{table}{0}
\setcounter{section}{0}
\setcounter{equation}{0}
\clearpage

\def\err#1#2{\lower2pt\hbox{ $\stackrel{\scriptstyle +#1}{\scriptstyle
-#2}$}}

\def\O{{$\slash\hskip -1pt {\rm O}$}}

\part{{\bf  Kaluza-Klein States of the Standard Model Gauge Bosons: 
Constraints From High Energy Experiments
} \\[0.5cm]\hspace*{0.8cm}
{\it K. Cheung and G. Landsberg
}}
\label{landsberg2sec}


\begin{abstract}
In theories with the standard model gauge bosons
propagating in TeV$^{-1}$-size extra dimensions, their Kaluza-Klein
states interact with the rest of the SM particles confined to the
3-brane.  We look for possible signals for this interaction in the present
high-energy collider data, and estimate the sensitivity offered by the
next generation of collider experiments. Based on the present data
from the LEP~2, Tevatron, and HERA experiments, we set a lower limit
on the extra dimension compactification scale $M_C > 6.8$~TeV at the
95\% confidence level (dominated by the LEP 2 results) and quote expected
sensitivities in the Tevatron Run 2 and at the LHC.
\end{abstract}


This contribution is a shortened version of the recent paper~\cite{Cheung:2001mq}, 
with the focus on future high-energy facilities. The details of the formalism 
used to obtain the results presented here can be found in~\cite{Cheung:2001mq}.

Recently, it has
been suggested that the Planck, string, and grand unification scales
can all be significantly lower than it was previously thought, perhaps
as low as a few 
TeV~\cite{Antoniadis:1990ew,Lykken:1996fj,Shiu:1998pa,Antoniadis:1998ax}. 
An interesting model was proposed 
\cite{Dienes:1998vg,Pomarol:1998sd,Masip:1999mk,Antoniadis:1999bq}, in 
which matter resides on a $p$-brane ($p>3$), with chiral fermions
confined to the ordinary three-dimensional world internal to the
$p$-brane and the SM gauge bosons also propagating in the extra $\delta >
0$ dimensions internal to the $p$-brane. (Gravity in the bulk is not
of direct concern in this model.) It was shown \cite{Dienes:1998vg} that in
this scenario it is possible to achieve the gauge coupling
unification at a scale much lower than the usual GUT scale, due to a
much faster power-law running of the couplings at the scales above
the compactification scale of the extra dimensions.
The SM gauge bosons that propagate in the extra dimensions
compactified on $S^1/Z_2$, in the four-dimensional point of view,
are equivalent to towers of Kaluza-Klein (KK) states with masses
$M_n = \sqrt{M_0^2 + n^2/R^2}$ ($n=1,2, ...$), where $R=M_C^{-1}$ 
is the size of the compact dimension, $M_C$ is the corresponding 
compactification scale, and $M_0$ is the mass of the corresponding SM 
gauge boson. 

There are two important consequences of the existence of 
the KK states of the gauge bosons in collider phenomenology. (i) Since 
the entire tower of KK states have the same quantum numbers as their
zeroth-state gauge boson, this gives rise to mixings among the
zeroth (the SM gauge boson) and the $n$th-modes ($n=1,2,3,...$) 
of the $W$ and $Z$ bosons. 
(The zero mass of the photon is protected by the U(1)$_{\rm EM}$
symmetry of the SM.) (ii) In addition to direct production and virtual
exchanges of the zeroth-state gauge bosons, both direct production and
virtual effects of the KK states of the $W,Z,\gamma$, and $g$ bosons
would become possible at high energies.

In this proceedings, we study the effects of virtual exchanges of the KK
states of the $W,Z,\gamma$, and $g$ bosons in high energy collider
processes.  While the effects on the low-energy precision measurements
have been studied in detail
\cite{Nath:1999aa,Nath:1999mw,Rizzo:1999br,Casalbuoni:1999ns,Strumia:1999jm,
Carone:1999nz,Delgado:1999sv,Cornet:1999im},
their high-energy
counterparts have not been systematically studied yet. We attempt to
bridge this gap by analyzing all the available high-energy collider
data including the dilepton, dijet, and top-pair production at the
Tevatron; neutral and charged-current deep-inelastic scattering at
HERA; and the precision observables in leptonic and hadronic
production at LEP~2.

We fit the observables in the above processes to the sum of the SM
prediction and the contribution from the KK states of the SM gauge
bosons. In all cases, the data do not require the presence of the
KK excitations, which is then translated to the limits on the
compactification scale $M_C$. The fit to the combined data set yields
a 95\% C.L. lower limit on $M_C$ of 6.8 TeV, which is substantially
higher than that obtained using only electroweak precision measurements.
In addition, we also estimate the expected reach on $M_C$ in
Run 2 of the Fermilab Tevatron and at the LHC, using dilepton
production.

\section{Interactions of the Kaluza-Klein States}

We use the formalism of 
Ref. \cite{Pomarol:1998sd,Masip:1999mk,Antoniadis:1999bq}, based on
an extension of the SM to five dimensions, with the fifth dimension,
$x^5$, compactified on the segment $S^1/Z_2$ (a circle of radius $R$
with the identification $x^5 \to -x^5$). This segment has the length
of $\pi R$. Two 3-branes reside at the fixed points $x^5=0$ and
$x^5=\pi R$. The SM gauge boson fields propagate in the 5D-bulk, while
the SM fermions are confined to the 3-brane located at $x^5=0$. 
The Higgs sector consists of two Higgs doublets, $\phi_1$ and $\phi_2$ 
(with the ratio of vacuum expectation values $v_2/v_1 \equiv \tan\beta$), 
which live in the bulk and on the SM brane, respectively.

In the case of SU(2)$_L \times$ U(1)$_Y$ symmetry,
the charged-current (CC) and neutral-current (NC) interactions,
after compactifying the fifth dimension,
are given by \cite{Casalbuoni:1999ns}:
\begin{eqnarray}
\label{P3_Landsberg_102901eq1} 
        {\cal L}^{\rm CC} &=& \frac{g^2 v^2}{8} \biggr [ W_1^2 +
        \cos^2\beta
        \sum_{n=1}^{\infty}( W_1^{(n)})^2 + 2\sqrt{2} \sin^2\beta W_1
        \sum_{n=1}^{\infty} W_1^{(n)} + 2 \sin^2\beta \left(
        \sum_{n=1}^{\infty} W_1^{(n)} \right )^2 \biggr ] \nonumber \\ &+&
        \frac{1}{2} \sum_{n=1}^{\infty} n^2 M_C^2 (W_1^{(n)})^2 - {g} (W_1^\mu
        + \sqrt{2} \sum_{n=1}^{\infty} W_1^{(n)\mu} ) J_\mu^1 +(1 \to 2) \;,
        \\
\label{P3_Landsberg_102901eq2} 
        {\cal L}^{\rm NC} &=& \frac{{g}v^2}{8 {c}_\theta^2} \biggr
        [ Z^2 + \cos^2\beta \sum_{n=1}^{\infty}( Z^{(n)})^2 + 2\sqrt{2}
        \sin^2\beta Z
        \sum_{n=1}^{\infty} Z^{(n)} + 2 \sin^2\beta \left (\sum_{n=1}^{\infty}
        Z^{(n)} \right )^2 \nonumber \\ &+& \frac{1}{2} \sum_{n=1}^{\infty}
        n^2 M_C^2 \biggr[ (Z^{(n)} )^2 + (A^{(n)})^2 \biggr] \nonumber \\ &-&
        \frac{{e}}{ {s}_\theta {c}_\theta } \left ( Z^\mu + \sqrt{2}
        \sum_{n=1}^{\infty} Z^{(n)\mu} \right ) J_\mu^Z - {e} \left ( A^\mu +
        \sqrt{2} \sum_{n=1}^{\infty} A^{(n)\mu} \right ) J_\mu^{\rm em} \;,
\end{eqnarray}

where the fermion currents are:
\[
        J_\mu^{1,2} = \bar \psi_L \gamma_\mu
        \left(\frac{\tau_{1,2}}{2} \right )
        \psi_L \;, \qquad
        J_\mu^Z = \bar \psi \gamma_\mu ( g_v - \gamma^5 g_a )
        \psi \;, \qquad
        J_\mu^{\rm em} = \bar \psi \gamma_\mu Q_\psi \psi \;,
\]
and $\langle \phi_1 \rangle = v \cos\beta, \langle
\phi_2 \rangle=v\sin\beta$; $g$ and $g'$ are the gauge couplings
of the SU(2)$_L$ and U(1)$_Y$, respectively; $g_v = T_{3L}/2 -
s_\theta^2 Q$ and $g_a = T_{3L}/2$. Here, we used the following short-hand
notations: $s_\theta \equiv \sin\theta_W$ and $c_\theta \equiv
\cos\theta_W$, where $\theta_W$ is the weak-mixing
angle. The tree-level (non-physical) $W$ and $Z$ masses are $M_W =
gv/2$ and $M_Z = M_W /c_\theta$. Since the compactification scale
$M_C$ is expected to be in the TeV range, we therefore ignore 
in the above equations the mass of the zeroth-state gauge boson 
in the expression for the mass of the $n$-th KK excitation: 
$M_n = \sqrt{M_0^2 + n^2 M_C^2} \approx n M_C$, $n=1,2,...$.

Using the above Lagrangians we can describe the two major effects of
the KK states: mixing with the SM gauge bosons and virtual exchanges
in high-energy interactions.

\subsection{Mixing with the SM Gauge Bosons}

The first few terms in the Eqs. (\ref{P3_Landsberg_102901eq1}) 
and (\ref{P3_Landsberg_102901eq2}) imply the
existence of mixings among the SM boson ($V$) and its KK excitations
($V^{(1)}$, $V^{(2)},\; ...$) where $V=W,Z$. There is no mixing for the 
$A^\mu$ fields because of the U(1)$_{\rm EM}$ symmetry. These mixings 
modify the electroweak observables (similar to the mixing between the 
$Z$ and $Z'$). The SM weak eigenstate of the $Z$-boson, $Z^{(0)}$, 
mixes with its excited KK states $Z^{(n)}$ ($n=1,2,...$) via a series 
of mixing angles, which depend on the masses of $Z^{(n)}, n=0,1,...$ 
and on the angle 
$\beta$. The $Z$ boson studied at LEP~1 is then the lowest mass 
eigenstate after mixing. The couplings of the $Z^{(0)}$ to fermions 
are also modified through the mixing angles. The observables at LEP~1 
can place strong constraints on the mixing, and thus on the 
compactification scale $M_C$. Similarly, the properties of the $W$ boson 
are also modified. 

The effects of KK excitations in the low-energy limit can be included
by eliminating their fields using equations of motion.  From 
the Lagrangians given by Eqs. (\ref{P3_Landsberg_102901eq1}) 
and (\ref{P3_Landsberg_102901eq2})
the $W,Z$ masses and the low-energy CC and NC interactions are given by
~\cite{Casalbuoni:1999ns}:
\begin{eqnarray} 
        M_W^2 &=& M_W^2 ( 1 - c^2_\theta \sin_\beta^4 X ) \;, \nonumber \\ 
        M_Z^2 &=& M_Z^2 ( 1 - \sin_\beta^4 X ) \;, \nonumber \\ 
        {\cal L}^{\rm CC}_{\rm int} &=& - g J^1_\mu W^{1\mu} (1-\sin^2\beta
        c^2_\theta X ) - \frac{g^2}{2M_Z^2} X J^1_\mu J^{1\mu} + (1\to 2) \;,
        \nonumber \\ 
        {\cal L}^{\rm NC}_{\rm int} &=& - \frac{e}{s_\theta
        c_\theta} J^Z_\mu Z^\mu (1-\sin^2\beta X ) - \frac{e^2}{2 s_\theta^2
        c_\theta^2 M_Z^2} X J^Z_\mu J^{Z\mu} \nonumber \\ 
        && - \; e J_\mu^{\rm
        em} A^{\mu} - \frac{e^2}{2 M_Z^2} X J^{\rm em}_\mu J^{{\rm em} \mu}
        \;, \nonumber \\ 
        X &=& \frac{\pi^2 M_Z^2}{3 M_C^2} \;.\nonumber
\end{eqnarray}

In the following, we summarize the results presented in
Refs. \cite{Nath:1999mw,Rizzo:1999br,Casalbuoni:1999ns,Strumia:1999jm,%
Carone:1999nz,Delgado:1999sv,Cornet:1999im}. Nath and
Yamaguchi~\cite{Nath:1999mw} used data on $G_F$, $M_W$, and $M_Z$ and set
the lower limit on $M_C \gsim 1.6$~TeV. Carone~\cite{Carone:1999nz} studied a
number of precision observables, such as $G_F$, $\rho$, $Q_W$, leptonic
and hadronic widths of the $Z$. The most stringent constraint on $M_C$
comes from the hadronic width of the $Z$: $M_C >3.85$ TeV. Strumia
\cite{Strumia:1999jm} obtained a limit $M_C > 3.4 - 4.3$ TeV from a set of
electroweak precision observables. Casalbuoni {\it et al.} 
\cite{Casalbuoni:1999ns} used the complete set of precision measurements, as well as
$Q_W$ and $R_\nu$'s from $\nu$-N scattering experiments, and obtained a
limit $M_C > 3.6$ TeV. Rizzo and Wells \cite{Rizzo:1999br} used the
same set of data as the previous authors and obtained a limit
$M_C > 3.8$ TeV. Cornet {\it et al.} \cite{Cornet:1999im} used the unitarity of
the CKM matrix elements and were able to obtain a limit $M_C > 3.3$
TeV. Delgado {\it et al.} \cite{Delgado:1999sv} studied a scenario in which
quarks of different families are separated in the extra spatial
dimension and set the limit $M_C > 5$ TeV in this scenario.

\subsection{Virtual Exchanges}

If the available energy is higher than the compactification scale the
on-shell production of the Kaluza-Klein excitations of the gauge
bosons can be observed~\cite{Rizzo:1999en,Davoudiasl:2000wi}. 
However, for the present
collider energies only indirect effects can be seen, as the 
compactification scale is believed to be at least a few TeV. These
indirect effects are due to virtual exchange of the KK-states.

When considering these virtual exchanges, we ignore a slight
modification of the coupling constants to fermions due to the mixings
among the KK states and so we use Eqs. (\ref{P3_Landsberg_102901eq1}) 
and (\ref{P3_Landsberg_102901eq2}) without 
the mixings.\footnote{For $M_C >> M_Z$ the mixings are very small.  
Furthermore, they completely vanish for  $\beta=0$.}
This implies that any Feynman 
diagram which has an exchange of a $W$, $Z$, $\gamma$, or $g$ will be 
replicated for every corresponding KK state with the masses $n M_C,$ 
where $n=1,2,...$.  Note that the coupling constant of the KK states 
to fermions is a factor of $\sqrt{2}$ larger than that for the 
corresponding SM gauge boson, due to the normalization of the KK 
excitations.

The effects of exchanges of KK states can be easily included by extending
reduced amplitudes.  In the limit $M_C \gg \sqrt{s}, \sqrt{|t|}, \sqrt{|u|}$, 
the reduced amplitudes take the form:
$$
        M^{\ell q}_{\alpha\beta}(s) = e^2 \Biggr \{ \frac{Q_\ell Q_q}{s} +
        \frac{g_\alpha^\ell g_\beta^q}{\sin^2\theta_W \cos^2 \theta_W } \;
        \frac{1}{s - M_Z^2 } - \left( Q_\ell Q_q + \frac{g_\alpha^\ell
        g_\beta^q} {\sin^2\theta_W \cos^2 \theta_W } \right ) \;
        \frac{\pi^2}{3 M_C^2 } \; \Biggr \} \;,
$$
based on which, the high energy processes can be described.  

\section{High Energy Processes and Data Sets}
\label{sec:data}

Before describing the data sets used in our analysis, let us first
specify certain important aspects of the analysis technique. Since the
next-to-leading order (NLO) calculations do not exist for the new
interactions yet, we use leading order (LO) calculations for 
contributions both from the SM and from new interactions, for 
consistency. However, in many cases, e.g. in the analysis of precision 
electroweak parameters, it is important to use the best available 
calculations of their SM values, as in many cases data is sensitive to 
the next-to-leading and sometimes even to higher-order corrections. Therefore, 
we normalize our leading order calculations to either the best
calculations available, or to the low-$Q^2$ region of the data set, where
the contribution from the KK states is expected to be vanishing. This
is equivalent to introducing a $Q^2$-dependent $K$-factor and using
the same $K$-factor for both the SM  contribution and the effects of 
the KK resonances, which is well justified by the similarity between these 
extra resonances and the corresponding ground-state gauge boson. The 
details of this procedure for each data set are given in the corresponding
section. Wherever parton distribution functions (PDFs) are needed, we
use the CTEQ5L (leading order fit) set \cite{Lai:1999wy}. 

\subsection{HERA Neutral and Charged Current Data}

ZEUS \cite{Breitweg:1999id,Breitweg:1999aa} and 
H1 \cite{Adloff:1999ah,Adloff:2000qj} have published results on
neutral-current (NC) and charged-current (CC) deep-inelastic
scattering (DIS) in $e^+ p$ collisions at $\sqrt{s} \approx 300$
GeV. The data sets collected by H1 and ZEUS correspond to
integrated luminosities of 35.6 and 47.7 pb$^{-1}$, respectively. H1
\cite{Adloff:1999ah,Adloff:2000qj} has also published 
a NC and CC analysis for the most recent data
collected in $e^- p$ collisions at $\sqrt{s} \approx 320 $ GeV with an
integrated luminosity of $16.4$ pb$^{-1}$.
We used single-differential cross sections $d\sigma/d Q^2$ presented by 
ZEUS \cite{Breitweg:1999id,Breitweg:1999aa} and 
double-differential cross sections $d^2 \sigma/dx dQ^2$
published by H1 \cite{Adloff:1999ah,Adloff:2000qj}. 

We normalize the tree-level SM cross section to that measured in the low-$Q^2$ 
data by a scale factor $C$ ($C$ is very close to 1 numerically). The cross 
section $\sigma$ used in the fitting procedure is given by
\begin{equation}
        \label{P3_Landsberg_102901eq3}
        \sigma = C \left( \sigma_{\rm SM} + \sigma_{\rm interf} +
        \sigma_{\rm KK} \right )\; ,
\end{equation} 
where $\sigma_{\rm interf}$ is the interference term between the SM and 
the KK states and $\sigma_{\rm KK}$ is the cross section due to the 
KK-state interactions only.

\subsection{Drell-Yan Production at the Tevatron}

Both CDF \cite{Abe:1997gt} and D\O\ \cite{Abbott:1998rr} measured the differential
cross section $d\sigma/dM_{\ell\ell}$ for Drell-Yan production, where
$M_{\ell\ell}$ is the invariant mass of the lepton pair. (CDF analyzed
data in both the electron and muon channels; D\O\ analyzed only the
electron channel.)

We scale this tree-level SM cross section 
by normalizing it to the $Z$-peak cross section measured with the data. 
The cross section used in the fitting procedure is then obtained 
similarly to that in Eq. (\ref{P3_Landsberg_102901eq3}).

\subsection{LEP~2 Data}

We analyze LEP~2 observables sensitive to the effects of the KK states
of the photon and $Z$, including hadronic and leptonic cross sections 
and forward-backward asymmetries. The LEP Electroweak Working
Group combined the $q \bar q, \mu^+ \mu^-$, and $\tau^+ \tau^-$ data 
from all four LEP collaborations \cite{LEPEWWG:2001xv} for the machine energies
between 130 and 202 GeV. We use the following quantities in our
analysis: (i) total hadronic cross sections; (ii) total $\mu^+ \mu^-,
\tau^+ \tau^-$ cross sections; (iii) forward-backward asymmetries in
the $\mu$ and $\tau$ channels; and (iv) ratio of $b$-quark and
$c$-quark production to the total hadronic cross section, $R_b$ and $R_c$. We 
take into account the correlations of the data points in
each data set as given by~\cite{LEPEWWG:2001xv}.

For other channels we use various data sets from individual experiments.
They are \cite{Barate:1999qx,Minard:1999ii,Barate:2000-047,
Barate:2001-019,Abreu:1999ug,Abreu:2000ap,Abreu:99-135,Abreu:2000-1328,
Acciarri:1996yr,Acciarri:1997iu,Acciarri:2398,Acciarri:1999rw,
Ackerstaff:1998nf,Abbiendi:1998ea,Abbiendi:1999wm,Abbiendi:424}: 
(i) Bhabha scattering cross section 
$\sigma(e^+ e^- \to e^+ e^-)$; (ii) angular distribution or 
forward-backward asymmetry in hadroproduction $e^+ e^- \to q \bar q$;
(iii) angular distribution or forward-backward asymmetry in the $e^+
e^-$, $\mu^+ \mu^-$, and $\tau^+ \tau^-$ production.

To minimize the uncertainties from higher-order corrections, we
normalize the tree-level SM calculations to the NLO cross section,
quoted in the corresponding experimental papers. We then scale our
tree-level results, including contributions from the KK states of the
$Z$ and $\gamma$, with this normalization factor, similar to
Eq. (\ref{P3_Landsberg_102901eq3}). When fitting angular distribution, we fit to the
shape only, and treat the normalization as a free parameter.

\subsection{Kaluza-Klein states of the Gluon in Dijet Production at
the Tevatron}

Since the gauge bosons propagate in extra dimensions, the Kaluza-Klein 
momentum conservation applies at their self-coupling vertices. Because 
of this conservation, the triple interaction vertex with two gluons on 
the SM 3-brane and one KK state of the gluon in the bulk vanishes. 
(However, the quartic vertex with two gluons on the SM 3-brane and two 
gluon KK states in the bulk does exist.) The cross sections for
dijet production, 
including the contributions from KK states of the gluon, are given in 
Ref.~\cite{Cheung:2001mq}.

Both CDF \cite{Abe:1996mj,Affolder:1999ua} and D\O~\cite{Abbott:1998nf,
Abbott:1998wh} published data on
dijet production, including invariant mass $M_{jj}$ and angular
distributions. In the fit, we take into account the full correlation
of data points in the data sets, as given by each experiment.

\subsection{Kaluza-Klein States of the Gluon in $t \bar t$ Production
at the Tevatron}

In Ref. \cite{Lykken:1999ms}, it was shown that $t\bar t$ production in
Run~2 of the Tevatron can be used to probe the compactification scales
up to $\sim 3$~TeV. In this paper, we consider the sensitivity from the 
existing
Run 1 data by using the tree-level $t\bar t$ production cross section,
including the contribution of the KK states of the gluon in the $q\bar q
\to t\bar t$ channel. (The $gg\to t\bar t$ channel does not have the
triple vertex interaction with two gluons from the SM 3-brane and one
KK state of the gluon in the bulk, as explained in the previous subsection.)

The latest theoretical calculations of the $t\bar t$ cross section,
including higher-order contributions, at $\sqrt{s}=1.8$ TeV correspond
to 4.7 -- 5.5 pb \cite{Berger:1996ad,Catani:1996dj}.  The present data 
on the $t\bar t$
cross sections are \cite{Blusk:ICHEP2000,Chakraborty:ICHEP2000}
\[
        \sigma_{t\bar t}\;({\rm CDF})= 6.5 \err{1.7}{1.4} \;\; {\rm pb;}
        \quad
        \sigma_{t\bar t}\;({\rm \mbox{D\O}})= 5.9 \pm{1.7}\;\; {\rm pb,}
\]
and the top-quark mass measurements are
\[
        m_{t}\;({\rm CDF}) = 176.1 \pm 6.6 \;\; {\rm GeV;} \quad
        m_{t}\;({\rm \mbox{D\O}}) = 172.1 \pm{7.1}\;\; {\rm GeV}.
\]
In our analysis, we normalize the tree-level SM cross
section to the mean of the latest theoretical predictions (5.1 pb),
and use this normalization coefficient to predict the cross section in
the presence of the KK states of the gluon (similar to 
Eq.~(\ref{P3_Landsberg_102901eq3})).

\section{Constraints from High Energy Experiments}
\label{sec:results}

Based on the above individual and combined data sets, we perform a fit to
the sum of the SM prediction and the contribution of the KK states of
gauge bosons, normalizing our tree-level cross section to the best
available higher-order calculations, as explained above. 
The effects of the KK states always enter the equations in the form 
$\eta = \pi^2/(3M_C^2)$~\cite{Cheung:2001mq}. 
Therefore, we parameterize these effects with a 
single fit parameter $\eta$. In most cases, 
the differential cross sections in the presence of the KK states 
of gauge bosons are bilinear in $\eta$.

The best-fit values of $\eta$ for each individual data set and their
combinations are shown in Table \ref{P3_Landsberg_102901table1}. 
In all cases, the preferred
values from the fit are consistent with zero, and therefore we proceed 
with setting limits on $\eta$. The one-sided 95\% C.L. upper limit on 
$\eta$ is defined as: 
\begin{equation}
        0.95 = \frac{\int_0^{\eta_{95}} d \eta \; P(\eta) }
        {\int_0^\infty d \eta \; P(\eta) } \;,\label{P3_Landsberg_102901eq4}
\end{equation}
where $P(\eta)$ is the fit likelihood function given by
$P(\eta)= \exp( -(\chi^2(\eta) - \chi^2_{\rm min})/2 )$.
The corresponding 
upper 95\% C.L. limits on $\eta$ and lower 95\% C.L. limits on $M_C$ are 
also shown in Table \ref{P3_Landsberg_102901table1}.

\begin{table}[htb]
\caption{Best-fit values of $\eta=\pi^2/(3M_C^2)$ 
and the 95\% C.L. upper limits on $\eta$ for individual data set 
and combinations. Corresponding 95\% C.L. lower limits on $M_C$ are 
also shown. \label{P3_Landsberg_102901table1} }
\medskip 
\centering
\begin{tabular}{lccc}
      & $\eta$ (TeV$^{-2}$)  & $\eta_{95}$ (TeV$^{-2}$) &  $M_C^{95}$ (TeV) \\
\hline
\hline
LEP~2:                       & & & \\
{} hadronic cross section, ang. dist., $R_{b,c}$
     & $-0.33 \err{0.13}{0.13}$ & 0.12  & 5.3 \\
{} $\mu,\tau$ cross section \& ang. dist.
     & $0.09\err{0.18}{0.18}$ & 0.42 & 2.8 \\
{} $ee$ cross section \& ang. dist.
     & $-0.62\err{0.20}{0.20}$ & 0.16 & 4.5\\
{} LEP combined          
     & $-0.28\err{0.092}{0.092}$ & 0.076 & 6.6 \\
\hline
HERA:     & & & \\
{} NC     
     & $-2.74\err{1.49}{1.51}$ & 1.59 & 1.4 \\
{} CC     
     & $-0.057\err{1.28}{1.31}$ & 2.45 & 1.2 \\
{} HERA combined 
     & $-1.23\err{0.98}{0.99}$ & 1.25  & 1.6 \\
\hline
TEVATRON:              & & & \\
{} Drell-Yan           
     & $-0.87\err{1.12}{1.03}$ & 1.96 & 1.3 \\       
{} Tevatron dijet      
     & $0.46\err{0.37}{0.58}$ & 1.0  & 1.8 \\
{} Tevatron top production 
     & $-0.53\err{0.51}{0.49}$ & 9.2 & 0.60 \\
{} Tevatron combined   
     & $-0.38\err{0.52}{0.48}$ & 0.65 & 2.3 \\
\hline
\hline
All combined 
     & $-0.29\err{0.090}{0.090}$ & 0.071 & 6.8 \\
\end{tabular}
\end{table}

\section{Sensitivity in Run~2 of the Tevatron and at the LHC}
\label{sec:sensitivity}

At the Tevatron, the best channel to probe the KK states of the photon or
the $Z$ boson is Drell-Yan production. In Ref. \cite{Cheung:1999wt}, we 
showed that using the double differential
distribution $d^2\sigma/ M_{\ell\ell} d \cos\theta$ can increase the
sensitivity to the KK states of the graviton compared to the use of
single-differential distributions.  Similarly, we expect this to be 
the case for the KK states of the photon and the $Z$ boson.  

We follow the prescription of Ref.~\cite{Cheung:1999wt} and use the Bayesian
approach, which correctly takes into account both the statistical and
systematic uncertainties, 
in the estimation of the sensitivity to 
$\eta \equiv \pi^2/(3M_C^2)$.\footnote{Note that the 
maximum likelihood method, as given by 
Eq.~(\protect\ref{P3_Landsberg_102901eq4}), 
artificially yields ~10\% higher sensitivity to $M_C$, 
as it does not properly treat the cases when the
likelihood maximum is found in the unphysical region $\eta<0$.}
Due to the 
high statistics in Run 2 and particularly at the LHC, the overall
systematics becomes dominated by the systematics on the $\hat s$-dependence
of the $K$-factor from the NLO corrections. (Systematic 
uncertainties on the integrated luminosity and efficiencies are not as
important as before, because they get canceled out when normalizing 
the tree level SM cross 
section to the $Z$-peak region in the data.) The uncertainty on the 
$K$-factor from the NLO calculations for Drell-Yan 
production~\cite{Hamberg:1991np} 
is currently known to a 3\% level, so we use this as the correlated 
systematics in our calculations on $M_C$. For the LHC we quote the limits 
for the same nominal 3\% uncertainty and also show how the sensitivity 
improves if the uncertainty on the $K$-factor shape is reduced to a 1\% 
level. It shows the importance of higher-order calculations of the 
Drell-Yan cross section, which we hope will become available by the time 
the LHC turns on.

In the simulation, we use a dilepton efficiency of 90\%, a rapidity
coverage of $|\eta| < 2.0$, and typical energy resolutions of the
Tevatron or LHC experiments. The simulation is done for a single
collider experiment in the combination of the dielectron and dimuon
channels.

As expected, the fit to double-differential cross sections yields a $\sim
10\%$ better sensitivity to $M_C$ than just using one-dimensional
differential cross sections.  We illustrate this by calculating the
sensitivity to $M_C$ in Run 1, which is slightly higher than the
result obtained from the fit to the invariant mass spectrum
from CDF and D\O. The sensitivity, at the 95\% C.L., to $M_C$ in Run 1 (120 pb$^{-1}$), 
Run 2a (2 fb$^{-1}$), Run 2b (15 fb$^{-1}$), and at the LHC (100 fb$^{-1}$)
is given in Table~\ref{P3_Landsberg_102901table2}. While the Run 2 sensitivity is somewhat 
inferior to the current indirect limits from precision electroweak data, 
LHC would offer a significantly higher sensitivity to $M_C$, well above 
10 TeV.

\begin{table}[htb]
\caption{
Sensitivity to the parameter $\eta=\pi^2/3M_C^2$ in Run 1, 
Run 2 of the Tevatron 
and at the LHC, using the dilepton channel. The corresponding 
95\% C.L. lower limits on $M_C$ are also shown.
\label{P3_Landsberg_102901table2} }
\medskip
\centering
\begin{tabular}{ccc}
     & $\eta_{95}$ (TeV$^{-2}$) & 95\% C.L. lower limit on $M_C$ (TeV) \\
\hline
\hline
Run 1 (120 pb$^{-1}$) & 1.62 & 1.4  \\
\hline
Run 2a (2 fb$^{-1}$) & 0.40 & 2.9 \\
\hline
Run 2b (15 fb$^{-1}$)& 0.19 & 4.2 \\
\hline
LHC (14 TeV, 100 fb$^{-1}$, 3\% systematics) & $1.81\times10^{-2}$ & 13.5 \\
\hline
LHC (14 TeV, 100 fb$^{-1}$, 1\% systematics) & $1.37\times10^{-2}$ & 15.5 \\
\end{tabular}
\end{table}


\section{Acknowledgments} We would like to thank Ignatios
Antoniadis, Keith Dienes, JoAnne Hewett, Steve Mrenna, Giacomo
Polesello, and Tom Rizzo for useful discussions. Many thanks to
the Les Houches organizers for making it a very productive scientific
workshop. This research was partially supported by the U.S.~Department 
of Energy under Grants
No. DE-FG02-91ER40688 and by A.P.~Sloan Foundation, and by the 
National Center for Theoretical Science under a grant from the 
National Science Council of Taiwan R.O.C.


%
\setcounter{figure}{0}
\setcounter{table}{0}
\setcounter{section}{0}
\setcounter{equation}{0}
\clearpage

\def\sg{$\tilde g$}
\def\sq{$\tilde q$}
\def\ptmiss{$P_{T}^{miss}$}
\def\Etmiss{$E_{T}^{miss}$}
\def\Pt{$p_{T}$}
\def\chione{\tilde \chi_1^0}
\def\chitwo{\tilde \chi_2^0}
\def\chithree{\tilde \chi_3^0}
\def\chifour{\tilde \chi_4^0}
\def\chii{\tilde \chi_i^0}
\def\chiipm{\tilde \chi_i^\pm}
\def\chionepm{\tilde \chi_1^\pm}
\def\chionemp{\tilde \chi_1^\mp}
\def\chionep{\tilde \chi_1^+}
\def\chitwopm{\tilde \chi_2^\pm}
\def\mgl{${\tilde{g}}$}
\def\msq{${\tilde{q}}$}
\def\mA{${{A}}$}
\def\mh{${{h}}$}
\def\mH{${{H}}$}
\def\m0{${0}$}
\def\mhf{${1/2}$}
\def\tanb{tan${\beta}$}
\def\tg{{\tilde g}}
\def\tq{{\tilde q}}
\def\tb{{\tilde b}}
\def\tchi{{\tilde\chi}}
\def\tl{{\tilde\ell}}
\def\sel{{\tilde e_L}}
\def\smul{{\tilde \mu_L}}
\def\sne{{\tilde\nu_e}}
\def\snm{{\tilde\nu_\mu}}
\def\lsp{{\tilde\chi_1^0}}
\def\GeV{{\rm GeV}}
\def\TeV{{\rm TeV}}
\def\Meff{{\rm eff}}
\def\wt{\widetilde}
\def\sgn{\mathop{\rm sgn}}
\def \sm {Standard Model }
\def \susy {supersymmetry }
\def \susyq {supersymmetric }
\def \mssm {minimal supersymmetric standard model }
\def \sugra {supergravity }
\def \fc {flavour changing }
\def \fcnc {flavour changing neutral current }
\def \brs {branching ratios }
\def \br {branching ratio }
\def\L {\Lambda }
\def\l {\lambda }
\def\ka {\kappa }
\def \t {\theta }
\def \vt {\vartheta }
\def\a {\alpha }
\def\dh {\partial }
\def \d {\delta }
\def \D {\Delta }
\def \bq {\bar q }
\def \bQ {\bar Q }
\def \g {\gamma }
\def \G {\Gamma }
\def \O {\Omega }
\def \o {\omega }
\def \b {\beta }
\def \S {\Sigma }
\def \s {\sigma }
\def \e {\epsilon }
\def \ud {{1 \over 2} }
\def \ut {{1 \over 3} }
\def \taud {{3 \over  2 } }
\def \bea {\begin{equation} }
\def \eea {\end{equation} }
\def \cc {coupling constant }
\def \ccs {coupling constants }
\def \tchi  {{\tilde \chi} }
\def \Eslash {E \kern-.9em\slash }
\def \pslash {p \kern-.5em\slash }
\def \kslash {k \kern-.5em\slash }
\def\rpv{\mbox{$\not \hspace{-0.10cm} R_p$ }}

\def\LV{\mbox{$\not \hspace{-0.15cm} L \hspace{0.1cm}$}}
\def\BV{\mbox{$\not \hspace{-0.15cm} B \hspace{0.1cm}$}}

\def\dofig#1#2{\epsfxsize=#1\centerline{\epsfbox{#2}}}
\def\dofigs#1#2#3{\centerline{\epsfxsize=#1\epsfbox{#2}%
   \hfil\epsfxsize=#1\epsfbox{#3}}}


%
\def\bentarrow{\:\raisebox{1.3ex}{\rlap{$\vert$}}\!\rightarrow}
\def\longbent{\:\raisebox{3.5ex}{\rlap{$\vert$}}\raisebox{1.3ex}%
        {\rlap{$\vert$}}\!\rightarrow}
\def\onedk#1#2{
        \begin{equation}
        \begin{array}{l}
         #1 \\
         \bentarrow #2
        \end{array}
        \end{equation}
                }
\def\dk#1#2#3{
        \begin{equation}
        \begin{array}{r c l}
        #1 & \rightarrow & #2 \\
         & & \bentarrow #3
        \end{array}
        \end{equation}
                }
\def\dksl#1#2#3#4{
        \begin{equation}
        \begin{array}{r c l }
        #1 & \rightarrow & #2  \\
         & & \bentarrow #3 \\
         & & \phantom{\bentarrow \;}  \bentarrow #4 
        \end{array}
        \label{eqq}
        \end{equation}
                }
\def\dkp#1#2#3#4{
        \begin{equation}
        \begin{array}{r c l}
        #1 & \rightarrow & #2#3 \\
         & & \phantom{\; #2}\bentarrow #4
        \end{array}
        \end{equation}
                }
\def\bothdk#1#2#3#4#5{
        \begin{equation}
        \begin{array}{r c l}
        #1 & \rightarrow & #2#3 \\
         & & \:\raisebox{1.3ex}{\rlap{$\vert$}}\raisebox{-0.5ex}{$\vert$}%
        \phantom{#2}\!\bentarrow #4 \\
         & & \bentarrow #5
        \end{array}
        \end{equation}
                }
%
%
\catcode`@=11 
\def \gsim{\mathrel{\mathpalette\@versim>}}
\def \lsim{\mathrel{\mathpalette\@versim<}}
\def \@versim#1#2{\lower0.4ex\vbox{\baselineskip\z@skip\lineskip\z@skip
     \lineskiplimit\z@\ialign{$\m@th#1\hfil##\hfil$%
     \crcr#2\crcr\sim\crcr}}}
\catcode`@=12 
%


\part{{\bf  Kaluza-Klein Excitations of Gauge Bosons in
the ATLAS Detector
} \\[0.5cm]\hspace*{0.8cm}
{\it G. Azuelos and G. Polesello
}}
\label{dysec}
%

\begin{abstract}
Kaluza-Klein excitations of the gauge bosons are a notable 
feature of theories with "small" ($\sim 1$~TeV) extra dimensions.  
The leptonic decays of the excitations of $\gamma$ and $Z$ bosons
provide a striking signature which can be detected at the
LHC.
We investigate the reach for these signatures 
through a parametrized simulation of the ATLAS detector. 
With an integrated luminosity of 100~fb$^{-1}$ a peak in 
the lepton-lepton invariant mass will be detected if 
the compactification scale ($M_c$) is below 5.8~TeV. If no peak is observed,
with an integrated luminosity of \mbox{300~fb$^{-1}$}
a limit of \mbox{$M_c<13.5$~Te}V can be obtained  
from a detailed study of the shape of the lepton-lepton invariant mass 
distribution. 
If a peak is observed, the study of the angular distribution 
of the two leptons will allow to distinguish the KK excitations 
from alternative models yielding the same signature.
\end{abstract}

\section{Introduction}\label{intro} 
%
In models with ``large"  Extra Dimensions, characterized by
compactification radii $\gg 1/{\mathrm TeV}$, gravity
propagates in the bulk, and the SM fields are confined to a
3-brane. The presence of the Extra Dimensions could be 
probed by searching for the Kaluza-Klein excitations 
of the gravitons at the future high energy accelerators, 
and these scenarios have been the subject of many phenomenological
studies.
An interesting variation of the  ADD model
\cite{Dienes:1998vg,Pomarol:1998sd,Masip:1999mk,Antoniadis:1999bq}
assumes that only the fermions are confined in the 3-brane, 
whereas the gauge fields propagate in a number of additional ``small"
extra dimensions orthogonal to the brane with 
compactification radius $\sim 1~$TeV.

For definiteness we concentrate here on a model 
with only one ``small" extra dimension, and  
where all of the SM fermions are on the same orbifold point ($D=0$).
The phenomenology of this model, which we will 
label as M1 is discussed in some detail in \cite{Rizzo:1999en}.
For compactification on $S^1/Z^2$ dimension, the model is completely
specified by a single parameter $M_c$, the compactification scale,
and the masses $M_n$ of the KK modes of the gauge bosons are
given by the relation $M^2_n= (nM_c)^2+M_0^2$, where $M_0$ is
the mass of the zero-mode excitation corresponding to the
Standard Model gauge boson.
The couplings are the same as the corresponding SM couplings,
scaled by a factor $\sqrt{2}$.
As an example of variation on our reference model 
we also briefly consider 
an alternative model, \cite{Arkani-Hamed:1999dc} (M2),
where quark and leptons are at opposite fixed points. 
The difference between the reference
model and M2 is that for model M2 the signs of the quark
couplings of the bosons are reversed for excitations with $n$ odd,
yielding a different interference pattern with the SM Drell-Yan production.

The constraints on the compactification scale from precision electroweak
measurements have been evaluated in a number of papers,
\cite{Rizzo:1999en,Nath:1999fs,Nath:1999mw,Casalbuoni:1999ns,%
Rizzo:1999br,Carone:1999nz,Delgado:1999sv,Cornet:1999im}.  These studies estimate 
an approximate lower limit of 4~TeV on the compactification
scale for the reference model considered in this analysis. 
A recent paper \cite{Cheung:2001mq,Cheunggreg2} calculates  the limits which can be extracted
from precision measurements at present high-energy accelerators. 
A 95\% limit of 6.8~TeV is obtained, dominated by the LEP 2 measurements.
The limit, however, comes from the fact that for two of the three fits
to LEP data, an unphysical  negative value for 
$M_c$ is obtained, 
with a significance of two to three standard deviations. In view of this fact, 
waiting for a clarification of the claimed discrepancy with the Standard Model, 
we conservatively study the performance of the ATLAS detector 
starting from $M_c=4$~TeV.

\section{Signal simulation and data analysis}
We simulate at particle level the production of the
gauge boson excitations, including the full
interference and angular information. We include the full Breit-Wigner shape
for the first two excitations of $\gamma$ and $Z$ \cite{RT}, and a resummed
expression for the higher lying states, for which the approximation
$M(i)\gg \hat{s}$ is used.
The expression with only  the first two resonances, 
does not alter the results for the reach in the peak region, 
but it significantly underestimates the deviation from the
Standard Model in the low mass off-peak region.
Since the dominant contributions to the low \mbox{$\hat s$} off-resonance region 
comes from the interference term between SM $\gamma/Z$ and 
the $KK$ excitations, the deviation from the SM is approximately proportional
to: 
\begin{equation}
\frac{1}{M_c^2}\sum_{n=1}^{\infty}\frac{1}{n^2}
\label{eq:eqsum}
\end{equation}
Therefore the deviation from the SM increases by 
$\frac{\pi^2/6}{1.25}-1 \sim 30\%$ 
when the full tower of resonances is considered instead of just the first two. 
If we consider model $M2$, the sum over the tower of resonances gives
a term proportional to 
$$
\frac{1}{M_c^2}\sum_{n=1}^{\infty}(-1)^n\frac{1}{n^2}\; = \;
-\frac{1}{2}\frac{1}{M_c^2}\sum_{n=1}^{\infty}\frac{1}{n^2}
$$
Therefore, the summed contribution of the interference terms in model M2 will be
of opposite sign and half of the one for the reference model.
The matrix elements are interfaced  to PYTHIA 6.125 \cite{PYTHIA}
event generator as an external process, and full events have been generated,
including the full PYTHIA machinery for QCD showering from the initial state
quarks, and for the hadronization.

\begin{figure}[tb]
\begin{center}
\dofig{0.7\textwidth}{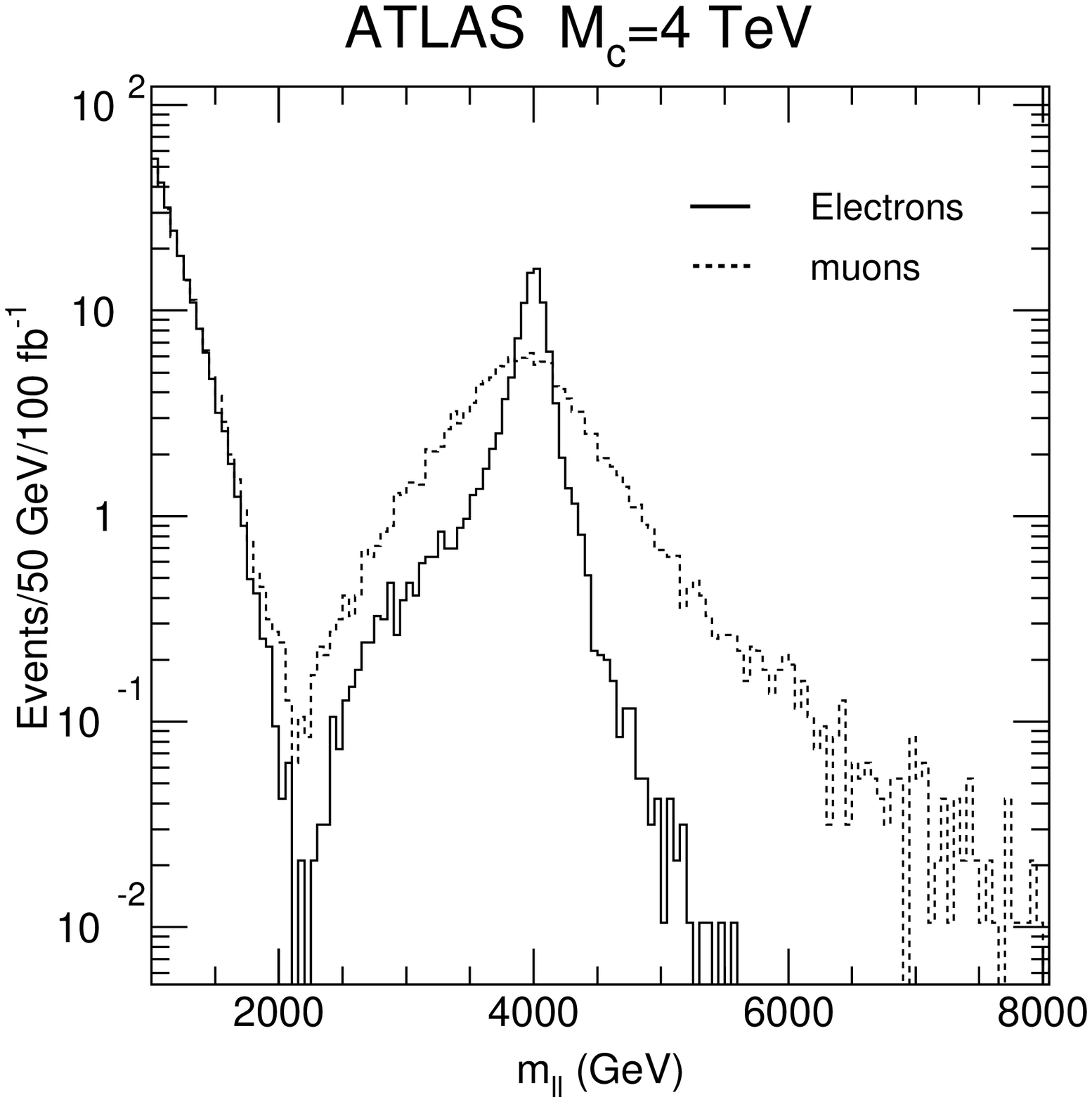}
\caption{\em
Distribution of the lepton lepton invariant mass for 
electrons (full line) and muons (dashed line). The distribution
assumes  4~TeV for the mass the lowest lying KK excitation.
}
\label{fig:emu}
\end{center}
\end{figure}

The events thus generated have been passed through
the  fast simulation of the ATLAS detector \cite{atlfast}.
As discussed in the introduction, the lowest $M_c$ considered in this
study is 4~TeV, consistent with precision electroweak measurements.
Therefore we need to detect and measure leptons with momenta in the few TeV range. 
In this range the energy resolution for electrons is dominated by a constant term 
due to the imperfect knowledge of the detector performance.
From studies performed on test beam data and on fully simulated 
events, for energies up to a few hundred GeV, 
this term has been evaluated as a few per mill. Detailed studies 
need to be performed to evaluate how well these results extrapolate
to the momentum range of interest for this analysis. 
With this caveat, we use here the 
standard parametrization included in the ATLFAST program which 
yields a resolution of $\sim$0.7\% for the energy measurement of 2~TeV electrons.
For muons the transverse momentum measurement of high $P_T$ muons 
is achieved through the sagitta measurement in the precision drift chambers,
and for a 2 TeV muon the resolution is of order 20\%. 
Considering the natural widths of the gauge excitations, 
the width of the lepton-lepton invariant mass distribution will be dominated 
by the natural width for electrons, and by the experimental momentum 
resolution for muons. 
This is illustrated in Figure \ref{fig:emu} where the invariant mass spectra for
a 4 TeV KK resonance is shown both for electrons (full line) and muons (dashed line). 
Although the muon peak is much broader, both lepton species can be
used in order to observe the existence of an excess in the peak region
with respect to the Standard Model.

\begin{figure}[htb]
\begin{center}
\dofig{0.7\textwidth}{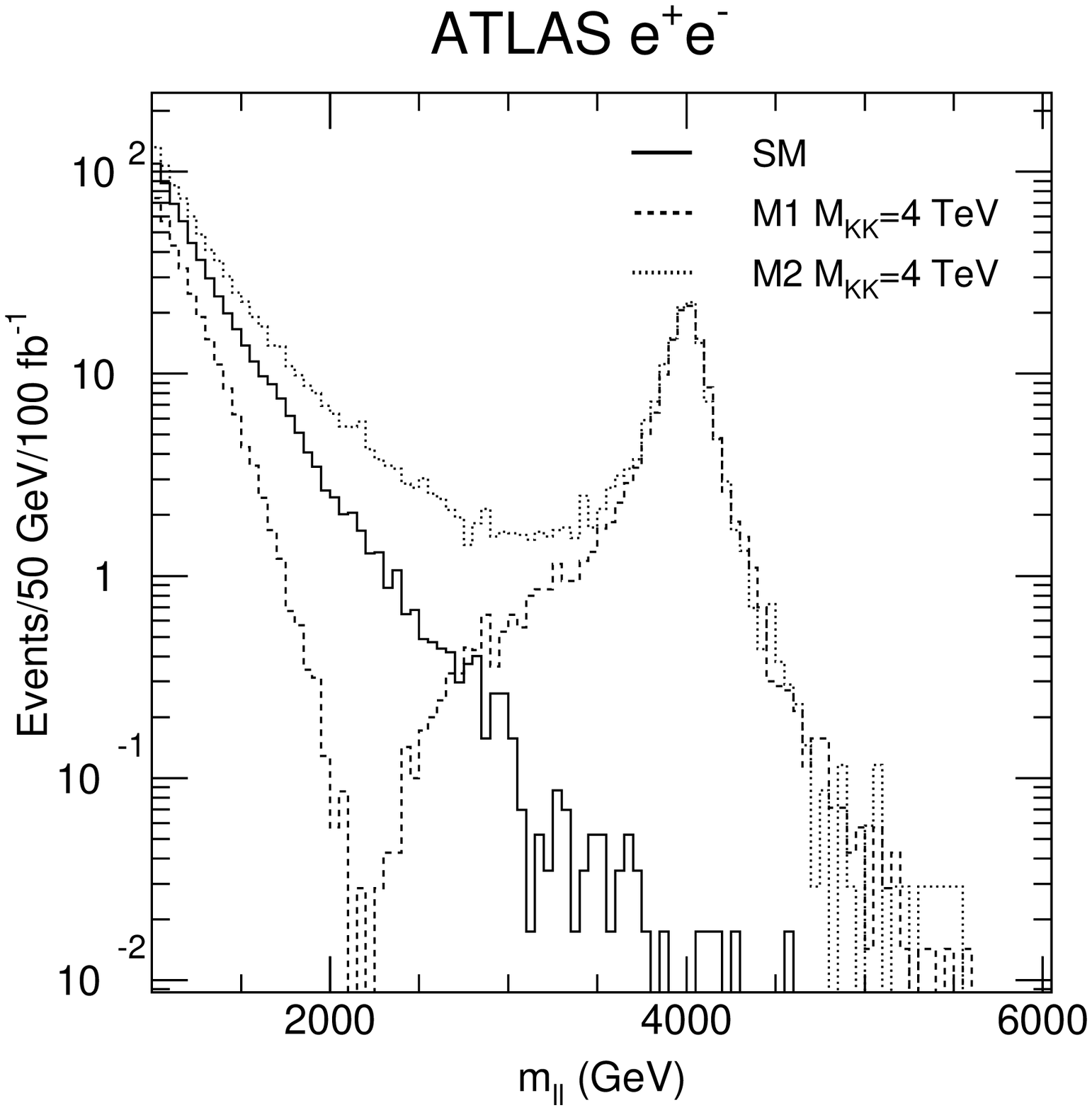}
\caption{\em  Invariant mass distribution of $e^+e^-$ pairs for the Standard
Model (full line) and for models M1 (dashed line) and M2 (dotted line).
The mass of
the lowest lying KK excitation is 4000 GeV. The histograms are
normalized to 100~fb$^{-1}$.
}
\label{fig:3mod}
\end{center}
\end{figure}

Comparing the two-lepton invariant mass spectrum for Standard
Drell-Yan production (full line), and for the reference model (dashed line) 
as shown  in Fig.\ref{fig:3mod}, 
two basic features can be observed: 
\begin{itemize}
\item
A peak centered around $M_c$, corresponding to the superposition
of the $\gamma(1)$ and $Z(1)$  Breit-Wigner shapes
\item
A suppression of the cross-section with respect to
the SM for masses below the resonance. This suppression is due
to the negative interference terms between the SM gauge bosons and
the whole tower of $KK$ excitations, and is sizable even for  compactification
masses well above the ones accessible to a direct detection of the mass peak.
This shape is the consequence of the model choices requiring both
the leptons and the quarks to be at the same orbifold point (D=0).
The different model choices corresponding to M2 would yield an 
enhancement of the off peak cross-section, as shown in the dotted line 
in Fig.~\ref{fig:3mod}.
\end{itemize}

We select events with two isolated opposite sign leptons, satisfying
the following requirements:
\begin{itemize}
\item
$m_{\ell^+\ell^-}>1000$ GeV ($\ell=e,\mu$)
\item
$P_T^{\ell}>20$~GeV,   $|\eta_{\ell}|<2.5$
\end{itemize}

The isolation criterion  consists in requiring a transverse energy 
deposition in the calorimeter smaller than 10~GeV in a ($\eta,\phi$) 
cone of radius 0.2 around the lepton direction.
In the absence of new physics, approximately 500 events 
survive these cuts for an integrated
luminosity of 100~fb$^{-1}$, corresponding to one year of high luminosity
LHC running for each of the lepton flavors.

The reach for the observation of a peak in the $m_{\ell^+\ell^-}$ distribution
can be naively estimated from Table~\ref{tab:peak},
which, for both electrons and muons gives the number of signal and 
background events for an integrated luminosity of 100 fb$^{-1}$
for different values of $M_c$.
As an arbitrary requirement
for discovery we ask for the detection above a given $M_{\ell\ell}$ of
10 events summed over the two lepton flavors,
and a statistical significance $S/\sqrt{B}>5$. The reach thus calculated
is $\sim$5.8~TeV for an integrated luminosity of 100~fb$^{-1}$.
In order to achieve this reach, a good control of high $m_{\ell\ell}$
background events which might be produced by the mismeasurement
of leptons is crucial. A handle on these events is however provided
by the consideration of the momentum balance of the event 
in the transverse plane, which will allow to reject events with one 
badly mismeasured lepton.

Unfortunately, even for the lowest allowed value of $M_c$, 4~TeV, no events would
be observed for the second resonance at 8~TeV, 
which would have been the most striking signature for this kind of model.
\begin{table}
\begin{center}
\begin{tabular}{|r|r|r|r|r|r|}
\hline
 $M_{c}$(GeV)  & Cut (GeV) & $N_{ev}(e)$ & $N_{ev}(\mu)$ & Background ($e$)& Background ($\mu$)\\
\hline
 4000 & 3000 & 172   & 156   & 1.45   & 1.8 \\
 5000 & 4000 & 23    &  20   & 0.15   & 0.22\\
 5500 & 4000 &  9    &   8   & 0.15   & 0.22\\
 6000 & 4500 & 3.3   &  2.8  & 0.05   & 0.1\\
 7000 & 5000 & 0.45  &  0.38  & 0.015  & 0.05 \\
 8000 & 6000 & 0.042 &  0.052 & 0.0015 & 0.012 \\
\hline
\end{tabular}
\caption{\em Expected number of events in the peak  for an integrated luminosity 
of 100~fb$^{-1}$, for different values of the mass  of the lowest lying
KK excitation, and Standard Model Drell-Yan background. The peak region
is defined by requiring a minimum  $\ell^+\ell^-$ invariant mass as shown in the second 
column. The results for electrons and muons are given separately.
}
\label{tab:peak}
\end{center}
\end{table}

\begin{figure}[htb]
\begin{center}
\dofig{0.7\textwidth}{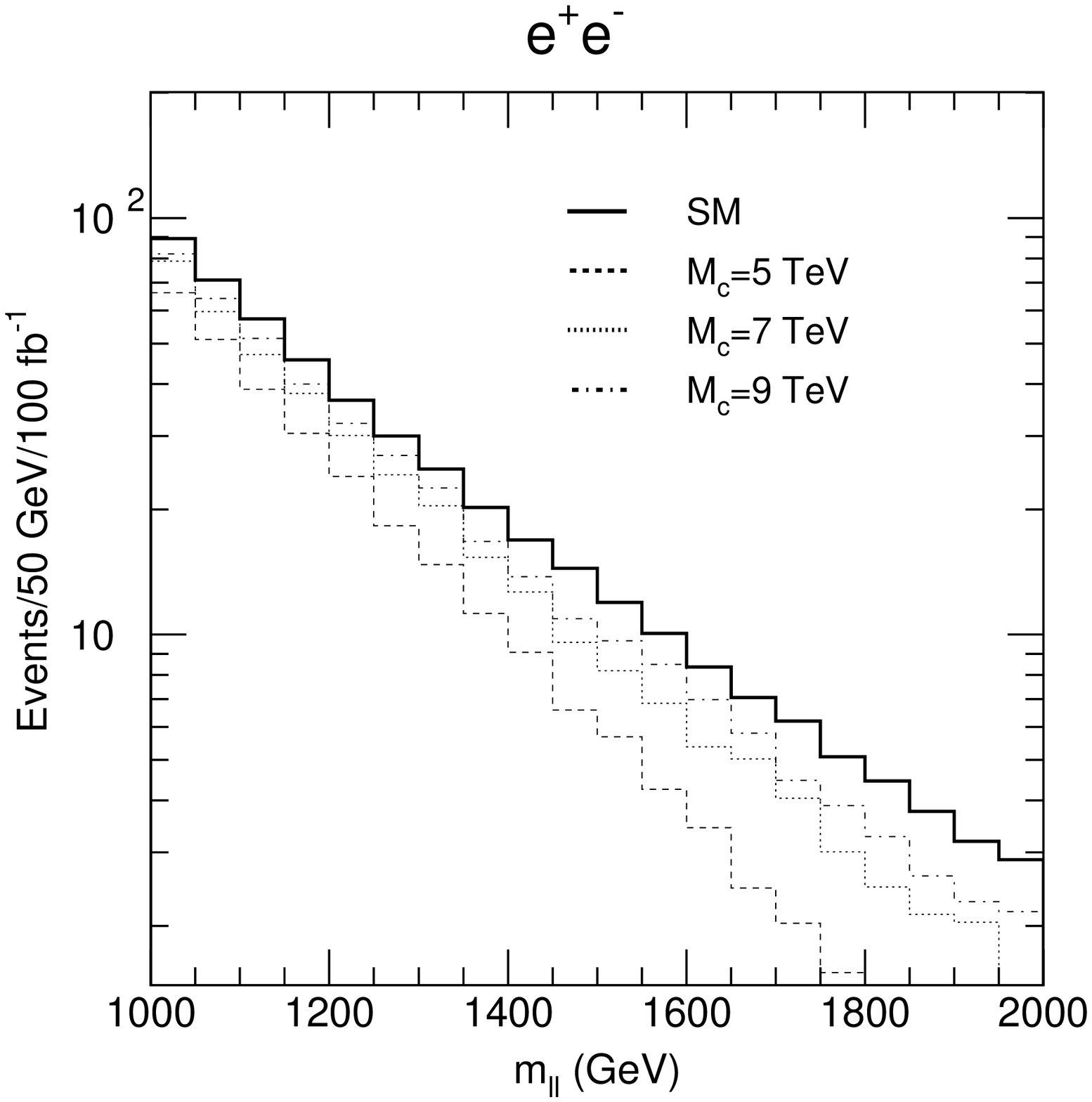}
\caption{\em  Invariant mass distribution of $e^+e^-$ pairs in the 
region below 2 TeV. The Standard Model contribution is shown as a thick line.
We show the reference model with three  different values for the compactification
scale $M_c$: 5000, 7000 and 9000 GeV as  dashed, dotted
and dash-dotted lines respectively.  The histograms are 
normalized to 100~fb$^{-1}$
}
\label{fig:int}
\end{center}
\end{figure}

In order to fully evaluate the sensitivity of the invariant mass spectrum
off-resonance to interference effects, a likelihood fit to the expected
spectrum can be performed, and will be discussed in the next section.
As a first approach, one can simply evaluate the variation 
in number of events within a given $m_{\ell\ell}$ range with respect 
to the SM, as a function of $M_c$. We show the invariant $e^+e^-$ 
mass spectrum between 1000 and 2000 GeV in Figure \ref{fig:int}
for Standard Model and for three choices of $M_c$.

A naive parameterization of the statistical  significance of 
the cross-section suppression is 
$$(N_{ev}(M_c)-N_{ev}(SM))/\sqrt{N_{ev}(SM)}.$$
A relevant variable which should also be considered 
is the ratio $N_{ev}(M_c)/N_{ev}(SM)$,
because the systematic uncertainty in our knowledge of the shape of $m_{\ell\ell}$
sets a limit on the detectable value of this ratio.
The choice of the mass interval is subject to the consideration 
of the systematical uncertainty, as the statistical significance
somewhat increases by lowering the lower limit of the
considered mass window, at the price of a worse $N_{ev}(M_c)/N_{ev}(SM)$.
We choose for this analysis a mass interval 
\mbox{$1000~<m_{\ell\ell}<2500$~GeV}. 

From the numbers in Table~\ref{tab:inter},
if we  consider both lepton flavors, the ATLAS 5~$\sigma$ reach is $\sim$8~TeV
for an integrated luminosity of 100 fb$^{-1}$ and $\sim$10.5~TeV
for 300 fb$^{-1}$. The deviation from the Standard Model 
will be 16\% for 8~TeV, and $\sim 10\%$ for 10.5~TeV,
defining in each case the level of systematic control 
on the  relevant region of the lepton-lepton invariant mass spectrum
we need to achieve to exploit the statistical power of the data.\\
\begin{table}
\begin{center}
\begin{tabular}{|r|r|}
\hline
 $M{c}$(GeV)  & $N_{ev}(\ell)$  \\
\hline
 SM    & 483  \\ 
 4000  & 210  \\
 5000  & 295  \\
 5500  & 324  \\
 6000  & 349  \\
 7000  & 381  \\
 8000  & 405  \\
 8500  & 413  \\
 9000  & 419  \\
 10000 & 432  \\
 11000 & 443  \\
 12000 & 450  \\
\hline
\end{tabular}
\caption{\em Expected number of events in the interference region for an integrated luminosity 
of 100~fb$^{-1}$, for different values of the compactification scale $M_c$
and Standard Model $e^+e^-$ background (1 lepton flavor). The considered 
mass interval is $1000 < {e^+e^-} < 2500$GeV 
}
\label{tab:inter}
\end{center}
\end{table}

\section{Optimal reach and mass measurement}
In the previous sections we have evaluated in a naive way
by simple event counting the $M_c$ range within which 
LHC will be able to observe a peak generated by the 
KK gauge excitations, and/or a deviation from the 
Standard Model in the $m_{\ell\ell}$ distribution off-peak.
An optimal estimate of the reach can be obtained by performing
a likelihood fit to the invariant mass shape expected for
different values of $M_c$. 
Instead of using just the invariant mass, we use the full 
information contained in the events. 
Ignoring the transverse momentum of the $\ell^+\ell^-$ system, the event 
kinematics for a given event $i$ is defined by the 
variables $x_1^i, x_2^i, cos\theta^i$. 
The values of $x_1$, $x_2$ have been evaluated
from the four-momenta of the detected electrons, according to the formulas:
$$
\frac{2\,P_L^{\ell^+\ell^-}}{\sqrt{s}}\, = \, x_1\, - \, x_2, \; \; \;
m^2_{\ell^+\ell^-}=x_1x_2s
$$
For the evaluation of  $\cos\theta$ we use the Collins-Soper
convention \cite{Collins:1977iv},
consisting in the equal sharing of the $\ell^+\ell^-$ system transverse momentum
between the two quarks.
A basic problem for the likelihood calculation is the fact that,
as the LHC is a $pp$ collider, it is not possible to know
from which direction the quark in the $\bar qq$ hard
scattering comes from, so only the absolute value of $\cos\theta$
can be measured, but not its sign.
Part of this information can however be recovered, by using
the knowledge of $x_1$ and $x_2$ and the fact that in the
proton the $x$ distribution for valence quarks is harder
than for anti-quarks. A detailed discussion of the experimental 
reconstruction of the three variables is given in \cite{rp}. 

For the processes under study, the initial state is $\bar qq$ for 
both signal and background, so the optimal result 
can be  obtained by just using the two physical variables 
sensitive to the dynamics of the hard-scattering processes, 
namely $m_{\ell\ell}=\sqrt{x_1x_2s}$ and $\cos\theta$.
However, since for electrons the effect of the experimental smearing is small,  
an effective approach to the problem is to use the theoretical cross-section 
expression to build an unbinned likelihood. In this approach, the use of 
only two variables would require an integration over the third 
one for each step in the likelihood calculation for each Monte Carlo experiment,
making the process unacceptably slow.
For muons the experimental smearing must 
be taken into account, and the fit can be performed by building an event  density grid
in the $m_{\ell\ell}-\cos\theta$ plane. 

In the following we will only perform the likelihood fit for electrons, 
calculating the unbinned likelihood functions on event samples
corresponding to an integrated luminosity of 100~fb$^{-1}$.
In order to evaluate the uncertainty on the $M_{c}$ measurement,
for each input $M_c$ value we generated an ensemble of Monte Carlo experiments 
(100~fb$^{-1}$ each) and for each of them we estimated 
$1/M_c^2$ by maximizing the likelihood function.\\
The likelihood fit is performed on the variable $1/M_c^2$, since 
for $m_{\ell\ell} \ll M_c$ it is the natural variable for describing 
the deviation of the cross-section from the Standard Model, as shown 
in Eq.~\ref{eq:eqsum}.
With the use of this test variable,
the Standard Model is the limit corresponding to $1/M_c^2=0$, 
and it is possible to build a continuous likelihood function extending
the evaluation to unphysical negative values of $1/M_c^2$.

\begin{figure}[tb]
\begin{center}
\dofig{\textwidth}{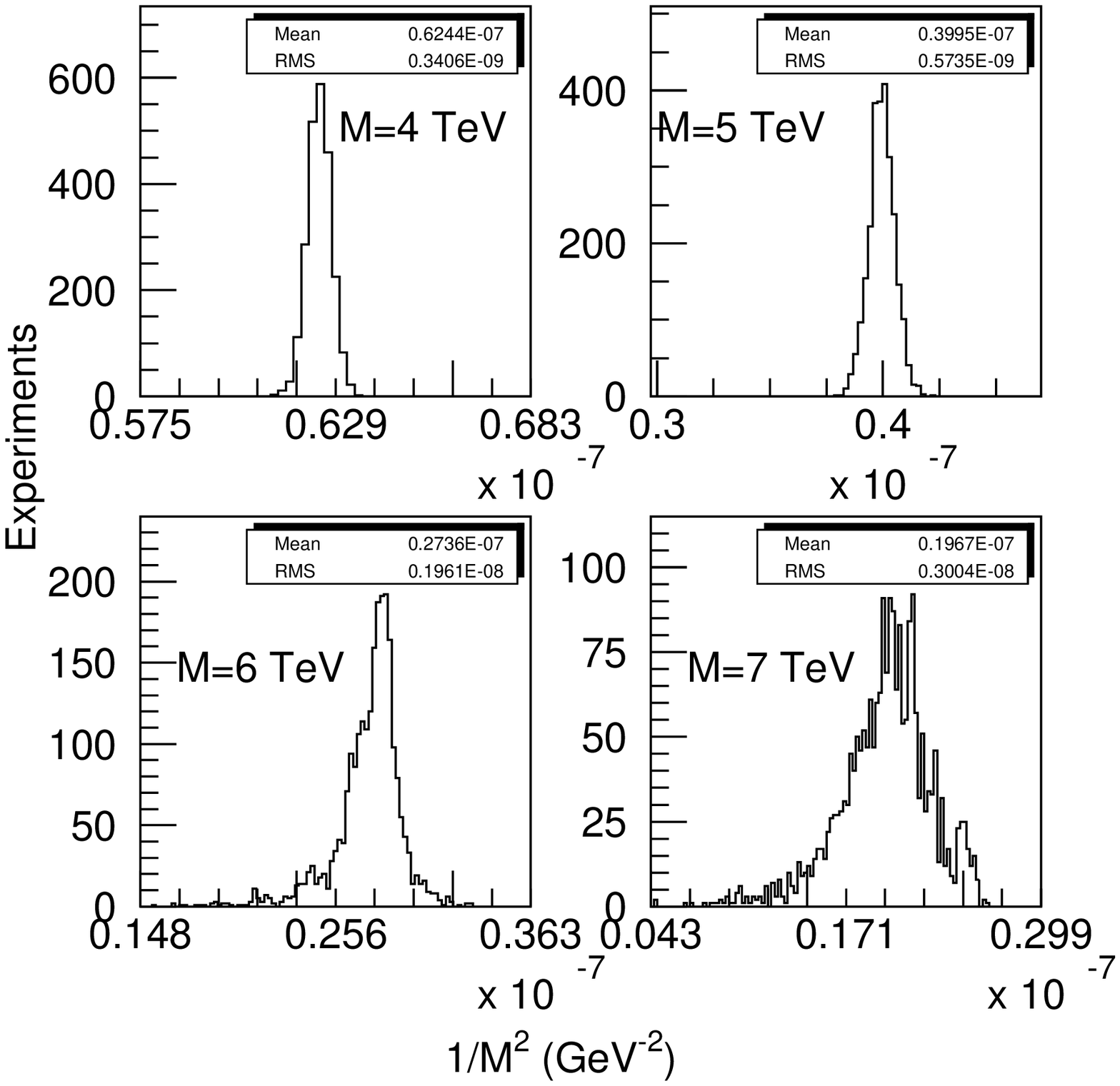}
\caption{\em Distribution of the value of $1/M_{c}^2$ estimated through
the maximization of the likelihood function for a set of $\sim$1500
Monte Carlo experiments for four $M_{c}$ values .
The input   $M_{c}$ are respectively 4, 5, 6 and 7 TeV.
The integrated luminosity is 100~fb$^{-1}$. }
\label{lik1}
\end{center}
\end{figure}

We show in Figure~\ref{lik1} the distributions of the estimated values  of
$1/M_{c}^2$ for four input values of $M_{c}$. As expected, the distributions are 
gaussian as long as events in the peak exist, and tails start to appear 
for $M_c$=6~TeV for which, on average, only three events appear in the peak 
region for the considered statistics. 
For  $M_c$=7~TeV, less than 1 event  is observed in the peak 
and the distribution becomes very broad, with an RMS corresponding 
to $\sim$600~GeV, and large tails.
The average and RMS of the estimated value of $M_c$ are given in Table
\ref{tab:lik}. 
The statistical error is below the percent level as long
as events are observed in the peak region. A small systematic shift 
in the average of the estimated $M_c$ is observed, due to the 
fact that the likelihood is built using analytical test functions neglecting
detector smearing and transverse motion of the $e^+e^-$ system. 

\begin{table}[tb]
\begin{center}
\begin{tabular}{|r|r|r|}
\hline
 $M{c}$(GeV)  & $M_{lik}$(GeV) & RMS  \\
\hline
     4000. &    4001.96 &      10.91 \\
     5000. &    5003.16 &      35.91 \\
     5500. &    5502.19 &      77.24 \\
     6000. &    6045.22 &     216.61 \\
     7000. &    7129.48 &     544.35 \\
\hline
\end{tabular}
\caption{\em Average estimated value ($M_{lik}$) and RMS of $M_c$ for $\sim2500$ experiments
and an integrated luminosity of of 100~fb$^{-1}$.}
\label{tab:lik}
\end{center}
\end{table}

The experimental sensitivity is defined in \cite{Feldman:1998qc} 
as the average upper limit that would be attained by an ensemble 
of experiments with the expected background and no true signal.
To evaluate the sensitivity, we therefore 
produced an ensemble of Monte Carlo experiments for which only 
SM Drell-Yan was generated. 
For the evaluation of the 95\% CL limit for each MonteCarlo experiment 
we use the following prescription.  
For each Monte Carlo experiment we build the likelihood  function ${\cal L}$ 
as a function of 1/$M_c^2$ as described above.
We then define as 95\% CL limit the value of $M_c$ such that the integral of ${\cal L}$
between zero  and 1/$M_c^2$ is 95\% of the integral between zero  and infinity. 
The experimental sensitivities  for one lepton flavor thus obtained
are respectively 9.5,  11 and 12~TeV 
for integrated luminosities of 100, 200 and 300 fb$^{-1}$.
These values are pessimistic, since they do not take
into account the systematic deviation from zero 
of the estimated 1/$M_c^2$ value due to the approximate test function 
used to perform the study.  
Correcting for the deviation from zero yields an improvement of 
approximately 200~GeV on these numbers. If we assume
similar sensitivity for electrons and muons, the sensitivity
is $\sim$13.5~TeV for 300~fb$^{-1}$ and both lepton species. 
These figures  only express
the statistical sensitivity of the ATLAS experiment, 
the possible sources of systematic uncertainty 
must be considered in order to evaluate the final ATLAS sensitivity.

\section{Systematic uncertainties}
As shown in the the previous section,  
the effect of KK excitation can be detected for $M_c$ well above the mass range
for which the direct observation of a peak is possible from a detailed study 
of the event shape in the interference region. The experimental sensitivity 
in this region crucially depends on our understanding of the kinematic
distributions of the lepton-lepton system both under the experimental and the 
theoretical point of view.

\begin{figure}[tb]
\begin{center}
\dofig{\textwidth}{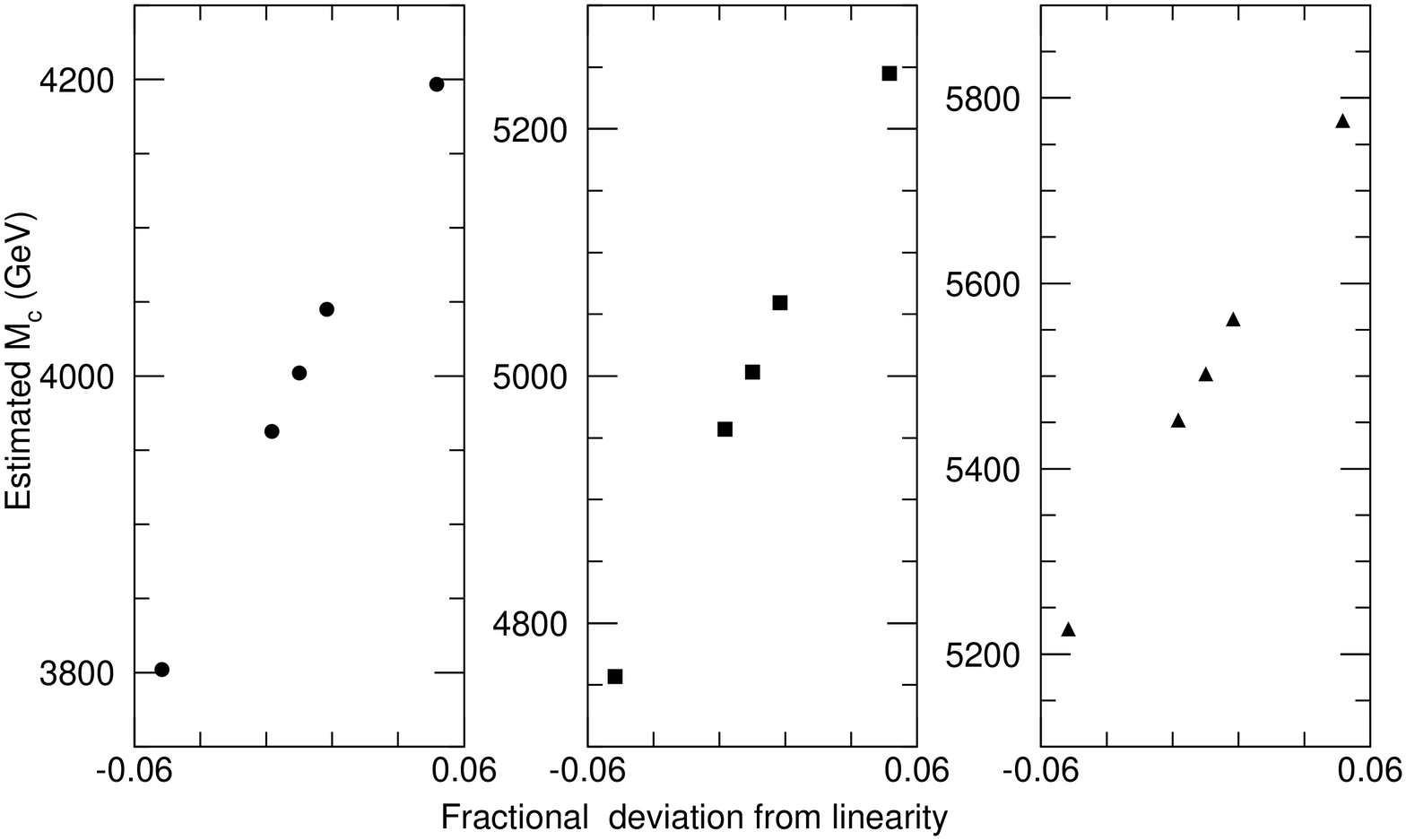}
\caption{\em Distribution of the estimated masses (100 fb$^{-1}$)
in GeV as a function of the allowed deviation from linearity for
electrons of 2~TeV momentum for three values of $M_c$: 4, 5 and 5.5 TeV
 (left, center and right plot respectively).}
\label{fig:linsig}
\end{center}
\end{figure}

\begin{figure}[tb]
\begin{center}
\dofig{0.7\textwidth}{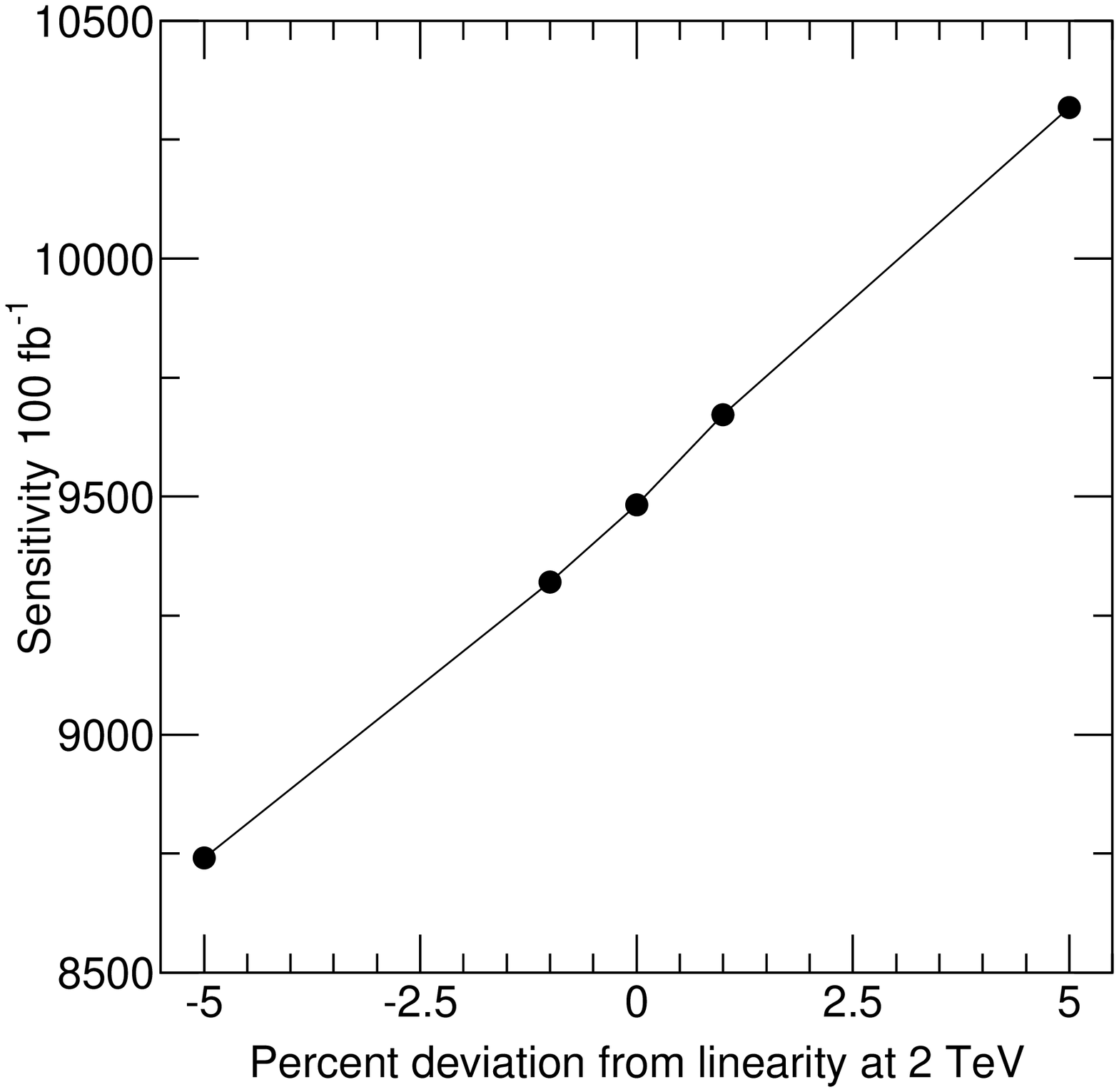}
\caption{\em Distribution of the expected sensitivity  (100 fb$^{-1}$)
in GeV as a function of the allowed deviation from linearity for 
electrons of 2~TeV momentum}
\label{fig:lin}
\end{center}
\end{figure}

As shown in Figure \ref{fig:int}, as $M_c$ increases, the difference in shape 
with respect to the standard model becomes less and less significant, and
systematic effects in the lepton-lepton invariant  mass measurement might affect the shape 
of the distribution, and destroy the experimental sensitivity.
We consider for this analysis electrons of very high momentum, around
1~TeV. At this energy scale the linearity of the lepton momentum measurement, as well 
as the momentum dependence of the acceptance are difficult to assess 
using the data. In fact very few of the  leptons from the decay of high 
momentum W and Z, which could in principle be used to perform the
measurement will have high enough momentum.   
From studies performed for lepton calibration in ATLAS, we know that the 
lepton energy scale will be known to 0.1\% at the $Z$ mass. We therefore 
parametrize the deviation from linearity as a logarithmic term 
which is zero for lepton momentum of 100~GeV, and $\pm$1 or $\pm$5\% for 
momenta of 2~TeV. We perform the likelihood analysis on all our simulated 
data samples, modifying event by event the reconstructed lepton energy  
according to the logarithmic formula.
For the evaluation of $M_c$ between 4 and 6~TeV, the 
relative deviation from the nominal $M_c$ approximately scales with 
the deviation from linearity for 2~TeV leptons, as shown in Fig.\ref{fig:linsig}
for 3 values of $M_c$: 4, 5 and 5.5 TeV. 
The variation of the sensitivity with the assumed value of the 
deviation from linearity is shown in Figure~\ref{fig:lin}. 
As discussed above, the systematic uncertainty is reflected in a systematic
shift of the average $M_c$ estimate, and an overestimate of the lepton 
calibration is going to yield an optimistic evaluation 
of the $M_c$ value excluded by the experiment.  
Taking the sensitivity values obtained with a negative deviation from linearity, 
the sensitivity for 100~fb$^{-1}$ and one lepton species is reduced from
9.5~TeV to 9.3~TeV  and 8.75~TeV for 1 and 5\% deviation respectively. 
As an approximate rule, the experimental limit should be reduced by 
$\sim$2\% for each percent of uncertainty on the energy calibration of 2~TeV
leptons.

An additional uncertainty factor is the theoretical systematics on the 
likelihood calculation.
The likelihood function is built by  weighting  real events 
according to a theoretical cross-section formula. Any 
discrepancy between the theoretical formula employed
and reality will induce an uncertainty on the measurement of 
$M_{c}$. In the likelihood analysis we are not sensitive to an
absolute $K$-factor,
since we do not use the absolute normalization, but only to 
distortions of the kinematic distributions of the  lepton-lepton
system. Three main sources of uncertainty can be identified:
\begin{itemize}
\item
QCD  higher order corrections;
\item
electroweak higher order corrections;  
\item
the parton distribution function (PDF) for the proton.
\end{itemize}
The main effect of QCD higher order corrections is the modification
of the $P_t$ distribution of the lepton-lepton pair, due to radiation
from initial state quarks. 
This effect is taken into account in a very 
pessimistic way in the study on fully generated events, 
where the likelihood is built from the leading order 2-to-2 Drell-Yan expression, 
and the events are generated with the full PYTHIA machinery for 
initial state radiation. Therefore, the experimental error
quoted in the previous section includes a very pessimistic 
estimate of this effect. In fact in a real experiment a  
more realistic theoretical modelling will probably be used to build the 
likelihood. 

Electroweak higher order corrections were recently 
calculated at NLO \cite{Baur:2001ze}, 
and shown to be sizable, leading to a reduction of the cross-section
which varies with the lepton-lepton invariant mass, and can be as large 
as 35\% for $pp\rightarrow\mu^+\mu^-$ and $m_{\mu^+\mu^-}$. 
The size of these corrections critically depends on the 
lepton identification and isolation criteria, as a substantial 
part of the higher order effects yield energetic photons produced
alongside the leptons. The evaluation of the uncertainties on these 
corrections is thus a complex interplay of experimental and theoretical
considerations which requires a dedicated study which is outside 
the scope of this analysis. 

\begin{figure}[tb]
\begin{center}
\dofig{0.7\textwidth}{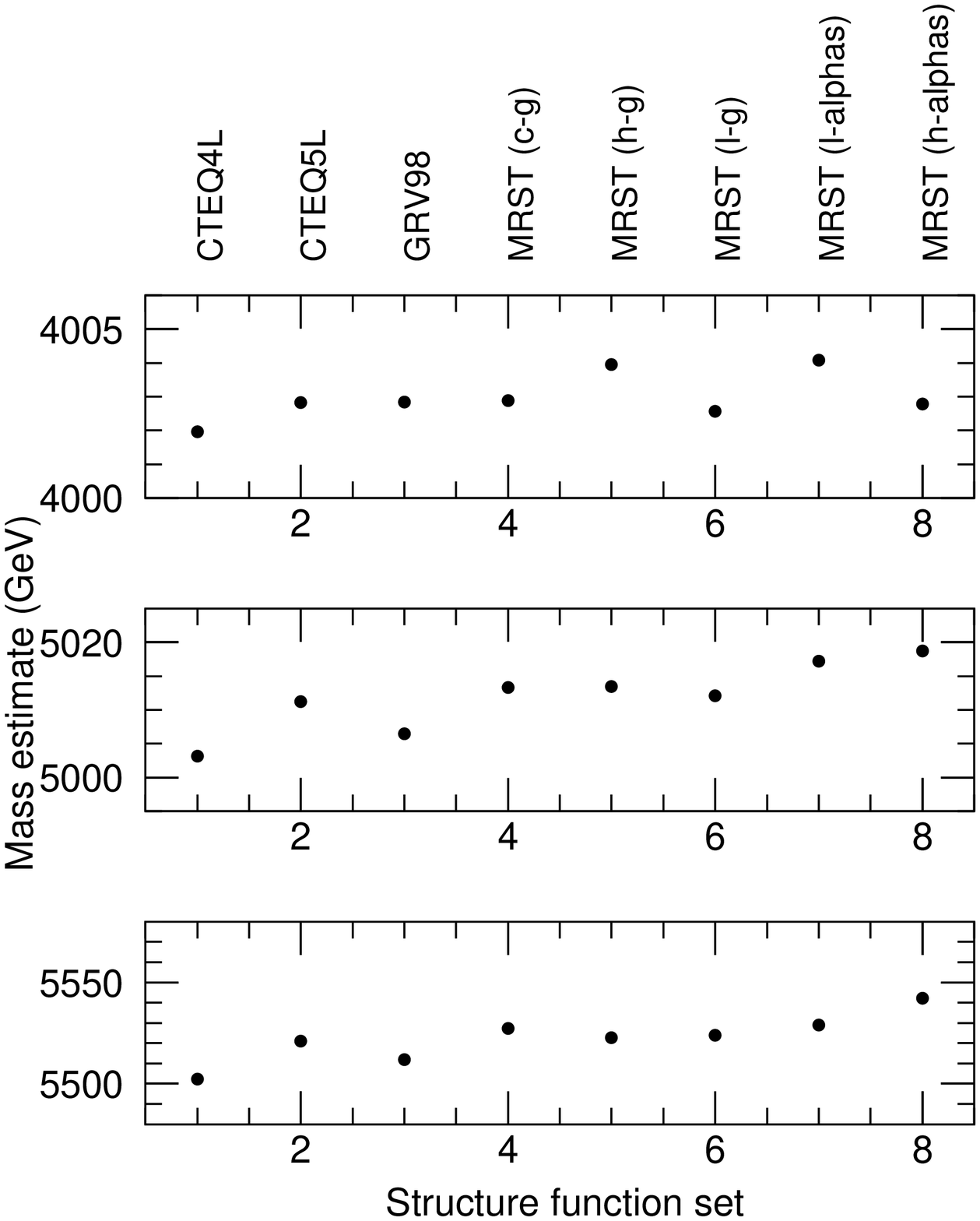}
\caption{\em Distribution of the measured values of $M_c$ in GeV 
as a function of the structure function set used for the likelihood 
fit. The events were generated using CTEQ4L. The input values for $M_c$ were 
respectively 4 TeV  (upper plot) 5~TeV (middle plot) and 5.5~TeV (lower plot).}
\label{fig:pdf}
\end{center}
\end{figure}

\begin{figure}[tb]
\begin{center}
\dofig{0.7\textwidth}{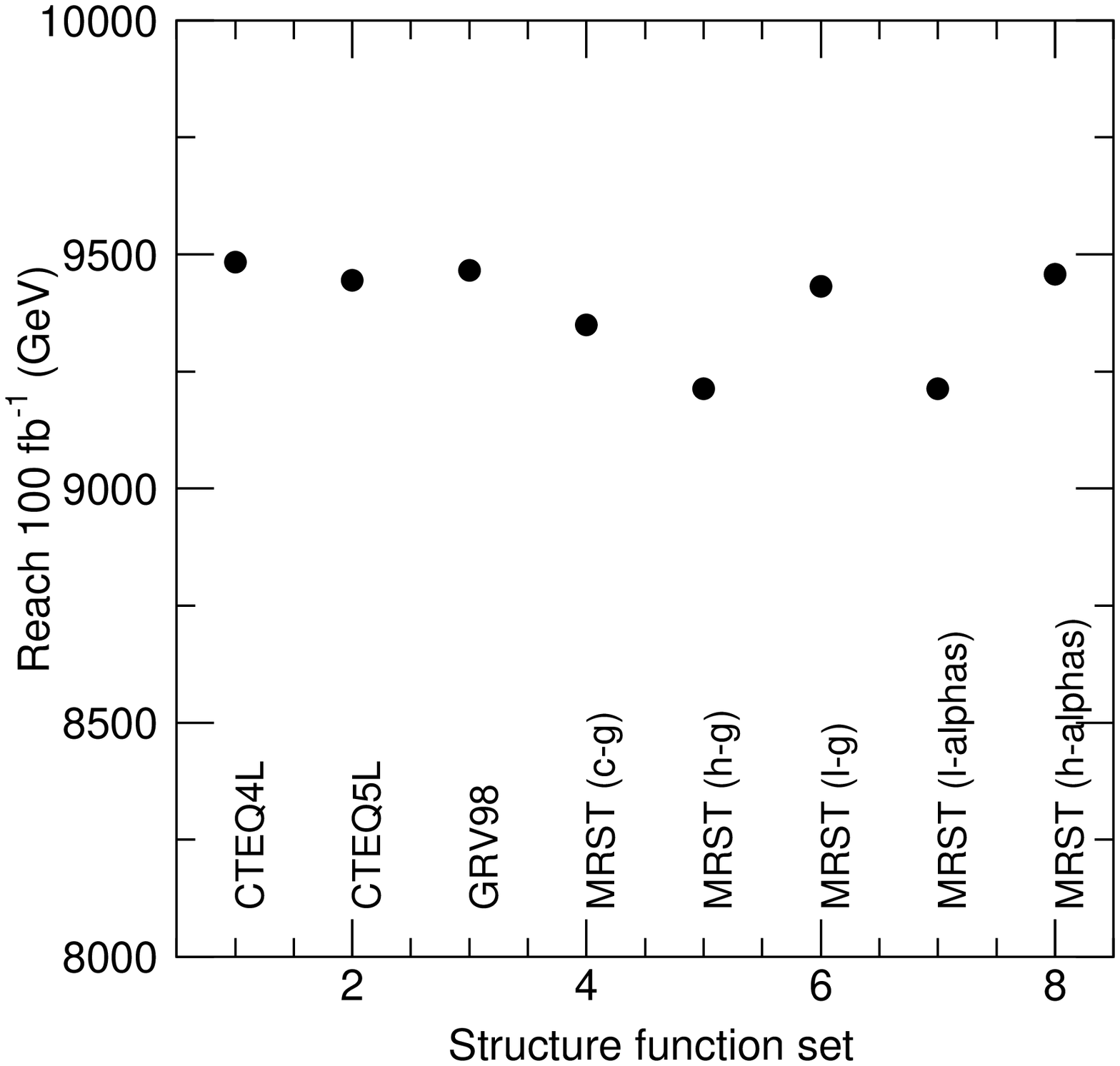}
\caption{\em Distribution of the expected sensitivity  (100 fb$^{-1}$)
in GeV as a function of the structure function set used for the likelihood 
fit. The events were generated using CTEQ4L.}
\label{fig:stfreach}
\end{center}
\end{figure}

The shape of the kinematic distributions of the lepton-lepton system,
in particular $m_{\ell\ell}$ has a strong dependence on the quark and 
antiquark PDF's in particular for high values of $x$.
All the events were generated with the
CTEQ4L PDF's.  In order to evaluate the effect of the uncertainty on the 
structure functions parametrization, the likelihood fit was performed 
on the data set thus generated, using 
a number of structure function sets. To this purpose we have selected the sets 
providing a leading order parametrization, and which are based on the latest 
available data sets.
The distributions of estimated masses are shown in Fig.\ref{fig:pdf} for 
the eight choices of structure function sets used for $M_c= 4, 5$ and 5.5~TeV.
The systematic displacement from the true value is between 3 and 4 GeV for 
4~TeV, increasing to 10-20~GeV for 5~TeV and 20-40~GeV for 5.5~TeV, 
and it is well below the RMS of the distributions given 
in Table~\ref{tab:lik}. Another notable effect is that the quality of the likelihood
fit is worse, giving rise to less Gaussian distributions, 
and sizable tails start to appear for $M_c=5.5$~TeV.
The experimental reach for 100~fb$^{-1}$ is shown in Fig.~\ref{fig:stfreach}, 
as a function of the structure function set. In the worst case the reach is reduced
by $\sim~200$~GeV with respect to CTEQ4L. 

\newcommand{\gexc} {\mbox{$Z^{(1)}/\gamma^{(1)}$}}

\section{Spin and Asymmetry Measurement} 
%
 If a Kaluza-Klein gauge excitation is discovered, one of the ways of 
distinguishing the signal from a $Z'$, predicted by GUT theories, or from
a narrow graviton resonance $G^*$ is by the angular distribution of the
decay products, which should be consistent with the spin 1 nature of the
excitation, and by the forward-backward asymmetry. By adjusting parameters of
the models, the cross sections can be made comparable, but, as
shown above and in~\cite{Rizzo:2001jj}, the shape
of the mass distribution can provide an additional distinguishing 
criterion. The present study
compares these distributions, but does not attempt to distinguish the resonances
by the shape of their mass distributions, by their relative
cross sections, nor by the branching ratios.

\subsection{Cases studied} 
%

 We studied the following cases

\renewcommand{\labelenumi}{\alph{enumi})}
\begin{enumerate}
        \item \gexc: this is the case of gauge excitations, 
model M1~\cite{Rizzo:1999en}, at mass 4 TeV. 
The process was implemented in PYTHIA~6.201.
        \item \gexc-M2: this case of gauge excitation is with
the alternative model M2~\cite{Arkani-Hamed:1999dc}, 
also at 4 TeV. The process was implemented in PYTHIA
        \item $Z'$: this is a standard model $Z'$. The same code as for case a) was 
used, but the first $\gamma$ excitation and higher excitations of $Z$ and $\gamma$ were removed.
        \item $G^*$: This is the case of a narrow graviton resonance, as was 
studied by ~\cite{Allanach:2000nr}. 
The process is implemented in PYTHIA. In order
to reproduce a resonance of width comparable to the \gexc\ of a) above, the 
dimensionless coupling which enters in the partial widths of the $G^*$ (PARP(50)
in PYTHIA) was set to 0.8. The reconstructed width is thus $\sigma \sim 82$ GeV.
 The angular distributions depend on the incoming partons. The two processes 
$ q q \to G^* \to \ell^+ \ell^-$ and $ g g \to G^* \to \ell^+ \ell^-$ 
were generated and added in proportion of their cross section. To their sum was added
the Standard Model Drell-Yan background $q q \to Z/\gamma \to \ell\ell$.
\end{enumerate}

\begin{figure}[tb]
 \begin{center}
      \mbox{\epsfig{file=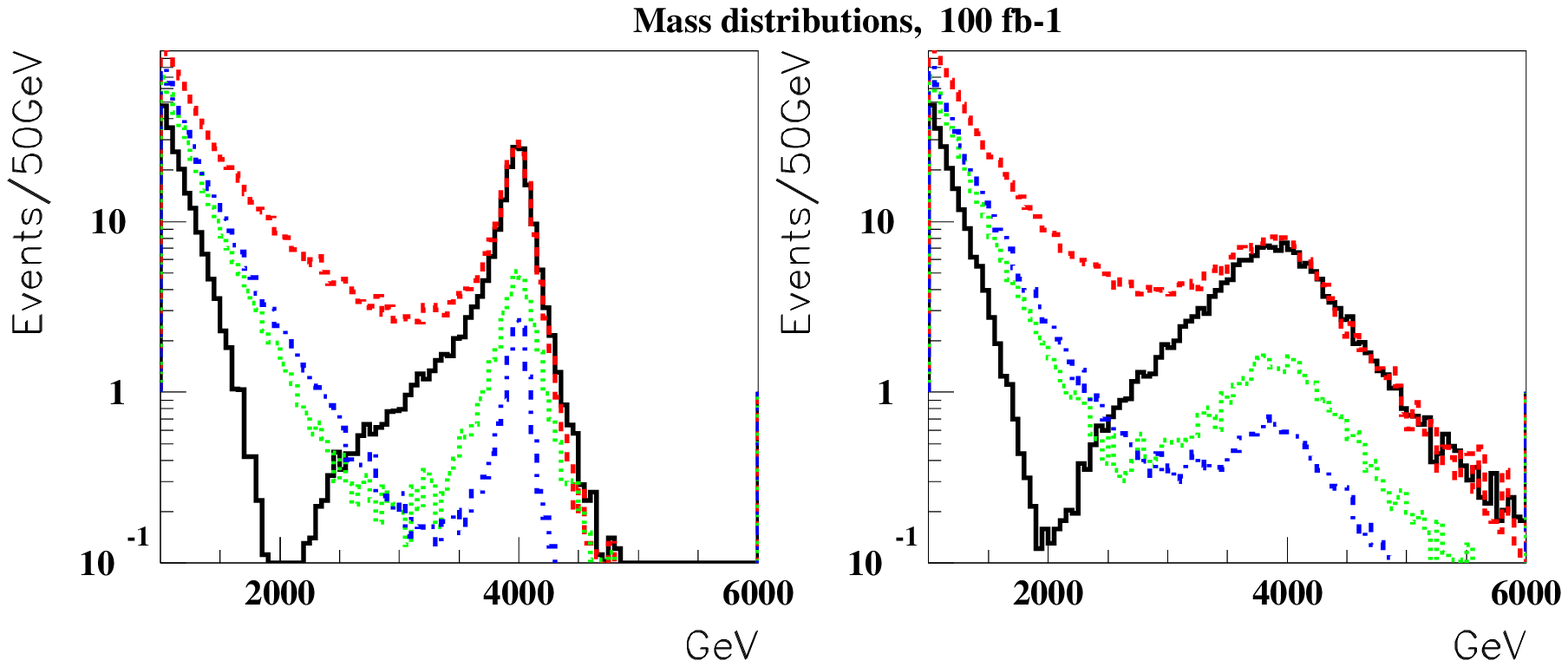,width=15cm}}
 \caption{ \em Mass distributions of the resonances considered: left: electron channel; 
          right: muon channel; (i) black full line, \gexc, (ii) red dashed line, \gexc\ model M2,
(iii) dotted green line, $Z'$, (iv) dash-dotted blue line, $G^*$+ SM Drell-Yan}
 \label{fig:massdist}
 \end{center}
 \end{figure}

  The mass distributions  normalized to a luminosity of 100 fb$^{-1}$ 
are displayed in Figs.~\ref{fig:massdist}  for the different cases.
The cross sections for the different processes are summarized in table~\ref{tab:sigma}.
 
\begin{table}[htb]
\begin{center}
\begin{tabular}{|l|c|} \hline
  process           & $\sigma\times BR(Z^*\to e^+e^-)$ (fb)\\ \hline
  \gexc             & 4.05                \\
  \gexc-M2          & 11.75                \\
  $Z'$              &  4.65   \\
  $qq\to G^*$       &  0.20  \\
  $gg\to G^*$       &  0.13  \\ 
  $q q \to e^+e^-$  &  4.83  \\ \hline
\end{tabular}
\end{center}
\caption{Nominal cross sections of the different processes, after a preselection 
$\sqrt{\hat s} >$ 1 TeV.}
\label{tab:sigma}
\end{table}

\subsection{Angular Distributions}
As mentioned above, because the colliding particles at LHC are both protons,
the forward-backward asymmetry
is measured with some ambiguity. Assuming that the resonances are produced 
by $q\bar q$ fusion, the third component of 
the reconstructed momentum of
the dilepton system is taken to be the quark direction, since the quark in the
proton is expected to have higher energy than an antiquark from the sea. 

\begin{figure}[tbp]
 \begin{center}
      \mbox{\epsfig{file=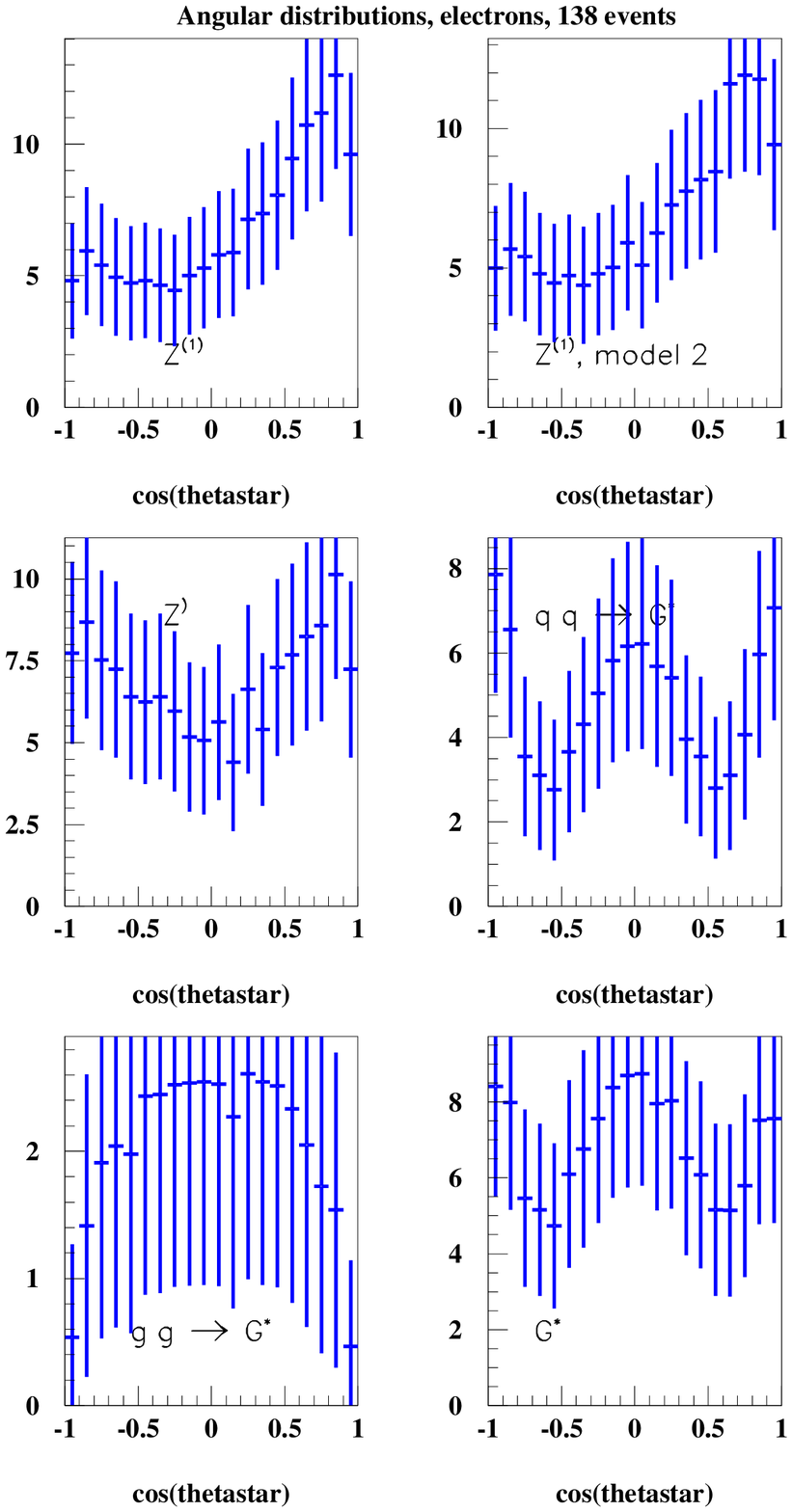,width=12cm}}
 \caption{ \em Electron channel: angular distributions for the different
types of resonances considered. The distributions
are normalized to a total of 138 events in the peak.}
 \label{fig:angdist_e_138ev}
 \end{center}
 \end{figure}

\begin{figure}[tbp]
 \begin{center}
      \mbox{\epsfig{file=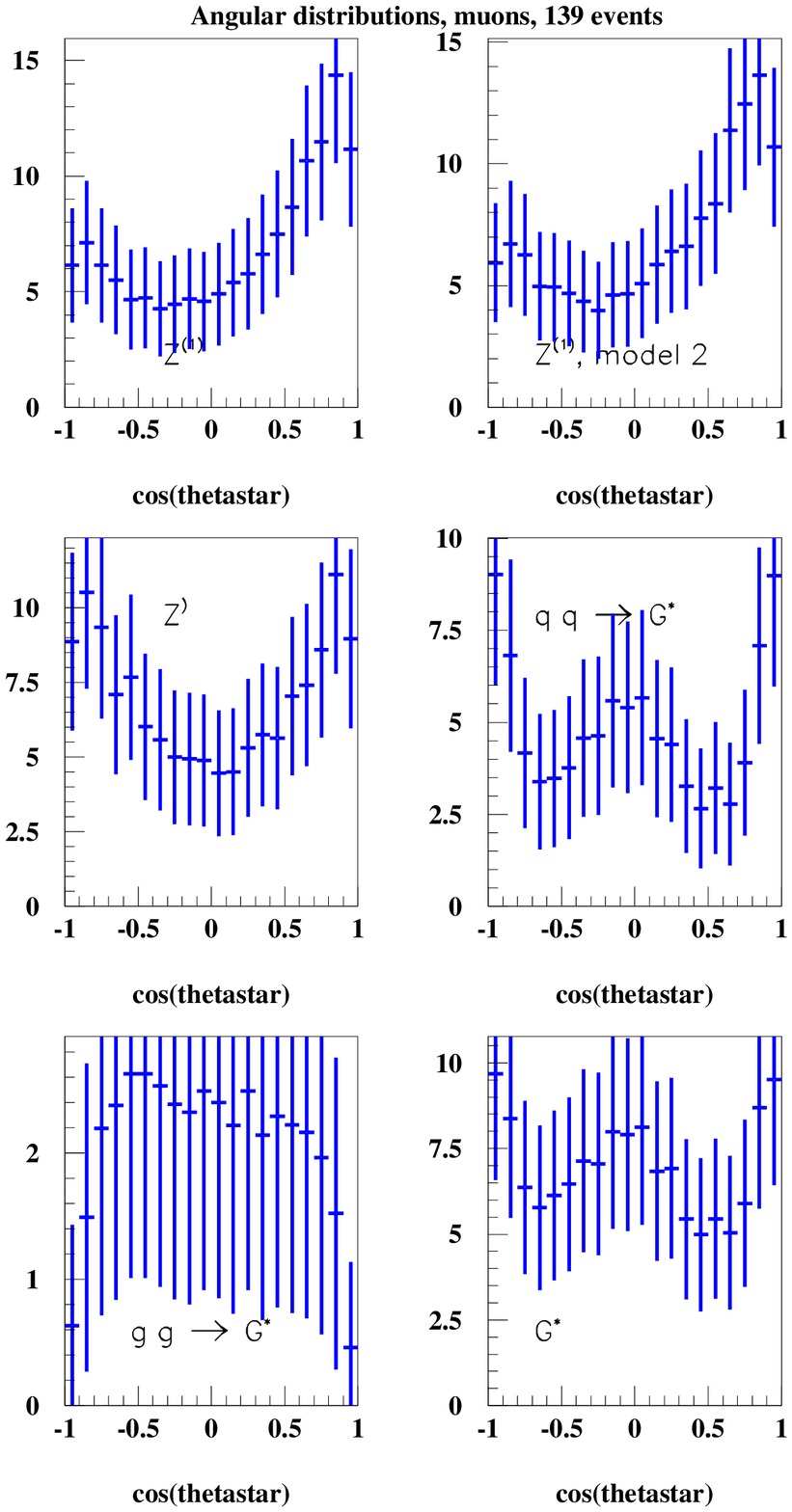,width=12cm}}
 \caption{ \em Muon channel: angular distributions for the different
types of resonances considered. The distributions
are normalized to a total of 138 events in the peak.}
 \label{fig:angdist_mu_138ev}
 \end{center}
 \end{figure}

Events around the peak of the resonance were selected: 3750 GeV $< m_{ee} <$ 4250 GeV
or 3250 GeV $< m_{\mu\mu} <$ 4750 GeV.
For these events, the
cosine of the angle of the lepton ($e^-$ or $\mu^-$), with respect to the beam direction,
in the frame of the decaying resonance, is shown in Figs.~\ref{fig:angdist_e_138ev}
and~\ref{fig:angdist_mu_138ev}.
The positive direction is defined by the sign of the reconstructed momentum of
the dilepton system. Since we will be interested only in the shape, and not in the cross sections,
the angular distribution histograms have been normalized, to a total
of 138 events, corresponding to the number of events predicted with an
integrated  luminosity of 100 fb$^{-1}$ for the reference case \gexc.
  
\begin{figure}[tb]
 \begin{center}
      \mbox{\epsfig{file=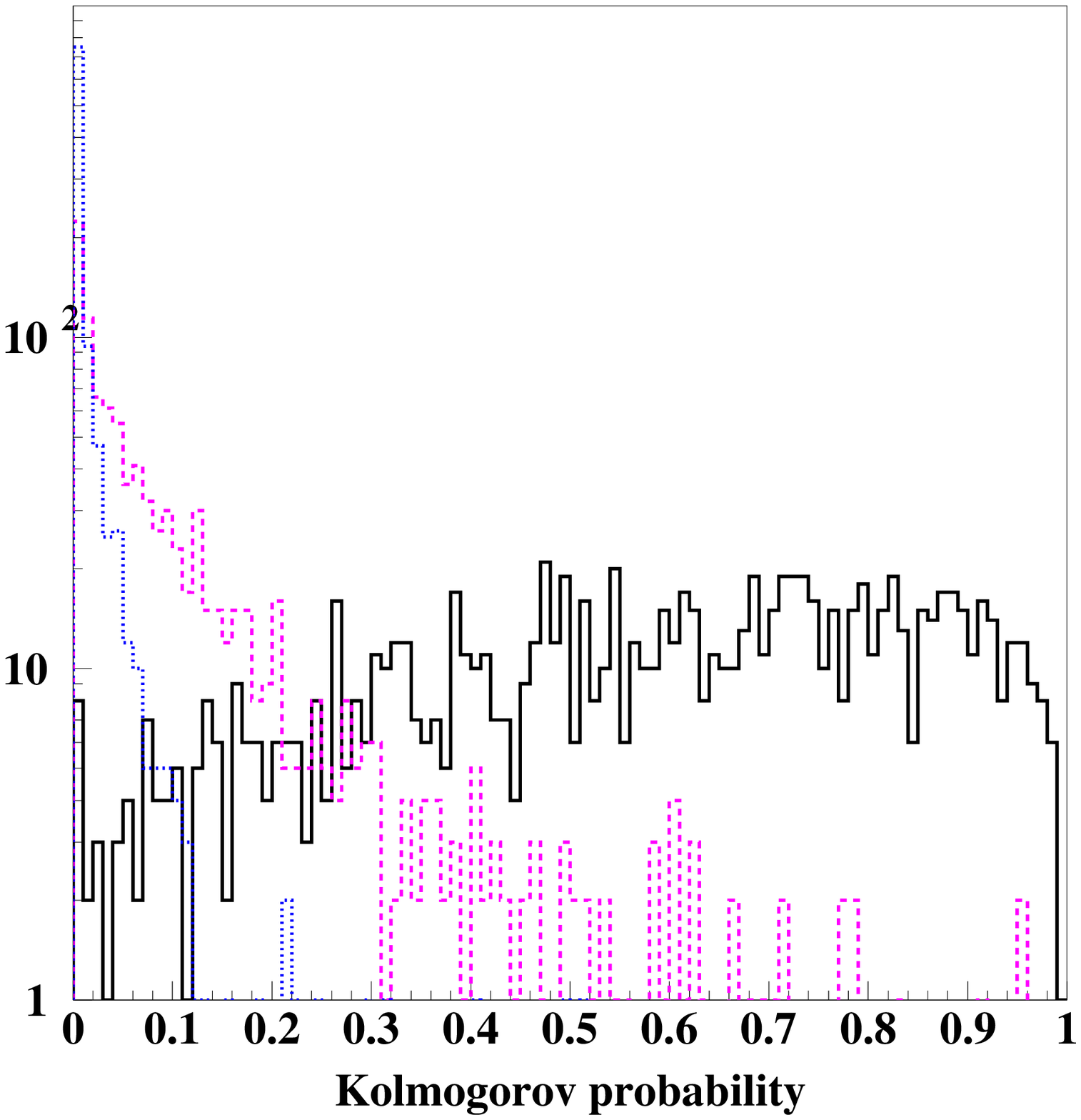,width=8cm}}
 \caption{ \em Kolmogorov probability from comparison of \gexc\ angular distribution with 
(i) black full line: model M2, (ii) pink dashed line: $Z'$ and (iii) blue dotted line: $G^*$. 
A histogram is constructed from 1000 pseudo samples of events.}
 \label{fig:Kolmogorov}
 \end{center}
 \end{figure}

 To compare the shape of these distributions, a set of 1000 angular 
distributions from the different
types of resonances was generated by sampling from the expected distributions 
of Figs.~\ref{fig:angdist_e_138ev} and~\ref{fig:angdist_mu_138ev}.
A Kolmogorov test was then 
applied\footnote{In principle, the Kolmogorov test should be applied on unbinned
data, but the application on binned data should still provide a valid test,
in the present case since the bins are narrower than the expected features}
between the expected \gexc\ distribution and distributions sampled from the
other resonances. 
The output of the test is expected to be a uniform distribution between 0
and 1 if they come from the same parent distribution. The histogram of the outputs
is displayed in Fig.~\ref{fig:Kolmogorov}. No significant difference is found between
models M1 and M2 of \gexc, as expected.  However, the
Kolmogorov test, applied to the distributions obtained for the $e^+e^-$ channel,
 will give an average probability of consistency between \gexc\ and 
$Z'$ or between \gexc\ and $G^*$ of 0.105 and 0.015 respectively and will reject,
at 95\% confidence level, the hypothesis that the distributions derive from the 
same parent distribution 52\% and 94\% of the times. For higher resonance masses
the statistical significance quickly decreases: at 5 TeV, with only 18 events in the
peak region, no discrimination becomes possible. However, for this mass but
with an integrated luminosity
of 300 fb$^{-1}$, the Kolmogorov test would 
reject the hypothesis, at 95\% CL, about 20\% of the times. Similar results are
obtained for the $\mu^+\mu^-$ channel.

 A $\chi^2$ test was also performed between these distributions, leading to
the same conclusions. Here, also, a histogram
of the calculated $\chi^2$ was produced from a sample of 1000 pseudo experiments
with 138 events each. 
The average 
$\chi^2$/d.f. are  0.998, 1.50 and 2.10 for the cases of model M2, $Z'$ and $G^*$ respectively.
The goodness of fit test between the \gexc\ and the $Z'$ or $G^*$ angular distributions
would yield a confidence level below 5\% respectively 38\% and 84\% of the times.


\subsection{Forward-backward asymmetry}

\begin{figure}[tb]
 \begin{center}
      \mbox{\epsfig{file=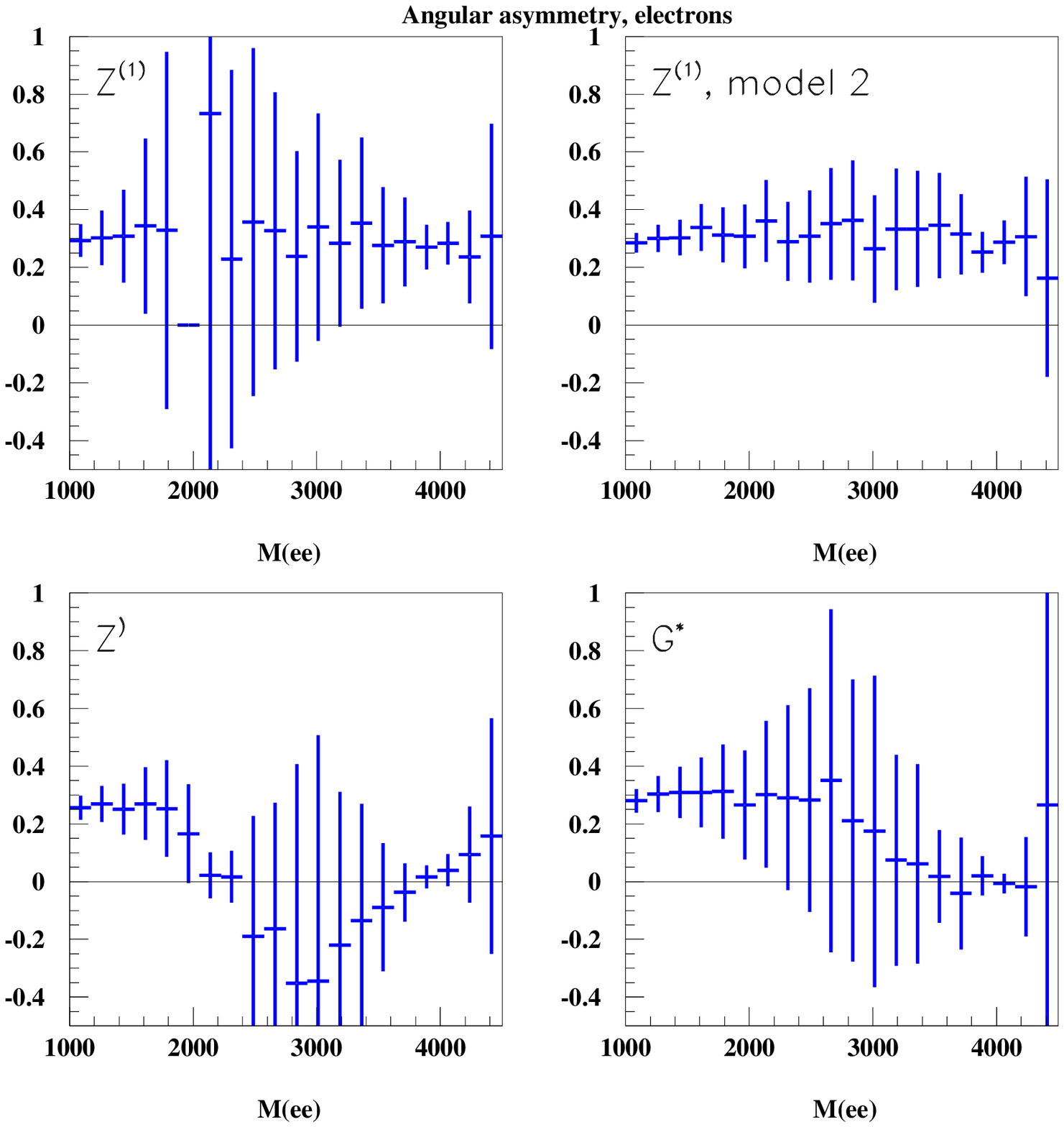,width=14cm}}
 \caption{ \em Electron channel: measured forward-backward asymmetry at LHC, 
for different types of resonances, centered at m = 4 TeV. The error
bars are representative of a sample having 138 events in the peak region, 
or 100 fb$^{-1}$ for \gexc.}
 \label{fig:fbasy_e}
 \end{center}
 \end{figure}

\begin{figure}[tb]
 \begin{center}
      \mbox{\epsfig{file=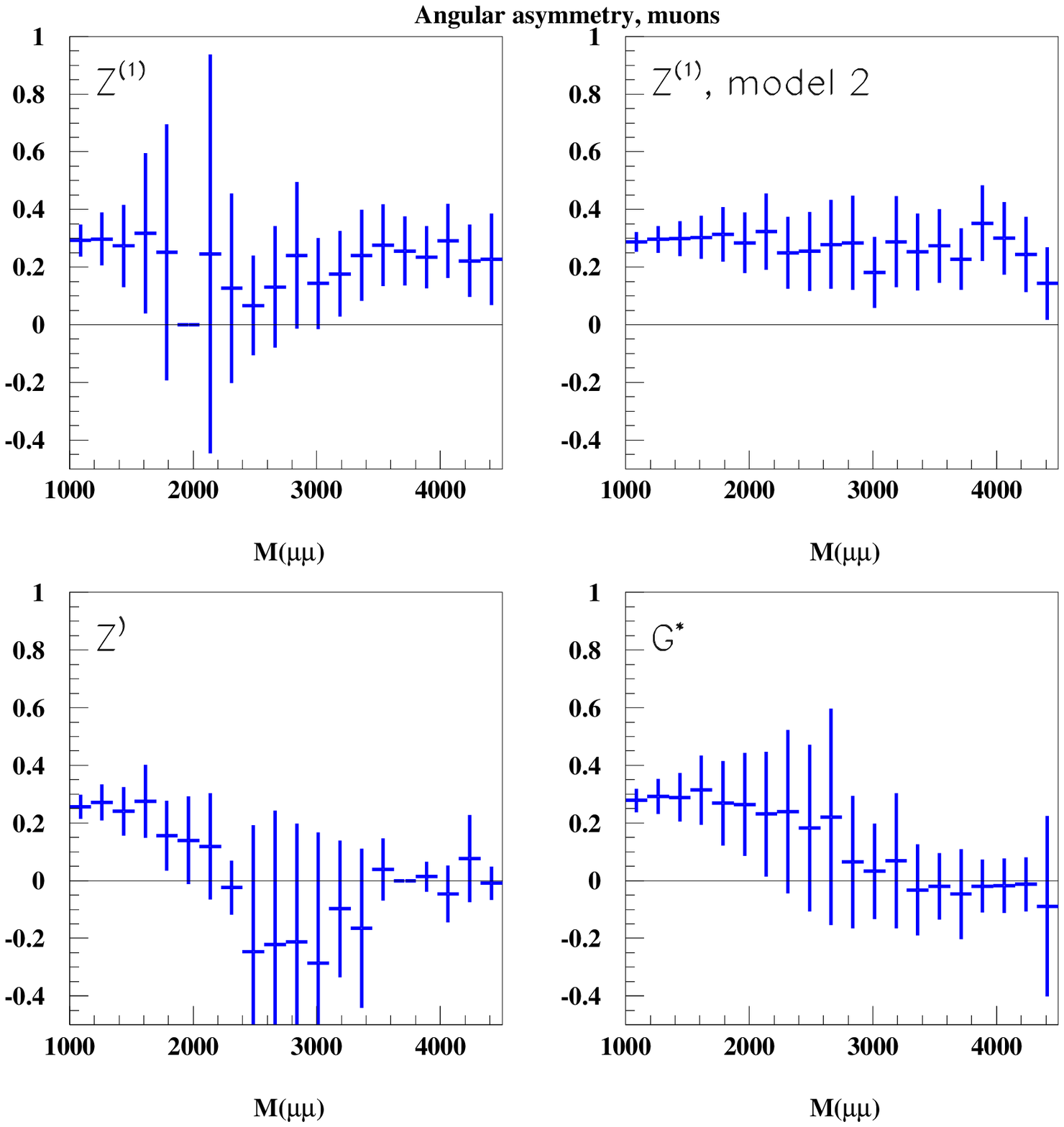,width=14cm}}
 \caption{ \em Muon channel: measured forward-backward asymmetry at LHC, 
for different types of resonances, centered at 4 TeV. The error
bars are representative of a sample having 138 events in the peak region, 
or 100 fb$^{-1}$ for \gexc.}
 \label{fig:fbasy_mu}
 \end{center}
 \end{figure}

From the angular distributions, the forward-backward asymmetry is obtained and shown 
in Figs.~\ref{fig:fbasy_e} and~\ref{fig:fbasy_mu} as a function of the reconstructed
dilepton mass. It allows a clear distinction between a resonance
due to  \gexc\ and either a $Z'$ or a  $G^*$ resonance.
Indeed, the asymmetry is expected to be close to $0$ at the mass 
peak of the $Z'$, if the couplings are
those of the SM, because $\sin^2\theta_W \sim 1/4$:

\begin{eqnarray}
  A^0_{FB} &=& \frac{3}{4} A_q A_\ell \\
\mbox{ with}\\
  A_\ell &=& \frac{2 v_\ell a_\ell}{v_\ell^2+a_\ell^2} 
              = \frac{2 (1 -4|Q_\ell| \sin^2\theta_W)} { 1 + (1-4|Q_\ell|\sin^2\theta_W)^2} \sim 0 \\
\end{eqnarray}

For masses below, but close to  the resonance, the FB asymmetry 
can also serve as a distinguishing criterion between the $Z'$ and the \gexc.
For large masses, however, 
the discrimination power becomes quickly limited by statistics.

\section{Conclusions} 
We have performed a detailed study of the leptonic signatures 
for the production of the KK excitations of the $\gamma$ and $Z$ 
in models with TeV-scale extra-dimensions.\par
The production and decay of the excitations were 
fully simulated, including initial state QCD radiation,
and the resulting particles were
passed through a parametrized simulation of the ATLAS detector.

We found that with an integrated luminosity of 100~$fb^{-1}$ 
ATLAS will be able to detect a peak in 
the lepton-lepton invariant mass if 
the compactification scale ($M_c$) is below 5.8~TeV. 
Even in the absence of a peak,
a detailed study of the shape of the lepton-lepton invariant mass 
will allow to observe a deviation from the standard model 
due to the interference of the KK excitations with the SM 
bosons.
From a study based on a maximum likelihood estimation 
of the compactification mass, ATLAS 
will be able to exclude at 95\% CL a signal from the models 
considered in this work for $M_c<13.5$~TeV 
with an integrated luminosity of 300~$fb^{-1}$.\\
We have performed an evaluation of the influence of
experimental and theoretical uncertainties on this result.
A 1\% deviation from linearity in lepton momentum measurement
yields a 2\% reduction in sensitivity. 
The maximum effect observed from the consideration of various
sets of PDF's is a reduction of order 200~GeV on the 
achievable limit.

Once a peak is observed, an important question is the assessment
of the model which has produced the signal.
We show that for resonances of mass up to $\lsim$ 5 TeV,
and with an integrated luminosity of 300 $fb^{-1}$, 
the KK excitations can be distinguished from mass peaks 
produced by SM-like $Z'$ or graviton resonances
from the study of the polar angle distribution of the leptons 
in the peak region.
The forward-backward lepton asymmetry as a function of invariant
mass can provide a more general
distinguishing criterion among the different models.
For invariant masses around the peak, it 
will allow  to distinguish the KK excitations 
from alternative models yielding the same signature.

\setcounter{figure}{0}
\setcounter{table}{0}
\setcounter{section}{0}
\setcounter{equation}{0}
\clearpage

\def\ptmiss{$P_{T}^{miss}$}
\def\Etmiss{$E_{T}^{miss}$}
\def\pT{$p_{T}$}
\def\GeV{{\rm GeV}}
\def\TeV{{\rm TeV}}
\def \sm {Standard Model }
\def \gsim{\mathrel{\mathpalette\@versim>}}
\def \lsim{\mathrel{\mathpalette\@versim<}}
\def \@versim#1#2{\lower0.4ex\vbox{\baselineskip\z@skip\lineskip\z@skip
     \lineskiplimit\z@\ialign{$\m@th#1\hfil##\hfil$%
     \crcr#2\crcr\sim\crcr}}}

\part{{\bf  Search for the Randall Sundrum Radion Using the ATLAS Detector
} \\[0.5cm]\hspace*{0.8cm}
{\it G. Azuelos, D. Cavalli, H. Przysiezniak, L. Vacavant
}}
\label{azuelossec_new}


\begin{abstract}
The possibility of observing the radion ($\phi$)
using the ATLAS detector at the LHC is investigated.
This scalar, postulated by 
Goldberger and Wise to stabilize brane fluctuations in the Randall-Sundrum model 
of extra dimensions, has Higgs-like couplings. Studies on searches 
for the Standard Model Higgs with the ATLAS detector are re-interpreted to obtain
limits on radion decay to $\gamma\gamma$ and ZZ$^{(*)}$.
The observability of radion decays into Higgs pairs, which subsequently decay into
$\mathrm{\gamma\gamma} + \mathrm{b \bar b}$  or $\tau\tau + \mathrm{b \bar b}$ is then
estimated.
\end{abstract}

\section{Introduction}

Theories with extra dimensions have recently received considerable attention.
One of the most interesting incarnations was formulated 
by Randall and Sundrum (RS)~\cite{Randall:1999ee}, 
who postulate a universe with two 4-d surfaces 
({\it branes}) bounding a slice of 5-d 
spacetime. The SM fields are assumed to be located on one of the branes (the TeV brane),
while gravity lives everywhere: on the TeV brane, on the Planck brane and in the bulk.
The metric is exponentially warped in the fifth dimension, 
allowing for a natural
resolution of the hierarchy problem.

The theory admits two types of massless excitations:
the usual 4-d graviton and a graviscalar.
In order to stabilize the size
of the extra dimension without fine tuning of parameters,
Goldberger and Wise (GW)~\cite{Goldberger:1999uk} have proposed
a mechanism which requires a massive bulk scalar $\phi$, the radion,
expected to be lighter than the J=2
Kaluza Klein excitations. The presence of the radion 
is one of the important phenomenological consequences of 
these theories of warped extra dimensions~\cite{Giudice:2000av,Bae:2000pk,%
Cheung:2000rw,Kribs:2001ic}. The study of this scalar
therefore constitutes a crucial probe of the model.

\subsection{Radion branching ratios and width}
\label{sec:br}

The radion couplings to fermions and bosons
are similar to those of the Standard Model (SM) Higgs~\cite{Giudice:2000av}.
They are expressed as a function of three parameters: 
the mass of the radion $m_\phi$,
the vacuum expectation value of the radion or scale, $\Lambda_\phi$, 
and $\xi$, the radion-SM Higgs mixing parameter~\cite{Giudice:2000av,Hewett:2002rz}. 

In the following study, 
we assume that the SM Higgs has been discovered and that its mass has been measured.
The branching ratios of the radion are calculated using
those of the SM Higgs as calculated in HDECAY~\cite{Djouadi:1998yw}, 
and using the ratio of the radion to Higgs branching ratios 
given by~\cite{Giudice:2000av}.

\begin{figure}[tbp]
\begin{center}
\includegraphics[height=18.cm,width=16.cm]{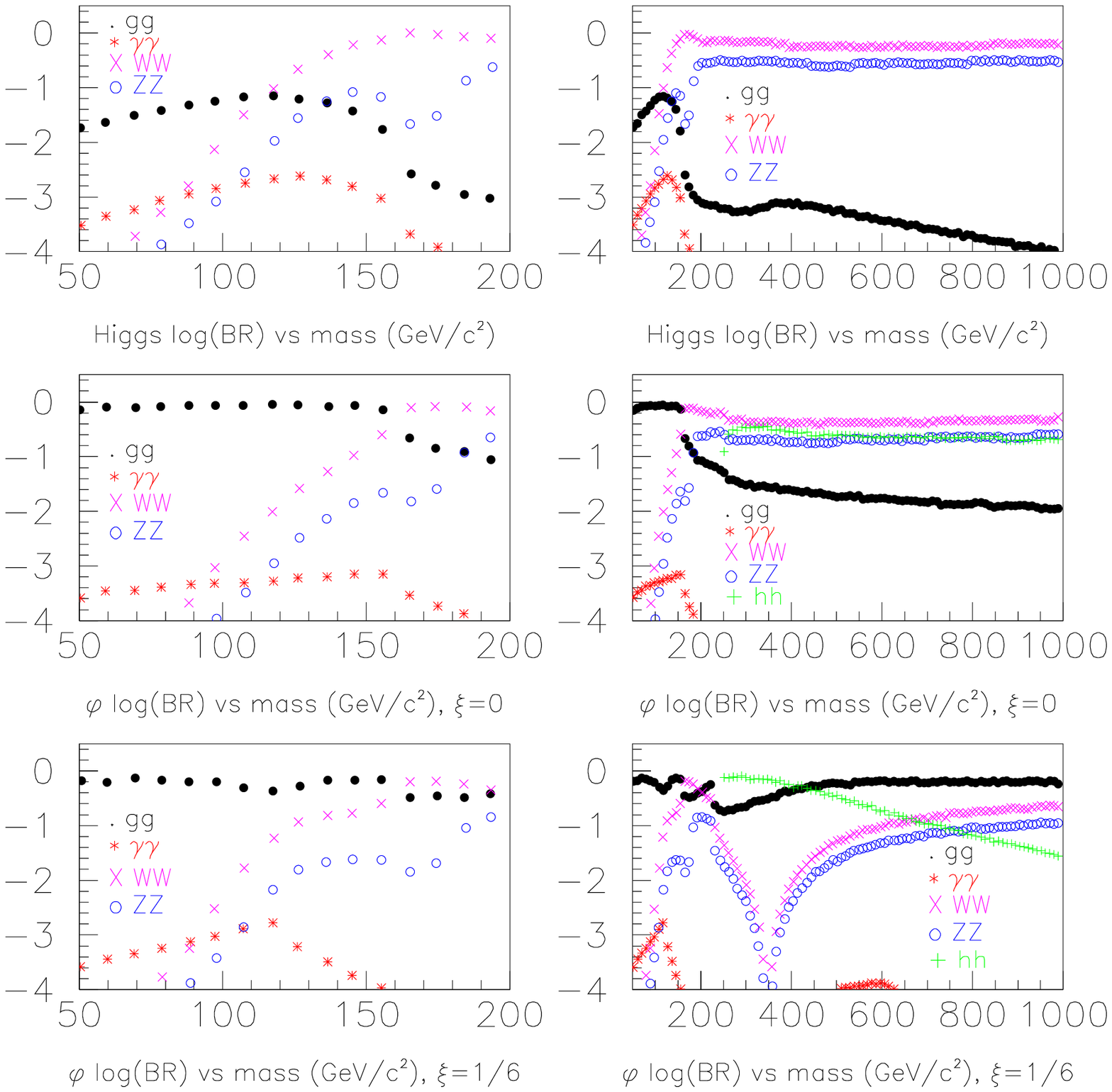}    
\caption {Log(BR) versus the mass of the scalar for the SM Higgs (top),
and for the radion when $\xi=0$ (middle) and $\xi=1/6$ (bottom)
when $\Lambda_\phi=1$ TeV. The Higgs mass in the lower curves is set to 
$m_{\mathrm h} = 125$GeV$/c^2$.
A smaller (larger) mass range is shown on the left(right)-hand side.}
\label{brxi0}
\end{center}
\end{figure}

Figure \ref{brxi0} 
shows the principal branching ratios as a function
of scalar mass for decays of the SM Higgs (top plots)
and of the radion when $m_{\mathrm h}=125$GeV$/c^2$ and $\Lambda_\phi=$1 TeV,
for $\xi=0$ when there is no $\phi$-h mixing (middle plots), 
and for $\xi=1/6$ when $\phi$ and h are heavily mixed (bottom plots).
We note the following:
\begin{itemize}
\item BR($\phi\rightarrow {\mathrm{gg}}$) is greatly enhanced with respect 
to the Higgs and is close to unity for $m_\phi>500$ GeV$/c^2$ and $\xi=1/6$
\item the radion decays into two SM Higgs for $m_\phi\geq 2m_{\mathrm h}$
\item BR($\phi\rightarrow \gamma\gamma$) is enhanced
for $\xi=1/6$ and $m_\phi\sim 600$ GeV$/c^2$.
\end{itemize}
For $\xi=1/6$ and for a radion with mass close to that of the Higgs, 
a strong interference produces a strong suppression of decays to vector bosons.

\begin{figure}[tb]
\begin{center}
\includegraphics[height=8.cm,width=8.cm]{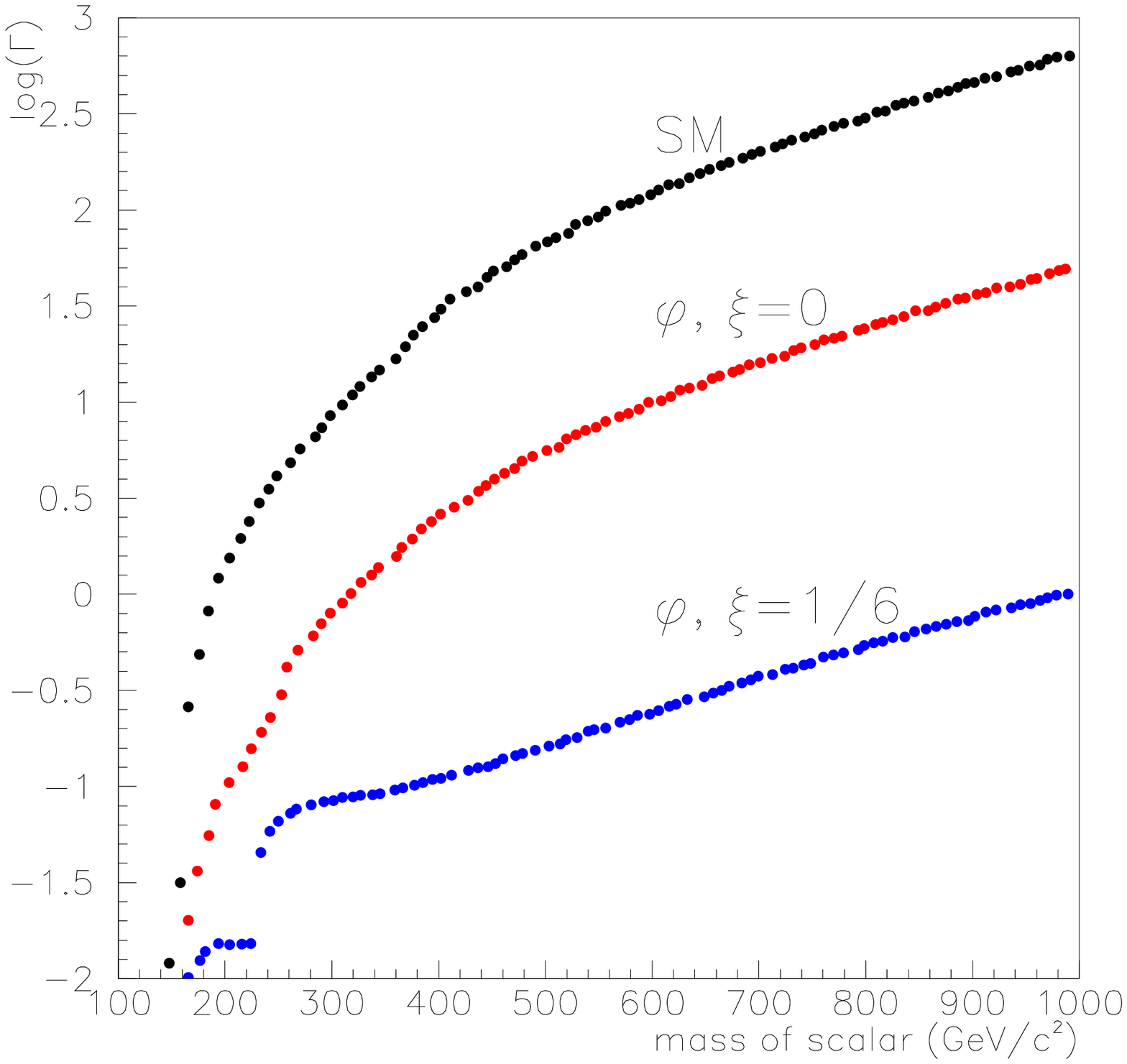}    
\caption {Log($\Gamma$) for the SM Higgs
and for the radion, for $\xi=$0 and 1/6 and for $\Lambda_\phi=1$ TeV.}
\label{width}
\end{center}
\end{figure}

The radion has a very narrow natural width.
Figure \ref{width} shows the total width as a function of mass, 
for the SM Higgs and for the radion with $\xi=0$ and 1/6, 
for $\Lambda_\phi=$1 TeV. The width is inversely
proportional to the square of $\Lambda_\phi$.

The aim of the present study is to 
investigate the possibility of observing a RS radion with the ATLAS detector 
through the following decays:
$\phi \rightarrow {\mathrm \gamma\gamma}$,
$\phi \rightarrow {\mathrm{ZZ}^{(*)}}\rightarrow 4\ell$,
$\phi \rightarrow {\mathrm{hh}}\rightarrow {\mathrm{b \bar b}}\gamma\gamma$
and
$\phi \rightarrow {\mathrm{hh}}\rightarrow {\mathrm{b \bar b}} \tau^+\tau^-$.
Only the direct production of the radion ${\mathrm{gg}}\rightarrow \phi$ is considered.

\section{$\phi\rightarrow\gamma\gamma$ and ${\mathrm{ZZ}^{(*)}}\rightarrow 4\ell$}

\begin{figure}[tb]
\begin{center}
\includegraphics[height=15.5cm,width=15.5cm]{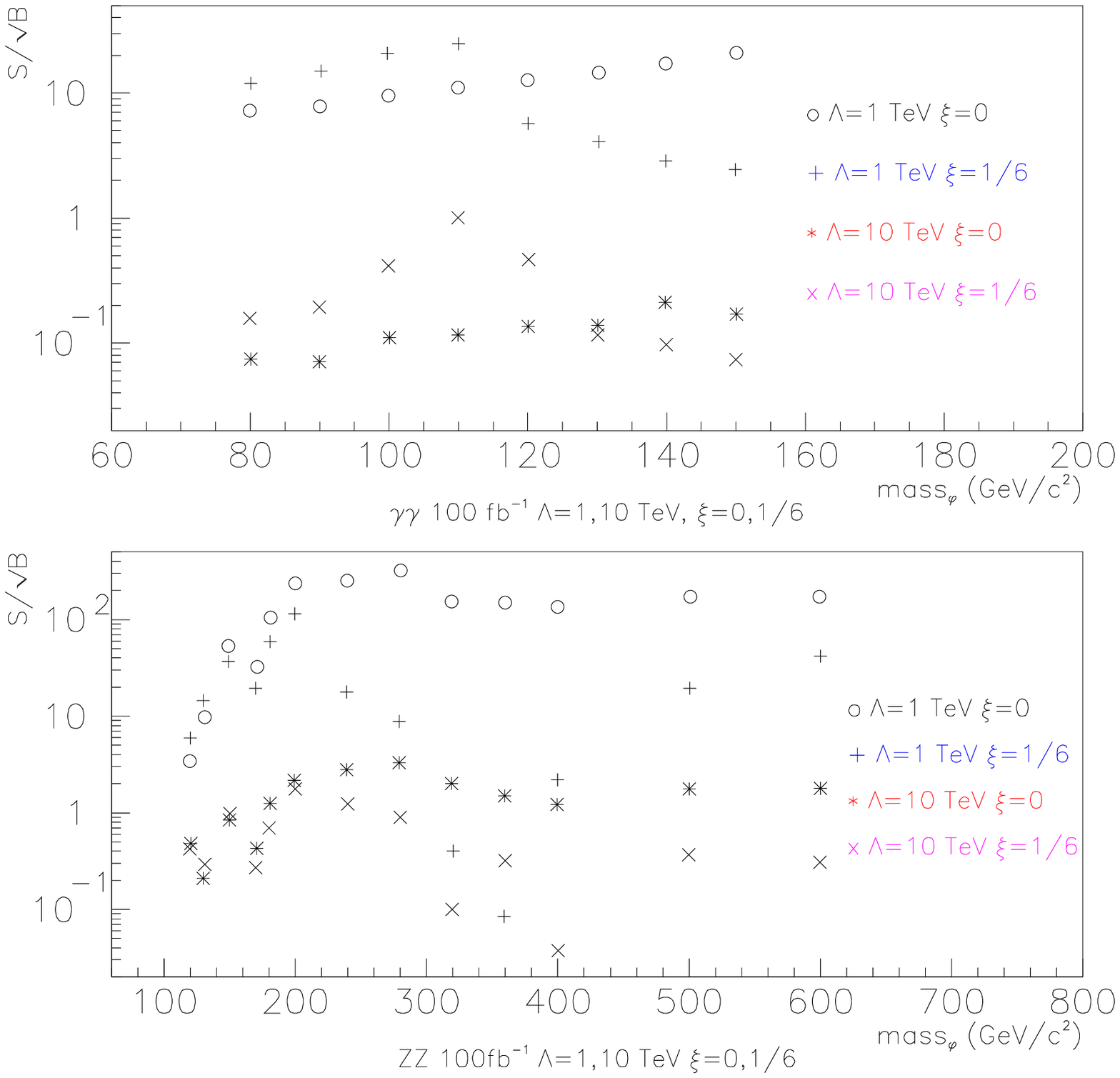}
\caption {Signal significance versus the mass of the radion,
for the $\gamma\gamma$ channel (top) and for the ZZ$^{(*)}$ channel (bottom).
In both plots, the values for $\Lambda_\phi=$1,10 TeV and 
$\xi=$0,1/6 are shown,
for an integrated luminosity of 100 $\mathrm{fb}^{-1}$.}
\label{sigbgd}
\end{center}
\end{figure}

For the $\gamma\gamma$ ($m_\phi<160\ {\mathrm{GeV}}/c^2$) and 
ZZ$^{(*)}$ ($m_\phi>100\ {\mathrm{GeV}}/c^2$) decay channels,
the radion signal significance is determined
from the SM Higgs results obtained in the ATLAS TDR~\cite{AtlasPhysTDR},
for 100 fb$^{-1 }$ (one year at high luminosity $\rm {10^{34}~cm^{-2} s^{-1}}$).
The ratio of the radion $\mathrm S/\sqrt{B}$ over that of the SM Higgs is given 
by~\cite{Giudice:2000av}:
$${\mathrm S/\sqrt{B}(\phi) \over S/\sqrt{B}(h)}=
{\Gamma(\phi\rightarrow {\mathrm{gg}}){\mathrm{BR}}(\phi\rightarrow \gamma\gamma,{\mathrm{ZZ}})\over
\Gamma({\mathrm h}\rightarrow {\mathrm{gg}}){\mathrm BR}({\mathrm h}\rightarrow 
\gamma\gamma,{\mathrm{ZZ}})}
\sqrt{{\mathrm{max}}(\Gamma_{\mathrm{tot}}^{\mathrm h},\sigma_m)\over 
{\mathrm{max}}(\Gamma_{\mathrm{tot}}^\phi,\sigma_m)}$$
where the mass resolutions are given by 
$\sigma_m^{\gamma\gamma}=0.10\sqrt{m}+0.005m$
and
$\sigma_m^{\mathrm{ZZ}}=
\sqrt{(\Gamma/2.36)^2+(0.02 m)^2}$.
Using the ATLAS TDR SM Higgs signal significance results,
the radion signal significance is determined and shown versus the mass of the radion,
in Figure \ref{sigbgd},
for the $\gamma\gamma$ channel (top) and for the ZZ$^{(*)}$ channel (bottom), 
for $\Lambda_\phi=$1,10 TeV, $\xi$=0,1/6,
and for an integrated luminosity of 100 $\mathrm{fb}^{-1}$.

\section{$\mathrm{\phi\rightarrow hh \rightarrow\gamma\gamma b\bar{b}}$}


The radion, unlike the SM Higgs but similarly to the ones in the 
Minimal Supersymmetric Standard Model (MSSM),
decays into Higgs pairs with relatively high BR (see Figure \ref{brxi0}).
As shown in Figure~\ref{width},
the total width of the radion is a factor of 10 (100) smaller for $\xi=$0 (1/6) than
that of the Higgs, such that it is completely negligible with respect to the
reconstructed mass resolutions.

The specific decay channel 
$\mathrm{\phi\rightarrow hh \rightarrow\gamma\gamma b\bar{b}}$
offers an interesting signature, 
with two high-$p_T$ isolated photons
and two b-jets. 
The background rate is expected to be very low for the relevant mass region 
$m_{\mathrm h} > 115$ GeV/$c^2$ and $m_\phi > 2m_{\mathrm h}$. 
In addition, 
triggering on such events is easy and the diphoton mass provides very good kinematical
constraints for the reconstruction of $m_\phi$.
 
The decay $\mathrm{ hh \rightarrow\gamma\gamma b\bar{b}}$ 
was studied in the context of the MSSM Higgs~\cite{mssmh}, 
although at that time the mass ranges investigated were lower. 
The approach and the selection we use in this study are very similar.

\subsection{Signal}

Signal events were generated with PYTHIA 6.158~\cite{PYTHIA}.
The modified version of HDECAY described in Section~\ref{sec:br}
is used  
at the initialization phase to input the correct parameters and at the end of 
the run to re-scale the cross-section. 
Note that the default PYTHIA parameters,
as opposed to the HDECAY ones, 
are used for the light Higgs couplings and parameters.

The heavy Higgs $H^0$ production process via gluon-gluon fusion
(in the framework of the Minimal 1-Higgs doublet Standard Model,
process {\tt ISUB=152})
is used to produce the radion.
The mass of the $H^0$ was set to reflect that of
the radion, and the light Higgs mass was set
to $m_{h}=125$ GeV$/c^2$.
In addition, since the width of the radion is much narrower than what is
usually
encountered in a Higgs scenario, a specific correction was
implemented~\cite{ts} and the branching ratio corrected appropriately.
Two samples of 100k events each were generated, 
for $m_{\phi}=300$ GeV/$c^2$ and for $m_{\phi}=600$ GeV/$c^2$.

\subsection{Background}

The backgrounds for this channel are $\mathrm{\gamma\gamma b\bar{b}}$ (irreducible), 
$\mathrm{\gamma\gamma c\bar{c}}$, $\mathrm{\gamma\gamma bj}$, $\mathrm{\gamma\gamma cj}$ 
and $\mathrm{\gamma\gamma jj}$ 
(reducible with b-tagging).
The events were generated with PYTHIA 6.158. The main production process is the box 
diagram ${\mathrm{gg}}\rightarrow\gamma\gamma$ (process {\tt ISUB=114}), where the jets
arise from initial state radiation, eventually combined with gluon splitting for heavy 
flavor jets. 
The rates are therefore very low. 
However large uncertainties apply to these backgrounds since the jet part comes only 
from radiation and not from the hard-scattering.
Generating a background sample of a sensible size turns out to be very CPU time 
consuming, and some cuts had to be applied at the event generation: 
the sample was generated in different bins of 
$\hat{p}_\perp$ (50, 100, 200, 400, 800, 1600 and 3200 GeV/$c$). 
For each bin, ten million events were generated.

Single photon production in the hard process $\gamma j$
where either the photon or jet is misidentified
represents another reducible background.
This background was studied in the context of the SM 
$\mathrm H \rightarrow \gamma\gamma$ channel,
and was found to increase the total background by a factor of two.
In the context of the radion where the backgrounds are negligible,
this would not affect the final results.

\subsection{Fast-simulation}

The detector effects on the signal and 
background events are simulated with ATLFAST 2.53~\cite{atlfast}. 
While most procedures and parameters are the standard ATLFAST ones for 
low luminosity operation ($\rm {10^{33}~cm^{-2} s^{-1}}$), 
a few improvements are applied for this study:
\begin{itemize}
\item jets are recalibrated using a detailed parameterization
\item the photon reconstruction efficiency is assumed to be 80\%
\item a $p_T$-dependent b-tagging parameterization is used with an average 
efficiency of $\epsilon_b=60\%$ and a rejection of approximately 93\%
for light-quark jets and 7\% for c-jets.
\end{itemize}

\subsection{Selection}

To extract the signal, two isolated photons with 
$p_T>20$ GeV$/c$ and $|\eta|<2.5$, and 
two jets with $p_T>15$ GeV$/c$, $|\eta|<2.5$
are required. At least one of the jets has to be tagged as a b.

\begin{figure}[tb]
\includegraphics[height=8cm]{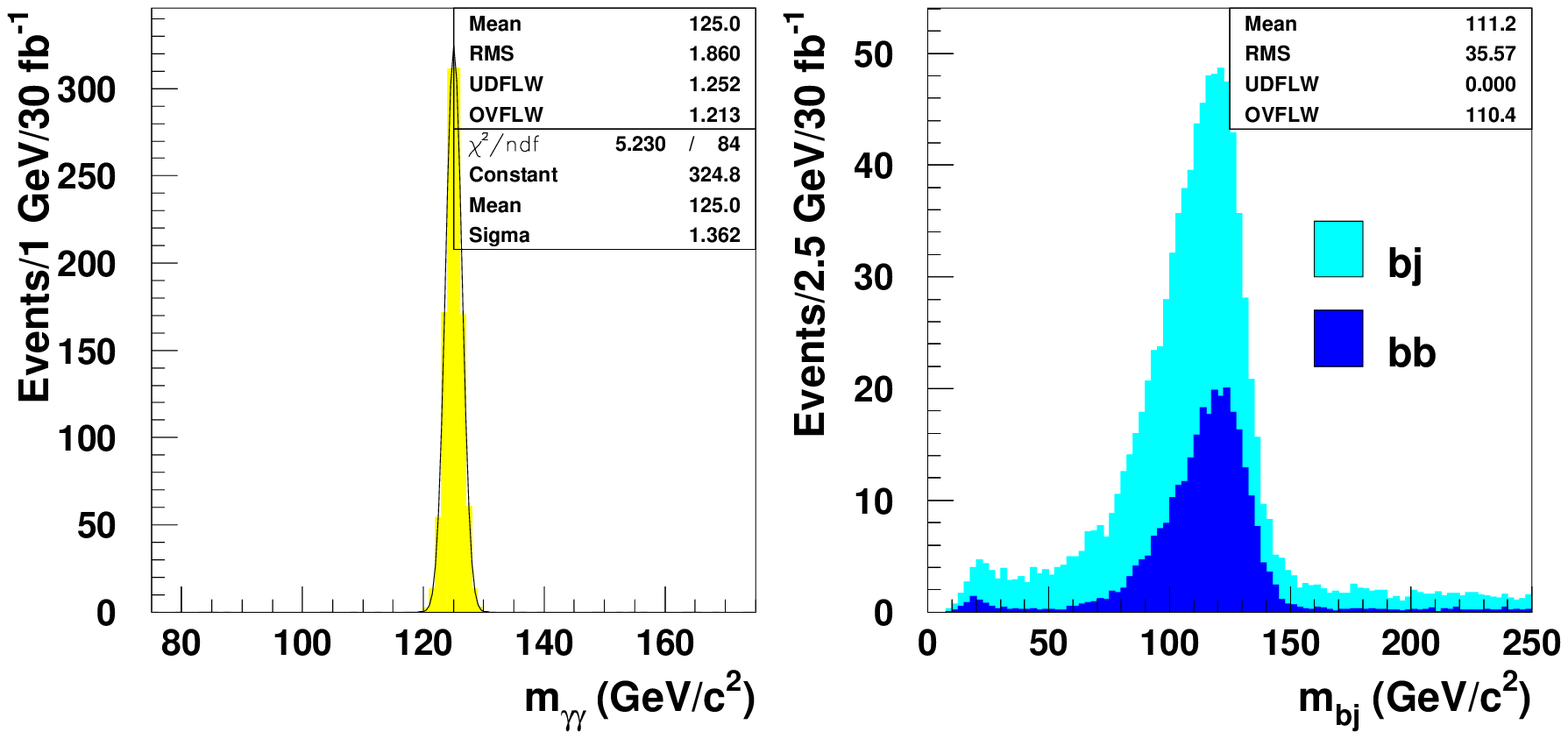}
\caption{Diphoton (left) and dijet (right) invariant mass distributions, 
for $m_{\phi}=300$ GeV/$c^2$,
$\xi=0$, $\Lambda_\phi=1$ TeV and 30 fb$^{-1}$ 
(three years at low luminosity $\rm {10^{33}~cm^{-2} s^{-1}}$). 
The right-hand plot shows the impact of requiring two b-tagged jets 
instead of one.}
\label{fig:mh}
\end{figure}

\begin{figure}[ht]
\includegraphics[height=8cm]{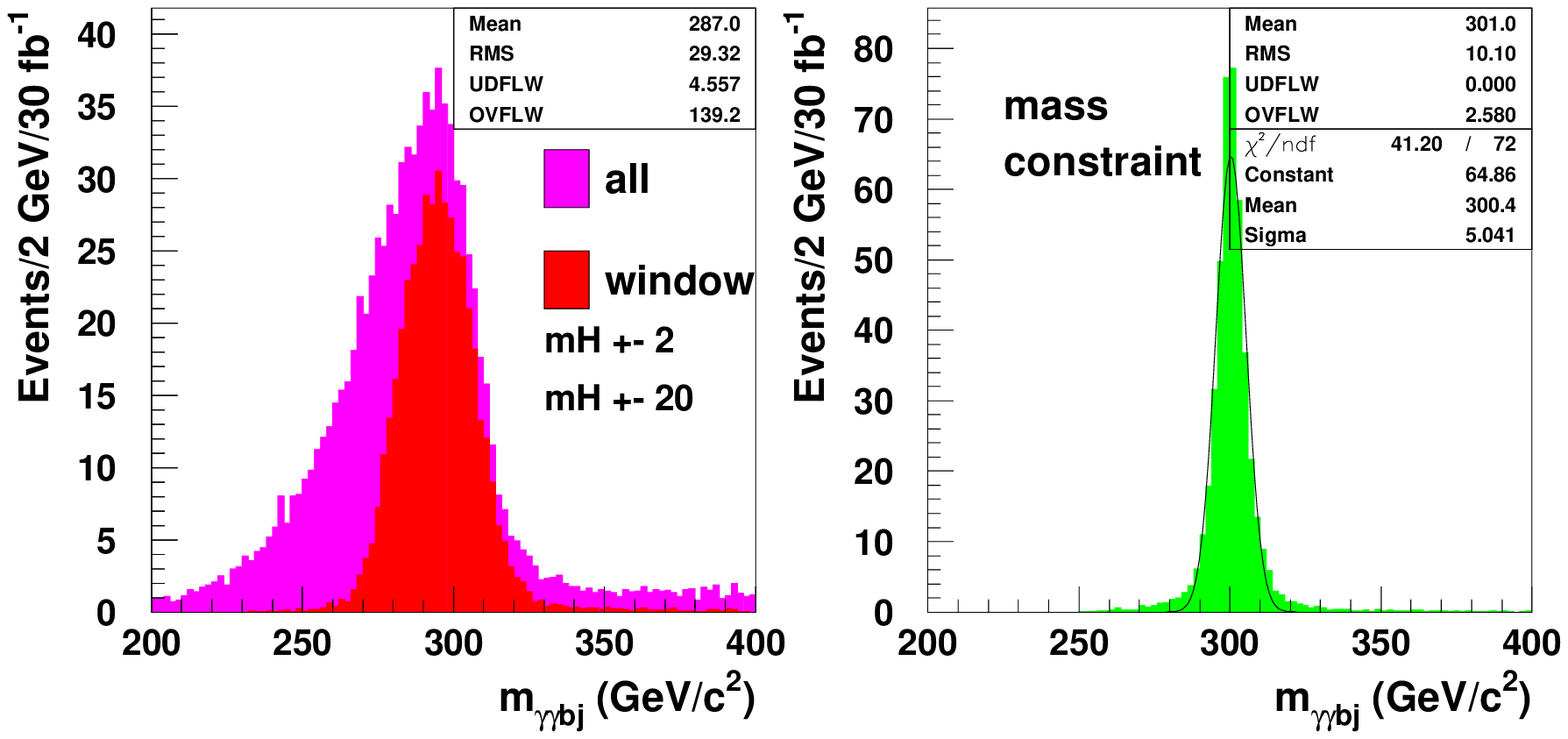}
\caption{Reconstructed $\mathrm{\gamma\gamma bj}$ invariant mass distribution, 
for $m_{\phi}=300$ GeV/$c^2$,
$\xi=0$, $\Lambda_\phi=1$ TeV and for three years at low luminosity.
The left-hand plot shows all the combinations and the ones fulfilling the mass window 
cuts (cf. text). 
The right-hand distribution is obtained by 
constraining the reconstructed masses $m_{\mathrm{bj}}$ and $m_{\gamma\gamma}$
to the light Higgs mass $m_{\mathrm h}$, after the mass window cuts.}
\label{fig:mphi}
\end{figure}

The diphoton and the dijet invariant masses are then formed.
Figure~\ref{fig:mh} shows the reconstructed invariant masses 
for $m_{\phi}=300$ GeV/$c^2$, $\xi=0$ and $\Lambda_\phi=1$ TeV.
Subsequently, two mass window cuts are applied by requiring that:
\begin{itemize}
\item $m_{\gamma\gamma} = m_{\mathrm h} \pm 2$ GeV$/c^2$
\item $m_{\mathrm{bj}} = m_{\mathrm h} \pm 20$ GeV$/c^2$.
\end{itemize}
The photons and jets fulfilling these requirements are combined to form the 
$m_{\mathrm{\gamma\gamma bj}}$ invariant mass as shown in Figure~\ref{fig:mphi}.
The mass resolution is improved down to a value of 5 GeV/$c^2$
by constraining the reconstructed masses $m_{\mathrm{bj}}$ and $m_{\gamma\gamma}$
to the light Higgs mass $m_{\mathrm h}$,
as shown on the right-hand plot of Figure~\ref{fig:mphi}. 
The signal acceptances after the various cuts described above are 
given in Table~\ref{tab:acc}.

\begin{table}[tb]
\begin{center}
\begin{tabular}{|c|c|c|}
\hline
Cuts & $m_\phi=300$ GeV/$c^2$ & $m_\phi=600$ GeV/$c^2$ \\
\hline
photons kinematics selection  & 46\% & 51\% \\
jets kinematics selection     & 36\% & 28\% \\
b-tagging                     & 76\% & 78\% \\
$m_{\gamma\gamma}$ window cut & 83\% & 85\% \\
$m_{\mathrm{bj}}$ window cut   & 49\% & 53\% \\
\hline
total                         &  5\% &  5\% \\
\hline
\end{tabular}
\caption{Acceptance for the signal, 
for $\xi=0$, $\Lambda_\phi=1$ TeV and for the two radion masses studied. 
For each cut the acceptance is defined with respect to the previous one.}
\label{tab:acc}
\end{center}
\end{table}

\begin{figure}[tb]
\includegraphics[height=8cm,width=\textwidth]{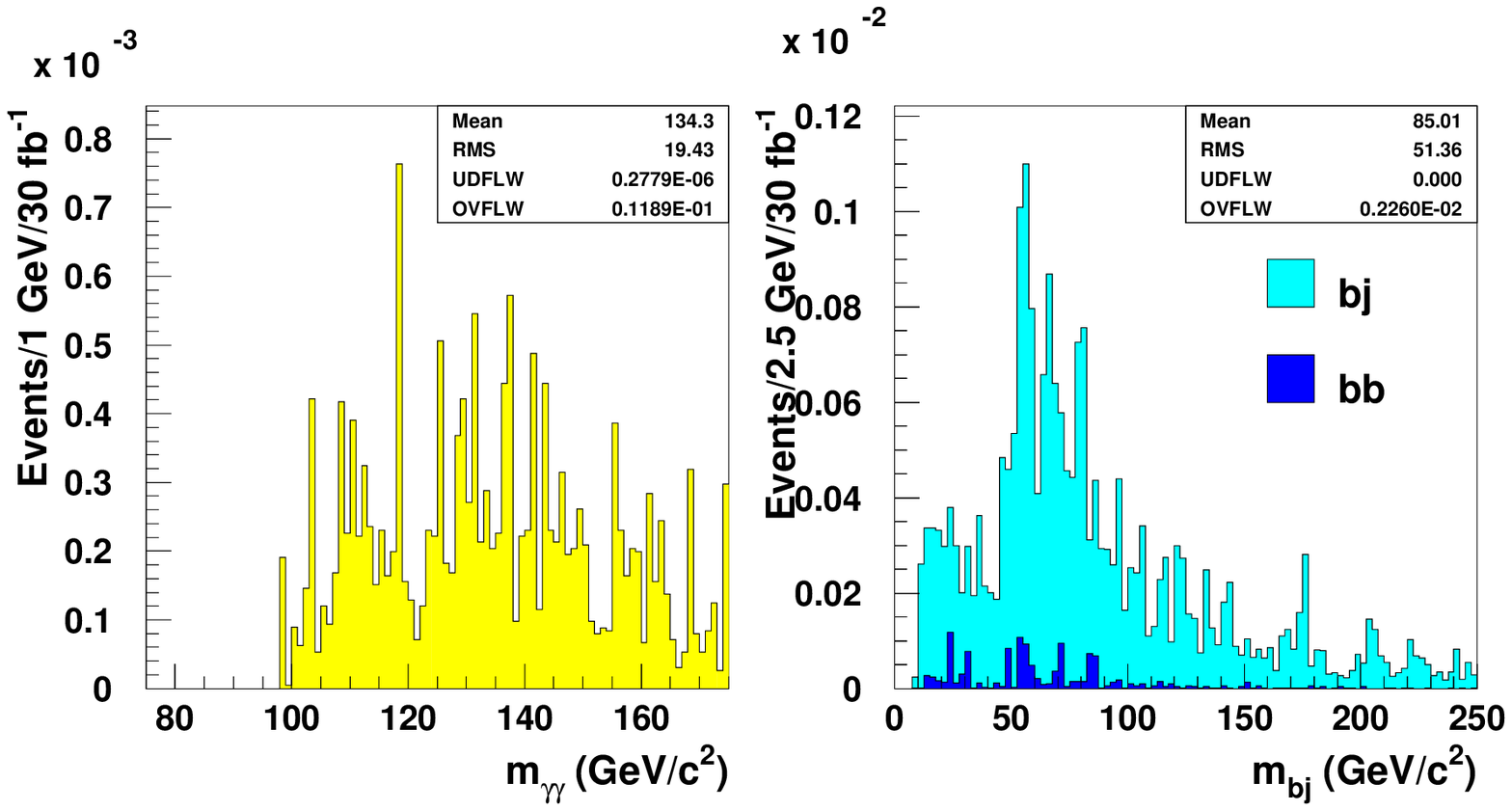}
\caption{Diphoton (left) and dijet (right) invariant mass distributions
for the background sample, for three years at low luminosity.
The right-hand plot shows the impact of requiring two b-tagged jets instead of one.}
\label{fig:mhb1}
\end{figure}

The same analysis procedure is applied to the background sample. 
The resulting $m_{\gamma\gamma}$ and $m_{\mathrm{bj}}$ distributions are shown in 
Figure~\ref{fig:mhb1}. 

\begin{figure}[tb]
\begin{center}
\includegraphics[height=15.cm,width=15.cm]{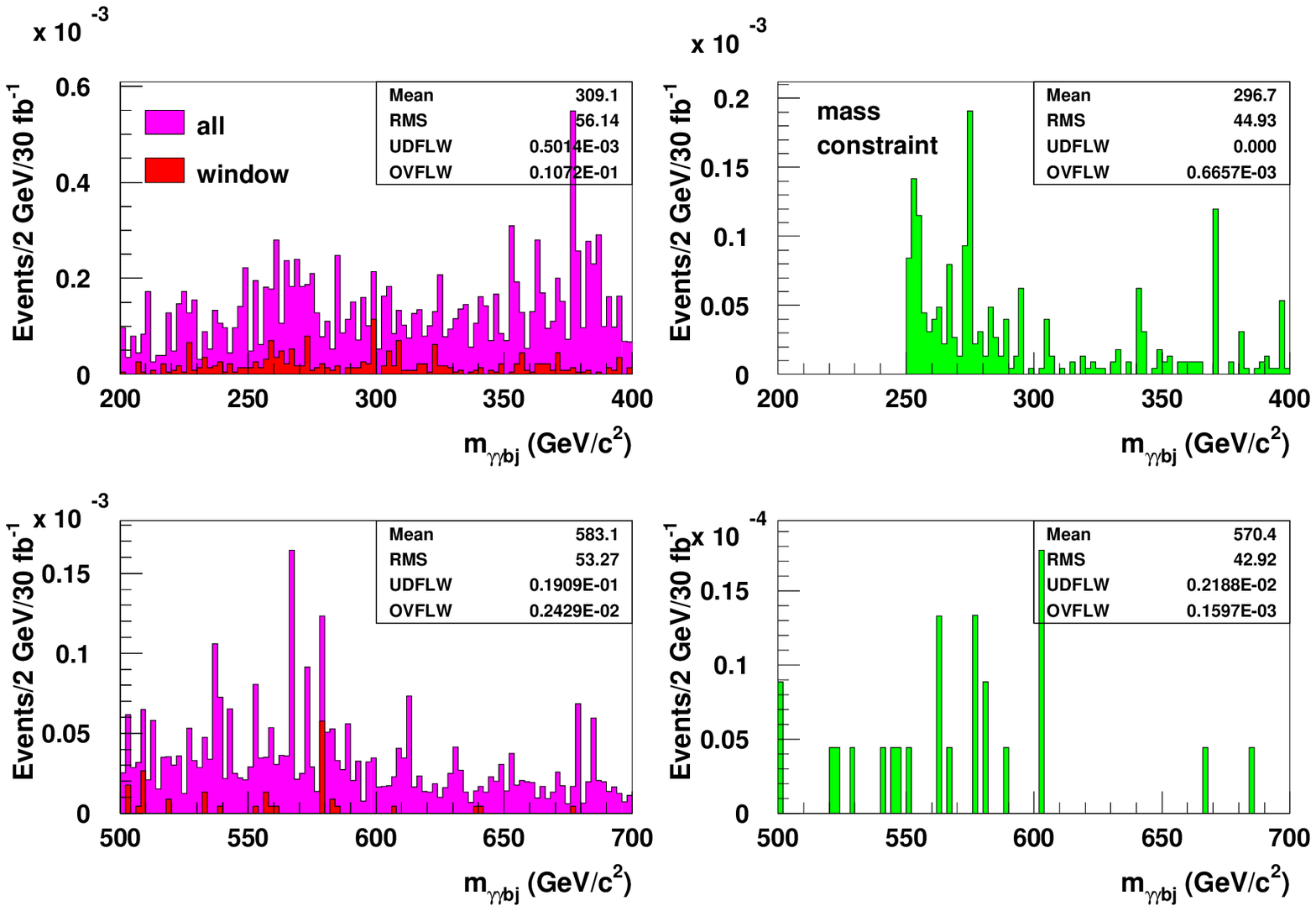}
\caption{Reconstructed $\mathrm{\gamma\gamma bj}$ invariant mass distribution for the 
background around the expected peak for the signal 
(top: $m_\phi=300$ GeV/$c^2$; bottom: $m_\phi=600$ GeV/$c^2$), 
for three years at low luminosity
and for the conservative approach (cf. text).
The left-hand plot shows all the combinations and the ones fulfilling the mass window cuts 
(cf. text). 
The right-hand distribution is obtained 
by constraining the reconstructed masses $m_{\mathrm{bj}}$ and $m_{\gamma\gamma}$
to the light Higgs mass $m_{\mathrm h}$,
after the mass window cuts.}
\label{fig:mphib300600b}
\end{center}
\end{figure}

Since there are some uncertainties concerning the level of the background, 
a more conservative approach is also tried: the mass window cuts are loosened 
to keep events fulfilling:
\begin{itemize}
\item $m_{\gamma\gamma} = m_{\mathrm h} \pm 30$ GeV/$c^2$
\item $m_{\mathrm{bj}} = m_{\mathrm h} \pm 40$ GeV/$c^2$
\end{itemize}
The $m_{\mathrm{\gamma\gamma bj}}$ invariant mass distributions for this conservative 
approach are shown in Figure~\ref{fig:mphib300600b}.

\subsection{Results}

The final number of events selected is obtained by counting the candidates
after all cuts within a mass window of
$\ <m_{\mathrm{\gamma\gamma bj}}>\pm 1.5\sigma_{m_{\mathrm{\gamma\gamma bj}}}\ $
for signal and background.
The results are shown in Table~\ref{tab:results}.
\begin{table}
\begin{center}
\begin{tabular}{||c|c||c|c||}
\hline
\multicolumn{2}{||c||}{$m_{\phi}=300$ \mbox{ GeV/$c^2$}} &
\multicolumn{2}{  c||}{$m_{\phi}=600$ \mbox{ GeV$/c^2$}} \\
\hline
\mbox{ background } & 0 &
\mbox{ background } & 0 \\
\mbox{ background (conserv.) } & $1.42\;10^{-4}$ &
\mbox{ background (conserv.) } & 0 \\
\hline
$\xi=0,   \Lambda_\phi=1$  \mbox{ TeV} & 380.1 &
$\xi=0,   \Lambda_\phi=1$  \mbox{ TeV} & 575.8 \\
\hline
$\xi=0,   \Lambda_\phi=10$ \mbox{ TeV} & 3.8 &
$\xi=0,   \Lambda_\phi=10$ \mbox{ TeV} & 5.9 \\
\hline
$\xi=1/6, \Lambda_\phi=1$  \mbox{ TeV} & 680.4 &
$\xi=1/6, \Lambda_\phi=1$  \mbox{ TeV} & 439.9 \\
\hline
$\xi=1/6, \Lambda_\phi=10$ \mbox{ TeV} & 5.5 &
$\xi=1/6, \Lambda_\phi=10$ \mbox{ TeV} & 5.9 \\
\hline
\end{tabular}
\caption{Number of events selected for background and for signal, 
for $m_\phi=$300 and 600 GeV$/c^2$,
for three years at low luminosity and for $m_{\mathrm h}=125$ GeV/$c^2$.}
\label{tab:results}
\end{center}
\end{table}
Since this channel is practically background free,
the usual significance $\mathrm S/\sqrt{B}$ is not relevant. 
A signal discovery, defined to be a minimum of ten events,
is straightforward for low values of $\Lambda_\phi$
early in the physics program of the LHC. 
This is shown in Table~\ref{tab:lmin} where the minimum integrated luminosities needed 
for discovery are shown. 
Approximately 1 $\mathrm{fb}^{-1}$ is needed
for $\Lambda_\phi\sim1$ TeV while requiring $\mathrm S>10$ and $\mathrm S/\sqrt{B}>5$.

In the special case where $\xi=0$, the cross-section is proportional to $\Lambda_\phi^{-2}$. 
Therefore the maximum reach in $\Lambda_\phi$ is derived from this study. 
This is obtained using the prescription of
~\cite{Feldman:1998qc}: 
for a known mean background of zero, 
the signal mean is larger than 10 with 95\% CL
if the number of observed events is larger than 18.
The corresponding reach in $\Lambda_\phi$ is 4.6 TeV for $m_\phi=300$ GeV$/c^2$ and 5.7 
TeV for $m_\phi=600$ GeV$/c^2$.

\begin{table}
\begin{center}
\[\begin{array}{|c|c|c|}
\hline
& m_{\phi}=300 \mbox{ GeV$/c^2$} & m_{\phi}=600 \mbox{ GeV$/c^2$} \\
\hline
\xi=0,   \Lambda_\phi=1  \mbox{ TeV} & 0.8 & 0.5 \\
\xi=0,   \Lambda_\phi=10 \mbox{ TeV} & 80  & 50  \\
\xi=1/6, \Lambda_\phi=1  \mbox{ TeV} & 0.4 & 0.7 \\
\xi=1/6, \Lambda_\phi=10 \mbox{ TeV} & 55  & 55  \\
\hline
\end{array}\]
\caption{Minimum integrated luminosity (fb$^{-1}$) needed for discovery.}
\label{tab:lmin}
\end{center}
\end{table}

\section{$\phi\rightarrow {\mathrm{hh}}\rightarrow {\mathrm{b \bar b}} \tau^+\tau^-$}

The channel 
$\phi\rightarrow {\mathrm{hh}}\rightarrow {\mathrm{b \bar b}} \tau^+\tau^-$
provides another potentially interesting signal for radion discovery,
although the background is higher and the reconstructed mass resolutions are poorer
than in the $\mathrm{\phi\rightarrow hh \rightarrow\gamma\gamma b\bar{b}}$ channel.

In order to provide a trigger, a leptonic decay of the $\tau$ is required.
Here, only the case when one $\tau$ decays leptonically 
and the other hadronically is considered.
As above, 
the events were generated by appropriately adapting the
process of MSSM decay of the heavy Higgs $\mathrm H^0$
into two light Higgs bosons (h) in Pythia 6.158~\cite{PYTHIA}.
The effect of the ATLAS detector on the resolution and
efficiency of reconstruction of these
events was simulated with the 
ATLAS fast simulation package (ATLFAST 2.53).
The efficiency for hadronic $\tau$ reconstruction is assumed to be 40\%.
For b-jet tagging, an efficiency of 60\% is assumed, with a rejection factor
of 10 for c jets and 100  for light jets\cite{AtlasPhysTDR}.
\subsection{Signals and backgrounds}

As in the previous section, the radion mass values investigated are 300 
and 600 GeV$/c^2$, while the Higgs mass is set to 125 GeV$/c^2$.

The fast simulated samples are:
\begin{itemize}
\item ${\mathrm{hh}}\rightarrow {\mathrm{b \bar b}}\ \ \tau^+\tau^-$
with one $\tau$ decaying leptonically and the other hadronically (10 000 events)
\item $\mathrm{t \overline{t} \rightarrow b W^+ \ \ \overline{b} W^- }$
with one W decaying leptonically and the other hadronically ($10^6$ events)
\item $\rm{Z \rightarrow \tau^+ \tau^-}$
with one $\tau$ decaying leptonically and the other hadronically ($10^6$ events).
Initial and final state radiation provide additional jets which can fake the signal.
\item inclusive W bosons decaying leptonically (2$\times 10^6$ events).

\end{itemize}

\subsection{The selection}

\begin{figure}
  \begin{minipage}[t]{0.48\textwidth} 
        \includegraphics[width=7cm]{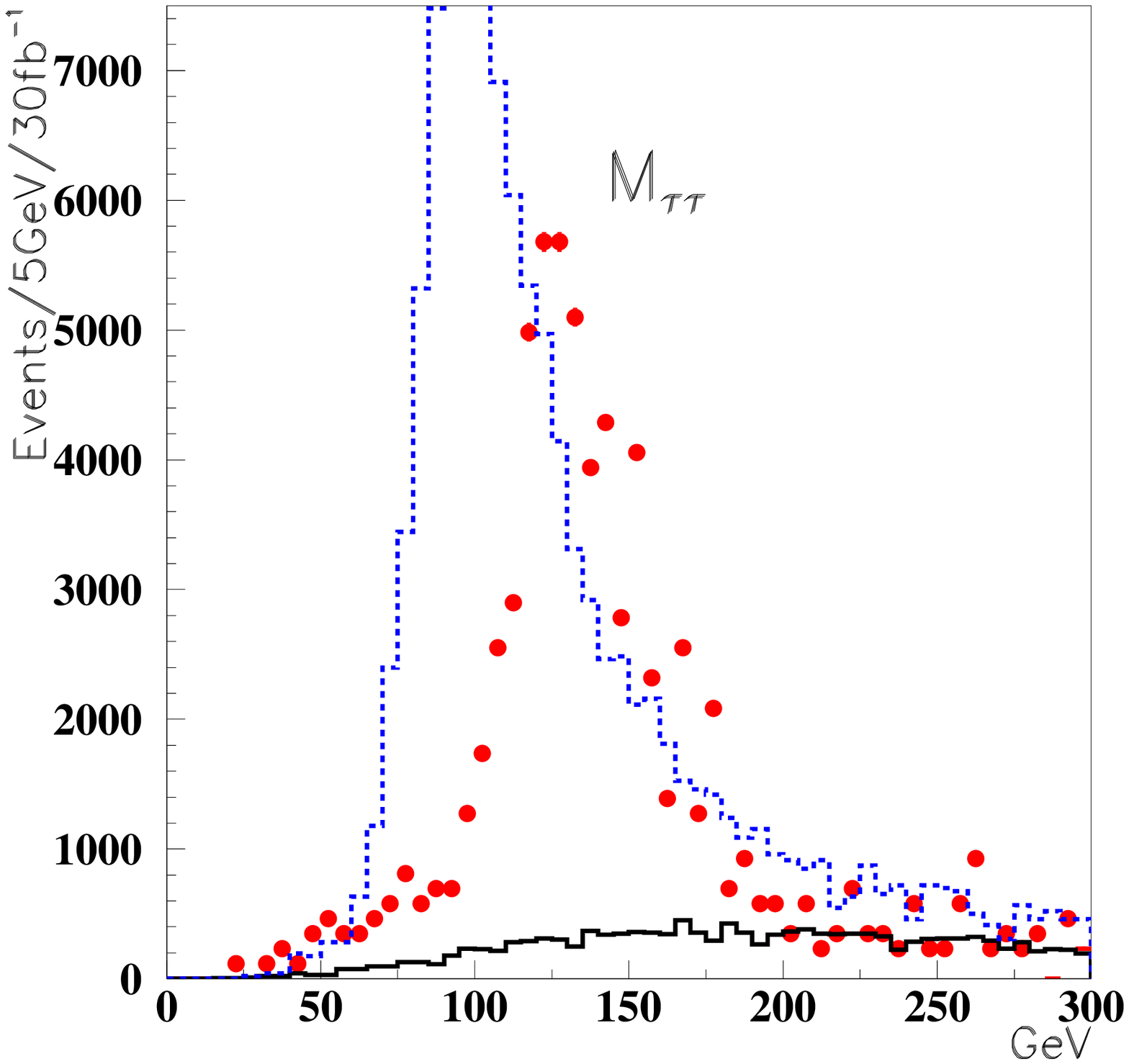}
        \caption{ Reconstructed $\tau\tau$ invariant mass for 
                  the signal (dots, arbitrary normalization)
                  and for the background:
                  $\mathrm{t\bar t}$ (full line, 30 fb$^{-1}$) and 
                  $\mathrm{Z\to \tau \tau}$ (dashed line, 30 fb$^{-1}$).} 
        \label{fig:tautaumass}
 \end{minipage} \hfill
  \begin{minipage}[t]{0.48\textwidth} 
        \includegraphics[width=7cm]{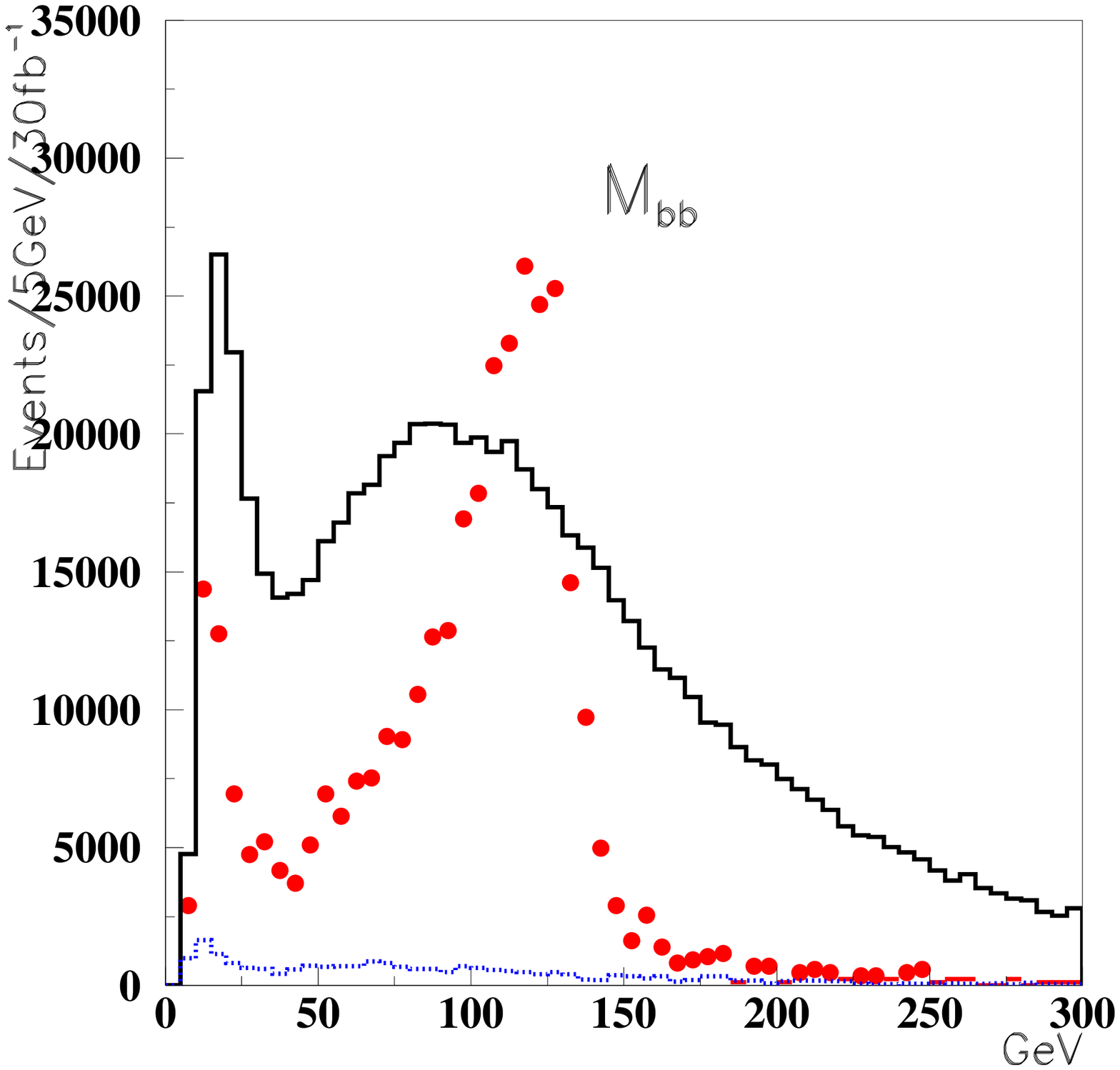}
        \caption{ Reconstructed $\mathrm{b\bar b}$ invariant mass for 
                  the signal (dots, arbitrary normalization)
                  and for the background:
                  $\mathrm{t\bar t}$ (full line, 30 fb$^{-1}$) and 
                  $\mathrm{Z\to \tau \tau}$ (dashed line, 30 fb$^{-1}$).}
        \label{fig:bbmass}
  \end{minipage}
\end{figure}

The study is performed assuming conditions of low luminosity 
($\rm {10^{33}~cm^{-2} s^{-1}}$) since, at
high luminosity
 ($\rm {10^{34}~cm^{-2} s^{-1}}$), the
reconstructed $\tau \tau$ mass resolution is seriously
compromised by pile-up effects.
\cite{AtlasPhysTDR}. 
The events are selected if they satisfy the following criteria:
\begin{itemize}
\item
 A lepton is required with $p_T^\ell>25$ GeV, $|\eta^\ell|<2.5$ 
(this lepton serves as a trigger).
\item
 The transverse mass $p_T^\ell-p_T^{\mathrm miss}$ is required to be $<40$ GeV$/c$.
  This cut rejects background events containing W bosons.
\item
 The $\tau \tau$ invariant mass is determined by combining the lepton with 
 a tagged $\tau$-jet having  $p_T^{\mathrm{jet}}>15$ GeV/$c$, $|\eta^{\mathrm{jet}}| <2.5$
(see Figure~\ref{fig:tautaumass}).
 If more than one jet is tagged as a tau-jet, the combination with the mass  
nearest to $m_{\rm{h}}$ is chosen.
\item
 A pair of b-tagged jets with $p_T>15$ GeV/$c$ and $|\eta| <2.5$ is
required and their jet-jet mass reconstructed
 (see Figure~\ref{fig:bbmass}). If more than two jets
are tagged as b-jets, the pair having the invariant mass closest to the
Higgs mass is chosen.
\item
   Cuts on the reconstructed  $\tau \tau$ mass and $\mathrm{b\bar b}$ mass are applied:\\
 $110<m_{\tau \tau}<140$ GeV$/c^2$ and $90<m_{\mathrm{b\bar b}}<140$ GeV$/c^2$ in the
case of the 300 GeV$/c^2$ radion, and 
 $110<m_{\tau \tau}<150$ GeV$/c^2$ and $85<m_{\mathrm{b\bar b}}<130$ GeV$/c^2$ in the
case of the 600 GeV$/c^2$ radion.

\end{itemize}

\subsection{Results}

Although the signal efficiency is low, the background rejection is high.
The expected cross sections for signal and background before the event selection
are given in Table \ref{tab:sig2} 
for $\Lambda_\phi=10$ TeV and $\xi=0$.
The branching ratios account for leptonic decays into a muon or an electron.

\begin{table}[tb]
\begin{center}
\begin{tabular}{|c|c|c|} 
\hline
 Signal                                                      &  $m_{\phi}$ = 300 GeV$/c^2$       &     $m_{\phi}$= 600 GeV$/c^2$\\
\hline
  $\mathrm{\sigma (gg \rightarrow \phi)}$   &       290 fb                      &   60 fb\\
  BR($\rm {\phi} \rightarrow$ hh)                            &       0.30                        &            0.25\\
  BR(hh $\rightarrow$ $\tau \tau$ bb)                        & $2\times 0.06 \times 0.8$         &  $2\times 0.06 \times 0.8$ \\
  BR($\rm{\tau \tau \rightarrow \ell + hadrons}$)            & $2\times 2\times 0.17\times 0.65$ & $2\times 2\times 0.17\times 0.65$\\
\hline
    $\sigma\times$BR =                                       &     3.98 fb                       &           0.652 fb\\
\hline
$\rm{t \bar{t} \to WbWb \to \ell + hadrons}$                 & \multicolumn{2}{c|}{$\sim$180 pb}     \\
$\rm{W \rightarrow \ell \nu}$                                & \multicolumn{2}{c|}{$\sim$40000 pb}  \\
$\rm{Z \to \tau \tau} \to \ell + hadrons$                    & \multicolumn{2}{c|}{$\sim$ 730 pb} \\
\hline
\end{tabular}
\caption{Expected cross sections for $\Lambda_\phi=$10 TeV and $\xi=0$ for signal and background before the event selection cuts. ($\ell = e, \mu$) }
\label{tab:sig2}
\end{center}
\end{table}

\begin{figure}[tb]
  \begin{minipage}[t]{0.48\textwidth} 
        \includegraphics[width=7cm]{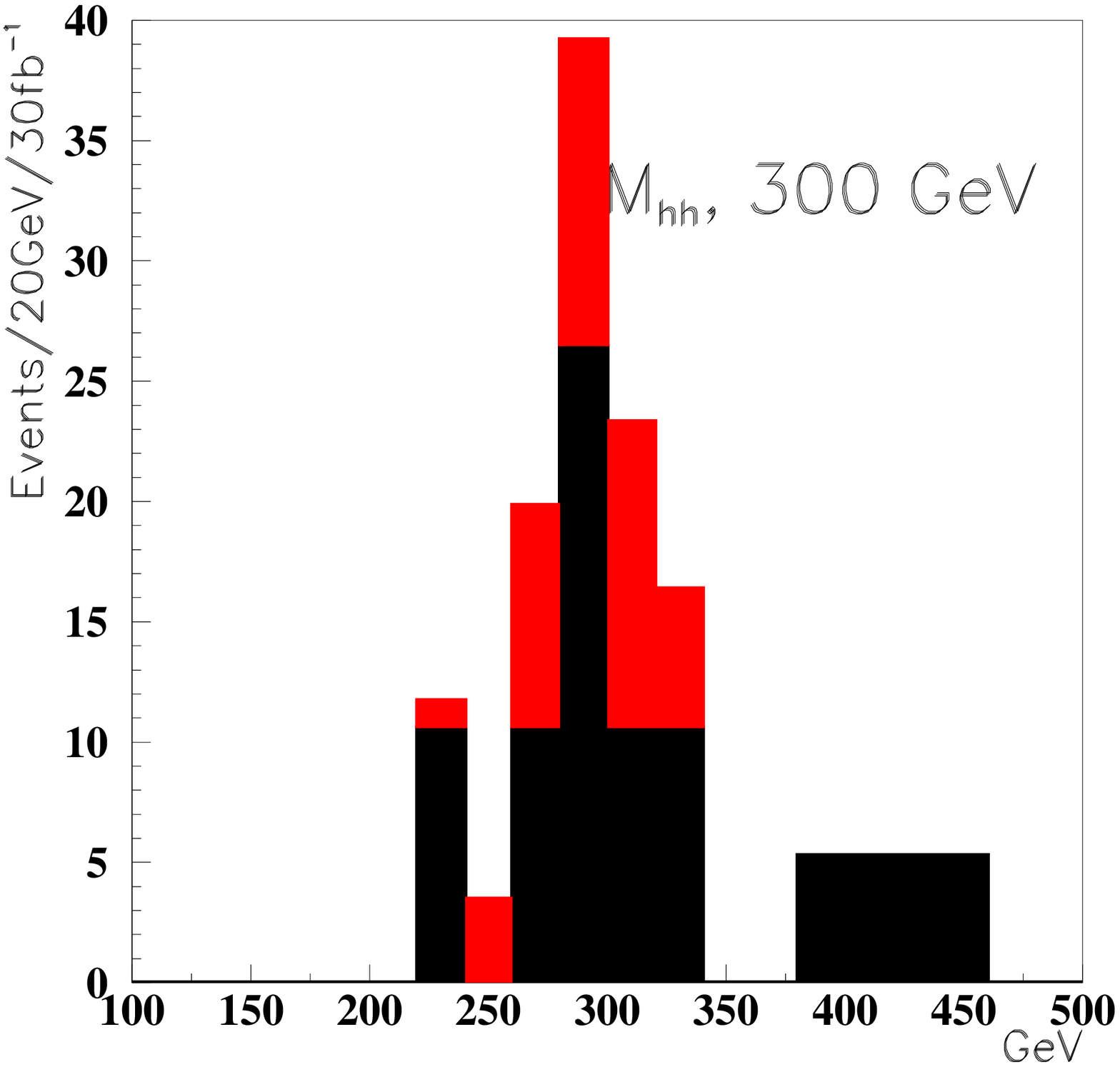}
        \caption{ Reconstructed mass of the radion ($m_\phi=$ 300 GeV$/c^2$) (dark: background and light: signal),
                  for 30 fb$^{-1}$, $\Lambda_\phi = 1 $ TeV and $\xi=0$.} 
        \label{fig:hhmass300}
 \end{minipage} \hfill
  \begin{minipage}[t]{0.48\textwidth} 
        \includegraphics[width=7cm]{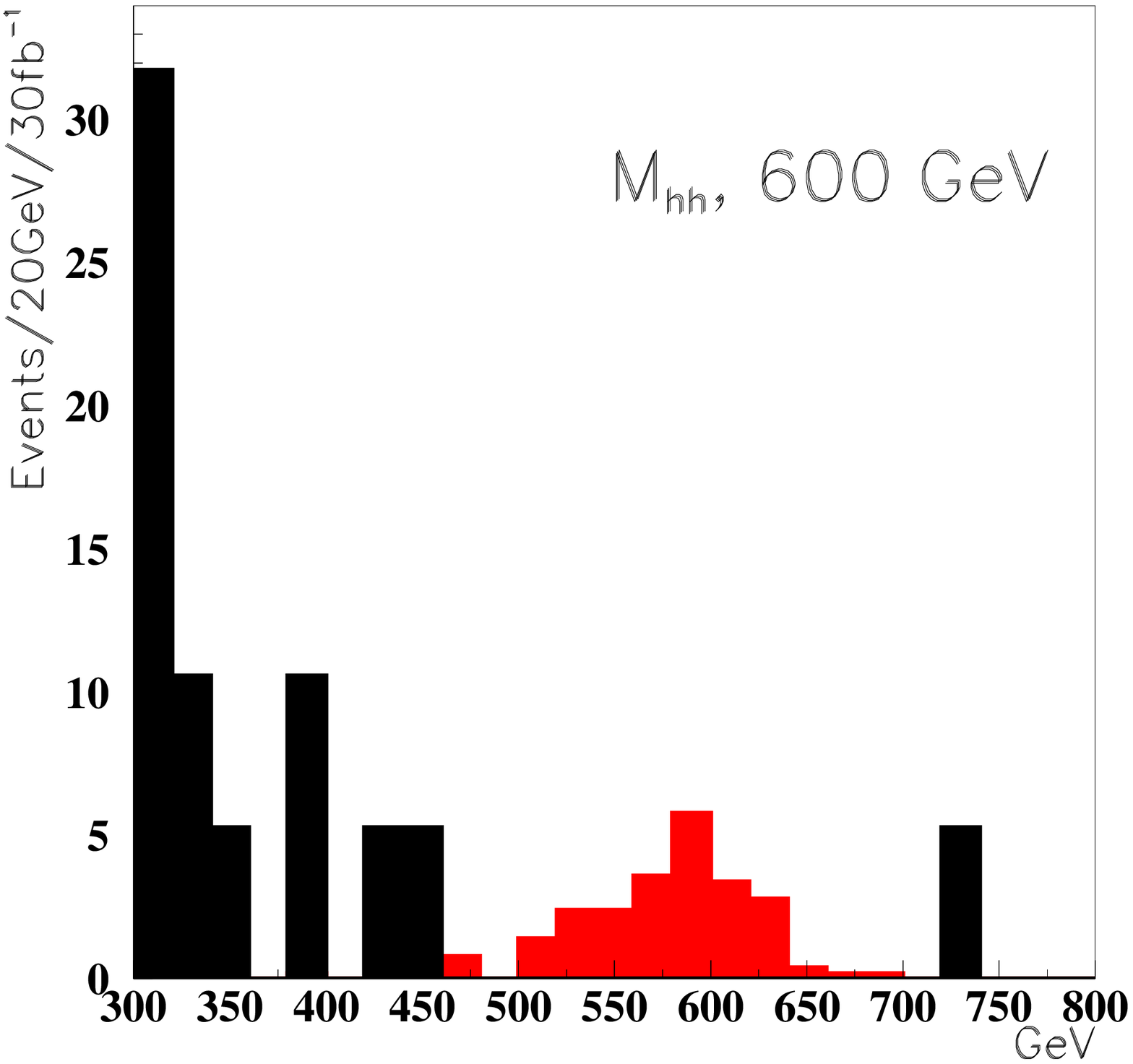}
        \caption{ Reconstructed mass of the radion ($m_\phi=$ 600 GeV$/c^2$) (dark: background and light: signal),
                  for 30 fb$^{-1}$, $\Lambda_\phi = 1 $ TeV and $\xi=0$.} 
        \label{fig:hhmass600}
  \end{minipage}
\end{figure}

Figures \ref{fig:hhmass300} and \ref{fig:hhmass600} show the reconstructed masses 
for signal when $m_{\phi}$=300 and 600 GeV$/c^2$ respectively, for 30 fb$^{-1}$,
$\Lambda_\phi = 1 $ GeV and $\xi=0$. The shape for a 300 GeV$/c^2$ radion is similar to 
the one for background (mostly $t\bar t$),
therefore systematic errors will most probably be dominated
by the understanding of the level of this background. 

The expected number of events for an integrated luminosity of 
30 $\rm {fb^{-1}}$ (three years at low luminosity) are
given in Table 5
for the two radion masses and for the 
backgrounds, when $\xi=0$ and $\Lambda_\phi =$ 1 TeV.
\begin{table}[tb]
\begin{center}
\begin{tabular}{|c|c|c|}  
\hline
                            & $m_{\phi}$=300 GeV$/c^2$        & $m_{\phi}$=600 GeV$/c^2$ \\
\hline
Signal                      & 43                              &   22 \\
\hline
 $\rm {t\bar t}$            & 58                              &  $\sim 6$  \\
 $\rm{Z \to \tau \tau}$     & $\sim 0$                        &  $\sim 0$ \\
 $\rm{W}$                   & $\sim 0$                        &   $\sim 0$  \\
\hline
$\mathrm S/\sqrt(B)$        &  5.6                            & 9.0 \\
\hline
\end{tabular}
\label{tab:sig1}
\caption{Expected number of events for signal and background,
for an integrated luminosity of 30 $\rm {fb^{-1}}$ for $m_\phi=$300 and 600 GeV$/c^2$,
$\xi=0$ and $\Lambda_\phi =$ 1 TeV, after all cuts.}
\end{center}
\end{table}

Requiring a minimum of 10 events and a $\mathrm  S/\sqrt{B} \ge  5$,
the maximum reach in $\Lambda_\phi$
is 1.05 and 1.4 TeV for $m_{\phi}$=300 GeV$/c^2$ and 
$m_{\phi}$=600 GeV$/c^2$ respectively, but the uncertainties in background subtraction
may affect considerably the observability of this channel in the first case.

\section{Conclusion}

We have studied the possibility of observing the radion using the ATLAS detector at the LHC.
The radion has couplings similar to those of the SM Higgs,
 and mixes with it, but it has also
a large effective coupling to gluons. 
Re-interpreting results of previous studies on the search for
a SM Higgs in ATLAS, a significance for observing a 
radion decaying into $\gamma\gamma$ or $\mathrm{ZZ^{(*)}}$
has been determined as a function of its mass (see ~Figure \ref{sigbgd}).
For an integrated luminosity of 100 $\mathrm{fb}^{-1}$,
the values $\mathrm S/\sqrt{B}\sim$10 (0.1) are obtained for
the $\gamma\gamma$ channel, with a mixing parameter $\xi$=0 and 
a scale $\Lambda_\phi$=1 (10)TeV, in the range 
$80~ {\mathrm{GeV}}/c^2 < m_\phi <160~ {\mathrm{GeV}}/c^2$.
For the $\mathrm{ZZ}^{(*)}$ channel,
$\mathrm S/\sqrt{B}\sim$100 (1) for 
$200~ {\mathrm{GeV}}/c^2 < m_\phi <600~ {\mathrm{GeV}}/c^2$
for the same conditions. Because the couplings are similar to those of the
SM Higgs, a good measurement of the branching ratios will be necessary to
discriminate between the two scalars. 

The radion can also decay into a pair of Higgs scalars, if the masses permit.
Two cases were examined: $\phi \to h h \to \gamma\gamma b \bar b$
and $\phi \to h h \to \tau\tau b \bar b$, for radion masses of 300 and 600 GeV$/c^2$,
for $m_h = 125$ GeV$/c^2$ and for an integrated luminosity of 30 fb$^{-1}$.
Limits on the maximal reach in $\Lambda_\phi$ were obtained for these two channels.
For the $\gamma\gamma b\bar b$ channel, the background is negligible and the 
reach in $\Lambda_\phi$ is
4.6 (5.7) TeV for $m_\phi=300$ (600) GeV$/c^2$, when $\xi$=0.
For the $\tau\tau b\bar b$ channel, the similarity between the signal and background shapes
make the observation of a radion of mass $m_{\phi}$=300 GeV$/c^2$
difficult, and the reach for
$\Lambda_\phi$ is about 1.4 TeV for $m_{\phi}$=600 GeV$/c^2$, when $\xi$=0.

\section{Acknowledgments}

We warmly thank the organizers of Les Houches 2001.
We thank T.Rizzo for useful discussions.


\setcounter{figure}{0}
\setcounter{table}{0}
\setcounter{section}{0}
\setcounter{equation}{0}
\clearpage

\def\ie{{\it i.e.}}
\def\eg{{\it e.g.}}
\def\etc{{\it etc}}
\def\etal{{\it et al.}}
\def\ibid{{\it ibid}.}
\def\mpl{\ifmmode \overline M_{Pl}\else $\overline M_{Pl}$\fi}
\def\to{\rightarrow}
\newskip\zatskip \zatskip=0pt plus0pt minus0pt
\def\matth{\mathsurround=0pt}
\def\lsim{\mathrel{\mathpalette\atversim<}}
\def\gsim{\mathrel{\mathpalette\atversim>}}
\def\atversim#1#2{\lower0.7ex\vbox{\baselineskip\zatskip\lineskip\zatskip
  \lineskiplimit 0pt\ialign{$\matth#1\hfil##\hfil$\crcr#2\crcr\sim\crcr}}}

\part{{\bf  Radion Mixing Effects on the Properties of the Standard Model 
Higgs Boson
} \\[0.5cm]\hspace*{0.8cm}
{\it J.L.Hewett and T.G. Rizzo
}}
\label{rizzo2sec}


\begin{abstract}
We examine how mixing between the Standard Model(SM) Higgs boson, $h$, and 
the radion 
of the Randall-Sundrum model modifies the expected properties of the Higgs 
boson. In particular we demonstrate that the total and 
partial decay widths of the 
Higgs, as well as the $h\to gg$ branching fraction,
 can be substantially altered from their SM expectations, while
the remaining branching fractions are modified less than $\lsim 5\%$
for most of the parameter region. 
\end{abstract}

The Randall-Sundrum(RS) model{\cite {Randall:1999ee}} 
offers a potential solution to the 
hierarchy problem that can be tested at present and future 
accelerators (for an overview of RS phenomenology, see
\cite{Davoudiasl:1999jd,Davoudiasl:2000my,Davoudiasl:2000wi}). 
In this model the SM fields lie on one of two 
branes that are embedded in 5-dimensional AdS space 
described by the metric 
$ds^2=e^{-2k|y|}\eta_{\mu\nu}dx^\mu dx^\nu-dy^2$, where $k$ is the 
5-d curvature 
parameter of order the Planck scale. To solve the hierarchy problem 
the separation between the two branes, $r_c$, must have a value of 
$kr_c \sim 11-12$. That this quantity can be stabilized and made natural has 
been demonstrated by a number 
of authors{\cite {Goldberger:1999uk,Goldberger:1999un,%
Csaki:1999mp,Csaki:2000zn,Charmousis:1999rg,Tanaka:2000er}} 
and leads directly to the existence of a radion ($r$), 
which corresponds to a quantum excitation of the brane separation. It can be 
shown that the radion couples to the trace of the 
stress-energy tensor with a strength $\Lambda$ of 
order the TeV scale, \ie, ${\cal L}_{eff}=-r~T^\mu_\mu /\Lambda$. 
(Note that $\Lambda= \sqrt 3 \Lambda_\pi$ in the notation of 
Ref.{\cite {Davoudiasl:1999jd,Davoudiasl:2000my,Davoudiasl:2000wi}}.)
This leads to gauge and 
matter couplings that are qualitatively similar to those of the SM 
Higgs boson. The radion mass ($m_r$) is expected to be significantly 
below the scale $\Lambda$ implying that the radion may be 
the lightest new field predicted by the RS model. One may expect on general 
grounds that this mass should lie in the range of  
a few $\times 10$ GeV $\leq m_r \leq \Lambda$. 
The phenomenology of the RS radion has been examined by a number of 
authors{\cite {Giudice:2000av,Mahanta:2000ci,Han:2001xs,%
Chaichian:2001rq,Chaichian:2001gr,Bae:2000pk,Bae:2000pd,Park:2001vk,%
Choudhury:2000fj,Cheung:2000rw}} and in particular has been reviewed 
by Kribs{\cite {Kribs:2001ic}}. 

\begin{figure}[tb]
\centerline{
\includegraphics[width=7cm,angle=90]{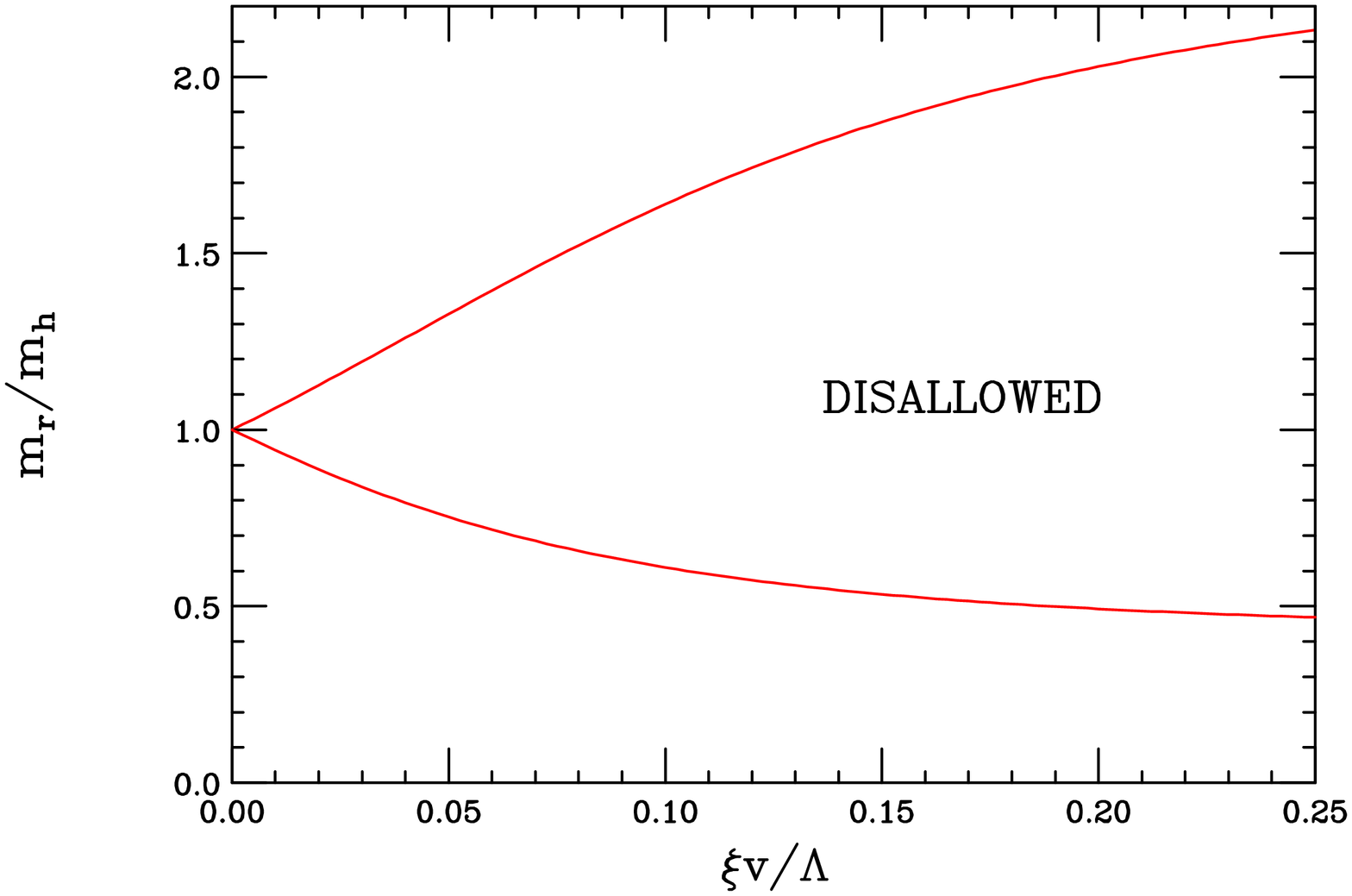}}
\vspace*{0.1cm}
\caption{Constraint on the ratio of the mass of the radion to that of 
the Higgs boson as a 
function of the product $\xi v/\Lambda$ as described in the text. The 
disallowed region lies between the curves.}
\label{p3-38_fig1}
\end{figure}

On general grounds of covariance, the radion may mix with the SM Higgs field on 
the TeV brane through an interaction term of the form 
\begin{equation}
S_{rH}=-\xi \int d^4x \sqrt{-g_w} R^{(4)}[g_w] H^\dagger H\,,
\end{equation}
where $H$ is the Higgs doublet field, 
$R^{(4)}[g_w]$ is the Ricci scalar constructed out of the induced metric $g_w$
on the SM brane,  and $\xi$ is a mixing parameter assumed to be 
of order unity and with unknown sign. The 
above action induces kinetic mixing between the `weak eigenstate' $r_0$ and 
$h_0$ fields which can be removed through a set of field redefinitions and 
rotations. Clearly, since the radion and Higgs boson 
couplings to other SM fields 
differ this mixing will induce modifications in the usual SM expectations for 
the Higgs decay widths. To make unique predictions in this scenario we need to 
specify four parameters: the masses of the {\it physical} Higgs and radion 
fields, $m_{h,r}$, the mixing 
parameter $\xi$ and the ratio $v/\Lambda$, where $v$ is 
the vacuum expectation value of the SM Higgs $\simeq 246$ GeV. Clearly the 
ratio $v/\Lambda$ cannot be too large as $\Lambda_\pi$ is already bounded 
from below by collider and electroweak precision 
data{\cite {Davoudiasl:1999jd,Davoudiasl:2000my,Davoudiasl:2000wi}}; 
for definiteness we will take $v/\Lambda \leq 0.2$ and 
$-1 \leq \xi \leq 1$ in what follows although larger absolute values of $\xi$ 
have been entertained in the literature.  
The values of the two physical masses themselves are not arbitrary. 
When we require that the weak basis mass-squared parameters of the radion and 
Higgs fields be real, as is required by hermiticity, we obtain an additional 
constraint on the ratio of the 
physical radion and Higgs masses which only depends on the product 
$|\xi| {v\over {\Lambda}}$. Explicitly one finds that 
either ${m_r^2\over {m_h^2}}\geq 
1+2\sin^2 \rho+2|\sin \rho| \sqrt{1+\sin^2 \rho}$ or ${m_r^2\over {m_h^2}}\leq 
1+2\sin^2 \rho-2|\sin \rho| \sqrt{1+\sin^2 \rho}$ where 
$\rho=\tan^{-1}(6\xi {v\over {\Lambda}})$. This disfavors the radion having a 
mass too close to that of the Higgs when there is significant mixing; the 
resulting excluded region is shown in Fig.~\ref{p3-38_fig1}. These 
constraints are somewhat restrictive; if we take $m_h=115$ GeV and 
$\xi {v\over {\Lambda}}=0.1(0.2)$ we find that either $m_r>189(234)$ GeV or 
$m_r<70(56)$ GeV. This lower mass range may be disfavored by direct 
searches. 

Following the notation of Giudice \etal {\cite {Giudice:1998ck}}, the coupling of the 
physical Higgs to the SM fermions and massive gauge bosons $V=W,Z$ 
is now given by
\begin{equation}
{\cal {L}}={-1\over {v}}(m_f\bar ff-m_V^2 V_\mu V^\mu)[\cos \rho \cos \theta +
{v\over {\Lambda}}(\sin \theta-\sin \rho \cos \theta)]h\,,
\end{equation}
where the angle $\rho$ is given above and $\theta$ can be calculated in 
terms of the 
parameters $\xi$ and $v/\Lambda$ and the physical Higgs and radion masses. 
Denoting the combinations $\alpha=\cos \rho \cos \theta$ and 
$\beta=\sin \theta-\sin \rho \cos \theta$, the corresponding Higgs 
coupling to gluons 
can be written as $c_g {\alpha_s\over {8\pi}}G_{\mu\nu}G^{\mu\nu}h$ with 
$c_g={-1\over {2v}}[(\alpha +{v\over {\Lambda}}\beta)F_g
-2b_3\beta {v\over {\Lambda}}]$ where $b_3=7$ is the $SU(3)$ $\beta$-function 
and $F_g$ is a well-known kinematic function of the ratio of masses of the  
top quark to the physical Higgs. 
Similarly the physical Higgs couplings to two photons is now given by 
$c_\gamma {\alpha_{em}\over {8\pi}}F_{\mu\nu}F^{\mu\nu}h$ where 
$c_\gamma={1\over {v}}[(b_2+b_Y)\beta {v\over {\Lambda}}-(\alpha 
+{v\over {\Lambda}}\beta)F_\gamma]$, where $b_2=19/6$ and $b_Y=-41/6$ are the 
$SU(2)\times U(1)$ $\beta$-functions and $F_\gamma$ is another well-known 
kinematic function of the ratios of the $W$ and top masses to the physical 
Higgs mass. (Note that in the simultaneous limits $\alpha \to 1,~\beta \to 0$ 
we recover the usual SM results.) From these expressions we can now compute  
the change of the various decay widths and branching fractions of the
SM Higgs due to mixing with the radion.

\begin{figure}[tb]
\centerline{
\includegraphics[width=5.4cm,angle=90]{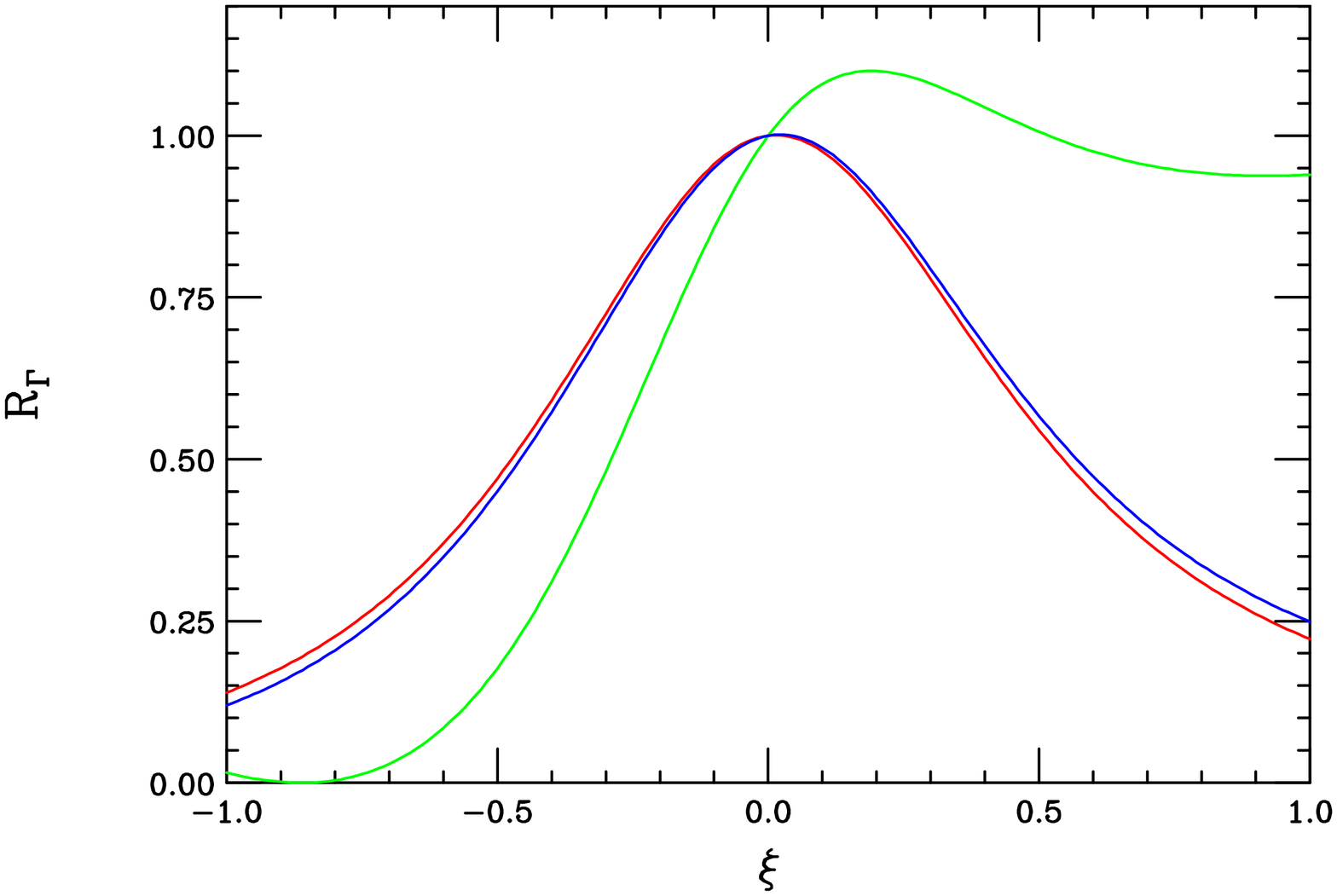}
\hspace*{5mm}
\includegraphics[width=5.4cm,angle=90]{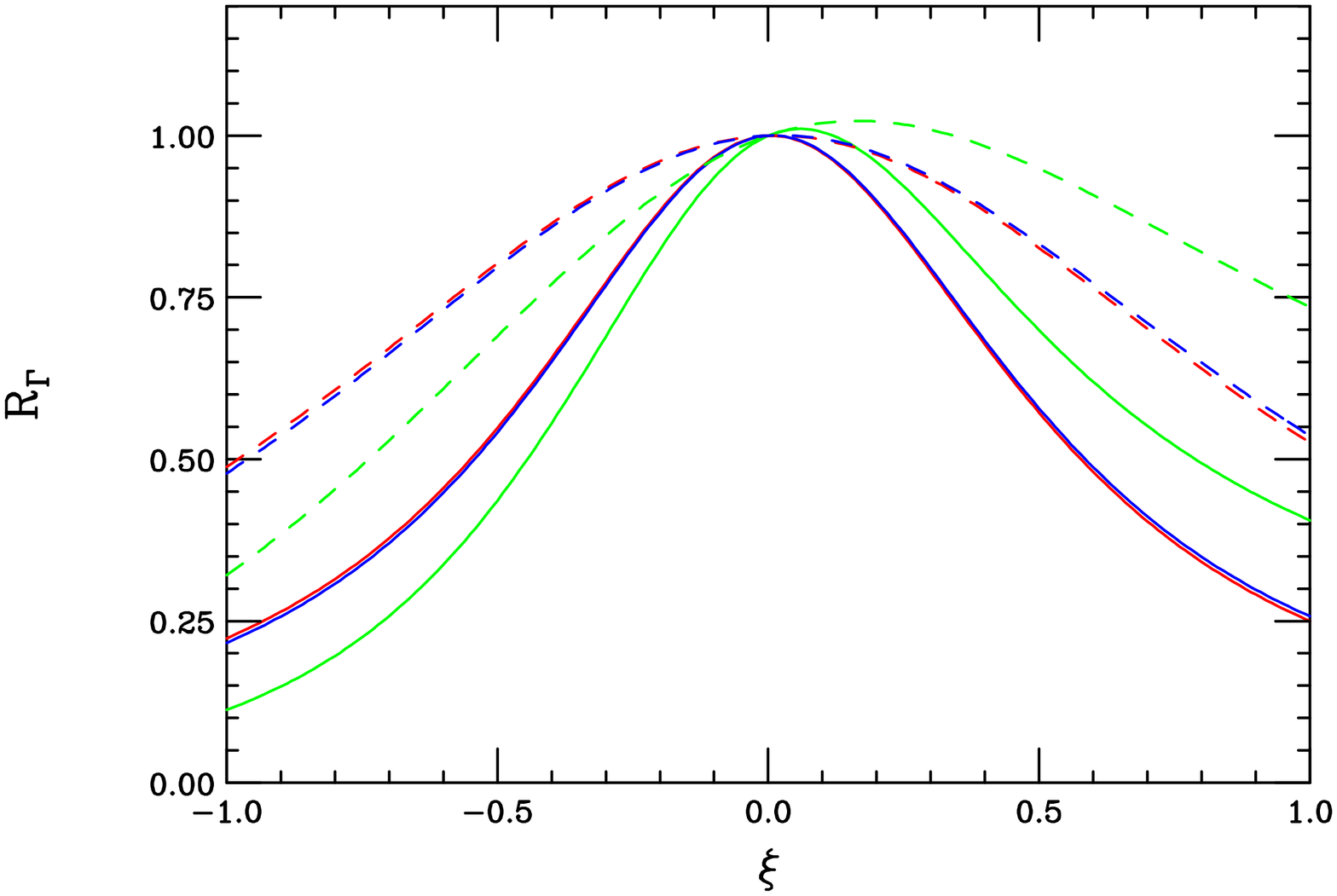}}
\vspace*{0.1cm}
\caption{Ratio of Higgs widths to their SM values, $R_\Gamma$, as a function 
of $\xi$ assuming a physical Higgs mass of 125 GeV: red for fermion pairs or 
massive gauge boson pairs, green for gluons and blue for photons. In the left 
panel we assume $m_r=300$ GeV and $v/\Lambda=0.2$. In the right panel the 
solid(dashed) curves are for $m_r=500(300)$ GeV and $v/\Lambda=0.2(0.1)$.}
\label{p3-38_fig2}
\end{figure}

Fig.~\ref{p3-38_fig2} shows the ratio of the various Higgs widths in 
comparison to their SM expectations as functions of the parameter $\xi$ 
assuming that $m_h=125$ GeV with different values of $m_r$ and 
${v\over {\Lambda}}$. We see several features right away: ($i$) the shifts in 
the widths to $\bar ff/VV$ and $\gamma \gamma$ final states are very similar; 
this is due to the relatively large magnitude of $F_\gamma$ while the 
combination $b_2+b_Y$ is rather small. ($ii$) On the other hand the shift for 
the $gg$ final state is quite different since $F_g$ is smaller than $F_\gamma$ 
and $b_3$ is quite large. ($iii$) For relatively light radions with a low 
value of $\Lambda$ the width into the $gg$ final state can come close to 
vanishing due to a strong 
destructive interference between the two contributions to the 
amplitude for values of $\xi$ near -1. ($iv$) Increasing the value of $m_r$ 
has less of an effect on the width shifts than does a decrease in the ratio 
${v\over {\Lambda}}$. 

\begin{figure}[tb]
\centerline{
\includegraphics[width=5.4cm,angle=90]{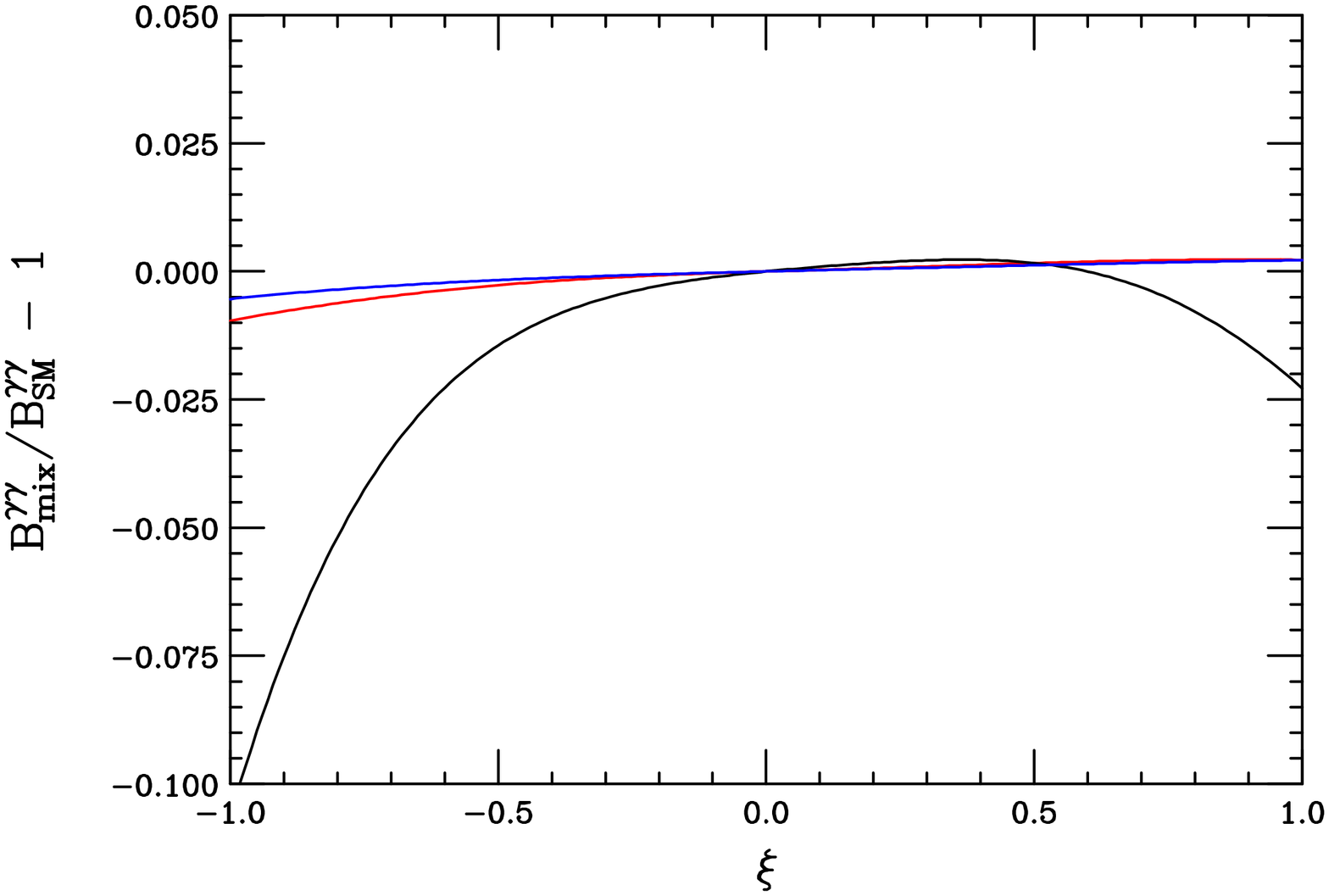}
\hspace*{5mm}
\includegraphics[width=5.4cm,angle=90]{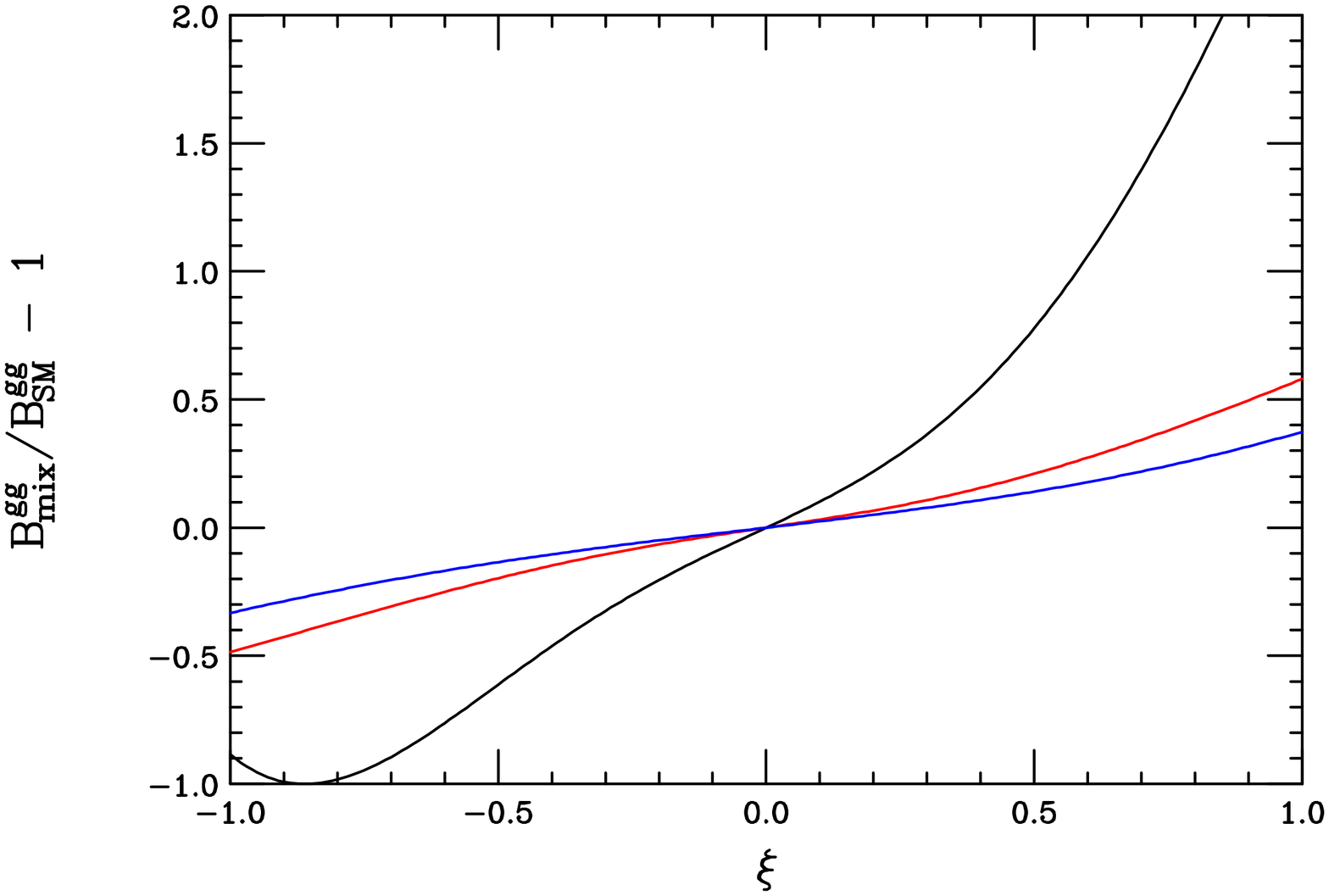}}
\vspace*{0.1cm}
\centerline{
\includegraphics[width=5.4cm,angle=90]{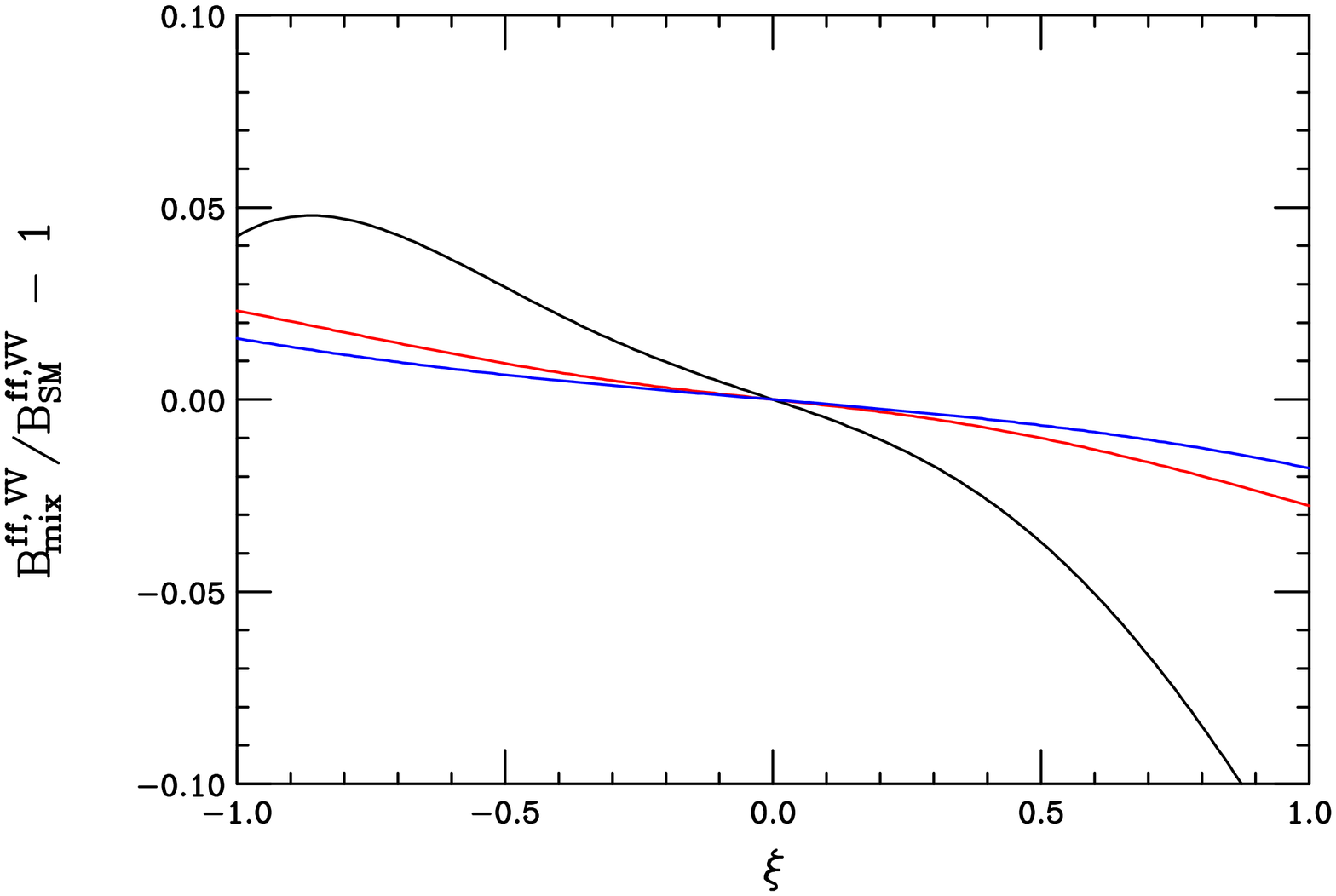}
\hspace*{5mm}
\includegraphics[width=5.4cm,angle=90]{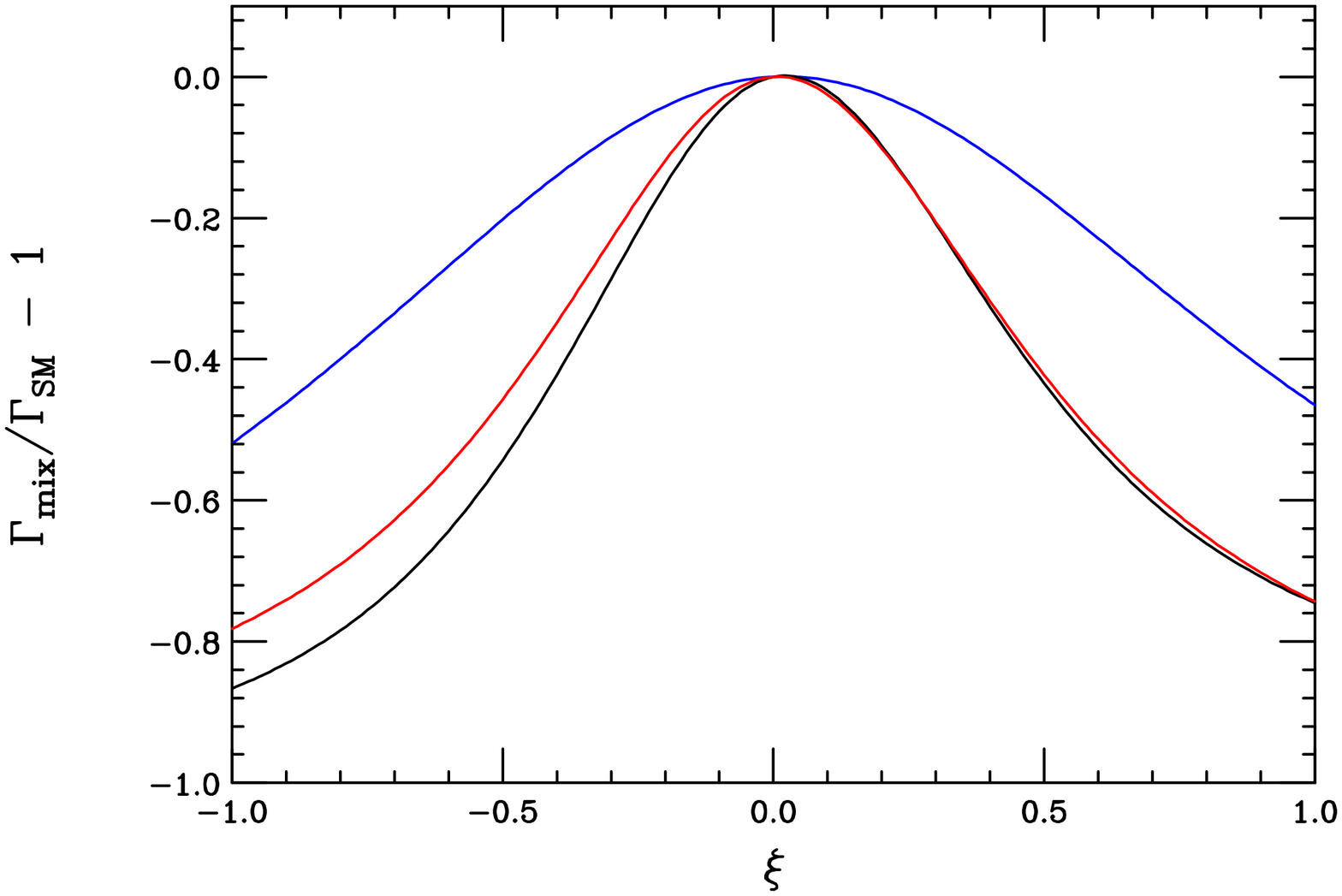}}
\vspace*{0.1cm}
\caption{The deviation from the SM expectations for the Higgs
branching fraction into $\gamma\gamma$, $gg$, $f\bar f$, and $VV$ final
states as labeled, as well as for the total width.  The black, red, and
blue curves correspond to the parameter choices $m_r=300, 500, 300$ GeV
with $v/\xi=0.2, 0.2, 0.1$, respectively.}
\label{p3-38_fig3}
\end{figure}

The  deviation from the SM expectations for
the various branching fractions, as well as the total width, of the 
Higgs are displayed in Fig. \ref{p3-38_fig3} as a function of the 
mixing parameter $\xi$.  We see that the gluon branching fraction and the
total width may be drastically different than that of the SM.  The former
will affect the Higgs production cross section at the LHC.  However, the
$\gamma\gamma$, $f\bar f$, and $VV$, where $V=W,Z$ branching fractions
receive small corrections to their SM values, of order $\lsim 5\%$
for most of the parameter region. Observation of these shifts 
will require the accurate determination
of the Higgs branching fractions available at an $e^+e^-$ Linear Collider.

In summary, we see that Higgs-radion mixing, which is present in some
extra dimensional scenarios, can have a substantial effect on the
properties of the Higgs boson.  These modifications affect the widths
and branching fractions of Higgs decay into various final states, which
in turn can alter the
Higgs production cross section at the LHC and may require the precision
of a Linear Collider to detect.

\setcounter{figure}{0}
\setcounter{table}{0}
\setcounter{section}{0}
\setcounter{equation}{0}
\clearpage

\def\nn{\noindent}
\def\Re{{\cal R \mskip-4mu \lower.1ex \hbox{\it e}\,}}
\def\Im{{\cal I \mskip-5mu \lower.1ex \hbox{\it m}\,}}
\def\ie{{\it i.e.}}
\def\eg{{\it e.g.}}
\def\etc{{\it etc}}
\def\etal{{\it et al.}}
\def\ibid{{\it ibid}.}
\def\sub#1{_{\lower.25ex\hbox{$\scriptstyle#1$}}}
\def\tev{\,{\ifmmode\mathrm {TeV}\else TeV\fi}}
\def\gev{\,{\ifmmode\mathrm {GeV}\else GeV\fi}}
\def\mev{\,{\ifmmode\mathrm {MeV}\else MeV\fi}}
\def\mpl{\ifmmode \overline M_{Pl}\else $\overline M_{Pl}$\fi}
\def\to{\rightarrow}
\def\slash{\not\!}
\def\subw{_{\rm w}}
\def\mh{\ifmmode m\sbl H \else $m\sbl H$\fi}
\def\mch{\ifmmode m_{H^\pm} \else $m_{H^\pm}$\fi}
\def\mt{\ifmmode m_t\else $m_t$\fi}
\def\mc{\ifmmode m_c\else $m_c$\fi}
\def\mz{\ifmmode M_Z\else $M_Z$\fi}
\def\mw{\ifmmode M_W\else $M_W$\fi}
\def\mws{\ifmmode M_W^2 \else $M_W^2$\fi}
\def\mhs{\ifmmode m_H^2 \else $m_H^2$\fi}   
\def\mzs{\ifmmode M_Z^2 \else $M_Z^2$\fi}
\def\mts{\ifmmode m_t^2 \else $m_t^2$\fi}
\def\mcs{\ifmmode m_c^2 \else $m_c^2$\fi}
\def\mchs{\ifmmode m_{H^\pm}^2 \else $m_{H^\pm}^2$\fi}
\def\ztwo{\ifmmode Z_2\else $Z_2$\fi}
\def\zone{\ifmmode Z_1\else $Z_1$\fi}
\def\mtwo{\ifmmode M_2\else $M_2$\fi}
\def\mone{\ifmmode M_1\else $M_1$\fi}
\def\tb{\ifmmode \tan\beta \else $\tan\beta$\fi}
\def\xw{\ifmmode x\subw\else $x\subw$\fi}
\def\ch{\ifmmode H^\pm \else $H^\pm$\fi}
\def\lum{\ifmmode {\cal L}\else ${\cal L}$\fi}
\def\inpb{\,{\ifmmode {\mathrm {pb}}^{-1}\else ${\mathrm {pb}}^{-1}$\fi}}
\def\infb{\,{\ifmmode {\mathrm {fb}}^{-1}\else ${\mathrm {fb}}^{-1}$\fi}}
\def\epem{\ifmmode e^+e^-\else $e^+e^-$\fi}
\def\ppb{\ifmmode \bar pp\else $\bar pp$\fi}
\def\bsg{\ifmmode B\to X_s\gamma\else $B\to X_s\gamma$\fi}
\def\bsll{\ifmmode B\to X_s\ell^+\ell^-\else $B\to X_s\ell^+\ell^-$\fi}
\def\bstt{\ifmmode B\to X_s\tau^+\tau^-\else $B\to X_s\tau^+\tau^-$\fi}
\def\lamt{\ifmmode \tilde\lambda\else $\tilde\lambda$\fi}
\def\shat{\ifmmode \hat s\else $\hat s$\fi}
\def\that{\ifmmode \hat t\else $\hat t$\fi}
\def\uhat{\ifmmode \hat u\else $\hat u$\fi}
\def\half{\textstyle{{1\over 2}}}
\def\elli{\ell^{i}}
\def\ellj{\ell^{j}}
\def\ellk{\ell^{k}} 
\newskip\zatskip \zatskip=0pt plus0pt minus0pt
\def\matth{\mathsurround=0pt}
\def\lsim{\mathrel{\mathpalette\atversim<}}
\def\gsim{\mathrel{\mathpalette\atversim>}}
\def\atversim#1#2{\lower0.7ex\vbox{\baselineskip\zatskip\lineskip\zatskip
  \lineskiplimit 0pt\ialign{$\matth#1\hfil##\hfil$\crcr#2\crcr\sim\crcr}}}
\def\undertext#1{$\underline{\smash{\vphantom{y}\hbox{#1}}}$}

\part{{\bf Probing Universal Extra Dimensions at Present and Future Colliders
} \\[0.5cm]\hspace*{0.8cm}
{\it Thomas G. Rizzo
}}
\label{rizzo1sec}


\begin{abstract}
In the Universal Extra Dimensions model of Appelquist, Cheng and Dobrescu, all 
of the Standard Model fields are placed in the bulk and thus have 
Kaluza-Klein (KK) excitations. These KK states can only be pair produced at 
colliders due to the tree-level conservation of KK number, with the lightest 
of them being stable and 
possibly having a mass as low as $\simeq 350-400$ GeV. 
We investigate the 
production cross sections and signatures for these particles at both hadron 
and lepton colliders. We demonstrate that these signatures critically depend 
upon whether the lightest KK states remain stable or are allowed to decay by 
any of a number of new physics mechanisms. These mechanisms which induce KK 
decays are studied in detail. 
\end{abstract}

\section{Introduction}

The possibility that the gauge bosons of the Standard Model (SM) may be 
sensitive to the existence of extra dimensions near the TeV scale has been 
known for some time {\cite{Antoniadis:1990ew,Antoniadis:1993fh,%
Antoniadis:1994jp,Antoniadis:2000vd,Antoniadis:1994yi,%
Benakli:1996ut,Accomando:1999sj}}. 
However, one finds that the phenomenology of these 
models is particularly sensitive to the manner in which the SM fermions (and 
Higgs bosons) are treated. 

Perhaps the most democratic possibility requires all of the SM fields to 
propagate in the $\sim$ TeV$^{-1}$ bulk {\cite {Appelquist:2000nn}}, \ie, Universal Extra 
Dimensions (UED). In this case,
the conservation of momentum in the extra dimensions is restored and 
one obtains interactions in the 4-d Lagrangian which take the form 
$\sim gC_{ijk}\bar f^{(i)} \gamma_\mu f^{(j)} G^\mu_{(k)}$, which for 
flat space metrics vanishes unless $i+j+k=0$, as a result of the afore 
mentioned momentum conservation. Although this momentum conservation is 
actually broken by 
orbifolding, one finds, at tree level, that KK number remains a 
conserved quantity. (As we will discuss below this conservation law is itself  
further 
broken at one loop order.) This implies that pairs of 
zero-mode fermions, which we identify with those of the SM, cannot directly 
interact singly with any of the excited modes in the gauge boson KK towers.  
Such a situation  
clearly limits any constraints arising from precision measurements since zero 
mode fermion fields can only interact with pairs of tower gauge boson fields. 
In addition, at colliders it now follows that 
KK states must be pair produced, thus significantly 
reducing the possible direct search reaches for these states. In fact, 
employing constraints from current experimental data, Appelquist, Cheng and 
Dobrescu (ACD) {\cite {Appelquist:2000nn}} find that the KK states in this 
scenario can be as light as $\simeq 350-400$ GeV. If these states 
are, in fact, nearby, they will be copiously produced at the LHC, and possibly 
also at the Tevatron, in a variety 
of different channels. It is the purpose of this paper to estimate the 
production rates for pairs of these particles in various channels and to 
discuss their possible production signatures. This is made somewhat difficult 
by the apparent conservation of KK number which appears to forbid the decay of 
heavier excitations into lighter ones.

Now although the KK number is conserved at the tree level it becomes apparent 
that it is no longer so at loop order {\cite {strumiapc}}. Consider a 
self-energy diagram  with a field that has KK number of $2n(2n+1)$ entering 
and a zero($n=1$) mode leaving 
the graph; KK number conservation clearly does not forbid such an amplitude 
and constrains the 2 particles in the intermediate state to both have KK number 
$n$($n$ and $n+1$). The existence of such amplitudes implies that all even 
and odd KK states mix separately so that the even KK excitations can 
clearly decay to zero modes while odd KK states can now 
decay down to the KK number=1 state. Thus it is KK {\it parity}, $(-1)^n$, 
which remains conserved while KK number 
itself is broken at one loop. Since the lightest KK excited states with 
$n=1$ have odd KK parity they remain stable unless new physics is introduced. 
As we are only concerned with the production of pairs of the lightest KK 
particles in our discussion below, we are faced with the possibility 
of producing heavy stable states at colliders.

\section{Collider Production}

\begin{figure}[ht]
\centerline{
\includegraphics[width=5.3cm,angle=90]{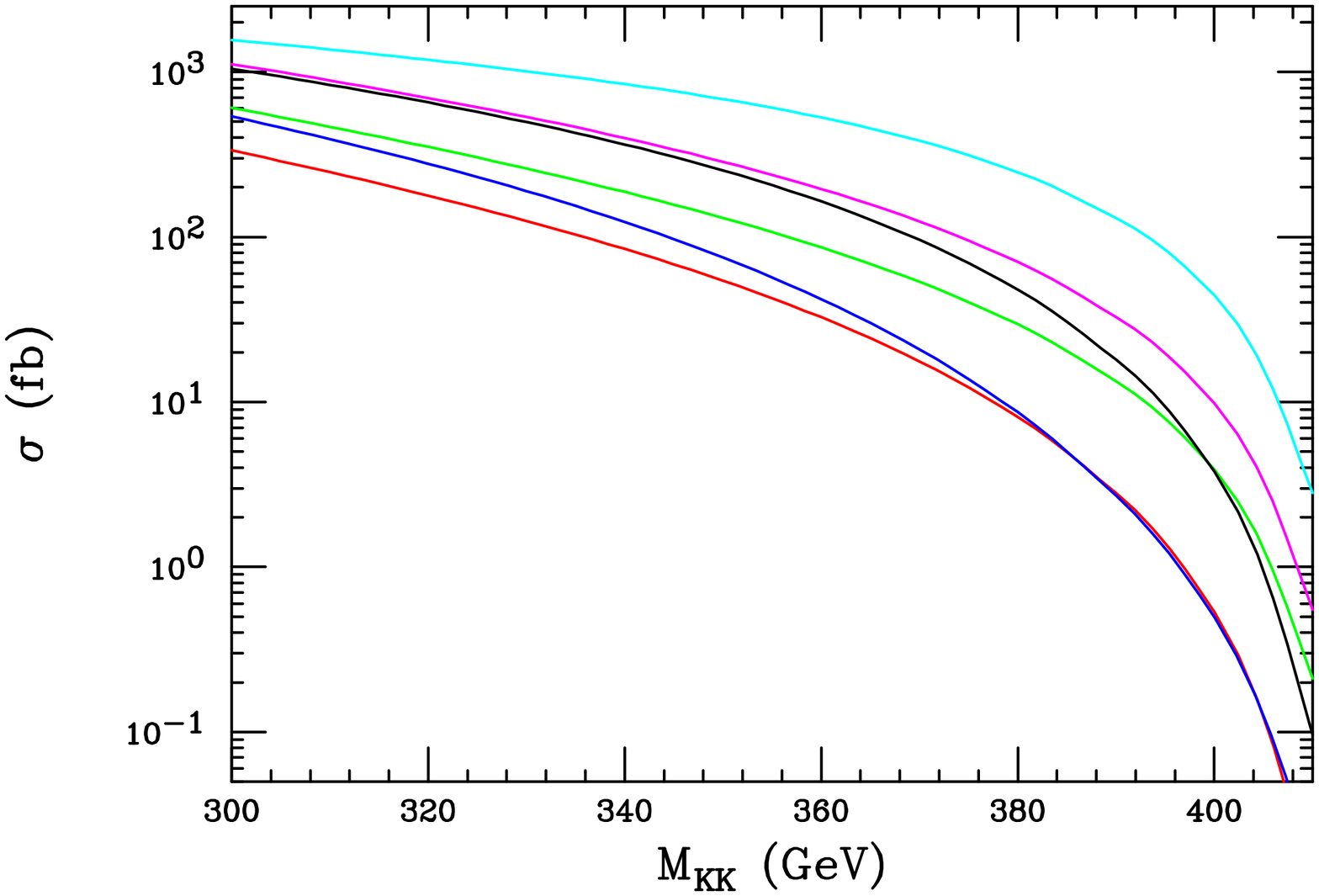}}
\vspace*{0.25cm}
\centerline{
\includegraphics[width=5.3cm,angle=90]{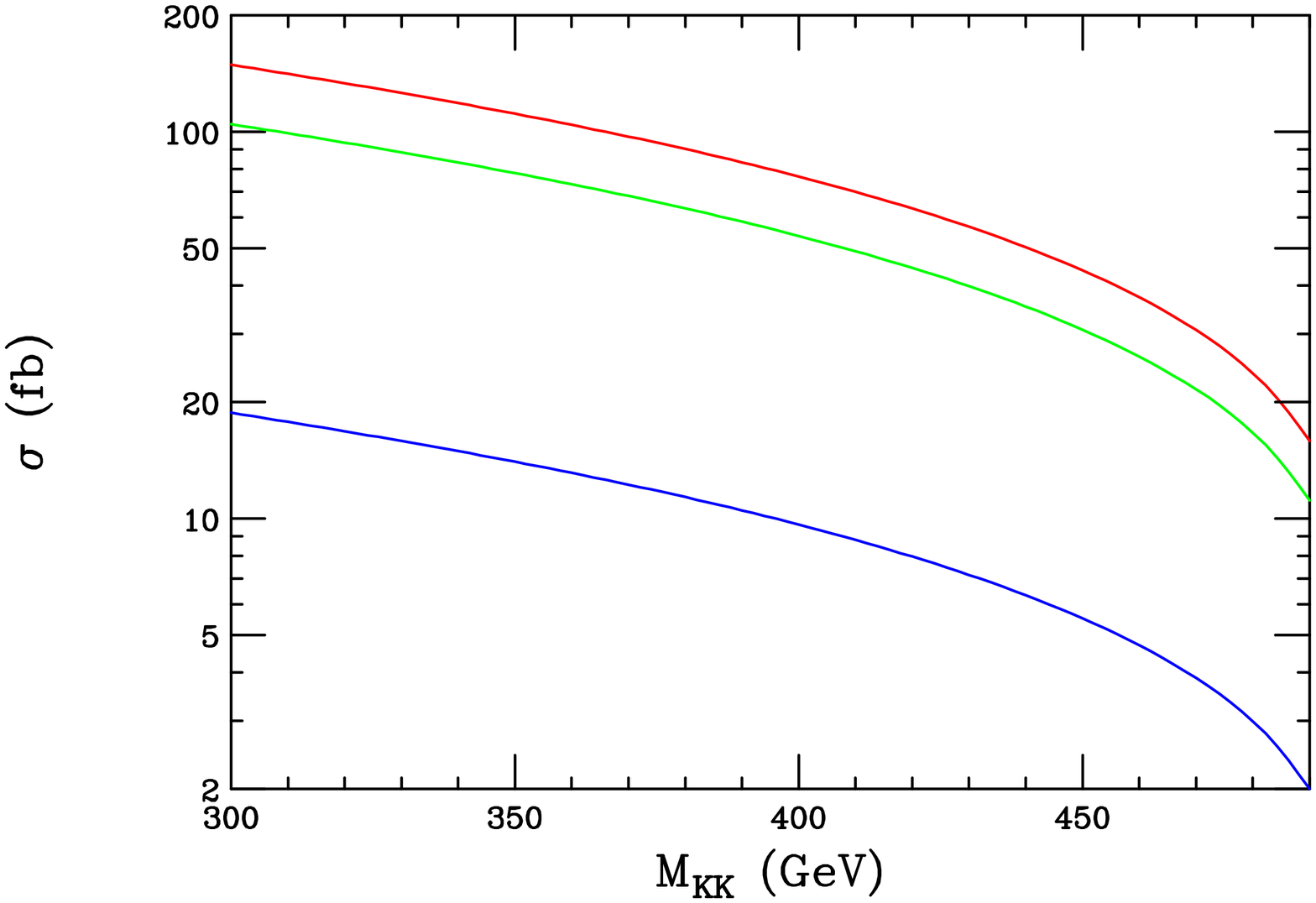}
\includegraphics[width=5.3cm,angle=90]{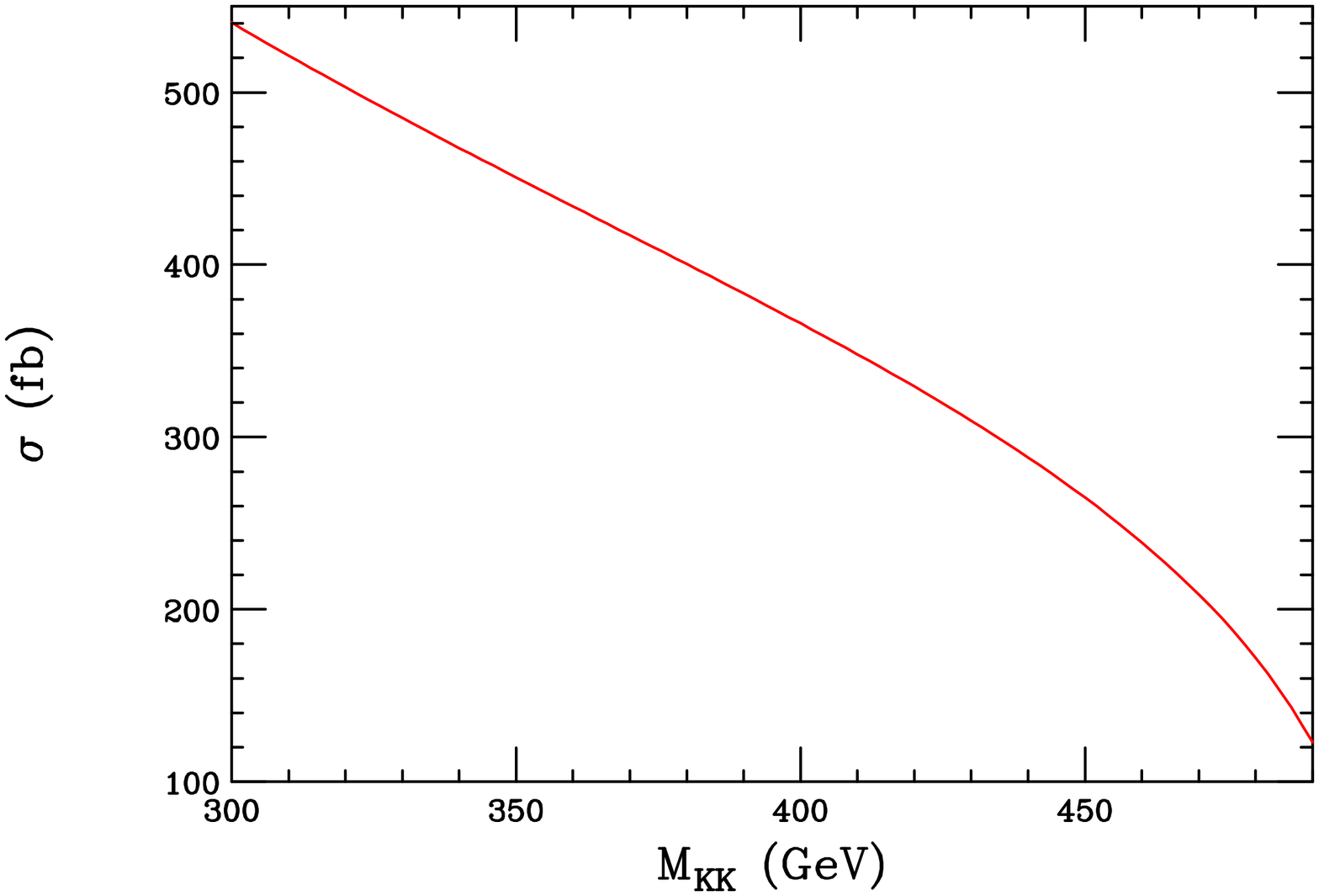}}
\vspace*{0.25cm}
\caption{Cross section for $\gamma\gamma \to W^{(1)+}W^{(1)-}$ (top panel) for 
different electron and laser polarizations for $\sqrt s_{ee}=1$ TeV. Cross 
section for $e^+e^- \to W^{(1)}W^{(1)}$ (lower right panel) for $\sqrt s=1$ 
TeV. Cross sections for (top to bottom, lower left panel) $e^+e^- \to 
2\gamma^{(1)}$, $Z^{(1)}\gamma^{(1)}$ and $2Z^{(1)}$ for $\sqrt s=1$ TeV.}
\label{fig1}
\end{figure}

Due to the conservation of KK parity, the first KK excitations of the SM 
fields must be pair-produced at colliders. At $\gamma \gamma$ and 
lepton colliders the production 
cross sections for all the kinematically accessible KK states will very  
roughly be of order 100 fb $(1~TeV/\sqrt s)^2$ which yields respectable event 
rates 
for luminosities in the $100-500 ~fb^{-1}$ range. A sample of relevant cross 
sections at both $\gamma\gamma$ and $e^+e^-$ colliders are shown in 
Fig. ~\ref{fig1}.   
In the case of $\gamma \gamma$ collisions we have chosen the process 
$\gamma \gamma \to W^{+(1)}W^{-(1)}$ as it the process which 
has the largest cross section for 
the production of the first KK state. Similarly, gauge boson pair production 
in $e^+e^-$ collisions naturally leads to a large cross section. Clearly, 
such states once produced would not be easily 
missed for masses up to close the kinematic limit of the machine independently 
of how they decayed or if they were stable. To directly probe heavier masses 
we must turn to hadron colliders.
\nn

\begin{figure}[tb]
\centerline{
\includegraphics[width=5.3cm,angle=90]{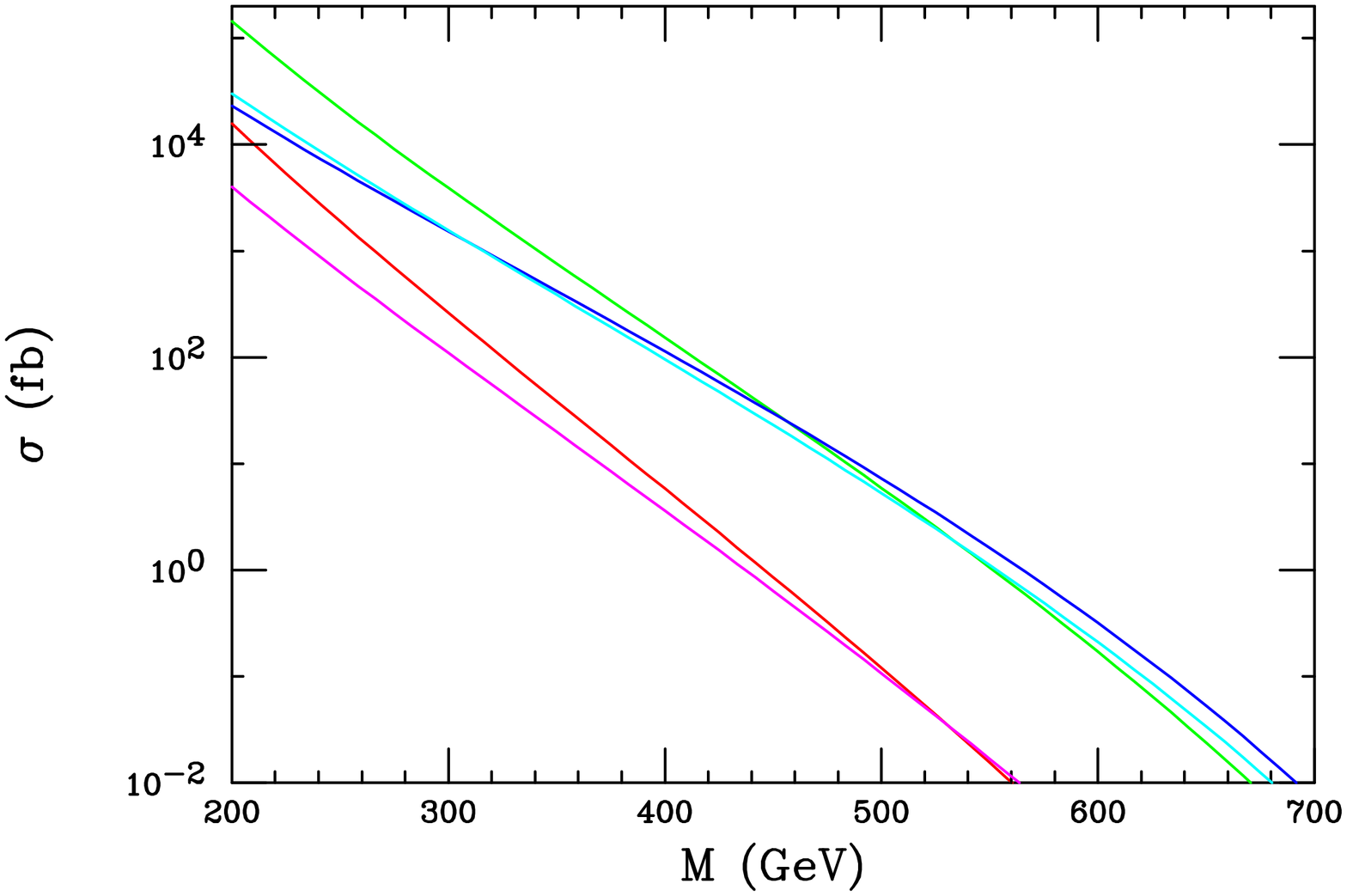}
\hspace*{-2mm}
\includegraphics[width=5.3cm,angle=90]{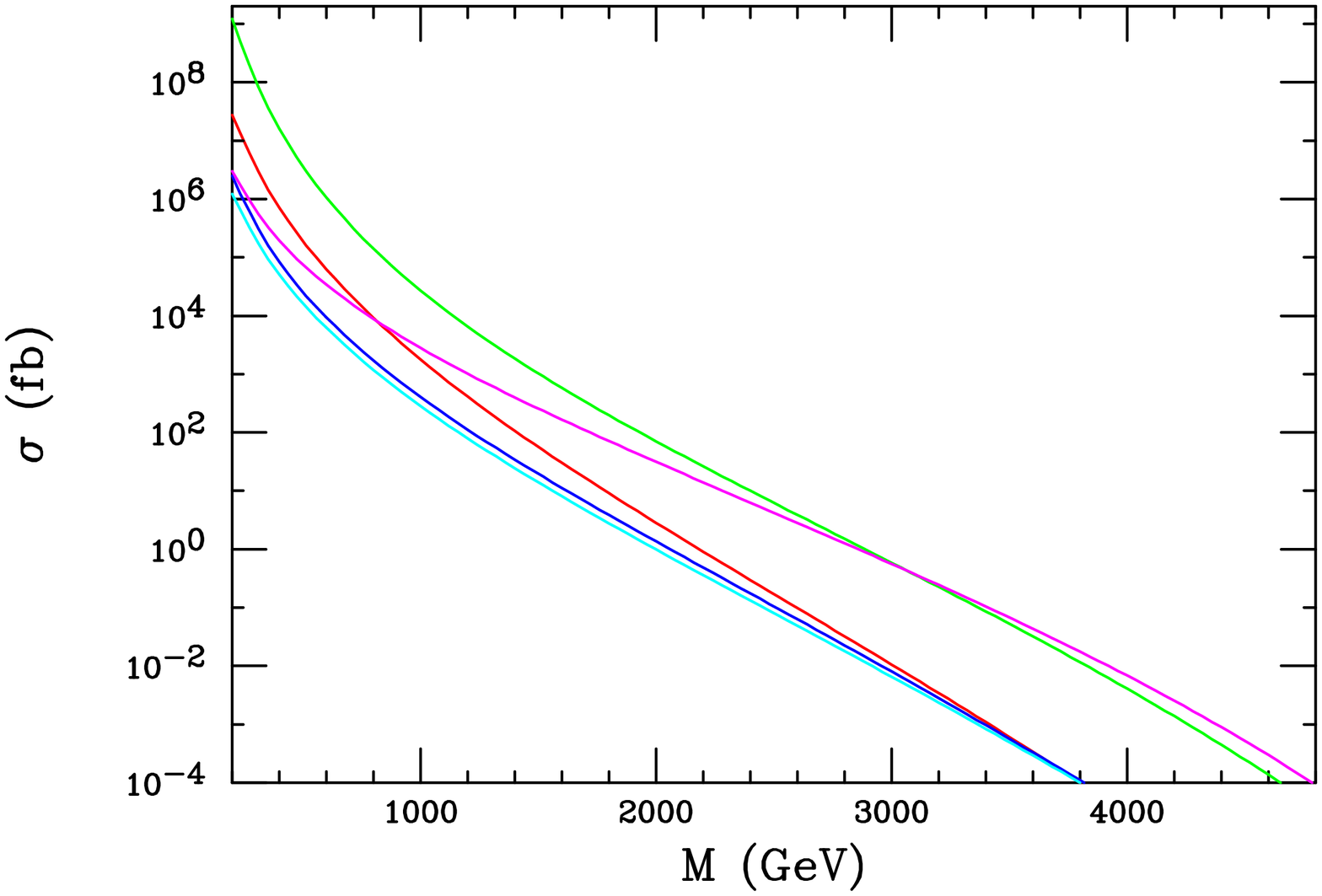}}
\caption{Cross section for the pair production of the lightest colored KK 
states at the $\sqrt s=2$ TeV Tevatron (left) and the LHC (right). In the left 
panel, from top to bottom on the left-hand side, the curves correspond to the 
processes $ii,v,iii,i$ and $iv$, respectively. In the right
panel, from top to bottom on the left-hand side, the curves correspond to the 
processes $ii,i,iv,iii$ and $v$, respectively. Antiquark contributions are 
included in reactions $ii$ and $iv$.}
\label{fig2}
\end{figure}

Since both QCD and electroweak exchanges can lead to KK pair production at 
hadron colliders there are three classes of basic processes to consider. 
Clearly the states with color 
quantum numbers will have the largest cross sections at hadron machines and 
there are a number of processes which can contribute to their production at 
order $\alpha_s^2$ {\cite {Dicus:2000hm}} several of which we list below:
\begin{eqnarray}
(i)~~gg &\to & g^{(1)}g^{(1)} \, \nonumber \\
(ii)~~qq'&\to & q^{(1)}q{'}^{(1)} \, \nonumber \\
(iii)~~gg+q\bar q &\to & q{'}^{(1)}\bar q{'}^{(1)} \, \nonumber \\
(iv)~~qq &\to & q^{(1)}q^{(1)} \, \nonumber \\
(v)~~q\bar q &\to & q^{(1)}q^{(1)} \,,
\end{eqnarray}
where the primes are present to denote flavor differences. 
Fig. ~\ref{fig2} shows the cross sections for these five processes at both the 
$\sqrt s=2$ TeV Tevatron and the LHC summed over flavors. It is clear that 
during the Tevatron Run II we should expect a reasonable yield of these 
KK particles for masses below $\simeq 600$ GeV if integrated luminosities in 
the range of 10-20 $fb^{-1}$ are obtained. Other processes that we have not 
considered may be able to slightly increase this reach. 
For larger masses we must turn to 
the LHC where we see that significant event rates should be obtainable for KK 
masses up to $\simeq 3$ TeV or so for an integrated luminosity of 100 
$fb^{-1}$. As one might expect we see that the most important QCD processes 
for the production of KK states are different at the two colliders.

\vspace*{-0.5cm}
\nn
\vspace*{0.4mm}

\begin{figure}[tb]
\centerline{
\includegraphics[width=5.3cm,angle=90]{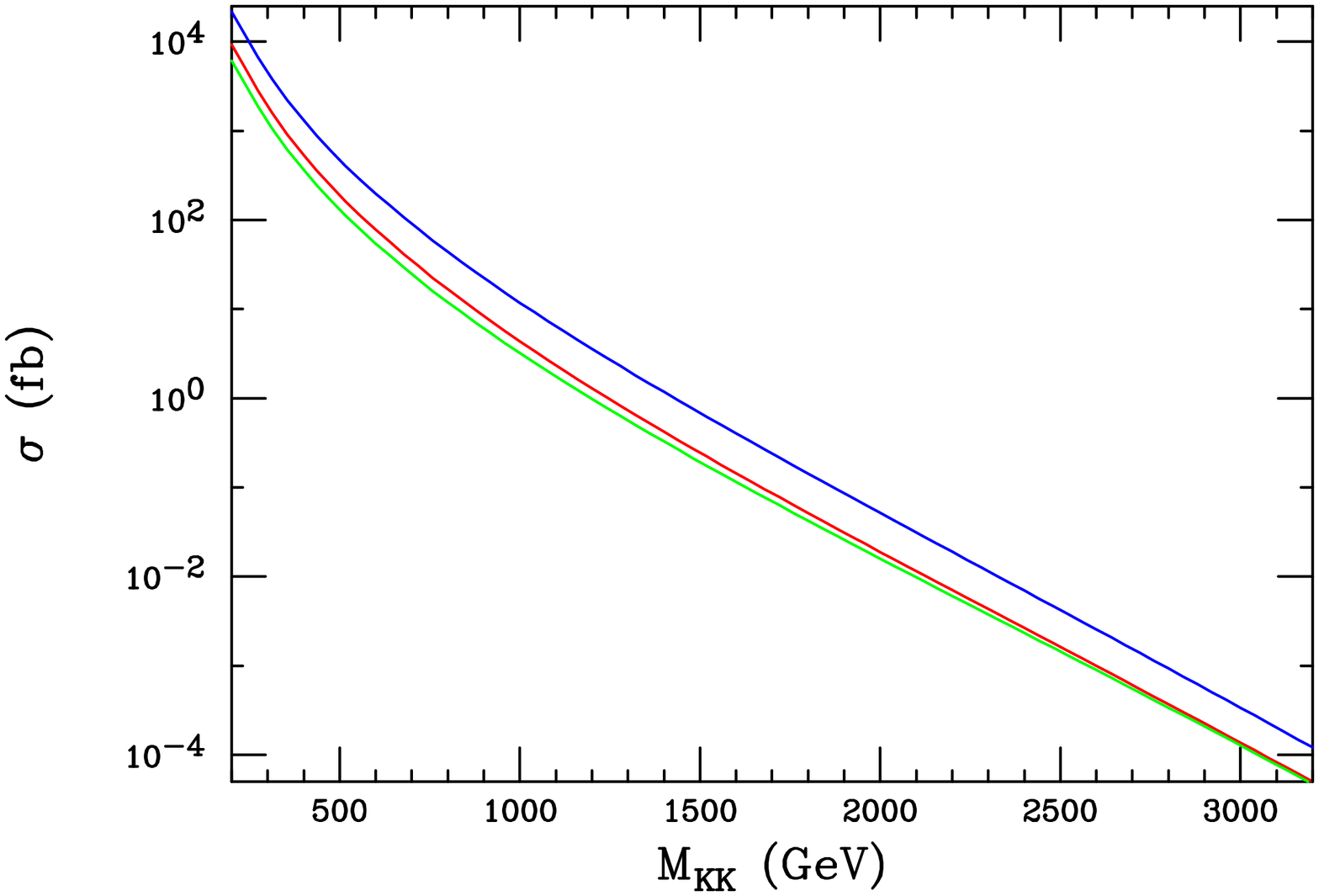}
\hspace*{5mm}
\includegraphics[width=5.3cm,angle=90]{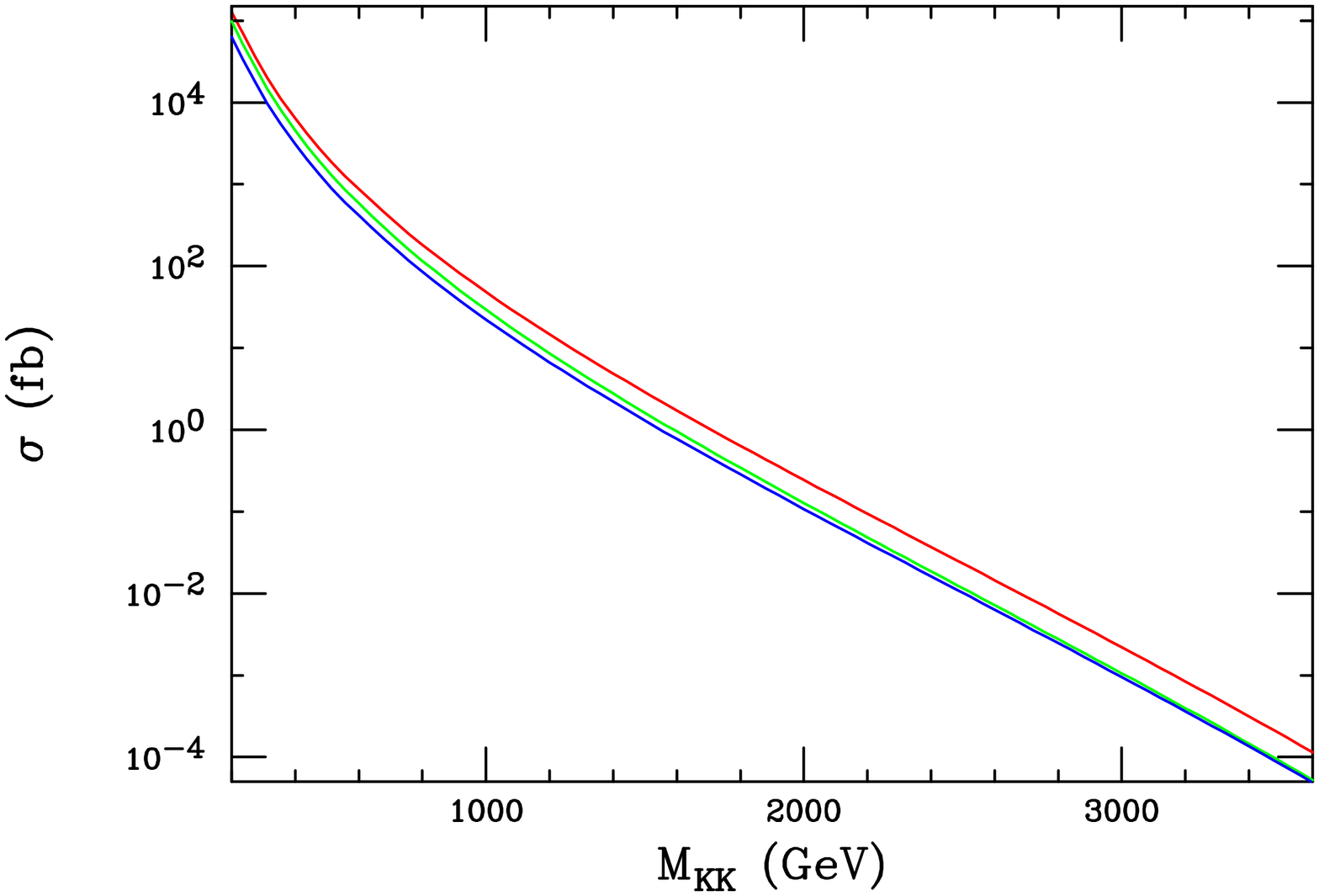}}
\caption{Cross sections for the associated production of the lightest color 
singlet KK states at the LHC: in the left panel, from top to bottom, for 
$g^{(1)}W^{(1)\pm},g^{(1)}Z^{(1)}$ and $g^{(1)}\gamma^{(1)}$ final states; in 
the right panel, from top to bottom, for $q^{(1)}W^{(1)\pm},q^{(1)}Z^{(1)}$ 
and $q^{(1)}\gamma^{(1)}$ final states. Anti-quark contributions are 
included.}
\label{fig3}
\end{figure}

\begin{figure}[tb]
\centerline{
\includegraphics[width=5.3cm,angle=90]{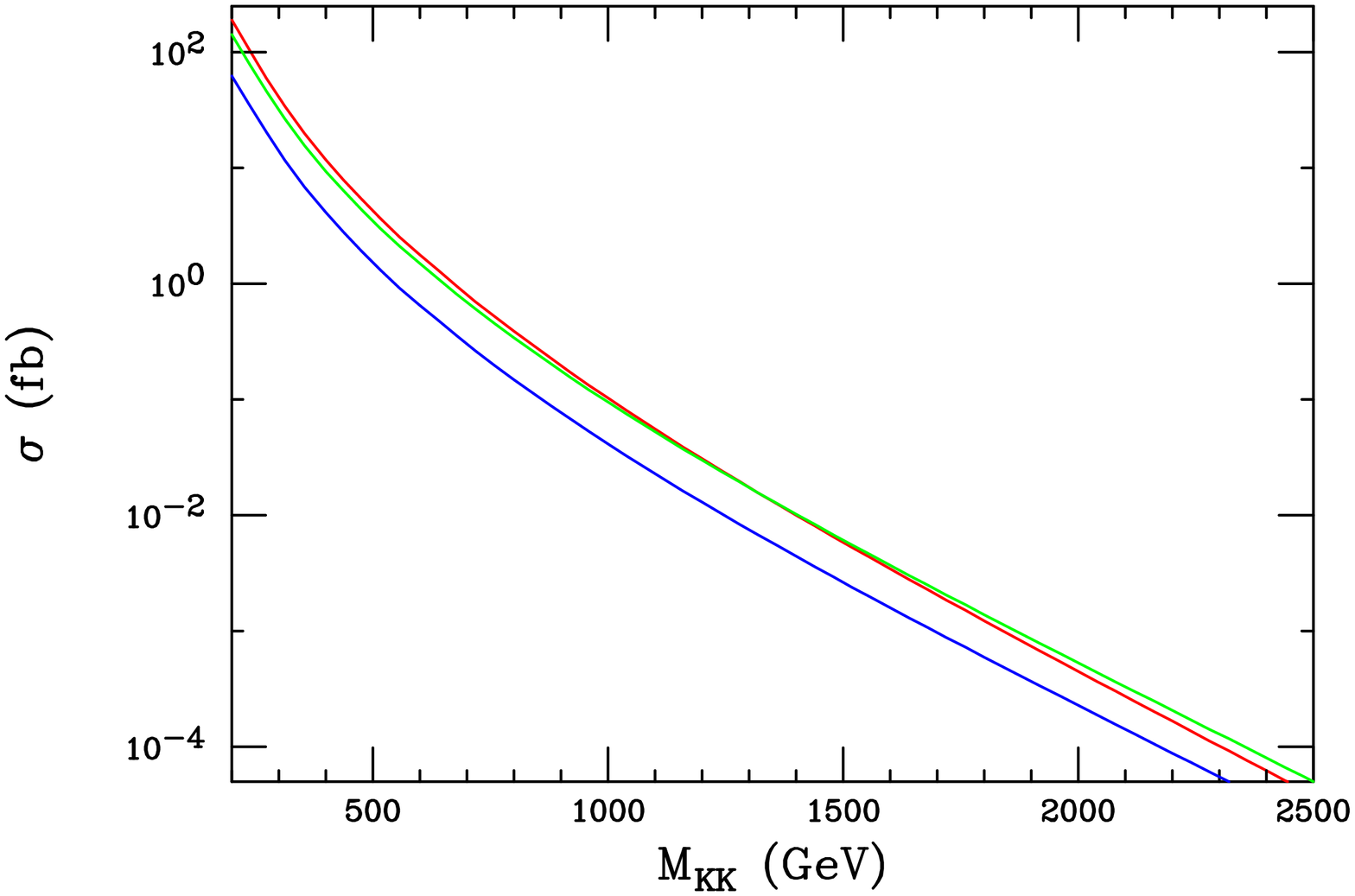}
\hspace*{-2mm}
\includegraphics[width=5.3cm,angle=90]{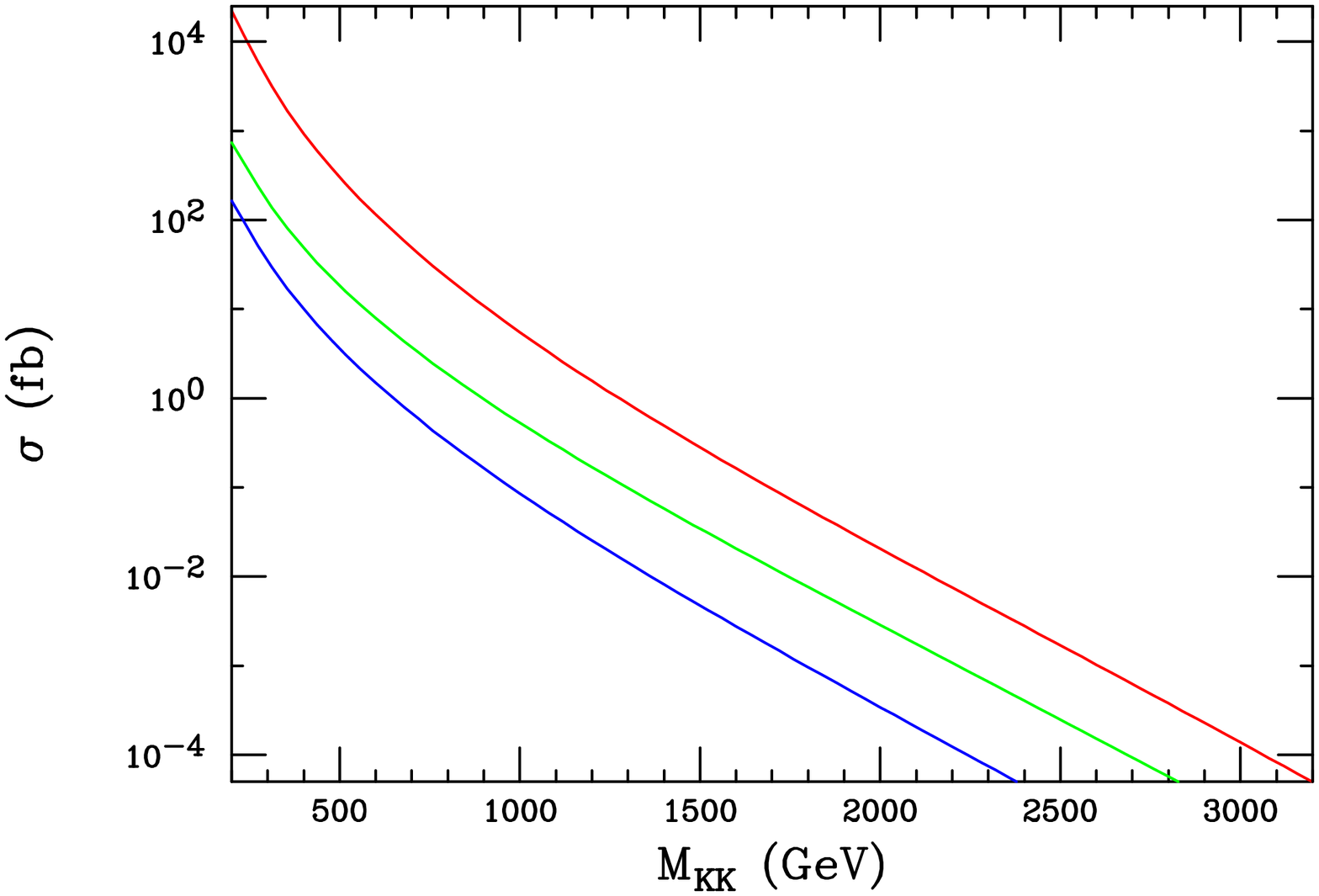}}
\caption{Cross sections for the production of the lightest color singlet KK 
states at the LHC: in the left panel, from top to bottom, for $2Z^{1)},
\gamma^{(1)}Z^{(1)}$ and $2\gamma^{(1)}$ final states; in the right panel, 
from top to bottom, for $W^{(1)+}W^{(1)-},W^{(1)\pm}Z^{(1)}$ and 
$W^{(1)\pm}\gamma^{(1)}$ final states.}
\label{fig4}
\end{figure}

The real signature of the UED scenario is that {\it all} of the SM fields 
have KK excitations. Thus we will also want to observe the production of the 
SM color singlet states. 
Of course color singlet states can also be produced, with the largest cross 
sections being for associated production with a colored state at order 
$\alpha \alpha_s$; these rates are of course smaller than for pairs of colored 
particles as can be seen in Fig.~\ref{fig3}. 
Here we see reasonable rates are obtained 
for KK masses in excess of $\simeq 1.8$ TeV or so. 
Lastly, it is possible to pair produce color singlets via electroweak 
interactions which thus lead to cross sections of order $\alpha^2$. Due to the 
large center of mass energy of the LHC these cross sections can also lead to 
respectable production rates for KK masses as great as $\simeq 1.5$ TeV 
as can be seen from Fig.~\ref{fig4}. 
It is clear from this analysis that the LHC will have 
a significant search reach 
for both colored and non-colored KK states provided that the production 
signatures are reasonably distinct.

\vspace*{-0.5cm}
\nn
\vspace*{0.4mm}
\vspace*{-0.5cm}
\nn
\vspace*{0.4mm}

\section{Signatures}

When examining collider signatures for KK pair production in the UED there are 
two important questions to ask: ($i$) Are the lightest KK modes stable and 
($ii$) if they are unstable what are their decay modes? From the discussion 
above it is clear that without introducing any new physics the $n=1$ KK states 
{\it are} stable so we must consider this possibility when looking at 
production signatures.

In their paper ACD {\cite {Appelquist:2000nn}} argue that cosmological constraints possibly 
suggest that KK states in the TeV mass range must be unstable on cosmological 
time scales. (Of course this does not mean that they would appear unstable 
on the time scale of a collider experiment in which case our discussion is 
the same as that above.) This would require the introduction of new physics 
beyond that contained in the original UED model. There are several possible 
scenarios for such new physics. Here we will discuss three possibilities in 
what follows, the first two of which 
were briefly mentioned by ACD {\cite {Appelquist:2000nn}}.

\begin{figure}[tb]
\centerline{
\includegraphics[width=5.3cm,angle=90]{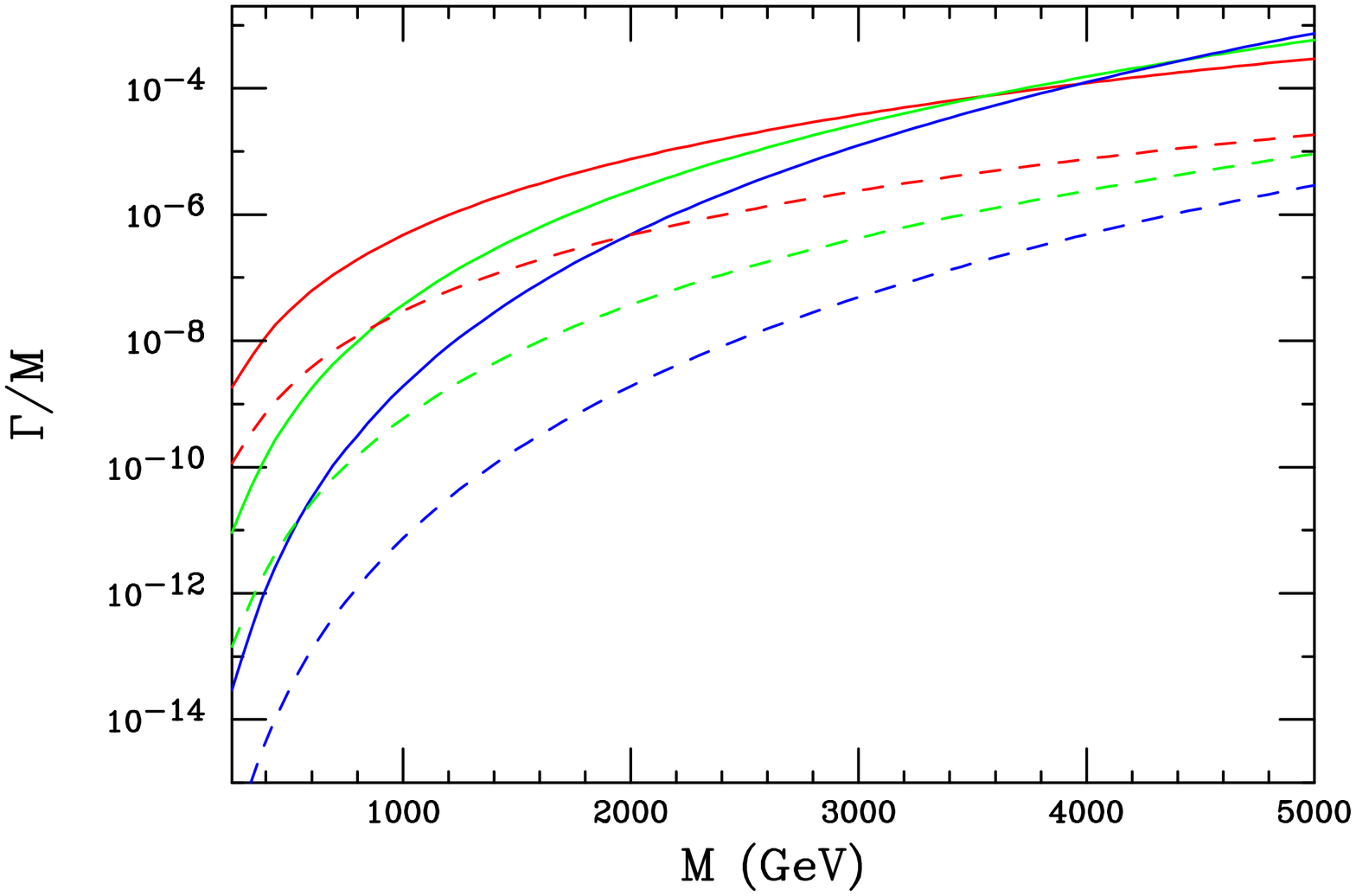}
\includegraphics[width=5.3cm,angle=90]{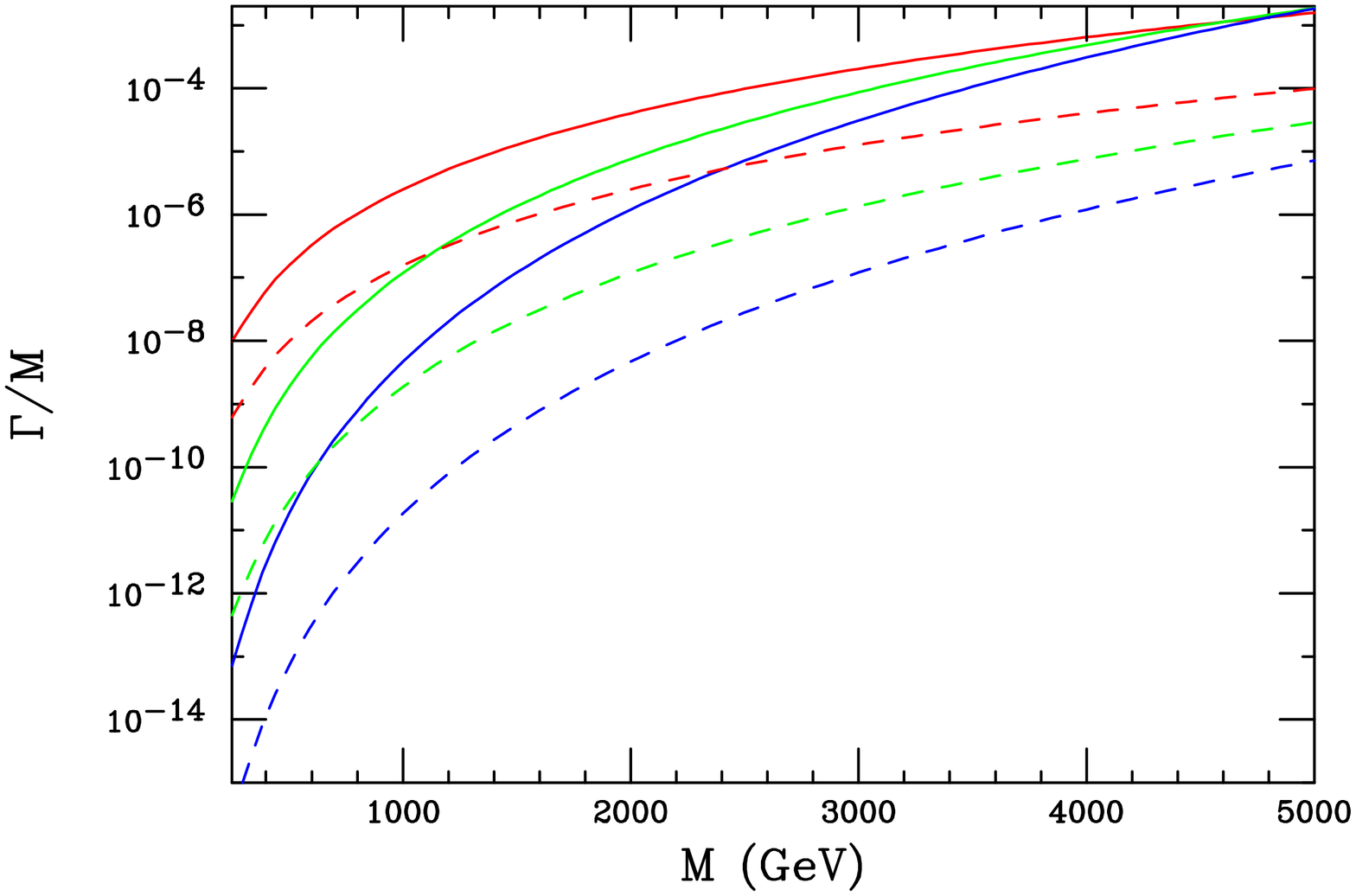}}
\vspace*{0.25cm}
\centerline{
\includegraphics[width=5.3cm,angle=90]{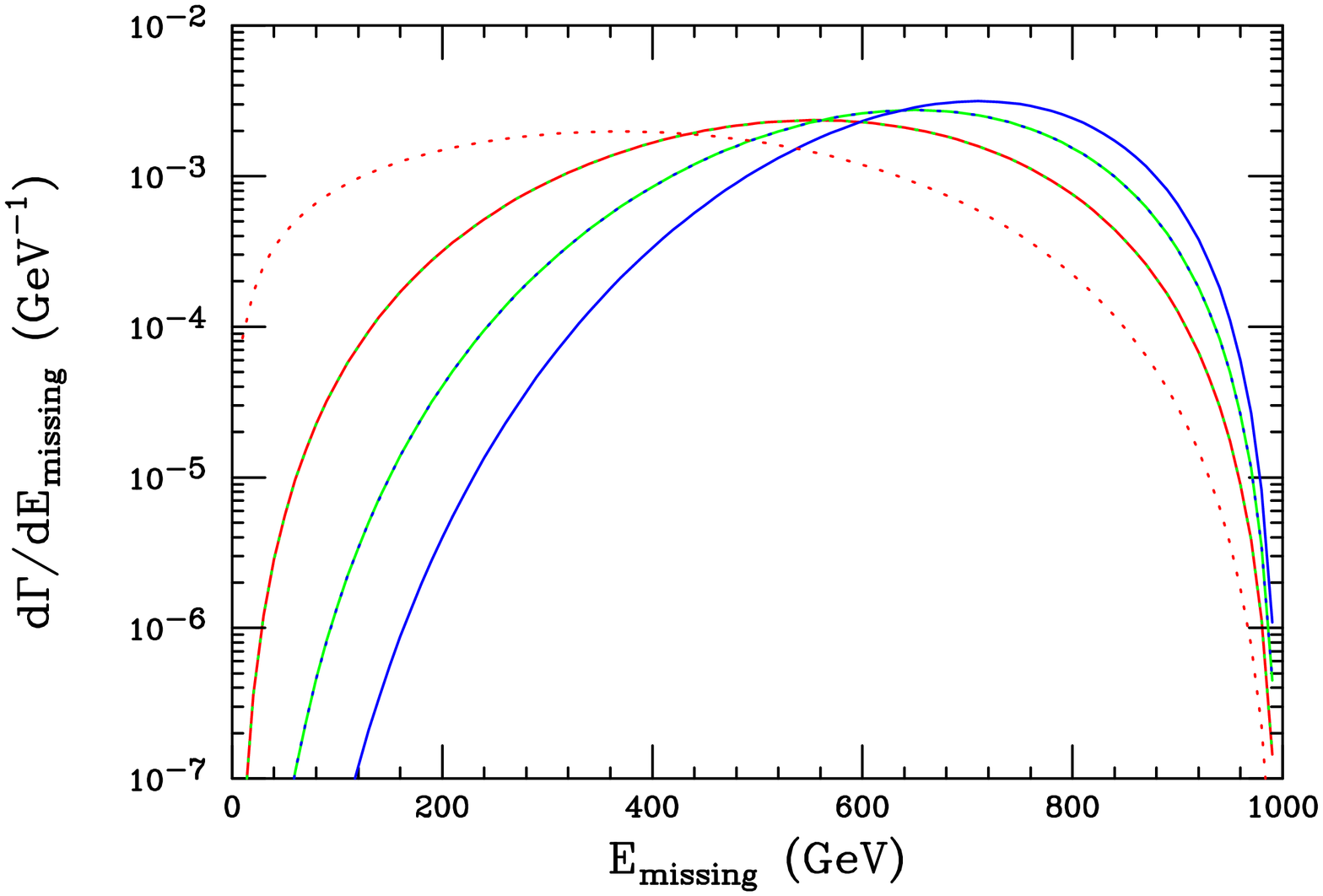}}
\vspace*{0.25cm}
\caption{Width for the decay of the first excited KK state (even-top left 
panel, 
odd-top right panel) into the corresponding zero mode and a graviton tower as a 
function of the mass of the KK state. The solid(dashed) lines are for 
$M_D=5(10)$ TeV and from top to bottom in each case the curves correspond to 
$\delta=2,4,6$, respectively. The lower panel shows the missing energy 
distribution for these decays for these same cases assuming a KK mass of 1 
TeV.}
\label{fig5}
\end{figure}

Scenario I: The TeV$^{-1}$-scale UED model is embedded inside a thick brane 
in a higher $(\delta+4)$-dimensional space, with a compactification scale 
$R_G>>R_c$, 
in which gravity is allowed to propagate in a manner similar to the model of 
Arkani-Hamed, Dimopoulos and 
Dvali {\cite{Arkani-Hamed:1998rs,Arkani-Hamed:1998nn,Antoniadis:1998ig}}. 
Since the graviton wave 
functions are normalized on a torus of volume $(2\pi R_G)^\delta$ while the KK 
states are normalized over $2\pi R_c$ the overlap of a KK zero mode with any 
even or odd KK tower state $n$ and a graviton will be non-zero. In a sense, 
the brane develops 
a transition form-factor analogous to that described in {\cite{DeRujula:2000he}}. 
This induces transitions of the form $KK(n=1)\to KK(n=0)+G$ where $G$ 
represents the graviton field which appears as missing energy in the collider 
detector. This means that production of a pair of KK excitations of, \eg,  
quarks or gluons would appear as two jets plus missing energy in the detector; 
the corresponding production of a KK excited pair of gauge bosons would appear 
as the pair production of the corresponding zero modes together with missing 
energy. We can express this form-factor simply as 
\begin{equation}
{\cal F}={\sqrt 2\over {\pi R_c}}\int_0^{\pi R_c}dy
e^{im_gy}(\cos ny/R_c,\sin ny/R_c)\,,
\end{equation}
for even and odd KK states, respectively, 
where $m_g$ is the graviton mass. Here we have assumed that the thick brane 
resides at $y_i=0$ for all $i\neq 1$. 
Given these form-factors we can calculate the actual decay rate for
$KK(n=1)\to KK(n=0)+G$, where we now must 
sum up the graviton towers by following the 
analyses in Ref.~{\cite {Giudice:1998ck,Han:1998sg}}; 
this result should be relatively independent 
of the spin of the original KK state. 
Performing the necessary integrations numerically we 
obtain the results shown in Fig.~\ref{fig5}. 
\nn
This figure shows that this mechanism 
provides for a very rapid decay over almost all of the parameter space. For 
light KK states with both $\delta$ and $M_D$ large the decay rate is suppressed 
and may lead to finite length charged tracks in the detector. (In particular 
the production of a charged KK state with a long lifetime would yield a 
kink-like track structure 
due to the decay to the graviton tower.) Although not a 
true two-body decay, Fig.~\ref{fig5} 
also shows that the typical missing energy in the 
gravitational decay of a KK state will be close to half its mass, which is 
quite significant for such heavy objects. It is clear that events with such a 
large fraction of missing energy should be observable above background given 
sufficient event rates. These events will not be confused with SUSY signals 
since they occur in every possible channel. 

Scenario II: KK decays can be induced in the UED model by adding a `benign' 
brane at some $y=y_0$ which induces new interactions. By `benign' we mean 
that these new interactions only do what we need them to do and do not alter 
the basic properties of the UED model. The simplest form of such interactions 
are just the four dimensional variants of the terms in the the 5-d UED 
action. For example, one might add a term such as 
\begin{equation}
\int d^4x \int dy ~\delta(y-y_0) ~{\lambda\over {M_s}} \bar \psi \gamma^A 
{\cal D}_A \psi \,,
\end{equation}
where $\lambda$ is some Yukawa-like coupling and $M_s$ is some large scale.  
Note that the brane is placed at some arbitrary position $y=y_0$ and {\it not} 
at the fixed points where only even KK modes would be effected. These new 
interactions result in a mixing of all KK states both even and odd and, in 
particular, with the zero mode. Thus we end up inducing decays of the form 
KK$^{(1)}\to$ KK$^{(0)}$ KK$^{(0)}$. For KK fermions the decay into a fermion 
plus gauge boson zero mode is found to be given by
\begin{equation}
\Gamma(f^{(1)}\to f^{(0)}V^{(0)})={g_V^2\over {8\pi}} s_\phi^2 M_c \cdot N_c 
\cdot PS
\,,
\end{equation}
where $s_\phi$ is the induced mixing angle, $N_c$ is a color factor, $g_V$, 
the relevant gauge coupling and PS is the phase space for the decay. It is 
assumed that the mixing angle is sufficiently small that single production 
of KK states at colliders remains highly suppressed but is large enough for 
the KK state to decay in the detector. For $\lambda \simeq 0.1$ and $M_s 
\simeq$ a few $M_c$ this level of suppression is quite natural. (Numerically, 
it is clear that the KK state will decay inside the detector unless the mixing 
angle is very highly suppressed.) The resulting branching fractions can 
be found in Table 1 where we see numbers that are not too different than those 
for excited fermions in composite models with similar decay signatures. 
However, unlike excited SM fields, single production modes are highly 
suppressed. For KK excitations of the gauge 
bosons, their branching fractions into zero mode fermions will be identical to 
those of the corresponding SM fields apart from corrections due to phase 
space, \ie, the first excited $Z$ KK state can decay to $t\bar t$ while the 
SM $Z$ cannot.

\begin{table}
\centering
\begin{tabular}{|l|c|c|c|c|} \hline\hline
       &  $g$  &  $\gamma$  &  $Z$  &  $W$  \\ \hline \hline
 $e^{(1)}$&   0   &   41.0&     14.4&   44.6\\
 $\nu^{(1)}$&   0   &   0&     39.1&   60.9\\
 $u^{(1)}$&   89.8   &    2.3&      2.1&    5.7\\
 $d^{(1)}$&   90.9   &    0.6&      2.7&    5.8\\ \hline\hline
\end{tabular}
\caption{Individual branching fractions in per cent 
for the first excited fermion KK modes when KK level mixing occurs as in 
Scenario II.}
\end{table}

\begin{figure}[tb]
\centerline{
\includegraphics[width=5.3cm,angle=90]{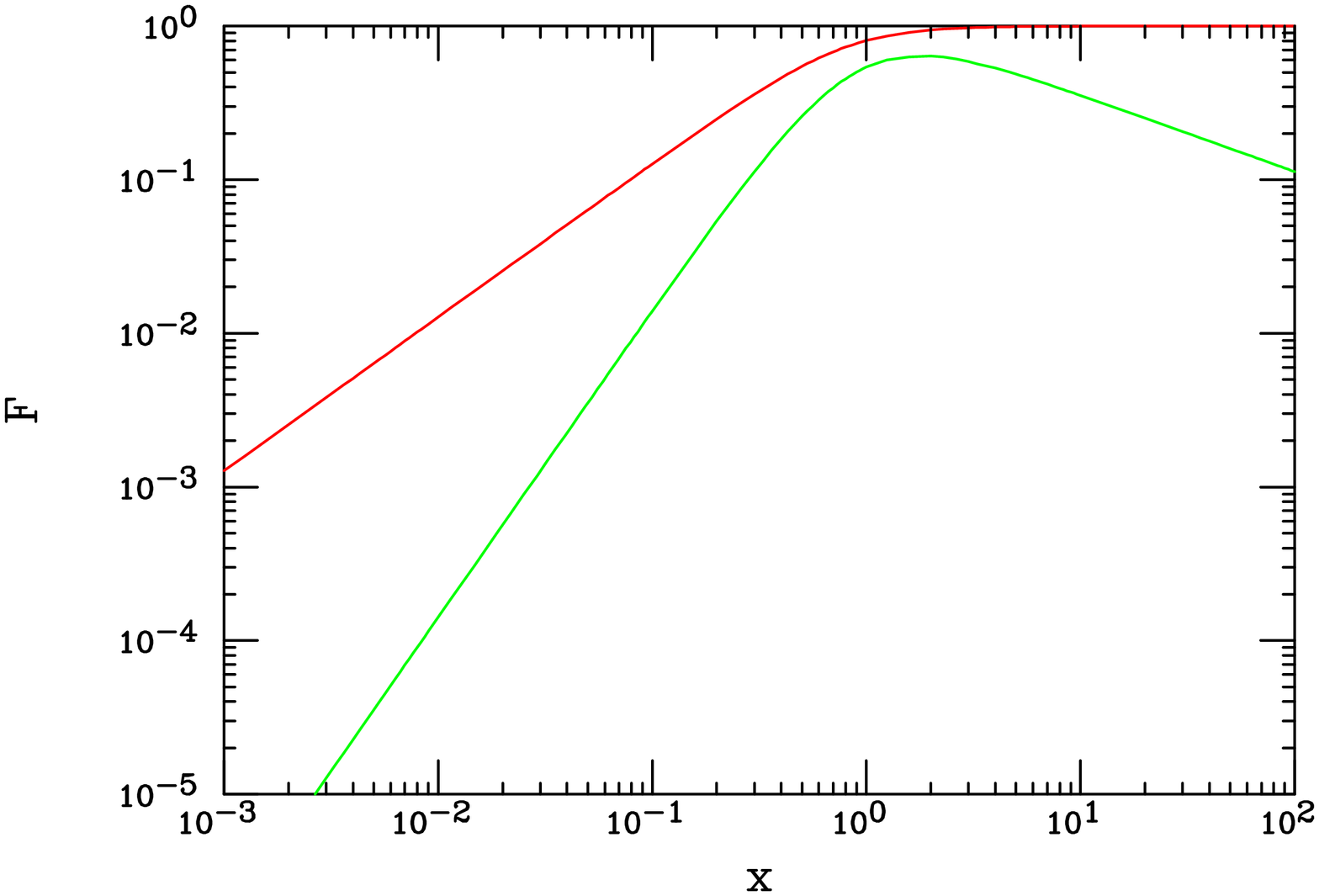}}
\vspace*{0.2cm}
\caption[*]{The form factors ${\cal G}$(upper curve) and ${\cal G}$'(lower 
curve) as discussed in the text for the case $n=1$.}
\label{fig6}
\end{figure}

Scenario III: We can add a common bulk mass term to the fermion action, \ie, 
a term of the form $m(\bar DD+\bar SS)$, where $D$ and $S$
represent the 5-dimensional isodoublet and isosinglet SM fields.
We chose a common mass both for simplicity and to avoid 
potentially dangerous flavor changing neutral currents. The largest influence 
of this new term is to modify the zero mode fermion wavefunction which is now 
no longer flat and takes the form $\sim e^{-m|y|}$
and thus remains $Z_2$-even. Clearly there is now a significant overlap in the 
5-d wavefunctions 
between pairs of fermion zero modes and any $Z_2$-even gauge KK mode which can  
be represented as another form-factor, ${\cal G}(x)$, where $x=mR_c$. 
This form factor then describes 
the decay $G^{(n)}\to \bar f^{(0)}f^{(0)}$ where $G$ represents a generic 
KK gauge 
field. Similarly we can obtain a form-factor that describes the corresponding 
decay $f^{(n)}\to G^{(0)}f^{(0)}$ which we denote by ${\cal G}'(x)$ 
where $x$ is as above. It is clear that the decays of KK states in this 
scenario will be essentially identical to Scenario II above although they are 
generated by a completely different kind of physics. 
Fig. ~\ref{fig6} shows the shape of these two form-factors as 
a function of the parameter $x$. The natural question to ask at this point 
is `what is the value of $m$ relative to $M_c$?'. It seems natural to imagine 
that the bulk mass would be of order the compactification scale, the only 
natural scale in the action, which would 
imply that $x\sim 1$ so that large form-factors would be obtained. While this 
scenario works extremely well for the decay of $Z_2$-even states it does not 
work at all for the case of $Z_2$-odd states.

\vspace*{-0.5cm}
\nn

\section{Summary and Conclusions}

In this paper we have begun a detailed examination of the predictions of the 
Universal Extra Dimensions model for future colliders. Since indirect 
searches for such states give rather poor reaches, direct 
searches are of greater importance in this model. To 
obtain interesting search reaches requires a hadron collider such as the 
Tevatron or LHC. Based on counting events we expect the reach at the Tevatron 
Run II (LHC) for KK states to be $\simeq 600(3000)$ GeV. Within the UED 
model itself these lightest KK states are stable even when loop corrections 
are included unless new interactions are introduced from elsewhere. If these 
states are indeed stable, the production of a large number of heavy stable 
charged
particles would not be missed at either collider. It is more likely, however, 
that new physics does indeed enter, rendering the KK modes unstable. In this 
paper we have examined three new physics scenarios that induce finite KK 
lifetimes and compared their decay signatures.

\setcounter{figure}{0}
\setcounter{table}{0}
\setcounter{section}{0}
\setcounter{equation}{0}
\clearpage

%
%
%
%
%
%
%

\def\MP{\mbox{$M_P$}}
\def\mp{\mbox{$M_P$}\ }
\def\TH{\mbox{$T_H$}\ }
\def\mbh{\mbox{$M_{\rm BH}$}\ }     
\def\MBH{\mbox{$M_{\rm BH}$}}         
\def\MET{\mbox{${\hbox{$E$\kern-0.6em\lower-.1ex\hbox{/}}}_T$}} 
\def\met{\mbox{${\hbox{$E$\kern-0.6em\lower-.1ex\hbox{/}}}_T$}\ } 
\def\ifb{fb$^{-1}$}                     
\def\etal{{\sl et al.}}                 
\def\vs{{\sl vs.}}                      
\def\et{\mbox{$E_T$}}




\part{{\bf Black Hole Production at Future Colliders
} \\[0.5cm]\hspace*{0.8cm}
{\it S. Dimopoulos and G. Landsberg
}}
\label{landsberg1sec}







\begin{abstract}
If the scale of quantum gravity is near a TeV, the 
CERN Large Hadron Collider will be producing one black hole (BH) 
about every second. The decays of the BHs into the final states 
with prompt, hard photons, electrons, or muons provide a clean signature 
with low background. The correlation between the BH mass and its 
temperature, deduced from the energy spectrum of the decay products, 
can test Hawking's evaporation law and determine the number of large 
new dimensions and the scale of quantum gravity. We also consider BH 
production at the proposed future high-energy colliders, such as CLIC 
and VLHC, and describe the Monte Carlo event generator that can be 
used to study BH production and decay.
\end{abstract}


%
%

%
%

\section{Introduction}

An exciting consequence of TeV-scale quantum gravity 
\cite{Arkani-Hamed:1998rs,Antoniadis:1998ig,Arkani-Hamed:1998nn} 
is the possibility of production of black holes 
(BHs)~\cite{Argyres:1998qn,Banks:1999gd,Emparan:2000rs,%
Dimopoulos:2001hw,Giddings:2001bu} at the LHC and beyond. 
This paper summarizes and extends our pioneer work 
on this subject~\cite{Dimopoulos:2001hw} to the post-LHC future 
and discusses additional aspects of black-hole 
phenomenology left out from~\cite{Dimopoulos:2001hw} due to lack of space. 
Since this work has been completed, numerous follow-up 
publications on this exciting subject have appeared in the archives, 
focusing on both the 
collider~\cite{Voloshin:2001vs,Dimopoulos:2001qe,%
Hossenfelder:2001dn,Cheung:2001ue} and cosmic 
ray~\cite{Feng:2001ib,Anchordoqui:2001ei,Emparan:2001kf} 
production. We hope that this new branch 
of phenomenology of extra dimensions will flourish in the months 
to come, as black hole production might be the very first 
evidence for the existence of large extra dimensions.

Black holes are well understood general-relativistic objects 
when their mass \mbh far exceeds the fundamental (higher dimensional) 
Planck mass $\MP \sim$TeV. As \mbh approaches \MP, the BHs 
become ``stringy'' and their properties complex. In what 
follows, we will ignore this obstacle and 
estimate the properties of light BHs by simple 
semiclassical arguments, strictly valid for $\MBH \gg \MP$. We expect that this 
will be an adequate approximation, since the important experimental
signatures rely on two simple qualitative properties: (i) the
absence of small couplings and (ii) the ``democratic" nature of BH 
decays, both of which may survive as average properties of the 
light descendants of BHs. Nevertheless, because of the 
unknown stringy corrections, our results are approximate estimates. 
For this reason, we will not attempt selective partial improvements -- 
such as time dependence, angular momentum, charge, hair, and other 
higher-order general relativistic refinements -- which, for light BHs, 
may be masked by larger unknown stringy effects. We will focus on 
the production and sudden decay of Schwarzschild black holes.

\section{Production} 

The Schwarzschild radius $R_S$ of an
$(4+n)$-dimensional black hole is given by \cite{Myers:1986un}, assuming 
that extra dimensions are large ($\gg$ $R_S$).

Consider two partons with the center-of-mass (c.o.m.) energy $\sqrt{\hat s} =
\MBH$ moving in opposite directions. Semiclassical reasoning
suggests that if the impact parameter is less than the (higher
dimensional) Schwarzschild radius, a BH with the mass \mbh forms.
Therefore the total cross section can be estimated from
geometrical arguments~\cite{footnote1}, and is of order
$$
    \sigma(\MBH) \approx \pi R_S^2 = \frac{1}{M_P^2}
    \left[
      \frac{\MBH}{\MP} 
      \left( 
        \frac{8\Gamma\left(\frac{n+3}{2}\right)}{n+2}
      \right)
    \right]^\frac{2}{n+1}
$$
(see Fig.~\ref{Landsberg1_fig1}a,d)~\cite{footnote0}.

This expression contains no small coupling constants; if the
parton c.o.m. energy $\sqrt{\hat s}$ reaches the fundamental
Planck scale $\MP \sim$~TeV then the cross section is of order
TeV$^{-2} \approx 400$~pb. At the LHC or VLHC, with the total c.o.m. energy 
$\sqrt{s}=14$~TeV or 100-200 TeV, respectively, BHs will 
be produced copiously. To calculate total production cross 
section, we need to take into account that only a
fraction of the total c.o.m. energy in a $pp$ collision is
achieved in a parton-parton scattering. We compute the full
particle level cross section using the parton luminosity
approach (see, e.g., Ref.~\cite{Eichten:1984eu}):
$$
    \frac{d\sigma(pp \to \mbox{BH} + X)}{d\MBH} = 
    \frac{dL}{dM_{\rm BH}} \hat{\sigma}(ab \to \mbox{BH})
    \left|_{\hat{s}=M^2_{\rm BH}}\right.,
$$
where the parton luminosity $dL/d\MBH$ is defined as the sum over
all the initial parton types:
$$
    \frac{dL}{dM_{\rm BH}} = \frac{2\MBH}{s} 
    \sum_{a,b} \int_{M^2_{\rm BH}/s}^1  
    \frac{dx_a}{x_a} f_a(x_a) f_b(\frac{M^2_{\rm BH}}{s x_a}),
$$
and $f_i(x_i)$ are the parton distribution functions (PDFs). We
used the MRSD$-'$~\cite{Martin:1993zi} PDF set with the $Q^2$ scale taken
to be equal to \MBH~\cite{footnote6}, which is within the allowed range for this
PDF set, up to the VLHC kinematic limit. Cross section dependence
on the choice of PDF is $\approx 10\%$.

\begin{figure}[tb]
\begin{center}
\includegraphics[width=3.2in]{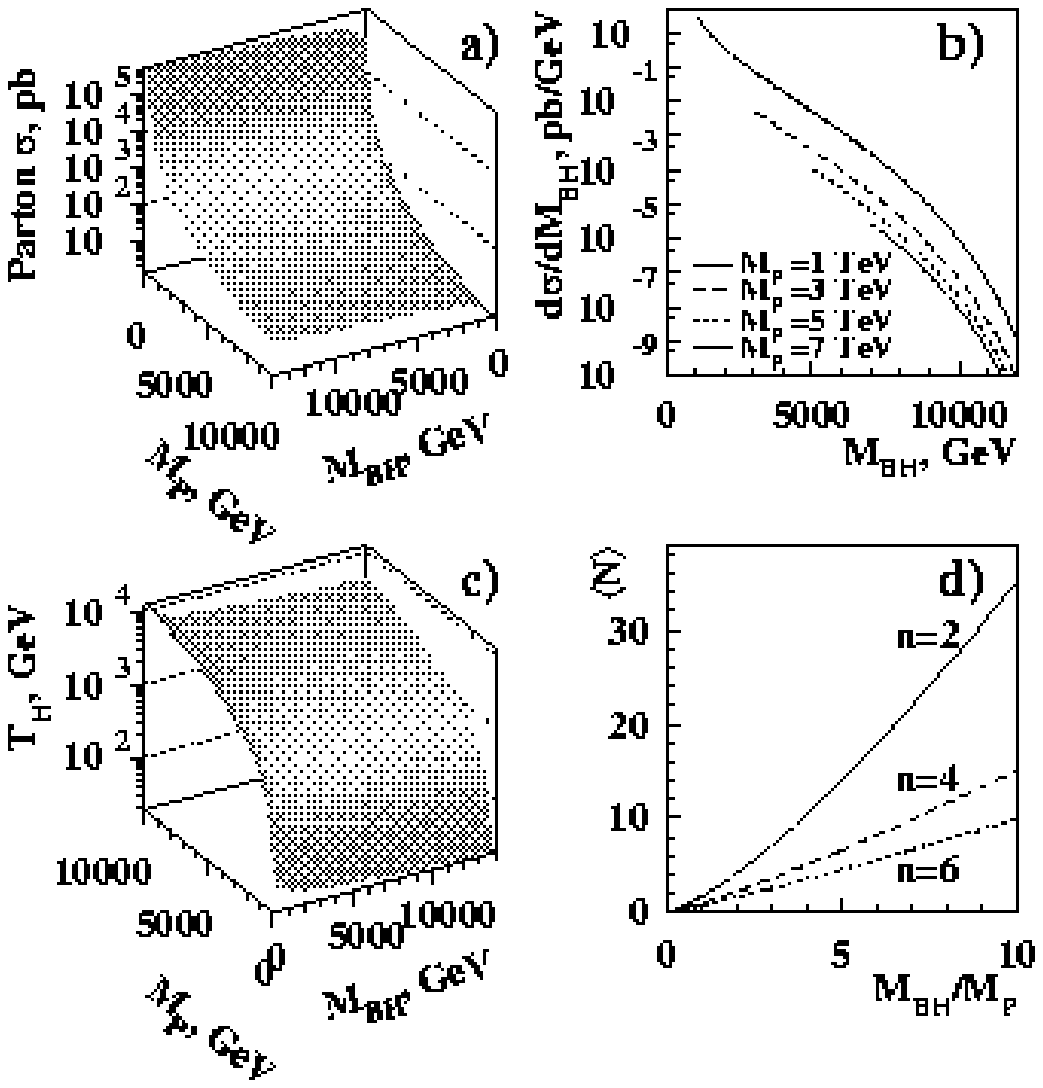}
\includegraphics[width=3.2in]{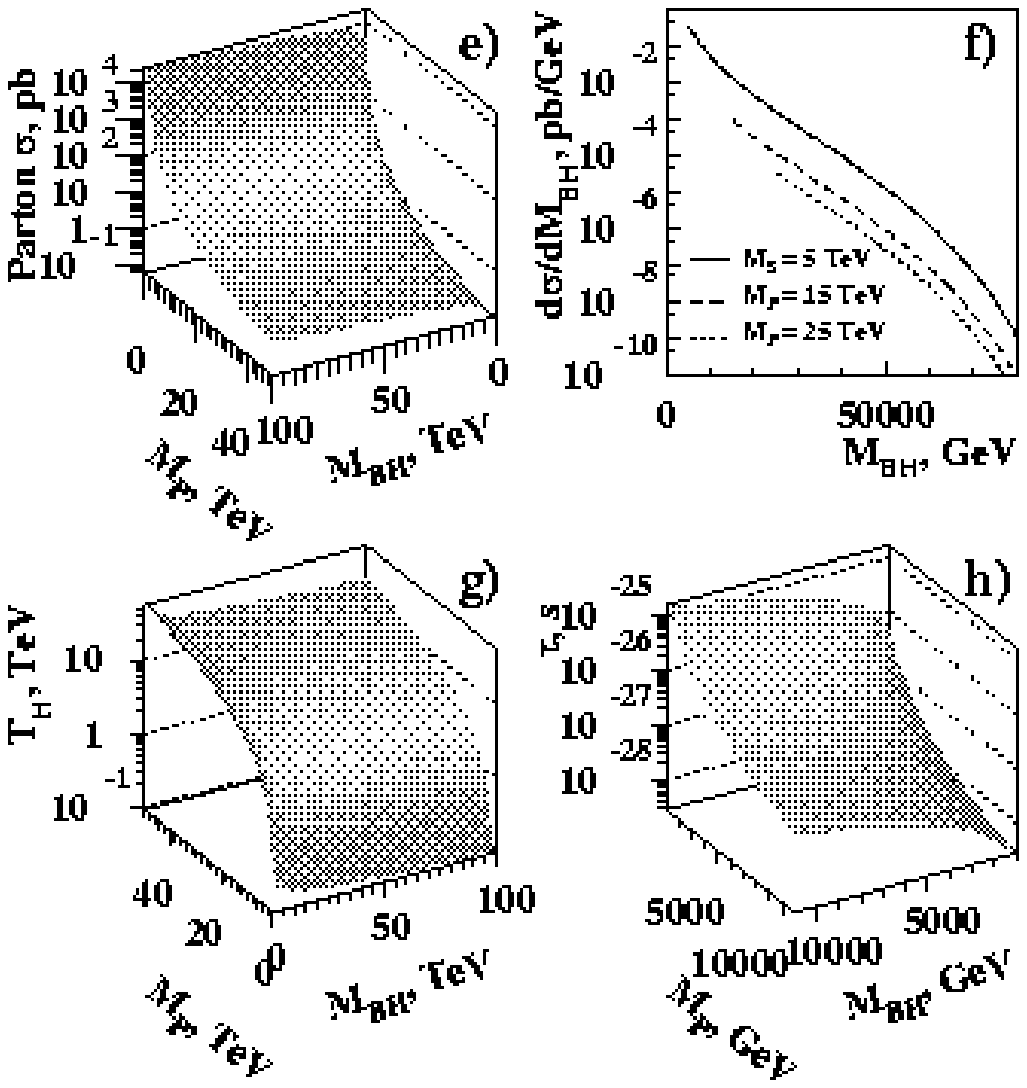}
\end{center}
\caption{Black-hole properties at the LHC a)-d),h) and VLHC d)-h). a,e) Parton-level production cross section; b,f) differential cross section $d\sigma/d\MBH$; c,g) Hawking 
temperature; d) average decay multiplicity for a Schwarzschild black hole; and h) black-hole lifetime. The number of extra spatial dimensions $n=4$ is used for a)-c), e)-h). The dependence of the cross section and the Hawking temperature on $n$ is weak and would be hardly noticeable on the logarithmic scale. The lifetime drops by about two orders of magnitude for $n$ increase from 2 to 7.}
\label{Landsberg1_fig1}
\end{figure}

\begin{figure}[tb]
\begin{center}
\includegraphics[width=3.in]{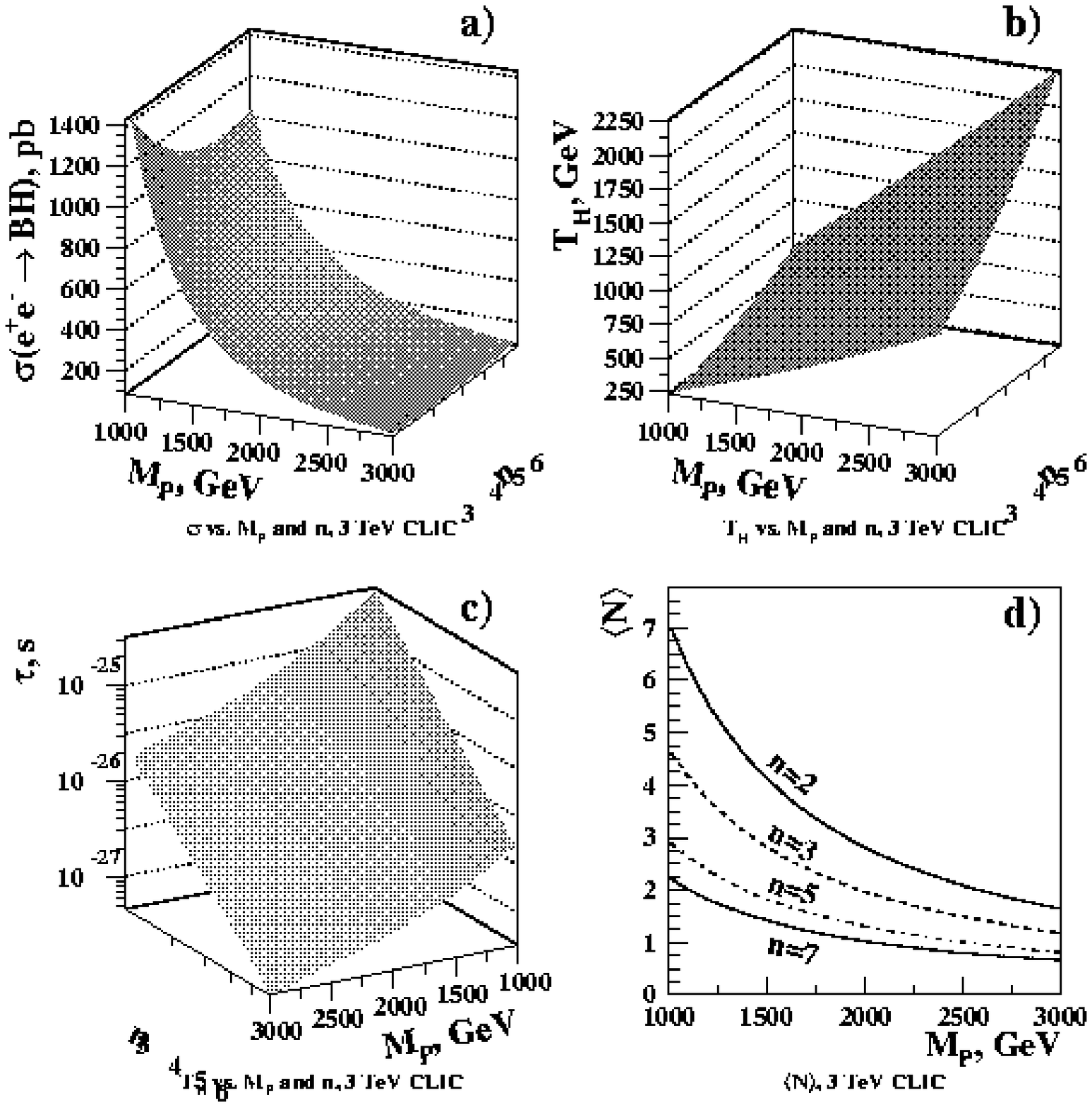}
\includegraphics[width=3.in]{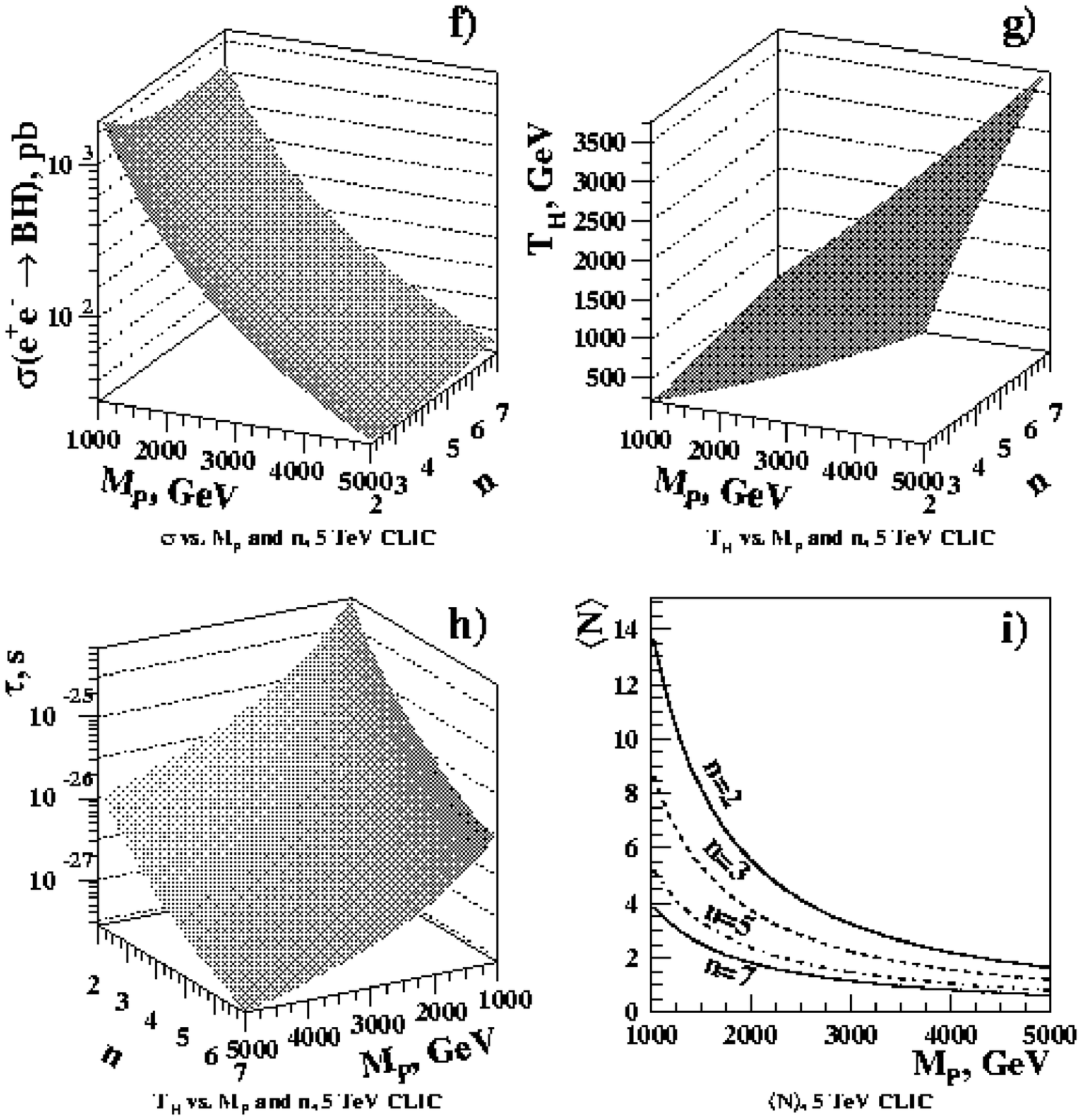}
\medskip
\includegraphics[width=3.in]{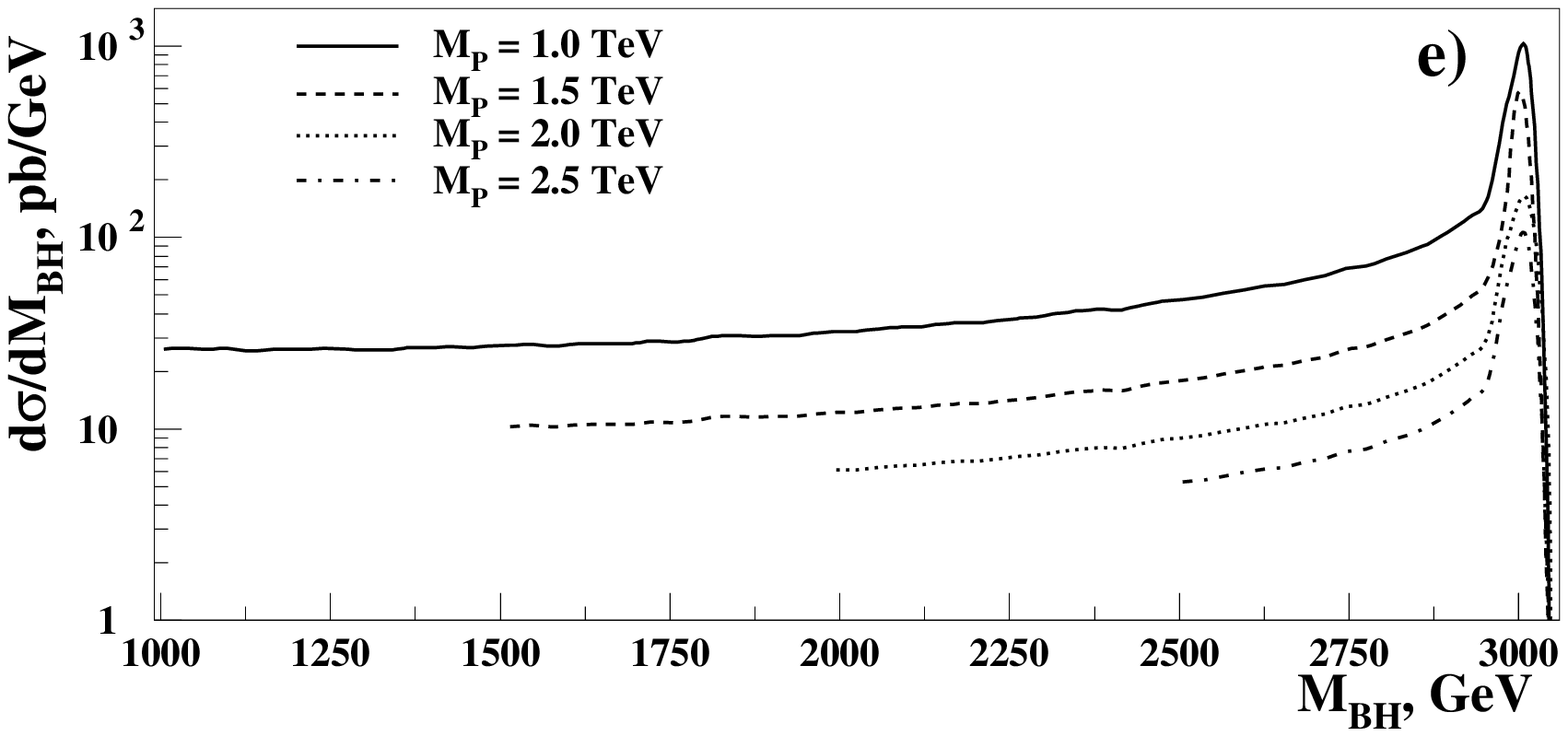}
\includegraphics[width=3.in]{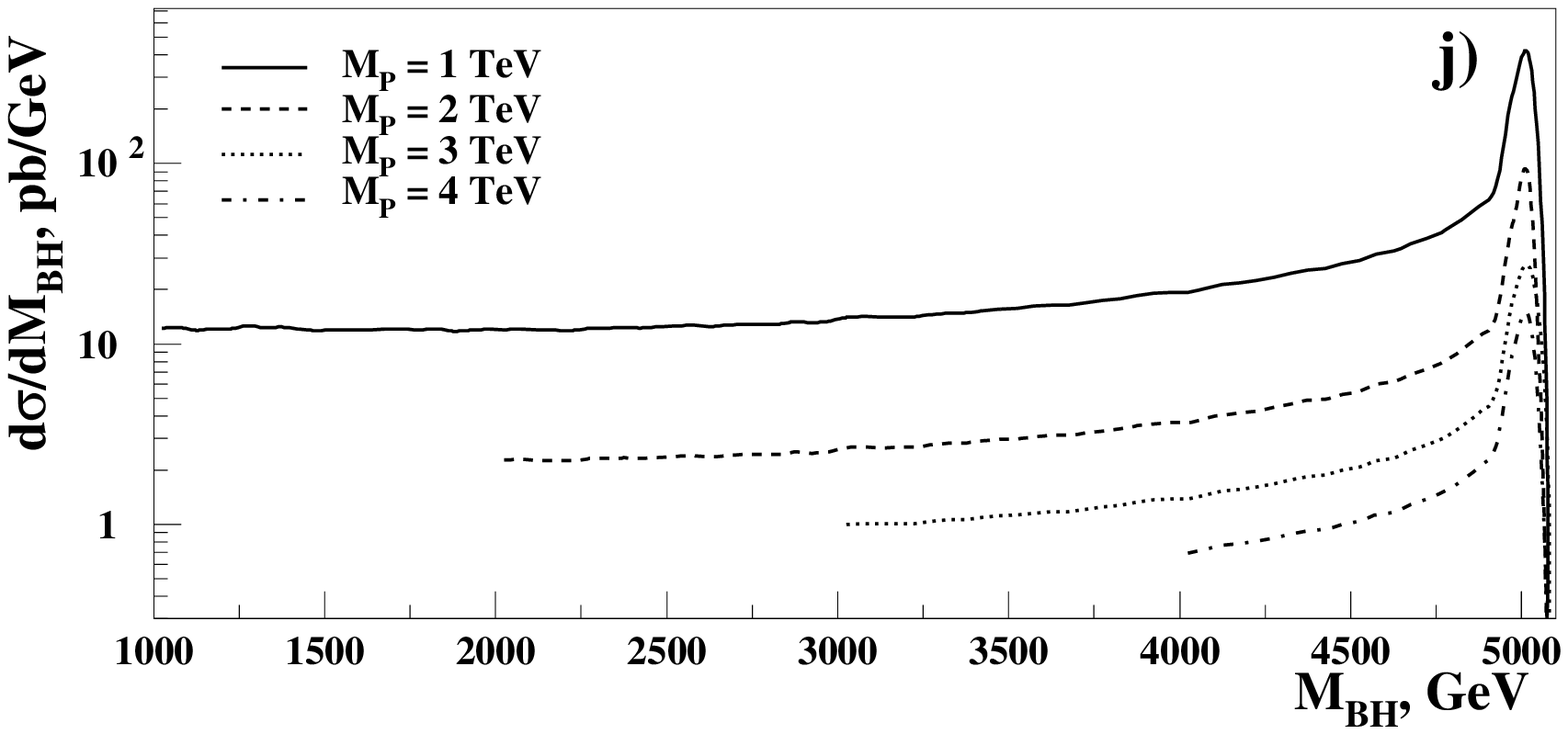}
\end{center}
\caption{Black hole properties at high-energy lepton colliders. Plots a)-d) and f)-i) correspond to the properties (production cross-section, temperature, lifetime, and average decay multiplicity) of a fixed-mass 3 TeV and 5 TeV black hole produced at a 3 TeV or a 5 TeV machine, respectively. Plots e),j) show the differential cross section of BH production for $n=4$, as a function of the BH mass at a 3 TeV or a 5 TeV CLIC $e^+e^-$-collider, respectively.}
\label{Landsberg1_fig2}
\end{figure}

The differential cross sections $d\sigma/d\MBH$ for the BH produced
at the LHC and a 200 TeV VLHC machines are shown in 
Figs.~\ref{Landsberg1_fig1}b and \ref{Landsberg1_fig1}f, 
respectively, for several choices of \MP. The total production cross 
section at the LHC for BH masses above \mp ranges from 0.5~nb 
for $\MP = 2$~TeV, $n=7$ to 120~fb for $\MP = 6$~TeV and $n=3$. 
If the fundamental Planck scale is
$\approx 1$~TeV, the LHC, with the peak luminosity of 30~\ifb/year
will produce over $10^7$ black holes per year. This is comparable to
the total number of $Z$'s produced at LEP, and suggests that we may
do high precision studies of TeV BH physics, as long as the backgrounds
are kept small. At the VLHC, BHs will be produced copiously for their masses and the value of the fundamental Planck scale as high as 25 TeV. The total production cross section is of the order of a millibarn for $\MP = 1$~TeV and of order a picobarn for $\MP = 25$~TeV.

Similarly, the black holes can be produced at future high-energy lepton colliders, such as CLIC or a muon collider. To a first approximation, such a machine produces black holes of a fixed mass, equal to the energy of the machine. The total cross section of such a BH produced at a 3~TeV and a 5 TeV machine, as a function of \mp and $n$, is shown in Fig~\ref{Landsberg1_fig2}a and Fig.~\ref{Landsberg1_fig2}f, respectively. For more elaborated studies of the BH production at electron colliders, one should take into account machine {\it beamstrahlung\/}. The beamstrahlung-corrected energy spectrum of the machine plays the same role as the parton luminosity at a hadron collider, except that for the $e^+e^-$ machine it is peaked at the nominal machine energy, rather than at small values of $\sqrt{\hat s}$, characteristic of a hadron collider. Using typical beamsstrahlung spectra expected for a 3 TeV or a 5 TeV CLIC machine, we show the differential cross section $d\sigma/d\MBH$ of the black hole production at a 3 and a 5 TeV CLIC in Fig.~\ref{Landsberg1_fig2}e and Fig.~\ref{Landsberg1_fig2}j, respectively.

\section{Decay} 

The decay of the BH is governed by its Hawking
temperature $T_H$, which is proportional to the inverse radius, and
given by~\cite{Myers:1986un}:
\begin{equation}
    T_H = \MP
    \left(
      \frac{\MP}{\MBH}\frac{n+2}{8\Gamma\left(\frac{n+3}{2}\right)}
    \right)^\frac{1}{n+1}\frac{n+1}{4\sqrt{\pi}} = \frac{n+1}{4\pi R_S}
\label{Landsberg1_eq1}
\end{equation}
(see Figs.~\ref{Landsberg1_fig1}c,g and \ref{Landsberg1_fig2}b,g). As the collision energy increases, 
the resulting BH gets heavier and its decay products get colder.

Note that the wavelength $\lambda = {2 \pi \over T_H}$
corresponding to the Hawking temperature is larger than the size
of the black hole. Therefore, the BH acts as a
point-radiator and emits mostly $s$-waves. This
indicates that it decays equally to a particle on the brane and in
the bulk, since it is only sensitive to the radial coordinate and
does not make use of the extra angular modes available in the
bulk. Since there are many more particles on our brane than in the
bulk, this has the crucial consequence that the BH decays
visibly to standard model (SM) particles~\cite{Emparan:2000rs,lenny}.

The average multiplicity of particles produced in the process of
BH evaporation is given by: $\langle N \rangle = \left\langle
\frac{\MBH}{E} \right\rangle$, where $E$ is the energy spectrum
of the decay products. In order to find $\langle N \rangle$, we note
that the BH evaporation is a blackbody radiation process, with the 
energy flux per unit of time given by Planck's formula:
$\frac{df}{dx} \sim \frac{x^3}{e^x + c}$, where $x \equiv E/T_H$, and
$c$ is a constant, which depends on the quantum statistics of the
decay products ($c = -1$ for bosons, $+$1 for fermions, and 0 for
Boltzmann statistics).

The spectrum of the BH decay products in the massless
particle approximation is given by: $\frac{dN}{dE} \sim
\frac{1}{E}\frac{df}{dE} \sim \frac{x^2}{e^x + c}$. For averaging
the multiplicity, we use the average of the distribution in the 
inverse particle energy:
\begin{equation}
    \left\langle \frac{1}{E} \right\rangle =
    \frac{1}{T_H}\frac{\int_0^\infty dx \frac{1}{x}
    \frac{x^2}{e^x + c}}{\int_0^\infty dx\frac{x^2}{e^x + c}}
    = a/T_H,
\label{Landsberg1_eq2}
\end{equation}
where $a$ is a dimensionless constant that depends on the type of
produced particles and numerically equals 0.68 for bosons, 0.46
for fermions, and $\frac{1}{2}$ for Boltzmann statistics. Since a
mixture of fermions and bosons is produced in the BH decay, we can
approximate the average by using Boltzmann statistics, which gives
the following formula for the average multiplicity: $\langle N
\rangle \approx \frac{\MBH}{2T_H}$. Using Eq. (\ref{Landsberg1_eq1}) for
Hawking temperature, we obtain:
\begin{equation}
    \langle N \rangle = \frac{2\sqrt{\pi}}{n+1}
    \left(\frac{\MBH}{\MP}\right)^\frac{n+2}{n+1}
    \left(\frac{8\Gamma\left(\frac{n+3}{2}\right)}{n+2}\right)^\frac{1}{n+1}\!\!\!\!\!\!\!\!\!\!\!\!.
\label{Landsberg1_eq3}
\end{equation}

Eq. (\ref{Landsberg1_eq3}) holds for $\MBH \gg T_H$, i.e. $\langle N \rangle \gg 1$; otherwise, the Planck spectrum is truncated at $E \approx \MBH/2$ 
by the decay kinematics~\cite{footnote2}. 
The average number of particles produced in the process of BH 
evaporation is shown in Figs.~\ref{Landsberg1_fig1}d 
and \ref{Landsberg1_fig2}d,i.

The lifetime of the BH can be estimated by using the Stefan's 
law of thermal radiation. Since BH evaporation occurs primarily 
in three spatial dimensions, the canonical 3-dimensional 
Stefan's law applies, and therefore the power dissipated 
by the Hawking's radiation per unit area of the BH event 
horizon is $p = \sigma T_H^4$, where $\sigma$ is the Stefan-Boltzmann 
constant and $T_H$ is the Hawking temperature of the BH. 
Since the effective evaporation area of the BH is the area 
of a 3-dimensional sphere with the radius equal to the BH 
Schwarzschild radius $R_S$, the total power dissipated by 
the BH is given by:
$$
        P = 4\pi R_S^2 p = 4\pi R_S^2 \sigma T_H^4 = \sigma T_H^2 \frac{(n+1)^2}{4\pi}.
$$
The BH lifetime $\tau$ is given by:
$$
        \tau = \MBH/P = \frac{4\pi\mbh}{\sigma T_H^2 (n+1)^2},
$$
and using Eq.~(\ref{Landsberg1_eq1}), as well as the expression for $\sigma$ in natural units ($\hbar=c=k=1$), $\sigma = \pi^2/60$~\cite{footnote5}, we find:
$$
  \tau = \frac{3840}{M_P (n+1)^4}\left(\frac{\mbh}{\mp}\right)^\frac{n+3}{n+1}
\left(\frac{8\Gamma\left(\frac{n+3}{2}\right)}{n+2}\right)^\frac{2}{n+1}.
$$
The lifetime of a black hole as a function of its mass and the fundamental Planck scale is shown in Figs.~\ref{Landsberg1_fig1}h and \ref{Landsberg1_fig2}c,h. A typical lifetime of a BH is $\sim 10^{-26}$ s, which corresponds to a rather narrow width of the BH state $\sim 10$~GeV, i.e. typical for, e.g., a $W'$ or $Z'$ resonance of a similar mass. 

We emphasize that, throughout this paper, we ignore time
evolution: as the BH decays, it gets lighter and hotter and its
decay accelerates. We adopt the ``sudden approximation'' in which
the BH decays, at its original temperature, into its decay
products. This approximation should be reliable as the BH
spends most of its time near its original mass and temperature,
because that is when it evolves the slowest; furthermore, that is
also when it emits the most particles. Later, when we test the
Hawking's mass-temperature relation by reconstructing Wien's
displacement law, we will minimize the sensitivity to the late and
hot stages of the BHs life by looking at only the soft
part of the decay spectrum. Proper treatment of time evolution,
for $\MBH \approx \MP$, is difficult, since it immediately takes us
to the stringy regime.

\section{Branching Fractions} 

The decay of a BH is thermal: it obeys all local conservation laws, but otherwise does not discriminate between particle species (of the same mass and spin). Theories with
quantum gravity near a TeV must have additional symmetries, beyond the
standard $SU(3) \times SU(2) \times U(1)$, to guarantee proton longevity,
approximate lepton number(s) and flavor conservation~\cite{footnote3}. 
There are many possibilities: discrete or continuous symmetries, four 
dimensional or higher dimensional ``bulk'' symmetries \cite{Arkani-Hamed:1998sj}. 
Each of these possible symmetries constrains the decays of the 
black holes. Since the typical decay involves a large number of  
particles, we will ignore the constraints imposed by the few conservation 
laws and assume that the BH decays with roughly equal probability to all 
of the $\approx 60$ particles of the SM. Since there are six charged leptons 
and one photon, we expect $\sim 10\%$  of the particles to be hard, 
primary leptons and $\sim 2\%$ of the particles to be hard photons, each 
carrying hundreds of GeV of energy. This is a very clean signal, with 
negligible background, as the production of SM leptons or photons in 
high-multiplicity events at the LHC occurs at a much smaller rate than 
the BH production (see Fig.~\ref{Landsberg1_fig3}). These events are also easy to 
trigger on, since they contain at least one prompt lepton or photon with 
the energy above 100 GeV, as well as energetic jets.

\section{Test of Hawking Radiation} 

Furthermore, since there
are three neutrinos, we expect only $\sim 5\%$ average missing
transverse energy (\MET) per event, which allows us to precisely
estimate the BH mass from the visible decay products. We can also
reconstruct the BH temperature by fitting the energy spectrum of
the decay products to Planck's formula. Simultaneous knowledge
of the BH mass and its temperature allows for a test of the Hawking
radiation and can provide evidence that the observed events come 
from the production of a BH, and not from some other new physics.

There are a few important experimental techniques that we will use
to carry out the numerical test. First of all, to improve
precision of the BH mass reconstruction we will use only the
events with \met consistent with zero. Given the small probability
for a BH to emit a neutrino or a graviton, total statistics won't
suffer appreciably from this requirement. Since BH decays have
large jet activity, the \mbh resolution will be dominated by the
jet energy resolution and the initial state radiation effects, and
is expected to be $\sim 100$~GeV for a massive BH. Second, we will
use only photons and electrons in the final state to reconstruct the
Hawking temperature. The reason is twofold: final states with
energetic electrons and photons have very low background at high
$\sqrt{\hat s}$, and the energy resolution for electrons and
photons remains excellent even at the highest energies achieved in
the process of BH evaporation. We do not use muons, as
their momenta are determined by the track curvature in
the magnetic field, and thus the resolution deteriorates fast 
with the muon momentum growth. We also ignore the $\tau$-lepton decay 
modes, as the final states with $\tau$'s have much higher background than inclusive
electron or photon final states, and also because their energies can
not be reconstructed as well as those for the electromagnetic objects.
The fraction of electrons and photons among the final state particles
is only $\sim 5\%$, but the vast amount of BHs produced at the LHC
allows us to sacrifice the rest of the statistics to allow for a
high-precision measurement. (Also, the large number of decay particles 
enhances the probability to have a photon or an electron in the event.) 
Finally, if the energy of a decay particle approaches the kinematic 
limit for pair production, $\MBH/2$, the shape of the energy spectrum 
depends on the details of the BH decay model. In order to eliminate 
this unwanted model dependence, we use only the low part of the energy 
spectrum with $E < \MBH/2$.

The experimental procedure is straightforward: we select the BH sample
by requiring events with high mass ($> 1$ TeV) and multiplicity of the
final state ($N \ge 4$), which contain electrons or photons with energy
$>100$~GeV. We smear the energies of the decay products with the 
resolutions typical of the LHC detectors. We bin the events in the invariant 
mass with the bin size (500 GeV) much wider than the mass resolution. 
The mass spectrum of the BHs produced at the LHC with 100~\ifb\
of integrated luminosity is shown in Fig.~\ref{Landsberg1_fig3} for several values 
of \mp and $n$. Backgrounds from the SM $Z(ee)+$ jets and 
$\gamma +$ jets production, as estimated with PYTHIA~\cite{PYTHIA}, 
are small (see Fig.~\ref{Landsberg1_fig3}).

\begin{figure}[tb]
\begin{center}
\includegraphics[width=5in]{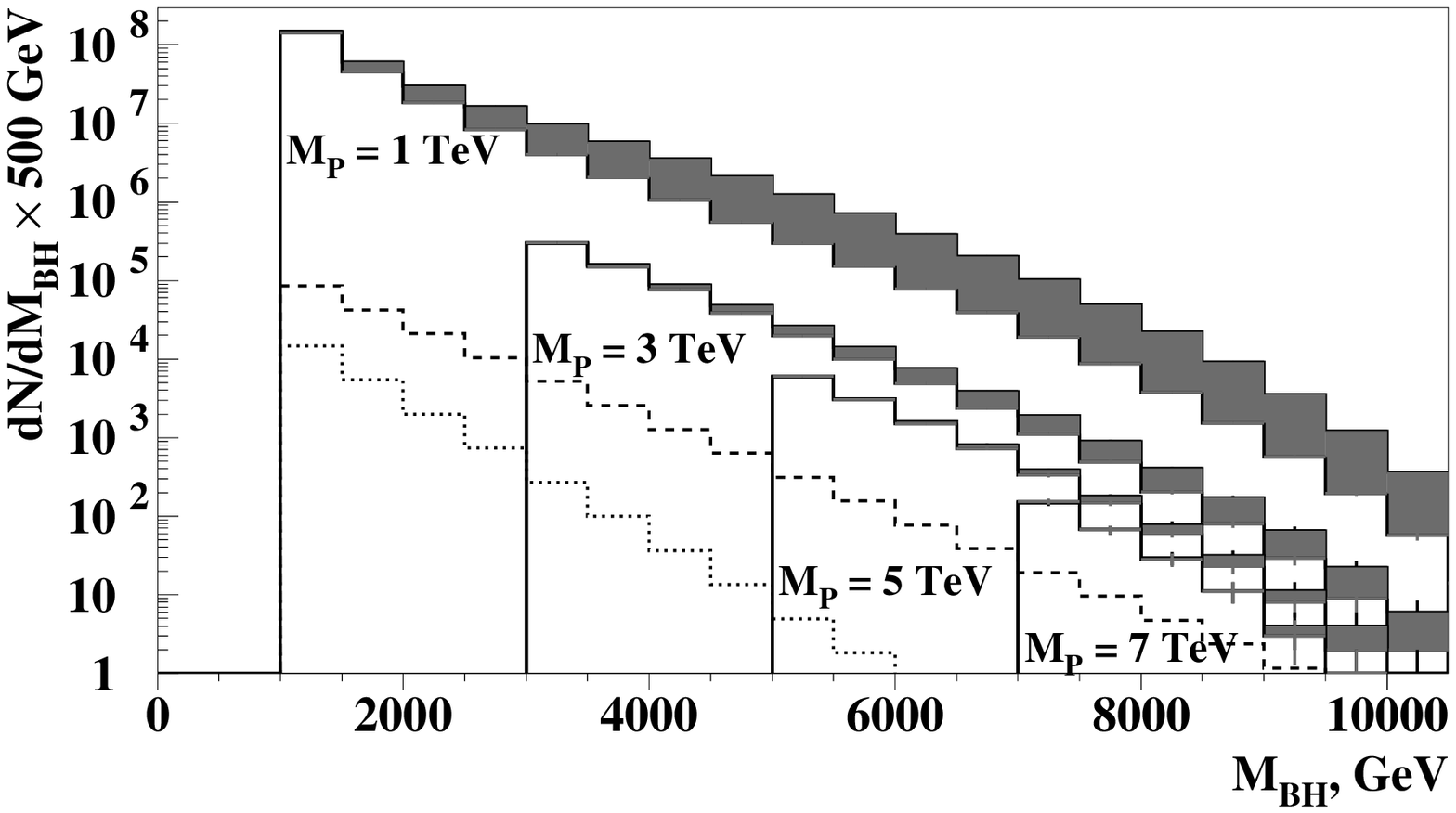}
\end{center}
\caption{Number of BHs produced at the LHC in the electron or photon decay 
channels, with 100~\protect\ifb of integrated luminosity, as a function of the BH 
mass. The shaded regions correspond to the variation in the number of events 
for $n$ between 2 and 7. The dashed line shows total SM background 
(from inclusive $Z(ee)$ and direct photon production). The dotted line 
corresponds  to the $Z(ee)+X$ background alone.}
\label{Landsberg1_fig3}
\end{figure}

To determine the Hawking temperature as a function of the BH mass, we perform 
a maximum likelihood fit of the energy spectrum of electrons and photons in the BH 
events to Planck's formula (with the coefficient $c$ determined by the
particle spin), below the kinematic cutoff (\MBH/2). This fit is performed using the 
entire set of the BH events (i.e., not on the event-by-event basis), separately in each of the \mbh bins. We then use the measured \mbh vs. \TH\ dependence and Eq.~(\ref{Landsberg1_eq1}) to determine the fundamental Planck scale \mp and the dimensionality of space $n$.
Note that to determine $n$ we can also take the logarithm of both
sides of Eq.~(\ref{Landsberg1_eq1}):
\begin{equation}
  \log(T_H) = \frac{-1}{n+1}\log(\MBH) + \mbox{const},
\label{Landsberg1_eq4}
\end{equation}
where the constant does not depend on the BH mass, but only
on \mp and on detailed properties of the bulk space, such as the shape of
extra dimensions. Therefore, the slope of a straight-line fit to
the $\log(\TH)$ vs. $\log(\MBH)$ data offers a direct way of determining
the dimensionality of space. This is a multidimensional analog of Wien's
displacement law. Note that Eq.~(\ref{Landsberg1_eq4}) is fundamentally different
from other ways of determining the dimensionality of space-time, e.g.
by studying a monojet signature or virtual graviton exchange processes,
also predicted by theories with large extra dimensions. (The properties of the 
latter two processes always depend on the volume of the extra-dimensional 
space, i.e. they cannot yield informaton on the number of extra dimensions 
without specific assumptions on their relative size.)

\begin{figure}[tb]
\begin{center}
\includegraphics[width=5in]{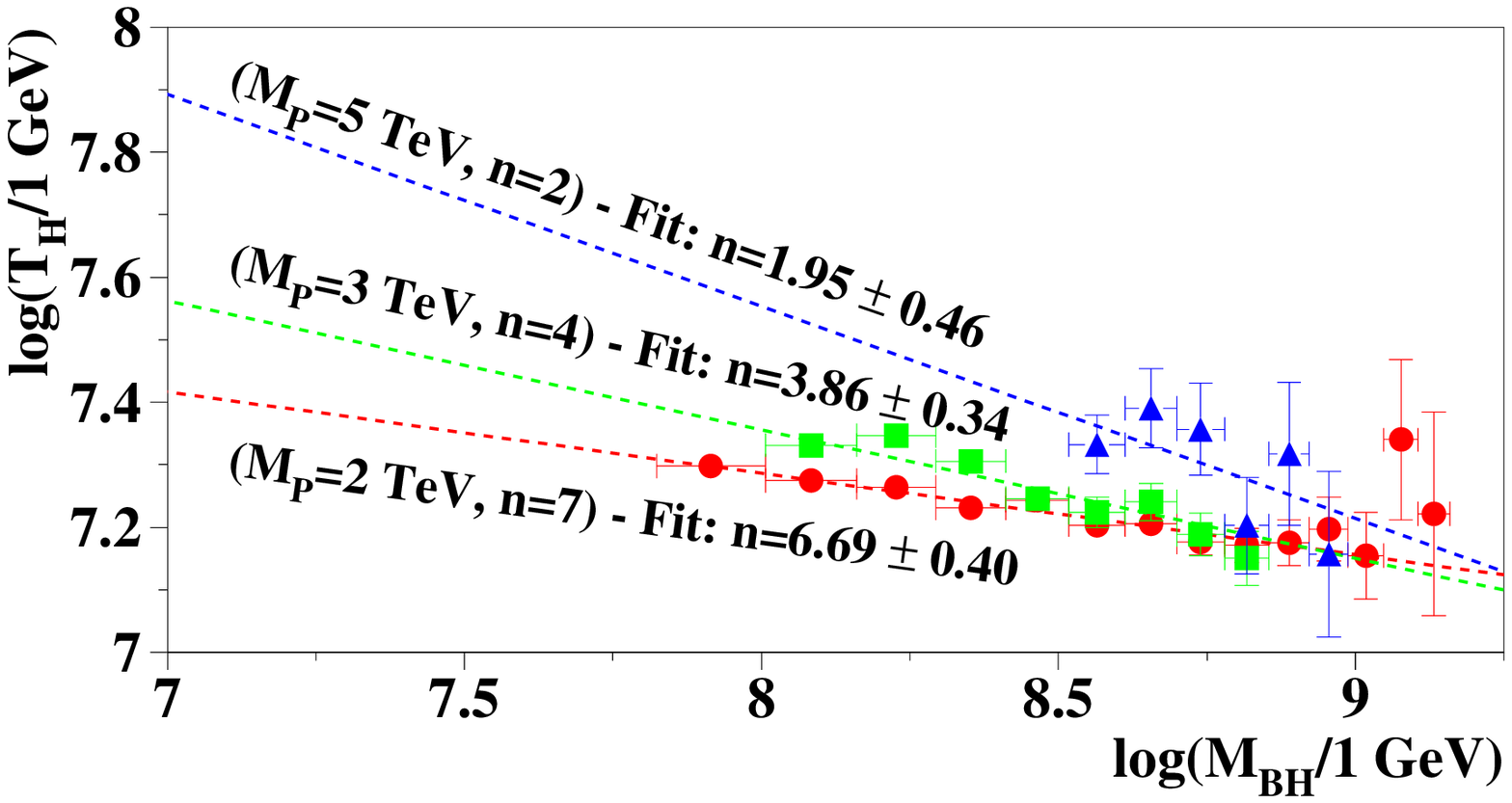}
\end{center}
\caption{Determination of the
dimensionality of space via Wien's displacement law at the LHC with
100~\protect\ifb\ of data.}
\label{Landsberg1_fig4}
\end{figure}

A test of Wien's law at the LHC would provide a confirmation
that the observed $e+X$ and $\gamma+X$ event excess is due to 
BH production. It would also be the first experimental test of
Hawking's radiation hypothesis. Figure~\ref{Landsberg1_fig4} shows typical fits
to the simulated BH data at the LHC, corresponding to 100~\ifb\ of
integrated luminosity, for the highest fundamental Planck scales
that still allow for determination of the dimensionality of space
with reasonable precision. The reach of the LHC for the
fundamental Planck scale and the number of extra dimensions via
Hawking radiation extends to $\MP \sim 5$~TeV and
is summarized in Table~\ref{bhtable}~\cite{footnote4}.

Similar tests can be performed at the VLHC and CLIC machines. 
While the VLHC case is identical to that at the LHC, with 
appropriately scaled energies, CLIC is complementary to the 
LHC in many ways, as the maximum number of BH produced at CLIC 
is found at the highest accessible masses. This has the
advantage that the stringy effects, as well as the kinematic distortion 
of the Planck black-body spectrum, decrease with the increase of the 
BH mass. Thus the \mbh vs. $T_H$ fit at CLIC is less affected by 
these unknown effects. Preliminary studies show that statistical 
sensitivity to the number of extra dimensions and the value of 
the fundamental Planck scale at CLIC is similar to that at the LHC.

\begin{table}[tb]
\begin{center}
\caption{Determination of \protect\mp and $n$ from Hawking radiation. 
The two numbers in each column correspond to fractional 
uncertainty in \protect\mp and absolute uncertainty in $n$, respectively.
\label{bhtable}
}
\begin{tabular}{cccccc}
\hline
~~\MP       &     1 TeV  &  2 TeV   & 3 TeV      &  4 TeV    & 5 TeV      \\
\hline
$n=2$     & 1\%/0.01 &  1\%/0.02& 3.3\%/0.10 & 16\%/0.35 & 40\%/0.46  \\
$n=3$     & 1\%/0.01 & 1.4\%/0.06 & 7.5\%/0.22 & 30\%/1.0 & 48\%/1.2  \\
$n=4$     & 1\%/0.01 & 2.3\%/0.13 & 9.5\%/0.34 & 35\%/1.5 & 54\%/2.0  \\
$n=5$     & 1\%/0.02 & 3.2\%/0.23 & 17\%/1.1 \\
$n=6$     & 1\%/0.03 & 4.2\%/0.34 & 23\%/2.5  & \multicolumn{2}{c}{Fit fails}\\
$n=7$     & 1\%/0.07 & 4.5\%/0.40 & 24\%/3.8 \\
\hline
\end{tabular}
\end{center}
\end{table}

Note, that the BH discovery potential at the LHC and VLHC is maximized in the
$e/\mu+X$ channels, where background is much smaller than that 
in the $\gamma+X$ channel (see Fig.~\ref{Landsberg1_fig3}). The 
reach of a simple counting experiment extends up to $\MP \approx 9$ 
TeV at the LHC and $\MP \approx 50$~TeV at the VLHC ($n=2$--7),
where one would expect to see a handful of BH events with negligible background.

\section{Black Hole Monte Carlo Event Generator}

A Monte Carlo package, TRUENOIR, has been developed for simulating 
production and decay of the black holes at high-energy colliders. 
This package is a plug-in module for the PYTHIA~\cite{PYTHIA}
Monte Carlo generator. It uses a heuristic algorithm and conservation 
of baryon and lepton numbers, as well as the QCD color, 
to simulate the decay of a black hole in a rapid-decay approximation.
While the limitations of such a simplistic approach are clear, 
further improvements to this generator are being worked on. 
In the meantime, it provides a useful qualitative tool to study 
the detector effects and other aspects of the BH event reconstruction.
Figure~\ref{Landsberg1_fig5} shows a display of a typical BH event 
at a 5 TeV CLIC collider, produced using the TRUENOIR code. 
A characteristic feature of this event is extremely large 
final state multiplicity, very atypical of the events 
produced in $e^+ e^-$ collisions.

\begin{figure}[tb]
\begin{center}
\includegraphics[width=3.0in]{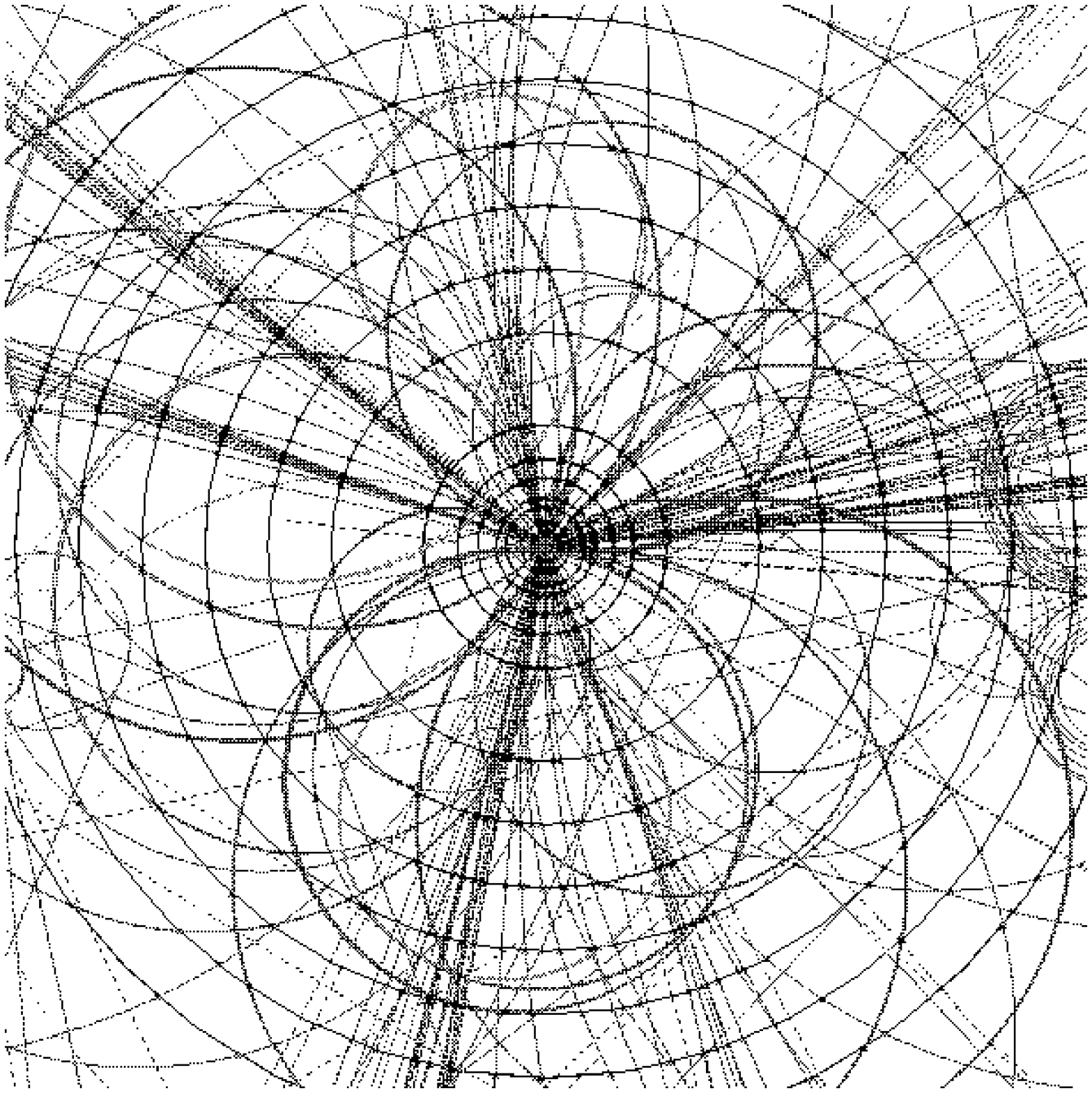}
\includegraphics[width=3.0in]{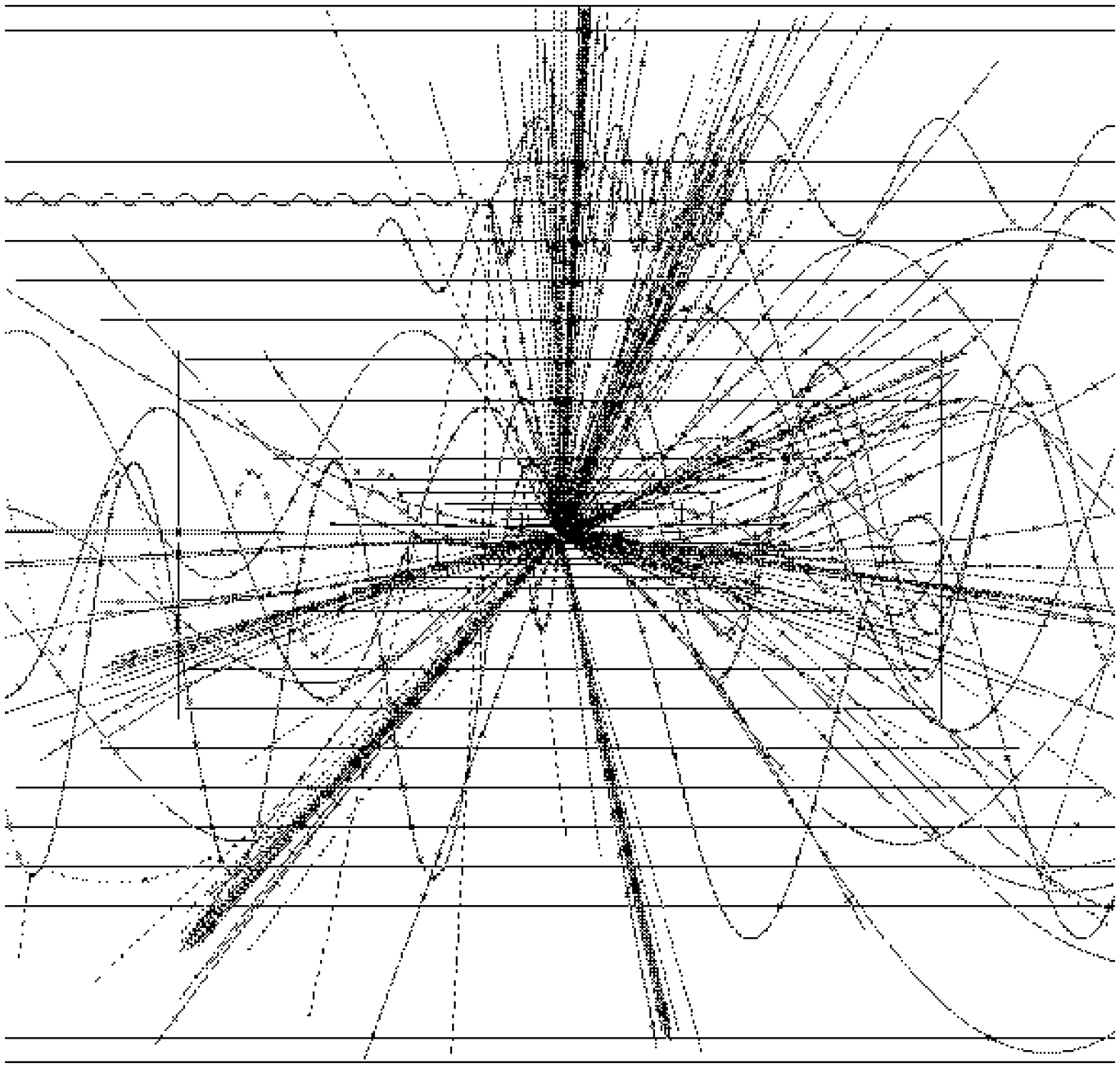}
\end{center}
\caption{A typical black-hole event at a 5 TeV CLIC accelerator. 
The two views correspond to the end and side-views of a CLIC detector. 
Detector simulation by Albert De Roeck.}
\label{Landsberg1_fig5}
\end{figure}

\section{Summary} 

Black hole production at the LHC and beyond may be one of the early
signatures of TeV-scale quantum gravity. It has three advantages:
\begin{itemize}
\item{(i)}
Large Cross Section: No small dimensionless coupling constants, 
analogous to $\alpha$, suppress the production of BHs. This leads to enormous rates.

\item{(ii)}
Hard, Prompt, Charged Leptons and Photons: Thermal decays are 
flavor-blind. This signature has practically vanishing SM background.

\item{(iii)}
Little Missing Energy: This facilitates the determination of 
the mass and the temperature of the black hole, and may lead 
to a test of Hawking radiation.
\end{itemize}

It is desirable to improve our primitive estimates, especially for the
light black holes ($\MBH \sim \MP$); this will involve string theory.
Nevertheless, the most telling signatures of BH production~-- large and
growing cross sections; hard leptons, photons, and jets~-- emerge from
qualitative features that are expected to be reliably estimated from the
semiclassical arguments of this paper.

Perhaps black holes will be the first signal of TeV-scale quantum gravity. This
depends on, among other factors, the relative magnitude of $M_P$ and the
(smaller) string scale $M_S$. For $M_S \ll \MP$, the vibrational modes of 
the string may be the first indication of the new physics.

Studies of the BH properties at future facilities, including very 
high-energy lepton and hadron colliders, would make it possible to map out 
properties of large extra dimensions, to measure the effects of 
quantum gravity, and to provide insight into other quantum 
phenomena, such as Hawking radiation, the information paradox, etc.

\section{Acknowledgments}
We would like to thank Gia Dvali, Veronika Hubeny, Nemanja Kaloper, 
Elias Kiritsis, Joe Lykken, Konstantin Matchev, Steve Mrenna, and 
Lenny Susskind for valuable conversations.
Many thanks to the organizers of the Les Houches workshop, where 
this work was started, for creating an excellent scientific atmosphere 
and productive environment.
This work was partially supported by the U.S. DOE Grant No. DE-FG02-91ER40688,
NSF Grant No. PHY-9870115, and A.P.~Sloan Foundation.


%
\setcounter{figure}{0}
\setcounter{table}{0}
\setcounter{section}{0}
\setcounter{equation}{0}
\clearpage

\bibliography{bsmreport}

\begin{thebibliography}{100}
\bibitem{RT}
We warmly thank Tom Rizzo for providing us with the FORTRAN code for $Z/\gamma$
  excitations production and for the help in implementing it.
\bibitem{footnote2}
To avoid this limitation when performing numerical tests of Wien's law, we
  truncate the integrals in Eq. (\protect\ref{Landsberg1_eq2}) at the kinematic
  limit (\MBH/2).
\bibitem{footnote5}
In the above calculatons we used the Stefan-Boltzmann constant, which is
  strictly speaking correct only for an evaporation via massless bosons.
  However, since the nature of our calculations is very approximate, this is a
  reasonable assumption.
\bibitem{footnote4}
We checked the effect of introducing an additional cutoff at $E < M_S$, where
  string dynamics is expected to affect the shape of the Planck spectrum, by
  taking $M_S =\MP/2$. As a result, the uncertainties in \mp and $n$ increased
  by about a factor of 2.
\end{thebibliography}
\end{document}